\documentclass[12pt,oneside,letterpaper]{report}

\newcommand{\thesistitle}{Magic zeroes in the black hole response problem and a Love symmetry resolution}
\newcommand{\thesisauthor}{Panagiotis Charalambous}
\newcommand{\thesisadvisor}{{\color{black}{Sergei Dubovsky}}}

\newcommand{\bibtitlename}{References} 

\newcommand{\thesisdept}{Physics}
\newcommand{\gradmonth}{May}
\newcommand{\gradyear}{2023}

\usepackage{style}

\usepackage[titletoc]{appendix}

\newcommand\listappendixname{List of Appendices}

\newcommand\appchapter[1]{%
	\chapter{#1}
	\addcontentsline{app}{chapter}{Appendix \thechapter:\hspace{0.3cm}#1}}
\newcommand\appsection[1]{%
	\section{#1}
	\addcontentsline{app}{section}{\thesection\hspace{0.5cm}#1}}
\makeatletter

\newcommand\listofappendices{%
	\chapter*{\listappendixname}\@starttoc{app}}
\makeatother

\makeatletter

\AtBeginEnvironment{appendices}{%
	\clearpage
	\addcontentsline{toc}{chapter}{Appendices}
	\write\@auxout{%
		\string\let\string\latex@tf@toc\string\tf@toc
		\string\let\string\tf@toc\string\tf@apc%
	}
	\titleformat
	{\chapter}
	[display]
	{\Large\bfseries} 
	{\Large \MakeUppercase{\@chapapp} \thechapter} 
	{1.5ex}{}
	{}
}

\AtEndEnvironment{appendices}{%
	\write\@auxout{%
		\string\let\string\tf@toc\string\latex@tf@toc%
	}
}

\makeatother

\usepackage[utf8]{inputenc}
\usepackage[T1]{fontenc}

\usepackage{subcaption}
\usepackage{tipa}
\usepackage{graphicx}
\usepackage[bottom]{footmisc}
\usepackage[titletoc]{appendix}
\usepackage{listings}
\usepackage{float}
\usepackage{amsmath}
\usepackage{mathrsfs}
\usepackage{amssymb}
\usepackage{mathtools}
\usepackage{amsfonts}
\usepackage{braket}
\usepackage{slashed}
\usepackage{dashbox}
\usepackage[colorinlistoftodos]{todonotes}
\usepackage{tikz-feynman}   
\usetikzlibrary{snakes}     
\usepackage[normalem]{ulem}
\usepackage{scalerel}
\usepackage{multirow}  
\usepackage{makecell}  

\usepackage{enumitem}																		 
\newlist{abbrv}{itemize}{1}																	 
\setlist[abbrv,1]{label=,labelwidth=0.8in,align=parleft,itemsep=0.01\baselineskip,leftmargin=!} %

\newcommand{\vev}[1]{\ensuremath{\left\langle #1 \right\rangle}}
\def\be{\begin{equation}}
\def\ee{\end{equation}}
\def\ba{\begin{aligned}}
\def\ea{\end{aligned}}

\newcommand{\SL}{\text{SL}\left(2,\mathbb{R}\right)}
\newcommand{\Lstr}{\text{\sout{$\mathcal{L}$}}}

\usepackage{setspace}

\begin{document}
	

\thispagestyle{empty}
%

\vspace*{25pt}
\begin{center}	
	{\Large
		\begin{doublespace}
			{\textcolor{SchoolColor}{\textsc{\thesistitle}}}
		\end{doublespace}
	}
	\vspace{.3in}
	
	by
	\vspace{.3in}
	
	\thesisauthor

	\vspace{.5in}
	\includegraphics[width=10cm]{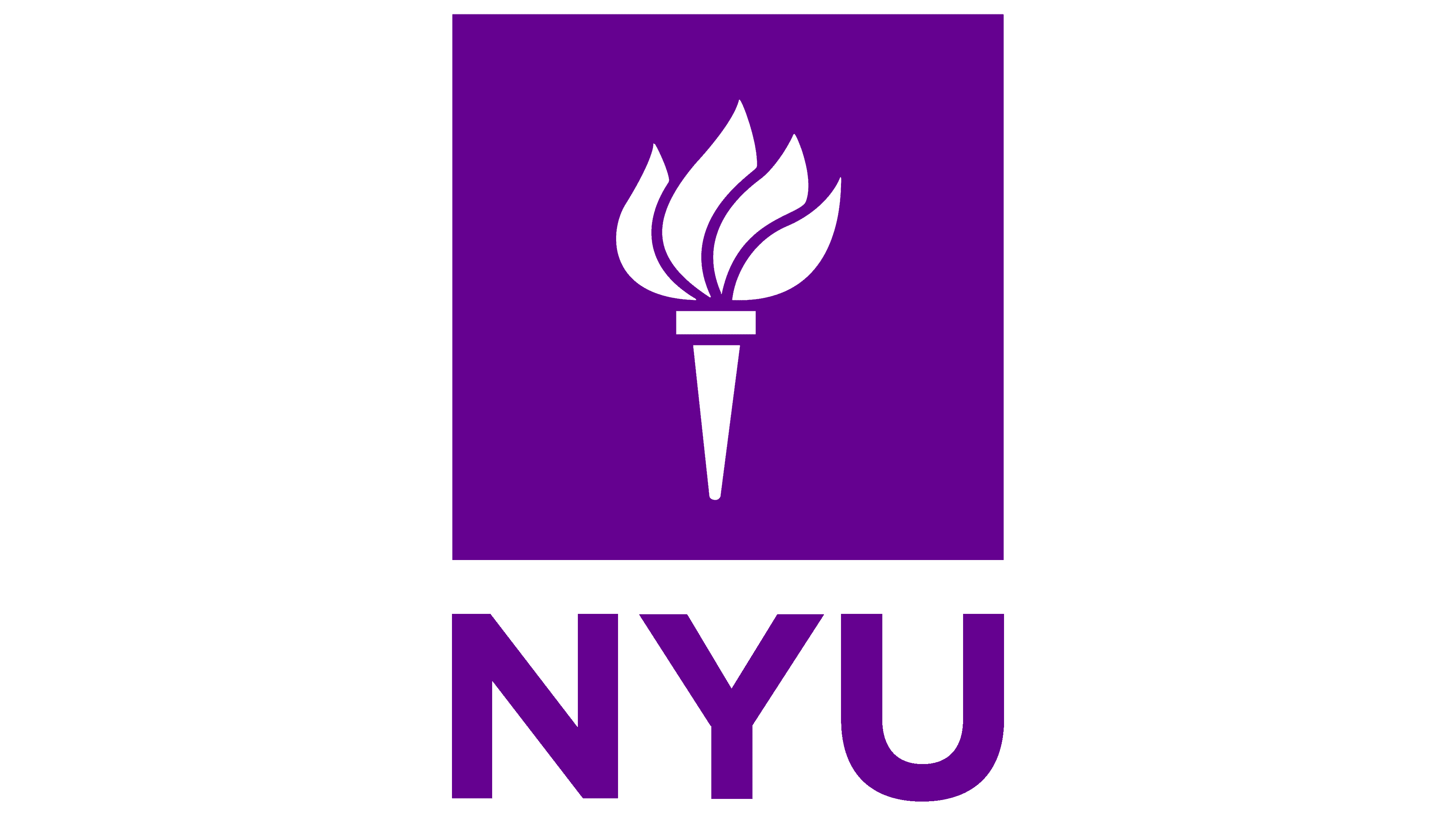}
	\vspace{.5in}

	\begin{doublespace}
		\textsc{
			A dissertation submitted in partial fulfillment\\
			of the requirements for the degree of\\
			Doctor of Philosophy\\
			Department of \thesisdept\\
			New York University\\
			\gradmonth, \gradyear}
	\end{doublespace}
\end{center}
\vfill

\newpage
	
	\pagenumbering{roman}


\thispagestyle{empty}
%

\vspace*{25pt}
\begin{center}	
	{\Large
		\begin{doublespace}
			{\textcolor{SchoolColor}{\textsc{\thesistitle}}}
		\end{doublespace}
	}
	\vspace{.7in}
	
	by
	\vspace{.7in}
	
	\thesisauthor
	\vfill
	
	\begin{doublespace}
		\textsc{
			A dissertation submitted in partial fulfillment\\
			of the requirements for the degree of\\
			Doctor of Philosophy\\
			Department of \thesisdept\\
			New York University\\
			\gradmonth, \gradyear}
	\end{doublespace}
\end{center}
\vfill

\noindent\makebox[\textwidth]{\hfill\makebox[2.5in]{\hrulefill}}\\
\makebox[\textwidth]{\hfill\makebox[2.5in]{\hfill\thesisadvisor}}

\newpage

\thispagestyle{empty}
\vspace*{25pt}
\begin{center}
	\scshape \noindent \small \copyright \  \small  \thesisauthor \\
	All rights reserved, \gradyear
\end{center}
\vspace*{0in}


	\chapter*{Acknowledgments}
	\addcontentsline{toc}{chapter}{Acknowledgments}
	I would like to start by expressing my sincere gratitude to my Thesis Advisor, Sergei Dubovsky, for his continuous supervision and guidance throughout the conduction of this research project, and for his genuine understanding and effort to try to accommodate both my academic as well as my non-academic requests. I am also grateful to Mikhail Ivanov who, besides an invaluable collaborator, has come to be for me for all intends and purposes a secondary advisor, patiently dealing with and addressing my inhibitions.

I would also like to thank my Academic Advisor, Prof. Massimo Porrati, who has always been standing by and ready to assist with my academic concerns. I would furthermore like to thank Barak Kol, Ramy Brustein and Zihan Zhou for fruitful discussions. My work at NYU has been financially supported from the 2018-2022 NYU MacCracken Fellowship, the GSAS Summer 2020 Relief funds, Summer 2021 and Summer 2022 RA appointments (NSF Grant No. PHY-1915219) and the 2022-2023 Dean's Dissertation Fellowship.

Last, I am deeply grateful to my family, especially my mother, Machi Philippou, and my grandmother, Stavroulla Philippou, for their continuous support and trust throughout my academic career. I am also thankful to my beloved partner, Georgia Antoniou, and I am deeply humbled by the support of her family, notably, the continuous emotional support from her mother, Phani Pilavaki, as well as the exceptional physical and mental mentoring guidance from her uncle, Dimitris Pilavakis.


	\chapter*{Abstract}
	\addcontentsline{toc}{chapter}{Abstract}	
	The tidal deformation problem has been an insightful setup for studying the nature of celestial bodies. On the one hand, it probes the internal structure of a compact body through the tidal response coefficients which are measurable quantities encoded in gravitational wave signals emitted during the inspiraling phase of a binary system. On the other hand, it poses a challenge to the naturalness of General Relativity, arising from the fact that the conservative parts of the static tidal response coefficients (also referred to as static tidal Love numbers) for asymptotically flat general-relativistic black holes are exactly zero in four spacetime dimensions. This behavior persists when studying linearized perturbations of black holes under external electromagnetic and scalar fields.

In this thesis, we present the emergence of an $\SL$ (``Love'') symmetry in the suitably defined near-zone region, relevant for studying the black hole response problem. This symmetry is globally defined and physical solutions of the black hole linearized field equations are closed under its action. The vanishing of static Love numbers is found to naturally arise as a selection rule following from the fact that the relevant solution belongs to a particular highest-weight representation of the Love symmetry. Interestingly, the Love symmetry appears to be connected to another well-known $\SL$ structure associated with black holes: the enhanced isometry of the near-horizon geometry of extremal black holes. This relation is simplest for extremal charged spherically symmetric (Reissner-Nordstr\"{o}m) black holes, where the Love symmetry exactly reduces to the isometry of the near-horizon $\text{AdS}_2$ throat. For rotating (Kerr-Newman) black holes, one is lead to consider an infinite-dimensional $\SL\ltimes\hat{U}\left(1\right)_{\mathcal{V}}$ extension of the Love symmetry. It contains three physically distinct $\SL$ subalgebras: the Love algebra, the Starobinsky near-zone algebra, and the near-horizon algebra whose extremal limit precisely recovers the Killing vectors of the corresponding $\text{AdS}_2$ throat.

Similar results persist when studying perturbations of general-relativistic black holes in higher spacetime dimensions. Even though the black hole static Love numbers in higher dimensions vanish only for a discrete set of resonant conditions that depends on the orbital number of the perturbation, Love symmetry still exists independently of the details of the perturbation. This hints at a geometric interpretation of the Love symmetry; a statement that becomes rigorous within the framework of ``subtracted geometries'', that is, effective black hole geometries that preserve the internal structure of the black hole but subtract information about the environment. However, Love symmetry appears to be theory-dependent and does not emerge in generic modifications of the Einstein-Hilbert gravitational action. We extract, in particular, a sufficient geometric condition for its existence. Based on this, Love symmetry does not appear to exist, for instance, for black holes in Riemann-cubed gravity or the low-energy effective Riemann-squared and Riemann-to-the-fourth gravitational actions of bosonic/heterotic string theory and type-II superstring theory respectively; a fact that is in accordance with explicit computations of the corresponding black hole static Love numbers.

	\newpage
	\tableofcontents

	\newpage
	\cleardoublepage
	\phantomsection
	\addcontentsline{toc}{chapter}{List of Figures}
	\listoffigures

	\newpage
	\cleardoublepage
	\phantomsection
	\addcontentsline{toc}{chapter}{List of Tables}
	\listoftables
	
	\newpage
	\cleardoublepage
	\phantomsection
	\addcontentsline{toc}{chapter}{List of Appendices}
	\listofappendices


	\newpage
	\pagenumbering{arabic} 


	\newpage
\chapter{Introduction}
\label{ch:Intro}

We have come a long way since the first (unsuccessful) attempts of Joseph Weber's resonant bar apparatus to detect gravitational waves~\cite{Weber:1960zz,Astone:2010mr}. It was only until September $14$, $2015$ at $09:50:45$ UTC that the LIGO detectors at Hanford, Washington and Livingston, Louisiana~\cite{Barish:1999vh} recorded the first ever confirmed observation of transient gravitational waves emitted during the final stages of the coalescence of a binary system of black holes~\cite{LIGOScientific:2016aoc}. Ever since, the worldwide network of detectors dedicated to searching gravitational waves has grown through the upgrade into Advanced LIGO detectors~\cite{LIGOScientific:2014pky}, the European Gravitational Observatory (EGO) and VIRGO collaboration in Italy, consisting of the upgraded Advanced VIRGO detector~\cite{VIRGO:2012dcp,VIRGO:2014yos}, and the KAGRA detector at the Kamioka Observatory, Japan~\cite{KAGRA:2020agh}. The current state-of-the-art third Gravitational-Wave Transient Catalog (GWTC-$3$)~\cite{KAGRA:2021vkt} of the recently formed LIGO-VIRGO-KAGRA collaboration has raised the total number of observations to $90$ candidate compact binary coalescences and will continue to improve in sensitivity in the future~\cite{Saleem:2021iwi,LIGOScientific:2016wof,Punturo:2010zza,2017arXiv170200786A,Reitze:2019iox}.

More notably, the space-based LISA~\cite{2017arXiv170200786A}, planned to lunch in $2037$ and operating in the low frequency range $10^{-4}-1\,\text{Hz}$, compared to LIGO's sensitivity in the $10-10^3\,\text{Hz}$ frequency range, will allow to observe gravitational waves from compact binary systems at much wider orbits and, hence, better study the early stages of the inspiraling phase of the binary system. This regime is particularly relevant for studying tidal effects. The tidal perturbations of the bodies are parameterized by the tidal Love numbers~\cite{Love:1912,PoissonWill2014}; these are response coefficients capturing the conservative response of a body when immersed in a tidal field, named after Augustus Edward Hough (A. E. H.) Love who first introduced them in his landmark mathematical study of the elastic properties of Earth~\cite{Love:1912}. As such, besides a source of playful puns, Love numbers are useful observables capable of probing the internal structure of the involved relativistic configurations, e.g. the elusive nuclear equation of state of Neutron Stars (NSs)~\cite{Chatziioannou:2020pqz}.

Their predictable imprint on gravitational wave signatures is a direct consequence of the employment of the worldline Effective Field Theory (EFT) as a toolkit for the construction of gravitational waveform templates~\cite{Goldberger:2004jt,Porto:2005ac,Porto:2016pyg,Levi:2015msa,Levi:2018nxp,Goldberger:2022ebt}. Within the worldline EFT, a compact body is approximated by its large-distance universal appearance as a point-particle propagating along a worldline. Finite-size effects are systematically treated by non-minimal couplings of the worldline with the curvature. The Love numbers, in particular, appear as Wilson coefficients for operators quadratic in the curvature and their computation reduces to a matching condition. It was Flanagan \& Hinderer~\cite{Flanagan:2007ix} who first pointed out the prospect of measuring Love numbers via the phase of gravitational waves emitted from compact binary coalescences. More explicitly, they found the following leading tidal contribution to the frequency space gravitational waveform phase $\Psi\left(f\right)$ for gravitational waves emitted by an isolated binary star system\footnote{This formula is accurate for orbital frequencies much smaller than the frequency of oscillation modes of NSs and for stars in quasi-Newtonian orbits.}
\[
	\Psi_{\text{tidal}}^{\text{LO}}\left(f\right)=-\frac{9}{16}\frac{\upsilon^5}{\mu \left(GM\right)^4}\left[\left(11\frac{m_2}{m_1}+\frac{M}{m_1}\right)\lambda_1 + 1\leftrightarrow2\right] \,,
\]
where $f$ is the frequency of the gravitational wave, equal to twice the orbital frequency of the binary, $m_1$ and $m_2$ are the masses of the two stars, $M=m_1+m_2$ and $\mu=\frac{m_1m_2}{M}$ are the total and reduced masses of the system respectively, $\upsilon=\left(\pi GMf\right)^{1/3}$ is the (dimensionless) Newtonian orbital velocity for circular orbits and $\lambda_1$ and $\lambda_2$ are the (dimensionful) quadrupolar Love numbers of the two stars. Compared to the leading order Newtonian contribution, $\Psi_{\text{N}}^{\text{LO}}\left(f\right)=\frac{3}{128\eta\upsilon^5}$, with $\eta=\frac{\mu}{M}$ the symmetric mass ratio, tidal effects enter at $\upsilon^{10}$, i.e. at $5$th Post-Newtonian ($5$PN) order. The dependence on the internal structure of the bodies is therefore quite weak and their effective description as point-particles propagating along geodesics is a surprisingly good approximation, a fact often referred to as the effacement theorem~\cite{Porto:2016pyg}. This effacement makes the measurement of Love numbers challenging, requiring high signal-to-noise ratio signals covering a large window of the coalescence~\cite{Chatziioannou:2020pqz}. Nevertheless, the quadrupolar tidal deformability parameter has been measured for the seminal GW170817 detection of a NS-NS merger~\cite{LIGOScientific:2017vwq}, already putting constraints on the possible nuclear equation of state~\cite{LIGOScientific:2017vwq,Raithel:2018ncd}.

Besides directly probing the internal structure of the involved bodies, Love numbers have been proposed as a mean to lift a degeneracy in measuring the luminosity distance and inclination plane. This degeneracy is due to the difficulty of differentiating the amplitudes between the $+$ and $\times$ polarizations of the detected gravitational waveforms. To get a sense of the magnitude of this degeneracy, the relative difference between the $+$ and $\times$ polarization amplitudes is less than $1\%$ for inclinations less than $30^{\circ}$ (or greater than $150^{\circ}$) and $5\%$ for inclinations less than $45^{\circ}$ (or greater than $135^{\circ}$)~\cite{Usman:2018imj}. As was suggested in~\cite{Xie:2022brn}, the distance-inclination degeneracy can be significantly reduced by incorporating ``I-Love-Q'' relations~\cite{Yagi:2013bca,Yagi:2013awa,Yagi:2015pkc,Yagi:2016qmr}; these are relations between the moment of inertia, the quadrupole moment and the quadrupolar tidal Love number that are surprisingly rigid, that is, they are quite insensitive to the equation of state of the neutron star.

The measurement of Love numbers has also been proposed as a testing arena for strong-field gravity~\cite{Cardoso:2017cfl,Cardoso:2018ptl}. And what better strong-field configuration than a black hole? The history of black holes is itself quite amusing. The first ever recorded reference to the concept of black holes, in the sense of a distribution whose escape velocity exceeds the speed of light, comes as early as $1784$ with John Michell's ``dark stars''~\cite{Michell1784}. However, the first black hole solution of Einstein's General theory of Relativity was initiated in the trenches of WWI by Karl Schwarzschild~\cite{Schwarzschild:1916uq} and was also the first exact solution of the theory beyond the trivial flat spacetime geometry. In fact, it took a couple of decades to realize that the Schwarzschild solution describes a black hole for sufficiently massive objects, in the sense that the apparent singularity of the metric at the event horizon was in fact just a coordinate, and not a physical, singularity~\cite{Eddington1924,Finkelstein:1958zz,Kruskal:1959vx}. The Schwarzschild solution describes an isolated black hole\footnote{It is usually often noted that the term ``black holes'' was coined by John Archibald Wheeler for what was previously referred to as ``gravitational completely collapsed objects'' during a talk at the NASA Goddard Institute of Space Studies (GISS) in Columbia, New York City in December 1967. In fact, the term is said to have been used four years earlier at the first-of-the-still-ongoing Texas Symposium of Relativistic Astrophysics, while it is rumored that the term had already been used sometime around 1960-1961 when the physicist Robert Dicke described gravitationally collapsed objects ``like the Black Hole of Calcutta''~\cite{BlackHoleNameScienceNews}. Regardless of the exact origin, it is truly John Wheeler who popularized the term and eventually made it part of the scientific terminology.} in asymptotically flat four-dimensional space with no other charges besides its mass. This was generalized to the astrophysically more relevant rotating black holes via the celebrated Kerr geometry solution~\cite{Kerr:1963ud}, while it was only a matter of a few years to equip the rotating black hole with an electric charge to construct what we now call the Kerr-Newman geometry~\cite{Newman:1965my}. It also took a while to accept such extreme configurations as potential astrophysical objects. Most notably, Roger Penrose's landmark $3$-page work~\cite{Penrose:1964wq} has changed the way we see black holes and he was awarded the $2020$ Nobel Prize for Physics ``for the discovery that black hole formation is a robust prediction of the general theory of relativity''. Beyond theoretical work on the existence of black holes, we have now come to the point where we can directly\footnote{``Directly'' here is a bit of misnomer. We can only observe the black hole's effect on the surroundings of course, but the EHT images have the novelty on focusing on single black holes rather than members of X-ray binaries, for which Cygnus X-$1$ is historically the first ever candidate astrophysical black hole~\cite{BOLTON1972}.} observe such objects by taking ``photos'' of them, namely, the momentous apparent black hole shadows of $\text{M}87^{\ast}$ and of our Milky Way's Sagittarius $\text{A}^{\ast}$, released on April $10$, $2019$~\cite{EventHorizonTelescope:2019dse} and on May $12$, $2022$~\cite{EventHorizonTelescope:2022wkp} respectively by the Event Horizon Telescope (EHT) collaboration.


Returning to the tidal response problem, general-relativistic black holes in four spacetime dimensions have a theoretically remarkable property: their static Love numbers vanish identically! This was first noticed for the quadrupolar tidal response of the four-dimensional Schwarzschild black hole in~\cite{Fang:2005qq} and was later confirmed for all multipolar orders by studying the limit of neutron star equation of states when their compactness approaches that of a black hole~\cite{Damour:2009vw,Binnington:2009bb}. The response problem has naturally been extended to rotating black holes in four-dimensional General Relativity and for responses beyond gravitational, namely, electromagnetic and scalar responses. The vanishing of the static Love numbers persists for these general black holes as well~\cite{Gurlebeck:2015xpa,Bicak:1977,Bicak:1976,Poisson:2014gka,Landry:2015zfa,Pani:2015hfa,LeTiec:2020spy,LeTiec:2020bos,Chia:2020yla,Charalambous:2021mea,Ivanov:2022hlo,Ivanov:2022qqt}. For one, these results tell us that a measurement of non-zero Love numbers for compact bodies with mass above the Tolman–Oppenheimer–Volkoff (TOV) limit of $\sim3\,M_{\odot}$~\cite{Tolman:1939jz,Oppenheimer:1939ne,Kalogera:1996ci} is a clear signature of physics beyond General Relativity.

On the more theoretical side, Love numbers appear as non-minimal coupling constants in the worldline EFT. The naturalness dogma \'{a} la 't Hooft~\cite{tHooft:1979rat} states that (dimensionless) physical observables are allowed to be small if setting them to zero enhances the symmetries of the system. The vanishing of static Love numbers cannot be assigned to any of the background symmetries of the corresponding black holes and is therefore an example of ``magic zeroes'' at first sight, raising naturalness concerns from the EFT point of view and calling upon the existence of an enhanced symmetry structure~\cite{Porto:2016zng}. There have been various works indicating a persisting hidden conformal structure of asymptotically flat black holes, with the classic paradigms being the extremal~\cite{Bardeen:1999px,Amsel:2009et,Guica:2008mu,Lu:2008jk} and the non-extremal~\cite{Castro:2010fd,Krishnan:2010pv} Kerr/CFT conjectures, while more efforts have recently been put forward to constructing holographic-like dictionaries between black hole geometries and conformal field theories~\cite{Bonelli:2021uvf,Consoli:2022eey}. It has also been suggested that conformal structures associated with black holes can leave distinct signatures on polarimteric observations, see e.g.~\cite{Johnson:2019ljv,Himwich:2020msm,Hadar:2022xag}. None of these, however, have been able to algebraically address the vanishing of static Love numbers.

This thesis is dedicated to the black hole response problem and the revealing of a new form of such conformal structures of black holes which can precisely be used to explain the vanishing of static Love numbers for general-relativistic black holes in four spacetime dimensions and is based on our works in Refs.~\cite{Charalambous:2021mea,Charalambous:2021kcz,Charalambous:2022rre}. This ``Love'' symmetry is an $\SL$ symmetry manifesting itself in the near-zone region, where perturbations have large wavelengths compared to the distance from the black hole. The Love symmetry outputs the vanishing of the static Love numbers as a selection rule following from the fact that the relevant perturbation solution belongs to a highest-weight representation. Furthermore, this thesis extends to the higher-dimensional black hole response problem and the relevant emergence of Love symmetries as found in our works in Refs.~\cite{Charalambous:2023jgq,Charalambous:2024tdj}.

The structure of this thesis is therefore the following. In Chapter~\ref{ch:TLNsDefinition}, we review the definition of Love numbers, starting with the tidal response problem in Newtonian gravity and extending to relativistic compact objects through the worldline EFT. In Chapter~\ref{ch:TLNsBlackHoles4d}, we employ the developed tools to compute the Love numbers of black holes in four spacetime dimensions. We will see, in particular, that the Kerr-Newman black hole exhibits no conservative static response under scalar~\cite{Charalambous:2021mea}, electromagnetic~\cite{Bicak:1977,Bicak:1976} or gravitational~\cite{Poisson:2014gka,LeTiec:2020spy,LeTiec:2020bos,Chia:2020yla} perturbations. Interestingly enough, the response will not be zero but will follow from purely dissipative effects surviving in the static limit due to frame dragging~\cite{Chia:2020yla,Charalambous:2021mea}. We will also explore black holes beyond General Relativity, namely, the modified Schwarzschild solution in Riemann-cubed gravity where we will find that the static Love numbers are non-zero and, in fact, exhibit the expected behavior based on power counting arguments within the worldline EFT~\cite{Charalambous:2022rre,DeLuca:2022tkm}.

In Chapter~\ref{ch:LoveSymmetry4d}, we present how Love symmetry naturally emerges as a near-zone symmetry of the equations of motion for perturbations of general-relativistic asymptotically flat black holes in four spacetime dimensions~\cite{Charalambous:2021kcz,Charalambous:2022rre}. This is an $\SL$ symmetry acting on perturbation of black holes with the property of being globally defined, in contrast to more conventional hidden $\SL$ structures associated with black holes~\cite{Bardeen:1999px,Amsel:2009et,Guica:2008mu,Lu:2008jk,Castro:2010fd,Krishnan:2010pv} whose generators are only locally realized and do not map physical solutions onto physical solutions. We will see that a particular representation of the Love symmetry, spanned by states regular at the (future) event horizon, contains the physical static solution. This is a highest-weight module of $\SL$ and the highest-weight property precisely outputs the vanishing of static Love numbers by dictating a (quasi-)polynomial profile of the solution with no response modes.

In Chapter~\ref{ch:Properties}, we will investigate various interesting properties of the Love symmetry~\cite{Charalambous:2022rre}. Most notably, we will construct an infinite-dimensional extension of the Love $\SL$ symmetry into $\SL\ltimes\hat{U}\left(1\right)_{\mathcal{V}}$ with $\mathcal{V}$ being a particular representation of the Love symmetry algebra. This infinite extension will turn out to contain all the possible globally defined $\SL$ structures of truncations of the equations of motion, up to local temporal translations that depend on the radial variable, which have the property of preserving the near-horizon dynamics. In particular, this contains the Starobinsky near-zone $\SL$ algebra, which is a second, different, globally defined $\SL$ symmetry emerging from a more conventional near-zone expansion of the black hole perturbation equations of motion~\cite{Starobinsky:1973aij,Starobinskil:1974nkd}. We will also see that the near-zone symmetries are approximate symmetries in a more rigorous sense, within the framework of ``subtracted geometries''~\cite{Cvetic:2011hp,Cvetic:2011dn}, that is, effective black hole geometries that preserve the internal structure of the black hole but subtract information associated with its environment. The two globally defined, Love and Starobinsky, near-zone symmetries will then acquire the geometric interpretation of being isometries of particular subtracted geometries of the Kerr-Newman black hole. We will also see that the existence of Love symmetry is rather special to General Relativity by studying the prospect of finding it in modified theories of gravity. This analysis will turn out to output a sufficient constraint on what kind of black hole geometries support a near-zone $\SL$ structure. Applying it to the example of Riemann-cubed gravity, we will see that the modified Schwarzschild solution does not satisfy this constraint in accordance with the corresponding Love numbers computed in Chapter~\ref{ch:TLNsBlackHoles4d}, which do not exhibit any form of fine-tuning. This serves as a preliminary justification of the name of Love symmetry, i.e. supports the fact that the existence of Love symmetry and the vanishing of static Love numbers are mutually inclusive.

Interestingly, the Love symmetry appears to be a cousin of another well-known $\SL$ symmetry associated with degenerate black holes: the enhanced isometry of the near-horizon geometry for extremal black holes~\cite{Bardeen:1999px,Amsel:2009et,Kunduri:2007vf}. Indeed, as we will see in Chapter~\ref{ch:NHE}, the infinite extension presented in Chapter~\ref{ch:Properties} contains a particular family of $\SL$ subalgebras whose extremal limit precisely recovers the Killing vectors of the $\SL$ isometry subgroup of the near-horizon extremal geometry. This hints at the interpretation of the near-zone symmetries as remnants of this enhanced isometry for extremal black holes.

The static Love numbers have also been studied for higher-dimensional spherically symmetric black holes in General Relativity~\cite{Kol:2011vg,Hui:2020xxx,Pereniguez:2021xcj,Charalambous:2024tdj}. In these examples, the static Love numbers are in general non-zero and exhibit no running in accordance with power counting arguments. However, there exist some resonant conditions between the multipolar order $\ell$ and the spacetime dimensionality $d$, e.g. whenever $\ell/\left(d-3\right)$ is an integer for the case of massless scalar perturbations of Schwarzschild black holes~\cite{Kol:2011vg,Hui:2020xxx}, for which the static Love numbers are indeed vanishing and hence also call upon an enhanced symmetry explanation. As we will see in Chapter~\ref{ch:LoveSymmetryDd}, despite this more intricate structure of the black hole static Love numbers, Love symmetry still exists for any multipolar order and spacetime dimensionality and is in accordance with these results~\cite{Charalambous:2021kcz,Charalambous:2022rre,Charalambous:2024tdj}. This appears to persist for higher-dimensional asymptotically flat, axisymmetric spinning (Myers-Perry) black holes~\cite{Myers:1986un} as we will see in Chapter~\ref{ch:5dMP}~\cite{Charalambous:2023jgq}.

We will finally close with a discussion of the results and prospect future directions in Chapter~\ref{ch:Discussion}, where we will also briefly comment on other attempts to infer the vanishing of Love numbers via hidden symmetry structures manifesting themselves directly in the IR static limit~\cite{Hui:2021vcv,Hui:2022vbh,Berens:2022ebl,Katagiri:2022vyz,BenAchour:2022uqo}.

In addition, this thesis contains supplementary material distributed in the appendices. In Appendix~\ref{app:Conventions}, we list the main symbols and abbreviations as well as the conventions employed throughout the chapters. In Appendix~\ref{app:2F1Gamma}, we enumerate various useful properties involving Euler's hypergeometric function and the $\Gamma$-function that are relevant when solving the near-zone equations of motion and extracting the Love numbers. In Appendix~\ref{app:SL2RRepresentations}, we present all the indecomposable standard $\SL$ modules, most of which turn out to be relevant when discussing how Love symmetry addresses the vanishing of Love numbers and/or the absence of logarithmic running. In Appendix~\ref{app:SL2RGenerators}, we set forth the derivation of the generators of the near-zone $\SL$ symmetries for the Kerr-Newman black hole. The involved steps are generic in nature and can be equally well applied for the extraction of the near-zone symmetries for black holes in higher spacetime dimensions. In Appendix~\ref{app:LieDerivative}, we investigate the extension of the Lie derivative into a generalized Lie derivative acting homogeneously on spin-weighted observables, which arise withing the Newman-Penrose (NP) or the Geroch-Held-Penrose (GHP) formalisms ~\cite{Newman:1961qr,Geroch:1973am} that are traditionally used to study electromagnetic and gravitational perturbations of Kerr-Newman black holes~\cite{Teukolsky:1972my,Teukolsky:1973ha}. Last, in Appendix~\ref{app:SphericalHarmonics}, we present an alternative basis for the spherical harmonic expansion in $4$ spatial dimensions that is most convenient when studying scalar perturbations of the $5$-dimensional doubly spinning Myers-Perry black hole~\cite{Charalambous:2023jgq} due its homogeneous transformations under the action of the $U\left(1\right)^2$ axisymmetry group of the black hole, rather than the $U\left(1\right)$ of the standard scalar spherical harmonics.
	\newpage
\chapter{The response problem for a relativistic compact body}
\label{ch:TLNsDefinition}

The response problem for a compact body is central to the study of its internal structure. A compact body in the current context is any finite-size, self-supported configuration of charges coupling to long-range forces. The most familiar such example is Earth's response to the tidal environment sourced by the Moon~\cite{Love:1912}. More generally, in Newtonian gravity, the tidal response problem consists of a self-gravitating body, configured of mass elements, which responds under the influence of the gravitational field of an external mass configuration~\cite{PoissonWill2014}. This idea can be generalized to relativistic configurations where one employs a geometric theory of gravity, e.g. General Relativity; gravity is then sourced by the density and flux of energy and momentum, as encoded in the stress-energy-momentum tensor~\cite{PoissonWill2014,Damour:2009vw,Binnington:2009bb}. In electromagnetism, one deals instead with a configuration of electric charges and currents which responds to the presence of electric and magnetic fields sourced by an external charge and current configuration, while one can similarly formulate a scalar response problem for a configuration of scalar charges interacting via forces mediated by a scalar field.

Before analyzing how a compact body responds to external forces, one first needs to describe the unperturbed compact body itself. This is achieved by the universal large-distance behavior of any finite-size body as a point-particle. To take into account its finite-size, this effective point-particle is dressed with charge and current multipole moments~\cite{Goldberger:2004jt}. Switching on an external source then induces a change in the body's multipole moments and these induced multipole moments encode all the accessible information about its internal structure as measured from far away. A distant observer can then set experimental apparatuses built of test objects in the exterior, interacting with the multipole moments and yielding field-gradient-induced effects. Such an effect would be, for example, the deviation of the trajectory of a point-like electric charge in the background of a dipole electric field. In a more modern experimental setup, field-gradient induced effects could be captured by interferometers detecting the radiation emitted from the compact-body--perturbing-source system~\cite{Flanagan:2007ix,Chatziioannou:2020pqz}.

In this chapter, we will review the ingredients needed to parameterize the linear response problem of a compact body. We will begin with the Newtonian definition of the linear response tensors in Section~\ref{sec:RCsNewtonian} and subsequently extend it to the relativistic definition via the worldline Effective Field Theory (EFT) in Section~\ref{sec:RCsGR}. In doing so, we will distinguish between conservative and dissipative responses, the former defining the ``Love'' part of the response. We will also present a matching condition of the worldline EFT definition onto GR calculations through the $1$-point function. Last, in Section~\ref{sec:NearZoneApproximation}, we will introduce the concept of the near-zone approximation as a necessary tool for performing such matchings between microscopic and EFT computations. This chapter is an adjustment of the background methods presented in our works in Refs~\cite{Charalambous:2021mea,Charalambous:2022rre,Charalambous:2023jgq}, but is otherwise based on the existing literature on the topic.

\section{Response coefficients in Newtonian Gravity}
\label{sec:RCsNewtonian}

Let us start with the standard formulation of the tidal response problem for a compact body in Newtonian gravity~\cite{Love:1912,PoissonWill2014}. This consists of solving the Poisson equation\footnote{Newton's gravitational constant $G$ is identified as the coupling constant appearing in the Einstein-Hilbert action $S=\frac{1}{16\pi G}\int d^{d}x\,\sqrt{-g}\,R$ such that the Einstein field equations $G_{\mu\nu}=8\pi G T_{\mu\nu}$ preserve their form in any number of spacetime dimensions.} for the Newtonian gravitation potential $\Phi_{\text{N}}$,
\be
	\nabla^2\Phi_{\text{N}} = \frac{d-3}{d-2}8\pi G\rho \,,
\ee
with $\rho=\rho\left(t,\mathbf{x}\right)$ the mass density of the mass configuration. In practice, this is accompanied by two field equations; the continuity equation and Euler's equations. Once supplemented with an equation of state, the problem is well posed.


The setup for introducing the tidal response coefficients begins with an unperturbed mass configuration at equilibrium, practically being hydrostatic equilibrium, which is adiabatically perturbed by a weak distant mass configuration sourcing tidal forces parameterized by its tidal moments $\bar{\mathcal{E}}_{L}\left(t\right)$. In response, the mass distribution of the body rearranges until a new equilibrium state is reached. The response is encoded in the induced mass multipole moments $\delta Q_{L}\left(t\right)$~\cite{PoissonWill2014}. The perturbation in the Newtonian gravitational potential in the exterior of the body is then given by, in frequency space,
\be
	\delta\Phi_{\text{N}}\left(\omega,\mathbf{x}\right) = \sum_{\ell=2}^{\infty}\frac{\left(\ell-2\right)!}{\ell!}\left[\bar{\mathcal{E}}_{L}\left(\omega\right) - N_{\ell}G\frac{\delta Q_{L}\left(\omega\right)}{r^{2\ell+d-3}}\right]x^{L} \,,
\ee
where we have defined the numerical constants
\be\label{eq:NnormDimLess}
	N_{\ell} \equiv  \frac{8\pi}{\left(d-2\right)\Omega_{d-2}}\frac{\left(2\ell+d-5\right)!!}{\left(\ell-2\right)!\left(d-5\right)!!} \,,
\ee
with $\Omega_{d-2}=2\pi^{\left(d-1\right)/2}/\Gamma\left(\frac{d-1}{2}\right)$ the surface area of the unit $\left(d-2\right)$-sphere $\mathbb{S}^{d-2}$. Both the tidal moments $\bar{\mathcal{E}}_{L}$ and the induced mass multipole moments $\delta Q_{L}$ are rank-$\ell$ symmetric tracefree (STF) spatial tensors by virtue of the Laplace equation in the exterior\footnote{If $V\left(\mathbf{x}\right)$ is a harmonic function, then $\partial_{L}V$ are rank-$\ell$ STF spatial tensors. In the same way, $\partial_{L}\frac{1}{r^{d-3}}$, which are involved in the multipole expansion defining the mass multipole moments, are STF tensors of rank-$\ell$, since the Green's function $\left|\mathbf{x}-\mathbf{x}^{\prime}\right|^{-\left(d-3\right)}$ is a harmonic function of $\mathbf{x}$ in the exterior.}. We also remark here that we are working in the body-centered frame where the induced dipole moment vanishes identically and all the sums start from $\ell=2$.

Assuming that the body exhibits no gravitational hysteresis, i.e. that it develops no new permanent multipole moments after the tidal source is switched off, we can apply linear response theory and define the dimensionful tidal response tensor $\lambda_{LL^{\prime}}$ in frequency space as the corresponding retarded Green's function\footnote{Upper and lower spatial indices in Newtonian gravity are lowered and raised with the flat space metric.},
\be
	\delta Q_{L}\left(\omega\right) = -\sum_{\ell^{\prime}=2}^{\infty}\lambda_{LL^{\prime}}\left(\omega\right)\bar{\mathcal{E}}^{L^{\prime}}\left(\omega\right) \,,
\ee
where the mixing of different $\ell$-modes is a necessary ingredient for rotating and non-spherically symmetric bodies and we are suppressing non-linear corrections. The tidal response tensor $\lambda_{LL^{\prime}}$ is STF with respect to the first multi-index $L$, while only $\lambda_{L\left\langle L^{\prime} \right\rangle}$ is physically relevant. Although we have only displayed an $\omega$-dependence, $\lambda_{LL^{\prime}}$ will strongly depend on other properties associated with the internal structure of the body, e.g. its background multipole moments, including its mass and angular momentum as well as other parameters entering its equation of state. As a last conventional step, we introduce the computationally favorable dimensionless tidal response tensor $k_{LL^{\prime}}\left(\omega\right)$, defined according to
\be\label{eq:DimensionlessResponse}
	\lambda_{LL^{\prime}}\left(\omega\right) \equiv k_{LL^{\prime}}\left(\omega\right)\frac{\mathcal{R}^{2\ell+d-3}}{N_{\ell}G} \,,
\ee
with $\mathcal{R}$ a scale associated to the unperturbed body's size, e.g. its radius if it is spherically symmetric, such that the frequency space gravitational potential perturbation in the exterior takes the form
\be
	\delta\Phi_{\text{N}}\left(\omega,\mathbf{x}\right) = \sum_{\ell,\ell^{\prime}=2}^{\infty}\frac{\left(\ell-2\right)!}{\ell!}\left[\delta_{L,L^{\prime}} + k_{LL^{\prime}}\left(\omega\right)\left(\frac{\mathcal{R}}{r}\right)^{2\ell+d-3}\right]\bar{\mathcal{E}}^{L^{\prime}}\left(\omega\right)x^{L} \,.
\ee
Although we have presented an analysis for the tidal deformation of a self-gravitating body, it can be extended to compact bodies supported by other types of forces as well. In particular, it can be extended to systems responding to spin-$1$ and spin-$0$ forces and define the corresponding response tensors. Then, the $\ell$-sums will start from $\ell=1$ or $\ell=0$ respectively, but the general prescription outlined above can be carried away unaffected, up to the conventional overall $\left(\ell-s\right)!/\ell!$ factor for a spin-$s$ force system, and define the spin-$s$ response tensors $k_{LL^{\prime}}^{\left(s\right)}$,
\be\label{eq:NewtonianResponseTensros}
	\delta\Phi^{\left(s\right)}\left(\omega,\mathbf{x}\right) = \sum_{\ell,\ell^{\prime}=s}^{\infty}\frac{\left(\ell-s\right)!}{\ell!}\left[\delta_{L,L^{\prime}} + k_{LL^{\prime}}^{\left(s\right)}\left(\omega\right)\left(\frac{\mathcal{R}}{r}\right)^{2\ell+d-3}\right]\bar{\mathcal{E}}^{\left(s\right)L^{\prime}}\left(\omega\right)x^{L} \,.
\ee
In this language, the above analysis of the tidal response of a gravitational system corresponds to the Newtonian definition of the $k_{LL^{\prime}}^{\left(2\right)}$ response tensor. For $s=1$, $\Phi^{\left(1\right)}$ is the electrostatic potential and $k_{LL^{\prime}}^{\left(1\right)}$ defines the electric susceptibility tensor of the body, which is now a self-supported configuration of electric charges and currents, while there is an analogous definition of the magnetic susceptibility tensor associated with induced electric current multipole moments in the vector potential profile which we do not write down here. Last, for $s=0$, $\Phi^{\left(0\right)}$ is a scalar field for the potential of a system interacting via scalar forces and $k_{LL}^{\left(0\right)}$ defines the scalar susceptibility tensor of a self-supported distribution of scalar charges.

\section{Response coefficients in General Relativity}
\label{sec:RCsGR}

When relativistic effects are taken into account, the definition of the associated response tensors is more subtle. To begin with, the response tensors are observables leaving detectable interferometric signatures and should therefore be gauge invariant under diffeomorphisms. In addition, the growing mode in the profiles of the potentials, the ``source'' part of the field, acquires relativistic corrections and results in an overlapping with the decaying mode, the ``response'' part of the field, thus, raising concerns for a source/response ambiguity~\cite{Gralla:2017djj}.

These concerns are all addressed within the framework of the worldline EFT~\cite{Goldberger:2004jt,Porto:2005ac} whose starting point is the leading order universal point-particle appearance of compact bodies from large distances. We will only briefly review the EFT definition of Love numbers here. A more complete review fitted to the response problem can be found in~\cite{Charalambous:2021mea,Ivanov:2022hlo}, while comprehensive reviews of the worldline EFT formalism for self-gravitating bodies can be found in~\cite{Porto:2016pyg,Levi:2018nxp}.

One is effectively integrating out the short-scale modes associated with the internal structure of the body. For instance, for a general-relativistic non-spinning neutron star of mass $M$ made up of $~10^{40}/\text{m}^3$ constituents, with the $I$'th constituent propagating along a worldline $x_{I}^{\mu}\left(\sigma_{I}\right)$ parameterized by an affine parameter $\sigma_{I}$, the full ``UV'' action looks like~\cite{Porto:2016pyg}
\be
	S_{\text{full}}\left[g,\left\{x_{I}\right\}\right] = S_{\text{EH}}\left[g\right] + S_{\text{int}}\left[\left\{x_{I}\right\},g^{\text{S}}\right] \,,
\ee
where $S_{\text{EH}}\left[g\right]$ is the Einstein-Hilbert action, $S_{\text{int}}\left[\left\{x_{I}\right\},g^{\text{S}}\right]$ describes the dynamics of the internal degrees of freedom and the metric tensor $g_{\mu\nu} = g^{\text{L}}_{\mu\nu} + g^{\text{S}}_{\mu\nu}$ has been split into a long-distance, $g^{\text{L}}_{\mu\nu}$, component and a short-distance, $g^{\text{S}}_{\mu\nu}$, component relative to the size $\mathcal{R}\sim 2GM/c^2$ of the neutron star\footnote{The gravitational radius, i.e. the Schwarzschild radius $r_{s}=2GM/c^2$, written here assumes we are in four spacetime dimensions. In $d=1+\left(d-1\right)$ spacetime dimensions, $r_{s}^{d-3}=\frac{16\pi GM}{\left(d-2\right)\Omega_{d-2}c^2}$.}. We note here that in this example we are neglecting referring to other types of long-range and short-range forces that might be involved in the internal dynamics of the star.

For an effective long-distance description we follow a bottom-up approach and adopt, besides the long-distance gravitational modes $g_{\mu\nu}^{\text{L}}$, the collective description of the neutron star constituents as a point-particle propagating along the center of mass of the body $x_{\text{cm}}\left(\sigma\right)$, parameterized by an affine parameter $\sigma$. Formally, the worldline EFT is constructed by integrating out the short-distance modes $g^{\text{S}}_{\mu\nu}$ and $\delta x_{I}^{\mu}=x_{I}^{\mu}-x_{\text{cm}}^{\mu}$,
\be
	e^{iS_{\text{EFT}}\left[x_{\text{cm}},g^{\text{L}}\right]} = \int Dg^{\text{S}}_{\mu\nu}\left(x\right)D\delta x_{I}^{\rho}\left(\sigma_{I}\right)e^{iS_{\text{EH}}\left[g\right] + iS_{\text{int}}\left[\left\{x_{I}\right\},g^{\text{S}}\right]} \,.
\ee
The resulting long-distance effective action will be equipped with the symmetries of diffeomorphism and worldline reparameterization invariance. At leading order in the point-particle approximation, it contains the ``minimal'' bulk and point-particle actions, while finite-size effects are accounted for by non-minimal couplings of the worldline with derivatives of the long-distance metric modes,
\be
	\begin{gathered}
		S_{\text{EFT}}\left[x_{\text{cm}},g^{\text{L}}\right] = S_{\text{EH}}\left[g^{\text{L}}\right] + S_{\text{pp}}\left[x_{\text{cm}},g^{\text{L}}\right] + S_{\text{finite-size}}\left[x_{\text{cm}},g^{\text{L}}\right] \,, \\
		S_{\text{pp}}\left[x_{\text{cm}},g^{\text{L}}\right] = - M\int d\sigma \sqrt{g^{\text{L}}_{\mu\nu}\left(x_{\text{cm}}\left(\sigma\right)\right)\frac{dx_{\text{cm}}^{\mu}\left(\sigma\right)}{d\sigma}\frac{dx_{\text{cm}}^{\nu}\left(\sigma\right)}{d\sigma}} \,.
	\end{gathered}
\ee

For a more general compact body, integrating out the short-scale modes leaves an effective action whose degrees of freedom are the worldline position $x_{\text{cm}}\left(\sigma\right)$ along which the center of mass of the body propagates and parameterized by an affine parameter $\sigma$, a set of vielbein vectors $e_{a}^{\mu}\left(\sigma\right)$ localized on the worldline and capturing rotational degrees of freedom in the case the body is spinning\footnote{If the body is spinning, the notion of a ``center of mass'' is not invariant and one needs to supplement with a spin gauge symmetry~\cite{Porto:2005ac} to compensate for this. The center of mass is then fixed via a ``Spin Supplementary Condition'' (SSC) corresponding to fixing the time-like vector of the worldline vielbein $e_{a=0}^{\mu}$~\cite{Porto:2016pyg,Levi:2018nxp}. Moreover, the symmetries of EFT are supplemented with $SO\left(d-1\right)$ rotational invariance of the worldline spatial vielbein.}, and the long-distance metric perturbations with respect to the Minkowski background, $h_{\mu\nu}=g^{\text{L}}_{\mu\nu}-\eta_{\mu\nu}$, for asymptotically flat spacetimes. These are supplemented with other types of long-distance bulk fields and symmetries in the case of non-pure gravity, e.g. with a $U\left(1\right)$ gauge field $A_{\mu}$ for the Einstein-Maxwell theory or a real scalar field $\Phi$ for systems interacting via scalar forces. The effective action then contains the minimal point-particle action, now augmented with minimal couplings of the rotational degrees of freedom to gravity~\cite{Porto:2005ac,Levi:2015msa,Porto:2016pyg,Levi:2018nxp}, while finite-size effects are captured by non-minimal couplings of the worldline with higher-derivative operators,
\be
	S_{\text{EFT}}\left[x_{\text{cm}},e,h,A,\Phi\right] = S_{\text{bulk}}\left[\eta+h,A,\Phi\right] + S_{\text{pp}}\left[x_{\text{cm}},e,h\right] + S_{\text{finite-size}}\left[x_{\text{cm}},e,h,A,\Phi\right] \,.
\ee

Love numbers are defined as particular Wilson coefficients in front of quadratic couplings of the worldline with field strength tensors. For the simplest case of a spherically symmetric, non-rotating body, for example, the static Love numbers are defined from\footnote{Let us briefly comment on the progress towards defining the multipole moments associated with a curved spacetime due to, for example, the presence of a compact body. The difficulty within the context of a geometric theory of gravity is that one needs a covariant definition of the multipole moments. This was first achieved through by Geroch~\cite{Geroch:1970cc,Geroch:1970cd} and Hansen~\cite{Hansen:1974zz} but only for stationary spacetime, and was later generalized by Thorne~\cite{Thorne:1980ru} via the notion of the astrophysically relevant ``asymptotically Cartesian and mass centered frame''. Within the worldline EFT, as we see here, the multipole moments are naturally defined covariantly as the variables conjugate to the multipole moments of the field strength tensors.}~\cite{Goldberger:2007hy,Porto:2016pyg,Levi:2018nxp}
\be\ba
	S_{\text{finite-size}} \supset S_{\text{Love}} &= \sum_{s=0}^{2}\sum_{\ell=s}^{\infty}\frac{C_{\text{el},\ell}^{\left(s\right)}}{2\ell!}\int d\tau\,\mathcal{E}_{L}^{\left(s\right)}\left(x_{\text{cm}}\left(\tau\right)\right)\mathcal{E}^{\left(s\right)L}\left(x_{\text{cm}}\left(\tau\right)\right) \\
	&\quad\quad + \left(\mathcal{E}\leftrightarrow \mathcal{B}\right) + \left(\mathcal{E}\leftrightarrow \mathcal{T}\right)\,,
\ea\ee
where we have chosen the affine parameter to be equal to the proper time along the worldline and $\mathcal{E}_{L}^{\left(s\right)}\equiv \mathcal{E}_{a_1\dots a_{\ell}}^{\left(s\right)}$ are the multipole moments of the electric-type field strength tensors projected onto spatial slices orthogonal to the $d$-velocity $u^{\mu}=\frac{dx^{\mu}}{d\tau}$ of the body. These are defined via a set of local spatial vielbein vectors $e_{a}^{\mu}$, satisfying $u_{\mu}e_{a}^{\mu}=0$\footnote{For a non-rotating body, for which $u_{\mu}u^{\mu} = -1$, this spatial projector is given by $e_{a}^{\mu} = \delta_{a}^{\mu} + \delta_{a}^{\nu}u_{\nu}u^{\mu}$.}. For $s=0,1,2$
\be\ba
	\mathcal{E}_{L}^{\left(s=0\right)} &= e_{a_1}^{\mu_1}\dots e_{a_{\ell}}^{\mu_{\ell}}\nabla_{\langle \mu_1}\dots \nabla_{\mu_{\ell}\rangle}\Phi \,, \\
	\mathcal{E}_{L}^{\left(s=1\right)} &= e_{a_1}^{\mu_1}\dots e_{a_{\ell-1}}^{\mu_{\ell-1}}\nabla_{\langle \mu_1}\dots \nabla_{\mu_{\ell-1}}E_{a_{\ell}\rangle} \,,\quad E_{a} = e^{\mu}_{a}u^{\nu}F_{\mu\nu} \,, \\
	\mathcal{E}_{L}^{\left(s=2\right)} &= e_{a_1}^{\mu_1}\dots e_{a_{\ell-2}}^{\mu_{\ell-2}}\nabla_{\langle \mu_1}\dots \nabla_{\mu_{\ell-2}}E_{a_{\ell-1}a_{\ell}\rangle} \,,\quad E_{ab} = e^{\mu}_{a}u^{\nu}e^{\rho}_{b}u^{\sigma}C_{\mu\nu\rho\sigma} \,,
\ea\ee
and they are by construction rank-$\ell$ STF spatial tensors. The Wilson coefficients $C_{\text{el},\ell}^{\left(s\right)}$ define the spin-$s$ electric-type static Love numbers, while there is a magnetic version of this interaction term defining the magnetic-type static Love numbers. In $d=4$ spacetime dimensions, this magnetic part is constructed from the odd-parity magnetic-type field strength tensors~\cite{Zhang:1986cpa,Binnington:2009bb},
\be\ba
	\mathcal{B}_{L}^{\left(s=1\right)} &= e_{a_1}^{\mu_1}\dots e_{a_{\ell-1}}^{\mu_{\ell-1}}\nabla_{\langle \mu_1}\dots \nabla_{\mu_{\ell-1}}B_{a_{\ell}\rangle} \,,\quad B_{a} = \frac{1}{2}e^{\mu}_{a}\varepsilon_{\mu\nu\rho\sigma}u^{\nu}F^{\rho\sigma} \,, \\
	\mathcal{B}_{L}^{\left(s=2\right)} &= e_{a_1}^{\mu_1}\dots e_{a_{\ell-2}}^{\mu_{\ell-2}}\nabla_{\langle \mu_1}\dots \nabla_{\mu_{\ell-2}}B_{a_{\ell-1}a_{\ell}\rangle} \,,\quad B_{ab} = \frac{1}{2}e^{\mu}_{a}e^{\nu}_{b}\varepsilon_{\mu\kappa\rho\sigma}u^{\kappa}C^{\rho\sigma}_{\,\,\,\,\,\,\,\,\nu\lambda}u^{\lambda} \,,
\ea\ee
with no notion of magnetic-type moments for $s=0$. In higher spacetime dimensions, one needs to perform a generalization of the electric/magnetic decomposition of the field-strength tensors for $s=1,2$~\cite{Hervik:2012jn,Hervik:2013nx}. For the purposes of this thesis, we use the equivalent definition~\cite{Hui:2020xxx,Charalambous:2021mea}
\be\ba
	\mathcal{B}_{L|b}^{\left(s=1\right)} &= e_{a_1}^{\mu_1}\dots e_{a_{\ell-1}}^{\mu_{\ell-1}}\nabla_{\langle \mu_1}\dots \nabla_{\mu_{\ell-1}}B_{a_{\ell}\rangle b} \,,\quad B_{ab} = e_{a}^{\mu}e_{b}^{\nu}F_{\mu\nu} \,, \\
	\mathcal{B}_{L|b}^{\left(s=2\right)} &= e_{a_1}^{\mu_1}\dots e_{a_{\ell-2}}^{\mu_{\ell-2}}\nabla_{\langle \mu_1}\dots \nabla_{\mu_{\ell-2}}B_{a_{\ell-1}a_{\ell}\rangle b} \,,\quad B_{abc} = u^{\mu}e_{a}^{\nu}e_{b}^{\rho}e_{c}^{\sigma}C_{\mu\nu\rho\sigma} \,.
\ea\ee
We note, in particular, that the magnetic-type moments that are typically employed in $d=4$ are the ones arising by dualizing $B_{ab}$ and $B_{abc}$. Furthermore, in $d>4$, one needs to include tensor-type moments for the $s=2$ case. These are defined as~\cite{Hui:2020xxx}
\be
	\mathcal{T}_{L|bc}^{\left(s=2\right)} = e_{a_1}^{\mu_1}\dots e_{a_{\ell-2}}^{\mu_{\ell-2}}\nabla_{\langle \mu_1}\dots \nabla_{\mu_{\ell-2}}T_{a_{\ell-1}|b|a_{\ell}\rangle c} \,,\quad T_{abcd} = e_{a}^{\mu}e_{b}^{\nu}e_{c}^{\rho}e_{d}^{\sigma}C_{\mu\nu\rho\sigma} \,,
\ee
and vanish identically for $d=4$. The tensor-type moments above give rise to new type of Love numbers, the tensor-type Love numbers, which have no $d=4$ analogue. More explicitly, the magnetic-type and tensor-type Love numbers are then defined as
\be\ba
	S_{\text{Love}}^{\text{magnetic}} &= \sum_{s=1}^{2}\sum_{\ell=s}^{\infty}\frac{C_{\text{mag},\ell}^{\left(s\right)}}{2\ell!}\frac{1}{2}\int d\tau\,\mathcal{B}_{L|b}^{\left(s\right)}\left(x_{\text{cm}}\left(\tau\right)\right)\mathcal{B}^{\left(s\right)L|b}\left(x_{\text{cm}}\left(\tau\right)\right) \,, \\
	S_{\text{Love}}^{\text{tensor}} &= \sum_{\ell=2}^{\infty}\frac{C_{\text{tensor},\ell}^{\left(s\right)}}{2\ell!}\frac{1}{4}\int d\tau\,\mathcal{T}_{L|bc}^{\left(s\right)}\left(x_{\text{cm}}\left(\tau\right)\right)\mathcal{T}^{\left(s\right)L|bc}\left(x_{\text{cm}}\left(\tau\right)\right) \,.
\ea\ee
In the rest of this chapter, we will focus to the electric-type responses for economy but all the analysis below can be straightforwardly applied for magnetic-type and tensor-type responses as well.

For a generic compact body, the Wilson coefficients $C_{\text{el},\ell}^{\left(s\right)}$ become ``Wilson tensors'' $C_{\text{el},LL^{\prime}}^{\left(s\right)}$ defining the static electric-type Love tensors\footnote{We are also assuming parity invariance of the geometry of the unperturbed body here, i.e. we are omitting mixing between electric and magnetic components which would otherwise be allowed.}~\cite{Porto:2016pyg,Levi:2018nxp,Charalambous:2021mea,LeTiec:2020spy,LeTiec:2020bos,Ivanov:2022hlo},
\be
	S_{\text{Love}}^{\text{electric}} = \sum_{s=0}^{2}\sum_{\ell,\ell^{\prime}=s}^{\infty}\int d\tau\,\frac{C_{\text{el},LL^{\prime}}^{\left(s\right)}}{2\ell!}\,\mathcal{E}^{\left(s\right)L}\left(x_{\text{cm}}\left(\tau\right)\right)\mathcal{E}^{\left(s\right)L^{\prime}}\left(x_{\text{cm}}\left(\tau\right)\right) \,.
\ee
Dynamic responses can also be defined in a similar fashion. These arise by operators of the form
\be
	S_{\text{dynamic Love}}^{\text{electric}} = \sum_{s=0}^{2}\sum_{\ell,\ell^{\prime}=s}^{\infty}\sum_{n=0}^{\infty}\int d\tau \left(-1\right)^{n}\frac{C^{\left(s\right)}_{\text{el},LL^{\prime};n}}{2\ell!}\,D^{n}\mathcal{E}^{\left(s\right)L}\left(x_{\text{cm}}\left(\tau\right)\right)D^{n}\mathcal{E}^{\left(s\right)L^{\prime}}\left(x_{\text{cm}}\left(\tau\right)\right) \,,
\ee
with $D = u^{\mu}\nabla_{\mu}$ the covariant ``time''-derivative. The $n=0$ term is of course the static response operator we just saw. It is more practical, however, to switch to frequency space where the dynamic responses become $\omega$-dependent and this interaction term is more conveniently written as
\be\label{eq:DynamicLoveFrequensy}
	S_{\text{dynamic Love}}^{\text{electric}} = \sum_{s=0}^{2}\sum_{\ell,\ell^{\prime}=s}^{\infty}\int\frac{d\omega}{2\pi}\,\frac{C_{\text{el},LL^{\prime}}^{\left(s\right)}\left(\omega\right)}{2\ell!}\,\mathcal{E}^{\left(s\right)L}\left(-\omega\right)\mathcal{E}^{\left(s\right)L^{\prime}}\left(\omega\right) \,.
\ee

To compute the Love tensors, one matches onto observables of the full theory. We will employ here the ``Newtonian matching'' condition consisting of inserting a pure $2^{\ell}$-pole background Newtonian source at large distances,
\be
	\Phi^{\left(s\right)}\left(\omega,\mathbf{x}\right) = \bar{\Phi}^{\left(s\right)}\left(\omega,\mathbf{x}\right) + \delta\Phi^{\left(s\right)}\left(\omega,\mathbf{x}\right) \,,\quad \bar{\Phi}^{\left(s\right)}\left(\omega,\mathbf{x}\right) = \frac{\left(\ell-s\right)!}{\ell!}\bar{\mathcal{E}}_{L}^{\left(s\right)}\left(\omega\right)x^{L} \,,
\ee
and matching EFT $1$-point functions onto microscopic computations of perturbation theory~\cite{Kol:2011vg,Charalambous:2021mea,Ivanov:2022hlo}. Alternative matching conditions involve matching onto the gauge invariant scattering and absorption cross-sections~\cite{Goldberger:2007hy,Porto:2016pyg,Ivanov:2022qqt}. Diagrammatically, the Newtonian matching procedure is given by\footnote{We are employing the dimensional regularization scheme which sets all the unphysical power-law divergences to zero.}
\be\label{eq:1ptNewtonianMatching}
	\vev{\delta\Phi^{\left(s\right)}\left(\omega,\mathbf{x}\right)} = \vcenter{\hbox{\begin{tikzpicture}
			\begin{feynman}
				\vertex (a0);
				\vertex[right=0.6cm of a0] (gblobaux);
				\vertex[left=0.00cm of gblobaux, blob] (gblob){};
				\vertex[below=1cm of a0] (p1);
				\vertex[above=1cm of a0] (p2);
				\vertex[right=1cm of p1] (a1);
				\vertex[right=1cm of p2] (a2);
				\vertex[right=0.69cm of p2] (a22){$\times$};
				\diagram*{
					(p1) -- [double,double distance=0.5ex] (p2),
					(a1) -- (gblob) -- (a2),
				};
			\end{feynman}
	\end{tikzpicture}}} =
	\underbrace{\vcenter{\hbox{\begin{tikzpicture}
				\begin{feynman}
					\vertex[dot] (a0);
					\vertex[below=1cm of a0] (p1);
					\vertex[above=1cm of a0] (p2);
					\vertex[right=0.4cm of a0, blob] (gblob){};							
					\vertex[right=1.5cm of p1] (b1);
					\vertex[right=1.5cm of p2] (b2);
					\vertex[right=1.19cm of p2] (b22){$\times$};
					\vertex[above=0.7cm of a0] (g1);
					\vertex[above=0.4cm of a0] (g2);
					\vertex[right=0.05cm of a0] (gdtos){$\vdots$};
					\vertex[below=0.7cm of a0] (gN);
					\diagram*{
						(p1) -- [double,double distance=0.5ex] (p2),
						(g1) -- [photon] (gblob),
						(g2) -- [photon] (gblob),
						(gN) -- [photon] (gblob),
						(b1) -- (gblob) -- (b2),
					};
				\end{feynman}
	\end{tikzpicture}}}}_{\text{source}} + 
	\underbrace{\vcenter{\hbox{\begin{tikzpicture}
				\begin{feynman}
					\vertex[dot] (a0);
					\vertex[left=0.00cm of a0] (lambda){$C_{\text{el},LL^{\prime}}^{\left(s\right)}\left(\omega\right)$};
					\vertex[below=1.6cm of a0] (p1);
					\vertex[above=0.4cm of a0] (p2);
					\vertex[right=1.5cm of p1] (b1);
					\vertex[right=1.5cm of p2] (b2);
					\vertex[right=1.19cm of p2] (b22){$\times$};
					\vertex[below=0.5cm of a0] (g1);
					\vertex[below=0.3cm of g1] (g2);
					\vertex[below=0.15cm of g2] (gdotsaux);
					\vertex[right=0.00cm of gdotsaux] (gdtos){$\vdots$};
					\vertex[below=0.5cm of gdotsaux] (gN);
					\vertex[below=0.7cm of a0] (gblobaux);
					\vertex[right=0.4cm of gblobaux, blob] (gblob){};
					\diagram*{
						(p1) -- [double,double distance=0.5ex] (p2),
						(b2) -- (a0) -- (gblob) -- (b1),
						(g1) -- [photon] (gblob),
						(g2) -- [photon] (gblob),
						(gN) -- [photon] (gblob),
					};
				\end{feynman}
	\end{tikzpicture}}}}_{\text{response}} \,,
\ee
where the double line represents the worldline, straight lines indicate propagators of the fields $\delta\Phi^{\left(s\right)}$, a ``$\times$'' represents a $\bar{\Phi}^{\left(s\right)}$ insertion and wavy lines correspond to interactions of the worldline with the graviton arising from the minimal point-particle action. We note here that we are not including dissipative effects which will be addressed shortly. In the above diagrammatic representation we have also demonstrated how the worldline EFT definition allows to unambiguously separate relativistic corrections in the ``source'' part of the field profile from tidal effects~\cite{Charalambous:2021mea,Ivanov:2022hlo}. This splitting is in fact equivalent to the method of analytically continuing the spacetime dimensionality $d$~\cite{Kol:2011vg} or the multipolar order $\ell$~\cite{LeTiec:2020bos,Charalambous:2021mea,Ivanov:2022hlo,Creci:2021rkz} as the ``source'' and ``response'' diagrams then have non-integer indicial powers $r^{\alpha}$ with $\alpha=\ell$ and $\alpha=-\left(\ell+d-3\right)$ respectively. These receive PN corrections from the interaction of the graviton with the worldline which have the form $r^{\alpha-n}$ with positive integer $n$ and, thus, the ``source'' and ``response'' modes do not overlap. If analytic continuation to real $\ell$ is not possible, then one has to systematically compute the contribution of the $\left(2\ell+1\right)$PN corrections in the ``source'' part and subtract this from the $1$-point function to be left with a purely response effect at $\left(2\ell+1\right)$PN order~\cite{Charalambous:2021mea,Ivanov:2022hlo}.

\begin{table}[t]
	\centering
	\begin{tabular}{|c||c|c||c|c||c|c|}
		\hline
		 & \multicolumn{2}{c||}{Electric-type response} & \multicolumn{2}{c||}{Magnetic-type response} & \multicolumn{2}{c|}{Tensor-type response} \\
		\hline
		$s$ & $\Phi^{\left(s\right)}$ & $N_{\text{prop}}^{\left(s\right),\mathcal{E}}$ & $\mathcal{A}^{\left(s\right)}_{a}$ & $N_{\text{prop}}^{\left(s\right),\mathcal{B}}$ & $\mathcal{H}^{\left(s\right)}_{ab}$ & $N_{\text{prop}}^{\left(s\right),\mathcal{T}}$ \\
		\hline\hline
		$0$ & $\Phi$ & $+1$ & $-$ & $-$ & $-$ & $-$ \\
		\hline
		$1$ & $u^{\mu}A_{\mu}$ & $-1$ & $e_{a}^{\mu}A_{\mu}^{\left(\text{V}\right)}$ & $+1$ & $-$ & $-$ \\
		\hline
		$2$ & $+\frac{1}{2}u^{\mu}u^{\nu}h_{\mu\nu}$ & $+8\pi G \frac{d-3}{d-2}$ & $+\frac{1}{2}u^{\mu}e_{a}^{\nu}h_{\mu\nu}^{\left(\text{V}\right)}$ & $-4\pi G$ & $+\frac{1}{2}e_{\langle a}^{\mu}e_{b \rangle}^{\nu}h_{\mu\nu}^{\left(\text{T}\right)}$ & $+8\pi G$ \\
		\hline
	\end{tabular}
	\caption[List of the fundamental fields in terms of which the response problem is defined in this thesis and their propagator normalization.]{List of the fundamental fields in terms of which the response problem is defined in this thesis and their propagator normalization. In the main text in this Chapter, we only analyze explicitly the case of electric-type responses. For magnetic-type responses, one employs the gauge-invariant transverse vector fields, indicated here by the superscript ``$\left(\text{V}\right)$'' (vector-modes). For tensor-type responses, which are present only in gravity for $d>4$, one instead employs the gauge-invariant transverse tracefree tensor field, indicated here by the superscript ``$\left(\text{T}\right)$'' (tensor-modes). The normalization of the propagator for vector and tensor modes should be understood in the sense that $\langle \mathcal{A}^{\left(s\right)}_{a}\mathcal{A}^{\left(s\right)}_{b} \rangle\left(p\right) = N_{\text{prop}}^{\left(s\right),\mathcal{B}}\delta_{ab}\frac{-i}{p^2}$ and $\langle \mathcal{H}^{\left(s\right)}_{ab}\mathcal{H}^{\left(s\right)}_{cd} \rangle\left(p\right) = N_{\text{prop}}^{\left(s\right),\mathcal{T}}P_{ab;cd}\frac{-i}{p^2}$, with $P_{ab;cd}=\delta_{a(c}\delta_{d)b} - \frac{1}{d-1}\delta_{ab}\delta_{cd}$.}
	\label{tbl:PhiATNprop}
\end{table}

In the Newtonian limit, in a gauge where the fields $\Phi^{\left(s\right)}$ are canonical variables up to an overall normalization constant $N^{\left(s\right),\mathcal{E}}_{\text{prop}}$ in momentum space (see Table~\ref{tbl:PhiATNprop}),
\be
	\vev{\Phi^{\left(s\right)}\Phi^{\left(s\right)}}\left(p\right) = N^{\left(s\right),\mathcal{E}}_{\text{prop}}\frac{-i}{p^2} \,,
\ee
and in the body centered frame where $x_{\text{cm}}=\left(t,\mathbf{0}\right)$ and $u^{\mu}=\left(1,\mathbf{0}\right)$, this gives
\be\ba
	{}&\vev{\delta\Phi^{\left(s\right)}\left(\omega,\mathbf{x}\right)} \rightarrow
	\vcenter{\hbox{\begin{tikzpicture}
			\begin{feynman}
				\vertex[dot] (a0);
				\vertex[below=1cm of a0] (p1);
				\vertex[above=1cm of a0] (p2);						
				\vertex[right=1cm of p1] (b1);
				\vertex[right=1cm of p2] (b2);
				\vertex[right=0.69cm of p2] (b22){$\times$};
				\diagram*{
					(p1) -- [double,double distance=0.5ex] (p2),
					(b1) -- (b2),
				};
			\end{feynman}
	\end{tikzpicture}}} + 
	\vcenter{\hbox{\begin{tikzpicture}
			\begin{feynman}
				\vertex[dot] (a0);
				\vertex[left=0.00cm of a0] (lambda){$C_{\text{el},LL^{\prime}}^{\left(s\right)}\left(\omega\right)$};
				\vertex[below=1cm of a0] (p1);
				\vertex[above=1cm of a0] (p2);
				\vertex[right=1.2cm of p1] (b1);
				\vertex[right=1.2cm of p2] (b2);
				\vertex[right=0.89cm of p2] (b22){$\times$};
				\diagram*{
					(p1) -- [double,double distance=0.5ex] (p2),
					(b2) -- (a0) -- (b1),
				};
			\end{feynman}
	\end{tikzpicture}}} \\
	&= \frac{\left(\ell-s\right)!}{\ell!} \sum_{\ell^{\prime}=s}^{\infty}\left[\delta_{L,L^{\prime}} + \frac{2^{\ell-2}\Gamma\left(\ell+\frac{d-3}{2}\right)}{\pi^{\left(d-1\right)/2}}N^{\left(s\right),\mathcal{E}}_{\text{prop}}\frac{\left[C_{\text{el},LL^{\prime}}^{\left(s\right)}\left(\omega\right)\right]_{\text{TRS}}}{r^{2\ell+d-3}}\right]\bar{\mathcal{E}}^{\left(s\right)L^{\prime}}\left(\omega\right)x^{L} \,,
\ea\ee
where,
\be
	\left[C_{\text{el},LL^{\prime}}^{\left(s\right)}\left(\omega\right)\right]_{\text{TRS}}\equiv\frac{1}{2}\left(C_{\text{el},LL^{\prime}}^{\left(s\right)}\left(\omega\right)+C_{\text{el},L^{\prime}L}^{\left(s\right)}\left(-\omega\right)\right) \,,
\ee
with ``TRS'' standing for time-reversal symmetric. From this, we identify the explicit correspondence between the electric-type Love tensor and the Wilson tensor for a compact body of size $\mathcal{R}$,
\be\label{eq:WTesnorConsResp}
	k_{LL^{\prime}}^{\left(s\right)\text{Love}}\left(\omega\right) = \frac{2^{\ell-2}\Gamma\left(\ell+\frac{d-3}{2}\right)}{\pi^{\left(d-1\right)/2}}N^{\left(s\right),\mathcal{E}}_{\text{prop}}\frac{\left[C_{\text{el},LL^{\prime}}^{\left(s\right)}\left(\omega\right)\right]_{\text{TRS}}}{\mathcal{R}^{2\ell+d-3}} \,.
\ee
We see therefore that the Love tensor is defined from the \textit{conservative} response, i.e. the part of the response tensor invariant under time-reversal transformations, which correspond to simultaneously flipping the sign of the frequency, $\omega\rightarrow-\omega$, and the exchange $L\leftrightarrow L^{\prime}$. This is implicit by the definition at the level of the action and the use of the in-out formalism since
\be\ba
	\sum_{\ell,\ell^{\prime}}\int\frac{d\omega}{2\pi}&\,\frac{\left[C_{\text{el},LL^{\prime}}^{\left(s\right)}\left(\omega\right)\right]_{\text{TRS}}}{2\ell!}\,\mathcal{E}^{\left(s\right)L}\left(-\omega\right)\mathcal{E}^{\left(s\right)L^{\prime}}\left(\omega\right) = \\
	&\sum_{\ell,\ell^{\prime}}\int\frac{d\omega}{2\pi}\,\frac{C_{\text{el},LL^{\prime}}^{\left(s\right)}\left(\omega\right)}{2\ell!}\,\mathcal{E}^{\left(s\right)L}\left(-\omega\right)\mathcal{E}^{\left(s\right)L^{\prime}}\left(\omega\right) \,.
\ea\ee

\subsection{Dissipation in EFT}
As we just saw, only $\left[C_{\text{el},LL^{\prime}}^{\left(s\right)}\left(\omega\right)\right]_{\text{TRS}}$ is relevant when computing $1$-point functions via the standard in-out formalism, i.e. local operators in the worldline EFT action capture only conservative effects. Dissipative effects are incorporated by introducing gapless internal degrees of freedom $X$. One then considers composite operators $Q_{L}^{\left(s\right)}\left(X\right)$ corresponding to the full multipole moments, including the dissipative multipole moments due to the internal degrees of freedom $X$, but whose exact dependence on $X$ is not known. These are then coupled to the field moments~\cite{Goldberger:2005cd,Goldberger:2019sya,Goldberger:2020fot,Goldberger:2020wbx},
\be
	S_{\text{diss}} = -\sum_{s=0}^{2}\sum_{\ell=s}^{\infty}\frac{1}{\ell!}\int d\tau\,Q_{L}^{\left(s\right),\mathcal{E}}\left(X\right)\mathcal{E}^{\left(s\right)L}\left(x_{\text{cm}}\left(\tau\right)\right) + \left(\mathcal{E}\leftrightarrow\mathcal{B},\mathcal{T}\right) \,.
\ee
In order to account for dissipative effects at the level of the $1$-point function, one then employs the in-in (Schwinger-Keldysh) formalism~\cite{Schwinger:1960qe,Keldysh:1964ud,Galley:2009px,Goldberger:2009qd,Goldberger:2005cd,Goldberger:2019sya,Goldberger:2020fot,Goldberger:2020wbx}. Within this framework\footnote{We assume that the operators are chosen in such a way that their vev's are zero, $\vev{Q}=0$.}~\cite{Ivanov:2022hlo},
\be\ba\label{eq:1ptInInResponse}
	{}&\vev{\delta\Phi^{\left(s\right)}\left(\omega,\mathbf{x}\right)} \supset
	\vcenter{\hbox{\begin{tikzpicture}
			\begin{feynman}
				\vertex[] (a0){};
				\vertex[left=0.2cm of a0] (a00);
				\vertex[above=0.4cm of a00] (a0t){};
				\vertex[below=0.4cm of a00] (a0b){};
				\vertex[left=0.4cm of a0] (p0);
				\vertex[above=0.4cm of p0] (p0t){$Q^{\left(s\right)}$};
				\vertex[below=0.4cm of p0] (p0b){$Q^{\left(s\right)}$};
				\vertex[below=1.5cm of a0] (p1);
				\vertex[above=1.5cm of a0] (p2);
				\vertex[below=0.6cm of a0] (p11);
				\vertex[above=0.6cm of a0] (p22);
				\vertex[below right=2cm of a0] (a1);
				\vertex[above right=2cm of a0] (a2);
				\vertex[above right=2cm of a0] (a22){$\times$};
				\diagram*{
					(p1) -- [double,double distance=0.5ex] (p2),
					(a1) -- (a0b) , (a0t) -- (a2),
					(p11) -- [ghost](p22)
				};
			\end{feynman}
	\end{tikzpicture}}} \\
	&= \frac{\left(\ell-s\right)!}{\ell!}\sum_{\ell^{\prime}=s}^{\infty}\left[\frac{2^{\ell-2}\Gamma\left(\ell+\frac{d-3}{2}\right)}{\pi^{\left(d-1\right)/2}}N^{\left(s\right),\mathcal{E}}_{\text{prop}}\frac{\vev{Q_{L}^{\left(s\right),\mathcal{E}}Q_{L^{\prime}}^{\left(s\right),\mathcal{E}}}\left(-\omega\right)}{r^{2\ell+d-3}}\right]\bar{\mathcal{E}}^{\left(s\right)L^{\prime}}\left(\omega\right)x^{L} \,,
\ea\ee
and the \textit{full} electric-type response tensors $k_{LL^{\prime}}^{\left(s\right)}\left(\omega\right)$ are matched onto the retarded $2$-point function~\cite{Ivanov:2022hlo},
\be\label{eq:WTesnorFullResp}
	k_{LL^{\prime}}^{\left(s\right)}\left(\omega\right) = \frac{2^{\ell-2}\Gamma\left(\ell+\frac{d-3}{2}\right)}{\pi^{\left(d-1\right)/2}}N^{\left(s\right),\mathcal{E}}_{\text{prop}}\frac{\vev{Q_{L}^{\left(s\right),\mathcal{E}}Q_{L^{\prime}}^{\left(s\right),\mathcal{E}}}\left(-\omega\right)}{\mathcal{R}^{2\ell+d-3}} \,.
\ee

This way, correlation functions of $Q$ can be extracted through matching to various observables such as low-energy graviton absorption cross-sections, see e.g.~\cite{Ivanov:2022hlo,Ivanov:2022qqt}. The latter determine the imaginary part of $QQ$ correlators. The real part can be reconstructed by making use of dispersion relations following from analyticity of the correlators. This reconstruction leaves undetermined a real polynomial piece of the $QQ$ correlators, which corresponds to the Love numbers.

Our discussion implies that, generically, the physical Love tensors are a sum of the Wilson tensors entering the worldline EFT in \eqref{eq:DynamicLoveFrequensy} and of the polynomial part of the $QQ$ correlators. However, by making use of the field redefinitions which shift $Q$'s by functions of the bulk metric evaluated at the origin, one may work in the operator basis where Love numbers are entirely associated with these Wilson tensors.

\section{From Love tensors to Love numbers}
\label{sec:LTs_LNs}

So far we have formulated the response problem in terms of the response tensors $k_{LL^{\prime}}^{\left(s\right)}$. In practice, one is interested in the spherical harmonic response coefficients, arising after performing a spherical harmonic expansion thanks to the $1$-to-$1$ correspondence between spatial STF tensors and spherical harmonics. From these, the Love numbers are identified as the conservative spherical harmonic response coefficients~\cite{Chia:2020yla,Charalambous:2021mea,Ivanov:2022hlo}. Isolating the conservative part of the spherical harmonic response coefficients is in general non-trivial, but for some particular configurations, e.g. a spherically and non-spinning body in $d$ spacetime dimensions or the remarkably integrable black hole perturbations in $4$ spacetime dimensions, this decomposition allows to identify the Love numbers as the real part of the spherical harmonic response coefficients, while the imaginary part is assigned to dissipative effects. We will demonstrate this here following the analysis done in~\cite{LeTiec:2020bos}, but extended to higher dimensions. For useful reviews of higher-dimensional spherical harmonics and their $1$-to-$1$ correspondence with spatial STF tensors, we refer to~\cite{Chodos:1983zi,Higuchi:1986wu,Hui:2020xxx}.

To this end, we begin by expanding the $\left(d-1\right)$-dimensional spatial STF tensors $\bar{\mathcal{E}}_{L}^{\left(s\right)}$ into spherical harmonics of orbital number $\ell\in\mathbb{N}$,
\be
	\bar{\mathcal{E}}^{\left(s\right)L} = \sum_{\mathbf{m}}\bar{\mathcal{E}}_{\ell,\mathbf{m}}^{\left(s\right)}\mathcal{Y}_{\ell,\mathbf{m}}^{L\ast} \,,
\ee
where the constant STF tensors $\mathcal{Y}_{\ell,\mathbf{m}}^{L}$ are given by
\be
	\mathcal{Y}_{\ell,\mathbf{m}}^{L} = \frac{1}{A_{\ell}}\oint_{\mathbb{S}^{d-2}}d\Omega_{d-2}\,n^{\left\langle L \right\rangle}Y_{\ell,\mathbf{m}}^{\ast}\left(\mathbf{n}\right) \,,\quad A_{\ell} \equiv \frac{\Omega_{d-2}\left(d-3\right)!!\ell!}{\left(2\ell+d-3\right)!!} \,,
\ee
with $Y_{\ell,\mathbf{m}}\left(\mathbf{n}\right)$ the scalar spherical harmonics on $\mathbb{S}^{d-2}$, $n^{i}\equiv x^{i}/r$ and asterisks indicate complex conjugation. For future reference, $\mathbf{m}=\left\{m_1,m_2,\dots,m_{d-3}\right\}$ is a multi-index consisting of $d-4$ polar number $\left\{m_2,m_3,\dots,m_{d-3}|m_{i}\in\mathbb{N}\right\}$ and 1 azimuthal number $m_{1}\in\mathbb{Z}$, ranging according to
\be
	0 \le \left|m_1\right| \le m_2 \le m_3 \le \dots \le m_{d-3} \le \ell
\ee
with unit steps.

Then, the response coefficients $k_{\ell,\mathbf{m};\ell^{\prime},\mathbf{m}^{\prime} }^{\left(s\right)}\left(\omega\right)$ are related to the response tensor $k_{LL^{\prime}}^{\left(s\right)}\left(\omega\right)$ according to
\be\label{eq:RNsToRTs}
	k_{\ell,\mathbf{m};\ell^{\prime},\mathbf{m}^{\prime}}^{\left(s\right)}\left(\omega\right) = A_{\ell}\,k_{LL^{\prime}}^{\left(s\right)}\left(\omega\right)\mathcal{Y}_{\ell,\mathbf{m}}^{L}\mathcal{Y}_{\ell^{\prime},\mathbf{m}^{\prime}}^{L^{\prime}\ast} \,.
\ee
Using the fact that the induced multipole moments $\delta Q_{L}^{\left(s\right)}\left(t\right)$ and source multipole moments $\mathcal{E}_{L}^{\left(s\right)}\left(t\right)$ are real in position space as well as the assumption that the response tensors $k_{LL^{\prime}}^{\left(s\right)}\left(\omega\right)$ are analytic in $\omega$, i.e. that
\be
	k_{LL^{\prime}}^{\left(s\right)}\left(\omega\right) = \sum_{n=0}^{\infty}k_{LL^{\prime};n}^{\left(s\right)}\left(i\omega\right)^{n} \,,
\ee
with real-valued modes $k_{LL^{\prime};n}^{\left(s\right)}$, we see that
\be
	k_{LL^{\prime}}^{\left(s\right)\ast}\left(\omega\right)=k_{LL^{\prime}}^{\left(s\right)}\left(-\omega\right) \,.
\ee
From $Y_{\ell,\mathbf{m}}^{\ast}=\left(-1\right)^{m_1}Y_{\ell,-m_1,m_2,\dots,m_{d-3}}$, we then deduce the following complex conjugacy relation for the response coefficients
\be\label{eq:klmCC}
	k_{\ell,\mathbf{m};\ell^{\prime},\mathbf{m}^{\prime}}^{\left(s\right)\ast}\left(\omega\right) = \left(-1\right)^{m_1+m_1^{\prime}} k_{\ell,-m_1,m_2,\dots,m_{d-3};\ell^{\prime},-m_1^{\prime},m_2^{\prime},\dots,m_{d-3}^{\prime}}^{\left(s\right)}\left(-\omega\right) \,.
\ee

We can now translate the conservative/dissipative decomposition of the response tensor (see \eqref{eq:WTesnorConsResp},\eqref{eq:WTesnorFullResp}),
\be\ba\label{eq:RTsConsDiss}
	k_{LL^{\prime}}^{\left(s\right)\text{Love}}\left(\omega\right) &= \frac{1}{2}\left(k_{LL^{\prime}}^{\left(s\right)}\left(\omega\right) + k_{L^{\prime}L}^{\left(s\right)}\left(-\omega\right)\right) \,, \\
	k_{LL^{\prime}}^{\left(s\right)\text{diss}}\left(\omega\right) &= \frac{1}{2}\left(k_{LL^{\prime}}^{\left(s\right)}\left(\omega\right) - k_{L^{\prime}L}^{\left(s\right)}\left(-\omega\right)\right) \,,
\ea\ee
at the level of the response coefficients $k_{\ell,\mathbf{m};\ell^{\prime},\mathbf{m}^{\prime}}^{\left(s\right)}\left(\omega\right)$. The definition \eqref{eq:RNsToRTs} and the complex conjugacy relation \eqref{eq:klmCC} immediately imply
\be\ba\label{eq:RCsConsDissGen}
	k_{\ell,\mathbf{m};\ell^{\prime},\mathbf{m}^{\prime}}^{\left(s\right)\text{Love}}\left(\omega\right) &= \frac{1}{2}\left(k_{\ell,\mathbf{m};\ell^{\prime}\mathbf{m}^{\prime}}^{\left(s\right)}\left(\omega\right) + k_{\ell^{\prime},\mathbf{m}^{\prime};\ell,\mathbf{m}}^{\left(s\right)\ast}\left(\omega\right)\right) \,, \\
	k_{\ell,\mathbf{m};\ell^{\prime},\mathbf{m}^{\prime}}^{\left(s\right)\text{diss}}\left(\omega\right) &= \frac{1}{2i}\left(k_{\ell,\mathbf{m};\ell^{\prime}\mathbf{m}^{\prime}}^{\left(s\right)}\left(\omega\right) - k_{\ell^{\prime},\mathbf{m}^{\prime};\ell,\mathbf{m}}^{\left(s\right)\ast}\left(\omega\right)\right) \,,
\ea\ee
such that
\be
	k_{\ell,\mathbf{m};\ell^{\prime},\mathbf{m}^{\prime}}^{\left(s\right)}\left(\omega\right) = k_{\ell,\mathbf{m};\ell^{\prime},\mathbf{m}^{\prime}}^{\left(s\right)\text{Love}}\left(\omega\right) + ik_{\ell,\mathbf{m};\ell^{\prime},\mathbf{m}^{\prime}}^{\left(s\right)\text{diss}}\left(\omega\right) \,.
\ee
At this point, however, $k_{\ell,\mathbf{m};\ell^{\prime},\mathbf{m}^{\prime}}^{\left(s\right)\text{Love}}\left(\omega\right)$ and $k_{\ell,\mathbf{m};\ell^{\prime},\mathbf{m}^{\prime}}^{\left(s\right)\text{diss}}\left(\omega\right)$ are in general complex numbers.

We now focus to two particular examples. First, for a spherically symmetric and non-spinning configuration, the background symmetries imply the decoupling of the $\ell$-modes and independence on the $\mathbf{m}$-modes. Then,
\be
	k_{\ell,\mathbf{m};\ell^{\prime},\mathbf{m}^{\prime}}^{\left(s\right)}\left(\omega\right) = k_{\ell}^{\left(s\right)}\left(\omega\right)\delta_{\ell\ell^{\prime}}\delta_{\mathbf{m};\mathbf{m}^{\prime}} \,,
\ee
and the spherical harmonic modes of the frequency space potential perturbation in the Newtonian limit simplify to
\be
	\delta\Phi^{\left(s\right)}_{\ell,\mathbf{m}}\left(\omega,r\right) = \frac{\left(\ell-s\right)!}{\ell!}\left[1 + k_{\ell}^{\left(s\right)}\left(\omega\right)\left(\frac{\mathcal{R}}{r}\right)^{2\ell+2}\right]r^{\ell}\bar{\mathcal{E}}_{\ell,\mathbf{m}}^{\left(s\right)}\left(\omega\right) \,.
\ee
The conservative/dissipative decomposition of the response coefficients \eqref{eq:RCsConsDissGen} for a spherically symmetric and non-spinning body is therefore trivially identical to a real/imaginary split,
\be\label{eq:RCsConsDiss_Spher}
	k_{\ell}^{\left(s\right)\text{Love}}\left(\omega\right) = \text{Re}\left\{ k_{\ell}^{\left(s\right)}\left(\omega\right) \right\} \,,\quad k_{\ell}^{\left(s\right)\text{diss}}\left(\omega\right) = \text{Im}\left\{ k_{\ell}^{\left(s\right)}\left(\omega\right) \right\} \,.
\ee
In the static limit, in particular, the entire response is conservative.

For the second example, we consider an axisymmetric and rotating configuration in $d=4$ spacetime dimensions. The axisymmetry of the background implies the decoupling of $m$-modes, while we further specialize here to the particular case where there is no $\ell$-mode mixing either, a case relevant for Kerr-Newman black holes~\cite{LeTiec:2020bos}. Then,
\be
	k_{\ell m;\ell^{\prime} m^{\prime}}^{\left(s\right)}\left(\omega\right) = k_{\ell m}^{\left(s\right)}\left(\omega\right)\delta_{\ell\ell^{\prime}}\delta_{mm^{\prime}}
\ee
and, in the Newtonian limit,
\be
	\delta\Phi^{\left(s\right)}_{\ell m}\left(\omega,r\right) = \frac{\left(\ell-s\right)!}{\ell!}\left[1 + k_{\ell m}^{\left(s\right)}\left(\omega\right)\left(\frac{\mathcal{R}}{r}\right)^{2\ell+2}\right]r^{\ell}\bar{\mathcal{E}}_{\ell m}^{\left(s\right)}\left(\omega\right) \,.
\ee

These response coefficients $k_{\ell m}^{\left(s\right)}\left(\omega\right)$ will in general be analytic functions in the angular momentum of the rotating body as well as the frequency $\omega$ with respect to an inertial observer. The complex conjugacy relation, which now reads $k_{\ell m}^{\left(s\right)\ast}\left(\omega\right) = k_{\ell,-m}^{\left(s\right)}\left(-\omega\right)$, then allows to explicitly separate the $m$-dependency of the response coefficients as
\be\label{eq:TLNs4d_mExpansion}
	k_{\ell m}^{\left(s\right)}\left(\omega\right) = k_{\ell}^{\left(0\right)}\left(\omega\right) + \chi\sum_{n=1}^{\infty}k_{\ell}^{(n)}\left(\omega,\chi\right) \left(im\right)^{n} \,,
\ee
with $\chi$ the dimensionless spin parameter associated with the angular momentum. All $k_{\ell}^{(n)}\left(\omega,\chi\right)$ are smooth functions of $\chi$, satisfying the complex conjugacy relation $k_{\ell}^{(n)\ast}\left(\omega,\chi\right)=k_{\ell}^{(n)}\left(-\omega,\chi\right)$.

Let us now extract a necessary condition for such a decoupling to occur~\cite{LeTiec:2020bos}. Starting from the physically relevant part of the response tensor,
\be
	k_{L\left\langle L^{\prime} \right\rangle}^{\left(s\right)}\left(\omega\right) = \frac{4\pi\ell!}{\left(2\ell+1\right)!!}\sum_{m}k_{\ell m}^{\left(s\right)}\left(\omega\right)\mathcal{Y}^{\ell m\ast}_{L}\mathcal{Y}^{\ell m}_{L^{\prime}} \,,
\ee
with $\ell^{\prime}=\ell$ understood, the expansion \eqref{eq:TLNs4d_mExpansion} implies
\be
	k_{L\left\langle L^{\prime} \right\rangle}^{\left(s\right)}\left(\omega\right) = k_{\ell}^{\left(0\right)}\left(\omega\right)\delta_{LL^{\prime}} + \chi\sum_{n=1}^{\infty} \left(-1\right)^{n} \left[k_{\ell}^{(2n)}\left(\omega,\chi\right)R_{LL^{\prime}}^{(2n)} + k_{\ell}^{(2n-1)}\left(\omega,\chi\right)I_{LL^{\prime}}^{(2n-1)}\right] \,,
\ee
and the tensorial structure of $k_{L\left\langle L^{\prime} \right\rangle}^{\left(s\right)}$ is completely determined by one real-valued symmetric and one real-valued antisymmetric STF tensors,
\be\ba
	R_{LL^{\prime}}^{(2n)} &\equiv \frac{8\pi\ell!}{\left(2\ell+1\right)!!}\sum_{m=1}^{\ell}m^{2n}\text{Re}\left\{\mathcal{Y}^{\ell m\ast}_{L}\mathcal{Y}^{\ell m}_{L^{\prime}}\right\} = +R_{L^{\prime}L}^{(2n)}\,, \\
	I_{LL^{\prime}}^{(2n-1)} &\equiv \frac{8\pi\ell!}{\left(2\ell+1\right)!!}\sum_{m=1}^{\ell}m^{2n-1}\text{Im}\left\{\mathcal{Y}^{\ell m\ast}_{L}\mathcal{Y}^{\ell m}_{L^{\prime}}\right\} = -I_{L^{\prime}L}^{(2n-1)} \,.
\ea\ee

Finally, let us write the conservative/dissipative decomposition of the response coefficients \eqref{eq:RCsConsDissGen} for the current special configuration,
\be\label{eq:RCsConsDiss}
	k_{\ell m}^{\left(s\right)\text{Love}}\left(\omega\right) = \text{Re}\left\{ k_{\ell m}^{\left(s\right)}\left(\omega\right) \right\} \,, \quad k_{\ell m}^{\left(s\right)\text{diss}}\left(\omega\right) = \text{Im}\left\{ k_{\ell m}^{\left(s\right)}\left(\omega\right) \right\} \,.
\ee
The Love numbers are therefore just the real part of the response coefficients, while the imaginary part encodes all the dissipative effects. We remark here that, in contrast to the case of a spherically symmetric and non-spinning body, dissipative effects can survive even in the static limit due to frame dragging~\cite{Chia:2020yla,Charalambous:2021mea,Ivanov:2022hlo}.

\section{Computing Love - Near-zone approximation}
\label{sec:NearZoneApproximation}

In order to match observables onto the $1$-body worldline EFT according to \eqref{eq:1ptNewtonianMatching}, we should solve the microscopic (``UV'' theory) equations of motion in the appropriate regime. The physical setup consists of a binary system of compact bodies during the early stages of their inspiraling phase where a Post-Newtonian expansion is accurate~\cite{Goldberger:2004jt,Porto:2005ac,Porto:2016pyg,Levi:2015msa,Levi:2018nxp}. Centering the body of interest at the origin, the companion sources perturbations with frequency equal to the orbital frequency of the system $\omega=\omega_{\text{orb}}$. The system loses energy by emitting radiation with frequency $\omega_{\text{rad}}\propto \omega_{\text{orb}}$ which is then detected by an observer located at infinity through, for example, an interferometer.

The worldline EFT arises after integrating out the short scale-internal degrees of freedom of the centered compact body, i.e. it is valid for low frequency perturbations relative to the inverse size of the body. Furthermore, the $1$-body worldline EFT ignores the dynamics of the companion body sourcing the perturbations and a second condition for its validity is that the wavelength of the perturbations is large with respect to the separation of the two bodies. This combination of conditions defines the \textit{near-zone region}. At the level of the microscopic equations of motion for a compact body of size $\mathcal{R}$, the near-zone approximation consists of working in the regime~\cite{Chia:2020yla,Charalambous:2021kcz,Starobinsky:1973aij,Starobinskil:1974nkd,Castro:2010fd,Maldacena:1997ih}
\be\label{eq:NZapprox}
	\omega \left(r-\mathcal{R}\right) \ll 1 \,,\quad \omega \mathcal{R} \ll 1\,.
\ee
In the near-zone region, one imposes the asymptotic boundary condition
\be\label{eq:Asympbc}
	\delta\Phi^{\left(s\right)}_{\ell,\mathbf{m}}\left(\omega,r\right) \xrightarrow{r\rightarrow\infty} \frac{\left(\ell-s\right)!}{\ell!}\bar{\mathcal{E}}^{\left(s\right)}_{\ell,\mathbf{m}}\left(\omega\right)\,r^{\ell} \,,
\ee
indicating the presence of a source at large distances with multipole moments $\bar{\mathcal{E}}^{\left(s\right)}_{\ell,\mathbf{m}}\left(\omega\right)$. This is supplemented with a boundary condition at the surface of the body. For example, for a star, this consists of the regularity of the wavefunction on the star's surface. For the compact bodies of interest in this thesis, i.e. for black holes, the boundary condition at the surface becomes an ingoing boundary condition at the event horizon.

An important remark here is that the near-zone approximation for black holes extends beyond the near-horizon or the low-frequency regimes. In particular, not only does it preserve the near-horizon dynamics in the radial operator for any frequency $\omega$, but it also overlaps with the asymptotically flat far-zone region $r\gg \mathcal{R}$ where outgoing boundary conditions are imposed. The overlapping intermediate region $\mathcal{R}\ll r \ll \omega^{-1}$ then serves as a matching region that probes the response of the centered body in the outgoing waves that are detected at infinity.

It should also be noted that the near-zone approximation is not unique as there are infinitely many ways to truncate the equations of motion as long as they differ by subleading terms. In practice, the truncation is done such that the equations of motion are exactly solvable in terms of elementary functions.

After solving the microscopic equations of motion, one can then match onto the worldline EFT $1$-point functions. In the rare, but relevant for general-relativistic black holes, case where the equations can be solved analytically for any orbital number $\ell$, this matching can be done directly at the level of the solution by analytically continuing $\ell$ to range in the field of real numbers. The corresponding response coefficient are then extracted from the decaying tail $r^{-\left(\ell+d-3\right)}$ in the wavefunction profile.

In practice, one is interested in the static ($\omega=0$) Love numbers for which the near-zone approximation becomes exact. The first analytic computation of the static Love numbers for neutron stars was carried out in~\cite{Hinderer:2007mb}, while the first analysis of the tidal perturbation of black holes can be traced back to~\cite{Fang:2005qq}, who studied the quadrupolar tidal response of a four-dimensional Schwarzschild black hole. This was generalized to higher multipolar orders in~\cite{Damour:2009vw,Binnington:2009bb} by extrapolating the behavior of the static Love numbers of spherically symmetric neutron stars in the limit where the compactness approaches that of a black hole. Computations of the static Love numbers for black holes have ever since flourished. For instance, static Love numbers have been computed by solving the static equations of motion for four-dimensional general-relativistic rotating black holes~\cite{Poisson:2014gka,LeTiec:2020spy,LeTiec:2020bos}, higher-dimensional spherically symmetric black holes~\cite{Kol:2011vg,Hui:2020xxx}, as well as for some beyond-general-relativistic black holes~\cite{Cardoso:2017cfl,Cardoso:2018ptl,Cai:2019npx,Charalambous:2022rre,DeLuca:2022tkm}.

However, it is instructive to study the near-zone equations of motion, rather their exact static versions. For one, this is in contact with the Newtonian matching procedure onto the worldline EFT. Furthermore, as we will see in Chapter~\ref{ch:LoveSymmetry4d}, it is within the near-zone approximation that an enhanced symmetry emerges to address some seemingly fine-tuned properties of the black hole Love numbers.
	\newpage
\chapter{Black hole Love numbers in four spacetime dimensions}
\label{ch:TLNsBlackHoles4d}

In the previous chapter, we saw how relativistic Love numbers are defined through local non-minimal couplings in the worldline EFT and how they can be matched onto $1$-point functions. Here, we will present the microscopic computation point-of-view and explicitly compute black hole Love numbers. This chapter is an adaptation of our work in Ref.~\cite{Charalambous:2021mea} which supplemented the literature by completing the analysis of the response problem for four-dimensional asymptotically flat black holes in General Relativity, via the computation of the scalar susceptibilities of rotating black holes, and contributed to resolving a dispute in the literature around the vanishing~\cite{Poisson:2014gka,Chia:2020yla} or not~\cite{LeTiec:2020spy,LeTiec:2020bos} of Love numbers, via the clarified definition through the worldline EFT. Furthermore, this chapter borrows original computations presented in our work in Ref.~\cite{Charalambous:2022rre} associated with the response problem of extremal general-relativistic black holes and also a more detailed and complete analysis of the response problem of black holes in modified theories of gravity.

We will begin by demonstrating the calculation for the simplest possible black hole response problem in four spacetime dimensions: the scalar susceptibility of the Schwarzschild black hole~\cite{Schwarzschild:1916uq}. As a sequel, we will study the scalar response problem for the most general asymptotically flat general-relativistic black hole in four spacetime dimension~\cite{Robinson:1975bv,Mazur1982}, i.e. the electrically charged and rotating Kerr-Newman black hole~\cite{Kerr:1963ud,Newman:1965my}. As was already touched upon in Section~\ref{sec:LTs_LNs}, the Kerr-Newman black hole is remarkably integrable, which will allow to separate the problem in a fashion similar to the Schwarzschild case.

We will then proceed to higher spin perturbations, namely, electromagnetic~\cite{Bicak:1977,Bicak:1976} and gravitational~\cite{Poisson:2014gka,Landry:2015zfa,Pani:2015hfa,LeTiec:2020spy,LeTiec:2020bos,Chia:2020yla} perturbations of the Kerr-Newman black hole, for the sake of which we will briefly review black hole perturbation theory in General Relativity using the Newman-Penrose formalism of null tetrads~\cite{Newman:1961qr,Geroch:1973am}. Within this formalism, the separability of linear perturbations around asymptotically flat general-relativistic black holes becomes manifest and is captured by the Teukolsky mater equation~\cite{Teukolsky:1972my,Teukolsky:1973ha,Dudley:1977zz,Kokkotas:1993ef,Kokkotas:1999bd,Berti:2005eb}. This master equation turns out to be exactly solvable in terms of elementary function for static perturbations~\cite{Bicak:1977,Bicak:1976,LeTiec:2020spy,LeTiec:2020bos,Chia:2020yla,Charalambous:2021mea}, while its exact solvability can be extended for non-static perturbations within the near-zone approximation~\cite{Chia:2020yla,Charalambous:2021kcz,Castro:2010fd,Starobinsky:1973aij,Starobinskil:1974nkd,Maldacena:1997ih}.

We will see that the leading order near-zone black hole response coefficients acquire a closed form expression and that the static Love numbers as defined in the previous chapter are in fact exactly zero~\cite{LeTiec:2020spy,LeTiec:2020bos,Chia:2020yla,Charalambous:2021mea}. This is to be contrasted with generic departures from General Relativity where the black hole Love numbers are no longer zero and exhibit a classical Renormalization Group (RG) flow~\cite{Cai:2019npx,Cardoso:2018ptl,DeLuca:2022tkm}. We will explicitly show this by studying perturbations of a scalar field in the background of the modified Schwarzschild black hole in Riemann-cubed gravity~\cite{Charalambous:2022rre}.

We will close by commenting on the behavior of the Love numbers in General Relativity and beyond. The general-relativist results are characterized with some seemingly fine-tuned properties~\cite{Porto:2016zng} calling upon an enhanced symmetry explanation~\cite{tHooft:1979rat}. This behavior appears to be theory-dependent since, for example, it is not present in the Riemann-cubed paradigm. We will study in more detail what kind of gravitational theories exhibit such behaviors in the next chapters where we will reveal the underlying enhanced symmetry reviving General Relativity's naturalness with respect to the response problem by outputting the superficially unnatural black hole Love numbers as selection rules~\cite{Charalambous:2021kcz,Charalambous:2022rre}.

\section{Scalar Love numbers of Schwarzschild black holes}
\label{sec:SLNs_Schwarzschild4d}

We begin by studying the linear scalar susceptibility problem of the Schwarzschild black hole in four spacetime dimensions. This is described by a massless scalar field $\Phi$ propagating in the background of the black hole, with action
\be
	S = -\frac{1}{2}\int d^4x\sqrt{-g}\,\left(\partial\Phi\right)^2 \,.
\ee
The geometry of the four-dimensional Schwarzschild black hole is read from the line element~\cite{Schwarzschild:1916uq}
\be
	ds^2 = -f\left(r\right)dt^2 + \frac{dr^2}{f\left(r\right)} + r^2d\Omega_2^2 \,,\quad f\left(r\right) = 1 - \frac{r_{s}}{r} \,,
\ee
where we have charted the manifold with Schwarzschild coordinates $\left(t,r,\theta,\phi\right)$ and $d\Omega_2^2 = d\theta^2 + \sin^2\theta\,d\phi^2$ is the metric on the unit-radius $2$-sphere, $\mathbb{S}^2$. This geometry turns out to be the most general static and asymptotically flat solution of Einstein's vacuum field equations equipped with a regular horizon~\cite{Israel:1967wq}. It describes an isolated, asymptotically flat, electrically neutral and non-rotating black hole in General Relativity. Its event horizon is located at the Schwarzschild radius $r_{s}$, which is related to the ADM mass $M$ of the black hole according to $r_{s}=2GM$. The event horizon is a Killing horizon relative to the Killing vector field $K=\partial_{t}$, i.e. $K$ becomes null at $r=r_{s}$.

The equation of motion we wish to solve is then the massless Klein-Gordon equation in the above curved background in the absence of any sources. Spherical symmetry ensures that the massless Klein-Gordon operator can be separated into a radial operator $\mathbb{O}_{\text{full}}$ and an angular operator $\mathbb{P}_{\text{full}}$,
\be
	\begin{gathered}
		\nabla^2\Phi = \frac{1}{r^2}\left[\mathbb{O}_{\text{full}} - \mathbb{P}_{\text{full}}\right]\Phi = 0 \,, \\
		\mathbb{O}_{\text{full}} = \partial_{r}\,\Delta\,\partial_{r} - \frac{r^4}{\Delta}\partial_{t}^2 \,,\quad \mathbb{P}_{\text{full}} = -\Delta_{\mathbb{S}^2} \,,
	\end{gathered}
\ee
where we have introduced the discriminant function $\Delta\left(r\right)\equiv r^2f\left(r\right)=r\left(r-r_{s}\right)$ for future convenience and $\Delta_{\mathbb{S}^2} = \frac{1}{\sin\theta}\partial_{\theta}\,\sin\theta\,\partial_{\theta} + \frac{1}{\sin^2\theta}\partial_{\phi}^2$ is the Laplace-Beltrami operator on $\mathbb{S}^2$. By exploiting the spherical and time translation symmetry to look for monochromatic scalar spherical harmonic modes on $\mathbb{S}^2$,
\be
	\Phi_{\omega\ell m}\left(t,r,\theta,\phi\right) = e^{-i\omega t}R_{\omega\ell m}\left(r\right)Y_{\ell m}\left(\theta,\phi\right) \,,
\ee
the Klein-Gordon equation decomposes to
\be
	\mathbb{O}_{\text{full}}\Phi_{\omega\ell m} = \ell\left(\ell+1\right)\Phi_{\omega\ell m} \,,\quad \mathbb{P}_{\text{full}}\Phi_{\omega\ell m} = \ell\left(\ell+1\right)\Phi_{\omega\ell m} \,.
\ee
with the angular problem of course already being solved by the spherical harmonic functions after identifying $\ell\in\mathbb{N}$ with the orbital number.

As already discussed in Section~\ref{sec:NearZoneApproximation}, in order to match onto the worldline EFT $1$-point function, we need to write down the equations of motion in a near-zone expansion,
\be
	\omega\left(r-r_{s}\right) \ll 1 \,,\quad \omega r_{s} \ll 1 \,.
\ee
We, therefore, split the radial operator as
\be\label{eq:NZsplitting_Schwarzschild4d}
	\begin{gathered}
		\mathbb{O}_{\text{full}} = \partial_{r}\,\Delta\,\partial_{r} + V_0 + \epsilon\,V_1 \,, \\
		V_0 = -\frac{r_{s}^4}{\Delta}\partial_{t}^2 \,,\quad V_1 = -\frac{r^4-r_{s}^4}{\Delta}\partial_{t}^2 \,,
	\end{gathered}
\ee
where $\epsilon$ is a formal parameter which is equal to unity for the full equations of motion and equal to zero for the near-zone approximation. As can be seen from these expressions, $V_1$ is indeed suppressed in the near-zone region compared to $V_0$. We remind here that there are infinitely many possible near-zone truncations of the equations of motion\footnote{There is actually a physical constraint on the allowed near-zone truncations coming from the fact that we want the solutions to behave like $r^{\ell}$ for large $r$ at leading order in the near-zone expansion, reflecting the fact that there is a source at large distances. This means that, whatever near-zone truncation we choose, it should be such that the point at infinity is a regular singular point of the near-zone radial differential equation with indicial powers $+\ell$ (and, hence, $-\ell-1$), i.e. that the leading behavior in the non-derivative terms comes from the $\ell\left(\ell+1\right)R_{\omega\ell m}$ contribution alone. More explicitly, the generic near-zone truncation consists of approximating
\[
	-\frac{r^4}{\Delta}\partial_{t}^2 \simeq -\frac{r_{s}^4g\left(r\right)}{\Delta}\partial_{t}^2
\]
at leading order, with $g\left(r=r_{s}\right)=1$ such that the near-horizon dynamics are preserved. The requirement that the indicial powers around the point at infinity are $+\ell$ and $-\ell-1$ then imposes the following constraint
\[
	\lim_{r\rightarrow\infty}\frac{g\left(r\right)}{r^2} = 0 \,.
\]
While there are infinitely many such functions $g\left(r\right)$, if one naturally focuses to polynomial functions, then the most general option is
\[
	g\left(r\right) = 1 + c_1\frac{r-r_{s}}{r_{s}} \,,
\]
for arbitrary constant $c_1$. Remarkably, even with this generic choice, the leading order near-zone equations of motion can still be solved analytically in terms of Euler's hypergeometric function.}. However, we have singled out the above one because it is the simplest possible that allows to analytically solve the resulting leading order near-zone equations of motion.

Indeed, after introducing the independent variable
\be
	x = \frac{r-r_{s}}{r_{s}} \,,
\ee
the leading order near-zone radial equation reads
\be
	\left[\frac{d}{dx}\,x\left(1+x\right)\frac{d}{dx} + \frac{\omega^2r_{s}^2}{x\left(1+x\right)}\right]R_{\omega\ell m} = \ell\left(\ell+1\right)R_{\omega\ell m} \,.
\ee
This is a Fuchsian differential equation and is solved in terms of Euler's hypergeometric function. The general solution is given by
\be\ba\label{eq:RNZGeneralSol_Schwarzschild}
	R_{\omega\ell m} &= \bar{R}_{\ell m}^{\text{in}}\left(\omega\right)\left(\frac{x}{1+x}\right)^{-i\omega r_{s}}{}_2F_1\left(\ell+1,-\ell;1-2i\omega r_{s};-x\right) \\
	&+ \bar{R}_{\ell m}^{\text{out}}\left(\omega\right)\left(\frac{x}{1+x}\right)^{+i\omega r_{s}}{}_2F_1\left(\ell+1,-\ell;1+2i\omega r_{s};-x\right) \,.
\ea\ee

Let us now impose the boundary conditions of the problem. First of all, there is an ingoing boundary condition at the event horizon, more specifically, the physical future event horizon. To understand what this looks like at the level of the radial wavefunction, we need to adapt null coordinates. For the case of Schwarzschild black hole, these are the advanced ($+$) and retarded ($-$) Eddington-Finkelstein coordinates $\left(t_{\pm},r,\theta,\phi\right)$, defined according to~\cite{Eddington1924,Finkelstein:1958zz}
\be\label{eq:EddingtonFinkelsteinCoord4d}
	dt_{\pm} = dt \pm \frac{dr}{f\left(r\right)} \Rightarrow t_{\pm} = t \pm\left\{r + r_{s}\ln\left|\frac{r-r_{s}}{r_{s}}\right|\right\} \,.
\ee
Ingoing waves at the future ($+$) or the past ($-$) event horizon are then imposed by the boundary condition
\be
	\Phi_{\omega\ell m} \xrightarrow{r\rightarrow r_{s}}T_{\ell mj}^{\left(\pm\right)}\left(\omega\right)e^{-i\omega t_{\pm}}Y_{\ell m}\left(\theta,\phi\right) \,,
\ee
with $T_{\ell mj}^{\left(\pm\right)}\left(\omega\right)$ the corresponding transmission amplitudes. At the level of the radial wavefunction in Schwarzschild coordinates, this translates to\footnote{We remark here that an ingoing wave at the future (past) event horizon looks like an outgoing wave at the past (future) event horizon. This degenerate behavior is special to non-rotating black holes due to time-reversal symmetry and will not be true once the black hole is equipped with a non-zero spin as we will see in the next section.}
\be
	R_{\omega\ell m} \sim \left(\frac{r-r_{s}}{r_{s}}\right)^{\mp i\omega r_{s}} \,\quad\text{as $r\rightarrow r_{s}$} \,.
\ee
As a result, requiring that the solution \eqref{eq:RNZGeneralSol_Schwarzschild} is ingoing at the future event horizon sets one of the integration constants to zero, namely, $\bar{R}_{\ell m}^{\text{out}}\left(\omega\right)=0$.

Next, the presence of a scalar source with multipole moments $\bar{\mathcal{E}}_{\ell m}\left(\omega\right)$ far away from the black hole is imposed by the asymptotic boundary condition
\be
	R_{\omega\ell m} \xrightarrow{r\rightarrow\infty} r^{\ell}\bar{\mathcal{E}}_{\ell m}\left(\omega\right) \,.
\ee
By analytically continuing the hypergeometric function at large distances\footnote{See Appendix~\ref{app:2F1Gamma} for a list of formulae around the hypergeometric and $\Gamma$ functions.}, we see that
\be\ba
	{}&R_{\omega\ell m} \xrightarrow{x\rightarrow\infty} \bar{R}_{\ell m}^{\text{in}}\left(\omega\right)\Gamma\left(1-2i\omega r_{s}\right) \\
	&\times\left[\frac{\Gamma\left(2\ell+1\right)}{\Gamma\left(\ell+1\right)\Gamma\left(\ell+1-2i\omega r_{s}\right)}x^{\ell} + \frac{\Gamma\left(-2\ell-1\right)}{\Gamma\left(-\ell\right)\Gamma\left(-\ell-2i\omega r_{s}\right)}x^{-\ell-1}\right] \,.
\ea\ee
Therefore, the second integration constant $\bar{R}_{\ell m}^{\text{in}}\left(\omega\right)$ is matched onto a multiple of the source strength multipole moments,
\be
	\bar{R}_{\ell m}^{\text{in}}\left(\omega\right) = \frac{\Gamma\left(\ell+1\right)\Gamma\left(\ell+1-2i\omega r_{s}\right)}{\Gamma\left(2\ell+1\right)\Gamma\left(1-2i\omega r_{s}\right)}r_{s}^{\ell}\bar{\mathcal{E}}_{\ell m}\left(\omega\right) \,.
\ee
As for the scalar response coefficients, these can also be directly read from this large distance expansion of the solution. We remind here that matching onto the worldline EFT $1$-point function can be done at the level of the above microscopic computation by analytically continuing the orbital number $\ell$ to range in the field of real numbers and then reading off the corresponding coefficient in a large distance expansion according to
\be
	R_{\omega\ell m} \xrightarrow{r\rightarrow\infty} \left[1 + k_{\ell}^{\left(0\right)}\left(\omega\right)\left(\frac{r_{s}}{r}\right)^{2\ell+1}\right]r^{\ell}\bar{\mathcal{E}}_{\ell m}\left(\omega\right) \,,\quad \ell\in\mathbb{R} \,,
\ee
where we have identified the characteristic size of the black hole with its gravitational radius and the response coefficients are independent of the azimuthal number by virtue of the spherical symmetry of the background geometry. The leading order near-zone scalar response coefficients are therefore extracted to be, before sending $\ell$ to take its physical natural-ranged values,
\be
	k_{\ell}^{\left(0\right)}\left(\omega\right) = \frac{\Gamma\left(-2\ell-1\right)\Gamma\left(\ell+1\right)\Gamma\left(\ell+1-2i\omega r_{s}\right)}{\Gamma\left(2\ell+1\right)\Gamma\left(-\ell\right)\Gamma\left(-\ell-2i\omega r_{s}\right)} \,.
\ee
After using the $\Gamma$-function mirror formula to massage this result, the conservative/dissipative split then tells us that,
\be\ba\label{eq:SLNsDissipativeRCs_Schwarzschild4d}
	k_{\ell}^{\left(0\right)\text{diss}}\left(\omega\right) &= A_{\ell}\left(\omega\right)\sinh2\pi\omega r_{s} \,, \\
	k_{\ell}^{\left(0\right)\text{Love}}\left(\omega\right) &= A_{\ell}\left(\omega\right)\tan\pi\ell\cosh2\pi\omega r_{s} \,,
\ea\ee
with $A_{\ell}\left(\omega\right)$ a real, positive and non-vanishing constant given by
\be
	A_{\ell}\left(\omega\right) = \frac{\Gamma^2\left(\ell+1\right)\left|\Gamma\left(\ell+1-2i\omega r_{s}\right)\right|^2}{2\pi\,\Gamma\left(2\ell+1\right)\Gamma\left(2\ell+2\right)} \,.
\ee

From this result, we see that sending $\ell$ to take its physical integer values results in vanishing near-zone scalar Love numbers for all frequencies,
\be
	k_{\ell}^{\left(0\right)\text{Love}}\left(\omega\right) = 0 \,.
\ee
Of course, this is accurate only to leading order in the near-zone approximation. In the current problem, this acquires corrections already at order $\mathcal{O}\left(\omega^2r_{s}^2\right)$. Therefore, the exact result from this analysis is actually the vanishing of only the static scalar Love numbers. Nevertheless, the above leading order vanishing for all frequencies is an interesting result hinting at an enhanced symmetry structure of the leading order near-zone equations of motion as well discuss in Section~\ref{sec:PowerCounting}.

\section{Scalar Love numbers of Kerr-Newman black holes}
\label{sec:SLNs_KerrNewman}

For the next example of Love numbers calculation, we study the scalar response of the most general four-dimensional asymptotically flat black hole solution solving the electrovacuum Einstein-Maxwell field equations~\cite{Robinson:1975bv,Mazur1982}. The background is now the Kerr-Newman black hole geometry~\cite{Kerr:1963ud,Newman:1965my},
\be\label{eq:KerrNewmanMetricBL}
	\begin{gathered}
		ds^2 = -dt^2 + \frac{r_{s}r-r_{Q}^2}{\Sigma}\left(dt-a\sin^2\theta\,d\phi\right)^2 + \frac{\Sigma}{\Delta}dr^2 + \Sigma\,d\theta^2 + \left(r^2+a^2\right)\sin^2\theta\,d\phi^2 \\
		\Delta = r^2 - r_{s}r + a^2 + r_{Q}^2 \,,\quad \Sigma = r^2 + a^2\cos^2\theta \,,
	\end{gathered}
\ee
and describes an isolated asymptotically flat, electrically charged and rotating black hole in General Relativity. The mass $M$, angular momentum $J$ and electric charge $Q$ of the black hole are encoded in the Schwarzschild radius, spin parameter and charge parameter as
\be
	r_{s} = 2GM \,,\quad a = \frac{J}{M} \quad\text{and}\quad r_{Q} = \sqrt{G}\left|Q\right|
\ee
respectively, where we have adopted CGS units for the electric charge. In the remaining of this thesis, we will adopt units with $G=1$. A new feature of this black hole solution compared to the Schwarzschild case is that there are now two horizons: the outer, event, horizon $r_{+}$ and the inner, Cauchy, horizon, related to the parameters of the geometry according to
\be
	r_{\pm} = \frac{1}{2}\left[r_{s} \pm \sqrt{r_{s}^2-4\left(a^2+r_{Q}^2\right)}\right] = M \pm \sqrt{M^2-\left(a^2+Q^2\right)} \,.
\ee
For a range of the spin and charge parameters, namely, for $a^2+Q^2>M^2$, the two horizons disappear and the geometry contains a naked singularity which is deemed improper as per the (weak) cosmic censorship hypothesis~\cite{Penrose:1969pc}. One then refers to non-extremal black holes if $a^2+Q^2<M^2$, for which $r_{+} \ne r_{-}$, or to extremal black holes if $a^2+Q^2=M^2$, for which the horizons degenerate to each other, $r_{+}=r_{-} = M$. The event ($+$) and Cauchy ($-$) horizons are Killing horizons with respect to the Killing vector fields $K^{\left(\pm\right)} = \partial_{t} + \Omega^{\left(\pm\right)}\,\partial_{\phi}$, with $\Omega^{\left(\pm\right)} = \frac{a}{r_{\pm}^2+a^2}$. In particular, $\Omega^{\left(+\right)}\equiv\Omega$ is the black hole angular velocity as realized by a static observer in the exterior.

Let us now write down the equations of motion we want to solve. Quite remarkably, the massless Klein-Gordon operator separates into a radial and angular operator~\cite{Press1972,Teukolsky:1972my,Teukolsky:1973ha,Starobinsky:1973aij,Starobinskil:1974nkd},
\be\label{eq:KleinGordonFull_KerrNewman}
	\begin{gathered}
		\nabla^2\Phi = \frac{1}{\Sigma}\left[\mathbb{O}_{\text{full}} - \mathbb{P}_{\text{full}}\right]\Phi = 0 \,, \\
		\mathbb{O}_{\text{full}} = \partial_{r}\,\Delta\,\partial_{r} - \frac{a^2}{\Delta}\partial_{\phi}^2 - \frac{\left(r^2+a^2\right)^2}{\Delta}\partial_{t}^2 - 2\frac{a\left(2Mr-Q^2\right)}{\Delta}\partial_{t}\partial_{\phi} \,, \\
		\mathbb{P}_{\text{full}} = -\left[\Delta_{\mathbb{S}^2} + a^2\sin^2\theta\,\partial_{t}^2\right] \,.
	\end{gathered}
\ee
Evidently, this large integrability of the Kerr-Newman black hole is owed to a higher-rank hidden symmetry, namely, the existence of a rank-$2$ Killing tensor~\cite{Walker:1970un}, a consequence of which is the conservation of the so-called Carter constant~\cite{Carter:1968rr}.

Similar to the Schwarzschild case, we look for separable solutions,
\be
	\begin{gathered}
		\Phi_{\omega\ell m}\left(t,r,\theta,\phi\right) = e^{-i\omega t}e^{im\phi}R_{\omega\ell m}\left(r\right)S_{\omega\ell m}\left(\theta\right) \,, \\
		\Rightarrow \mathbb{O}_{\text{full}}\Phi_{\omega\ell m} = \ell\left(\ell+1\right)\Phi_{\omega\ell m} \,,\quad \mathbb{P}_{\text{full}}\Phi_{\omega\ell m} = \ell\left(\ell+1\right)\Phi_{\omega\ell m} \,.
	\end{gathered}
\ee
A significant difference compared to the spherically symmetric Schwarzschild example is that the angular problem is not solved by spherical harmonics. Rather it is solved by spheroidal harmonics which have no analytic expressions in terms of elementary functions. Furthermore, despite appearance, the angular eigenvalues are not integers for non-zero $\omega a$, but can be treated, for instance, as a perturbative series in $\omega^2a^2$ around the integer spherical harmonics eigenvalues~\cite{Teukolsky:1972my,Teukolsky:1973ha}. In the above separated full problem, therefore, the physical values of $\ell$ are \textit{not} integers.

The next step in extracting the scalar susceptibilities of the Kerr-Newman black hole is to near-zone split the equations of motion, with the near-zone approximation now being described by
\be
	\omega\left(r-r_{+}\right) \ll 1 \,,\quad \omega r_{+} \ll 1 \,.
\ee
For the angular problem, we can near-zone approximate the angular operator as
\be\label{eq:NZSplitAngular_KerrNewman}
	\mathbb{P}_{\text{full}} = -\left[\Delta_{\mathbb{S}^2} + \epsilon\,a^2\sin^2\theta\,\partial_{t}^2\right] \,,
\ee
where, as in the previous section, $\epsilon$ is a formal order parameter. Within this regime, the leading order near-zone angular solutions are then the scalar spherical harmonics on $\mathbb{S}^2$ and the physical angular eigenvalues become truly integers. As for the radial problem, we will use here the following near-zone splitting
\be\label{eq:NZSplitRadial_KerrNewman}
	\begin{gathered}
		\mathbb{O}_{\text{full}} = \partial_{r}\,\Delta\,\partial_{r} + V_0 + \epsilon\,V_1 \,, \\
		\ba
			V_0 = &-\frac{\left(r_{+}^2+a^2\right)^2}{\Delta}\left[ \left(\partial_{t} +\Omega\,\partial_{\phi}\right)^2 + 4\Omega\frac{r-r_{+}}{r_{+}-r_{-}}\partial_{t} \partial_{\phi} \right] \,, \\
			V_1 = &-\frac{\left(r+r_{+}\right)\left(r^2 + r_{+}^2 +2a^2\right)}{r-r_{-}}\partial_{t}^2 + 2\frac{r_{+}^2+a^2}{r-r_{-}}\left(\beta-2M\right)\Omega\,\partial_{t}\partial_{\phi} \,,
		\ea
	\end{gathered}
\ee
where we have introduce the (inverse of the) Hawking temperature of the Kerr-Newman black hole\footnote{More concretely, $\beta$ is the inverse surface gravity at the event horizon since we are in the purely classical regime. However, we will keep referring to $\beta$ as ``the (inverse of the) Hawking temperature'', even though there are no $\hbar$'s attached to the problem we are solving, to make contact with the Kerr/CFT correspondence~\cite{Castro:2010fd,Guica:2008mu}.},
\be\label{eq:betaKerrNewman}
	\beta \equiv \frac{1}{2\pi T_{H}} = 2\frac{r_{+}^2+a^2}{r_{+}-r_{-}} \,.
\ee

To study the ingoing boundary condition at the future/past event horizon, we need to adopt the advanced ($+$)/retarded ($-$) null coordinates $\left(t_{\pm},r,\theta,\varphi_{\pm}\right)$, defined by\footnote{We have fixed the integration constants such that we obtain a smooth extremal limit of these expressions, namely,
\[
	a^2 + Q^2 \rightarrow M^2 \Rightarrow t_{\pm} \rightarrow t \pm \left\{r+2r_{+}\ln\left|\frac{r-r_{+}}{r_{+}}\right| - \frac{r_{+}^2+a^2}{r-r_{+}}\right\} \,,\quad \varphi_{\pm} \rightarrow \phi \mp \frac{a}{r-r_{+}} \,,
\]
with $r_{-}=r_{+}=M$ at extremality.}
\be\ba\label{eq:NullCoordiantes_KerrNewman}
	dt_{\pm} &= dt \pm \frac{r^2+a^2}{\Delta}dt \Rightarrow t_{\pm} = t \pm \left\{r + \frac{1}{2\kappa_{+}}\ln\left|\frac{r-r_{+}}{r_{+}}\right| + \frac{1}{2\kappa_{-}}\ln\left|\frac{r-r_{-}}{r_{+}}\right|\right\} \,, \\
	d\varphi_{\pm} &= d\phi \pm \frac{a}{\Delta}dr \quad\,\,\,\,\,\Rightarrow \varphi_{\pm} = \phi \pm \frac{a}{r_{+}-r_{-}}\ln\left|\frac{r-r_{+}}{r-r_{-}}\right| \,,
\ea\ee
where $\kappa_{+} = \frac{1}{2}\frac{r_{+}-r_{-}}{r_{+}^2+a^2}$ and $\kappa_{-} = \frac{1}{2}\frac{r_{-}-r_{+}}{r_{-}^2+a^2}$ are the surface gravities\footnote{The surface gravity $\kappa$ of a Killing horizon, at which a Killing vector $K=K^{\mu}\partial_{\mu}$, normalized such that $K_{\mu}K^{\mu}\rightarrow-1$ at infinity, becomes null, is defined by $K^{\nu}\nabla_{\nu}K^{\mu} = \kappa K^{\mu}$ and measures the acceleration as realized by a distant observer to keep an object at the horizon. For the surface gravities $\kappa_{\pm}$ at the outer ($+$) and the inner ($-$) horizons of the Kerr-Newman black hole, the relevant properly normalized Killing vectors are the Killing vectors $K^{\left(\pm\right)}=\partial_{t}+\Omega^{\left(\pm\right)}\partial_{\phi}$ with respect to which the horizons are Killing horizons.} at the outer and inner horizons respectively. In particular, $\kappa_{+}=\beta^{-1}$ in our current notation. Then, an ingoing wave at the future/past event horizon looks like
\be
	\Phi_{\omega\ell m} \xrightarrow{r\rightarrow r_{+}} T_{\ell m}^{\left(\pm\right)}\left(\omega\right)e^{-i\omega t_{\pm}}e^{im\varphi_{\pm}}S_{\omega\ell m}\left(\theta\right)
\ee
or, at the level of the radial wavefunction,
\be
	R_{\omega\ell m} \sim \left(r-r_{+}\right)^{\pm iZ_{+}\left(\omega\right)}\left(r-r_{-}\right)^{\pm iZ_{-}\left(\omega\right)} \quad \text{as $r\rightarrow r_{+}$} \,,
\ee
where have defined
\be\label{eq:Zpm_KerrNewman}
	Z_{\pm}\left(\omega\right) \equiv \frac{1}{2\kappa_{\pm}}\left(m\Omega^{\left(\pm\right)}-\omega\right) \,.
\ee
This is to be supplemented with the worldline EFT matching asymptotic boundary conditions
\be\label{eq:ScalarEFTmatching_KerrNewman}
	R_{\omega\ell m}\xrightarrow{r\rightarrow\infty} \left[1 + k_{\ell m}^{\left(0\right)}\left(\omega\right)\left(\frac{r_{s}}{r}\right)^{2\ell+1}\right]r^{\ell}\bar{\mathcal{E}}_{\ell m}\left(\omega\right) \,,
\ee
which is to be understood as arising from the large-$r$ expansion of the solution \textit{after} analytically continuing $\ell\in\mathbb{N}\rightarrow\ell\in\mathbb{R}$.

Performing the independent variable transformation
\be
	x = \frac{r-r_{+}}{r_{+}-r_{-}} \,,
\ee
the leading order near-zone radial equation of motion reads
\be\label{eq:RNZEq_KerrNewman}
	\left[\frac{d}{dx}\,x\left(1+x\right)\frac{d}{dx} + \frac{Z_{+}^2\left(\omega\right)}{x} - \frac{\tilde{Z}_{-}^2\left(\omega\right)}{1+x}\right]R_{\omega\ell m} = \ell\left(\ell+1\right)R_{\omega\ell m} \,,
\ee
with $\tilde{Z}_{-}\left(\omega\right)\equiv\frac{\beta}{2}\left(m\Omega+\omega\right)$. This is again a second order Fuchsian differential equation that is solved by Euler's hypergeometric functions. The solution that is ingoing at the future event horizon is then found to be
\be\ba\label{eq:RNZSol_KerrNewman}
	R_{\omega\ell m} &= \bar{R}_{\ell m}^{\text{in}}\left(\omega\right)\left(\frac{x}{1+x}\right)^{iZ_{+}\left(\omega\right)} \\
	&\times\left(1+x\right)^{im\beta\Omega}{}_2F_1\left(\ell+1+im\beta\Omega,-\ell+im\beta\Omega;1+2iZ_{+}\left(\omega\right);-x\right) \,,
\ea\ee
while the asymptotic boundary condition \eqref{eq:ScalarEFTmatching_KerrNewman} matches the integration constants $\bar{R}_{\ell m}^{\text{in}}\left(\omega\right)$ and the leading order near-zone scalar response coefficients $k_{\ell m}^{\left(0\right)}\left(\omega\right)$ to
\be\ba\label{eq:SRCs_KerrNewman}
	\bar{R}_{\ell m}^{\text{in}}\left(\omega\right) &= \frac{\Gamma\left(\ell+1+im\beta\Omega\right)\Gamma\left(\ell+1-i\beta\omega\right)}{\Gamma\left(2\ell+1\right)\Gamma\left(1+2iZ_{+}\left(\omega\right)\right)}\left(r_{+}-r_{-}\right)^{\ell}\bar{\mathcal{E}}_{\ell m}\left(\omega\right) \,, \\
	k_{\ell m}^{\left(0\right)}\left(\omega\right) &= \frac{\Gamma\left(-2\ell-1\right)\Gamma\left(\ell+1+im\beta\Omega\right)\Gamma\left(\ell+1-i\beta\omega\right)}{\Gamma\left(2\ell+1\right)\Gamma\left(-\ell+im\beta\Omega\right)\Gamma\left(-\ell-i\beta\omega\right)}\left(\frac{r_{+}-r_{-}}{r_{s}}\right)^{2\ell+1} \,.
\ea\ee
As already prescribed in Section~\ref{sec:LTs_LNs}, the Love numbers and dissipative response coefficients are extracted from the real and imaginary parts of the response coefficients respectively. This can be worked out be
\be\ba\label{eq:SLNsDissipativeRCs_KerrNewman}
	k_{\ell m}^{\left(0\right)\text{diss}}\left(\omega\right) &= -A_{\ell m}\left(\omega\right)\sinh2\pi Z_{+}\left(\omega\right) \,, \\
	k_{\ell m}^{\left(0\right)\text{Love}}\left(\omega\right) &= A_{\ell m}\left(\omega\right)\bigg[ \tan\pi\ell\,\cosh\pi m\beta\Omega\,\cosh\pi\beta\omega \\
	&\quad\quad\quad\quad\,\,\,+ \cot\pi\ell\,\sinh\pi m\beta\Omega\,\sinh\pi\beta\omega \bigg] \,,
\ea\ee
with the real and positive constant $A_{\ell m}\left(\omega\right)$ given by
\be\label{eq:AConstantKerrNewman}
	A_{\ell m}\left(\omega\right) = \frac{\left|\Gamma\left(\ell+1+im\beta\Omega\right)\right|^2\left|\Gamma\left(\ell+1-i\beta\omega\right)\right|^2}{2\pi\,\Gamma\left(2\ell+1\right)\Gamma\left(2\ell+2\right)}\left(\frac{r_{+}-r_{-}}{r_{s}}\right)^{2\ell+1} \,.
\ee

The qualitative behavior of the Kerr-Newman black hole scalar Love numbers at leading order in the near-zone expansion can be found in Table~\ref{tbl:NZLNs_KerrNewmanS}. In what follows, we will investigate in detail the various cases associated with the running of the scalar Love numbers and their vanishing and/or non-running.

\begin{table}[t]
	\centering
	\begin{tabular}{|l|l|l|}
		\hline
		\multicolumn{2}{|c||}{\Gape[8pt]{Range of parameters}} & \multicolumn{1}{c|}{Behavior of $k^{\left(0\right)\text{Love}}_{\ell m}\left(\omega\right)$} \\
		\hline\hline
		\multicolumn{2}{|c||}{\Gape[8pt]{$im\beta\Omega\notin\mathbb{Z}$ AND $i\beta\omega\notin\mathbb{Z}$}} & \multicolumn{1}{c|}{Running} \\
		\hline
		\multirow{2}{*}{$\begin{matrix} im\beta\Omega = k\in\mathbb{Z} \\ \text{OR} \\ -i\beta\omega = k\in\mathbb{Z} \end{matrix}$} & \multicolumn{1}{c||}{\Gape[9pt]{$-\ell\le k \le\ell$}} & \multicolumn{1}{c|}{Vanishing} \\
		\cline{2-3}
		& \multicolumn{1}{c||}{\Gape[10pt]{$k>\ell$ OR $k<-\ell$}} & \multicolumn{1}{c|}{Non-vanishing and Non-running} \\
		\hline
	\end{tabular}
	\caption[Behavior of the Kerr-Newman scalar Love numbers at leading order in the near-zone expansion.]{Behavior of the Kerr-Newman scalar Love numbers at leading order in the near-zone expansion, Eq.~\eqref{eq:SLNsDissipativeRCs_KerrNewman}, after analytically continuing the frequency $\omega$ and azimuthal number $m$ to range in the field of complex numbers. For general black hole and perturbation parameters, the leading order near-zone scalar Love numbers exhibit a classical RG flow, while there is a discrete set of resonant conditions associated with imaginary frequencies and azimuthal numbers for which the Love numbers are vanishing and/or non-running.}
	\label{tbl:NZLNs_KerrNewmanS}
\end{table}

\subsection{Running Love}
The expression in Eq.~\eqref{eq:SLNsDissipativeRCs_KerrNewman} for the scalar Love numbers of the Kerr-Newman black hole at leading order in the near-zone expansion has some new features compared to the Schwarzschild case. First of all, we see that sending $\ell$ to take its physical integer values results in seemingly diverging Love numbers for generic black hole angular momentum and perturbation frequency and azimuthal number. As we will see right away, this is indicative of a classical RG flow exhibited by the Wilson coefficients in the worldline EFT~\cite{Kol:2011vg,Porto:2016pyg,Porto:2016zng,Hui:2020xxx,Ivanov:2022hlo}.

Let us first demonstrate this at the level of the microscopic computation. The diverging behavior of the Love numbers comes from a simple pole in Eq.~\eqref{eq:SRCs_KerrNewman}, arising from evaluating $\Gamma\left(-2\ell-1\right)$ at negative integers. To tackle this, we will be using an ``orbital-number regularization'' scheme, corresponding to replacing $\ell\rightarrow\ell-\varepsilon$ in Eq.~\eqref{eq:SLNsDissipativeRCs_KerrNewman}, sending $\ell$ to take its physical whole integer values and then studying the $\varepsilon\rightarrow0$ behavior. From the residue of the $\Gamma$-function near negative integers, $\Gamma\left(-n+\varepsilon\right) = \frac{\left(-1\right)^{n}}{n!\varepsilon}+\mathcal{O}\left(\varepsilon^0\right)$, the developed pole in the response coefficients therefore reads
\be\label{eq:RunningSRCs_KerrNewman}
	k_{\ell-\varepsilon,m}^{\left(0\right)}\left(\omega\right) = -\frac{1}{2\varepsilon}\frac{\Gamma\left(\ell+1+im\beta\Omega\right)\Gamma\left(\ell+1-i\beta\omega\right)}{\left(2\ell\right)!\left(2\ell+1\right)!\Gamma\left(-\ell+im\beta\Omega\right)\Gamma\left(-\ell-i\beta\omega\right)}\left(\frac{r_{+}-r_{-}}{r_{s}}\right)^{2\ell+1} + \mathcal{O}\left(\varepsilon^0\right) \,.
\ee
The full scalar field profile, of course, does not suffer from any divergence. This can be traced back to a compensating divergence in the ``source'' part of the scalar field profile. Indeed, according to our matching procedure of performing the source/response splitting before sending $\ell$ to take its physical values, the ``source'' and ``response'' contributions to the radial wavefunction at leading order in the near-zone expansion are
\be\ba
	R_{\omega\ell m}\left(r\right) &= \left[Z_{\omega\ell m}^{\text{source}}\left(r\right) + k_{\ell m}^{\left(0\right)}\left(\omega\right)\left(\frac{r_{s}}{r}\right)^{2\ell+1}Z_{\omega\ell m}^{\text{response}}\left(r\right)\right]r^{\ell}\bar{\mathcal{E}}^{\left(0\right)}_{\ell m}\left(\omega\right) \,, \\
	Z_{\omega\ell m}^{\text{source}}\left(r\right) &= \left(1-\frac{r_{+}}{r}\right)^{\ell}\left(\frac{r-r_{-}}{r-r_{+}}\right)^{i\beta\left(m\Omega+\omega\right)/2} \\
	&\quad\times{}_2F_1\left(-\ell+im\beta\Omega,-\ell+i\beta\omega;-2\ell;\frac{r_{+}-r_{-}}{r_{+}-r}\right) \,, \\
	Z_{\omega\ell m}^{\text{response}}\left(r\right) &= \left(1-\frac{r_{+}}{r}\right)^{-\ell-1}\left(\frac{r-r_{-}}{r-r_{+}}\right)^{i\beta\left(m\Omega+\omega\right)/2} \\
	&\quad\times{}_2F_1\left(\ell+1+im\beta\Omega,\ell+1+i\beta\omega;2\ell+2;\frac{r_{+}-r_{-}}{r_{+}-r}\right) \,.
\ea\ee
In the limit $\varepsilon\rightarrow0$, the ``source'' part $Z_{\omega,\ell-\varepsilon,m}^{\text{source}}$ also develops a simple pole, this time coming from evaluating the hypergeometric function with the third parameter argument being a negative integer. From the residue formula for the hypergeometric function,
\be\ba\label{eq:2F1Residue}
	{}_2F_1\left(a,b;-n+\varepsilon;z\right) &= \Gamma\left(-n+\varepsilon\right)\frac{\Gamma\left(a+n+1\right)\Gamma\left(b+n+1\right)}{\Gamma\left(a\right)\Gamma\left(b\right)\left(n+1\right)!} \\
	&\quad\times{}_2F_1\left(a+n+1,b+n+1;n+2;z\right) + \mathcal{O}\left(\varepsilon^0\right) \,,
\ea\ee
we see that
\be\ba
	Z_{\omega,\ell-\varepsilon,m}^{\text{source}}\left(r\right) &= -\frac{1}{2\varepsilon}\frac{\Gamma\left(\ell+1+im\beta\Omega\right)\Gamma\left(\ell+1+i\beta\omega\right)}{\left(2\ell\right)!\left(2\ell+1\right)!\Gamma\left(-\ell+im\beta\Omega\right)\Gamma\left(-\ell+i\beta\omega\right)}\left(\frac{r_{+}-r_{-}}{r_{s}}\right)^{2\ell+1} \\
	&\quad\times\left(\frac{r_{s}}{r}\right)^{2\ell+1}Z_{\omega\ell m}^{\text{response}}\left(r\right) + \mathcal{O}\left(\varepsilon^0\right) \,.
\ea\ee
By employing the $\Gamma$-function mirror formula, this turns out to exactly cancel the divergence in the scalar response coefficients whenever $\ell$ is an integer.

From the worldline EFT point of view, this diverging behavior of the scalar Love numbers is interpreted as a classical RG flow. More specifically, as we will see in detail in the Section~\ref{sec:PowerCounting}, power counting arguments reveal that the Wilson coefficients defining the Love numbers should get renormalized from an overlapping with PN corrections of the ``source'' part of the $1$-point function, namely, from the following type of diagrams~\cite{Kol:2011vg,Charalambous:2022rre,Ivanov:2022hlo}
\be
	\Phi_{\omega\ell m} \supset \vcenter{\hbox{\begin{tikzpicture}
			\begin{feynman}
				\vertex[dot] (a0);
				\vertex[below=1cm of a0] (p1);
				\vertex[above=1cm of a0] (p2);
				\vertex[right=0.4cm of a0, blob] (gblob){};							
				\vertex[right=1.5cm of p1] (b1);
				\vertex[right=1.5cm of p2] (b2);
				\vertex[right=1.19cm of p2] (b22){$\times$};
				\vertex[above=0.7cm of a0] (g1);
				\vertex[above=0.4cm of a0] (g2);
				\vertex[right=0.05cm of a0] (gdtos){$\vdots$};
				\vertex[below=0.7cm of a0] (gN);
				\vertex[left=0.2cm of gN] (gN1);
				\vertex[left=0.2cm of g1] (g11);
				\diagram*{
					(p1) -- [double,double distance=0.5ex] (p2),
					(g1) -- [photon] (gblob),
					(g2) -- [photon] (gblob),
					(gN) -- [photon] (gblob),
					(b1) -- (gblob) -- (b2),
				};
				\draw [decoration={brace}, decorate] (gN1.south west) -- (g11.north west)
				node [pos=0.55, left] {\(2\ell+1\)};
			\end{feynman}
	\end{tikzpicture}}} \,.
\ee
From the theory of differential equations point of view, whenever $2\ell+1\in\mathbb{N}$, the series solutions of the radial differential equation~\eqref{eq:RNZEq_KerrNewman} fall into the degenerate case where the characteristic exponents near $x\rightarrow\infty$ differ by an integer number and, thus, according to Fuchs' theorem, only one of the independent solutions can be written as a Frobenius series around there, while the second independent solution will unavoidably contain logarithms. More explicitly, the solution regular at the future event horizon is still given by Eq.~\eqref{eq:RNZSol_KerrNewman}, but its analytic continuation at large distances must be taken as a limiting case with the end result being~\cite{NIST}
\be\ba
	{}&R_{\omega\ell m j} = \bar{\mathcal{E}}_{\ell m}^{\left(0\right)}\left(\omega\right) r_{s}^{\ell}\left(\frac{r-r_{-}}{r-r_{+}}\right)^{i\beta\left(m\Omega+\omega\right)/2} \\
	&\times\bigg\{ \left(\frac{r-r_{+}}{r_{s}}\right)^{\ell}\sum_{k=0}^{\ell}\frac{\left(-\ell+im\beta\Omega\right)_{k}}{\left(-k+\ell+1-i\beta\omega\right)_{k}}\frac{\left(\ell-k\right)!}{\ell!\,k!}\left(-x\right)^{-k} \\
	&\,\,\,+\underset{\varepsilon\rightarrow0}{\text{Res}}\left\{k_{\ell-\varepsilon,m}^{\left(0\right)}\left(\omega\right)\right\} \left(\frac{r-r_{+}}{r_{s}}\right)^{-\ell-1}\sum_{k=0}^{\infty}\frac{\left(\ell+1+im\beta\Omega\right)_{k}}{\left(-k-\ell-i\beta\omega\right)_{k}}\frac{\left(\ell+1\right)!}{\left(k+\ell+1\right)!\,k!}\left(+x\right)^{-k} \\
	{}&\,\,\,\times\bigg[\log x + \psi\left(k+1\right) + \psi\left(k+\ell+2\right) - \psi\left(k+\ell+1+im\beta\Omega\right) - \psi\left(-k-\ell-i\beta\omega\right)\bigg] \bigg\} \,,
\ea\ee
where we have identified that the coefficient multiplying the second term is the residue of the response coefficients as $\ell$ approaches an integer number, Eq.~\eqref{eq:RunningSRCs_KerrNewman}. From the EFT point of view, this residue is precisely the $\beta$-function dictating the classical RG flow of the Love numbers~\cite{Kol:2011vg,Ivanov:2022hlo,Charalambous:2022rre},
\be
	L\frac{d k_{\ell m}^{\left(0\right)}}{dL} = -\frac{\Gamma\left(\ell+1+im\beta\Omega\right)\Gamma\left(\ell+1-i\beta\omega\right)}{2\left(2\ell\right)!\left(2\ell+1\right)!\Gamma\left(-\ell+im\beta\Omega\right)\Gamma\left(-\ell-i\beta\omega\right)}\left(\frac{r_{+}-r_{-}}{r_{s}}\right)^{2\ell+1} \,.
\ee
This $\beta$-function is evidently real and is therefore entirely associated with the running of the Love numbers, while the dissipative response exhibits no RG flow.

\subsection{Vanishing static Love}
The scalar Love numbers for the Kerr-Newman black hole given in Eq.~\ref{eq:SLNsDissipativeRCs_KerrNewman} are at leading order in the near-zone expansion. In fact, by inspecting the near-zone splitting in Eqs.~\eqref{eq:NZSplitAngular_KerrNewman}-\eqref{eq:NZSplitRadial_KerrNewman}, we see that the scalar Love numbers admit corrections already at $\mathcal{O}\left(\omega a, \beta^2\omega^2\right)$. Therefore, the only exact result we obtain from this analysis is the response to static perturbations. In the static limit, the scalar response coefficients \eqref{eq:SRCs_KerrNewman} become~\cite{LeTiec:2020spy,LeTiec:2020bos,Chia:2020yla,Charalambous:2021mea}
\be
	k_{\ell m}^{\left(0\right)}\left(\omega=0\right) = -\frac{i m\beta\Omega}{2}\left(\frac{r_{+}-r_{-}}{r_{s}}\right)^{2\ell+1}\frac{\left(\ell!\right)^2}{\left(2\ell\right)!\left(2\ell+1\right)!}\prod_{n=1}^{\ell}\left(n^2+\left(m\beta\Omega\right)^2\right) \,,
\ee
where we have used the $\Gamma$-function formula $\left|\Gamma\left(\ell+1+ib\right)\right|^2 = \frac{\pi b}{\sinh\pi b}\prod_{n=1}^{\ell}\left(n^2+b^2\right)$, holding for $\ell\in\mathbb{N}$, $b\in\mathbb{R}$.

For zero black hole angular momentum, we retrieve the corresponding vanishing result for the static scalar susceptibility of the four-dimensional Schwarzschild black hole. Furthermore, for non-zero black hole spin parameter, the response coefficients also vanish for axisymmetric ($m=0$) perturbations~\cite{Gurlebeck:2015xpa,Poisson:2014gka,Landry:2015zfa}. Beyond these special cases, we find the above non-vanishing result for the static scalar susceptibility of the Kerr-Newman black hole. This was first noted explicitly in~\cite{LeTiec:2020spy,LeTiec:2020bos} for spin-$2$ (tidal) perturbations.

However, as per our definition of the Love numbers as Wilson coefficients in front of particular \textit{local} operators in the worldline EFT, the actual ``Love'' part of the response is encoded in its conservative component. According to the prescription extracted in Section~\ref{sec:LTs_LNs}, the conservative component of the response coefficients is given by their real part. The purely imaginary static scalar susceptibilities above are therefore purely dissipative in nature and the corresponding static scalar Love numbers are exactly zero~\cite{Chia:2020yla,Charalambous:2021mea},
\be
	k_{\ell m}^{\left(0\right)}\left(\omega=0\right) = ik_{\ell m}^{\left(0\right)\text{diss}}\left(\omega=0\right) \,,\quad k_{\ell m}^{\left(0\right)\text{Love}}\left(\omega=0\right) = 0 \,.
\ee

In fact, the static dissipation arises from frame-dragging effects~\cite{Chia:2020yla,Charalambous:2021mea}. Another way to see this purely frame-dragging effect is to work directly in advanced null coordinates $\left(t_{+},r,\theta,\varphi_{+}\right)$, which become co-rotating coordinates at the future event horizon. Then, the corresponding static radial wavefunction becomes purely polynomial in the radial variable and the absence of a decaying mode is indicative of the vanishing of the gauge invariant conservative response~\cite{Chia:2020yla,Charalambous:2021mea,Charalambous:2021kcz,Charalambous:2022rre}.

\subsection{Non-running Love}
Even though the leading order near-zone scalar Love numbers are not accurate beyond the static case, it is still interesting to study the resonant conditions under which Eq.~\eqref{eq:SLNsDissipativeRCs_KerrNewman} exhibits no logarithmic running for non-zero frequencies. This happens whenever $\sinh\pi\beta\omega=0$ or $\sinh\pi m\beta\Omega=0$, i.e. whenever
\be
	\beta\omega \in i\mathbb{Z} \quad\text{OR}\quad m\beta\Omega \in i\mathbb{Z} \,.
\ee
The first cases correspond to modes that are purely growing or decaying in time and can be regarded as approximate in their nature. The second case corresponds to purely imaginary azimuthal numbers for which the scalar field profile acquires conical singularities coming from the periodic identification of the azimuthal angle, $\phi\sim\phi+2\pi$, and are, hence, unphysical. Despite their unphysical nature, as we will see in the next chapter, they still admit an elegant interpretation in terms of representation theory arguments from an enhanced symmetry arising in the near-zone approximation~\cite{Charalambous:2023jgq}.

Let us focus to the resonant conditions associated with purely imaginary azimuthal numbers. These can be broken down into two classes,
\be
	im\beta\Omega = k\in\mathbb{Z} \quad \begin{cases}
		-\ell\le k \le \ell \\
		k>\ell \quad\text{OR}\quad k<-\ell
	\end{cases} \,.
\ee
The corresponding radial wavefunction reads
\be\ba
	R_{\omega\ell m}\bigg|_{im\beta\Omega=k\in\mathbb{Z}} &= \bar{R}_{\ell m}^{\text{in}}\left(\omega\right)\left(\frac{x}{1+x}\right)^{-i\beta\omega/2} \\
	&\quad\times\left[x\left(1+x\right)\right]^{k/2}{}_2F_1\left(\ell+1+k,-\ell+k;	1+k-i\beta\omega;-x\right)
\ea\ee
For $k\le\ell$, these acquire an interesting quasi-polynomial form analogous to the static case,
\be\ba\label{eq:RadialWFkL_KerrNewman}
	R_{\omega\ell m}\bigg|_{im\beta\Omega=k\le\ell} &= \bar{R}_{\ell m}^{\text{in}}\left(\omega\right)\left(\frac{x}{1+x}\right)^{-i\beta\omega/2}\left[x\left(1+x\right)\right]^{k/2}\sum_{n=0}^{\ell-k}\begin{pmatrix} \ell-k \\ n \end{pmatrix} \frac{\left(\ell+1+k\right)_{n}}{\left(1+k-i\beta\omega\right)_{n}}x^{n} \,.
\ea\ee
However, for $k<-\ell$, one starts encountering decaying, $r^{-\ell-1}$, modes which persist even when working in advanced null coordinates and, therefore, cannot be attributed to purely frame-dragging effects. The absence of logarithms indicates that this is not a PN correction to the growing (``source'') mode and these decaying modes are, hence, interpreted as non-vanishing and non-running Love numbers. Nevertheless, the intermediate range $-\ell \le k\le \ell$ contains perturbations with vanishing scalar Love numbers to leading-order in the near-zone expansion. For $k>\ell$, the radial wavefunction does not acquire any particular quasi-polynomial form but the absence of RG flowing is still present in these situations as well. These cases are all captured by the formula for the scalar Love numbers, Eq.~\eqref{eq:SLNsDissipativeRCs_KerrNewman}. For the resonant conditions associated with purely imaginary frequencies, this analysis can be similarly carried through for $-i\beta\omega=k\in\mathbb{Z}$ and gives with same results, i.e. that there is no running for any $k$, while the scalar Love numbers vanish for $-\ell \le k \le \ell$.

\section{Spin-$s$ Love numbers of Kerr-Newman black holes}
\label{sec:LNs_KerrNewmanS}

The separability of Kerr-Newman black hole perturbations turns out to persist for electromagnetic as well as gravitational perturbations. This becomes manifest when working with field-strength tensors, rather than gauge field and metric perturbations, which leads to the traditional setup of black hole perturbation theory in General Relativity within the Newman-Penrose formalism~\cite{Newman:1961qr,Geroch:1973am}.

\subsection{Newman-Penrose formalism and the Teukolsky equation}
Let us briefly review some basics of the Newman-Penrose (NP) formalism that will be important for our purposes~\cite{Newman:1961qr}. The NP formalism is a particular case of the tetrad formalism, with the local tetrads chosen to be null. The relevant four local tetrads are composed of two real vectors, $\ell^{\mu}$ and $n^{\mu}$, and two complex-valued vectors, $m^{\mu}$ and its complex conjugate\footnote{Throughout this thesis, we are denoting complex conjugation by asterisks. However, here, and only here, we denote the complex conjugate of $m^{\mu}$ as $\bar{m}^{\mu}$ instead of $m^{\mu\ast}$ to make contact with the standard conventions in the NP formalism.} $\bar{m}^{\mu}$, normalized according to
\be
	\ell_{\mu}n^{\mu} = -1\,,\quad m_{\mu}\bar{m}^{\mu} = 1
\ee
and with all other inner products of two tetrad vectors being zero. These tetrads play the role of the ``square root'' of the metric tensor, which can be reconstructed according to
\be
	g_{\mu\nu} = 2\left(-\ell_{(\mu}n_{\nu)} + m_{(\mu}\bar{m}_{\nu)}\right) \,.
\ee
Within the NP formalism, non-scalar degrees of freedom are encoded into complex spacetime scalars obtained by projecting the relevant tensors on the local null tetrads. In particular, the $10$ independent components of the Weyl tensor are cast into the $5$ complex Weyl scalars,
\be
	\begin{split}
		\psi_0 &=  C_{\mu\nu\rho\sigma}\ell^{\mu}m^{\nu}\ell^{\rho}m^{\sigma} \,,\quad \psi_1 = C_{\mu\nu\rho\sigma}\ell^{\mu}n^{\nu}\ell^{\rho}m^{\sigma} \,, \quad \psi_2 = C_{\mu\nu\rho\sigma}\ell^{\mu}m^{\nu}\bar{m}^{\rho}n^{\sigma} \,, \\ 
		\psi_3 & = C_{\mu\nu\rho\sigma}\ell^{\mu}n^{\nu}\bar{m}^{\rho}n^{\sigma} \,,\quad \psi_4 = C_{\mu\nu\rho\sigma}n^{\mu}\bar{m}^{\nu}n^{\rho}\bar{m}^{\sigma} \,,
	\end{split}
\ee
while the $6$ independent components of the Maxwell field strength tensor are rearranged into the $3$ complex Maxwell-NP scalars,
\be
	\phi_0 =  F_{\mu\nu}\ell^{\mu}m^{\nu} \,,\quad \phi_1 =  \frac{1}{2}F_{\mu\nu}\left(\ell^{\mu}n^{\nu} + \bar{m}^{\mu}m^{\nu}\right) \,,\quad \phi_2 =  F_{\mu\nu}\bar{m}^{\mu}n^{\nu} \,.
\ee
and the NP scalar of a scalar field is the scalar field itself. This process is of course reversible and one can reconstruct, for example, the Maxwell field strength tensor according to
\be
	F_{\mu\nu} = 4\,\text{Re}\left\{ - \phi_0\,n_{[\mu}m_{\nu]} - \phi_1\left(\ell_{[\mu}n_{\nu]} - m_{[\mu}\bar{m}_{\nu]}\right) + \phi_2\,\ell_{[\mu}m_{\nu]} \right\} \,.
\ee

All of the above NP scalars transform homogeneously under two particular local Lorentz transformations known as local rotations and local boosts\footnote{These are also known as type III transformations in the traditional language of the NP formalism~\cite{Newman:1961qr}.},
\begin{itemize}
	\item \underline{Local rotations}: These are complex local $SO\left(2\right)$ rotations of the tetrad $m^{\mu}$, keeping the real vectors $\ell^{\mu}$ and $n^{\mu}$ invariant,
	\be
		\begin{pmatrix}
			\ell^{\mu} \\
			n^{\mu} \\
			m^{\mu} \\
			\bar{m}^{\mu}
		\end{pmatrix} \xrightarrow{\chi}
		\begin{pmatrix}
			\ell^{\mu} \\
			n^{\mu} \\
			e^{i\chi}m^{\mu} \\
			e^{-i\chi}\bar{m}^{\mu}
		\end{pmatrix} \,.
	\ee
	
	\item \underline{Local boosts}: These are real local rescalings of the real tetrads $\ell^{\mu}$ and $n^{\mu}$, keeping the complex vector $m^{\mu}$ invariant,
	\be
		\begin{pmatrix}
			\ell^{\mu} \\
			n^{\mu} \\
			m^{\mu} \\
			\bar{m}^{\mu}
		\end{pmatrix} \xrightarrow{\lambda}
		\begin{pmatrix}
			\lambda\ell^{\mu} \\
			\lambda^{-1}n^{\mu} \\
			m^{\mu} \\
			\bar{m}^{\mu}
		\end{pmatrix} \,.
	\ee
\end{itemize}

NP scalars are labeled according to their transformation properties w.r.t. local rotations and boosts via a spin-weight\footnote{Unfortunately, the standard notation for the spin-weight of an NP scalar and the spin of a field within the context of Chapter~\ref{ch:TLNsDefinition} involve using exactly the same symbol ``$s$''. However, these two are not equivalent and should not be confused. Nevertheless, as we will see shortly, the absolute values of the spin-weights of the particular NP scalars that enter the relevant equations of motion turn out to match the actual spins of the perturbation.} $s$ and a boost-weight $b$. In particular, a NP scalar $\Psi$ transforming as
\be
	\Psi \xrightarrow{\lambda,\chi}\lambda^{b}e^{is\chi}\Psi
\ee
is said to have weights $\left\{b,s\right\}$. These are related to the GHP weights $p$ and $q$~\cite{Geroch:1973am} according to,
\be
	b = \frac{p+q}{2}\,,\quad s=\frac{p-q}{2} \,.
\ee
The weights associated with the scalar field, Maxwell-NP scalars and Weyl scalars can be found in Table~\ref{tbl:NPweights}.

\begin{table}
	\centering
	\begin{tabular}{|c|c|c|}
		\hline
		NP scalar & $\left\{b,s\right\}$ & $\left\{p,q\right\}$ \\
		\hline
		\hline
		$\psi_0$ & $\left\{+2,+2\right\}$ & $\left\{+4,0\right\}$ \\
		\hline
		$\psi_1$ & $\left\{+1,+1\right\}$ & $\left\{+2,0\right\}$ \\
		\hline
		$\psi_2$ & $\left\{0,0\right\}$ & $\left\{0,0\right\}$ \\
		\hline
		$\psi_3$ & $\left\{-1,-1\right\}$ & $\left\{-2,0\right\}$ \\
		\hline
		$\psi_4$ & $\left\{-2,-2\right\}$ & $\left\{-4,0\right\}$ \\
		\hline
		\hline
		$\phi_0$ & $\left\{+1,+1\right\}$ & $\left\{+2,0\right\}$ \\
		\hline
		$\phi_1$ & $\left\{0,0\right\}$ & $\left\{0,0\right\}$ \\
		\hline
		$\phi_2$ & $\left\{-1,-1\right\}$ & $\left\{-2,0\right\}$ \\
		\hline
		\hline
		$\Phi$ & $\left\{0,0\right\}$ & $\left\{0,0\right\}$ \\
		\hline
	\end{tabular}
	\caption[Spin weights, boost weights and GHP weights of the Weyl scalars $\psi_{a}$, the Maxwell-NP scalars $\phi_{a}$ and the scalar field $\Phi$.]{Spin weights, boost weights and GHP weights of the Weyl scalars $\psi_{a}$, the Maxwell-NP scalars $\phi_{a}$ and the scalar field $\Phi$.}
	\label{tbl:NPweights}
\end{table}

In the tetrad formalism, the Christoffel symbols are repackaged into the Ricci rotation coefficients, known as spin coefficients when the tetrad vectors are null as in the NP formalism. In total, there are $12$ spin coefficients. These can be categorized into the $8$ so-called ``good'' spin coefficients,
\be\ba\label{eq:NPSpinCoefficientsGood}
	& \kappa = - m^{\mu}D\ell_{\mu} \,,\quad \tau = - m^{\mu}\triangle\ell_{\mu} \,,\quad \sigma = - m^{\mu}\delta\ell_{\mu} \,,\quad \rho = - m^{\mu}\bar{\delta}\ell_{\mu} \,, \\
	& \pi =  \bar{m}^{\mu}Dn_{\mu} \,,\quad \nu =  \bar{m}^{\mu}\triangle n_{\mu} \,,\quad \mu =  \bar{m}^{\mu}\delta n_{\mu} \,,\quad \lambda = \bar{m}^{\mu}\bar{\delta}n_{\mu} \,,
\ea\ee
which transform covariantly under local rotation and boost transformations, and the $4$ so-called ``bad'' spin coefficients,
\be\label{eq:NPSpinCoefficientsBad}
	\begin{split}
		\varepsilon &= -\frac{1}{2}\left(n^{\mu}D\ell_{\mu}-\bar{m}^{\mu}D m_{\mu}\right) \,,\quad \gamma = -\frac{1}{2}\left(n^{\mu}\triangle\ell_{\mu}-\bar{m}^{\mu}\triangle m_{\mu}\right) \,, \\
		\beta &= -\frac{1}{2}\left(n^{\mu}\delta\ell_{\mu}-\bar{m}^{\mu}\delta m_{\mu}\right) \,,\quad \alpha = -\frac{1}{2}\left(n^{\mu}\bar{\delta}\ell_{\mu}-\bar{m}^{\mu}\bar{\delta}m_{\mu}\right) \,,
	\end{split}
\ee
which do not have definite spin and boost weights. In the above expressions, the NP directional derivatives $D$, $\triangle$, $\delta$ and $\bar{\delta}$ are the covariant derivatives projected onto the null tetrads,
\be
	D \equiv \ell^{\mu}\nabla_{\mu} \,,\quad \triangle \equiv n^{\mu}\nabla_{\mu} \,,\quad \delta \equiv m^{\mu}\nabla_{\mu} \,,\quad \bar{\delta} \equiv \bar{m}^{\mu}\nabla_{\mu} \,,
\ee
whose action on a NP scalar also does not transform homogeneously (see Appendix~\ref{app:LieDerivative}).

The advantage of working with the spin-weighted scalars is that for some of them the relevant dynamical equations fully factorize in the generic Kerr-Newman black hole metric. The corresponding separable master equation was obtained by Teukolsky in Ref.~\cite{Teukolsky:1972my}\footnote{Historically, the Teukolsky equation describes linear perturbations around the electrically neutral Kerr black hole~\cite{Teukolsky:1972my,Teukolsky:1973ha}. The master equation for a more general class of spacetimes, known as type-$D$ spacetimes in the Petrov classification (see~\cite{Petrov:2000bs} for an English translation of the relevant work), which also contains the Kerr-Newman black geometry as a special case, was derived later in~\cite{Dudley:1977zz}, see also~\cite{Kokkotas:1993ef,Kokkotas:1999bd,Berti:2005eb}, and sometimes goes by the name ``the Dudley-Finley master equation''. However, we will adopt the standard terminology of referring to this generalized Teukolsky-like equation as ``the Teukolsky master equation''.}. A spin-$s$ NP scalar $\Psi_{s}$ can be factorized as follows
\be
	\Psi_{s\omega\ell m}\left(t,r,\theta,\phi\right) =  e^{-i\omega t}e^{im\phi}R_{s\omega\ell m}\left(r\right){}_{s}S_{\omega\ell m}\left(\theta\right) \,.
\ee
It satisfies the Teukolsky master equation~\cite{Teukolsky:1972my,Teukolsky:1973ha,Dudley:1977zz} which, in the Kinnersley tetrad~\cite{Kinnersley1969}
\be\ba\label{eq:KinnersleyTetrad}
	\ell &= \frac{r^2+a^2}{\Delta}\left(\partial_{t}+\frac{a}{r^2+a^2}\,\partial_{\phi}\right) + \partial_{r} \,, \\
	n &= \frac{\Delta}{2\Sigma}\left[\frac{r^2+a^2}{\Delta}\left(\partial_{t}+\frac{a}{r^2+a^2}\,\partial_{\phi}\right) - \partial_{r} \right] \,, \\
	m &= \frac{1}{\sqrt{2}\left(r+ia\cos\theta\right)}\left[ia\sin\theta\,\partial_{t} + \partial_{\theta} + \frac{i}{\sin\theta}\,\partial_{\phi}\right] \,,
\ea\ee
takes the following form in the absence of sources
\be\label{eq:teuk}
	\begin{split}
		\mathbb{O}^{\left(s\right)}_{\text{full}}\Psi_{s\omega\ell m} = \ell\left(\ell+1\right)\Psi_{s\omega\ell m} \,,\quad \mathbb{P}^{\left(s\right)}_{\text{full}}\Psi_{s\omega\ell m} = \ell\left(\ell+1\right)\Psi_{s\omega\ell m} \,.
	\end{split}
\ee
The separation constant $\ell\left(\ell+1\right)$ is identified with the angular problem eigenvalue, which is now the spectrum of the spin-weighted spheroidal harmonics. As also discussed in the previous section, the orbital number $\ell$ above \textit{is not an integer} in general.

The explicit forms of the radial and angular differential operators are
\be\label{eq:teukExplicit}
	\begin{split}
		\mathbb{O}^{\left(s\right)}_{\text{full}} &= \Delta^{-s}\partial_{r}\,\Delta^{s+1}\partial_{r} - \frac{a^2}{\Delta}\partial_{\phi}^2  -\frac{\left(r^2+a^2\right)^2}{\Delta}\partial_{t}^2 - 2\frac{a\left(2Mr-Q^2\right)}{\Delta}\partial_{t}\partial_{\phi} \\
		&\quad+ s\left(s+1\right) + s\frac{a\Delta^{\prime}}{\Delta}\partial_{\phi}	+ 2s\left[\frac{M\left(r^2-a^2\right)-Q^2r}{\Delta}-r\right]\partial_{t} \,, \\
		\mathbb{P}^{\left(s\right)}_{\text{full}} &= -\left[ \Delta_{\mathbb{S}^2}^{\left(s\right)} - s\left(s+1\right) + a^2\sin^2\theta\,\partial_{t}^2 -2isa\cos\theta\,\partial_{t}\right] \,,
	\end{split}
\ee
with $\Delta_{\mathbb{S}^2}^{\left(s\right)}$ the spin-weighted Laplace-Beltrami operator on the $2$-sphere,
\be
	\Delta_{\mathbb{S}^2}^{\left(s\right)} = \frac{1}{\sin\theta}\partial_{\theta}\,\sin\theta\,\partial_{\theta}+\frac{\left(\partial_{\phi}+is\cos\theta\right)^2}{\sin^2\theta} + s\,.
\ee
For $s=0$ the spacetime scalar function $\Psi_{s}$ is just the massless scalar field, for $s=+1$ ($s=-1$) this is the transverse ingoing (outgoing) radiation Maxwell-NP scalar and for $s=+2$ ($s=-2$) this is the transverse ingoing (outgoing) radiation Weyl scalar~\cite{Teukolsky:1972my,Teukolsky:1973ha},
\be\label{eq:NPScalarsTeukolsky}
	\begin{gathered}
		\Psi_{0} = \Phi \,, \\
		\Psi_{+1} = \phi_0 \,,\quad \Psi_{-1} = \rho^{-2}\phi_2 \,, \\
		\Psi_{+2} = \psi_0 \,,\quad \Psi_{-2} = \rho^{-4}\psi_4 \,,
	\end{gathered}
\ee
where $\rho=\left(r+ia\cos\theta\right)^{-1}$ is one of the background spin coefficients.

\subsection{Master formula for spin-$s$ near-zone Love numbers}
With the Teukolsky master equation at hand, we can go ahead and compute the spin-$s$ Love numbers of the Kerr-Newman black hole. In principle, one should first reconstruct the gauge field and metric perturbations profile from the spin-weighted NP scalars before matching onto the worldline EFT as prescribed in Section~\ref{sec:RCsGR}. However, as we will see, the solutions for $\Psi_{s}$ at leading order in the near-zone approximation can be found analytically for any $\ell$. Analytically continuing $\ell$ to range in the field of real numbers is then sufficient to do this matching directly at the level of $\Psi_{s}$. This is because the spin-weighted NP scalars are built from derivatives of the gauge field or the metric perturbations and this operation preserves the source/response splitting procedure since the indicial powers of the growing (source) and the decaying (response) modes only shift by integer numbers and remain non-overlapping for real $\ell$. More explicitly, taking into account the $\sim r^{-1}$ scaling of the spin coefficient $\rho$ in the definition \eqref{eq:NPScalarsTeukolsky} of the NP scalars entering the Teukolsky equation, the only thing that changes in the Newtonian matching is that the asymptotic behavior now takes the form
\be
	\Psi_{s\omega\ell m} \xrightarrow{r\rightarrow\infty} \left[1 + k_{\ell m}^{\mathcal{E},\left(s\right)}\left(\omega\right)\left(\frac{r_{s}}{r}\right)^{2\ell+1}\right]r^{\ell-s}\bar{\Psi}_{\ell m}^{\mathcal{E},\left(s\right)}\left(\omega\right) + \left(\mathcal{E}\leftrightarrow\mathcal{B}\right) \,,
\ee
where $\bar{\Psi}_{\ell m}^{\mathcal{E},\left(s\right)}$ ($\bar{\Psi}_{\ell m}^{\mathcal{B},\left(s\right)}$) is proportional to the electric-type (magnetic-type) source multipole moments $\bar{\mathcal{E}}^{\left(\left|s\right|\right)}_{\ell m}$ ($\bar{\mathcal{B}}^{\left(\left|s\right|\right)}_{\ell m}$). Similarly, $k_{\ell m}^{\mathcal{E},\left(s\right)}$ ($k_{\ell m}^{\mathcal{B},\left(s\right)}$) is proportional to the electric-type (magnetic-type) response coefficients $k_{\ell m}^{\mathcal{E},\left(\left|s\right|\right)}$ ($k_{\ell m}^{\mathcal{B},\left(\left|s\right|\right)}$). The electric-type and magnetic-type responses are in fact exactly equal to each other~\cite{LeTiec:2020spy,LeTiec:2020bos,Charalambous:2021mea}. This is the statement of the isospectrality of the two types of response following from the electric-magnetic duality that emerges in four spacetime dimensions~\cite{Chandrasekhar:1985kt}, see also~\cite{Porto:2007qi,Glampedakis:2017rar,Hui:2020xxx}. It is therefore sufficient to consider the matching condition
\be\label{eq:KerrNewmanS_EFTMatching}
	\Psi_{s\omega\ell m} \xrightarrow{r\rightarrow\infty} \left[1 + k_{\ell m}^{\left(s\right)}\left(\omega\right)\left(\frac{r_{s}}{r}\right)^{2\ell+1}\right]r^{\ell-s}\bar{\Psi}_{\ell m}^{\left(s\right)}\left(\omega\right) \,.
\ee
To be more accurate, the response coefficients $k_{\ell m}^{\left(s\right)}$ above are equal to the actual response coefficients $k^{\Phi^{\left(\left|s\right|\right)}}_{\ell m}$ associated with the gauge field or metric perturbations up to an overall factor arising from the derivatives involved in the definition of the NP scalars,
\be
	k_{\ell m}^{\left(s\right)}\left(\omega\right) = \left(-1\right)^{s}\frac{\left(\ell-s\right)!\left(\ell+s\right)!}{\left(\ell!\right)^2} k_{\ell m}^{\Phi^{\left(\left|s\right|\right)}}\left(\omega\right) \,.
\ee

To complete our analysis on the boundary conditions, we need to see how the ingoing boundary condition at the horizon is imposed onto the NP scalars $\Psi_{s}$. Again, due to the fact that the spin-weighted NP scalars $\Psi_{s}$ are constructed from derivatives of the gauge field or metric perturbations, the only modification compared to the scalar field example we saw will come from near-horizon contributions of the tetrad vector fields. The Kinnersley tetrad \eqref{eq:KinnersleyTetrad} is itself singular at the event horizon but one can transit to a regular one by performing a local boost with the boost factor having the near-horizon behavior $\lambda \sim \Delta$. The ingoing boundary condition at the horizon for the boosted NP scalars, with near-horizon behavior $\sim \Delta^{s}\Psi_{s}$, will then look like that for a scalar field. Consequently, the ingoing wave condition at the future ($+$)/past ($-$) event horizon at the level of the radial wavefunction in the current Kinnersley tetrad will have the form~\cite{Teukolsky:1972my}
\be
	R_{s\omega\ell m} \sim \left(r-r_{+}\right)^{\pm iZ_{+}\left(\omega\right)-s}\left(r-r_{-}\right)^{\pm iZ_{-}\left(\omega\right)-s} \quad\text{as $r \rightarrow r_{+}$} \,,
\ee
with the scalar field near-horizon characteristic exponents $Z_{\pm}\left(\omega\right)$ given in \eqref{eq:Zpm_KerrNewman}.

We now proceed to apply the near-zone approximation to the Teukolsky radial and angular operators. As before, we introduce a formal near-zone order parameter $\epsilon$. We choose the following near-zone splitting~\cite{Charalambous:2021kcz,Charalambous:2022rre}
\be\ba\label{eq:NZSplitOP_KerrNewmanS}
	\mathbb{O}^{\left(s\right)}_{\text{full}} &= \Delta^{-s}\partial_{r}\,\Delta^{s+1}\partial_{r} + V_0 +\epsilon V_1 + s\left(s+1\right) \,, \\
	\mathbb{P}^{\left(s\right)}_{\text{full}} &= -\left[ \Delta_{\mathbb{S}^2}^{\left(s\right)} - s\left(s+1\right) + \epsilon\left(a^2\sin^2\theta\,\partial_{t}^2 -2isa\cos\theta\,\partial_{t}\right) \right] \,,
\ea\ee
with
\be\ba\label{eq:NZSplitV0V1_KerrNewmanS}
	V_0 = &-\frac{\left(r_{+}^2+a^2\right)^2}{\Delta}\left[(\partial_t +\Omega\,\partial_\phi)^2 + 4\Omega\frac{r-r_{+}}{r_{+}-r_{-}}\partial_t \partial_\phi - 2s\,\beta^{-1}\partial_{t}\right] +s\frac{a\Delta^{\prime}}{\Delta}\partial_\phi \,,\\
	V_1 = &-\frac{\left(r+r_{+}\right)\left(r^2 + r_{+}^2 +2a^2\right)}{r-r_{-}}\partial_t^2 + 2\frac{r_{+}^2+a^2}{r-r_{-}}\left(\beta-2M\right)\Omega\,\partial_t\partial_\phi \\
	&+ 2s\left[\frac{M\left(r+r_{+}\right)-Q^2}{r-r_{-}}-r\right]\partial_t \,.
\ea\ee
As far as the angular problem is concerned, the leading near-zone approximation is chosen to coincide with the static approximation. The leading angular eigenfunctions are then given by the standard spin-weighted spherical harmonics (see Ref.~\cite{Charalambous:2021kcz} for our conventions), 
\be
	{}_{s}S_{\omega=0,\ell m}\left(\theta\right) = e^{-im\phi} {}_{s}Y_{\ell m}\left(\theta,\phi\right) \,.
\ee
where the factor $e^{-im\phi}$ is introduced to strip off the $\phi$-dependence from ${}_{s}Y_{\ell m}\left(\theta,\phi\right)$. The orbital number $\ell\ge\left|s\right|$ is always an integer in the leading near-zone approximation.

We can then go ahead and solve the leading order near-zone radial equation of motion. By working with the same independent variable $x=\frac{r-r_{+}}{r_{+}-r_{-}}$ as before, the leading order radial wavefunction that satisfies ingoing boundary conditions at the future event horizon is given by
\be\ba\label{eq:RNZSol_KerrNewmanS}
	R_{\omega\ell m} &= \bar{R}_{\ell m}^{\text{in}}\left(\omega\right)x^{-s}\left(\frac{x}{1+x}\right)^{iZ_{+}\left(\omega\right)} \\
	&\times\left(1+x\right)^{im\beta\Omega}{}_2F_1\left(\ell+1+im\beta\Omega,-\ell+im\beta\Omega;1-s+2iZ_{+}\left(\omega\right);-x\right) \,.
\ea\ee
In this expression, the information that $\ell$ is a whole number has not been used at all. Therefore, we can analytically continuing $\ell\in\mathbb{N}\rightarrow\ell\in\mathbb{R}$ to perform the matching \eqref{eq:KerrNewmanS_EFTMatching}. More explicitly, expanding around large $x$, we find that the leading order spin-$s$ response coefficients for the Kerr-Newman black hole are
\be\label{eq:ResponseCoefficients_KerrNewmanS}
	k_{\ell m}^{\left(s\right)}\left(\omega\right) = \frac{\Gamma\left(-2\ell-1\right)\Gamma\left(\ell+1+im\beta\Omega\right)\Gamma\left(\ell+1-s-i\beta\omega\right)}{\Gamma\left(2\ell+1\right)\Gamma\left(-\ell+im\beta\Omega\right)\Gamma\left(-\ell-s-i\beta\omega\right)}\left(\frac{r_{+}-r_{-}}{r_{s}}\right)^{2\ell+1} \,.
\ee
Decomposing this into its real and imaginary pieces as prescribed in Section~\ref{sec:LTs_LNs}, we, hence, find the following dissipative responses and Love numbers
\be\ba\label{eq:LNs_KerrNewmanS}
	k_{\ell m}^{\left(s\right)\text{diss}}\left(\omega\right) &= B_{\ell}^{\left(s\right)\text{Re}}\left(\omega\right)k_{\ell m}^{\left(0\right)\text{diss}}\left(\omega\right) + B_{\ell}^{\left(s\right)\text{Im}}\left(\omega\right)k_{\ell m}^{\left(0\right)\text{Love}}\left(\omega\right) \,, \\
	\quad k_{\ell m}^{\left(s\right)\text{Love}}\left(\omega\right) &= B_{\ell}^{\left(s\right)\text{Re}}\left(\omega\right)k_{\ell m}^{\left(0\right)\text{Love}}\left(\omega\right) - B_{\ell}^{\left(s\right)\text{Im}}\left(\omega\right)k_{\ell m}^{\left(0\right)\text{diss}}\left(\omega\right) \,,
\ea\ee
with $k_{\ell m}^{\left(0\right)\text{Love}}\left(\omega\right)$ and $k_{\ell m}^{\left(0\right)\text{diss}}\left(\omega\right)$ the scalar Love numbers and scalar dissipative response coefficients respectively, given in \eqref{eq:SLNsDissipativeRCs_KerrNewman}, and $B_{\ell}^{\left(s\right)\text{Re}}\left(\omega\right)$ and $B_{\ell}^{\left(s\right)\text{Im}}\left(\omega\right)$ the real and imaginary parts of
\be
	B_{\ell}^{\left(s\right)\text{Re}}\left(\omega\right)+iB_{\ell}^{\left(s\right)\text{Im}}\left(\omega\right) = \left(-1\right)^{s}\prod_{n=1}^{\left|s\right|}\frac{\ell+n+\text{sign}\left\{s\right\}i\beta\omega}{\ell+1-n-\text{sign}\left\{s\right\}i\beta\omega} \,.
\ee

It is important to remember though that \eqref{eq:ResponseCoefficients_KerrNewmanS} is \textit{exact} only for static perturbations ($\omega=0$). For physical orbital numbers $\ell\in\mathbb{N}$, in particular, we find a purely dissipative response~\cite{Chia:2020yla,Charalambous:2021mea}
\be\label{eq:ResponseCoefficientsStatic_KerrNewmanS}
	k_{\ell m}^{\left(s\right)}\left(\omega=0\right) = \frac{i m\beta\Omega}{2}\left(\frac{r_{+}-r_{-}}{r_{s}}\right)^{2\ell+1}\left(-1\right)^{s+1}\frac{\left(\ell-s\right)!\left(\ell+s\right)!}{\left(2\ell\right)!\left(2\ell+1\right)!}\prod_{n=1}^{\ell}\left(n^2+\left(m\beta\Omega\right)^2\right) \,,
\ee
Therefore, the spin-$s$ static Love numbers for the Kerr-Newman black hole vanish identically,
\be
	k_{\ell m}^{\left(s\right)\text{Love}}\left(\omega=0\right) = 0 \,.
\ee

\subsection{NLO Love numbers}
We have repeatedly stressed out that the frequency-dependent part of the response coefficients at leading order in the near-zone expansion is \textit{approximate} and does not capture the full black hole response at non-zero $\omega$. For completeness, we will compute here the accurate linear order in $\omega a$ correction to the response coefficients for the Kerr ($Q=0$) black hole. The first thing to note in the full Teukolsky equation in Eqs.~\eqref{eq:teuk}-\eqref{eq:teukExplicit} is that the angular eigenvalues $\ell\left(\ell+1\right)$ are \text{not} integers. In fact, one can use standard time-independent perturbation theory to show that~\cite{Teukolsky:1972my,Teukolsky:1973ha}
\be
	\text{$\ell\left(\ell+1\right)$ in Eq.~\eqref{eq:teuk}} \rightarrow \ell\left(\ell+1\right) -\omega a\frac{2ms^2}{\ell\left(\ell+1\right)} + \mathcal{O}\left(\left(\omega a\right)^2\right) \,,
\ee
where $\ell\ge\left|s\right|$ is now an integer and $\left|m\right|\le\ell$.

To compute the Next-to-Leading-Order (NLO) Love numbers as an expansion in frequency, we need to use an accurate solution of the Teukolsky equation in the form of a series over hypergeometric functions, obtained in Refs.~\cite{Mano:1996vt,Mano:1996mf,Mano:1996gn,Sasaki:2003xr}. Taking the solution that is regular at the horizon from these references and focusing on the part which reproduces the multipole expansion in the low-frequency limit, we find~\cite{Charalambous:2021mea}
\be\ba
	{}&R_{s\omega\ell m} \supset \bar{R}_{s\ell m}\left(\omega\right)e^{-i\frac{\beta\omega r_{s}}{2r_{+}}x} x^{-s}\left(\frac{x}{1+x}\right)^{-iZ_{-}\left(\omega\right)} \\
	&\times\bigg[x^{\nu}{}_2F_1\left(-\nu+2iZ_{-}\left(\omega\right)-i\omega r_{s},-\nu+s+i\omega r_{s};-2\nu;-x^{-1}\right) \\
	&+\varkappa_{\nu m}^{\left(s\right)}\left(\omega\right) x^{-\nu-1}{}_2F_1\left(\nu+1+2iZ_{-}\left(\omega\right)-i\omega r_{s},\nu+1+s+i\omega r_{s};2\nu+1;-x^{-1}\right)\bigg] \,,
\ea\ee
where $\bar{R}_{s\ell m}\left(\omega\right)$ is an integration constant proportional to the source multipole moments, the coefficients in front of the decaying mode is given by
\be
	\varkappa_{\nu m}^{\left(s\right)}\left(\omega\right) = \frac{\Gamma\left(-2\nu-1\right)\Gamma\left(\nu+1-s-i\omega r_{s}\right)\Gamma\left(\nu+1+2iZ_{-}\left(\omega\right) - i\omega r_{s}\right)}{\Gamma\left(2\nu+1\right)\Gamma\left(-\nu-s-i\omega r_{s}\right)\Gamma\left(-\nu+2iZ_{-}\left(\omega\right)-i\omega r_{s}\right)}
\ee
and the ``renormalized orbital number'' $\nu$ is a series expansion in powers of the frequency around the integer-valued orbital number $\ell$~\cite{Mano:1996vt,Mano:1996mf,Mano:1996gn,Sasaki:2003xr},
\be\ba
	\nu &= \ell + \Delta\ell \,, \\
	\Delta\ell &= \frac{\omega^2r_{s}^2}{2\ell+1}\left[-2-\frac{s^2}{\ell\left(\ell+1\right)}+\frac{\left(\left(\ell+1\right)^2-s^2\right)^2}{\left(2\ell+1\right)\left(2\ell+2\right)\left(2\ell+3\right)} - \frac{\left(\ell^2-s^2\right)^2}{\left(2\ell-1\right)2\ell\left(2\ell+1\right)}\right] \\
	&\quad + \mathcal{O}\left(\left(\omega r_{s}\right)^3\right) \,.
\ea\ee
Expanding around small frequencies, we encounter a pole in the coefficients $\varkappa_{\nu m}^{\left(s\right)}\left(\omega\right)$,
\be
	\varkappa_{\nu m}^{\left(s\right)}\left(\omega\right) = -\frac{1}{2\Delta\ell}\left(\omega r_{s}\right)\left(m\beta\Omega\right)\left(-1\right)^{s+1}\frac{\left(\ell-s\right)!\left(\ell+s\right)!}{\left(2\ell\right)!\left(2\ell+1\right)!} \prod_{n=1}^{\ell}\left(n^2 + \left(m\beta\Omega\right)^2\right) + \mathcal{O}\left(\left(\omega r_{s}\right)^0\right) \,.
\ee
When we match the worldline EFT result to the above microscopic calculation, we regularize such singularities by introducing counterterms in the EFT. In the end, we use only finite parts that depend on the employed regularization scheme, and hence this singular contribution can be ignored. However, it is important to note that this term also generates a finite logarithmic contribution coming from
\be
	x^{-2\nu-1} = x^{-2\ell-1}\left[1-2\Delta\ell\,\ln x + \mathcal{O}\left(\left(\omega r_{s}\right)^3\right)\right] \,.
\ee
After taking this into account, as well as an overall $\left(\frac{r_{+}-r_{-}}{r_{s}}\right)^{2\ell+1}$ factor entering the matching onto the worldline EFT by identifying the size of the black hole with its gravitational radius, we find that the NLO spin-$s$ Love numbers are corrected by
\be\label{eq:ksf}
	\begin{split}
		\delta  k^{\left(s\right)\text{Love}}_{\ell m} = & \left(\omega r_{s}\right)\left(m\beta\Omega\right) \ln\frac{r-r_+}{r_{+}-r_{-}} \\
		&\times \left(\frac{r_{+}-r_{-}}{r_s}\right)^{2\ell+1}\left(-1\right)^{s+1}\frac{\left(\ell-s\right)!\left(\ell+s\right)!}{\left(2\ell\right)!\left(2\ell+1\right)!} \prod_{n=1}^{\ell}\left(n^2 + \left(m\beta\Omega\right)^2\right) \,,
	\end{split} 
\ee
where we retained only the scheme-independent logarithmic part. This result indicates that non-static (``dynamical'') 
black hole Love numbers are in general non-zero and exhibit logarithmic running, which, as we will discuss later in this chapter, is in agreement with expectations of Wilsonian naturalness.

\subsection{Love numbers of extremal black holes}
The near-zone expansions employed so far do not apply for extremal black holes with $a^2+Q^2=M^2$, for which $r_{+}=r_{-}=M$. This happens because the near-horizon dynamics change at extremality and the near-zone split used above is no longer capable of capturing the full near-horizon behavior. More explicitly, the correction to the potential $V_1$, given in \eqref{eq:NZSplitV0V1_KerrNewmanS} and neglected in the leading order near-zone approximation, develops a singularity at the horizon in the extremal limit. Nevertheless, there exists a consistent near-zone truncation of the Teukolsky equation even in the extremal limit as we will demonstrate here.

The full radial Teukolsky equation for extremal Kerr-Newman black holes reads
\be\ba\label{eq:TeukolskyExtremalFull}
	{}&\bigg[ \left(r-M\right)^{-2s}\partial_{r}\,\left(r-M\right)^{2\left(s+1\right)}\partial_{r} + \frac{\left(M^2+a^2\right)^2}{\left(r-M\right)^2}\left(m\Omega-\omega\right)^2 - \frac{4Ma}{r-M} m\omega \\
	&\quad + \frac{r+M}{r-M}\left(r^2+M^2+2a^2\right)\omega^2 + 2is\frac{M^2+a^2}{r-M}\left(m\Omega-\omega\right) \\
	&\quad + 2is\left(r-M\right)\omega + s\left(s+1\right) \bigg] R_{s\omega\ell m} = \ell\left(\ell+1\right)R_{s\omega\ell m} \,.
\ea\ee
The ingoing boundary condition at the future event horizon now becomes
\be
	R_{s\omega\ell m} \sim \left(r-M\right)^{-2s+2iM\left(m\Omega-\omega\right)}e^{-\frac{M^2+a^2}{r-M}\left(m\Omega-\omega\right)} \quad \text{as $r\rightarrow M$} \,.
\ee
Introducing the dimensionless parameter
\be
	\tilde{\alpha} = \frac{a}{M}\left(1-\frac{\omega}{m\Omega}\right)
\ee
and dimensionless independent variable
\be
	x = \frac{r-M}{M} \,,
\ee
the full radial equation of motion can be rewritten as
\be\ba
	{}&\bigg[ x^{-2s}\partial_{x}\,x^{2\left(s+1\right)}\partial_{x} + \frac{m^2\tilde{\alpha}^2}{x^2} -2\left(2\omega M-is\right)\frac{m\tilde{\alpha}}{x} + 2is \omega Mx \\
	&\quad + 2\omega^2 M^2\left(\frac{x^2}{2}+2x+3+\frac{a^2}{M^2}\right) + s\left(s+1\right) - \ell\left(\ell+1\right)\bigg] R_{s\omega\ell m} = 0 \,.
\ea\ee
In the near-zone regime, $2\omega M\ll1$, $\omega\left(r-M\right)\ll1$, a consistent near-zone truncation is the following~\cite{Charalambous:2022rre}
\be
	\bigg[ x^{-2s}\partial_{x}\,x^{2\left(s+1\right)}\partial_{x} + \frac{m^2\tilde{\alpha}^2}{x^2} -2\left(2\omega M-is\right)\frac{m\tilde{\alpha}}{x} + s\left(s+1\right) - \ell\left(\ell+1\right) \bigg] R_{s\omega\ell m} = 0 \,,
\ee
which is solved in terms of confluent hypergeometric functions. The solution regular at the future event horizon, in particular, is found to be~\cite{Charalambous:2022rre}
\be
	R_{s\omega\ell m} = \bar{R}_{s\ell m}^{\text{in}}\left(\omega\right)e^{-im\tilde{\alpha}/x}\left(\frac{2im\tilde{\alpha}}{x}\right)^{\ell+1+s} \Psi\left(\ell+1-s-2i\omega M;2\ell+2;\frac{2im\tilde{\alpha}}{x} \right) \,,
\ee
where $\Psi\left(a;b;z\right)$ is the confluent hypergeometric function of the second kind~\cite{Bateman:100233}. To find the Love numbers we expand $\Psi\left(a,n+1,y\right)$ at spatial infinity $x\rightarrow\infty$ and find logarithms in the solution profile,
\be 
	\begin{split}
		& \Psi\left(\ell + 1-s-2i\omega M,2\ell+2;\frac{2im\tilde{\alpha}}{x} \right) \xrightarrow{x\rightarrow\infty} \\
		&\frac{\left(2\ell\right)!}{\Gamma\left(\ell+1-s-2i\omega M\right)}\left(\frac{2im\tilde{\alpha}}{x}\right)^{-2\ell-1} +\frac{\log\left(\frac{2im\tilde{\alpha}}{x}\right)}{\left(2\ell+1\right)!\Gamma\left(-\ell-s-2i\omega M\right)} \,,
	\end{split}
\ee
which gives the following expression for the response coefficients
\be 
	\begin{split}
		k^{\left(s\right)}_{\ell m}\left(\omega\right) &= \left(\log\left(\frac{2m\tilde{\alpha}}{x}\right)+i\frac{\pi}{2}\right) \frac{\left(im\tilde{\alpha}\right)^{2\ell+1}\Gamma\left(\ell+s-s-2i\omega M\right)}{\left(2\ell\right)!\left(2\ell+1\right)!\Gamma\left(-\ell-s-2i\omega M\right)} \\
		&= \left(2\omega M\right)\left(m\tilde{\alpha}\right)^{2\ell+1} \left(\log\frac{2m\tilde{\alpha}M}{r-M}+i\frac{\pi}{2}\right) \left(-1\right)^{s}\frac{\left(\ell-s\right)!\left(\ell+s\right)!}{\left(2\ell\right)!\left(2\ell+1\right)!} \,,
	\end{split}
\ee
where in the last line we kept terms linear in $\omega M$. We see that frequency-dependent response coefficients are not zero and run with the distance. However, the exact static Love numbers do vanish,
\be 
	k^{\left(s\right)\text{Love}}_{\ell m}\left(\omega=0\right) = 0\,,\quad a^2+Q^2=M^2 \,.
\ee

\section{Black hole Love numbers beyond General Relativity}
\label{sec:SLNs_R34d}

A simple important example of non-vanishing and running static Love numbers is given by the black hole scalar response in the Riemann-cubed gravity in four dimensions~\cite{Charalambous:2022rre}.

General effective field theory arguments suggest the presence of higher-order curvature corrections in the gravity action~\cite{Donoghue:2017pgk}. In the pure gravity case the Riemann-squared corrections to the Einstein-Hilbert action can be eliminated by means of the leading (general-relativistic) equations of motion in the vacuum and the Gauss-Bonnet identity. Hence the first non-trivial contribution appears at the cubic order in curvature. In perturbation theory this contribution becomes important at the two-loop order~\cite{Goroff:1985th}. The action of the Riemann-cubed gravity is given by,
\be\label{eq:R3act}
	S_{\text{gr}}\left[g\right] = -\frac{1}{16\pi G}\int d^{4}x\,\sqrt{-g}\,\left(R + \alpha\,R^{\mu\nu}_{\quad\rho\sigma}R^{\rho\sigma}_{\quad\kappa\lambda}R^{\kappa\lambda}_{\quad\mu\nu}\right) \,.
\ee
This modified gravitational action leads to the vacuum field equations
\be
	G_{\mu\nu} + \alpha K_{\mu\nu}=0 \,,
\ee
where $G_{\mu\nu}=R_{\mu\nu}-\frac{1}{2}Rg_{\mu\nu}$ is the Einstein tensor and
\be
	K_{\mu\nu} \equiv -\frac{1}{2}g_{\mu\nu}R^{\alpha\beta}_{\quad\rho\sigma}R^{\rho\sigma}_{\quad\kappa\lambda}R^{\kappa\lambda}_{\quad\alpha\beta} + 3R_{\mu\alpha\kappa\lambda}R^{\kappa\lambda\rho\sigma}R_{\nu\,\,\,\,\rho\sigma}^{\,\,\,\,\alpha} - 6\nabla_{\sigma}\nabla_{\rho}\left(R_{(\mu|}^{\,\,\,\,\,\,\,\,\sigma\kappa\lambda}R_{\kappa\lambda\,\,\,\,|\nu)}^{\,\,\,\,\,\,\,\rho}\right)\,.
\ee

The first step in calculating the black hole Love numbers is to find the background geometry. We focus on the modified Schwarzschild solution describing an asymptotically flat and spherically symmetric black hole in the vacuum of this theory of gravity. As such, the general background metric can be written as
\be
	ds^2 = -f_{t}\left(r\right)dt^2 + \frac{dr^2}{f_{r}\left(r\right)} + r^2 d\Omega_{2}^2 \,.
\ee
We perturbatively expand the problem around the Schwarzschild results at $\alpha=0$. To keep track of the boundary conditions at the horizon, it is convenient to introduce the dimensionless variable\footnote{This is not to be confused with the dimensionless variable $x=\frac{r-r_{+}}{r_{+}-r_{-}}$ used in the previous sections.}
\be
	x = \frac{r_{\text{h}}}{r} \,,
\ee
where $r=r_{\text{h}}$ is the radial position of the event horizon. We keep $r_{\text{h}}$ fixed at all values of $\alpha$. In this parameterization the ADM mass of the resulting solutions is a non-trivial function of $\alpha$. Up to linear order in $\alpha$, we then find~\cite{Charalambous:2022rre}
\be\ba
	f_{t} &= \left(1-x\right)\left(1 + \frac{\alpha}{r_{\text{h}}^4}\left(-5x\frac{1-x^5}{1-x} - 5x^6\right) + \mathcal{O}\left(\alpha^2\right) \right) \,, \\
	f_{r} &= \left(1-x\right)\left(1 + \frac{\alpha}{r_{\text{h}}^4}\left(-5x\frac{1-x^5}{1-x} + 49x^6\right) + \mathcal{O}\left(\alpha^2\right) \right) \,.
\ea\ee

Let us calculate now the black hole static response at leading order in $\alpha$. For simplicity, we restrict to the scalar field response, so that we need to compute the static scalar field profile by solving the Klein-Gordon equation
\be\label{eq:R3KG}
	\left[f_{r} \partial_{x}^2 + \frac{\partial_{x}\left(f_{t}f_{r}\right)}{2f_{t}}\partial_{x}\right]\Phi_{\ell m} = \frac{\ell\left(\ell+1\right)}{x^2}\Phi_{\ell m} \,,
\ee
where an expansion into spherical harmonics has been used. We look for a perturbative solution of \eqref{eq:R3KG} as a series in $\alpha$,
\be
	\Phi_{\ell m}\left(x\right) = r_{\text{h}}^{\ell}\bar{\mathcal{E}}_{\ell m}^{\left(0\right)}\left[ \Phi_{\ell}^{\left(0\right)}\left(x\right) + \frac{\alpha}{r_{\text{h}}^4}\Phi_{\ell}^{\left(1\right)}\left(x\right) + \mathcal{O}\left(\alpha^2\right) \right] \,,
\ee
where at zeroth-order in $\alpha$ one gets the conventional static general-relativistic solution
\be
	\Phi_{\ell}^{\left(0\right)}\left(x\right) = \frac{\left(\ell!\right)^2}{\left(2\ell\right)!}P_{\ell}\left(\frac{2-x}{x}\right) \,,
\ee
corresponding to vanishing static Love numbers. Higher-order terms $\Phi_{\ell}^{\left(n\right)}$ are all regular at the horizon at $x=1$ and grow at infinity slower than the leading order solution,
\be
	\lim_{x\rightarrow0}x^{\ell}\Phi_{\ell}^{\left(n\right)}\left(x\right) = 0 \,\quad\text{for $n>0$} \,.
\ee

The leading corrections $\Phi_{\ell}^{\left(1\right)}$ are then extracted by solving the resulting linear inhomogeneous ordinary differential equation. For $\ell=1,2$, we find
\be\ba
	\Phi_{1}^{\left(1\right)} &={7\over 2}+{2\over 5}x^2+{2\over 5}x^3+{9\over 25}x^4+{83\over 25} x^5 \,, \\
	\Phi_{2}^{\left(1\right)} &= -{5\over x}+{5\over3}+{236\over 75}x^3+{118\over 25}x^4-{83\over 25} x^5 \,.
\ea\ee
We see that these corrections are no longer polynomial in $r$ and, in particular, terms proportional to $x^{\ell+1}\propto r^{-\ell-1}$ are arising, but there is still no sign of logarithmic running. This changes for higher multipole numbers, $\ell\ge3$. For example, for $\ell=3$, we find
\begin{gather}
	\Phi_{3}^{\left(1\right)}={230385\over 2x^2}-{143994\over x}+{154877\over  4}-480  x+48x^2-5x^3-{1851x^4\over 250}+{249x^5\over  125} \\
	-{1920\left( 11x^2-60x+60\right)\log  x\over x^2}+{5760(x-2)(x^2-10x+10)\left(\log x\log(1-x)+\text{Li}_2\left(x\right)\right)\over x^3} \,,
	\nonumber
\end{gather}
where $\text{Li}_{2}\left(x\right)$ is the dilogarithm function. At face value, the coefficients in front of the $x^{\ell+1}$ terms that correspond to the Love numbers read, for $\ell=1,2,3,4$,
\be\label{eq:varkappas}
	\begin{gathered}
		\varkappa_1 = \frac{2}{5} \frac{\alpha}{r_{\text{h}}^4} \,\,\,,\,\,\, \varkappa_2 = \frac{236}{75} \frac{\alpha}{r_{\text{h}}^4} \,, \\
		\varkappa_3 = \frac{\alpha}{r_{\text{h}}^4}\left(\frac{450501}{12250} + \frac{288}{7}\log x\right)\,\,\,,\,\,\,\varkappa_4 = \frac{\alpha}{r_{\text{h}}^4}\left(\frac{1540202}{11025} + \frac{736}{7} \log x\right) \,,
	\end{gathered}
\ee
where we denoted them as $\varkappa_{\ell}$ in order to emphasize that it would be premature to identify these with the Love numbers at this stage.

Indeed, as we discussed in Section~\ref{sec:RCsGR}, in order to rigorously identify the Love numbers through the matching procedure, we need to compute the graviton corrections to the source term and subtract them from the full GR solution. An alternative to this procedure is to perform an analytic continuation $\ell\in\mathbb{N}\rightarrow\ell\in\mathbb{R}$, which removes an overlap between the source and the response. Unfortunately, the complexity of the equations of motion in Riemann-cubed gravity does not allow us to construct a solution for generic $\ell$. Therefore, a more accurate calculation involving a systematic expansion of the bulk action into interaction vertices and calculations of the corresponding loop corrections is required. However, as we will see in the next section, for many purposes, it is possible to bypass a complete calculation by making use of the straightforward dimensional analysis alone.

In particular, the situation simplifies when the logs appear. Logarithmic corrections are present both in the IR and UV theories. In the IR, the logarithms are associated with divergences in the EFT loop integrals~\cite{Kol:2011vg}. The finite logarithmic part is associated with the RG flow of the physical Love number. This is exactly the situation that we have in Riemann-cubed gravity with $\ell\geq3$, and we conclude that the $\ell\geq3$ Love numbers exhibit log running there, with $\varkappa_{\ell}$ being the corresponding leading log coefficients.

\section{A nod to the naturalness dogma}
\label{sec:PowerCounting}

Let us summarize the results of this section. We have found that the spin-$s$ static Love numbers for Kerr-Newman black holes are exactly zero, even though the response coefficients are in fact non-zero but assigned to purely dissipative (frame-dragging) effects~\cite{LeTiec:2020spy,LeTiec:2020bos,Chia:2020yla,Charalambous:2021mea}. At leading order in the near-zone expansion \eqref{eq:NZSplitOP_KerrNewmanS}-\eqref{eq:NZSplitV0V1_KerrNewmanS}, the $\omega$-dependent Love numbers acquire a richer structure; for spinning black holes and for non-axisymmetric perturbations, they exhibit a classical RG flow while, for non-rotating black holes (including the spherically symmetric and electrically charged Reissner-Nordstr\"{o}m black hole), they turn out to vanish for all frequencies for $s=0$. If we take the expressions for the leading order spin-$s$ Love numbers \eqref{eq:LNs_KerrNewmanS} and analytically continue $\omega\in\mathbb{R}\rightarrow\omega\in\mathbb{C}$, then one can extract more resonant conditions at which these Love numbers vanish. A similar approach can be applied by analytically continuing the azimuthal number $m$ to range in the field of complex numbers, although the resulting radial wavefunctions will develop conical singularities following from the periodic identification of the azimuthal angle. A summary of this analysis can be found in Table~\ref{tbl:NZLNs_KerrNewmanS} for the case of the scalar ($s=0$) response.

We furthermore observe that, in addition to vanishing of the static Love numbers, the black hole response exhibits an additional surprising feature. Namely, at non-integer $\ell$ the static radial wavefunction can be rewritten as
\be\ba
	R_{s,\omega=0,\ell m} &= \bar{R}_{s\ell mj}^{\text{in}}\left(\omega=0\right)x^{-s}\left(\frac{x}{1+x}\right)^{im\beta\Omega/2} \\
	&\quad\times\left(1+x\right)^{-s}{}_2F_1\left(\ell+1-s,-\ell-s;1+im\beta\Omega;-x\right) \,.
\ea\ee
Expanding at large distances, the hypergeometric function above splits into a sum of non-overlapping source and response contributions of the form
\[
	{}_2F_1\sim r^\ell\left(1+\dots\right)+k r^{-\ell-1}\left(1+\dots\right) \,,
\]
where dots stand for two infinite series of power law corrections $\sum  a_n/r^n$ with positive integer $n$'s. 
At the (physical) integer values of $\ell$ the two series overlap and conspire to produce a polynomial answer for the full static solution, even though both source and response are still given by non-trivial infinite series as reflected by the presence of non-trivial static response coefficients \eqref{eq:ResponseCoefficientsStatic_KerrNewmanS}. Hence, the actual puzzle to explain is the (quasi-)polynomial form of the spin-$s$ radial wavefunction. We will see in the next chapter that it arises as a consequence of the highest-weight property of the corresponding representation of an $\SL$ symmetry emerging in the near-zone regime.

On the other hand, once we depart from General Relativity, like in the Riemann-cubed gravity paradigm we saw, the generic behavior for the static Love numbers is governed by a classical RG flow. This is true at least for multipolar order $\ell\ge3$. For $\ell=1,2$ instead we have found no running. As we will learn right away, the absence of running for $\ell=1,2$ is actually expected and, in fact, follows from the vanishing of the static Love numbers in General Relativity.

\subsection{Power counting argument in worldline EFT}
Let us see now that the presence of this classical RG running for $\ell\geq 3$ scalar Love numbers in Riemann-cubed gravity 
can be understood based on a power counting argument. In fact, the argument applies for a general theory of gravity with the gravitational action being of the following schematic form 
\be
	S_{\text{gr}} \sim \frac{1}{16\pi G}\int d^{d}x\,\sqrt{-g}\,\sum_{k=1}^{\infty}\alpha_{k}\left(R_{\mu\nu\rho\sigma}\right)^{k} \,,
\ee
where the coupling constants $\alpha_{k}$ have mass dimensions $2\left(1-k\right)$. As we discussed, the source/response ambiguity and associated running of Love numbers occurs when gravitational non-linearities caused by the source contribute to the total field at the same order, $r^{-\ell-1}$, as the response. A typical EFT diagram representing these non-linearities has the following schematic form
\be\label{eq:1pointSource}
\vcenter{\hbox{\begin{tikzpicture}
			\begin{feynman}
				\vertex[dot] (a0);
				\vertex[below=1.5cm of a0] (p1);
				\vertex[above=1.5cm of a0] (p2);
				\vertex[above=1cm of a0] (a1);
				\vertex[above=0.5cm of a0] (a2);
				\vertex[below=1cm of a0] (a3);
				\vertex[left=0.2cm of a1] (a11);
				\vertex[left=0.2cm of a3] (a31);
				\vertex[right=0.1cm of a0] (a10){$\vdots$};
				\vertex[right=0.7cm of a0, blob] (b0){};
				\vertex[right=0.425cm of a0] (a00);
				\vertex[right=1cm of a00, dot] (b00){};
				\vertex[below right=2cm of a00] (b1);
				\vertex[above right=2cm of a00] (b2);
				\vertex[above right=1.55cm of a00] (b22){$\times$};
				\diagram*{
					(p1) -- [double,double distance=0.5ex] (p2),
					(a1) -- [photon] (b0),
					(a2) -- [photon] (b0),
					(a3) -- [photon] (b0),
					(b1) -- [plain] (b00) -- (b2),
				};
				\draw [decoration={brace}, decorate] (a31.south west) -- (a11.north west)
				node [pos=0.5, left] {\(N\)};
			\end{feynman}
\end{tikzpicture}}}
\sim \bar{\mathcal{E}}_{\ell m}r^{\ell}\left(\frac{GM}{r}\right)^{N}\times
\vcenter{\hbox{\begin{tikzpicture}
			\begin{feynman}
				\vertex[blob] (a0){};
				\vertex[above left=0.37cm and 0.147986cm of a0] (a1){$\times$};
				\vertex[above left=0.24cm and 0.32cm of a0] (a2){$\times$};
				\vertex[below left=0.0cm and 0.5cm of a0] (d1){$\cdot$};
				\vertex[below left=0.15cm and 0.47697cm of a0] (d2){$\cdot$};
				\vertex[below left=0.28612cm and 0.41cm of a0] (d3){$\cdot$};
				\vertex[below left=0.37cm and 0.147986cm of a0] (a3){$\times$};
				\vertex[right=0.38cm of a0, dot] (b0){};
				\diagram*{
				};
			\end{feynman}
\end{tikzpicture}}} \,,
\ee
where $\bar{\mathcal{E}}_{\ell m}r^{\ell}$, $M^{N}$ and $\left(G/r\right)^{N}$ come from the asymptotic source insertion, graviton-worldline vertices and graviton propagators attached to the worldline respectively. The remaining blob diagram contributes a factor of 
\be\label{eq:NDA}
	\left({G\over r^2}\right)^L\prod_k \left(\alpha_k\over r^{2(k-1)}\right)^{n_k} \,,
\ee 
where $L$ is a number of loops in the blob and $n_k$  is the multiplicity of the $\alpha_k$ vertex. The classical contribution to the response, which is our focus here, corresponds to $L=0$. In these expressions we reconstructed the powers of $r$ using the ``naive dimensional analysis" (NDA). NDA correctly reproduces powers of $r$ (but misses the logarithms, which generically arise whenever the EFT diagrams are divergent) provided one is using a mass-independent scheme, such as dimensional regularization, to regulate the EFT infinities. To give rise to a non-trivial contribution into the corresponding Love number, the diagram should scale as $r^{-\ell-1}$, which implies that
\be\label{eq:Neq}
	N=2\ell+1-2\sum_{k}n_k(k-1) \,.
\ee
A specific case considered in the last section---the leading order calculation in Riemann-cubed gravity---corresponds to $n_3=1$ with $n_2=0$ and $n_{k>3}=0$. In this case one also finds $N\geq 2$, given that the $\alpha_3$ vertex has at least three graviton legs so that for tree level diagrams the number of gravitons attached to the worldline cannot be smaller than two (this accounts for the fact that one graviton connects to the bulk scalar field vertex). This leaves as with the condition
\[
	\ell \geq  3
\]
for the possibility of the source responce mixing, in agreement with the presence of logarithmic running as observed in the microscopic calculation in the last section. As a byproduct, this argument indicates also that there is no source-response mixing at $\ell = 1,2$, so that the values of $\varkappa_{1,2}$ given by \eqref{eq:varkappas} are the actual Love numbers\footnote{Let us recall, however, that in applications of these expressions it may be necessary to account for the non-trivial relation between $r_\text{h}$ and the black hole ADM mass in our conventions.}. It is worth mentioning that the case of gravitational perturbations in Riemann-cubed gravity was studied in Ref.~\cite{Cai:2019npx}, which showed that the $\ell=2$ Love numbers are not zero without addressing an issue of the source/response separation. The property of RG flowing Love numbers for black holes in Riemann-cubed gravity was also noted in Ref.~\cite{DeLuca:2022tkm}, who studied vector and tensor Love numbers after our work on the scalar Love numbers~\cite{Charalambous:2022rre}. A similar analysis and conclusions were made in the gravity theory with $R^4$ corrections in Ref.~\cite{Cardoso:2018ptl}.

It is instructive to contrast our results in Riemann-cubed gravity with properties of Love numbers in the Einstein theory in $d$ dimensions. There, the analog of \eqref{eq:Neq} is
\be\label{eq:dNeq}
	N\left(d-3\right)=2\ell+d-3\;,
\ee
where we accounted for the fact that Newton's constant $G$ in $d$ dimensions has mass dimension $\left[\text{mass}\right]^{d-2}$ and that the decaying solution of the Laplace equation scales as 
\[
	\Phi_\ell\propto r^{-\ell-d+3} \,,.
\]
Hence, based on the dimensional analysis presented above, one expects to find logarithmic running of the Love numbers whenever \eqref{eq:dNeq} can be satisfied with integer $N$, {\it i.e.} for integer and half-integer values of $\hat{\ell}=\frac{\ell}{d-3}$. In particular, at $d=4$ one expects the static Love numbers to exhibit running at all values of $\ell$; instead they all vanish identically.

On naturalness grounds~\cite{tHooft:1979rat}, the dimensionless static Love numbers are expected to be RG-flowing $\mathcal{O}\left(1\right)$ numbers. The above general-relativistic results then raise naturalness concerns since there are no background symmetries that output these vanishings as selection rules~\cite{tHooft:1979rat,Porto:2016zng}. Therefore, it is only ``natural'' to seek for an enhanced symmetry structure associated with general-relativistic black holes in four spacetime dimensions. In fact, as we will see in the Chapter~\ref{ch:LoveSymmetryDd} and Chapter~\ref{ch:5dMP}, this seemingly fine-tuned behavior persists in higher spacetime dimensions as well. There, we will find that general-relativistic black hole static Love numbers do exhibit the expected RG flow pattern for half-integer $\hat{\ell}$ and also do acquire the expected non-running and non-vanishing values for $2\hat{\ell}\notin\mathbb{N}$. However, there exist resonant conditions, e.g. whenever $\hat{\ell}\in\mathbb{N}$ for the case of the static scalar Love numbers of the higher-dimensional Schwarzschild black hole, for which one again encounters such ``magic zeroes''~\cite{Kol:2011vg,Hui:2020xxx}. Technically, for spin-$0$ and spin-$2$ fields, the absence of logarithmic running for integer values of $\hat{\ell}$ can be understood within EFT from the special structure of nonlinear vertices in the Einstein action in the conveniently chosen ADM gauge~\cite{Kol:2011vg,Ivanov:2022hlo}. However, this still leaves one wonder whether this fact itself is indicative of an additional symmetry structure in the microscopic theory, which turns out to be the case as we will see in the next chapter.

\section{Love numbers Vs no-hair theorems}
\label{sec:LNsVsNoHair}

Vanishing of Love numbers is often linked to another famous property of black holes---no-hair theorems. Let us briefly compare these phenomena, the main point being that the two are in fact quite distinct and different, at least as far as the discussion of EFT naturalness and fine-tuning go. To make a proper comparison it is important to be precise about what black hole hair are. Note also that the no-hair property is often used to distinguish black holes from conventional objects such as stars, rocky planets or scalar field solutions. Again, we will see that to draw such a distinction it is important to be precise in defining the notion of hair.

A simple observation showing that the vanishing of Love numbers is not a consequence of no-hair theorems is based on the Bekenstein argument~\cite{Bekenstein:1972ky,Cardoso:2016ryw}. The no-hair theorem in the Bekenstein sense is the statement that one cannot ``anchor'' a regular and decaying at infinity scalar field profile to the black hole geometry. This statement is true both in four and in higher dimensions. However, we know that generically Love numbers do not vanish in higher dimensions~\cite{Kol:2011vg,Hui:2020xxx}.

Let us now define hair more systematically. 
Possible black hole hair can be divided into three broad categories. The first type of hair is a situation when a black hole is a member of a family of solutions which have additional continuous parameters on top of mass, spin and gauge charge(s). This type of hair is sometimes called ``primary hair''. This primary hair match the perturbative notion of Bekenstein, as its presence implies the presence of a zero mode, i.e. the possibility to anchor an additional field to a black hole, see e.g.~\cite{Dubovsky:2007zi}.

Primary hair have to be distinguished from ``secondary hair'', which refer to a situation when black holes support additional fields (typically, scalar fields), which are not associated to any conserved gauge charges. An example of this situation is a theory with a non-minimally coupled scalar field (e.g.~\cite{Berti:2015itd,Herdeiro:2015waa}), which give new black hole solutions that are different from the ones existing in the absence of such couplings. Black holes do not acquire any new continuous parameters in this case, so that the corresponding field profiles, if they exist, are still uniquely fixed by the values of a black hole mass, spin and charge(s).

Finally, for a proper comparison of black holes to conventional objects, such as rocky planets, it is important to consider also a possibility of ``discrete hair''. This is a situation when at fixed values of mass, spin and gauge charge(s) one finds a (potentially very large) ``discretuum'' of distinct black hole solutions. Similarly to the secondary hair, no new continuum parameter is present in this case. In the context of rocky planets, there are new parameters characterizing the multipolar structure such as e.g. ``mountain height''. Another example is given by higher-dimensional black holes, which are characterized by a ``discrete'' set of topologies~\cite{Hollands:2012xy}.

Let us see now that these three scenarios have very different properties both from the worldline EFT viewpoint and in the microscopic theory, and also as far as the relation to Love numbers is concerned. To check the absence (or presence) of primary hair at the level of the microscopic theory one needs to perform a study which is very similar to the calculation of the static Love numbers. Indeed, primary hair correspond to time-independent solutions to the Teukolsky equation, which are regular both at the horizon and at the spatial infinity.

Importantly, the presence of primary hair would be a fine-tuning and would require a symmetry explanation, similarly to the vanishing of static Love numbers. In the worldline EFT description, primary hair give rise to additional gapless degrees of freedom on the worldline. Also in this description their presence would be indicative of fine-tuning unless some additional symmetry is present. This implies that the absence of black hole primary hair (contrary to the vanishing of Love numbers) is not surprising, and is not that special to black holes. Primary hair are generically absent also for conventional objects, such as rocky planets and solitons. The only peculiar feature of black holes in this respect, is that, unlike for conventional objects, their continuum parameters can only be {\it gauge} charges, {\it i.e.}, black holes are neutral w.r.t. global charges. 

In spite of these differences, the absence of primary hair still has some relevance for the properties of Love numbers. Namely, if primary hair were present in a certain sector, it would be impossible to define the corresponding tidal/electromagnetic/scalar response. Indeed, Love numbers are defined by the decaying tail of the black hole perturbation, which is regular at the horizon. The presence of primary hair would allow to change the decaying tail arbitrarily by adding the corresponding solution. A physical realization of an object with primary hair would be a planet made of something like plastic, which can be deformed continuously. An object like this may exhibit a hysteresis property, preventing a definition of the response.

One may be puzzled by this discussion given that the absence of hair is often used to distinguish black holes from conventional objects. We already saw one sense in which this is correct, namely, unlike black holes, conventional objects may carry primary hair corresponding to global charges. In addition, it is the possibility of discrete hair which distinguishes conventional objects from black holes in four-dimensional General Relativity. Thus, unlike a black hole, a piece of rock of fixed mass and angular momentum (and other charges) may take many different shapes, which cannot be smoothly deformed into each other without breaking the rock into smaller pieces and gluing them back together. However, the absence or presence of such discrete hair is invisible from the viewpoint of the worldline EFT describing each individual shape, and has no bearing on the properties of  Love numbers.

Finally, secondary hair are associated to tadpole couplings in the worldline EFT. Similarly to the vanishing of Love numbers, their absence also appears as tuning in the worldline EFT unless additional symmetries are present. However, secondary hair again has little to do with the properties of Love numbers. Secondary hair is a property of the background solution, while the Love numbers are determined by the behavior of small perturbations.

	\newpage
\chapter{Love symmetry}
\label{ch:LoveSymmetry4d}

As we just saw, general-relativistic black hole Love numbers in four spacetime dimensions exhibit some superficially fine-tuning properties. In this chapter, we will present the emergence of an enhanced symmetry which governs the dynamics of black hole perturbations in the near-zone region and precisely revives the naturalness of the theory. This chapter is an adaptation of results from our original work in Ref.~\cite{Charalambous:2021kcz} where we revealed the discovery of this symmetry. We call it ``Love symmetry'' because it is this symmetry that forces the black hole Love numbers to vanish. We also borrow elements from our works in Refs.~\cite{Charalambous:2022rre,Charalambous:2023jgq} around the more detailed study of the structure of the representations of the Love symmetry relevant for the vanishing of the black hole Love numbers and the absence of RG flow.

We will describe this symmetry following a similar structure as in the previous chapter. First, we will start with the simple case of a massless scalar field in the four-dimensional Schwarzschild black hole background, which had already be known to posses an enhanced $\SL$ symmetry in the near-zone approximation~\cite{Bertini:2011ga}, although no connection with the response problem had been made in~\cite{Bertini:2011ga}. We will explicitly show that static scalar Schwarzschild black hole perturbations belong to highest-weight representations of $\SL$, which implies the vanishing of scalar Love numbers. We will extend this argument to a massless scalar field in the Kerr-Newman black hole background, which posses a similar globally defined $\SL$ structure. Then, we will generalize our results to generic spin-$s$ fields by presenting a general $\SL$ symmetry which addresses the vanishing of scalar ($s=0$), electromagnetic ($s=\pm1$), and gravitational ($s=\pm2$) Love numbers of Kerr-Newman black holes in four spacetime dimensions.

We will furthermore comment on the absence of RG flow which manifests itself at the level of the Love symmetry by offering an algebraic local criteria of distinguishing solutions regular at the future and past event horizons~\cite{Charalambous:2021kcz,Charalambous:2022rre}. Finally, we will investigate in more detail the structure of the Love symmetry highest-weight representation and show that this vector space is precisely spanned by states with vanishing and/or non-running Love numbers at leading order in the near-zone expansion.

\section{Scalar perturbations of Schwarzschild black holes}
\label{sec:SL2R_Schwarzschild4d}

We begin with the scalar response problem for the Schwarzschild black hole. As was already pointed out in~\cite{Bertini:2011ga}, the leading order near-zone equations of motion \eqref{eq:NZsplitting_Schwarzschild4d} acquire a conformal structure generated by the vector fields
\be\label{eq:SL2RSchwarzschild4d}
	\begin{split}
		L_0 = -\beta\,\partial_{t} \,,\quad L_{\pm1} = e^{\pm t/\beta}\left[\mp\sqrt{\Delta}\,\partial_{r} + \partial_{r}\left(\sqrt{\Delta}\right)\beta\,\partial_{t}\right] \,,
	\end{split}
\ee
where $\beta = 2r_{s}$ is the inverse Hawking temperature of the four-dimensional Schwarzschild black hole. By transforming these vectors into advanced ($+$)/retarded ($-$) Eddington-Finkelstein coordinates, see Eq.~\eqref{eq:EddingtonFinkelsteinCoord4d},
\be
	\begin{gathered}
		L_0 = -\beta\,\partial_{t_{+}} = -\beta\,\partial_{t_{-}} \,, \\
		\ba
			L_{+1} &= e^{+\left(t_{+}-r\right)/\beta}\sqrt{\frac{r_{s}}{r}}\left[-r\,\partial_{r}-\left(r-r_{s}\right)\,\partial_{t_{+}}\right] \\
			&= e^{+\left(t_{-}+r\right)/\beta}\sqrt{\frac{r}{r_{s}}}\left[-\left(r-r_{s}\right)\partial_{r} + \frac{r^2+2r_{s}r-r_{s}^2}{r}\,\partial_{t_{-}}\right] \,,
		\ea \\
		\ba
			L_{-1} &= e^{-\left(t_{+}-r\right)/\beta}\sqrt{\frac{r}{r_{s}}}\left[+\left(r-r_{s}\right)\partial_{r} + \frac{r^2+2r_{s}r-r_{s}^2}{r}\,\partial_{t_{+}}\right] \\
			&= e^{-\left(t_{-}+r\right)/\beta}\sqrt{\frac{r_{s}}{r}}\left[+r\,\partial_{r}-\left(r-r_{s}\right)\,\partial_{t_{+}}\right] \,,
		\ea
	\end{gathered}
\ee
it is clear that they are regular both at the future and at the past event horizons. Also, they obey the $\SL$ algebra commutation relations,
\be\label{eq:SL2RAlgebra}
	\left[L_{m},L_n\right] = \left(m-n\right)L_{m+n} \,,\quad m,n=-1,0,+1 \,.
\ee
That this is indeed a symmetry of the leading order near-zone equations of motion can be seen from the fact that the quadratic Casimir of this algebra exactly reproduces the differential operator of the Klein-Gordon equation in the near-zone approximation ($\epsilon=0$ in \eqref{eq:NZsplitting_Schwarzschild4d}),
\be
	\mathcal{C}_2 = L_0^2 - \frac{1}{2}\left(L_{+1}L_{-1}+L_{-1}L_{+1}\right) = \partial_{r}\,\Delta\,\partial_{r} - \frac{r_{s}^4}{\Delta}\,\partial_{t}^2 \,.
\ee
More importantly, the separable solution $\Phi_{\omega\ell m}$ of this equation is then an eigenvector of both $L_0$ and $\mathcal{C}_2$
\be
	\begin{split}
		L_0 \Phi_{\omega\ell m} = h\,\Phi_{\omega\ell m}= i\beta\omega\,\Phi_{\omega\ell m} \,,\quad \mathcal{C}_2\Phi_{\omega\ell m} = \ell\left(\ell+1\right)\Phi_{\omega\ell m}\,,
	\end{split}
\ee
where $\ell$ is an integer number by virtue of the static angular eigenvalue problem, namely, $\ell$ is the orbital number label of the spherical harmonics. Therefore, all separable solutions $\Phi_{\omega\ell m}$ furnish representations of this $\SL$, ``Love'', symmetry. Moreover, as a consequence of the regularity of the vector fields $L_{m}$, solutions regular at the horizon are closed under the $\SL$ action.

This is a powerful statement, because now we can derive many properties of black hole perturbations from group theory arguments.
In particular, let us demonstrate how the Love symmetry implies the vanishing of static scalar Love numbers, $k_{\ell}^{\left(0\right)}\left(\omega=0\right)$, of the four-dimensional Schwarzschild black hole~\cite{Charalambous:2021kcz,Charalambous:2022rre}. These are independent of the azimuthal number $m$ due to the spherical symmetry of the background, so in this section we set $m=0$ without loss of generality. They are extracted from the static solution, which is a null-vector of $L_0=-\beta\,\partial_{t}$. This solution belongs to a highest-weight representation of $\SL$\footnote{See Appendix~\ref{app:SL2RRepresentations} for a brief review of the standard indecomposable $\SL$ representations.}. To see this, let us explicitly construct this representation. It is generated by a highest-weight (primary) vector $\upsilon_{-\ell,0}$ with weight $h=-\ell$,
\be
	L_{+1}\upsilon_{-\ell,0} = 0 \,,\quad L_0\upsilon_{-\ell,0} = -\ell\,\upsilon_{-\ell,0} \,.
\ee
Up to an overall normalization factor, this highest-weight state is given by
\be
	\upsilon_{-\ell,0} = \left(-e^{+t/\beta}\sqrt{\Delta}\right)^{\ell} \,.
\ee
This function solves the near-zone massless Klein-Gordon equation for multipolar order $\ell$ with an imaginary frequency $\omega_{-\ell,0}=i\ell/\beta$ and is regular both at the future and at the past event horizons. Since the generators $L_{\pm1}$ are regular on the horizon, all the descendants,
\be
	\upsilon_{-\ell,n} = \left(L_{-1}\right)^{n}\upsilon_{-\ell,0}
\ee
are also regular solutions of the massless Klein-Gordon equation, now with frequency $\omega_{-\ell,n}=i\left(\ell-n\right)/\beta$. In particular, we immediately see that the physical static solution with zero frequency is an element of this highest-weight representation,
\be
	\Phi_{\omega=0,\ell,m=0} \propto \upsilon_{-\ell,\ell} = \left(L_{-1}\right)^{\ell}\upsilon_{-\ell,0} \,.
\ee
As such, it must be annihilated by $\left(L_{+1}\right)^{\ell+1}$. From the explicit action of the vector fields \eqref{eq:SL2RSchwarzschild4d}, we observe that, for an arbitrary radial function $F\left(r\right)$,
\be
	\left(L_{+1}\right)^{n}F\left(r\right) = \left(-e^{+t/\beta}\sqrt{\Delta}\right)^{n}\frac{d^{n}}{dr^{n}}F\left(r\right) \,.
\ee
For the static solution $\upsilon_{-\ell,\ell} = F\left(r\right)$, the annihilation condition then reads,
\be
	\left(L_{+1}\right)^{\ell+1}\upsilon_{-\ell,\ell} = \left(-e^{+t/\beta}\sqrt{\Delta}\right)^{\ell+1}\frac{d^{\ell+1}}{dr^{\ell+1}}\upsilon_{-\ell,\ell} = 0 \,,
\ee
implying that the physical static solution must be a degree-$\ell$ polynomial in $r$,
\be
	\upsilon_{-\ell,\ell} = \sum_{n=0}^{\ell}c_{n}r^{n} = c_{\ell}r^{\ell} + \dots + c_1r + c_0 \,.
\ee
Clearly, this solution does not have terms with decaying powers $\propto r^{-\ell-1}$, which is precisely indicative for the vanishing of static Love numbers.

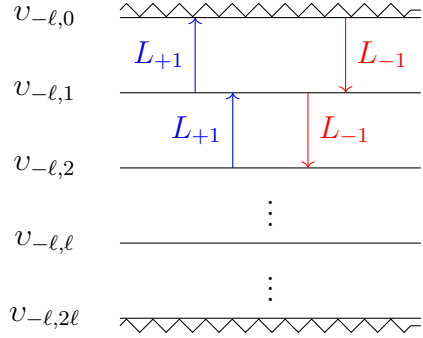
\begin{figure}[t]
	\centering
	\begin{tikzpicture}
		\node at (0,0) (uml4) {$\upsilon_{-\ell,2\ell}$};
		\node at (0,1) (uml3) {$\upsilon_{-\ell,\ell}$};
		\node at (0,2) (uml2) {$\upsilon_{-\ell,2}$};
		\node at (0,3) (uml1) {$\upsilon_{-\ell,1}$};
		\node at (0,4) (uml0) {$\upsilon_{-\ell,0}$};
		
		\draw [snake=zigzag] (1,-0.1) -- (5,-0.1);
		\draw (1,0) -- (5,0);
		\draw (1,1) -- (5,1);
		\node at (3,1.5) (up) {$\vdots$};
		\node at (3,0.5) (um) {$\vdots$};
		\draw (1,2) -- (5,2);
		\draw (1,3) -- (5,3);
		\draw (1,4) -- (5,4);
		\draw [snake=zigzag] (1,4.1) -- (5,4.1);
		
		\draw[blue] [->] (2.5,2) -- node[left] {$L_{+1}$} (2.5,3);
		\draw[blue] [->] (2,3) -- node[left] {$L_{+1}$} (2,4);
		\draw[red] [<-] (4,3) -- node[right] {$L_{-1}$} (4,4);
		\draw[red] [<-] (3.5,2) -- node[right] {$L_{-1}$} (3.5,3);
	\end{tikzpicture}
	\caption[The finite-dimensional highest weight representation of $\SL$ whose elements solve the leading order near-zone equations of motion for a massless scalar field in the four-dimensional Schwarzschild black hole background with multipolar index $\ell$ and contains the regular static solution.]{The finite-dimensional highest weight representation of $\SL$ whose elements solve the leading order near-zone equations of motion for a massless scalar field in the four-dimensional Schwarzschild black hole background with multipolar index $\ell$ and contains the regular static solution.}
	\label{fig:HWSL2RSchwarzschild}
\end{figure}

We note that, in the case of a scalar field in the four-dimensional Schwarzschild black hole background, one can arrive at the same conclusion starting from a lowest-weight state. Indeed, we can construct the lowest-weight vector of weight $h=+\ell$,
\be
	L_{-1}\bar{\upsilon}_{+\ell,0} = 0 \,,\quad L_0\bar{\upsilon}_{+\ell,0} = +\ell\,\bar{\upsilon}_{+\ell,0} \quad \Rightarrow \quad \bar{\upsilon}_{+\ell,0} = \left(+e^{-t/\beta}\sqrt{\Delta}\right)^{\ell} \,,
\ee
which is also a solution of the leading order near-zone massless Klein-Gordon equation with multipolar index $\ell$ that is regular on both the future and the past event horizons, but this time with frequency $\bar{\omega}_{+\ell,0}=-i\ell/\beta$. A regular static solution would then be identified as the particular ascendant with zero $L_0$-eigenvalue,
\be
	\Phi_{\omega=0,\ell,m=0} \propto \bar{\upsilon}_{+\ell,-\ell} = \left(-L_{+1}\right)^{\ell}\bar{\upsilon}_{+\ell,0}
\ee
By the uniqueness of the regular solution, this implies that the highest-weight and lowest-weight representations are actually the same, i.e. this is a finite $\left(2\ell+1\right)$-dimensional representation of $\SL$, and consequently
\be
	\bar{\upsilon}_{+\ell,0} = \upsilon_{-\ell,2\ell} \,.
\ee
Our construction of the highest-weight representation of $\SL$ in the leading order near-zone equations of motion for the scalar perturbations of the Schwarzschild black hole is summarized in Figure~\ref{fig:HWSL2RSchwarzschild}.

The fact that the highest-weight and lowest-weight representations of $\SL$ coincide is unique to the Schwarzschild background and to spin-$0$ fields. This property can be traced back to the time-reversal symmetry of the background. Solutions of the leading order near-zone equations of motion regular at the future event horizon belong to a highest-weight representation, while the corresponding solutions regular at the past event horizon belong to a lowest-weight representation. Indeed, the $t\rightarrow-t$ symmetry ensures that static scalar perturbations regular at the future event horizon will \textit{also} be regular at the past event horizon and therefore, the two representations overlap to furnish the finite-dimensional representation of the Love $\SL$ symmetry we just saw.

\begin{figure}[t]
	\centering
	\begin{tikzpicture}
		\node at (0,-1) (uml8) {$\tilde{\upsilon}_{-\ell-1,2\ell+3}$};
		\node at (0,0) (uml7) {$\tilde{\upsilon}_{-\ell-1,2\ell+2}$};
		\node at (0,1) (uml6) {$\tilde{\upsilon}_{-\ell-1,2\ell+1}$};
		\node at (0,2) (uml5) {$\tilde{\upsilon}_{-\ell-1,2\ell}$};
		\node at (0,3) (uml4) {$\tilde{\upsilon}_{-\ell-1,\ell+1}$};
		\node at (0,4) (uml3) {$\tilde{\upsilon}_{-\ell-1,2}$};
		\node at (0,5) (uml2) {$\tilde{\upsilon}_{-\ell-1,1}$};
		\node at (0,6) (uml1) {$\tilde{\upsilon}_{-\ell-1,0}$};
		\node at (0,7) (uml0) {$\tilde{\upsilon}_{-\ell-1,-1}$};
		
		\node at (3,-1.5) (umm) {$\vdots$};
		\draw (1,-1) -- (5,-1);
		\draw [snake=zigzag] (1,0.1) -- (5,0.1);
		\draw (1,0) -- (5,0);
		\draw (1,1) -- (5,1);
		\draw (1,2) -- (5,2);
		\node at (3,2.5) (um) {$\vdots$};
		\draw (1,3) -- (5,3);
		\node at (3,3.5) (up) {$\vdots$};
		\draw (1,4) -- (5,4);
		\draw (1,5) -- (5,5);
		\draw (1,6) -- (5,6);
		\draw [snake=zigzag] (1,5.9) -- (5,5.9);
		\draw (1,7) -- (5,7);
		\node at (3,7.5) (upp) {$\vdots$};
		
		\draw[blue] [->] (2,-1) -- node[left] {$L_{+1}$} (2,0);
		\draw[red] [<-] (4,-1) -- node[right] {$L_{-1}$} (4,0);
		\draw[red] [<-] (3,0) -- node[right] {$L_{-1}$} (3,1);
		\draw[blue] [->] (2,1) -- node[left] {$L_{+1}$} (2,2);
		\draw[red] [<-] (4,1) -- node[right] {$L_{-1}$} (4,2);
		\draw[blue] [->] (2,4) -- node[left] {$L_{+1}$} (2,5);
		\draw[red] [<-] (4,4) -- node[right] {$L_{-1}$} (4,5);
		\draw[blue] [->] (3,5) -- node[left] {$L_{+1}$} (3,6);
		\draw[blue] [->] (2,6) -- node[left] {$L_{+1}$} (2,7);
		\draw[red] [<-] (4,6) -- node[right] {$L_{-1}$} (4,7);
	\end{tikzpicture}
	\caption[The infinite-dimensional representation of $\SL$ whose elements solve the leading order near-zone equations of motion for a massless scalar field in the four-dimensional Schwarzschild black hole background with multipolar index $\ell$ and contains the singular static solution.]{The infinite-dimensional representation of $\SL$ whose elements solve the leading order near-zone equations of motion for a massless scalar field in the four-dimensional Schwarzschild black hole background with multipolar index $\ell$ and contains the singular static solution.}
	\label{fig:SingSL2RSchwarzschild}
\end{figure}
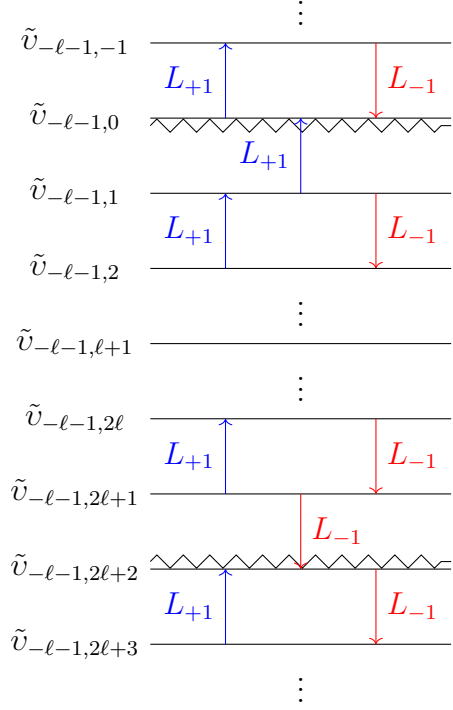

We will see momentarily that for the stationary, but not static, Kerr-Newman black holes, and also for spin-$s$ fields in the Schwarzschild metric, this is not the case. In these more general cases the highest-weight representation contains solutions that are regular on the physical future event horizon and always singular at the past event horizon, while the lowest-weight representation contains physically irrelevant solutions that are regular on the past event horizon but always singular on the future event horizon. In the four-dimensional Schwarzschild $s=0$ case, instead, a static solution that is singular on the event horizon belongs to the representation shown in Figure~\ref{fig:SingSL2RSchwarzschild}. This is spanned by vectors $\tilde{\upsilon}_{-\ell,j}$, $j\in\mathbb{Z}$. The upper part of the ladder is constructed by ascendants of the lowest-weight state $\tilde{\upsilon}_{-\ell-1,0}$ with weight $h=-\ell-1$,
\be
	L_{-1}\tilde{\upsilon}_{-\ell-1,0} = 0 \,,\quad L_0\tilde{\upsilon}_{-\ell-1,0} = -\left(\ell+1\right)\tilde{\upsilon}_{-\ell-1,0} \,\, \Rightarrow \,\, \tilde{\upsilon}_{-\ell-1,0} = \left(+e^{-t/\beta}\sqrt{\Delta}\right)^{-\left(\ell+1\right)} \,.
\ee
All the elements in this upper ladder are in fact singular (regular) at the future (past) event horizon with frequencies $\tilde{\omega}_{-\ell-1,-n} = i\left(\ell+1+n\right)/\beta$. Nevertheless, solutions singular at both the future and the past event horizons exist in the middle part of the ladder, constructed by requiring that $\tilde{\upsilon}_{-\ell-1,0}$ is itself an ascendant of $\tilde{\upsilon}_{-\ell-1,1}$, a condition that gives an inhomogeneous first-order differential equation to solve,
\be
	\tilde{\upsilon}_{-\ell-1,0} \propto L_{+1}\tilde{\upsilon}_{-\ell-1,1} \ne 0 \quad \Rightarrow \quad \tilde{\upsilon}_{-\ell-1,1} \propto \left(+e^{-t/\beta}\sqrt{\Delta}\right)^{-\ell} \ln\frac{r-r_{s}}{r} \,.
\ee
Clearly, $\tilde{\upsilon}_{-\ell-1,1}$ is singular at the horizon\footnote{We are ignoring here an irrelevant additive piece which is regular at the event horizon and is annihilated by $L_{+1}$. This freedom reflects the fact that, starting from one particular singular solution, we can always construct another (linearly dependent) singular solution by adding to the profile a regular solution.}. The subsequent descendants will then also be singular, up until $\tilde{\upsilon}_{-\ell-1,2\ell+2}$ beyond which we enter the lower part of the ladder. At that step, $\tilde{\upsilon}_{-\ell-1,2\ell+2}$ becomes a highest-weight vector of weight $h=+\ell+1$. Then, all the descendants $\tilde{\upsilon}_{-\ell-1,2\ell+2+n}$, with $n\ge0$, are regular (singular) at the future (past) event horizon with frequencies $\tilde{\omega}_{-\ell-1,2\ell+2+n} = -i\left(\ell+1+n\right)/\beta$. The region of interest of course is the middle part of the ladder, spanned by the singular vectors $\tilde{\upsilon}_{-\ell-1,n+1}$, $n=0,\dots,2\ell$ which have frequencies $\omega_{-\ell-1,n+1} = i\left(\ell-n\right)/\beta$. The singular static solution is then identified as the state $\tilde{\upsilon}_{-\ell-1,\ell+1}$ and the structure of the representation implies the selection rule
\be
	L_{-1}\left(L_{+1}\right)^{\ell+1}\tilde{\upsilon}_{-\ell-1,\ell+1} = 0 \Rightarrow \frac{d}{dr}\left(\Delta^{\ell+1}\frac{d^{\ell+1}}{dr^{\ell+1}}\tilde{\upsilon}_{-\ell-1,\ell+1}\right) = 0 \,.
\ee
This is indeed the condition satisfied by the singular at the horizon static solution as can be checked by its explicit expression in terms of Legendre polynomials of the second kind~\cite{Binnington:2009bb,Kol:2011vg}.

It is natural to ask whether other solutions which enter in the same multiplets with the static ones have any physical interpretation. In the Schwarzschild case the corresponding non-vanishing imaginary frequencies are given by
\be
	\omega_n= i\,2\pi T_H n \,.
\ee
Interestingly, the spacing between these modes matches the spacing of highly-damped quasinormal modes (QNMs) given by~\cite{Nollert:1993zz}
\be
	\omega_k =\frac{i}{4M}\left(k+{1}/{2}\right) = i\,2\pi T_H (k +1/2) \,. 
\ee
As we will see later in this chapter, there is an alternative interpretation of the vectors spanning the highest-weight representation of the Love symmetry as states with vanishing near-zone Love numbers.

Finally, let us briefly comment on the massive wave equation with mass $\mu$. In the regime $\mu\left(r-r_s\right)\ll 1$, and $\mu M\ll 1$, it has the following form in the Schwarzschild black hole near-zone approximation~\cite{Bertini:2011ga},
\be
	\left[\partial_{r}\,\Delta\,\partial_{r}-\frac{r_{s}^4}{\Delta}\,\partial_{t}^2\right]\Phi_{\omega\ell m} = \left[\ell\left(\ell+1\right)+\mu^2 r_{s}^2\right]\Phi_{\omega\ell m} \,.
\ee
Even though the Love symmetry is still present, the mass changes the eigenvalue of the Casimir operator such that the physical static solution $\Phi_{\omega=0,\ell m}$ regular at the horizon does not belong to the highest-weight $\SL$ representation anymore. A direct calculation shows that Love numbers are not zero. As can be seen by replacing $\ell \rightarrow \ell + \frac{\mu^2r_{s}^2}{2\ell+1} + \mathcal{O}\left(\mu^4r_{s}^4\right)$ in \eqref{eq:SLNsDissipativeRCs_Schwarzschild4d}, the static scalar Love numbers at first order in $\mu^2r_s^2$ take constant values without any running. This example illustrates that the presence of the Love symmetry alone is not enough to ensure the vanishing of Love numbers. The crucial role is played by the highest-weight property of the corresponding representations.

\section{Scalar perturbations of Kerr-Newman black holes}
\label{sec:SL2R_KerrNewman}

For a massless scalar field in the background of the Kerr-Newman black hole, the generators of the Love symmetry are~\cite{Charalambous:2021kcz,Charalambous:2022rre}
\be\label{eq:SL2RKerrNewman}
	L_0 = -\beta\,\partial_{t} \,,\quad L_{\pm1} = e^{\pm t/\beta}\left[\mp\sqrt{\Delta}\,\partial_{r} + \partial_{r}\left(\sqrt{\Delta}\right)\beta\,\partial_{t} + \frac{a}{\sqrt{\Delta}}\,\partial_{\phi}\right]\,,
\ee
with the inverse Hawking temperature $\beta$ of the Kerr-Newman black hole given in \eqref{eq:betaKerrNewman}. The new generators also satisfy the $\SL$ algebra \eqref{eq:SL2RAlgebra}, while the resulting Casimir operator,
\be
	\mathcal{C}_2 = \partial_{r}\,\Delta\,\partial_{r} - \frac{\left(r_{+}^2+a^2\right)^2}{\Delta}\left[\left(\partial_{t}+\Omega\,\partial_{\phi}\right)^2 + 4\Omega\frac{r-r_{+}}{r_{+}-r_{-}}\partial_{t}\partial_{\phi}\right] \,,
\ee
again coincides with the leading order near-zone radial differential operator, Eq.~\eqref{eq:NZSplitRadial_KerrNewman}. As in the Schwarzschild case, the Love vector fields are regular on both future and past event horizons as can be seen by transforming into advanced ($+$)/retarded ($-$) null coordinates $\left(t_{\pm},r,\theta,\varphi_{\pm}\right)$,
\be
	\begin{gathered}
		L_0 = -\beta\,\partial_{t_{+}} = -\beta\,\partial_{t_{-}} \,, \\
		\ba
			L_{\pm1} &= e^{\pm\left(t_{+}-r\right)/\beta}\left(\frac{r-r_{-}}{r_{+}}\right)^{\mp r_{s}/\beta} \\
			&\mkern-36mu \left[\mp\left(r-r_{\mp}\right)\partial_{r} + \frac{r-r_{\mp}}{r-r_{-}}\left(\beta\mp\left(r+r_{+}\right)\right)\partial_{t_{+}} + \left(1\mp1\right)\frac{r_{+}^2+a^2}{r-r_{-}}\left(\partial_{t_{+}}+\Omega\,\partial_{\varphi_{+}}\right)\right] \\
			&= e^{\pm\left(t_{-}+r\right)/\beta}\left(\frac{r-r_{-}}{r_{+}}\right)^{\pm r_{s}/\beta} \\
			&\mkern-36mu \left[\mp\left(r-r_{\pm}\right)\partial_{r} + \frac{r-r_{\pm}}{r-r_{-}}\left(\beta\pm\left(r+r_{+}\right)\right)\partial_{t_{-}} + \left(1\pm1\right)\frac{r_{+}^2+a^2}{r-r_{-}}\left(\partial_{t_{-}}+\Omega\,\partial_{\varphi_{-}}\right)\right] \,. \\
		\ea
	\end{gathered}
\ee

Supplementary to the Schwarzschild black hole, now we need to take into account that the separable solution $\Phi_{\omega\ell m}$ is charged under the $U\left(1\right)$ axial rotation symmetry,
\be
	J_0\Phi_{\omega\ell m}	= -i\partial_{\phi}\Phi_{\omega\ell m} = m\,\Phi_{\omega\ell m} \,.
\ee
We note here that the azimuthal $U\left(1\right)$ commutes with the Love symmetry. Now we can show that the static scalar Love numbers of the Kerr-Newman black hole also vanish as a result of the Love symmetry. Most of the argument can be straightforwardly borrowed from the Schwarzschild case. The highest-weight vector with $h=-\ell$ is defined by,
\be\label{eq:HWkerr}
	L_{+1}\upsilon_{-\ell,0}^{\left(m\right)}=0 \,,\quad L_0\upsilon_{-\ell,0}^{\left(m\right)} = -\ell\,\upsilon_{-\ell,0}^{\left(m\right)} \,\quad J_0\upsilon_{-\ell,0}^{\left(m\right)} = m\,\upsilon_{-\ell,0}^{\left(m\right)} \,.
\ee
Solving Eq.~\eqref{eq:HWkerr}, we obtain, up to an arbitrary normalization constant,
\be
	\upsilon_{-\ell,0}^{\left(m\right)} = e^{im\phi}\left(\frac{r-r_{+}}{r-r_{-}}\right)^{im\beta\Omega/2}\left(-e^{+t/\beta}\sqrt{\Delta}\right)^{\ell} \,.
\ee
This solves the leading order near-zone (radial) scalar Teukolsky equation for multipolar order $\ell$, with imaginary frequency $\omega_{-\ell,0}=i\ell/\beta$. An important difference with respect to the Schwarzschild case is the regularity condition. The solution $\upsilon_{-\ell,0}^{\left(m\right)}$ is still regular on the physical future event horizon, but it is now singular on the past event horizon. As a result all its descendants,
\be
	\upsilon_{-\ell,n}^{\left(m\right)} = \left(L_{-1}\right)^{n}\upsilon_{-\ell,0}^{\left(m\right)} \,,
\ee
are also solutions of the leading order near-zone equations of motion that are regular on the future event horizon. On the other hand, for a lowest-weight vector with $\bar{h}=+\ell$ we have
\be
	\begin{gathered}
		L_{-1}\bar{\upsilon}_{+\ell,0}^{\left(m\right)} = 0 \,,\quad L_0\bar{\upsilon}_{+\ell,0}^{\left(m\right)} = +\ell\,\bar{\upsilon}_{+\ell,0}^{\left(m\right)} \,\quad J_0\bar{\upsilon}_{+\ell,0}^{\left(m\right)} = m\,\bar{\upsilon}_{+\ell,0}^{\left(m\right)} \,, \\
		\Rightarrow \bar{\upsilon}_{+\ell,0}^{\left(m\right)} = e^{im\phi}\left(\frac{r-r_{+}}{r-r_{-}}\right)^{-im\beta\Omega/2}\left(+e^{-t/\beta}\sqrt{\Delta}\right)^{\ell} \,.
	\end{gathered}
\ee
This solution is regular at the past event horizon, but singular at the future event horizon. It gives rise to an ascending tower of solutions that are regular on the past event horizon.

\begin{figure}[t]
	\centering
	\begin{subfigure}[b]{0.49\textwidth}
		\centering
		\begin{tikzpicture}
			\node at (0,1) (uml3) {$\upsilon_{-\ell,\ell}^{\left(m\right)}$};
			\node at (0,2) (uml2) {$\upsilon_{-\ell,2}^{\left(m\right)}$};
			\node at (0,3) (uml1) {$\upsilon_{-\ell,1}^{\left(m\right)}$};
			\node at (0,4) (uml0) {$\upsilon_{-\ell,0}^{\left(m\right)}$};
			
			\draw (1,1) -- (5,1);
			\node at (3,1.5) (up) {$\vdots$};
			\node at (3,0.5) (um) {$\vdots$};
			\draw (1,2) -- (5,2);
			\draw (1,3) -- (5,3);
			\draw (1,4) -- (5,4);
			\draw [snake=zigzag] (1,4.1) -- (5,4.1);
			
			\draw[red] [<-] (2.5,2) -- node[left] {$L_{-1}$} (2.5,3);
			\draw[red] [<-] (2,3) -- node[left] {$L_{-1}$} (2,4);
			\draw[blue] [->] (4,3) -- node[right] {$L_{+1}$} (4,4);
			\draw[blue] [->] (3.5,2) -- node[right] {$L_{+1}$} (3.5,3);
		\end{tikzpicture}
		\caption{A highest-weight $\SL$ representation that contains scalar perturbations of the Kerr-Newman black hole that are regular on the future event horizon.}
	\end{subfigure}
	\hfill
	\begin{subfigure}[b]{0.49\textwidth}
		\centering
		\begin{tikzpicture}
			\node at (0,3) (upll) {$\bar{\upsilon}_{+\ell,\ell}^{\left(m\right)}$};
			\node at (0,2) (upl2) {$\bar{\upsilon}_{+\ell,2}^{\left(m\right)}$};
			\node at (0,1) (upl1) {$\bar{\upsilon}_{+\ell,1}^{\left(m\right)}$};
			\node at (0,0) (upl0) {$\bar{\upsilon}_{+\ell,0}^{\left(m\right)}$};
			
			\draw [snake=zigzag] (1,-0.1) -- (5,-0.1);
			\draw (1,0) -- (5,0);
			\draw (1,1) -- (5,1);
			\draw (1,2) -- (5,2);
			\node at (3,2.5) (up) {$\vdots$};
			\draw (1,3) -- (5,3);
			\node at (3,3.5) (um) {$\vdots$};
			
			\draw[blue] [->] (2.5,0) -- node[left] {$L_{+1}$} (2.5,1);
			\draw[blue] [->] (2,1) -- node[left] {$L_{+1}$} (2,2);
			\draw[red] [<-] (4,1) -- node[right] {$L_{-1}$} (4,2);
			\draw[red] [<-] (3.5,0) -- node[right] {$L_{-1}$} (3.5,1);
		\end{tikzpicture}
		\caption{A lowest-weight $\SL$ representation that contains scalar perturbations of the Kerr-Newman black hole that are regular on the past event horizon.}
	\end{subfigure}
	\caption[The infinite-dimensional highest-weight and lowest-weight representations of $\SL$ whose elements solve the leading order near-zone equations of motion for a massless scalar field in the Kerr-Newman black hole background with multipolar index $\ell$ and contain the static solutions.]{The infinite-dimensional highest-weight and lowest-weight representations of $\SL$ whose elements solve the leading order near-zone equations of motion for a massless scalar field in the Kerr-Newman black hole background with multipolar index $\ell$ and contain the static solutions.}
	\label{fig:HWSL2RKerr}
\end{figure}
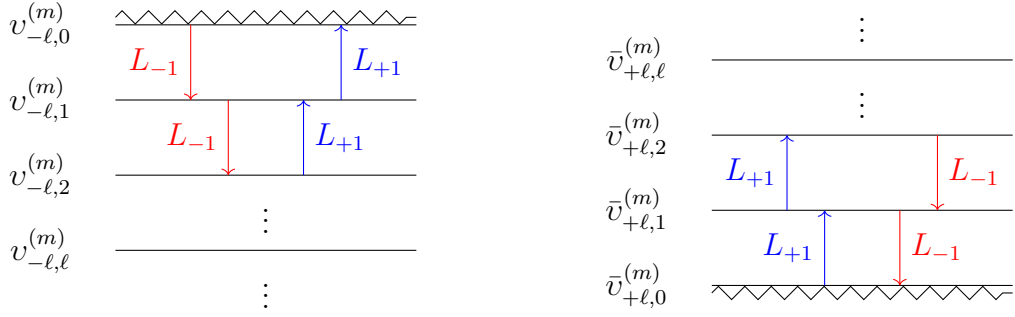

A physical black hole formed as a result of a collapse does not exhibit the past event horizon. Hence, we are interested in the solutions which are regular at the future event horizon, and their singularity at the past horizon does not pose a problem. This singles out the highest-weight representation. As follows from the above discussion, it is now infinite-dimensional, falling into the general category of Verma modules, see Fig.~\ref{fig:HWSL2RKerr}.

As before, the physical static solution is identified with the null $L_0$-eigenstate in the highest-weight representation,
\be
	\Phi_{\omega=0,\ell m} \propto \upsilon_{-\ell,\ell}^{\left(m\right)} = \left(L_{-1}\right)^{\ell}\upsilon_{-\ell,0}^{\left(m\right)} \,.
\ee
Observing that
\be
	\left(L_{+1}\right)^{n}\left[ e^{im\phi}\left(\frac{r-r_{+}}{r-r_{-}}\right)^{im\beta\Omega/2}F\left(r\right)\right] = e^{im\phi}\left(\frac{r-r_{+}}{r-r_{-}}\right)^{im\beta\Omega/2} \left(-e^{+t/\beta}\sqrt{\Delta}\right)^{n} \frac{d^{n}}{dr^{n}}F\left(r\right)
\ee
for arbitrary $F\left(r\right)$ and $n$, the annihilation condition for the static case ($n=\ell+1$) implies a polynomial form for $F\left(r\right)$,
\be
	\begin{gathered}
		\upsilon_{-\ell,\ell}^{\left(m\right)} = e^{im\phi}\left(\frac{r-r_{+}}{r-r_{-}}\right)^{im\beta\Omega/2}F\left(r\right) \,, \\
		\left(L_{+1}\right)^{\ell+1}\upsilon_{-\ell,\ell}^{\left(m\right)} = e^{im\phi}\left(\frac{r-r_{+}}{r-r_{-}}\right)^{im\beta\Omega/2}\left(-e^{+t/\beta}\sqrt{\Delta}\right)^{\ell+1}
		\frac{d^{\ell+1}}{dr^{\ell+1}}F\left(r\right) = 0 \,, \\
		\Rightarrow \quad \Phi_{\omega=0,\ell m} = e^{im\phi}\left(\frac{r-r_{+}}{r-r_{-}}\right)^{im\beta\Omega/2}\sum_{n=0}^{\ell}c_{n}r^{n} \,.
	\end{gathered}
\ee
While this does not look like a pure polynomial solution due the overall $r$-dependent factor, as discussed in the previous chapter, this form-factor can be attributed to frame-dragging and does not affect Love numbers. A polynomial form of the solution apart of this factor indicates that the highest-weight property ensures the vanishing of the Love numbers also in this case.

It is worth mentioning that the spacing of states in the Kerr-Newman Love multiplet $2\pi T_H$ matches the Hod conjecture~\cite{Hod:2003jn,Berti:2005eb,Berti:2009kk} for the asymptotic frequencies of highly-damped Kerr quasinormal modes. This conjecture was addressed analytically only in the cases of small rotation $a\to 0$ and near extremity $a\to M$~\cite{Berti:2009kk,Hod:2013fea}. In contrast, our expression is formally valid for all values of black hole spin. Note however that numerical studies do not support the original Hod conjecture~\cite{Berti:2009kk}, which implies that corrections to the leading damping frequency are not negligible. Nevertheless, our preliminary considerations suggest that the Love symmetry may become a powerful tool for an analytic study of Kerr quasinormal modes.

On the other hand, the spacing of the so-called anomalous total transmission modes is indeed equal to $2\pi T_H$ for the Kerr black holes~\cite{Cook:2016fge,Cook:2016ngj}. These modes require a source at the spatial infinity pumping energy into the system~\cite{MaassenvandenBrink:2000iwh,Cook:2016fge,Cook:2016ngj}, a lot like how the response problem requires a perturbing source at large distances.

As a last remark, we note that the Love symmetry generators are singular in the extremal limit $r_{+}-r_{-} \rightarrow 0$, which also corresponds to the vanishing of the Hawking temperature. This limit is very interesting on its own will be discussed in detail in a separate chapter where it will hint at a connection of the near-zone symmetries with a potential holographic description of asymptotically flat black holes.

\section{General perturbations of Kerr-Newman black holes}
\label{sec:SL2R_KerrNewmanS}

Let us now present the most general case of the four-dimensional Kerr-Newman black hole perturbations with an arbitrary integer spin-weight $s$. The generators of the Love $\SL$ algebra are not spacetime vectors anymore: they acquire a scalar part proportional to the spin-weight of the perturbation~\cite{Charalambous:2021kcz,Charalambous:2022rre},
\be\label{eq:SL2RsKerrNewman}
	\begin{gathered}
		L_0^{\left(s\right)} = -\beta\,\partial_{t} + s \\
		L_{\pm1}^{\left(s\right)} = e^{\pm t/\beta}\left[\mp\sqrt{\Delta}\,\partial_{r} + \partial_{r}\left(\sqrt{\Delta}\right)\beta\,\partial_{t} + \frac{a}{\sqrt{\Delta}}\,\partial_{\phi} - s\left(1\pm1\right)\partial_{r}\left(\sqrt{\Delta}\right) \right] \,.
	\end{gathered}
\ee
The Casimir of the algebra spanned by the above generators correctly reproduces the leading order near-zone truncation of the radial Teukolsky operator, Eqs.~\eqref{eq:NZSplitOP_KerrNewmanS}-\eqref{eq:NZSplitV0V1_KerrNewmanS}. These additional $s$-dependent parts are singular at the horizon. However, this singular behavior is an artifact related to the use of the Kinnersley tetrads~\cite{Kinnersley1969} that are singular at the horizon themselves. The relevant spin-$s$ scalars are regular at the horizon for the appropriate well-behaved choice of coordinates and tetrads~\cite{Teukolsky:1972my}. Another important difference with respect to the spin-$0$ case is that the static solution now has a non-zero $L_0$-eigenvalue,
\be
	L_0^{\left(s\right)} \Psi_{s,\omega=0,\ell m} = s\,\Psi_{s,\omega=0,\ell m} \,.
\ee
Let us show now that this state belongs to a highest-weight representation with weight $-\ell$ (recall that  $\ell\geq |s|$).
We start with a highest-weight vector $\upsilon_{-\ell,0}^{\left(m,s\right)}$ of azimuthal number $m$, satisfying
\be 
	L_{+1}^{\left(s\right)}\upsilon_{-\ell,0}^{\left(m,s\right)} = 0 \,,\quad L_0^{\left(s\right)}\upsilon_{-\ell,0}^{\left(m,s\right)} = -\ell\,\upsilon_{-\ell,0}^{\left(m,s\right)} \,,\quad J_0\upsilon_{-\ell,0}^{\left(m,s\right)} = m\,\upsilon_{-\ell,0}^{\left(m,s\right)} \,.
\ee
It is straightforward to integrate these equations and to obtain
\be
	\upsilon_{-\ell,0}^{\left(m,s\right)} = e^{im\phi}\Delta^{-s}\left(\frac{r-r_{+}}{r-r_{-}}\right)^{im\beta\Omega/2}\left(-e^{+t/\beta}\sqrt{\Delta}\right)^{\ell+s} \,.
\ee
By transforming this vector into the advanced null coordinates we see that it is regular on the future horizon, but singular on the past one. Similarly, the lowest-weight state,
\be
	\begin{gathered}
		L_{-1}^{\left(s\right)}\bar{\upsilon}_{+\ell,0}^{\left(m,s\right)} = 0 \,,\quad L_0^{\left(s\right)}\bar{\upsilon}_{+\ell,0}^{\left(m,s\right)} = +\ell\,\bar{\upsilon}_{+\ell,0}^{\left(m,s\right)} \,,\quad J_0\bar{\upsilon}_{+\ell,0}^{\left(m,s\right)} = m\,\bar{\upsilon}_{+\ell,0}^{\left(m,s\right)} \,, \\
		\Rightarrow \bar{\upsilon}_{+\ell,0}^{\left(m,s\right)} = e^{im\phi}\left(\frac{r-r_{+}}{r-r_{-}}\right)^{-im\beta\Omega/2}\left(+e^{-t/\beta}\sqrt{\Delta}\right)^{\ell-s} \,,
	\end{gathered}
\ee
is singular at the future event horizon, but regular at the past one. Now it is instructive to rewrite $L^{\left(s\right)}_{+1}$ as
\be
	L^{\left(s\right)}_{+1}= L_{+1} - 2s\,e^{+t/\beta}\partial_{r}\left(\sqrt{\Delta}\right) \,,
\ee
where $L_{+1}$ is the vector in the scalar field case, which is manifestly regular at the horizon.

\begin{figure}[t]
	\centering
	\begin{subfigure}[b]{0.49\textwidth}
		\centering
		\begin{tikzpicture}
			\node at (0,1) (uml3) {$\upsilon_{-\ell,\ell+s}^{\left(m,s\right)}$};
			\node at (0,2) (uml2) {$\upsilon_{-\ell,2}^{\left(m,s\right)}$};
			\node at (0,3) (uml1) {$\upsilon_{-\ell,1}^{\left(m,s\right)}$};
			\node at (0,4) (uml0) {$\upsilon_{-\ell,0}^{\left(m,s\right)}$};
			
			\draw (1,1) -- (5,1);
			\node at (3,1.5) (up) {$\vdots$};
			\node at (3,0.5) (um) {$\vdots$};
			\draw (1,2) -- (5,2);
			\draw (1,3) -- (5,3);
			\draw (1,4) -- (5,4);
			\draw [snake=zigzag] (1,4.1) -- (5,4.1);
			
			\draw[red] [<-] (2.5,2) -- node[left] {$L_{-1}$} (2.5,3);
			\draw[red] [<-] (2,3) -- node[left] {$L_{-1}$} (2,4);
			\draw[blue] [->] (4,3) -- node[right] {$L_{+1}$} (4,4);
			\draw[blue] [->] (3.5,2) -- node[right] {$L_{+1}$} (3.5,3);
		\end{tikzpicture}
		\caption{A highest-weight $\SL$ representation that contains spin-$s$ perturbations of the Kerr-Newman black hole that are regular on the future event horizon.}
	\end{subfigure}
	\hfill
	\begin{subfigure}[b]{0.49\textwidth}
		\centering
		\begin{tikzpicture}
			\node at (0,3) (upll) {$\bar{\upsilon}_{+\ell,\ell-s}^{\left(m,s\right)}$};
			\node at (0,2) (upl2) {$\bar{\upsilon}_{+\ell,2}^{\left(m,s\right)}$};
			\node at (0,1) (upl1) {$\bar{\upsilon}_{+\ell,1}^{\left(m,s\right)}$};
			\node at (0,0) (upl0) {$\bar{\upsilon}_{+\ell,0}^{\left(m,s\right)}$};
			
			\draw [snake=zigzag] (1,-0.1) -- (5,-0.1);
			\draw (1,0) -- (5,0);
			\draw (1,1) -- (5,1);
			\draw (1,2) -- (5,2);
			\node at (3,2.5) (up) {$\vdots$};
			\draw (1,3) -- (5,3);
			\node at (3,3.5) (um) {$\vdots$};
			
			\draw[blue] [->] (2.5,0) -- node[left] {$L_{+1}$} (2.5,1);
			\draw[blue] [->] (2,1) -- node[left] {$L_{+1}$} (2,2);
			\draw[red] [<-] (4,1) -- node[right] {$L_{-1}$} (4,2);
			\draw[red] [<-] (3.5,0) -- node[right] {$L_{-1}$} (3.5,1);
		\end{tikzpicture}
		\caption{A lowest-weight $\SL$ representation that contains spin-$s$ perturbations of the Kerr-Newman black hole that are regular on the past event horizon.}
	\end{subfigure}
	\caption[The infinite-dimensional highest-weight and lowest-weight representations of $\SL$ whose elements solve the leading order near-zone spin-$s$ Teukolsky equation for the Kerr-Newman black hole with multipolar index $\ell$ and contain the static solutions.]{The infinite-dimensional highest-weight and lowest-weight representations of $\SL$ whose elements solve the leading order near-zone spin-$s$ Teukolsky equation for the Kerr-Newman black hole with multipolar index $\ell$ and contain the static solutions.}
	\label{fig:HWSL2RKerrS}
\end{figure}
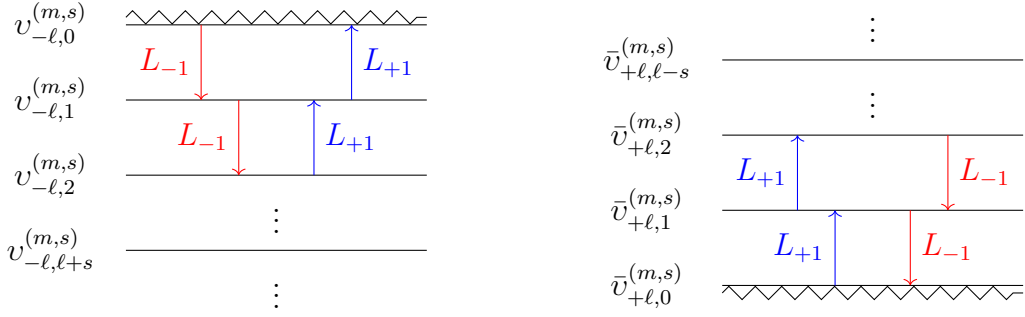

A Newman-Penrose scalar $\Psi_s$ that is regular at the future event horizon must have the form $\Delta^{-s}f$ with $f$ being a function that is regular on the future event horizon~\cite{Teukolsky:1972my,Teukolsky:1973ha}. For any function $f$ we have
\be
	L^{\left(s\right)}_{+1}\left(\Delta^{-s}f\right)  = \Delta^{-s}L_{+1}f \,,\quad L^{\left(s\right)}_{-1}\left(\Delta^{-s}f\right) = \Delta^{-s} \left[L_{-1} - 2s\,e^{-t/\beta}\partial_{r}\left(\sqrt{\Delta}\right)\right]f \,.
\ee
As $L_{\pm1} f$ is regular for regular $f$, any vector of the form $\Delta^s(L^{\left(s\right)}_{\pm 1})^n\left[\Delta^{-s}f\right]$ is regular too. This way, acting on $\upsilon_{-\ell,0}^{\left(m,s\right)}$ with $L^{\left(s\right)}_{\pm1}$, we will get new regular solutions thereby generating a multiplet containing solutions to the leading order near-zone Teukolsky equation that are regular at the future event horizon. The descendant with $L_0$-eigenvalue equal to $s$ is then identified with the regular static solution,
\be
	\Psi_{s,\omega=0,\ell m} \propto \upsilon_{-\ell,\ell+s}^{\left(m,s\right)} = \left(L^{\left(s\right)}_{-1}\right)^{\ell+s}\upsilon_{-\ell,0}^{\left(m,s\right)} \,.
\ee
This state satisfies the annihilation condition $(L_{+1}^{\left(s\right)})^{\ell+s+1}\upsilon_{-\ell,\ell+s}^{\left(m,s\right)} \propto L_{+1}^{\left(s\right)}\upsilon_{-\ell,0}^{\left(m,s\right)}=0$. Noting that, for any $n$ and $F\left(r\right)$,
\be
	\begin{split}
		&\left(L^{(s)}_{+1}\right)^{n}\left[e^{im\phi}\Delta^{-s}\left(\frac{r-r_{+}}{r-r_{-}}\right)^{im\beta\Omega/2} F\left(r\right)\right] \\
		&= e^{im\phi}\Delta^{-s}\left(\frac{r-r_{+}}{r-r_{-}}\right)^{im\beta\Omega/2}\left(-e^{+t/\beta}\sqrt{\Delta}\right)^{n} \frac{d^{n}}{dr^{n}}F\left(r\right) \,,
	\end{split}
\ee
the physical static solution $\upsilon_{-\ell,\ell+s}^{\left(m,s\right)}=e^{im\phi}\Delta^{-s}\left(\frac{r-r_{+}}{r-r_{-}}\right)^{im\beta\Omega/2}F\left(r\right)$ satisfies
\be
	\left(L_{+1}^{\left(s\right)}\right)^{\ell+s+1}\upsilon_{-\ell,\ell+s}^{\left(m,s\right)} = e^{im\phi} \Delta^{-s}\left(\frac{r-r_{+}}{r-r_{-}}\right)^{im\beta\Omega/2}\left(-e^{+t/\beta}\sqrt{\Delta}\right)^{\ell+s+1} \frac{d^{\ell+s+1}}{dr^{\ell+s+1}}F\left(r\right) = 0 \,,
\ee
which implies that $F\left(r\right)$ is a polynomial of order $\ell+s$. Consequently, the static solution is given by
\be\label{eq:QuasiPolynomialSL2R_KerrNewmanS}
	\Psi_{s,\omega=0,\ell m} = e^{im\phi}\Delta^{-s}\left(\frac{r-r_{+}}{r-r_{-}}\right)^{im\beta\Omega/2}\sum_{n=0}^{\ell+s}c_{n}r^{n} \,.
\ee
From this quasi-polynomial expression, we conclude that black hole Love numbers with respect to general spin-$s$ static perturbations vanish, in accordance with the results of the previous chapter. The construction of the highest-weight and lowest-weight Verma modules of Love symmetry for spin-$s$ perturbations of the Kerr-Newman black hole is depicted in Figure~\ref{fig:HWSL2RKerrS}.

\section{Running Vs Non-running}
\label{sec:SL2RZeroBeta_KerrNewmanS}

Another feature of the Love symmetry is that it offers an algebraic explanation of the non-running of Love numbers. We have already touched upon the algebraic manifestation of RG flow for the static Love numbers. Namely, the static solutions that are regular or singular at the future event horizon belong to locally distinguishable representations of the Love $\SL$ symmetry. For the case of a scalar field in the background of the four-dimensional Schwarzschild black hole, the static solution regular at the event horizon belongs to the finite-dimensional highest-weight representation in Figure~\ref{fig:HWSL2RSchwarzschild}, while the singular static solution belongs to the infinite-dimensional representation in Figure~\ref{fig:SingSL2RSchwarzschild}. These two satisfy the selection rules
\be\ba
	\text{Regular static solution: }&\left(L_{+1}\right)^{\ell+1}\Phi_{\omega=0,\ell m}^{\text{regular}} = 0 \,, \\
	\text{Singular static solution: }&L_{-1}\left(L_{+1}\right)^{\ell+1}\Phi_{\omega=0,\ell m}^{\text{singular}} = 0 \,,
\ea\ee
which can indeed be distinguished at any point in spacetime. In the notation of~\cite{Howe1992}, the representations in Figure~\ref{fig:HWSL2RSchwarzschild} and Figure~\ref{fig:SingSL2RSchwarzschild} are of type-``$[\circ]$'' and type-``$\circ]\circ[\circ$'' respectively.

For the case of spin-$s$ perturbations of the Kerr-Newman black hole, the relevant representations are the highest-weight and lowest-weight Verma modules of $\SL$; these are depicted in Figure~\ref{fig:HWSL2RKerr} for $s=0$ and in Figure~\ref{fig:HWSL2RKerrS} for $s\ne0$. In the notation of~\cite{Howe1992}, the highest-weight and lowest-weight Verma modules are of type-``$[\circ$'' and type-``$\circ]$'' respectively. The regular and singular branches of the static solution now satisfy the annihilation conditions
\be\ba
	\text{Regular static solution: }&\left(L_{+1}^{\left(s\right)}\right)^{\ell+s+1}\Psi_{s,\omega=0,\ell m}^{\text{regular}} = 0 \,, \\
	\text{Singular static solution: }&\left(L_{-1}^{\left(s\right)}\right)^{\ell-s+1}\Psi_{s,\omega=0,\ell m}^{\text{singular}} = 0 \,,
\ea\ee
which are again locally distinguishable and, hence, the absence of logarithmic running is outputted as a selection rule. 

This might sound a bit tautological since we have already shown that the highest-weight property implies vanishing static Love numbers without referring to any energy scale at all, but the above algebraic local criteria are in fact important to infer the absence of RG flow for the other resonant conditions in Table~\ref{tbl:NZLNs_KerrNewmanS}, and their $s\ne0$ generalizations. Indeed, we will see now how the Love symmetry can address the non-running for all the resonant conditions associated with purely imaginary frequencies, $-i\beta\omega = k \in \mathbb{Z}$. To do this, we will first show that the highest-weight and lowest-weight Verma modules with weights $h=-\ell$ and $\bar{h}=+\ell$ are in fact extendable to the reducible type-``$\circ[\circ[\circ$'' and type-``$\circ]\circ]\circ$'' $\SL$ modules respectively.

\subsection{The type-``$\circ[\circ[\circ$'' and type-``$\circ]\circ]\circ$'' modules}
In the previous analyses, we only focused on showing that the static solution regular (singular) at the future event horizon was an element of the highest-weight (lowest-weight) Verma module with weight $h=-\ell$ ($\bar{h}=+\ell$) of the Love $\SL$ symmetry. Here, we will see in that the type-``$[\circ$'' and type-``$\circ]$'' representations encountered in the previous sections are in fact only a subspace of the type-``$\circ[\circ[\circ$'' and type-``$\circ]\circ]\circ$'' representations respectively.

To demonstrate this, let us first count how many highest-weight representations of $\SL$ one can construct. A primary state $\upsilon_{h,0}$ of weight $h$ satisfies the following conditions
\be
	L_0\upsilon_{h,0} = h\,\upsilon_{h,0} \,,\quad L_{+1}\upsilon_{h,0} = 0
\ee
and, therefore,
\be
	\mathcal{C}_2\upsilon_{h,0} = h\left(h-1\right)\upsilon_{h,0} \,.
\ee
To make contact with the problem at hand, it is customary to label the Casimir eigenvalues as $\ell\left(\ell+1\right)$. As a result, there are in fact two possible highest-weight $\SL$ representations, with weights
\be\label{eq:hHW12}
	h^{\text{hw}}_1 = -\ell \quad\text{OR}\quad h^{\text{hw}}_2 = +\ell+1 \,.
\ee
The descendants of these primary states will then have $L_0$-eigenvalues
\be
	\upsilon_{h,n} = \left(L_{-1}\right)^{n}\upsilon_{h,0} \quad\Rightarrow\quad L_0\upsilon_{h,n} = \left(h+n\right)\upsilon_{h,n} \,.
\ee
Consequently, as long as the states in the two highest-weight representations satisfy the same boundary conditions and the weights differ by an integer number, one highest-weight representation forms an invariant subspace of the other and together they furnish the larger reducible, but indecomposable, representation of type-``$[\circ[\circ$''. In particular, the primary state with the larger weight is a descendant of the primary state with the lower weight. In the notation of Eq.~\eqref{eq:hHW12} with $2\ell\in\mathbb{N}$,
\be
	\upsilon_{+\ell+1,0} = \upsilon_{-\ell,2\ell+1} = \left(L_{-1}\right)^{2\ell+1}\upsilon_{-\ell,0} \,,\quad 2\ell\in\mathbb{N} \,.
\ee
This representation can also be augmented to a type-``$\circ[\circ[\circ$'' representation by appending to the multiplet vectors $\upsilon_{-\ell,-n-1}$, $n\ge0$, spanning the ladder above the primary state with the lower weight. These are all ascendants of $\upsilon_{-\ell,-1}$,
\be
	\upsilon_{-\ell,-n-1} = \frac{1}{n!\left(2\ell+2\right)_{n}}\left(L_{+1}\right)^{n}\upsilon_{-\ell,-1} \,,
\ee
which satisfies the inhomogeneous equation
\be
	L_{-1}\upsilon_{-\ell,-1} = \upsilon_{-\ell,0} \,.
\ee

As for the possible lowest-weight $\SL$ representations, the lowest-weight vector $\bar{\upsilon}_{\bar{h},0}$ of weight $\bar{h}$ satisfies the following conditions
\be
	L_0\bar{\upsilon}_{\bar{h},0} = \bar{h}\,\bar{\upsilon}_{\bar{h},0} \,,\quad L_{-1}\bar{\upsilon}_{\bar{h},0} = 0
\ee
and, hence,
\be
	\mathcal{C}_2\bar{\upsilon}_{\bar{h},0} = \bar{h}\left(\bar{h}+1\right)\bar{\upsilon}_{\bar{h},0} \equiv \ell\left(\ell+1\right)\bar{\upsilon}_{\bar{h},0} \quad\Rightarrow\quad \bar{h}^{\text{lw}}_1 = +\ell \quad\text{OR}\quad \bar{h}^{\text{lw}}_2 = -\ell-1 \,.
\ee
The $L_0$-spectrum of the corresponding ascendants is
\be
	\bar{\upsilon}_{\bar{h},n} = \left(-L_{+1}\right)^{n}\bar{\upsilon}_{\bar{h},0} \quad\Rightarrow\quad L_0\bar{\upsilon}_{\bar{h},n} = \left(\bar{h}-n\right)\bar{\upsilon}_{\bar{h},n}
\ee
and, therefore, as long as the states in the two lowest-weight representations satisfy the same boundary conditions and the weights differ by an integer number, one lowest-weight representation forms an invariant subspace of the other and together they furnish the larger reducible, but indecomposable, representation of type-``$\circ]\circ]$''. The lowest-weight state with the lower weight is then identified as an ascendant of the lowest-weight state with the larger weight,
\be
	\bar{\upsilon}_{-\ell-1,0} = \bar{\upsilon}_{+\ell,2\ell+1} = \left(-L_{+1}\right)^{2\ell+1}\bar{\upsilon}_{+\ell,0} \,,\quad 2\ell\in\mathbb{N} \,.
\ee
Appending with descendants $\bar{\upsilon}_{+\ell,-n-1}$, $n\ge0$, of a vector $\bar{\upsilon}_{+\ell,-1}$ satisfying an inhomogeneous equation,
\be
	L_{+1}\bar{\upsilon}_{+\ell,-1} = -\bar{\upsilon}_{+\ell,0} \,,\quad \bar{\upsilon}_{+\ell,-n-1} = \frac{\left(-1\right)^{n}}{n!\left(2\ell+2\right)_{n}}\left(L_{-1}\right)^{n}\bar{\upsilon}_{+\ell,-1} \,,
\ee
the vectors $\bar{\upsilon}_{+\ell,j}$, $j\in\mathbb{Z}$, will span the entire representation of type-``$\circ]\circ]\circ$''. These constructions are depicted in Figure~\ref{fig:FullHWSL2R}.

\begin{figure}[t]
	\centering
	\begin{subfigure}[b]{0.49\textwidth}
		\centering
		\begin{tikzpicture}
			\node at (0,-1) (uml5) {$\upsilon_{-\ell,2\ell+2}$};
			\node at (0,0) (uml4) {$\upsilon_{-\ell,2\ell+1}$};
			\node at (0,1) (uml3) {$\upsilon_{-\ell,2\ell}$};
			\node at (0,2) (uml2) {$\upsilon_{-\ell,2}$};
			\node at (0,3) (uml1) {$\upsilon_{-\ell,1}$};
			\node at (0,4) (uml0) {$\upsilon_{-\ell,0}$};
			\node at (0,5) (umlm1) {$\upsilon_{-\ell,-1}$};
			\node at (0,6) (umlm2) {$\upsilon_{-\ell,-2}$};
			
			\node at (3,-1.5) (um) {$\vdots$};
			\draw (1,-1) -- (5,-1);
			\draw [snake=zigzag] (1,0.1) -- (5,0.1);
			\draw (1,0) -- (5,0);
			\draw (1,1) -- (5,1);
			\node at (3,1.5) (up) {$\vdots$};
			\draw (1,2) -- (5,2);
			\draw (1,3) -- (5,3);
			\draw (1,4) -- (5,4);
			\draw [snake=zigzag] (1,4.1) -- (5,4.1);
			\draw (1,5) -- (5,5);
			\draw (1,6) -- (5,6);
			\node at (3,6.5) (up2) {$\vdots$};
			
			\draw[red] [<-] (2,-1) -- node[left] {$L_{-1}$} (2,0);
			\draw[red] [<-] (3,0) -- node[left] {$L_{-1}$} (3,1);
			\draw[red] [<-] (2.5,2) -- node[left] {$L_{-1}$} (2.5,3);
			\draw[red] [<-] (2,3) -- node[left] {$L_{-1}$} (2,4);
			\draw[red] [<-] (3,4) -- node[left] {$L_{-1}$} (3,5);
			\draw[red] [<-] (2,5) -- node[left] {$L_{-1}$} (2,6);
			\draw[blue] [->] (4,5) -- node[right] {$L_{+1}$} (4,6);
			\draw[blue] [->] (4,3) -- node[right] {$L_{+1}$} (4,4);
			\draw[blue] [->] (3.5,2) -- node[right] {$L_{+1}$} (3.5,3);
			\draw[blue] [->] (4,-1) -- node[right] {$L_{+1}$} (4,0);
		\end{tikzpicture}
		\caption{The type-``$\circ[\circ[\circ$'' representation a subspace of which is the highest-weight Verma module of $\SL$ with weight $h=-\ell$.}
	\end{subfigure}
	\hfill
	\begin{subfigure}[b]{0.49\textwidth}
		\centering
		\begin{tikzpicture}
			\node at (0,5) (upl5) {$\bar{\upsilon}_{+\ell,2\ell+2}$};
			\node at (0,4) (upl4) {$\bar{\upsilon}_{+\ell,2\ell+1}$};
			\node at (0,3) (upl3) {$\bar{\upsilon}_{+\ell,2\ell}$};
			\node at (0,2) (upl2) {$\bar{\upsilon}_{+\ell,2}$};
			\node at (0,1) (upl1) {$\bar{\upsilon}_{+\ell,1}$};
			\node at (0,0) (upl0) {$\bar{\upsilon}_{+\ell,0}$};
			\node at (0,-1) (uplm1) {$\bar{\upsilon}_{+\ell,-1}$};
			\node at (0,-2) (uplm2) {$\bar{\upsilon}_{+\ell,-2}$};
			
			\node at (3,-2.5) (um2) {$\vdots$};
			\draw (1,-2) -- (5,-2);
			\draw (1,-1) -- (5,-1);
			\draw [snake=zigzag] (1,-0.1) -- (5,-0.1);
			\draw (1,0) -- (5,0);
			\draw (1,1) -- (5,1);
			\draw (1,2) -- (5,2);
			\node at (3,2.5) (up) {$\vdots$};
			\draw (1,3) -- (5,3);
			\draw (1,4) -- (5,4);
			\draw [snake=zigzag] (1,3.9) -- (5,3.9);
			\draw (1,5) -- (5,5);
			\node at (3,5.5) (um) {$\vdots$};
			
			\draw[blue] [->] (2,-2) -- node[left] {$L_{+1}$} (2,-1);
			\draw[blue] [->] (3,-1) -- node[left] {$L_{+1}$} (3,0);
			\draw[blue] [->] (2,0) -- node[left] {$L_{+1}$} (2,1);
			\draw[blue] [->] (2.5,1) -- node[left] {$L_{+1}$} (2.5,2);
			\draw[blue] [->] (3,3) -- node[left] {$L_{+1}$} (3,4);
			\draw[blue] [->] (2,4) -- node[left] {$L_{+1}$} (2,5);
			\draw[red] [<-] (4,4) -- node[right] {$L_{-1}$} (4,5);
			\draw[red] [<-] (3.5,1) -- node[right] {$L_{-1}$} (3.5,2);
			\draw[red] [<-] (4,0) -- node[right] {$L_{-1}$} (4,1);
			\draw[red] [<-] (4,-2) -- node[right] {$L_{-1}$} (4,-1);
		\end{tikzpicture}
		\caption{The type-``$\circ]\circ]\circ$'' representation a subspace of which is the lowest-weight Verma module of $\SL$ with weight $\bar{h}=+\ell$.}
	\end{subfigure}
	\caption[The type-``$\circ\mathrm{[}\circ\mathrm{[}\circ$'' and type-``$\circ\mathrm{]}\circ\mathrm{]}\circ$'' representations of which the highest-weight and lowest-weight Verma modules of $\SL$ with weights $h=-\ell$ and $\bar{h}=+\ell$ are invariant subspaces respectively, under the conditions that $2\ell\in\mathbb{N}$ and that the elements satisfy the same boundary conditions.]{The type-``$\circ[\circ[\circ$'' and type-``$\circ]\circ]\circ$'' representations of which the highest-weight and lowest-weight Verma modules of $\SL$ with weights $h=-\ell$ and $\bar{h}=+\ell$ are invariant subspaces respectively, under the conditions that $2\ell\in\mathbb{N}$ and that the elements satisfy the same boundary conditions.}
	\label{fig:FullHWSL2R}
\end{figure}

We now focus to the $\SL$ generators \eqref{eq:SL2RsKerrNewman} relevant for spin-$s$ perturbations of the Kerr-Newman black hole. The corresponding highest-weight states with definite azimuthal number $m$ can be easily be solved for and are given up to a normalization constant by
\be
	\upsilon_{h,0}^{\left(m,s\right)} = e^{im\phi}\Delta^{-s}\left(\frac{r-r_{+}}{r-r_{-}}\right)^{im\beta\Omega/2}\left(-e^{+t/\beta}\sqrt{\Delta}\right)^{s-h} \,.
\ee
These states are regular at the future event horizon and singular at the past one for \text{any} value of $h$. Consequently, the actual highest-weight Verma module we encountered in Section~\ref{sec:SL2R_KerrNewmanS} is of type-``$[\circ[\circ$''. We can also construct the ladder spanned by states $\left\{\upsilon_{-\ell,-n-1}^{\left(m,s\right)}|n\ge0\right\}$ above the highest-weight state $\upsilon_{-\ell,0}^{\left(m,s\right)}$. The starting point is to solve the inhomogeneous first order differential equation
\be
	\begin{gathered}
		L_{-1}^{\left(s\right)}\upsilon_{-\ell,-1}^{\left(m,s\right)} = \upsilon_{-\ell,0}^{\left(m,s\right)} \,, \\
		\ba
			\Rightarrow\quad \upsilon_{-\ell,-1}^{\left(m,s\right)} &= -e^{im\phi}\Delta^{-s}\left(\frac{r-r_{+}}{r-r_{-}}\right)^{im\beta\Omega/2}\left(-e^{+t/\beta}\sqrt{\Delta}\right)^{\ell+s+1} \\
			&\quad\times\frac{{}_2F_1\left(1+,2\ell+2;2+\ell+im\beta\Omega;-\frac{r-r_{+}}{r_{+}-r_{-}}\right)}{\left(r_{+}-r_{-}\right)\left(\ell+1+im\beta\Omega\right)} \,.
		\ea
	\end{gathered}
\ee
The state $\upsilon_{-\ell,-1}^{\left(m,s\right)}$ is also regular (singular) at the future (past) event horizon and, therefore, the entire upper ladder is spanned by vectors
\be
	\upsilon_{-\ell,-\ell-n-1}^{\left(m,s\right)} = \frac{1}{n!\left(2\ell+2\right)_{n}}\left(L_{+1}^{\left(s\right)}\right)^{n}\upsilon_{-\ell,-1}^{\left(m,s\right)}
\ee
sharing the same near-horizon boundary conditions. As a result, the solutions regular at the future event horizon that solve the leading order near-zone equations of motion and have an integer $L_0$-spectrum are all elements of the type-``$\circ[\circ[\circ$'' reducible representation of the Love $\SL$ symmetry. Similarly, the corresponding solutions singular at the future event horizon can be showed to span the entire type-``$\circ]\circ]\circ$'' representation.

Looking at conditions associated with imaginary frequencies, $-i\beta\omega=k\in\mathbb{Z}$, for which the leading order near-zone Love numbers do not exhibit running, the corresponding solutions have the form
\be
	\Psi_{s\omega\ell m}\bigg|_{-i\beta\omega=k} = e^{kt/\beta}e^{im\phi}R_{s\omega\ell m}\left(r\right)\bigg|_{-i\beta\omega=k} \,,
\ee
where we are suppressing the $\theta$-dependence, and therefore satisfy
\be
	L_0^{\left(s\right)}\Psi_{s\omega\ell m}\bigg|_{-i\beta\omega=k} = \left(s-k\right)\,\Psi_{s\omega\ell m}\bigg|_{-i\beta\omega=k} \,.
\ee
Consequently, they cover the entire integer-valued $L_0$-spectrum. As we just saw, the branch regular at the future event horizon spans the type-``$\circ[\circ[\circ$'' module of the Love $\SL$ symmetry, while the singular branch spans the locally distinguishable type-``$\circ]\circ]\circ$'' representation; this is the algebraic manifestation of the absence of RG flow for the resonant conditions $-i\beta\omega=k\in\mathbb{Z}$.

If $k\le\ell+s$, the regular at the future event horizon solution is, in particular, recognized to be the $n=\ell+s-k$ descendant in the highest-weight multiplet that the static solution belongs to,
\be
	\Psi_{s\omega\ell m}\bigg|_{-i\beta\omega=k} \propto \upsilon_{-\ell,\ell+s-k}^{\left(m,s\right)} = \left(L_{-1}^{\left(s\right)}\right)^{\ell+s-k}\upsilon_{-\ell,0}^{\left(m,s\right)} \,.
\ee
It satisfied the annihilation condition
\be
	\left(L_{+1}^{\left(s\right)}\right)^{\ell+s-k+1}\Psi_{s\omega\ell m}\bigg|_{-i\beta\omega=k} = 0 \,,
\ee
while the quasi-polynomial form, Eq.~\eqref{eq:QuasiPolynomialSL2R_KerrNewmanS}, of the static descendant can be used to infer that all these states are quasi-polynomial as well. However, only the vectors belonging to the quotient representation sandwiched by the two primary states, i.e. for $-\ell \le k \le \ell$, will yield polynomials whose large distance expansion does not contain a decaying, $\propto r^{-\ell-1}$, term and, therefore, we arrive at the algebraic interpretation of these states as the near-zone-approximated solutions with vanishing spin-$s$ Love numbers.

As for all the other non-resonant conditions, $-i\beta\omega\notin\mathbb{Z}$, for which the leading order near-zone Love numbers exhibit an RG flow, this can be algebraically seen from the fact that the solutions regular and singular at the future event horizon belong to the same $\SL$ representation. More specifically, they are both elements of the $W\left(4\ell\left(\ell+1\right),0\right)$ standard module\footnote{We refer to Appendix~\ref{app:SL2RRepresentations} for a list of the standard modules of the $\SL$ algebra.}, and, hence, there is no local algebraic criteria for distinguishing the regular and singular branches.

There are two features of the Love numbers that we have not addressed yet via Love symmetry representation theory arguments. One concerns the resonant conditions corresponding to imaginary azimuthal numbers, $im\beta\Omega\in\mathbb{Z}$. As we will see in the next chapter, these arise as a selection rules of a second commuting $\SL$ structure manifesting itself in the near-zone region. This second $\SL$ is in fact only \textit{locally} defined which makes sense once we recall that imaginary azimuthal numbers give rise to conical singularities in the perturbation field profiles.

The second feature we have not addressed here is the peculiar vanishing of the leading order near-zone scalar Love numbers of the Schwarzschild black hole at \textit{all} orders in frequency, see Eq.~\eqref{eq:SLNsDissipativeRCs_Schwarzschild4d}. We defer this analysis for the next chapter where we will reveal a hyperparameter extension of the Love symmetry capable of capturing this situation and hence explaining the corresponding vanishing of the near-zone-approximated Love numbers for spherically symmetric black holes via highest-weight arguments similar to the ones presented here.
	\newpage
\chapter{Properties and Generalizations of Love symmetry}
\label{ch:Properties}

In the previous chapter we demonstrated the existence of an $\SL$ structure for the particular near-zone splitting \eqref{eq:NZSplitOP_KerrNewmanS}-\eqref{eq:NZSplitV0V1_KerrNewmanS}, which we dubbed ``Love symmetry'' based on its capability of explaining the vanishing of the spin-$s$ static Love numbers for the Kerr-Newman black hole~\cite{Charalambous:2021kcz,Charalambous:2022rre}. In this chapter, we will explore various properties and generalizations of this near-zone conformal structure by adjusting results from our work in Ref.~\cite{Charalambous:2022rre}, starting with the observation that there is a second particularly interesting near-zone splitting of the Teukolsky equation which is equipped with a globally defined $\SL$ structure: the Starobinsky near-zone approximation~\cite{Starobinsky:1973aij,Starobinskil:1974nkd,Charalambous:2021kcz,Charalambous:2022rre}. At leading order in the Starobinsky near-zone expansion, the equations of motion for the Kerr-Newman black hole have the remarkable property of outputting vanishing Love numbers for all frequencies and angular momenta~\cite{Charalambous:2022rre}. We will see that this vanishing of the dynamical Love numbers at leading order in the near-zone expansion can be algebraically explained as selection rules arising from a continuous set of $\SL$ algebras constructed via a hyperparameter-extension of the Love symmetry~\cite{Charalambous:2022rre}.

We will next complete the enhanced symmetry algebraic explanation of the non-running/vanishing of the near-zone Love numbers for spin-$s$ perturbations of four dimensional black holes in General Relativity by demonstrating that the globally defined Love $\SL$ symmetry associated with the near-zone splitting \eqref{eq:NZSplitOP_KerrNewmanS}-\eqref{eq:NZSplitV0V1_KerrNewmanS} is in fact a local $\SL\times\SL$ symmetry, with the second $\SL$ factor being \text{non-globally} defined. This second $\SL$ factor turns out to exactly capture all the remaining resonant conditions in Table~\ref{tbl:NZLNs_KerrNewmanS} associated with imaginary azimuthal numbers which the globally defined Love $\SL$ symmetry was not able to address. The algebraic argument will be identical in nature as the one used to explain the resonant conditions associated with imaginary frequencies: the leading order near-zone solutions with $im\beta\Omega\in\mathbb{Z}$ span the entire type-``$\circ[\circ[\circ$'' representation of the new non-globally defined $\SL$ factor~\cite{Charalambous:2023jgq}.

After these analyses, we will present a remarkable infinite extension of the Love symmetry $\SL$ algebra into $\SL\ltimes\hat{U}\left(1\right)$ which contains both the Love symmetry and the Starobinsky near-zone algebra as two particular subalgebras~\cite{Charalambous:2021kcz,Charalambous:2022rre}. This infinite extension will turn out to play an important role when studying the extremal limit in the next chapter, to find a relation with the isometries of the near-horizon throat of extremal black holes. We will furthermore see that there are actually infinitely many near-zone truncations of the Teukolsky equation that give rise to globally defined $\SL$ symmetries; these will turn out be classifiable into two towers of near-zone $\SL$ structures, arising from local temporal translations onto the Love and Starobinsky near-zone symmetries~\cite{Charalambous:2022rre}.

Moving forwards, we will demonstrate that the near-zone symmetries acquire a geometric interpretation as isometries of effective black hole geometries, known as ``subtracted geometries''~\cite{Cvetic:2011dn,Cvetic:2011hp}. These are geometries that preserve the internal structure of the black hole but approximate (``subtract'') the information associated with the environment. This observation will be particularly useful in attempting to give a geometric interpretation of the scalar pieces for the $s\ne0$ $\SL$ algebras via spin-weighted Lie derivatives~\cite{Ludwig2000,Ludwig:2001hx,Charalambous:2022rre}.

In Section~\ref{sec:LoveBeyondGR4d}, we will study the prospect of near-zone $\SL$ structures arising in a general metric theory of gravity, not necessarily General Relativity. We will do this by studying the scalar response problem of a generic static, spherically symmetric, asymptotically flat and non-extremal black hole in four spacetime dimensions~\cite{Charalambous:2022rre}. We will extract a sufficient geometric constraint for the existence of Love symmetry and show that such black holes have vanishing static scalar Love numbers explained by the highest-weight property of the representation that the corresponding static solution belongs to. The Riemann-cubed modification of the Schwarzschild black hole we encountered in Chapter~\ref{ch:TLNsBlackHoles4d} will turn out \textit{not} to satisfy this geometric constraint and, hence, Love symmetry appears not to exist for this type of theories of gravity giving stronger evidence in favor of the connection between vanishing Love numbers and existence of near-zone symmetries~\cite{Charalambous:2022rre}.

Last, in Section~\ref{sec:TTMs4d}, we will infer the states of vanishing Love numbers as total transmission modes~\cite{Cook:2016fge,Cook:2016ngj,MaassenvandenBrink:2000iwh,Hod:2013fea}. This will be achieved by matching the Love numbers onto the gauge invariant scattering cross-sections and then showing that vanishing Love numbers imply reflectionless modes~\cite{Ivanov:2022qqt}.

\section{Starobinsky near-zone splitting}
\label{sec:SL2R_Starobinsky}

As already mentioned, there are infinitely many possible near-zone truncations of the equations of motion. So far, we have been utilizing the particular near-zone splitting given in Eqs.~\eqref{eq:NZSplitOP_KerrNewmanS}-\eqref{eq:NZSplitV0V1_KerrNewmanS} of the Teukolsky equation, which had the special property of admitting the globally defined Love $\SL$ symmetry we saw in the previous chapter.

There is a second particularly interesting near-zone splitting that was historically the first one to be applied to the Teukolsky equation. This is the Starobinsky near-zone approximation and is given by~\cite{Starobinsky:1973aij,Starobinskil:1974nkd}
\be\label{eq:V0V1_StarobinskyS}
	\begin{gathered}
		\mathbb{O}_{\text{full}}^{\left(s\right)} = \Delta^{-s}\partial_{r}\,\Delta^{s+1}\,\partial_{r} + V_0^{\text{Star}} + \epsilon\,V_1^{\text{Star}} + s\left(s+1\right) \,, \\\\
		\ba
			V_0^{\text{Star}} &= - \frac{\left(r_{+}^2+a^2\right)^2}{\Delta}\left(\partial_{t}+\Omega\,\partial_{\phi}\right)^2 + s\frac{\left(r_{+}^2+a^2\right)\Delta^{\prime}}{\Delta}\left(\partial_{t}+\Omega\,\partial_{\phi}\right) \,, \\
			V_1^{\text{Star}} &= -\frac{\left(r+r_{+}\right)\left(r^2+r_{+}^2+2a^2\right)}{r-r_{-}}\,\partial_{t}^2 - \frac{4Ma}{r-r_{-}}\,\partial_{t}\partial_{\phi} + 2s\left[M\frac{r-r_{+}}{r-r_{-}}-r\right]\,\partial_{t} \,.
		\ea
	\end{gathered}
\ee
Following the procedure prescribed in Chapter~\ref{ch:TLNsBlackHoles4d}, the radial wavefunction that is ingoing at the future event horizon at leading order in this near-zone expansion can be found to be~\cite{Charalambous:2022rre}
\be\ba
	R_{s\omega\ell m} &= \bar{R}_{\ell m}^{\text{in}}\left(\omega\right)\Delta^{-s}\left(\frac{r-r_{+}}{r-r_{-}}\right)^{iZ_{+}\left(\omega\right)}{}_2F_1\left(-\ell-s,\ell+1-s;1+2iZ_{+}\left(\omega\right);-\frac{r-r_{+}}{r_{+}-r_{-}}\right) \,,
\ea\ee
with the near-horizon exponent $Z_{+}\left(\omega\right)$ given in Eq.~\eqref{eq:Zpm_KerrNewman} and we have yet to use the fact that the physical values of the orbital number are integers. Expanding around large distances while analytically continuing $\ell\in\mathbb{N}\rightarrow\ell\in\mathbb{R}$, the response coefficients are extracted to be, after massaging via the $\Gamma$-function mirror formula and eventually sending the orbital number to range in its physical integer values $\ell\ge\left|s\right|$~\cite{Chia:2020yla,Charalambous:2021mea,Charalambous:2022rre},
\be
	k_{\ell m}^{\left(s\right)}\left(\omega\right) = \frac{i\sinh2\pi Z_{+}\left(\omega\right)}{2\pi}\left|\Gamma\left(\ell+1-2iZ_{+}\left(\omega\right)\right)\right|^2\left(-1\right)^{s+1}\frac{\left(\ell+s\right)!\left(\ell-s\right)!}{\left(2\ell\right)!\left(2\ell+1\right)!}\left(\frac{r_{+}-r_{-}}{r_{s}}\right)^{2\ell+1} \,.
\ee
These are purely imaginary and therefore correspond to dissipative effects, while the conservative Love numbers are zero for \textit{all} values of the perturbation frequency in the Starobinsky near-zone approximation. Of course, the above response coefficients are only accurate within the leading order near-zone regime but, nevertheless, this observation calls for a symmetry explanation analogous to the static case. As we saw in Chapter~\ref{ch:TLNsBlackHoles4d}, this property is shared by scalar perturbations of the four-dimensional Schwarzschild black hole. In fact, for $s=0$, the only qualitative change to the near-zone splitting is the replacement $\partial_{t}\rightarrow\partial_{t}+\Omega\,\partial_{\phi}$ in Eq.~\eqref{eq:NZsplitting_Schwarzschild4d} which is a purely frame dragging effect that ensures that the near-horizon dynamics are preserved.

\subsection{Starobinsky near-zone algebra}
There is another reason why the Starobinsky near-zone approximation is interesting: it is also equipped with a globally defined $\SL$ structure, now generated by
\be\label{eq:SL2RStarobinsky}
	\begin{gathered}
		L_0^{\text{Star},\left(s\right)} = -\beta\left(\partial_{t}+\Omega\partial_{\phi}\right) \,,\\
		L_{\pm1}^{\text{Star},\left(s\right)} = e^{\pm t/\beta}\left[\mp\sqrt{\Delta}\,\partial_{r} + \partial_{r}\left(\sqrt{\Delta}\right)\beta\left(\partial_{t}+\Omega\partial_{\phi}\right) \mp s\frac{r-r_{\mp}}{\sqrt{\Delta}}\right] \,,
	\end{gathered}
\ee
with $\beta$ the inverse Hawking temperature \eqref{eq:betaKerrNewman} of the Kerr-Newman black hole. We will denote this algebra as $\SL_{\text{Star}}$. The Casimir of $\SL_{\text{Star}}$ does indeed reproduce the $\epsilon=0$ Teukolsky differential equation in Eq.~\eqref{eq:V0V1_StarobinskyS}. The charge of a separable solution $\Psi_{s\omega\ell m}$ under the action of $L_0^{\text{Star},\left(s\right)}$ is
\be
	L_0^{\text{Star},\left(s\right)} \Psi_{s\omega\ell m} = i\beta\left(\omega-m\Omega\right)\Psi_{s\omega\ell m}\,,
\ee
from which we see that black hole perturbations at the locking frequency,
\[
	\omega = m\Omega	\,,
\]
play a special role in the Starobinsky near-zone approximation. Namely, these black hole perturbations form highest-weight $\SL_{\text{Star}}$ representations as can be shown completely analogously to the static case for the Love $\SL$ symmetry. This highest-weight Verma module, along with its extended type-``$\circ[\circ[\circ$'' structure, is spanned by states with spectrum $\omega = m\Omega + i2\pi T_{H} k$, $k\in\mathbb{Z}$, which matches the ``near-horizon quasinormal modes'' studied in~\cite{Zimmerman:2011dx}, and correspond to solutions regular at the future event horizon with vanishing/non-running response coefficients in whole. As far as static Love numbers are concerned, the Starobinsky near-zone algebra allows us to make only one exact statement: Love numbers vanish for axisymmetric static perturbations, $m=0$, $\omega=0$~\cite{Poisson:2014gka,Landry:2015zfa,Gurlebeck:2015xpa,LeTiec:2020spy,LeTiec:2020bos}.

Another interesting property of the Starobinsky near-zone regime is the emergence of a modified rotational symmetry, produced by the following generators
\be
	J_0 = -i\,\partial_{\phi} \,,\quad J_{\pm1} = e^{\pm i\left(\phi-\Omega t\right)}\left[\partial_{\theta} \pm i\cot\theta\,\partial_{\phi} \mp \frac{s}{\sin\theta}\right] \,,
\ee
such that the full near-zone symmetry gets enhanced to $\SL_{\text{Star}}\times SO\left(3\right)$. This can be contrasted with the usual Love near-zone $\SL$ which does not commute with the $SO\left(3\right)$ of the near-zone-approximated angular equations of motion in the general Kerr-Newman background. The emergence of the modified spherical symmetry is an intriguing fact hinting at a more general symmetry structure of the near-zone Teukolsky equation.

\subsection{A finite frequency generalization}
We will now present a symmetry explanation of the vanishing of the spin-$s$ Love numbers for Kerr-Newman black holes at leading order in the Starobinsky near-zone expansion, starting with the $s=0$ scalar perturbations for simplicity. This is achieved by the following generators
\be\label{eq:SL2RStarobinskyOmega}
	\begin{gathered}
		L_0^{\left(\tilde{\omega}\right)} = -\beta\left(\partial_{t}+i\tilde{\omega}\right) \,,\\
		L_{\pm1}^{\left(\tilde{\omega}\right)} = e^{\pm t/\beta}\left[\mp\sqrt{\Delta}\partial_{r} + \partial_{r}\left(\sqrt{\Delta}\right)\beta\,\partial_{t} + \frac{a}{\sqrt{\Delta}}\,\partial_{\phi} + i\beta \tilde{\omega}\sqrt{\frac{r-r_{+}}{r-r_{-}}}\right] \,,
	\end{gathered}
\ee
defined for some arbitrary parameter $\tilde{\omega}$. These generators are regular in advanced/retarded null coordinates \eqref{eq:NullCoordiantes_KerrNewman}. In fact, the vector parts of the generators are just the Love symmetry vector fields \eqref{eq:SL2RKerrNewman}, while the supplementary scalar pieces are themselves regular at both the future and the past event horizon. The modified generators form an $\SL$ algebra which we dub $\SL_{\tilde{\omega}}$. Its quadratic Casimir is given by
\be\label{eq:wtilde}
	\mathcal{C}_2^{\left(\tilde{\omega}\right)} = \partial_{r}\,\Delta\,\partial_{r} - \frac{\left(r_{+}^2+a^2\right)^2}{\Delta}\left[\left(\partial_{t}+\Omega\,\partial_{\phi}\right)^2 + 4\frac{r-r_{+}}{r_{+}-r_{-}}\left(\partial_{t}+i\tilde{\omega}\right)\left(\Omega\,\partial_{\phi}-i\tilde{\omega}\right)\right] \,.
\ee
A separable scalar black hole perturbation $\Phi_{\omega\ell m}$ of frequency $\omega$ is an eigenstate of $L_0^{\left(\tilde \omega\right)}$ with eigenvalue
\be
	L_0^{\left(\tilde \omega\right)} \Phi_{\omega\ell m} = i\beta\left(\omega - \tilde{\omega}\right)\Phi_{\omega\ell m} \,,
\ee
so that the null $L_0^{\left(\tilde{\omega}\right)}$-eigenstate is a monochromatic black hole perturbation with frequency $\omega=\tilde\omega$. By comparing \eqref{eq:wtilde} and \eqref{eq:V0V1_StarobinskyS}, we observe that null $L_0^{\left(\tilde{\omega}\right)}$-eigenstates satisfy the $s=0$ Teukolsky differential equation in the Starobinsky near-zone approximation. Hence, analogously to the arguments of Chapter~\ref{ch:LoveSymmetry4d}, one concludes that any regular black hole perturbation of finite frequency $\omega=\tilde\omega$ in the Starobinsky near-zone approximation belongs to a highest-weight representation of the $\SL_{\tilde \omega}$ algebra \eqref{eq:SL2RStarobinskyOmega}. The corresponding highest-weight vector, satisfying
\be
	L_{+1}^{\left(\tilde{\omega}\right)}\upsilon_{-\ell,0}^{\left(m,\tilde{\omega}\right)} = 0 \,,\quad L_0^{\left(\tilde{\omega}\right)}\upsilon_{-\ell,0}^{\left(m,\tilde{\omega}\right)} = -\ell\,\upsilon_{-\ell,0}^{\left(m,\tilde{\omega}\right)} \,,\quad J_0\upsilon_{-\ell,0}^{\left(m,\tilde{\omega}\right)} = m\,\upsilon_{-\ell,0}^{\left(m,\tilde{\omega}\right)} \,,
\ee
has the following form
\be
	\upsilon_{-\ell,0}^{\left(m,\tilde{\omega}\right)} = e^{-i\tilde{\omega}t}e^{im\phi}\left(\frac{r-r_{+}}{r-r_{-}}\right)^{i Z_{+}\left(\tilde{\omega}\right)}\left(-e^{+t/\beta}\sqrt{\Delta}\right)^{\ell} \,.
\ee
It describes a regular at the future event horizon solution with frequency $\omega_{-\ell,0} = \tilde{\omega} +i\ell/\beta$. The $L^{\left(\tilde{\omega}\right)}_0$-null vector $\upsilon_{-\ell,\ell}^{\left(m,\tilde{\omega}\right)}$ is then a regular solution of the scalar Teukolsky equation at frequency $\omega_{-\ell,\ell}=\tilde\omega$ in the Starobinsky near-zone approximation. It satisfies
\be
	\left(L^{\left(\tilde{\omega}\right)}_{+1}\right)^{\ell+1}\upsilon_{-\ell,\ell}^{\left(m,\tilde{\omega}\right)} = 0 \,,
\ee
which implies the following quasi-polynomial form of the solution
\be
	\Phi_{\omega=\tilde{\omega},\ell m} \propto \upsilon_{-\ell,\ell}^{\left(m,\tilde{\omega}\right)} = e^{-i\tilde{\omega}t}e^{im\phi}\left(\frac{r-r_{+}}{r-r_{-}}\right)^{iZ_{+}\left(\tilde{\omega}\right)}\sum_{n=0}^{\ell}c_{n}r^{n} \,.
\ee
The overall form-factor can be removed by a transition into the advanced null coordinates and plays no role in the worldline EFT matching calculation of Love numbers. The remaining polynomial does not contain negative powers in $r$ and hence the frequency-dependent Love numbers vanish identically.

The extension of the $\SL_{\tilde{\omega}}$ algebra to spin-$s$ fields is straightforward. One has to start with the usual spin-$s$ Love generators~\eqref{eq:SL2RsKerrNewman} and add the same $\tilde{\omega}$-dependent pieces as in Eq.~\eqref{eq:SL2RStarobinskyOmega}. For monochromatic perturbations $\Psi_{s\omega\ell m}$ with $\omega=\tilde{\omega}$, the corresponding action of the Casimir reduces to the spin-$s$ Teukolsky equation in the Starobinsky near-zone approximation.

This proves the very intriguing property of the Starobinsky near-zone approximation of vanishing linear conservative response at all frequencies. It is important to stress again that, unlike for static Love numbers, this property does not hold for the full response, i.e. beyond the leading order near zone approximation. Indeed, as already examined in Section~\ref{sec:LNs_KerrNewmanS}, using the full solution to the Teukolsky equation~\cite{Mano:1996vt,Mano:1996gn}, one can see that the conservative black hole response does not vanish already at the linear order in $\omega$~\cite{Charalambous:2021mea}.

Also, it is currently unclear whether this algebraic argument is a qualitatively new piece of information or simply a restatement of the result obtained by a direct solution of the corresponding differential equation. The problem is the interpretation of the hyperparameter $\tilde{\omega}$ in the general case of non-monochromatic solutions, and the geometric meaning of the scalar $\tilde\omega$-dependent generators appearing in Eq.~\eqref{eq:SL2RStarobinskyOmega}.

\section{Local $\SL\times\SL$ Love symmetry}
\label{sec:LocalSL2RLove4d}

For the near-zone truncation \eqref{eq:NZSplitOP_KerrNewmanS}-\eqref{eq:NZSplitV0V1_KerrNewmanS} for which the Love symmetry becomes manifest, we were so far able to address the vanishing of the leading order near-zone spin-$s$ Love numbers of the Kerr-Newman black hole only for the resonant conditions associated with imaginary perturbation frequencies. Somewhat surprisingly, we can also address the remaining resonant conditions $im\beta\Omega=k\in\mathbb{Z}$, even for $\omega\ne0$. This is ought to the observation that the particular near-zone truncation is equipped with a larger $\SL_{\text{L}}\times\SL_{\text{R}}$ structure. The first $\SL$ factor is the Love symmetry $\SL$,
\be
	\SL_{\text{L}}=\SL_{\text{Love}} \,,
\ee
whose globally defined generators are given in Eq.~\eqref{eq:SL2RsKerrNewman}. The generators of the second, $\SL_{\text{R}}$, factor are
\be\label{eq:SL2RR_KerrNewmanS}
	\begin{gathered}
		L_0^{\text{R},\left(s\right)} = -\beta\Omega\,\partial_{\phi} \,, \\
		L_{\pm1}^{\text{R},\left(s\right)} = e^{\pm\phi/\left(\beta\Omega\right)}\left[\mp\sqrt{\Delta}\partial_{r}+\partial_{r}\left(\sqrt{\Delta}\right)\beta\Omega\,\partial_{\phi} + \frac{r_{+}-r_{-}}{2\sqrt{\Delta}}\beta\,\partial_{t} \mp s\frac{r-r_{\mp}}{\sqrt{\Delta}} \right] \,.
	\end{gathered}
\ee
The Casimirs of the two commuting $\SL$'s are exactly the same,
\be
	\mathcal{C}_2^{\text{R},\left(s\right)} = \mathcal{C}_2^{\text{L},\left(s\right)} \,,
\ee
and equal to near-zone truncation \eqref{eq:NZSplitOP_KerrNewmanS}-\eqref{eq:NZSplitV0V1_KerrNewmanS}. Transforming the vector parts into advanced ($+$)/retarded ($-$) null coordinates, reveals that the vector fields $L_{m}^{\text{R},\left(s=0\right)}$ do not develop poles as $r\rightarrow r_{+}$, while the corresponding singular behavior of the scalar pieces for $s\ne0$ is assigned to the employment of the Kinnersley tetrad~\cite{Kinnersley1969,Teukolsky:1972my,Teukolsky:1973ha}. However, only the $U\left(1\right)_{\text{R}}$ subgroup of $\SL_{\text{R}}$ is globally defined by virtue of the periodic identification of the azimuthal angle, $\phi\sim\phi+2\pi$, under which $L_{\pm1}^{\text{R},\left(s\right)}$ develop conical deficits,
\be
	L_{\pm1}^{\text{R},\left(s\right)} \xrightarrow{\phi\rightarrow\phi+2\pi}e^{\pm2\pi/\left(\beta\Omega\right)}L_{\pm1}^{\text{R},\left(s\right)} \,.
\ee
As such, this local $\SL_{\text{R}}$ maps regular solutions onto solutions with conical singularities. Furthermore, the generators of $\SL_{\text{R}}$ do not have a smooth spinless, $\Omega\rightarrow0$ limit.

Despite this rather unphysical property of $\SL_{\text{R}}$, it can still be used to explain the vanishing/non-running associated to the situations where the azimuthal number of the perturbation is imaginary, $im\beta\Omega=k\in\mathbb{Z}$, in a similar fashion as in the previous chapter. These resonant conditions share the same degree of ``non-physicality'' in the sense that the corresponding solutions also suffer from conical singularities. Nevertheless, generic solutions of the near-zone equations of motion will furnish representations labeled by the Casimir and $L_0$-eigenvalues,
\be
	\mathcal{C}_2^{\text{R},\left(s\right)}\Psi_{s\omega\ell m}=\ell\left(\ell+1\right)\Psi_{s\omega\ell m} \,,\quad L_0^{\text{R},\left(s\right)}\Psi_{s\omega\ell m} = -im\beta\Omega\,\Psi_{s\omega\ell m} \,.
\ee

Let us construct the analogous highest-weight representation of $\SL_{\text{R}}$. The primary state with highest-weight $h^{\text{R}}=-\ell$, satisfying
\be
	L_0^{\text{R},\left(s\right)}\upsilon_{-\ell,0}^{\left(\omega,s\right)} = -\ell\,\upsilon_{-\ell,0}^{\left(\omega,s\right)} \,,\quad 	L_{+1}^{\text{R},\left(s\right)}\upsilon_{-\ell,0}^{\left(\omega,s\right)} = 0 \,,
\ee
and having definite frequency $\omega$ is given by, up to an overall normalization constant,
\be
	\upsilon_{-\ell,0}^{\left(\omega,s\right)} = e^{-i\omega t}\left(r-r_{+}\right)^{-s}\left(\frac{r-r_{+}}{r-r_{-}}\right)^{-i\beta\omega/2}\left(-e^{+\phi/\left(\beta\Omega\right)}\sqrt{\Delta}\right)^{\ell} \,.
\ee
This state is singular at the past event horizon and has a conical singularity as we go around the azimuthal circle, but develops no pole at the future event horizon with respect to the radial variable. The descendants,
\be
	\upsilon_{-\ell,n}^{\left(\omega,s\right)} = \left(L_{+1}^{\text{R},\left(s\right)}\right)^{n}\upsilon_{-\ell,0}^{\left(\omega,s\right)} \,,
\ee
share the same boundary conditions, with the conical deficit measured by their charge under $L_0^{\text{R},\left(s\right)}$,
\be
	L_0^{\text{R},\left(s\right)}\upsilon_{-\ell,n}^{\left(\omega,s\right)} = \left(n-\ell\right)\upsilon_{-\ell,n}^{\left(\omega,s\right)} \,.
\ee

However, there exists a particular descendant that has no conical singularity and is therefore truly regular at the future event horizon. This is the $n=\ell$ descendant which is neutral under the action of $L_0^{\text{R},\left(s\right)}$ and corresponds to the axisymmetric, $m=0$, regular near-zone solution. Noticing that
\be\ba
	\left(L_{+1}^{R,\left(s\right)}\right)^{n}&\left(e^{-i\omega t}\left(r-r_{+}\right)^{-s}\left(\frac{r-r_{+}}{r-r_{-}}\right)^{-i\beta\omega/2}F\left(r\right)\right) = \\
	&e^{-i\omega t} \left(r-r_{+}\right)^{-s} \left(\frac{r-r_{+}}{r-r_{-}}\right)^{-i\beta\omega/2} \left[-e^{+\phi/\left(\beta\Omega\right)}\sqrt{\Delta}\right]^{n}\frac{d^{n}}{dr^{n}}F\left(r\right) \,,
\ea\ee
for generic $F\left(r\right)$, we see that the highest-weight property,
\be
	\left(L_{+1}^{\text{R},\left(s\right)}\right)^{\ell+1}\Psi_{s\omega\ell,m=0} = 0 \,,
\ee
implies the following quasi-polynomial form for the radial wavefunction
\be
	R_{s\omega\ell,m=0} = e^{-i\omega t} \left(r-r_{+}\right)^{-s} \left(\frac{r-r_{+}}{r-r_{-}}\right)^{-i\beta\omega/2}\sum_{n=0}^{\ell}c_{n}r^{n} \,,
\ee
which exactly matches the one from \eqref{eq:RadialWFkL_KerrNewman} for $s=0$, $k=0$ we wanted to address. The absence of RG flow is also encoded in the representation theory analysis of $\SL_{\text{R}}$, with the solution singular at the future event horizon (and regular at the past event horizon) being the $\ell$'th ascendant of the locally distinguishable lowest-weight representation of $\SL_{\text{R}}$ with weight $\bar{h}^{\text{R}}=+\ell$.

For the other situations of vanishing/non-running scalar Love numbers for which the relevant near-zone solutions develop conical singularities, $im\beta\Omega=k\ne0$, we can follow the same procedure as in the Section~\ref{sec:SL2RZeroBeta_KerrNewmanS} and show that all the relevant regular (singular) at the future event horizon near-zone solutions span the entire representation of type-``$\circ[\circ[\circ$'' (type-``$\circ]\circ]\circ$'').

This completes the algebraic explanation of vanishing/non-running Love numbers at leading order in the near-zone expansion for four-dimensional asymptotically flat and non-extremal black holes in General Relativity. It is worth noting that similar local $\SL\times\SL$ structures have been previously constructed in the context of the Kerr/CFT proposal~\cite{Castro:2010fd,Chen:2010zwa,Lowe:2011aa}. However, in general, neither of the two $\SL$ factors is in globally defined, again due to conical deficits accompanying the action of the generators. As we will soon see in this chapter, the Love and Starobinsky near-zone symmetries are in fact the best one can do, i.e. they are the \textit{only} globally defined near-zone $\SL$ structures.

\section{Extremal middle zones}
\label{sec:ExtremalMiddleZones}

At the end of Section~\ref{sec:LNs_KerrNewmanS}, we saw that the static Love numbers of extremal Kerr-Newman black holes are also zero. In this section, we will see that in the extremal case there is another set of generators that allows us to derive the vanishing of Love numbers from group theory arguments. Let us first focus on the Reissner-Nordstr{\"o}m case. Consider the following set of vector fields regular at the future horizon~\cite{Charalambous:2022rre}
\be
	\begin{split}
		& L_0 = -2M\partial_{t} +s \,, \\ 
		&L_{\pm 1}= e^{\pm \left(\frac{t}{2M} -\frac{M}{2\left(r-M\right)}\right)} \left[\mp \left(r-M\right)\partial_{r}  + 2M\left(1\pm \frac{M}{2\left(r-M\right)}\right)\partial_{t} -s\left(1\pm 1\right)\right] \,.
	\end{split}
\ee
The corresponding Casimir is given by
\be
	\begin{split}
		\mathcal{C}_2=&\left(r-M\right)^{-2s}\partial_{r}\left(r-M\right)^{2\left(s+1\right)}\partial_{r} + \frac{M^4}{\left(r-M\right)^2}\,\partial_{t}^2 \\
		&-2M^2\,\partial_{t} \partial_{r} - 2s\frac{M^2}{r-M}\,\partial_{t} + s\left(s+1\right) \,.
	\end{split} 
\ee
Strictly speaking, it matches the Teukolsky equation \eqref{eq:TeukolskyExtremalFull} only in the static limit; the $\partial_{t}^2$ term has a wrong sign, and hence the Casimir provides a good approximation to the physical Teukolsky operator only in the regime
\be\label{eq:mz}
	\omega M^2 \ll r-M \ll 1/\omega \,.
\ee
We call this region the ``middle zone'' in what follows. The appearance of the middle zone is natural because the extremal near-horizon patch decouples from the asymptotically flat region. The near-horizon region, in fact, becomes an infinite throat that is described by an $\text{AdS}_2$ geometry. As we will see in detail in the next chapter, the Killing vectors of this near-horizon geometry will be capable of reproducing the static solution and deriving the properties of static Love numbers~\cite{Charalambous:2022rre}. However, the Love numbers themselves are defined through the matching in the asymptotically flat space patch, and the fact that they can be fully extracted from the near-horizon approximation looks like a miracle. This might simply be a result of some accidental degeneracy that takes place in the Reissner-Nordstr{\"o}m case.

In contrast to the near-horizon region, the middle zone interpolates between the asymptotic infinity and the near-horizon region. Hence, it is natural to expect that it is this symmetry that should be important for Love numbers of extremal black holes. Indeed, we can use this symmetry for an alternative derivation of the vanishing of Love numbers of extremal four-dimensional black holes. However, the near-horizon symmetry is a symmetry of the background geometry itself,
while right now we do not have a similar interpretation for the ``middle zone'' symmetry. It remains to be seen if there is a deeper reason behind the appearance of this symmetry.

Importantly, there is an analog of the middle zone for the extremal rotating black holes, which captures solutions that are not even present in the $\text{AdS}_2$ near-horizon throat. The static part of the extremal Kerr-Newman black hole Teukolsky equation can be reproduced with the following $\SL$ generators,
\be\label{eq:algmid}
	\begin{split}
		L_0 &= -2M\partial_{t} + s \,, \\ 
		L_{\pm 1} &= e^{\pm \left(\frac{t}{2M} -\frac{M^2+a^2}{2M\left(r-M\right)}\right)} \bigg[\mp \left(r-M\right)\partial_{r}  + 2M \left(1\pm \frac{M^2+a^2}{2M\left(r-M\right)}\right)\partial_{t} \\
		&\qquad\qquad\qquad\qquad\,\,\,\, +\frac{a}{r-M}\,\partial_{\phi} - s\left(1\pm 1\right)\bigg] \,,
	\end{split} 
\ee
which are regular in advanced null coordinates. The Casimir of this algebra,
\be
	\begin{split}
		&\mathcal{C}_2= \left(r-M\right)^{-2s}\partial_{r} \left(r-M\right)^{2\left(s+1\right)} \partial_{r} +\frac{\left(M^2+a^2\right)^2}{\left(r-M\right)^2} \left(\partial_{t}^2 - \Omega^2\,\partial_{\phi}^2\right) \\
		&-\frac{4M a}{r-M}\,\partial_{t}\partial_{\phi} - 2\left(M^2+a^2\right) \partial_{t} \partial_{r} - 2s\frac{M^2+a^2}{r-M}\left(\partial_{t} - \Omega\,\partial_{\phi}\right) + s\left(s+1\right) \,,
	\end{split}
\ee
is again somewhat different from the full extremal Kerr-Newman low-frequency Teukolsky operator \eqref{eq:TeukolskyExtremalFull}, but it is still accurate provided that Eq.~\eqref{eq:mz} holds true. Importantly, the middle zone symmetry also captures states with mode frequencies different from the locking one, and hence it is suitable to describe dynamics outside the throat.

We can use the middle zone symmetry for an algebraic derivation of the vanishing of Love numbers for general static non-axisymmetric perturbations. Unlike the Reissner-Nordsr{\"o}m case, this statement cannot be made within the non-extremal Love symmetry or the near-horizon symmetry. The proof goes essentially the same way as in the non-extremal Kerr-Newman case. The crucial observation is that the static solution belongs to the highest-weight $\SL$ representation with $h=-\ell$, which dictates its polynomial form in $r$.

\section{An infinite-dimensional extension}
\label{sec:InfiniteExtension4d}

Very general arguments~\cite{Hofman:2011zj} suggest that the $\SL$ symmetry discussed so far is just a small part of a full infinite dimensional algebra. We note that the proof of~\cite{Hofman:2011zj} does not apply here directly, because it relies on unitarity, and the representations encountered so far are all non-unitary. Nevertheless, there are indications that the Love $\SL$ symmetry is indeed a part of a much larger algebraic structure, for instance we have seen in Section~\ref{sec:SL2R_Starobinsky} that, for non-extremal black holes, in addition to the near-zone approximation leading to the Love symmetry, the Starobinsky near-zone approximation also has very special and interesting properties.

We will show now that these two near-zone approximations can be naturally combined into a single algebraic structure. The construction is based on the observation that, for any regular at the horizon $\SL$ representation $\mathcal{V}$, the Love algebra can be extended into a semidirect product $\SL\ltimes U\left(1\right)_{\mathcal{V}}$. Focusing to the $s=0$ case to begin with, the $U\left(1\right)_{\mathcal{V}}$ factor is generated by vector fields of the form $\upsilon\,\beta\Omega\,\partial_{\phi}$ with $\upsilon\in \mathcal{V}$.

A representation $\mathcal{V}$, which leads to a unified description of the $s=0$ Love symmetry and Starobinsky near-zone $\SL$, can be constructed as follows. We start with $\upsilon_{0,0}=-1$, which is a single element of the one-dimensional unitary $\SL$ representation (``singleton'') of the Love symmetry~\cite{Barut1965}. It has $h=0$ and also satisfies $L_{\pm1}\upsilon_{0,0} = 0$. Then, we solve for the vectors $\upsilon_{0,\pm 1}$ satisfying 
\be
	L_{0}\upsilon_{0,\pm1} = \mp\upsilon_{0,\pm1} \,,\quad L_{\mp 1}\upsilon_{0,\pm 1}=\mp \upsilon_{0,0} \,.
\ee 
These give us
\be
	\upsilon_{0,\pm1} = e^{\pm t/\beta}\sqrt{\frac{r-r_{+}}{r-r_{-}}} \,, \quad \upsilon_{0,0}=-1 \,.
\ee
All these functions are regular at the future and past event horizons. By definition, starting with $\upsilon_{0,\pm1}$ one can reach $\upsilon_{0,0}$ by acting on them with $L_{\pm1}$. However, one cannot reach $\upsilon_{0,\pm1}$ starting from $\upsilon_{0,0}$. A (reducible) representation which contains all three vectors $\upsilon_{0,0}$, $\upsilon_{0,\pm1}$ is spanned by the following $\upsilon_{0,n}$ vectors
\be
	\upsilon_{0,\pm n} = \left(L_{\pm1}\right)^{n-1}\upsilon_{0,\pm1} = \left(\pm1\right)^{n-1}\left(n-1\right)! \,e^{\pm n t/\beta}\left(\frac{r-r_{+}}{r-r_{-}}\right)^{n/2} \,,
\ee
with $n\in\mathbb{N}$. We present a graphical representation of this construction in Fig.~\ref{fig:VSL2R4d}.

\begin{figure}
	\centering
	\begin{tikzpicture}
		\node at (0,0) (um3) {$\upsilon_{0,-3}$};
		\node at (0,1) (um2) {$\upsilon_{0,-2}$};
		\node at (0,2) (um1) {$\upsilon_{0,-1}$};
		\node at (0,3) (u0) {$\upsilon_{0,0}$};
		\node at (0,4) (up1) {$\upsilon_{0,+1}$};
		\node at (0,5) (up2) {$\upsilon_{0,+2}$};
		\node at (0,6) (up3) {$\upsilon_{0,+3}$};
		
		\node at (3,-0.4) (um) {$\vdots$};
		\draw (1,0) -- (5,0);
		\draw (1,1) -- (5,1);
		\draw (1,2) -- (5,2);
		\draw [snake=zigzag] (1,2.9) -- (5,2.9);
		\draw (1,3) -- (5,3);
		\draw [snake=zigzag] (5,3.1) -- (1,3.1);
		\draw (1,4) -- (5,4);
		\draw (1,5) -- (5,5);
		\draw (1,6) -- (5,6);
		\node at (3,6.4) (up) {$\vdots$};
		
		\draw[blue] [->] (2,0) -- node[left] {$L_{+1}$} (2,1);
		\draw[blue] [->] (2.5,1) -- node[left] {$L_{+1}$} (2.5,2);
		\draw[blue] [->] (3,2) -- node[left] {$L_{+1}$} (3,3);
		\draw[blue] [->] (2.5,4) -- node[left] {$L_{+1}$} (2.5,5);
		\draw[blue] [->] (2,5) -- node[left] {$L_{+1}$} (2,6);
		\draw[red] [<-] (4,0) -- node[right] {$L_{-1}$} (4,1);
		\draw[red] [<-] (3.5,1) -- node[right] {$L_{-1}$} (3.5,2);
		\draw[red] [<-] (3,3) -- node[right] {$L_{-1}$} (3,4);
		\draw[red] [<-] (3.5,4) -- node[right] {$L_{-1}$} (3.5,5);
		\draw[red] [<-] (4,5) -- node[right] {$L_{-1}$} (4,6);
	\end{tikzpicture}
	\caption[A representation $\mathcal{V}$ of $\SL$ used to construct the $\SL\ltimes U\left(1\right)_{\mathcal{V}}$ extension of the Love algebra in four spacetime dimensions.]{A representation $\mathcal{V}$ of $\SL$ used to construct the $\SL\ltimes U\left(1\right)_{\mathcal{V}}$ extension of the Love algebra in four spacetime dimensions.}
	\label{fig:VSL2R4d}
\end{figure}

This way, we obtain the general $U\left(1\right)_{\mathcal{V}}$ representation that extends the Love $\SL$ symmetry into $\SL\ltimes \hat{U}\left(1\right)_{\mathcal{V}}$,
\be\label{eq:U1V}
	\upsilon = \sum_{n=-\infty}^\infty \alpha_{n} \upsilon_{0,n}\,\beta\Omega\,\partial_\phi \,.
\ee 

A one-parameter family of the $\SL$ subalgebras from this $\SL\ltimes U\left(1\right)_{\mathcal{V}}$ is of a particular interest,
\be\label{eq:StarLove}
	L_{m}\left(\alpha\right) = L_{m} + \alpha\,\upsilon_{0,m}\,\beta\Omega\,\partial_{\phi} \,, \quad m=0,\pm1 \,.
\ee
The Casimir of this algebra is given by 
\be 
	\mathcal{C}_2\left(\alpha\right) = \partial_{r}\,\Delta\,\partial_{r} - \frac{\left(r_{+}^2+a^2\right)^2}{\Delta}\left(\partial_{t}+\Omega\,\partial_{\phi}\right)^2 + 2\frac{r_{+}^2+a^2}{r-r_{-}}\left(\partial_{t}+\alpha\,\Omega\,\partial_{\phi}\right)\left(\alpha-1\right)\beta\Omega\,\partial_{\phi} \,.
\ee 
In general, this Casimir does not capture the physical near-zone limit: its static part does not match that of the scalar Teukolsky equation unless $\alpha=0$ or $\alpha = 1$. The first choice, $\alpha=0$, corresponds to the Love symmetry, see Eq.~\eqref{eq:NZSplitRadial_KerrNewman}. The second special choice $\alpha = 1$ reduces this Casimir to the one matching the Starobinsky near-zone split, see Eq.~\eqref{eq:V0V1_StarobinskyS} with $s=0$.

Nevertheless, it does capture the near-horizon characteristic exponents of the Teukolsky equation. In fact, as we show explicitly in Appendix~\ref{app:SL2RGenerators}, this one-parameter family of $\SL$ subalgebras precisely contains \textit{all} possible globally defined and ``time-reversal'' symmetric\footnote{``Time-reversal'' here refers to the simultaneous time-reversal transformation $t\rightarrow-t$ and the flip of the angular momentum of the black hole, $a\rightarrow-a$.} approximations with an $\SL$ enhancement that preserve these near-horizon dynamics. This property will prove crucial in the next chapter where a connection with enhanced isometries of near-horizon geometries of extremal black holes will be discussed.

For non-zero spin-weight $s$, the above construction goes through identically, with the exception that the $U\left(1\right)$ vector in \eqref{eq:U1V} is supplemented with a scalar piece,
\be\label{eq:U1Vs}
	\beta\Omega\,\partial_{\phi} \rightarrow \beta\Omega\,\partial_{\phi} + s \,.
\ee
The one-parameter family of $\SL_{\left(\alpha\right)}$ subalgebras of this $s$-extended $\SL\ltimes U\left(1\right)_{\mathcal{V}}$ are then generated by
\be\label{eq:SL2Ralpha_s}
	L_{m}^{\left(s\right)}\left(\alpha\right) = L_{m}^{\left(s\right)} + \alpha\,\upsilon_{0,m}\left(\beta\Omega\,\partial_{\phi} + s\right) \,,\quad m=0,\pm1 \,,
\ee
where $L_{m}^{\left(s\right)}$ are the $s$-extended Love generators \eqref{eq:SL2RsKerrNewman} and $\upsilon_{0,m}$ belong to the representation $\mathcal{V}$ constructed from $L_{m}^{\left(s=0\right)}$ (see Figure~\ref{fig:VSL2R4d}). The associated Casimir is given by,
\be
	\begin{aligned}
		\mathcal{C}_2^{\left(s\right)}\left(\alpha\right) &= \Delta^{-s}\partial_{r}\Delta^{s+1}\partial_{r} - \frac{\left(r_{+}^2+a^2\right)^2}{\Delta}\left(\partial_{t}+\Omega\,\partial_{\phi}\right)^2 \\
		&\quad+ s\frac{\left(r_{+}^2+a^2\right)\Delta^{\prime}}{\Delta}\left(\partial_{t}+\Omega\,\partial_{\phi}\right) + s\left(s+1\right) \\
		&\quad+ 2\frac{r_{+}^2+a^2}{r-r_-}\left(\partial_{t}+\alpha\,\Omega\,\partial_{\phi} + \alpha\,\frac{s}{\beta}\right)\left(\alpha-1\right)\beta\Omega\,\partial_{\phi} \,.
	\end{aligned}
\ee
Again, even though all of these operators preserve the near-horizon characteristic exponents of the non-extremal Teukolsky equation, only the choices $\alpha=0$ and $\alpha=1$ give rise to valid near-zone approximations, corresponding to the $s$-extended Love, Eqs.~\eqref{eq:NZSplitOP_KerrNewmanS}-\eqref{eq:NZSplitV0V1_KerrNewmanS}, and Starobinsky, Eq.~\eqref{eq:V0V1_StarobinskyS}, near-zones respectively.

\section{Infinite zones of Love from local time translations}
\label{sec:SL2RtTranslations4d}

Interestingly, by carefully solving the constraints that need to be satisfied for a near-zone $\SL$ symmetry to exist as is sketched in Appendix~\ref{app:SL2RGenerators}, the most general forms of the generators come into two classes. These are generalizations of the Love and Starobinsky near-zones and are controlled by an arbitrary radial function $g\left(r\right)$ that is regular at the horizon, $g\left(r_{+}\right)=\text{finite}$, according to
\be\label{eq:SL2RExtension4d}
	\begin{gathered}
		L_0\left[g\left(r\right)\right] = L_0\left[0\right] \,, \\
		L_{\pm1}\left[g\left(r\right)\right] = e^{\pm g\left(r\right)/\beta}L_{\pm1}\left[0\right] \pm e^{\pm\left(t+g\left(r\right)\right)/\beta}\sqrt{\Delta}g^{\prime}\left(r\right)\partial_{t} \,,
	\end{gathered}
\ee
where primes denote radial derivatives and $L_0\left[0\right]$ and $L_{\pm1}\left[0\right]$ are the already found expressions for the Love and Starobinsky near-zone $\SL$ generators in \eqref{eq:SL2RsKerrNewman} and \eqref{eq:SL2RStarobinsky} respectively. These generalizations are not simple rescalings of $L_{\pm1}$, which trivially leave the algebra and the Casimir unchanged. They are \textit{local} rescalings accompanied with an additional $t$-component which are realized as local, $r$-dependent, translations of the temporal coordinate,
\be
	t \rightarrow \tilde{t} = t + g\left(r\right) \,.
\ee
This is seen more transparently by comparing \eqref{eq:SL2RsKerrNewman} and \eqref{eq:SL2RStarobinsky} in $\left(t,r,\phi\right)$ coordinates with the explicit expressions for the generalized generators in $\left(\tilde{t},r,\phi\right)$ coordinates,
\begin{itemize}
	\item \underline{Generalized Love near-zone}:
	\be
		\begin{gathered}
			L_0\left[g\right] = -\beta\partial_{\tilde{t}} + s \\
			L_{\pm1}\left[g\right] = e^{\pm\tilde{t}/\beta} \left[ \mp\sqrt{\Delta}\,\partial_{r} + \partial_{r}\left(\sqrt{\Delta}\right)\beta\,\partial_{\tilde{t}} + \frac{a}{\sqrt{\Delta}}\,\partial_{\phi} - s\left(1\pm1\right)\partial_{r}\left(\sqrt{\Delta}\right) \right]
		\end{gathered}
	\ee
	
	\item \underline{Generalized Starobinsky near-zone}:
	\be
		\begin{gathered}
			L_0^{\text{Star}}\left[g\right] = -\beta\left(\partial_{\tilde{t}}+\Omega\partial_{\phi}\right) \\
			L_{\pm1}^{\text{Star}}\left[g\right] = e^{\pm\tilde{t}/\beta} \left[ \mp\sqrt{\Delta}\,\partial_{r} + \partial_{r}\left(\sqrt{\Delta}\right)\beta\left(\partial_{\tilde{t}}+\Omega\,\partial_{\phi}\right) \mp s\frac{r-r_{\mp}}{\sqrt{\Delta}} \right]
		\end{gathered}
	\ee
\end{itemize}
The corresponding equations of motion arising from the Casimirs of the above two generalized near-zone $\SL$'s are the same as the ``fundamental'' ($g=0$) ones but with $t$ replaced by $\tilde{t}$.

The Love symmetry argument implying vanishing/non-running Love numbers can still be applied in the same way as before but with this translated temporal coordinate $\tilde{t}$ appearing in place of $t$ in the elements of highest-weight representations. This does not alter the conclusion that static Love numbers vanish due to the polynomial form, up to overall irrelevant form-factors, of the static solution.

\section{Near-zone symmetries as isometries of subtracted geometries}
\label{sec:SubtractedGeometries4d}

The Love symmetry \eqref{eq:SL2RsKerrNewman} or the $s=0$ Starobinsky near-zone $\SL$ symmetry \eqref{eq:SL2RStarobinsky} are not exact isometries of the Kerr-Newman black hole. In a sense, they are approximate symmetries manifesting themselves in the near-zone region. This statement can be made geometrically rigorous thanks to the fact that there exists a framework where the vector fields generating the Love or the Starobinsky near-zone $\SL$'s are realized as Killing vectors of effective geometries, which are in turn realized as relatives of subtracted geometries of the Kerr-Newman black hole geometry~\cite{Cvetic:2011hp,Cvetic:2011dn}.

Let us briefly review how the construction of subtracted geometries is performed, focusing to the geometry of interest of a Kerr-Newman black hole. In the notation of Ref.~\cite{Cvetic:2011dn}, the full Kerr-Newman geometry is written as,
\be
	ds^2 = -\Delta_0^{-1/2}G\left(dt+\mathcal{A}\right)^2 +\Delta_0^{1/2}\left(\frac{dr^2}{X}+d\theta^2+\frac{X}{G}\sin^2\theta\,d\phi^2\right) \,,
\ee
where
\be
	G=X-a^2\sin^2\theta \,.
\ee
In the standard notations of \eqref{eq:KerrNewmanMetricBL}, the warp factor $\Delta_0$, the discriminant $X$ and the angular potential $\mathcal{A}$ are given by,
\be
	\Delta_0 = \Sigma^2 \,,\quad X=\Delta \,,\quad \mathcal{A} = \frac{a\sin^2\theta}{G}\left(2Mr-Q^2\right)\text{d}\phi \,.
\ee
The main observation then is that the thermodynamic variables of the black hole are completely independent of the warp factor $\Delta_0$, a fact suggesting that $\Delta_0$ encodes information about the environment around the black hole rather than its interior~\cite{Cvetic:2011dn}. Furthermore, the location of the ergosphere only depends on $G$, corresponding to the locus $G=0$. A subtracted geometry as introduced in~\cite{Cvetic:2011hp} then corresponds to modifying the warp factor $\Delta_0$, which preserves the internal structure of the black hole. We will more loosely refer to a subtracted geometry of the Kerr-Newman black hole, to whatever geometry which preserves its thermodynamic properties and the location of the ergosphere. In other words, besides arbitrary modifications of the warp factor $\Delta_0$, we will also allow alterations of the angular potential $\mathcal{A}$ such that its near-horizon behavior is preserved,
\be\label{eq:SubtractedGeometry}
	ds^2_{\text{sub}} = -\Delta_{0,\text{sub}}^{-1/2}G\left(dt+\mathcal{A}_{\text{sub}}\right)^2 +\Delta_{0,\text{sub}}^{1/2}\left(\frac{dr^2}{\Delta}+d\theta^2+\frac{\Delta}{G}\sin^2\theta\,d\phi^2\right) \,,
\ee
with arbitrary $\Delta_{0,\text{sub}}$ and,
\be\ba
	{}&\lim\limits_{r\rightarrow r_{+}}\mathcal{A}_{\text{sub}} = \lim\limits_{r\rightarrow r_{+}}\mathcal{A} = -\left(r_{+}^2+a^2\right)\text{d}\phi \,.
\ea\ee

Within this framework, both the Love symmetry as well as the Starobinsky near-zone $\SL$ vector fields can be realized as isometries of the relevant subtracted geometries. For the $s=0$ Starobinsky near-zone $\SL$ generators \eqref{eq:SL2RStarobinsky}, the relevant subtracted geometry that ensures separability of the massless Klein-Gordon operator in Boyer-Lindquist coordinates is given by \eqref{eq:SubtractedGeometry} with~\cite{Charalambous:2022rre}
\be\label{eq:SubtractedGeometry_Starobinsky}
	\Delta_{0,\text{Star}} = \left(r_{+}^2+a^2\right)^2 \,,\quad \mathcal{A}_{\text{Star}} = \frac{a\sin^2\theta}{G}\left(r_{+}^2+a^2\right)\text{d}\phi \,.
\ee
This geometry is what~\cite{Hui:2022vbh} refers to as the ``effective near-zone geometry'' of the Kerr-Newman black hole, here connected to the earlier notion of subtracted geometries~\cite{Cvetic:2011hp,Cvetic:2011dn}. Interestingly, this geometry is conformally flat and, therefore, besides the $6$ Killing vectors generating the $\SL\times SO\left(3\right)$ symmetry of the Teukolsky equation in the Starobinsky near-zone approximation, it admits $9$ additional conformal Killing vectors, whose explicit form can be found in~\cite{Hui:2022vbh}.

For the Love symmetry $s=0$ vector fields \eqref{eq:SL2RKerrNewman}, the corresponding subtracted geometry is given by \eqref{eq:SubtractedGeometry} with~\cite{Charalambous:2022rre}
\be\label{eq:SubtractedGeometry_Love}
	\begin{gathered}
		\Delta_{0,\text{Love}} = \left(r_{+}^2+a^2\right)^2\left(1+\beta^2\Omega^2\sin^2\theta\right) \,, \\
		\mathcal{A}_{\text{Love}} = \frac{a\sin^2\theta}{G}\left(r_{+}^2+a^2+\beta\left(r-r_{+}\right)\right)\text{d}\phi \,.
	\end{gathered}
\ee
In contrast to the Starobinsky near-zone, this subtracted geometry does not appear to exhibit any interesting new features.

\section{Spin-weighted generators of isometries}
\label{sec:SpinWeightedGenerators}

We have just saw that the $s=0$ vector fields generating the globally defined near-zone symmetries acquire a geometric interpretation in terms of subtracted black hole geometries. We will now attempt to infer the $s\ne0$ scalar pieces within the NP formalism. For the special case of a Killing vector field $\xi=\xi^{\mu}\partial_{\mu}$, there exists a canonical generalization $\Lstr_{\xi}$ of the usual Lie derivative $\mathcal{L}_{\xi}$ that acts on a NP scalar of spin-weight $s$ and boost-weight $b$~\cite{Ludwig2000,Ludwig:2001hx},
\be\label{eq:LieLudwig}
	\Lstr_{\xi} = \mathcal{L}_{\xi} + \left(b\,n_{\mu}\mathcal{L}_{\xi}\ell^{\mu} - s\,\bar{m}_{\mu}\mathcal{L}_{\xi}m^{\mu}\right) \,.
\ee
These are improvements of the traditional GHP Lie derivatives such that the information that $\xi$ is a Killing vector can be read off from the tetrad vectors themselves, namely, if there exists a null tetrad vector which is 
annihilated by $\Lstr_{\xi}$, then $\xi$ is a Killing vector field~\cite{Ludwig2000,Ludwig:2001hx}. We will refer to this generalized Lie derivative as the spin-weighted Lie derivative in what follows. The spin-weighted Lie derivative operator is covariant with respect to local spin and boost transformations and does not change the spin-weight or boost-weight of the NP scalar it acts on.

For the spin-weighted NP scalars entering the Teukolsky equation, their boost-weight is equal to their spin-weight, $b=s$ (see Table~\ref{tbl:NPweights}). In a coordinate basis, these modified derivatives modify the isometry generators by scalar $s$-dependent pieces. The spin-weighted generators turn out to inherit the algebraic structure of the underlying isometries, i.e. the spin-weighted Lie derivatives satisfy the same algebra as the usual Lie derivatives. This fact only holds whenever Lie dragging along a Killing vector. Indeed, as is analyzed in Appendix~\ref{app:LieDerivative}, once we attempt to construct the most general generalized Lie derivative with respect to a vector field $\xi$ acting on NP scalars, there is an unambiguous choice only when $\xi$ is a Killing vector, while there are infinitely main options when $\xi$ is not an element of an isometry; the algebra preserving property we just mentioned turns out to be unique to generalized Lie derivatives along Killing vectors.

In order to apply this prescription to attempt to infer the $s\ne0$ pieces of Love and Starobinsky near-zone symmetries, we therefore need to use their geometric interpretation as isometries of subtracted geometries we saw in the previous section. This is important because applying the spin-weighted Lie derivative \eqref{eq:LieLudwig} requires to have a basis of tetrad fields reproducing the corresponding geometry for which $\xi$ is a Killing vector.

We start with the $s=0$ Starobinsky near-zone $\SL$ generators \eqref{eq:SL2RStarobinsky}. We first need to identify a proper set of tetrad vectors that reconstructs the subtracted geometry, Eq.~\eqref{eq:SubtractedGeometry} with the choices \eqref{eq:SubtractedGeometry_Starobinsky}. This is achieved by
\be\ba
	\ell_{\text{Star}} &= \frac{r_{+}^2+a^2}{\Delta}\left(\partial_{t}+\Omega\partial_{\phi}\right) + \partial_{r} \,, \\
	n_{\text{Star}} &= \frac{\Delta}{2\left(r_{+}^2+a^2\right)}\left[\frac{r_{+}^2+a^2}{\Delta}\left(\partial_{t}+\Omega\partial_{\phi}\right) - \partial_{r}\right] \,, \\
	m_{\text{Star}} &= \frac{1}{\sqrt{2}\left(r_{+}+ia\right)}\left[\partial_{\theta} + \frac{i}{\sin\theta}\partial_{\phi}\right] \,,
\ea\ee
which preserves the algebraic classification of the Kerr-Newman black hole, i.e. the above subtracted geometry is still a Petrov type-D spacetime. A direct calculation then reveals that the $s\ne0$ pieces in the Starobinsky near-zone $\SL$ generators \eqref{eq:SL2RStarobinsky} are exactly reproduced by the spin-weighted Lie derivative \eqref{eq:LieLudwig},
\be
	\begin{gathered}
		\Lstr_{L_{0}^{\text{Star}}} = L_{0}^{\text{Star}} = L_{0}^{\text{Star},\left(s\right)} \,, \\
		\Lstr_{L_{\pm1}^{\text{Star}}} = L_{\pm1}^{\text{Star}} \mp s\,e^{\pm t/\beta}\frac{r-r_{\mp}}{\sqrt{\Delta}} = L_{\pm1}^{\text{Star},\left(s\right)} \,,
	\end{gathered}
\ee
where $L_{m}^{\text{Star}}$ are the $s=0$ vector fields generating the Starobinsky near-zone $\SL$ symmetry of the massless Klein-Gordon equation. This shows that the $s\ne0$ pieces in the Starobinsky near-zone $\SL$ generators can indeed be assigned a geometric interpretation in terms of spin-weighted Lie derivatives in a particular subtracted geometry.

For the Love symmetry $s=0$ vector fields \eqref{eq:SL2RKerrNewman}, the corresponding subtracted geometry is given in Eq.~\eqref{eq:SubtractedGeometry} with the choice \eqref{eq:SubtractedGeometry_Love}. The associated tetrad vectors are, up to local boosts and rotations,
\be\ba
	\ell_{\text{Love}} &= \frac{r_{+}^2+a^2}{\Delta}\left(\partial_{t}+\frac{\Delta^{\prime}}{r_{+}-r_{-}}\Omega\partial_{\phi}\right) + \partial_{r} \,, \\
	n_{\text{Love}} &= \frac{\Delta}{2\Delta_{0,\text{Love}}^{1/2}}\left[\frac{r_{+}^2+a^2}{\Delta}\left(\partial_{t}+\frac{\Delta^{\prime}}{r_{+}-r_{-}}\Omega\partial_{\phi}\right) - \partial_{r}\right] \,, \\
	m_{\text{Love}} &= \frac{1}{\sqrt{2}\mathcal{M}_{0,\text{Love}}}\left[\partial_{\theta} + i\sqrt{\frac{1}{\sin^2\theta}+\beta^2\Omega^2}\,\partial_{\phi}\right] \,,
\ea\ee
with $\left|\mathcal{M}_{0,\text{Love}}\right|^2 = \Delta_{0,\text{Love}}^{1/2}$. These would imply the following $s\ne0$ extensions of the Love symmetry vector fields,
\be
	\begin{gathered}
		\Lstr_{L_{0}^{\text{Love}}} = L_{0}^{\text{Love}} = L_{0}^{\text{Love},\left(s\right)}-s \,, \\
		\Lstr_{L_{\pm1}^{\text{Love}}} = L_{\pm1}^{\text{Love}} \mp s\,e^{\pm t/\beta}\frac{r-r_{\mp}}{\sqrt{\Delta}} = L_{\pm1}^{\text{Love},\left(s\right)} + s\,e^{\pm t/\beta}\sqrt{\frac{r-r_{+}}{r-r_{-}}} \,,
	\end{gathered}
\ee
which do not reproduce the actual $s\ne0$ pieces involved in the Love symmetry generators \eqref{eq:SL2RsKerrNewman}. As a result a geometric interpretation in terms of spin-weighted Lie derivatives for the subtracted geometry does not seem to be possible in this case\footnote{As a disclaimer here, there actually exists a possible choice of tetrads that ensures that the spin-weighted Lie derivative does indeed capture the correct $s\ne0$ pieces of the Love symmetry generators in \eqref{eq:SL2RsKerrNewman}. This is given by a particular local boost,
	\begin{equation*}
		\ell_{\text{Love}} \rightarrow \lambda \ell_{\text{Love}} \,,\quad n_{\text{Love}}\rightarrow \lambda^{-1}n_{\text{Love}} \,,\quad m_{\text{Love}}\rightarrow m_{\text{Love}} \,,
	\end{equation*}
	with $\lambda = e^{\left(t-\phi/\Omega\right)\beta}$. However, this local boost is not globally defined and does not have a smooth spinless limit.}. The corresponding Casimir associated with this spin-weighted Lie derivative along the Love symmetry vector fields is,
\be
	\text{\sout{$\mathcal{C}$}}_2^{\text{Love}} = \mathcal{C}_2^{\text{Love},\left(s\right)} + s \frac{r_{+}-r_{-}}{r-r_{-}}\left(\partial_{t}+\Omega\,\partial_{\phi}\right) \,,
\ee
which fails to be a valid near-zone truncation due to the additional static contribution for $\Omega\ne0$, while, for $\Omega=0$, this is just the Starobinsky near-zone approximation \eqref{eq:V0V1_StarobinskyS}.

Nevertheless, the Love symmetry generators have the peculiar property that the corresponding $s$-dependent pieces can be completely gauged away through a particular globally defined local boost transformation of the NP scalars involved in the Teukolsky equation,
\be
	\Psi_{s} \rightarrow \tilde{\Psi}_{s} = \left(e^{-t/\beta}\sqrt{\Delta}\right)^{s}\Psi_{s} \Rightarrow L_{m}^{\text{Love},\left(s\right)}\Psi_{s} = \left(e^{-t/\beta}\sqrt{\Delta}\right)^{-s}L_{m}^{\text{Love}}\tilde{\Psi}_{s} \,.
\ee
In other words, the Love near-zone truncation of the spin-$s$ Teukolsky equation \eqref{eq:NZSplitOP_KerrNewmanS}-\eqref{eq:NZSplitV0V1_KerrNewmanS} is effectively a near-zone truncation of the $s=0$ Teukolsky equation for the boosted NP scalars $\tilde{\Psi}_{s}$. We note, however, that the required boost has a non-trivial time dependence, so that it turns static perturbations into time-dependent ones.

Last, it is instructive to ask whether the $\SL_{\left(\alpha\right)}$ subalgebras of the infinite extension $\SL\ltimes U\left(1\right)_{\mathcal{V}}$ admit similar subtracted geometry analyses as with the near-zone $\SL$'s. The would-be subtracted geometry associated with $\SL_{\left(\alpha\right)}$ is given by \eqref{eq:SubtractedGeometry}, now with,
\be\ba
	\Delta_{0,\left(\alpha\right)} &= \left(r_{+}^2+a^2\right)^2\left(1+\left(\alpha-1\right)^2\beta^2\Omega^2\sin^2\theta\right) \\
	\mathcal{A}_{\left(\alpha\right)} &= \frac{a\sin^2\theta}{G_{\left(\alpha\right)}}\left[r_{+}^2+a^2-\left(\alpha-1\right)\beta\left(r-r_{+}\right)\right] \,,
\ea\ee
but also with a modified function $G$,
\be
	G_{\left(\alpha\right)} = G + 4\,\alpha\left(\alpha-1\right)\frac{r-r_{+}}{r_{+}-r_{-}}a^2\sin^2\theta \,.
\ee
The $\SL_{\left(\alpha\right)}$ generators \eqref{eq:StarLove} are then Killing vectors of this geometry. Even though this effective geometry preserves the thermodynamic properties of the Kerr-Newman black hole, it fails to capture properties that extend beyond the near-horizon behavior. For instance, the ergosphere of this black hole geometry is now located at $G_{\left(\alpha\right)}=0$ which does not match the original locus condition $G=0$ unless $\alpha=0$ or $\alpha=1$, that is unless the algebra is the Love or Starobinsky near-zone $\SL$ algebra respectively. This observation also demonstrates how the near-zone approximation captures information beyond the near-horizon regime at the level of the geometry itself.

Let us finish this section by examining whether the $s\ne0$ pieces of these $\SL$ subalgebras, obtained via the substitution \eqref{eq:U1Vs} in \eqref{eq:StarLove}, can be inferred from the spin-weighted Lie derivative \eqref{eq:LieLudwig}. The corresponding tetrad vectors are, up to local boosts and rotations,
\be\ba
	\ell_{\left(\alpha\right)} &= \frac{r_{+}^2+a^2}{\Delta}\left(\partial_{t}+\frac{\Delta^{\prime}-2\alpha\left(r-r_{+}\right)}{r_{+}-r_{-}}\Omega\partial_{\phi}\right) + \partial_{r} \,, \\
	n_{\left(\alpha\right)} &= \frac{\Delta}{2\Delta_{0,\left(\alpha\right)}^{1/2}}\left[\frac{r_{+}^2+a^2}{\Delta}\left(\partial_{t}+\frac{\Delta^{\prime}-2\alpha\left(r-r_{+}\right)}{r_{+}-r_{-}}\Omega\partial_{\phi}\right) - \partial_{r}\right] \,, \\
	m_{\left(\alpha\right)} &= \frac{1}{\sqrt{2}\mathcal{M}_{0,\left(\alpha\right)}}\left[\partial_{\theta} + i\sqrt{\frac{1}{\sin^2\theta}+\left(\alpha-1\right)^2\beta^2\Omega^2}\,\partial_{\phi}\right] \,,
\ea\ee
with $\left|\mathcal{M}_{0,\left(\alpha\right)}\right|^2=\Delta_{0,\left(\alpha\right)}^{1/2}$. The spin-weighted Lie derivative \eqref{eq:LieLudwig} then outputs
\be
	\begin{gathered}
		\Lstr_{L_{0}\left(\alpha\right)} = L_{0}\left(\alpha\right) = L_{0}^{\left(s\right)}\left(\alpha\right) - \left(1-\alpha\right)s \,, \\
		\Lstr_{L_{\pm1}\left(\alpha\right)} = L_{\pm1}\left(\alpha\right) \mp s\,e^{\pm t/\beta}\frac{r-r_{\mp}}{\sqrt{\Delta}} = L_{\pm1}^{\left(s\right)}\left(\alpha\right) +\left(1-\alpha\right)s\,e^{\pm t/\beta}\sqrt{\frac{r-r_{+}}{r-r_{-}}} \,,
	\end{gathered}
\ee
and the corresponding spin-weighted Casimir is given by,
\be
	\text{\sout{$\mathcal{C}$}}_2^{\left(s\right)}\left(\alpha\right) = \mathcal{C}_2^{\left(s\right)}\left(\alpha=1\right) + 2\frac{r_{+}^2+a^2}{r-r_{-}}\left(\partial_{t}+\alpha\,\Omega\,\partial_{\phi}+\frac{s}{\beta}\right)\left(\alpha-1\right)\beta\Omega\,\partial_{\phi} \,,
\ee
where $\mathcal{C}_2^{\left(s\right)}\left(\alpha=1\right)$ is the Starobinsky near-zone $\SL$ Casimir, \eqref{eq:V0V1_StarobinskyS}.

Just like in the analysis for the Love symmetry generators, the spin-weighted Lie derivative does not agree with the actual $s\ne0$ pieces of $L_{m}^{\left(s\right)}\left(\alpha\right)$\footnote{Similar to the Love symmetry example, there exists a particular local boost on the tetrad vectors with the same boost parameter $\lambda=e^{\left(t-\phi/\Omega\right)/\beta}$ whose implementation in the spin-weighted Lie derivative correctly reproduces the appropriate scalar pieces in $L_{m}^{\left(s\right)}\left(\alpha\right)$, but is not globally defined and does not have a smooth $\Omega\rightarrow0$ limit. Interestingly, this boost parameter is independent of $\alpha$, but the Starobinsky near-zone algebra is in fact invariant under such transformations involving the co-rotating azimuthal angle, which allows to always gauge away such pathological factors.}. In fact, the $s\ne0$ corrections predicted from the spin-weighted Lie derivative are independent of $\alpha$ and they are always equal to the $s\ne0$ pieces of the Starobinsky near-zone $\SL$ algebra generators in Eq.~\eqref{eq:SL2RStarobinsky}. However, in contrast to the Love symmetry example, the $s\ne0$ pieces of $L_{m}^{\left(s\right)}\left(\alpha\right)$ cannot be gauged away by any globally defined local boost or rotation.

\section{Love symmetry beyond General Relativity}
\label{sec:LoveBeyondGR4d}

In this section, we investigate general conditions for the existence of the Love symmetry in modifications of General Relativity in four spacetime dimensions. Let us focus on a simple example of a massless scalar field in the background of a generalized spherically symmetric black hole geometry which can always be brought to the form,
\be
	ds^2 = -f_{t}\left(r\right)dt^2 + \frac{dr^2}{f_{r}\left(r\right)} + r^2d\Omega_{2}^2\,.
\ee
The functions $f_{t}\left(r\right)$ and $f_{r}\left(r\right)$ are arbitrary at this point. In vacuum General Relativity, $f_{r}\left(r\right)=f_{t}\left(r\right)$ but in general modified gravity this needs not be true anymore. The preliminary assumptions we impose on these radial functions are that there exists a horizon at $r=r_{\text{h}}$, which is a root of both $f_{t}\left(r\right)$ and $f_{r}\left(r\right)=0$ with multiplicity one,
\be
	f_{t}\left(r_{\text{h}}\right) = f_{r}\left(r_{\text{h}}\right) = 0 \,,\quad f_{t}^{\prime}\left(r_{\text{h}}\right) \ne 0\,,\quad f_{r}^{\prime}\left(r_{\text{h}}\right) \ne 0\,,
\ee
that is, the geometry describes a non-extremal spherically symmetric black hole. The full radial Klein-Gordon equation after expanding over monochromatic spherical harmonic modes of orbital number $\ell$ reads,
\be
	\begin{gathered}
		\mathbb{O}_{\text{full}}\Phi_{\omega\ell m} = \ell\left(\ell+1\right)\Phi_{\omega\ell m} \,, \\
		\mathbb{O}_{\text{full}} = \partial_{r}\,\Delta_{r}\,\partial_{r} + \frac{\Delta_{r}^2}{2\Delta_{t}}\left(\frac{\Delta_{t}}{\Delta_{r}}\right)^{\prime}\partial_{r} - \frac{r^4}{\Delta_{t}}\,\partial_{t}^2
	\end{gathered}
\ee
where $\Delta_{t} \equiv r^2f_{t}$, $\Delta_{r} \equiv r^2f_{r}$ and primes denote derivatives with respect to $r$.

We now explore whether there exist near-zone truncations that are equipped with an $\SL$ structure. Of particular interest is the following near-zone approximation of the radial Klein-Gordon operator,
\be\label{eq:NZModGR}
	\mathbb{O}_{\text{NZ}} = \partial_{r}\,\Delta_{r}\,\partial_{r} + \frac{\Delta_{r}^2}{2\Delta_{t}}\left(\frac{\Delta_{t}}{\Delta_{r}}\right)^{\prime}\partial_{r} - \frac{r_{\text{h}}^4}{\Delta_{t}}\,\partial_{t}^2 \,.
\ee
This differential operator can be represented as a Casimir of a certain regular and globally defined $\SL$ algebra if and only if the following condition is satisfied
\be\label{eq:SL2RModGRConstraint}
	\frac{\Delta_{r}}{\Delta_{t}}\Delta_{t}^{\prime\prime} + \frac{1}{2}\Delta_{t}^{\prime}\left(\frac{\Delta_{r}}{\Delta_{t}}\right)^{\prime} = 2 \,.
\ee
The generators of the $\SL$ near-zone symmetry are then given by
\be\label{eq:SL2RModGR}
	\begin{gathered}
		L_0 = -\beta\,\partial_{t} \,,\quad
		L_{\pm1} = e^{\pm t/\beta}\left[\mp\sqrt{\Delta_{r}}\,\partial_{r} + \sqrt{\frac{\Delta_{r}}{\Delta_{t}}}\partial_{r}\left(\sqrt{\Delta_{t}}\right)\beta\,\partial_{t}\right] \,,
	\end{gathered}
\ee
with $\beta = 2/\sqrt{f_{t}^{\prime}\left(r_{\text{h}}\right)f_{r}^{\prime}\left(r_{\text{h}}\right)}$ the inverse Hawking temperature of this general spherically symmetric black hole, and are regular at both the future and the past event horizons. The geometric constraint \eqref{eq:SL2RModGRConstraint} can be solved explicitly,
\be
	\Delta_{r}\left(r\right) = \Delta_{t}\left(r\right)\frac{4\Delta_{t}\left(r\right)+\left(\frac{\beta_{s}}{\beta}r_{\text{h}}\right)^2}{\Delta_{t}^{\prime2}\left(r\right)} \,,
\ee
with $\beta_{s}=2r_{\text{h}}$ the inverse Hawking temperature of the four-dimensional Schwarzschild black hole. At the level of the functions $f_{t}\left(r\right)$ and $f_{r}\left(r\right)$ themselves, the above condition reads,
\be
	f_{r}\left(r\right) = f_{t}\left(r\right) \frac{1}{\left(r^2f_{t}\left(r\right)\right)^{\prime2}} \left[4r^2f_{t}\left(r\right) + \left(\frac{\beta_{s}}{\beta}r_{\text{h}}\right)^2 \right] \,.
\ee
As can be checked explicitly, the asymptotic flatness condition is automatically imposed by the above condition if either $f_{t}$ or $f_{r}$ is asymptotically flat. For the case where $f_{r}\left(r\right)=f_{t}\left(r\right)$, we get that the most general such geometry is the four-dimensional Reissner-Nordstr\"{o}m black hole. Following the procedure outlined in Appendix~\ref{app:SL2RGenerators}, one can in fact show that the near-zone truncation \eqref{eq:NZModGR} employed here is the only one that is a candidate of being equipped with an $\SL$ structure. Indeed, the above results set absolute geometric constraints on the existence of Love symmetry beyond General Relativity as long as we ignore possible scalar field redefinitions and only consider fields minimally coupled to gravity.

One can also check that the above near-zone $\SL$ implies the vanishing of static Love numbers. Using the same symmetry argument of the regular static solution being an element of a highest-weight representation of this $\SL$, we obtain $\left(L_{+1}\right)^{\ell+1}\Phi_{\omega=0,\ell m} = 0$. We now get a modified ``polynomial'' requirement. In particular, noticing that
\be
	\left(L_{+1}\right)^{n}F\left(r\right) = \left(-e^{+t/\beta}\sqrt{\Delta_{t}}\right)^{n}\left[\sqrt{\frac{\Delta_{r}}{\Delta_{t}}}\frac{d}{dr}\right]^{n}F\left(r\right)
\ee
for an arbitrary purely radial function $F\left(r\right)$, we get that the physical static radial wavefunction $R_{\omega=0,\ell m}$ belongs to a highest-weight representation and that it is again a polynomial, this time not in the radial variable $r$, but in the variable $\tilde{r}$, defined as
\be
	d\tilde{r} \equiv \sqrt{\frac{\Delta_{t}}{\Delta_{r}}} \,dr \Rightarrow \tilde{r} = \sqrt{\Delta_{t}+\left(\frac{\beta_{s}}{2\beta}r_{\text{h}}\right)^2} + \tilde{r}_{\text{h}} - \frac{\beta_{s}}{2\beta}r_{\text{h}} \,,
\ee
where $\tilde{r}_{\text{h}}$ is an integration constant indicating the location of the event horizon in this new radial coordinate,
\be
	R_{\omega=0,\ell m}\left(r\right) = \sum_{n=0}^{\ell}c_{n}^{\left(m\right)}\tilde{r}^{n}\left(r\right) \,.
\ee
We note that, asymptotically, $\tilde{r}\rightarrow r$ due to the asymptotic flatness of $f_{t}$. Expanding this polynomial in $\tilde{r}$ at large distance in the initial radial variable $r$, one observes the appearance of an $r^{-\ell-1}$ term. However, this term is a relativistic correction in the profile of the ``source'' part of the solution, rather than a response effect from induced multipole moments. Indeed, if the geometric condition \eqref{eq:SL2RModGRConstraint} for the existence of a near-zone $\SL$ symmetry is satisfied, we arrive at a situation practically identical to the case of scalar perturbations of the four-dimensional Schwarzschild black hole, Eq.~\eqref{eq:NZsplitting_Schwarzschild4d}, when working with the variable $\tilde{r}$. More explicitly, the full radial Klein-Gordon operator reads
\be
	\mathbb{O}_{\text{full}} = \partial_{\tilde{r}}\,\Delta_{t}\,\partial_{\tilde{r}} - \frac{r^4}{\Delta_{t}}\,\partial_{t}^2 \,,
\ee
and $\Delta_{t}$ is a quadratic polynomial in $\tilde{r}$,
\be
	\Delta_{t} = \left(\tilde{r}-\tilde{r}_{\text{h}}\right)\left(\tilde{r}-\tilde{r}_{\text{h}}+\frac{\beta_{s}}{\beta}r_{\text{h}}\right) \,.
\ee
Matching onto the worldline EFT is equivalent to solving the equations motion after analytically continuing the orbital number to perform the source/response split of the scalar field, and only in the end sending $\ell$ to take its physical integer values. Doing this, we see that the ``response'' part of the static scalar field is singular at the horizon when sending $\ell\in\mathbb{N}$ and is therefore absent, while the ``source'' part becomes a polynomial of degree $\ell$ in $\tilde{r}$. Consequently, the corresponding static Love numbers vanish identically and we see again how a polynomial form of the solution is indicative of this vanishing.

All in all, we observe that the Love symmetry can be present beyond General Relativity. We have derived a generic class of geometries enjoying the near-zone $\SL$ symmetries. All these geometries must have exactly zero static scalar Love numbers in four spacetime dimensions as a result of the highest weight property.

An instructive application of our construction is the non-vanishing of Love numbers in Riemann-cubed gravity considered in Sec.~\ref{sec:SLNs_R34d}. In this case, the Klein-Gordon equation does not posses an $\SL$ symmetry, which can be seen
from the fact that the geometric constraint \eqref{eq:SL2RModGRConstraint} is violated for any non-zero Riemann-cubed coupling constant $\alpha$. Thus, Riemann-cubed gravity gives an explicit example where the absence of the Love symmetry is accompanied by running Love numbers.

\section{Vanishing Love numbers at total transmission modes}
\label{sec:TTMs4d}

In this last section, we will shine more light on the connection between the states of the Love symmetry highest-weight multiplets and total transmission modes of black holes~\cite{Cook:2016fge,Cook:2016ngj,MaassenvandenBrink:2000iwh,Hod:2013fea}. To do this, let us for simplicity focus to the scattering of a monochromatic scalar field from a spherically symmetric black hole in four spacetime dimensions. Using the following field redefinition of the separable solution
\be
	\Phi_{\omega\ell m}\left(t,r,\theta\right) = \frac{R_{\omega\ell m}\left(r\right)}{r}e^{-i\omega t}Y_{\ell m}\left(\theta\right) \,,
\ee
the equation of motion for the radial wavefunction can be brought to canonical form, i.e. a Schr\"{o}dinger-like equation, when working with the tortoise coordinate,
\be
	\left[\frac{d^2}{dr_{\ast}^2} + \omega^2-f_{t}\left(r\right)V_{\ell}^{\left(0\right)}\left(r\right)\right]R_{\omega\ell m} = 0 \,,
\ee
with the scalar potential given by
\be
	V_{\ell}^{\left(0\right)}\left(r\right) = \frac{1}{r^2}\left[\ell\left(\ell+1\right) + \frac{r\left(f_{t}\left(r\right)f_{r}\left(r\right)\right)^{\prime}}{2f_{t}\left(r\right)}\right] \,.
\ee
For asymptotically flat black hole geometries, one can then solve for the radial wavefunction in the far-zone region $r\gg r_{\text{h}}$,
\be\label{eq:RadialScalarAsymptotic4d}
	R_{\omega\ell m}^{\infty} = \sqrt{\frac{\pi\omega r}{2}} e^{-i\frac{\pi}{2}\left(\ell+1\right)}\mathcal{I}_{\ell m}\left(\omega\right)\left[H^{\left(2\right)}_{\ell+\frac{1}{2}}\left(\omega r\right) + \left(-1\right)^{\ell+1}\mathcal{R}_{\ell m}\left(\omega\right)H^{\left(1\right)}_{\ell+\frac{1}{2}}\left(\omega r\right) \right] \,,
\ee
where the integration constants were fixed such that
\be
	\Phi_{\omega\ell m} \xrightarrow{r\rightarrow\infty} \frac{\mathcal{I}_{\ell m}\left(\omega\right)}{r}\left[e^{-i\omega t_{+}}+\mathcal{R}_{\ell m}\left(\omega\right)e^{-i\omega t_{-}}\right] \,,
\ee
that is, $\left|\mathcal{I}_{\ell m}\left(\omega\right)\right|^2$ is the incoming flux and $\mathcal{R}_{\ell m}\left(\omega\right)$ is the reflection amplitude.

The canonical radial equation of motion is particularly useful to apply standard scattering theory via the partial wave method. In particular, the total scattering cross-section can be decomposed into elastic (conservative) and absorption (dissipative) cross-sections. In terms of the real-valued conservative phase-shifts $\delta_{\ell}\left(\omega\right)$ and dissipative transmission factors $\eta_{\ell}\left(\omega\right)$~\cite{Matzner1978ApJS,Futterman:1988ni,Dolan:2008kf,Ivanov:2022qqt}:
\be\ba
	\sigma_{\text{elastic}} &= \frac{4\pi}{\omega^2}\sum_{\ell=0}^{\infty}\left(2\ell+1\right)\sin^2\delta_{\ell}\left(\omega\right) \,, \\
	\sigma_{\text{absorption}} &= \frac{\pi}{\omega^2}\sum_{\ell=0}^{\infty}\left(2\ell+1\right)\left(1-\eta_{\ell}^2\left(\omega\right)\right) \,.
\ea\ee

These can be extracted from the reflection amplitude according to
\be
	\eta_{\ell}\left(\omega\right)e^{2i\delta_{\ell}\left(\omega\right)} = \left(-1\right)^{\ell+1}\mathcal{R}_{\ell0}\left(\omega\right) \,.
\ee
Let us now see what happens in the near-zone region, $\omega r\ll 1$. For the asymptotic wavefunction \eqref{eq:RadialScalarAsymptotic4d} to be supported in the near-zone region, we must of course also have $\omega r_{\text{h}} \ll 1$ such that the intermediate region $r_{\text{h}} \ll r \ll\omega^{-1}$ is non-empty. From the asymptotic behaviors of the Hankel functions for small arguments, we see that it takes the form
\be
	R^{\infty}_{\omega\ell m} \xrightarrow{\omega r\ll1} \bar{\mathcal{E}}^{\left(0\right)}_{\ell m}\left(\omega\right) r^{\ell+1}\left[1+k_{\ell}\left(\omega\right)\left(\frac{r_{\text{h}}}{r}\right)^{2\ell+1}\right] \,,
\ee
Therefore, we can find a matching prescription of the response coefficients with the reflection amplitude. In particular, to linear order in the response coefficients, we see that the phase-shifts and transmission factors are matched onto~\cite{Ivanov:2022qqt}
\be
	\eta_{\ell}\left(\omega\right)e^{2i\delta_{\ell}\left(\omega\right)} = 1 + i\frac{\left(\ell!\right)^2}{\left(2\ell\right)!\left(2\ell+1\right)!}\left(2\omega r_{\text{h}}\right)^{2\ell+1}\,k_{\ell}\left(\omega\right) \,.
\ee
This reveals that
\be\ba
	\eta_{\ell}\left(\omega\right) &= 1 - \frac{\left(\ell!\right)^2}{\left(2\ell\right)!\left(2\ell+1\right)!}\left(2\omega r_{\text{h}}\right)^{2\ell+1}\,\text{Im}\left\{k_{\ell}\left(\omega\right)\right\} \,, \\
	\delta_{\ell}\left(\omega\right) &= \frac{1}{2}\frac{\left(\ell!\right)^2}{\left(2\ell\right)!\left(2\ell+1\right)!}\left(2\omega r_{\text{h}}\right)^{2\ell+1}\,\text{Re}\left\{k_{\ell}\left(\omega\right)\right\} \,,
\ea\ee
which is an alternative way to see that the Love numbers only enter the conservative sector, directly at the level of gauge invariant observables such as scattering cross-sections. The upshot of the analysis done here is that we can see explicitly that vanishing Love numbers immediately imply vanishing elastic scattering cross-section for each partial wave~\cite{Ivanov:2022qqt},
\be
	\sigma_{\text{elastic},\ell}\left(\omega\right) = 0 \quad \text{if} \quad k_{\ell}^{\text{Love}}\left(\omega\right) = 0 \,,
\ee
while the corresponding partial absorption cross-section is maximized. Vanishing Love numbers, are thus interpreted as reflectionless, total transmission modes~\cite{Cook:2016fge,Cook:2016ngj,MaassenvandenBrink:2000iwh,Hod:2013fea}. This supports the interpretation of the states in the Love symmetry highest-weight multiplets with total transmission modes. However, it is still not clear whether this connection is exact. For instance, Figure 19 in \cite{Cook:2016fge} shows that the Love symmetry highest-weight multiplet over-counts the number of total transmission modes.

	\newpage
\chapter{Relation to near-horizon isometries of extremal black holes}
\label{ch:NHE}

In Section~\ref{sec:RCsGR}, we saw how dissipative effects are captured by the worldline EFT by introducing gapless internal degrees of freedom $X$ localized on the horizon. This description of the black hole EFT is holographic in its nature~\cite{Goldberger:2005cd}. Namely, the idea of holography is essentially the statement that $X$'s are more than an EFT bookkeeping device, but represent an actual quantum mechanical system describing the black hole dynamics. Remarkably, the AdS/CFT correspondence~\cite{Aharony:1999ti} tells us exactly what this system is for certain classes of black holes in string theory. For instance, black $3$-branes in type-IIB string theory are described in this way by the maximally supersymmetric Yang--Mills theory. The discovery of this correspondence~\cite{Maldacena:1997re} was guided by the matching calculations~\cite{Klebanov:1997kc,Gubser:1997yh} of the type described in Section~\ref{sec:RCsGR}.

A promise of the Kerr/CFT correspondence~\cite{Guica:2008mu,Castro:2010fd} is that one day a similar success may be achieved for the actual real world black holes. It is natural to ask what the role of the Love numbers is in this story. The discussion above implies that they do not have an intrinsic CFT interpretation. Rather, they describe how the CFT (describing a near-horizon throat emerging in the extremal limit) is glued to the rest of the spacetime. 

Along these lines, from the holographic viewpoint, it is natural to think about black holes with different values of a mass as excitations of a single quantum mechanical system rather than representing genuinely different systems. Namely, an extremal black hole of a minimal possible mass at the fixed values of gauge charges (and zero Hawking temperature) corresponds to the ground state. Non-extremal black holes correspond instead to excited finite temperature states. As a result, symmetries of the system become more manifest as one approaches the extremal limit. Geometrically, a near-extremal charged black hole develops a near-horizon anti de Sitter (AdS) throat as we will see in this chapter. Isometries of this region correspond to the conformal symmetry present in the holographic description. This makes it natural to study the fate of the Love symmetry in the near extremal limit.

The near-zone $\SL$ symmetries generators are formally singular in the extremal limit, as the Hawking temperature approaches a vanishing value. However, extremal black holes have the astonishing property of developing an $\text{AdS}_2$ near-horizon throat that completely decouples from the asymptotically flat region~\cite{Bardeen:1999px,Amsel:2009et}. The isometry group of $\text{AdS}_2$ is precisely $\SL$ which leaves one to wonder whether the algebraic structure of the near-zone symmetries can be related to these enhanced symmetries of the near-horizon geometry of extremal black holes.

In this chapter, we will explore this possibility. We will begin with a review of how the near-horizon geometry of extremal black holes decouples from the far-horizon region and develops an infinite throat with enhanced symmetries. Following this, we will attempt to infer the near-zone symmetries as remnants of this near-horizon extremal isometry that survive for non-extremal black holes. This will be achieved by appropriately taking the extremal limit of a particular family of $\SL$ subalgebras of the infinite extension $\SL\ltimes U\left(1\right)_{\mathcal{V}}$ we presented in Section~\ref{sec:InfiniteExtension4d}, as was first pointed out in our work in Ref.~\cite{Charalambous:2022rre}.

\section{Review of near-horizon geometries}
\label{sec:NHEReview4d}

The extremal black hole limit has many peculiarities. For simplicity, we start with the four-dimensional extremal Reissner-Nordstr\"{o}m black hole,
\be
	ds ^2 = -f\left(r\right)dt^2 + \frac{dr^2}{f\left(r\right)} + r^2d\Omega_2^2 \,,\quad f\left(r\right) = \left(1-\frac{M}{r}\right)^2 \,.
\ee
The horizon is now a double root of the discriminant function and corresponds to the degeneracy of the inner and outer horizons, $r_{+}=r_{-}=M$, for which the Hawking temperature vanishes. To obtain the near-horizon geometry, we employ the scaling transformations
\be\label{eq:scalRN}
	r = M +  \lambda \rho \,,\quad t = \frac{\tau}{\lambda} \,,
\ee
and zoom-in on the horizon by taking the $\lambda\to 0$ limit. The resulting geometry,
\be
	ds^2 =  -\frac{\rho^2}{M^2} d\tau^2 + M^2\frac{d\rho^2}{\rho^2} + M^2d\Omega^2_2 \,,
\ee
is an $\text{AdS}_2\times \mathbb{S}^2$ manifold. Besides the $SO\left(3\right)$ symmetry that is shared with the non-extremal configuration, it acquires the following additional Killing vectors
\be\label{eq:isoRN}
	\xi_0 = \tau\,\partial_{\tau} - \rho\,\partial_{\rho} \,, \quad 
	\xi_{+1} = \partial_{\tau}\,, \quad 
	\xi_{-1} = \left(\frac{M^4}{\rho^2} + \tau^2\right)\partial_{\tau} -2\tau\rho\,\partial_{\rho}  \,,
\ee
which satisfy the $\SL$ algebra, appropriate for the $2$-dimensional maximally symmetric $\text{AdS}_2$ submanifold. In the original Schwarzschild coordinates $\left(t,r,\theta,\phi\right)$, they take the following form
\be\label{eq:isoRN2}
	\begin{split}
		\xi_0 &= t\,\partial_{t} - \left(r-M\right)\partial_{r} \,, \quad \xi_{+1} = \lambda^{-1}\partial_{t}\,, \\ 
		\xi_{-1} & = \lambda \left[\left( \frac{M^4}{(r-M)^2} + t^2\right)\partial_{t} - 2t\left(r-M\right)\partial_{r} \right] \,.
	\end{split}
\ee
The Casimir of this $\SL$ near-horizon isometry is given by, in Schwarzschild coordinates,
\be\label{eq:C2NHE_RN4d}
	\mathcal{C}_2 = \xi_0^2 - \frac{1}{2}\left(\xi_{+1}\xi_{-1}+\xi_{-1}\xi_{+1}\right) = \partial_{r}\,\left(r-M\right)^2\partial_{r} -\frac{M^4}{\left(r-M\right)^2}\,\partial_t^2 \,.
\ee

A massless scalar field $\Phi$ propagating in the background of an extremal Reissner-Nordstr\"{o}m black hole satisfies the following Klein-Gordon equation,
\be
	\begin{gathered}
		\mathbb{O}_{\text{full}}\Phi_{\omega\ell m} = \ell\left(\ell+1\right)\Phi_{\omega\ell m} \,, \\
		\mathbb{O}_{\text{full}} = \partial_{r}\,\left(r-M\right)^2\partial_{r} - \frac{r^4}{\left(r-M\right)^2}\,\partial_{t}^2 \,,
	\end{gathered}
\ee
where we have expanded into spherical harmonic modes of integer orbital number $\ell$. Zooming-in on the horizon via the $\lambda\rightarrow0$ limit of the scaling transformation \eqref{eq:scalRN}, the radial Klein-Gordon operator becomes
\be\ba
	\mathbb{O}_{\text{full}} &= \partial_{\rho}\,\rho^2\,\partial_{\rho} - \frac{M^4}{\rho^2}\,\partial_{\tau}^2 + \mathcal{O}\left(\lambda\right) \\
	&=  \partial_{r}\,\left(r-M\right)^2\partial_{r} - \frac{M^4}{\left(r-M\right)^2}\,\partial_{t}^2 + \mathcal{O}\left(\lambda\right) \,,
\ea\ee
where in the second line we switched back to the original Schwarzschild coordinates. As expected, the leading order part exactly matches the Casimir operator of the near-horizon $\SL$ algebra. The Klein-Gordon equation in the near-horizon scaling limit can then be written as
\be
	\mathcal{C}_2 \Phi_{\omega\ell m} =  \ell\left(\ell+1\right)\Phi_{\omega\ell m} \,,
\ee
with integer $\ell$ by virtue of the spherical symmetry of the background. Furthermore, according to the scaling transformation \eqref{eq:scalRN}, the throat frequency $\varpi$ of the scalar field perturbation is related to the frequency $\omega$ as measured by an asymptotic observer in the exterior as
\be
	\omega = \lambda \varpi \,.
\ee
Therefore, as $\lambda\rightarrow0$, all frequencies in the throat correspond to the single frequency $\omega=0$ in the exterior. In other words, from the perspective of the full geometry, only static perturbations survive in the throat and this enhanced $\SL$ isometry fully constrains their wave dynamics. Surprisingly, the near-horizon Klein-Gordon operator turns out to coincide with the Klein-Gordon differential operator in the near-zone approximation, see Eq.~\eqref{eq:NZSplitRadial_KerrNewman} with $a=0$ and $r_{+}=r_{-}=M$.

The situation is more intricate in the extremal Kerr-Newman case~\cite{Bardeen:1999px,Amsel:2009et}. One obtains the near-horizon extremal geometry by taking the $\lambda\rightarrow0$ limit of the following rescaled co-rotating coordinates\footnote{The angular velocity of the extremal Kerr-Newman black hole is equal to $\Omega=\frac{a}{r_{+}^2+a^2} = \frac{a}{M^2+a^2}$ and, therefore, $\varphi = \phi - \frac{a}{M^2+a^2}\, t = \phi - \Omega\,t$ is indeed the co-rotating azimuthal angle.}
\be\label{eq:limKerr}
	r =M + \lambda\rho \,,\quad t = \frac{\tau}{\lambda} \,,\quad \phi = \varphi + \frac{a}{M^2+a^2}\frac{\tau}{\lambda} \,,
\ee
which generates the metric 
\be
	ds^2 = \left(1-\frac{a^2}{\rho^2_0}\sin^2\theta\right) \left[ -\frac{\rho^2}{\rho_0^2}d\tau^2 + \frac{\rho_0^2}{\rho^2}d\rho^2 + \rho_0^2 d\theta^2 \right] + \frac{\rho_0^2\sin^2\theta}{1-\frac{a^2}{\rho^2_0}\sin^2\theta} \left( d\varphi + \frac{2Ma\rho}{\rho_0^4}d\tau\right)^2 \,,
\ee
with $\rho_0^2\equiv M^2 + a^2$. This metric possesses the azimuthal symmetry $U\left(1\right)$ generated by the vector field $J_0 = -i\,\partial_{\varphi}$, as well as additional $\SL$ Killing vectors of the $\text{AdS}_2$ factor,
\be\ba\label{eq:SL2RNHEKerrNewman}
	\xi_0 &= \tau\,\partial_{\tau} - \rho\,\partial_{\rho} \,,\quad \xi_{+1} = \partial_{\tau} \,, \\
	\xi_{-1} &= \left( \frac{\left(M^2+a^2\right)^2}{\rho^2} + \tau^2\right)\partial_{\tau} -2\tau\rho\,\partial_{\rho} - \frac{4M a}{\rho}\,\partial_{\varphi} \,,
\ea\ee
or, in Boyer-Lindquist coordinates,
\be\ba\label{eq:SL2RNHEKerrNewmanBL}
	\xi_0 &= t\left(\partial_{t}+\Omega\,\partial_{\phi}\right) - \left(r-M\right) \partial_{r} \,,\quad \xi_{+1} = \lambda^{-1}\left(\partial_{t} +\Omega\,\partial_{\phi}\right) \,, \\
	\xi_{-1} &= \lambda \left[ \left( \frac{\left(M^2+a^2\right)^2}{\left(r-M\right)^2}+t^2 \right)\left(\partial_{t} + \Omega\,\partial_{\phi}\right) - 2t\left(r-M\right)\partial_{r} - \frac{4M a}{r-M}\partial_{\phi} \right] \,.
\ea\ee
The full isometry group is, hence, $\SL\times U\left(1\right)$. The Casimir of the $\SL$ factor expressed in the original Boyer-Lindquist coordinates is given by
\be\label{eq:KNNHE}
	\mathcal{C}_2 = \partial_{r}\left(r-M\right)^2\partial_{r} - \left(\frac{M^2+a^2}{r-M}\right)^2\left(\partial_{t}+\Omega\,\partial_{\phi}\right)^2 + \frac{4Ma}{r-M}\left(\partial_{t}+\Omega\,\partial_{\phi}\right)\partial_{\phi} \,.
\ee
Performing the scaling transformation \eqref{eq:limKerr} at the level of the radial Klein-Gordon equation in the extremal Kerr-Newman black hole background, Eq.~\eqref{eq:KleinGordonFull_KerrNewman} with $r_{+}=r_{-}=M$, and taking the $\lambda\rightarrow0$ limit reveals that the near-horizon $s=0$ Teukolsky equation for an extremal Kerr-Newman black hole can be rewritten as
\be\label{eq:C2SL2RExtremalKerr}
	\mathcal{C}_2\Phi_{\omega\ell m} = \left[\ell\left(\ell+1\right)-2\left(3M^2+a^2\right)m^2\Omega^2\right]\Phi_{\omega\ell m} \,,
\ee
where the angular eigenvalues $\ell\left(\ell+1\right)$ are no longer integers, unless $m=0$. The explicit range of values of $\ell$ for non-zero azimuthal numbers is known only numerically and can be found, for example, in~\cite{Bardeen:1999px}. Similarly to the Reissner-Nordstr\"{o}m case, not all frequencies of scalar perturbations propagating in the full geometry survive in the throat. In the rotating case, solutions which can be extended to smooth perturbations of the full Kerr-Newman geometry can only have the locking frequency $\omega = m\Omega$. However, in contrast to the extremal Reissner-Nordstr\"{o}m case, the near-horizon Teukolsky equation for extremal Kerr-Newman black holes does not coincide with the extremal limit of any of the near-zone truncations.

\section[Near-horizon isometries from near-zone symmetries of scalar perturba-\linebreak tions]{Near-horizon isometries from near-zone symmetries of scalar perturbations}
\label{sec:NHEfromNZ4d}
We will now explore whether the near-horizon Killing vectors of extremal black holes can be retrieved by appropriately taking extremal limits of the $s=0$ Love $\SL$ generators that exist for non-extremal black holes. We will first do this for Reissner-Nordstr\"{o}m black holes where we will see that the near-horizon $\SL$ algebra can be obtained from the extremal limit of a particular combination of the near-zone Love algebra~\cite{Charalambous:2022rre}. We will then attempt to do the same for the Love symmetry manifested in the near-zone region for non-extremal Kerr-Newman black holes; this will not quite work as straightforwardly, but utilizing the infinite extension $\SL\ltimes U\left(1\right)_{\mathcal{V}}$ we found in Section~\ref{sec:InfiniteExtension4d} will turn out to precisely recover the desired Killing vectors~\cite{Charalambous:2022rre}.

\subsection{Reissner-Nordstr{\"o}m case}
\label{subsec:NHE_RN4d}
Let us take the extremal limit of the $s=0$ Love generators in the spherically symmetric Reissner-Nordstr\"{o}m case, Eq.~\eqref{eq:SL2RKerrNewman} with $a=0$. We parameterize the near-extremal expansion via the Hawking temperature $T_{H}$ of the black hole. Then,
\be
	Q^2=\frac{1-8\pi T_{H}M}{8\pi^2T_{H}^2}\left[\frac{1-4\pi T_{H}M}{\sqrt{1-8\pi T_{H}M}} - 1\right] = M^2\left(1-4\pi^2 T_{H}^2M^2\right) + \mathcal{O}\left(T_{H}^3\right) \,.
\ee
Taylor expanding the Love symmetry generators \eqref{eq:SL2RKerrNewman} up to $\mathcal{O}(T_{H}^2)$ we obtain
\be\ba
	L_0 &= -\frac{1}{2\pi T_{H}}\,\partial_{t} \\
	L_{\pm1} &= -\left(r-M\right)\left(\pm1 + t \,2\pi T_{H}\right)\partial_{r} + \left[\frac{1}{2\pi T_{H}} \pm t +\left(\frac{M^4}{\left(r-M\right)^2} + t^2\right)2\pi T_{H}\right] \partial_{t} \,.
\ea\ee
These generators are singular at $T_H=0$, so that taking the extremal limit of the Love symmetry is somewhat non-trivial. To achieve this, we consider three linear combinations of generators with $T_{H}$-dependent coefficients which have a finite non-trivial $T_H=0$ limit. These linear combinations are 
\be\ba
	\xi_{+1} &= \lambda^{-1}\lim_{T_{H}\to 0}\left(-2\pi T_{H} L_0\right) = \lambda^{-1} \partial_{t} \,,\\
	\xi_{0} &= \lim_{T_{H}\to 0}\frac{L_{+1}-L_{-1}}{2} = t\,\partial_{t} -\left(r-M\right)\partial_{r} \,,\\
	\xi_{-1} &= \lambda\lim_{T_{H}\to 0}\frac{L_1+L_{-1}+2L_0}{2\pi T_{H}} = \lambda\left[ \left(t^2 +\frac{M^4}{(r-M)^2}\right)\partial_{t} - 2t\left(r-M\right)\partial_{r} \right] \,.
\ea\ee
They exactly reproduce the generators of the $\SL$ near-horizon isometry of extremal charged black holes \eqref{eq:isoRN2} after identifying the parameter $\lambda$ with the scaling parameter associated with the near-horizon limit\footnote{This overall constant reciprocal rescaling of the $\xi_{+1}$ and $\xi_{-1}$ generators is in fact irrelevant as it falls into the automorphisms of $\SL$, i.e. its choice does not change the algebra or the Casimir, and merely serves to demonstrate explicitly the exact recovering of the $\SL$ Killing vectors of the near-horizon geometry of extremal Reissner-Nordstr\"{o}m black holes.}.

This limiting procedure is somewhat similar to the Wigner-In\"{o}n\"{u} contraction~\cite{Inonu:1953sp}.
Unlike the latter case though, here the limiting algebra of the near-horizon isometries is again $\SL$, so it is isomorphic to the original Love symmetry algebra. However, the limit itself is algebraically non-trivial. Namely, the resulting mapping of the generators does not correspond to the algebra automorphism.

To see this, it is instructive to investigate the algebraic structure of solutions of the Klein-Gordon equation for extremal Reissner-Nordstr\"{o}m black holes as dictated by this $\SL$ near-horizon isometry. In particular, we are interested in the representation that the static solution belongs to. By definition, the static solution is annihilated by $\xi_{+1}$ and is, therefore, a {\it primary} vector $\upsilon_{-\ell,0}$ of the highest-weight representation,
\be 
	\xi_{+1}\upsilon_{-\ell,0} = 0\,,\quad \xi_0\upsilon_{-\ell,0} = -\ell\,\upsilon_{-\ell,0} \,,\quad \Rightarrow 
	\quad \upsilon_{-\ell,0} = \left(r-M\right)^{\ell} \,.
\ee
Other solutions in the highest-weight $\SL$ multiplet are then obtained as descendants,
\be\ba
	\upsilon_{-\ell,1} &= \xi_{-1} \upsilon_{-\ell,0} = -2\lambda\ell\,t \left(r-M\right)^{\ell} \,, \\
	\upsilon_{-\ell,2} &= \xi_{-1} \upsilon_{-\ell,1} = -2\lambda^{2}\ell\left[ \left(1-2\ell\right) t^2 +\frac{M^4}{\left(r-M\right)^2} \right] \left(r-M\right)^{\ell} \,,
\ea\ee
and so on. From all of these elements of the highest-weight representation, $\upsilon_{-\ell,0}$ and $\upsilon_{-\ell,1}$ are exact solutions of the full Klein-Gordon equation for an extremal Reissner-Nordstr\"{o}m black hole, while $\upsilon_{-\ell,n}$ with $n>1$ are only approximate solutions valid in the near-horizon region. A change of the algebraic structure manifests itself in the fact that the static solution is a primary $\SL$ vector now, which is not the case before taking the extremal limit. The corresponding solution still has a polynomial dependence on $r$, indicative of the vanishing Love numbers. However, in the extremal limit this polynomial form does not link directly to the highest-weight property of the corresponding representation.

\subsection{Kerr-Newman case}
\label{subsec:NHE_KerrNewman}
Analogously to the Reissner-Nordstr\"{o}m black hole, the extremal limit for the case of Kerr-Newman black holes is realized as the $T_{H}\rightarrow0$ limit of vanishing Hawking temperature, in terms of which,
\be
	a^2+Q^2 = M^2\left(1 - 4\pi^2T_{H}^2M^2\right) + \mathcal{O}\left(T_{H}^3\right) \,.
\ee
However, an attempt to take linear combinations of the Love symmetry generators analogously to the non-rotating paradigm does not recover the near-horizon $\SL$ Killing vectors in this case. This may not come as a surprise since the Casimir of the $\SL$ Love symmetry does not capture the full near-horizon behavior in the extremal case, in contrast to the Reissner-Nordstr\"{o}m black hole.

Nevertheless, the near-horizon $\SL$ Killing vector can still be recovered in the rotating case as well by making use of the infinite extension $\SL\ltimes U\left(1\right)_{\mathcal{V}}$ constructed in Section~\ref{sec:InfiniteExtension4d}. Namely, consider a family of $\SL$ subalgebras corresponding to the choice
\be\label{eq:NHa}
	\alpha = 1+4\pi T_{H}M + \mathcal{O}\left(T_{H}^2\right)
\ee
in Eq.~\eqref{eq:StarLove}, for arbitrary subleading contributions in the extremal limit. The corresponding generators are given by
\be\label{eq:SL2RNHne}
	\begin{gathered}
		L_0^{\text{NHE}} = -\frac{1}{2\pi T_{H}}\left[\partial_{t}+\left(1+4\pi T_{H}M\right)\Omega\,\partial_{\phi}\right] + \mathcal{O}\left(T_{H}\right) \,, \\
		\ba
			L_{\pm 1}^{\text{NHE}} = e^{\pm 2\pi T_{H}t} &\bigg[\mp\sqrt{\Delta}\,\partial_{r} + \partial_{r}\left(\sqrt{\Delta}\right)\frac{1}{2\pi T_{H}}\left(\partial_{t}+\Omega\,\partial_{\phi}\right) \\
			&+ 2M\sqrt{\frac{r-r_{+}}{r-r_{-}}} \Omega\,\partial_{\phi} + \mathcal{O}\left(T_{H}\right) \bigg] \,,
		\ea
	\end{gathered}
\ee
and the associated quadratic Casimir operator is
\be\label{eq:C2nhe2}
	\mathcal{C}_2^{\text{NHE}} = \partial_{r}\Delta\partial_{r} - \frac{\left(r_{+}^2+a^2\right)^2}{\Delta}\left(\partial_{t}+\Omega\,\partial_{\phi}\right)^2 + \frac{4Ma}{r-r_{-}}\left(\partial_{t}+\Omega\,\partial_{\phi}\right)\partial_{\phi} + \mathcal{O}\left(T_{H}\right) \,,
\ee
which indeed reduces to the appropriate near-horizon Teukolsky differential operator \eqref{eq:KNNHE} in the extremal limit.

Reshuffling the vector fields $L_{m}^{\text{NHE}}$ as before and taking the $T_{H}\to 0$ limit, we arrive at the exact near-horizon extremal Kerr-Newman $\SL$ Killing vectors \eqref{eq:SL2RNHEKerrNewmanBL},
\be\ba
	\xi_{+1} &= \lambda^{-1}\lim_{T_{H}\to 0} \left(-2\pi T_{H} L_0^{\text{NHE}}\right) = \lambda^{-1}\left(\partial_{t} +\Omega\,\partial_{\phi}\right) \,, \\
	\xi_0 &= \lim_{T_{H}\to 0} \frac{L_{+1}^{\text{NHE}}-L_{-1}^{\text{NHE}}}{2}= t\left(\partial_{t}+\Omega\,\partial_{\phi}\right) - \left(r-M\right) \partial_{r} \,, \\
	\xi_{-1} &= \lambda\lim_{T_{H}\to 0} \frac{L_{+1}^{\text{NHE}}+L_{-1}^{\text{NHE}}+2 L_0^{\text{NHE}}}{2\pi T_{H}} \\
	&= \lambda\left[ \left( \frac{\left(M^2+a^2\right)^2}{\left(r-M\right)^2}+t^2 \right)\left(\partial_{t} + \Omega\,\partial_{\phi}\right) - 2t\left(r-M\right)\partial_{r} - \frac{4M a}{r-M}\partial_{\phi} \right] \,,
\ea\ee
after identifying $\lambda$ with the near-horizon scaling parameter as before.

A test field separable solution $\Phi_{\omega\ell m}$ must belong to a certain representation of $\SL$, which is characterized by eigenvalues of the Casimir \eqref{eq:C2SL2RExtremalKerr} and the $\xi_0$ operator. For the static axisymmetric mode $\Phi_{\omega=0,\ell,m=0}$, the orbital number $\ell$ is integer and
\be
	\mathcal{C}_2\Phi_{\omega=0,\ell,m=0} = \ell\left(\ell+1\right)\Phi_{\omega=0,\ell,m=0} \,,\quad \xi_{+1}\Phi_{\omega=0,\ell,m=0} = 0\,.
\ee
This mode is clearly the primary state $\upsilon_{-\ell,0}^{\left(m=0\right)}$ of a highest-weight representation with weight $h=-\ell$ and azimuthal number $m=0$. Integrating $\xi_0 \upsilon_{-\ell,0}^{\left(m=0\right)} = -\ell\,\upsilon_{-\ell,0}^{\left(m=0\right)}$, we obtain
\be
	\upsilon_{-\ell,0}^{\left(m=0\right)} = \left(r-M\right)^{\ell} \,.
\ee
The rest of the representation is then built by acting with $\xi_{-1}$,
\be\ba
	\upsilon_{-\ell,1}^{\left(m=0\right)} &= \xi_{-1} \upsilon_{-\ell,0}^{\left(m=0\right)} = -2\lambda\ell\,t \left(r-M\right)^{\ell} \,, \\
	\upsilon_{-\ell,2}^{\left(m=0\right)} &= \xi_{-1} \upsilon_{-\ell,1}^{\left(m=0\right)} = -2\lambda^{2}\ell\left[ \left(1-2\ell\right) t^2 + \frac{\left(M^2+a^2\right)^2}{\left(r-M\right)^2} \right] \left(r-M\right)^{\ell} \,,
\ea\ee
and so on. Similarly to the extremal Reissner-Nordstr\"{o}m black hole, only the static solution $\upsilon_{-\ell,0}^{\left(m=0\right)}$ satisfies the full extremal $s=0$ Teukolsky equation and can be extended beyond the near-horizon approximation; its polynomial form implies the vanishing of the Love numbers for static axisymmetric perturbations. However, this polynomial form is related to the special form of the generator $\xi_0$, rather than the highest-weight property.

For a generic perturbation with magnetic number $m$ the corresponding frequency is fixed to be $\omega=m\Omega$,
in which case the Casimir eigenvalue is not an integer. These solutions do not belong to highest-weight $\SL$ representations and their response coefficients are not zero.

\section{Near-horizon isometries from near-zone symmetries of general perturbations}
\label{sec:NHEfromNZS4d}
We will finish this chapter by supplementing with the corresponding results for the $s\ne0$ extremal radial Teukolsky equation in the near-horizon limit,
\be\ba\label{eq:NHTeukolskyS}
	{}&\mathbb{O}_{\text{full}}^{\left(s\right)} = \rho^{-2s}\partial_{\rho}\,\rho^{2\left(s+1\right)}\partial_{\rho} - \frac{\left(M^2+a^2\right)^2}{\rho^2}\,\partial_{\tau}^2 + \frac{4Ma}{\rho}\,\partial_{\tau}\partial_{\varphi} + 2s\frac{M^2+a^2}{\rho}\,\partial_{\tau} \\
	&\quad\quad\quad- 2\left(3M^2+a^2\right)\Omega^2\partial_{\varphi}^2 + \mathcal{O}\left(\lambda\right) \\
	&= \left(r-M\right)^{-2s}\partial_{r}\,\left(r-M\right)^{2\left(s+1\right)}\partial_{r} - \frac{\left(M^2+a^2\right)^2}{\left(r-M\right)^2}\left(\partial_{t}+\Omega\,\partial_{\phi}\right)^2 \\
	&\quad+ \frac{4Ma}{r-M}\left(\partial_{t}+\Omega\,\partial_{\phi}\right)\partial_{\phi} + 2s\frac{M^2+a^2}{r-M}\left(\partial_{t}+\Omega\,\partial_{\phi}\right) - 2\left(3M^2+a^2\right)\Omega^2\partial_{\phi}^2 + \mathcal{O}\left(\lambda\right) \,.
\ea\ee
This can be reproduced by the spin-weighted Killing vectors of the near-horizon isometry $\SL$ subgroup, constructed via the spin-weighted Lie derivative we introduced in Section~\ref{sec:SpinWeightedGenerators}. Expanding the Kinnersley tetrad \eqref{eq:KinnersleyTetrad} at leading order in the near-horizon scaling limit,
\be\ba
	\ell &= \frac{\lambda^{-1}}{\rho}\left(\frac{M^2+a^2}{\rho}\,\partial_{\tau} - \frac{2Ma}{M^2+a^2}\,\partial_{\varphi} + \rho\,\partial_{\rho}\right) + \mathcal{O}\left(\lambda^0\right) \,, \\
	n &= \frac{\lambda}{2\rho\left(M^2+a^2\cos^2\theta\right)}\left(\frac{M^2+a^2}{\rho}\,\partial_{\tau} - \frac{2Ma}{M^2+a^2}\,\partial_{\varphi} - \rho\,\partial_{\rho}\right) + \mathcal{O}\left(\lambda^2\right)  \,, \\
	m &= \frac{1}{\sqrt{2}\left(M+ia\cos\theta\right)}\left[\partial_{\theta} + \frac{i}{\sin\theta}\left(1 - \frac{a^2\sin^2\theta}{M^2+a^2}\right)\partial_{\varphi}\right] + \mathcal{O}\left(\lambda\right) \,,
\ea\ee
and employing the spin-weighted Lie derivative construction \eqref{eq:LieLudwig},
\be
	\xi_{m}^{\left(s\right)} = \Lstr_{\xi_{m}} \,,\quad m=0,\pm1 \,,
\ee
we find
\be
	\xi_0^{\left(s\right)} = \xi_0 - s \,,\quad \xi_{+1}^{\left(s\right)} = \xi_{+1} \,,\quad \xi_{-1}^{\left(s\right)} = \xi_{-1} -2s\left(\frac{M^2}{\rho} + \tau\right) \,,
\ee
or, in the original Boyer-Lindquist coordinates,
\be\ba\label{eq:SL2RNHEKerr_s}
	\xi_0^{\left(s\right)} &= t\left(\partial_{t}+\Omega\,\partial_{\phi}\right) - \left(r-M\right) \partial_{r} - s \,,\quad \xi_{+1}^{\left(s\right)} = \lambda^{-1}\left(\partial_{t} +\Omega\,\partial_{\phi}\right) \,, \\
	\xi_{-1}^{\left(s\right)} &= \lambda \bigg[ \left( \frac{\left(M^2+a^2\right)^2}{\left(r-M\right)^2}+t^2 \right)\left(\partial_{t} + \Omega\,\partial_{\phi}\right) - 2t\left(r-M\right)\partial_{r} - \frac{4M a}{r-M}\partial_{\phi} \\
	&\quad\quad-2s\left(\frac{M^2}{r-M} + t\right) \bigg] \,.
\ea\ee
The Casimir operator of this algebra correctly captures the near-horizon Teukolsky equation \eqref{eq:NHTeukolskyS} after supplementing with the appropriate $U\left(1\right)$ factor, namely, the separable NP scalar $\Psi_{s\omega\ell m}$ in the near-horizon regime will satisfy \eqref{eq:C2SL2RExtremalKerr}.

As in the scalar example analyzed in the previous section, the non-extremal $\SL$ algebras from which to retrieve the above spin-weighted near-horizon Killing vectors are subalgebras of the infinite extension $\SL\ltimes U\left(1\right)_{\mathcal{V}}$ found in Section~\ref{sec:InfiniteExtension4d}. Choosing $\alpha$ as in equation \eqref{eq:NHa} for arbitrary sub-extremal contributions, we obtain the $\SL$ algebra,
\be\label{eq:SL2RNHne_s}
	\begin{gathered}
		L_0^{\text{NHE},\left(s\right)} = -\frac{1}{2\pi T_{H}}\left[\partial_{t}+\left(1+4\pi T_{H}M\right)\Omega\,\partial_{\phi}\right] + \mathcal{O}\left(T_{H}\right) \,, \\
		\ba
			L_{\pm 1}^{\text{NHE},\left(s\right)} = e^{\pm 2\pi T_{H}t} &\bigg[\mp\sqrt{\Delta}\,\partial_{r} + \partial_{r}\left(\sqrt{\Delta}\right)\frac{1}{2\pi T_{H}}\left(\partial_{t}+\Omega\,\partial_{\phi}\right) + s\frac{r-r_{\mp}}{\sqrt{\Delta}} \\
		&+ 2M\sqrt{\frac{r-r_{+}}{r-r_{-}}} \Omega\,\partial_{\phi} + \mathcal{O}\left(T_{H}\right) \bigg] \,.
		\ea
	\end{gathered}
\ee
The Casimir of this algebra is explicitly given by
\be\ba\label{eq:CasNH1_s}
	\mathcal{C}_2^{\text{NHE},\left(s\right)} &= \Delta^{-s}\partial_{r}\Delta^{s+1}\partial_{r} - \frac{\left(r_{+}^2+a^2\right)^2}{\Delta}\left(\partial_{t}+\Omega\,\partial_{\phi}\right)^2 + \frac{4Ma}{r-r_{-}}\left(\partial_{t}+\Omega\,\partial_{\phi}\right)\partial_{\phi} \\
	&\quad+ s\frac{\left(r_{+}^2+a^2\right)\Delta^{\prime}}{\Delta}\left(\partial_{t}+\Omega\,\partial_{\phi}\right) + s\left(s+1\right)  \\
	&\quad+ \frac{r_{+}-r_{-}}{r-r_{-}}\left(2M\Omega\,\partial_{\phi}+s\right)2M\Omega\,\partial_{\phi} + \mathcal{O}\left(T_{H}\right) \,,
\ea\ee
which reduces to the appropriate near-horizon extremal Teukolsky operator \eqref{eq:NHTeukolskyS} in the extremal limit.

We can then follow the same prescription we used for $s=0$ to recover the spin-weighted Killing vectors \eqref{eq:SL2RNHEKerr_s},
\be\ba
	\xi_{+1}^{\left(s\right)} &= \lambda^{-1}\lim_{T_{H}\to 0} \left(-2\pi T_{H} L_0^{\text{NHE},\left(s\right)}\right) \,, \\
	\xi_0^{\left(s\right)} &= \lim_{T_{H}\to 0} \frac{L_{+1}^{\text{NHE},\left(s\right)}-L_{-1}^{\text{NHE},\left(s\right)}}{2} \,, \\
	\xi_{-1}^{\left(s\right)} &= \lambda\lim_{T_{H}\to 0} \frac{L_{+1}^{\text{NHE},\left(s\right)}+L_{-1}^{\text{NHE},\left(s\right)}+2 L_0^{\text{NHE},\left(s\right)}}{2\pi T_{H}} \,.
\ea\ee

An important remark here that was not relevant in the scalar example is that the infinite-dimensional extension $\SL\ltimes U\left(1\right)_{\mathcal{V}}$ is now needed not only in the Kerr-Newman case, but also in the spherically symmetric Reissner-Nordstr\"{o}m case in order to correctly reproduce the full spin-weighted Killing vectors \eqref{eq:SL2RNHEKerr_s}.

\section{Summary of Chapters~\ref{ch:TLNsBlackHoles4d}-\ref{ch:NHE}}
Let us summarize the results around the study of the black hole response problem in four spacetime dimensions. In Chapter~\ref{ch:TLNsBlackHoles4d}, we found that asymptotically flat general-relativistic black holes in four spacetime dimensions have exactly zero static Love numbers~\cite{Fang:2005qq,Damour:2009vw,Binnington:2009bb,Gurlebeck:2015xpa,Poisson:2014gka,Landry:2015zfa,Pani:2015hfa,LeTiec:2020spy,LeTiec:2020bos,Chia:2020yla,Charalambous:2021mea,Ivanov:2022hlo,Ivanov:2022qqt}.

From the point of view of the worldline EFT, this raises naturalness concerns~\cite{tHooft:1979rat,Porto:2016zng} (see Section~\ref{sec:PowerCounting}) which are resolved by the manifestation of the Love $\SL$ symmetry in the near-zone region~\cite{Charalambous:2021kcz,Charalambous:2022rre} as we saw in Chapter~\ref{ch:LoveSymmetry4d}. This is achieved by the fact that the physical static solution belongs to a particular highest-weight representation of the Love symmetry. The highest-weight property then implies a selection rule of the form
\be
	\left(L_{+1}\right)^{\ell+1}\vev{\Phi_{\omega=0,\ell m}} = 0
\ee
for the case of, for instance, a scalar field perturbation, which dictates a (quasi-)polynomial form of the static solution regular at the future event horizon with no response modes.

In Chapter~\ref{ch:Properties}, we investigated various properties of the Love symmetry. We saw, to begin with, that the Teukolsky equation admits a second near-zone splitting which also acquires a globally defined near-zone $\SL$ structure; the Starobinsky near-zone algebra~~\cite{Charalambous:2021kcz,Charalambous:2022rre}. Furthermore, we saw that representation theory arguments analogous to those for the static case can be used to infer \textit{all} the interesting properties of the Love numbers at leading order in the near-zone expansion, even for non-zero frequencies despite the fact that these are not accurate within the near-zone regime~\cite{Charalambous:2022rre}. The states in the highest-weight Love multiplet were interpreted via scattering theory observables as reflectionless, total transmission modes~\cite{Hod:2013fea,Cook:2016fge,Cook:2016ngj,MaassenvandenBrink:2000iwh}, although the exact connection remains unclear. We also attempted to infer the form of the generators for electromagnetic and gravitational ($s\ne0$) perturbations by introducing the notion of the spin-weighted Lie derivative acting on NP scalars~\cite{Ludwig2000,Ludwig:2001hx} in Section~\ref{sec:SpinWeightedGenerators}. This was motivated by the fact that the near-zone symmetries are approximate in a quite rigorous sense, i.e. in the sense that they are isometries of subtracted geometries~\cite{Cvetic:2011hp,Cvetic:2011dn} which have the property of preserving the internal structure of the black hole while subtracting information about the environment. Moreover, in Section~\ref{sec:LoveBeyondGR4d}, we investigated the possibility of Love symmetry arising in beyond-general-relativistic configurations and extracted a sufficient geometric condition for its existence; this was successfully contrasted with explicit computations of the static Love numbers for black holes in Riemann-cubed gravity in Section~\ref{sec:SLNs_R34d} which do not exhibit any seemingly fine-tuning behaviors.

More importantly, we presented in Section~\ref{sec:InfiniteExtension4d} an infinite-dimensional extension of the Love $\SL$ symmetry into $\SL\ltimes U\left(1\right)_{\mathcal{V}}$~\cite{Charalambous:2021kcz,Charalambous:2022rre}. This infinite extension contains a particularly interesting family of $\SL$ subalgebras whose extremal limit precisely recovers the Killing vectors that generate the enhanced $\SL$ isometry subgroup of the near-horizon geometry of extremal black holes~\cite{Charalambous:2022rre} as we saw in detail in this chapter.

A natural next step is to study the black hole response problem in higher spacetime dimensions~\cite{Kol:2011vg,Hui:2020xxx,Pereniguez:2021xcj,Charalambous:2024tdj}. This will be the topic of interest in the next two chapters.
	\newpage
\chapter{Black hole Love numbers and Love symmetry in higher spacetime dimensions}
\label{ch:LoveSymmetryDd}

We will now go beyond the four-dimensional examples we analyzed in the previous chapters and investigate the black hole response problem in higher spacetime dimensions. We will start by presenting in Section~\ref{sec:BHs_SphericallySymmetricDd} the main features of the geometry of an asymptotically flat and spherically symmetric black hole in a general geometric theory of gravity, allowed to carry an electric charge to begin with. In Section~\ref{sec:2pDm2Decomposition}, we will set the arena of studying perturbations of such static background geometries in a practical and covariant way as prescribed in \cite{Ishibashi:2003ap,Martel:2005ir}. We will then derive the equations of motion for the scalar (spin-$0$) response of such a generic black hole in Section~\ref{sec:EOM_Seq0}. Moving on, we will focus to electrically neutral black holes and extract the equations of motion relevant for computing the electric and magnetic (spin-$1$) susceptibilities of the black hole in Section~\ref{sec:EOM_Seq1}. This will be generalized to $p$-form perturbations in Section~\ref{sec:EOM_Seqp}~\cite{Yoshida:2019tvk}. As a last example in the category of spherical symmetric black holes, we will specialize to general-relativistic black holes and built the gravitational (spin-$2$) equations of motion for the tensor-mode gravitational perturbations of the higher-dimensional Reissner-Nordstr\"{o}m black hole~\cite{Pereniguez:2021xcj} as well as the equations of motion for the vector-mode (Regge-Wheeler) and scalar-mode (Zerilli) gravitational perturbations of the higher-dimensional Schwarzschild black hole~\cite{Hui:2020xxx} in Section~\ref{sec:EOM_Seq2}.

With this setup, we will go ahead and compute the leading order near-zone Love numbers for general-relativistic black holes in Sections~\ref{sec:LNs_SchwarzschildDd}-\ref{sec:LNs_RNDd}. We will find a much richer structure compared to the four-dimensional situation. In particular, the Love numbers will be controlled by the rescaled orbital number $\hat{\ell}\equiv\frac{\ell}{d-3}$~\cite{Kol:2011vg,Hui:2020xxx}. As per the discussion of Section~\ref{sec:PowerCounting} on the expected behavior of the static Love numbers based on power counting arguments~\cite{Charalambous:2022rre}, we will find that the static Love numbers are non-zero and exhibit no logarithmic running for generic values of $\hat{\ell}$. For half-integer rescaled orbital numbers, $\hat{\ell}\in\mathbb{N}+\frac{1}{2}$, we will also find the expected RG-flowing behavior. This persists for the magnetic susceptibilities and the magnetic-type tidal Love numbers, as well as $p$-form modes and $\left(p-1\right)$-form modes with $2\le p\le d-4$ at integer $\hat{\ell}$. However, we will find the surprising vanishing of the general-relativistic black hole static Love numbers under a discrete set of resonant conditions with the characteristic example being the vanishing of the electric-type Love numbers for integer $\hat{\ell}$.

As in the four-dimensional paradigms, we will see in Section~\ref{sec:LoveSymmetryDd} that these magic zeroes can be addressed by the existence of a globally defined near-zone $\SL$ symmetry. This is rather special to general-relativistic configurations as we will see in Section~\ref{sec:LoveSymmetryModGR} where we will also derive a sufficient geometric condition for the existence of Love symmetry. This chapter is an adaptation of our work in Ref.~\cite{Charalambous:2024tdj}, with elements of our works in Refs.~\cite{Charalambous:2021kcz,Charalambous:2022rre} around the scalar response problem, where we infer the ``magic zeroes'' of black hole Love numbers in higher spacetime dimensions, originally found in Refs.~\cite{Kol:2011vg,Hui:2020xxx,Pereniguez:2021xcj}, by revealing the emergence of the Love symmetry acting on $p$-form and gravitational perturbations.

\section{Spherically symmetric black holes in higher spacetime dimensions}
\label{sec:BHs_SphericallySymmetricDd}

We start by presenting the background geometry. We will be dealing with a general asymptotically flat, spherically symmetric and non-extremal black holes, which can always be brought to the form
\be\label{eq:SphericallySymemtricBHDd}
	ds^2 = -f_{t}\left(r\right)dt^2 + \frac{dr^2}{f_{r}\left(r\right)} + r^2d\Omega_{d-2}^2 \,,
\ee
where $d\Omega_{d-2}^2 = \Omega_{AB}\left(\theta\right)d\theta^{A}d\theta^{B}$ is the metric on $\mathbb{S}^{d-2}$, with angular coordinates labeled by capital Latin indices, $A,B=1,\dots,d-2$, and the argument ``$\theta$'' in $\Omega_{AB}\left(\theta\right)$ collectively indicating all the angular coordinates\footnote{In spherical coordinates $\theta^{A}$, $A=1,\dots,d-2$, with $\theta^{1},\theta^{2},\dots,\theta^{d-3}\in\left[0,\pi\right]$ the $d-3$ polar angles and $\theta^{d-2}\in\left[0,2\pi\right)$ the periodically identified azimuthal angle, $\Omega_{AB}\left(\theta\right) = \delta_{AB}\prod_{C=1}^{A-1}\sin^2\theta^{C}$.}. The asymptotic flatness and non-extremality conditions are imposed by the requirements
\be
	\begin{gathered}
		\lim_{r\rightarrow\infty}f_{t,r}\left(r\right)=1 \,, \\
		f_{t}\left(r_{\text{h}}\right)=f_{r}\left(r_{\text{h}}\right)=0 \quad \text{and} \quad f_{t}^{\prime}\left(r_{\text{h}}\right),f_{r}^{\prime}\left(r_{\text{h}}\right)\ne0 \,.
	\end{gathered}
\ee
Similar to the $4$-dimensional Schwarzschild black hole, the event horizon is a Killing horizon with respect to the Killing vector $K=\partial_{t}$.

To analyze the behavior of the perturbations near the event horizon, it is necessary to employ the tortoise coordinate,
\be
	dr_{\ast} = \frac{dr}{\sqrt{f_{t}\left(r\right)f_{r}\left(r\right)}} \,,
\ee
in terms of which the advanced ($+$) and retarded ($-$) null coordinates $\left(t_{\pm},r,\theta^{A}\right)$ are defined as
\be\label{eq:AdvancedRetarded_SchwazrschildDd}
	dt_{\pm} = dt \pm dr_{\ast} \,.
\ee
In particular, the near-horizon behavior of the tortoise coordinate can be extracted explicitly to be
\be\label{eq:Tortoise_SchwazrschildDd}
	r_{\ast} \sim \frac{\beta}{2}\ln\left|\frac{r-r_{\text{h}}}{r_{\text{h}}}\right| \quad \text{as $r\rightarrow r_{\text{h}}$} \,,
\ee
with $\beta$ the inverse Hawking temperature,
\be\label{eq:betaDd}
	\beta = \frac{1}{2\pi T_{H}} = \frac{2}{\sqrt{f_{t}^{\prime}\left(r_{\text{h}}\right)f_{r}^{\prime}\left(r_{\text{h}}\right)}} \,.
\ee
Then, monochromatic waves of frequency $\omega$ that are ingoing at the future ($+$)/past ($-$) event horizon behave as
\be\label{eq:NHbehavior_SphericalSymmetric}
	e^{-i\omega t_{\pm}} \sim e^{-i\omega t}\left(\frac{r-r_{\text{h}}}{r_{\text{h}}}\right)^{\mp i\beta\omega/2} \,.
\ee

For General Relativity, the most general spherically symmetric and asymptotically flat black hole geometry is the higher-dimensional Reissner-Nordstr\"{o}m(-Tangherlini) solution~\cite{Tangherlini:1963bw,Hollands:2012xy},
\be\ba\label{eq:RNDdGeometry}
	f_{t}\left(r\right) = f_{r}\left(r\right) &= 1 - \left(\frac{r_{s}}{r}\right)^{d-3} + \left(\frac{r_{Q}}{r}\right)^{2\left(d-3\right)} \\
	&=  \left[1-\left(\frac{r_{+}}{r}\right)^{d-3}\right]\left[1-\left(\frac{r_{-}}{r}\right)^{d-3}\right] \,.
\ea\ee
where the Schwarzschild radius $r_{s}$ and the charge parameter $r_{Q}$ are related to the ADM mass $M$ and the electric charge $Q$ (in CGS units) of the black hole according to
\be
	r_{s}^{d-3} = \frac{16\pi GM}{\left(d-2\right)\Omega_{d-2}} \,,\quad r_{Q}^{2\left(d-3\right)} = \frac{32\pi^2 GQ^2}{\left(d-2\right)\left(d-3\right)\Omega_{d-2}^2} \,,
\ee
while the inner and outer horizons are expressed in terms of $r_{s}$ and $r_{Q}$ as
\be
	r_{\pm}^{d-3} = \frac{1}{2}\left[r_{s}^{d-3} \pm \sqrt{r_{s}^{2\left(d-3\right)}-4r_{Q}^{2\left(d-3\right)}}\right] \,.
\ee
The essential singularity at $r\rightarrow0$ is hidden behind an event horizon as along as the magnitude of the electric charge is bounded from above from the mass of the black hole,
\be
	Q^2 \le 2\frac{d-3}{d-2}GM^2 \,,
\ee
with the saturation of the inequality indicating the extremality condition.

\section{$2+\left(d-2\right)$ decomposition}
\label{sec:2pDm2Decomposition}

Let us now develop a covariant formalism for studying the perturbations of the above spherically symmetric black hole. A higher-dimensional version of the Newman-Penrose formalism is possible~\cite{Pravda:2004ka,Coley:2004jv,Pravdova:2008gp,Durkee:2010xq} and the separability of perturbations around an algebraically special background geometry relevant for black holes has been shown explicitly for the class of the so-called Kundt spacetimes~\cite{Durkee:2010qu,Godazgar:2011sn}. Even though this class includes the higher-dimensional Schwarzschild black hole, we choose here to work in a less involved formalism that is also more reminiscent of the early days of studying the stability of the four-dimensional Schwarzschild black hole by Regge and Wheeler~\cite{Regge:1957td} and Zerilli~\cite{Zerilli:1970se}.

The background geometry is of the form of a generic $\mathcal{M}^{\left(2\right)}\times\mathbb{S}^{d-2}$ manifold equipped with a time-like Killing vector $t^{a}$,
\be
	\begin{gathered}
		ds^2 = g_{ab}\left(x\right)dx^{a}dx^{b} + r^2\left(x\right)\Omega_{AB}\left(\theta\right)d\theta^{A}d\theta^{B} \,, \\
		\mathcal{L}_{t}g_{ab} = 0 \,,\quad t^{a}\nabla_{a}\Omega_{AB} = 0 \,,\quad t^{a}\nabla_{a}r = 0 \,,
	\end{gathered}
\ee
where small Latin indices run over\footnote{We employ a slight change of notation in this chapter relative to the rest of the thesis to make contact with the literature~\cite{Ishibashi:2003ap,Martel:2005ir}. In particular, instead of referring to the tensorial structure on spatial slices, small Latin indices here refer to the transformation properties under diffeomorphisms on $\mathcal{M}^{\left(2\right)}$.} $\mathcal{M}^{\left(2\right)}$, $a,b=0,1$, and capital Latin indices run over the spherical coordinates $\theta^{A}$ of $\mathbb{S}^{d-2}$, $A=1,2,\dots,d-2$. With respect to $\mathcal{M}^{\left(2\right)}$, $r\left(x\right)$ is a scalar. In order to perform a covariant $2+\left(d-2\right)$ decomposition, we follow~\cite{Ishibashi:2003ap,Martel:2005ir} (see also~\cite{Kodama:2003jz,Kodama:2003kk,Ishibashi:2011ws}) and introduce the $\mathcal{M}^{\left(2\right)}$ co-vector normal to surfaces of constant $r\left(x\right)$,
\be
	r_{a} \equiv \nabla_{a}r \,.
\ee
In Schwarzschild coordinates, $r_{a}=\left(0,1\right)$. This allows to covariantly define $f_{t}\left(r\right)$ and $f_{r}\left(r\right)$ as
\be
	f_{t}\left(r\right) = -t_{a}t^{a} \,,\quad f_{r}\left(r\right) = r_{a}r^{a} \,.
\ee
The time-like Killing vector $t^{a}$ and the $2$-vector $r^{a}$ are orthogonal to each other, $t_{a}r^{a}=0$, and serve as a basis for $\mathcal{M}^{\left(2\right)}$. For example, the metric tensor $g_{ab}$ and the Levi-Civita tensor $\varepsilon_{ab}$ on $\mathcal{M}^{\left(2\right)}$ can be written as
\be
	g_{ab} =  -\frac{1}{f_{t}}t_{a}t_{b} + \frac{1}{f_{r}}r_{a}r_{b} \,,\quad \varepsilon_{ab} = -\frac{1}{\sqrt{f_{t}f_{r}}}\left(t_{a}r_{b}-r_{a}t_{b}\right) \,.
\ee
A zweibein for $\mathcal{M}^{\left(2\right)}$ would then be $\ell^{a} = t^{a}/\sqrt{f_{t}}$ and $n^{a} = r^{a}/\sqrt{f_{r}}$, such that $g_{ab} = -\ell_{a}\ell_{b} + n_{a}n_{b}$.

Let us now see how to decompose covariant derivatives. First of all, for spacetime scalar functions,
\be
	\nabla_{a}\phi = D_{a}\phi = -\frac{1}{f_{t}}t^{b}D_{b}\phi\,t_{a} + \frac{1}{f_{r}}r^{b}D_{b}\phi\,r_{a} \,,\quad \nabla_{A}\phi = D_{A}\phi \,,
\ee
where $D_{a}$ and $D_{A}$ are the covariant derivatives compatible with $g_{ab}$ and the unit-sphere metric $\Omega_{AB}$ respectively. For higher-spin fields, we need the $2+\left(d-2\right)$ decomposition of Christoffel symbols,
\be\ba
	\Gamma^{a}_{bA} &= 0 \,,\quad \Gamma^{a}_{AB} = -rr^{a}\Omega_{AB} \,, \\
	\Gamma^{A}_{ab} &= 0 \,,\quad \Gamma^{A}_{aB} = \frac{1}{r}r_{a}\delta^{A}_{B} \,.
\ea\ee
Then, we can see that, for a dual vector field $V_{\mu}$,
\be
	\begin{gathered}
		\nabla_{b}V_{a} = D_{b}V_{a} \,,\quad \nabla_{A}V_{a} = D_{A}V_{a} - \frac{1}{r}r_{a}V_{A} \\
		\nabla_{a}V_{A} = rD_{a}\left(\frac{V_{A}}{r}\right) \,,\quad \nabla_{B}V_{A} = D_{B}V_{A} + rr^{a}A_{a}\Omega_{AB} \,,
	\end{gathered}
\ee
while, for a rank-$2$ co-tensor $T_{\mu\nu}$,
\be
	\begin{gathered}
		\nabla_{c}T_{ab} = D_{c}T_{ab} \,,\quad \nabla_{A}T_{ab} = D_{A}T_{ab} - \frac{1}{r}\left(r_{a}T_{Ab}+r_{b}T_{aA}\right) \,, \\
		\nabla_{b}T_{aA} = rD_{b}\left(\frac{T_{aA}}{r}\right) \,,\quad \nabla_{B}T_{aA} = D_{B}T_{aA} - \frac{1}{r}r_{a}T_{BA} + rr^{b}T_{ab}\Omega_{AB} \,, \\
		\nabla_{a}T_{AB} = r^2 D_{a}\left(\frac{T_{AB}}{r^2}\right) \,,\quad  \nabla_{C}T_{AB} = D_{C}T_{AB} + rr^{a}\left(T_{aB}\Omega_{AC}+T_{Aa}\Omega_{BC}\right) \,.
	\end{gathered}
\ee

Furthermore, it will be useful to have explicit formulas for the covariant derivatives of the vectors $t^{a}$ and $r^{a}$. It is straightforward to show that
\be
	D_{a}t_{b} = -\frac{f_{t}^{\prime}}{2f_{t}}\left(t_{a}r_{b}-r_{a}t_{b}\right) \,,\quad D_{a}r_{b} = f_{r}\left[-\frac{f_{t}^{\prime}}{2f_{t}^2}\,t_{a}t_{b} + \frac{f_{r}^{\prime}}{2f_{r}^2}\,r_{a}r_{b}\right] \,,
\ee
where primes denote derivatives with respect to $r$ and we used that $r^{a}D_{a}F\left(r\right)=F^{\prime}\left(r\right)r_{a}r^{a} = F^{\prime}\left(r\right)f_{r}\left(r\right)$.

\section{Equations of motion for spin-$0$ perturbations}
\label{sec:EOM_Seq0}

We now begin extracting the equations of motion governing perturbations of the background spherical symmetric black hole geometry, starting from the action for a free scalar field minimally coupled to gravity,
\be
	S^{\left(0\right)} = \int d^{d}x\sqrt{-g}\left[-\frac{1}{2}\left(\nabla\Phi\right)^2 - \frac{1}{2}m^2\Phi^2\right] \,.
\ee
The scalar field is $\left(2+\left(d-2\right)\right)$-decomposed in spherical harmonic modes according to
\be
	\Phi\left(x\right) = \sum_{\ell,\mathbf{m}}\frac{\Psi^{\left(0\right)}_{\ell,\mathbf{m}}\left(t,r\right)}{r^{\left(d-2\right)/2}}Y_{\ell,\mathbf{m}}\left(\theta\right) \,.
\ee
The resulting reduced action then describes a scalar field minimally coupled to $2$-d gravity,
\be
	\begin{gathered}
		S^{\left(0\right)} = \sum_{\ell,\mathbf{m}}S^{\left(0\right)}_{\ell,\mathbf{m}} \,, \\
		S^{\left(0\right)}_{\ell,\mathbf{m}} = \int d^2x\sqrt{-g^{\left(2\right)}}\,\left[-\frac{1}{2}D_{a}\bar{\Psi}^{\left(0\right)}_{\ell,\mathbf{m}}D^{a}\Psi^{\left(0\right)}_{\ell,\mathbf{m}}-\frac{1}{2}V^{\left(0\right)}_{\ell}\left(r\right)\left|\Psi^{\left(0\right)}_{\ell,\mathbf{m}}\right|^2\right] \,,
	\end{gathered}
\ee
with the potential given by
\be\ba\label{eq:V0S}
	V^{\left(0\right)}_{\ell}\left(r\right) &= \frac{\ell\left(\ell+d-3\right)}{r^2} + \frac{\left(d-2\right)\left(d-4\right)}{4r^2}r_{a}r^{a} + \frac{d-2}{2r}D_{a}r^{a} + m^2 \\
	&=\frac{\ell\left(\ell+d-3\right)}{r^2} + \frac{\left(d-2\right)\left(d-4\right)}{4r^2}f_{r} + \frac{d-2}{2r}\frac{\left(f_{t}f_{r}\right)^{\prime}}{2f_{t}} + m^2 \,.
\ea\ee

Working with the tortoise coordinate, this reduces to the action for a scalar field propagating in $2$-d flat spacetime under the influence of a potential,
\be
	S^{\left(0\right)}_{\ell,\mathbf{m}} = \frac{1}{2}\int dtdr_{\ast} \left[\frac{1}{2}\left|\partial_{t}\Psi^{\left(0\right)}_{\ell,\mathbf{m}}\right|^2-\frac{1}{2}\left|\partial_{r_{\ast}}\Psi^{\left(0\right)}_{\ell,\mathbf{m}}\right|^2 - \frac{1}{2}f_{t}\left(r\right)V^{\left(0\right)}_{\ell}\left(r\right)\left|\Psi^{\left(0\right)}_{\ell,\mathbf{m}}\right|^2\right] \,,
\ee
and the equation of motion for scalar field perturbations reduces to a Shr\"{o}dinger equation,
\be
	\left[\partial_{r_{\ast}}^2-\partial_{t}^2-f_{t}\left(r\right)V^{\left(0\right)}_{\ell}\left(r\right)\right]\Psi^{\left(0\right)}_{\ell,\mathbf{m}} = 0 \,.
\ee

\section{Equations of motion for spin-$1$ perturbations}
\label{sec:EOM_Seq1}

Next, for electromagnetic perturbations, we focus to an electrically neutral black hole background such that there is no background electric field\footnote{This simply ensures that we will not need to deal with coupled equations of motion involving gravitational perturbation modes.}. To treat the Maxwell action,
\be
	S^{\left(1\right)} = \int d^{d}x\sqrt{-g}\left[-\frac{1}{4}F_{\mu\nu}F^{\mu\nu}\right] \,,\quad F_{\mu\nu} = \partial_{\mu}A_{\nu}-\partial_{\nu}A_{\mu} \,,
\ee
the $2+\left(d-2\right)$ decomposition involves first decomposing the components of the gauge field into irreducible representations of $SO\left(d-1\right)$,
\be
	A_{\mu}\left(x\right) =
	\begin{pmatrix}
		A_{a}\left(x\right) \\
		D_{A}A^{\left(\text{L}\right)}\left(x\right) + A_{A}^{\left(\text{T}\right)}\left(x\right)
	\end{pmatrix} \,.
\ee
With respect to $SO\left(d-1\right)$ transformations, $A_{a}$ and $A^{\left(\text{L}\right)}$ are scalars, while $A_{A}^{\left(\text{T}\right)}$ is a transverse co-vector, $D^{A}A_{A}^{\left(\text{T}\right)}=0$. Under gauge transformations $\delta_{\Lambda}A_{\mu}=\partial_{\mu}\Lambda$, the $SO\left(d-1\right)$-decomposed components transform according to
\be
	\delta_{\Lambda}A_{a} = \partial_{a}\Lambda \,,\quad \delta_{\Lambda}A^{\left(\text{L}\right)} = \Lambda \,,\quad \delta_{\Lambda}A_{A}^{\left(\text{T}\right)} = 0 \,.
\ee
Notably, the transverse vectors are gauge invariant, while we also see how the longitudinal modes $A^{\left(\text{L}\right)}$ are redundant degrees of freedom; they are pure gauge. A second gauge invariant quantity can then be constructed as
\be
	\mathcal{A}_{a} = A_{a} - D_{a}A^{\left(\text{L}\right)}
\ee
and the $\left(2+\left(d-2\right)\right)$-decomposed field strength tensor reads
\be\ba
	F_{ab} &= D_{a}\mathcal{A}_{b} - D_{b}\mathcal{A}_{a} \\
	F_{aA} &= D_{a}A^{\left(\text{T}\right)}_{A} - D_{A}\mathcal{A}_{a} \\
	F_{AB} &= D_{A}A^{\left(\text{T}\right)}_{B} - D_{B}A^{\left(\text{T}\right)}_{A} \,.
\ea\ee

The next step is to expand the scalars into scalar spherical harmonic modes $Y_{\ell,\mathbf{m}}\left(\theta\right)$ and the transverse vector into transverse vector spherical harmonic modes\footnote{For more information on the scalar, vector and tensor spherical harmonics in higher dimensions, we refer to~\cite{Hui:2020xxx,Chodos:1983zi,Higuchi:1986wu}.} $Y^{\left(\text{T}\right)A}_{\ell,\mathbf{m}}\left(\theta\right)$,
\be\ba
	\mathcal{A}^{a}\left(x\right) &= \sum_{\ell,\mathbf{m}}\mathcal{A}^{a}_{l,\mathbf{m}}\left(t,r\right)Y_{\ell,\mathbf{m}}\left(\theta\right) \,, \\
	A^{\left(\text{T}\right)}_{A}\left(x\right) &= \sum_{\ell,\mathbf{m}}A^{\left(\text{V}\right)}_{l,\mathbf{m}}\left(t,r\right)Y^{\left(\text{T}\right)}_{A;\ell,\mathbf{m}}\left(\theta\right) \,.
\ea\ee
Spherical symmetry of the background ensures that the scalar modes $\mathcal{A}^{a}_{\ell,\mathbf{m}}$ and the vector modes $A^{\left(\text{V}\right)}_{\ell,\mathbf{m}}$ will completely decouple from each other. Indeed, the Maxwell action after this expansion reads
\be
	\begin{gathered}
		S^{\left(1\right)} = \sum_{\ell,\mathbf{m}}\left(S^{\left(\text{V}\right)}_{\ell,\mathbf{m}} + S^{\left(\text{S}\right)}_{\ell,\mathbf{m}}\right) \,, \\
		\ba
			S^{\left(\text{V}\right)}_{\ell,\mathbf{m}} &= \int d^2x\sqrt{-g^{\left(2\right)}}\,r^{d-4} \left[-\frac{1}{2}D_{a}\bar{A}^{\left(\text{V}\right)}_{\ell,\mathbf{m}}D^{a}A^{\left(\text{V}\right)}_{\ell,\mathbf{m}} - \frac{1}{2}\frac{\left(\ell+1\right)\left(\ell+d-4\right)}{r^2}\left|A^{\left(\text{V}\right)}_{\ell,\mathbf{m}}\right|^2 \right] \,, \\
			S^{\left(\text{S}\right)}_{\ell,\mathbf{m}} &= \int d^2x\sqrt{-g^{\left(2\right)}}\,r^{d-2} \left[-\frac{1}{4}\bar{\mathcal{F}}_{ab;\ell,\mathbf{m}}\mathcal{F}^{ab}_{\ell,\mathbf{m}} - \frac{1}{2}\frac{\ell\left(\ell+d-3\right)}{r^2}\bar{\mathcal{A}}_{a;\ell,\mathbf{m}}\mathcal{A}^{a}_{\ell,\mathbf{m}}\right] \,,
		\ea
	\end{gathered}
\ee
where
\be
	\mathcal{F}^{ab}_{\ell,\mathbf{m}} \equiv D^{a}\mathcal{A}^{b}_{\ell,\mathbf{m}} - D^{b}\mathcal{A}^{a}_{\ell,\mathbf{m}} \,.
\ee

\subsection{Vector modes}
We begin by studying the decoupled vector modes. Similar to the scalar field case, these will be governed by the action for a scalar field minimally coupled to $2$-d gravity propagating under the influence of a potential. More explicitly, writing
\be
	A^{\left(\text{V}\right)}_{\ell,\mathbf{m}}\left(t,r\right) = \frac{\Psi^{\left(\text{V}\right)}_{\ell,\mathbf{m}}\left(t,r\right)}{r^{\left(d-4\right)/2}} \,,
\ee
the reduced action for the vector modes reads
\be
	S_{\ell,\mathbf{m}}^{\left(\text{V}\right)} = \int d^2x\sqrt{-g^{\left(2\right)}}\left[-\frac{1}{2}D_{a}\bar{\Psi}^{\left(\text{V}\right)}_{\ell,\mathbf{m}}D^{a}\Psi^{\left(\text{V}\right)}_{\ell,\mathbf{m}} - \frac{1}{2}V^{\left(\text{V}\right)}_{\ell}\left(r\right)\left|\Psi^{\left(\text{V}\right)}_{\ell,\mathbf{m}}\right|^2\right]
\ee
with the potential given by
\be\ba\label{eq:V1V}
	V^{\left(\text{V}\right)}_{\ell}\left(r\right) &= \frac{\left(\ell+1\right)\left(\ell+d-4\right)}{r^2} + \frac{\left(d-4\right)\left(d-6\right)}{4r^2}r_{a}r^{a} + \frac{d-4}{2r}D_{a}r^{a} \\
	&= \frac{\left(\ell+1\right)\left(\ell+d-4\right)}{r^2} + \frac{\left(d-4\right)\left(d-6\right)}{4r^2}f_{r} + \frac{d-4}{2r}\frac{\left(f_{t}f_{r}\right)^{\prime}}{2f_{t}} \,.
\ea\ee
Working with the tortoise coordinate, this becomes the action for a scalar field in $2$-d flat spacetime,
\be
	S^{\left(\text{V}\right)}_{\ell,\mathbf{m}} = \frac{1}{2}\int dtdr_{\ast} \left[\frac{1}{2}\left|\partial_{t}\Psi^{\left(\text{V}\right)}_{\ell,\mathbf{m}}\right|^2-\frac{1}{2}\left|\partial_{r_{\ast}}\Psi^{\left(\text{V}\right)}_{\ell,\mathbf{m}}\right|^2 - \frac{1}{2}f_{t}\left(r\right)V^{\left(\text{V}\right)}_{\ell}\left(r\right)\left|\Psi^{\left(\text{V}\right)}_{\ell,\mathbf{m}}\right|^2\right] \,,
\ee
with Schr\"{o}dinger-like equations of motion,
\be
	\left[\partial_{r_{\ast}}^2-\partial_{t}^2-f_{t}\left(r\right)V^{\left(\text{V}\right)}_{\ell}\left(r\right)\right]\Psi^{\left(\text{V}\right)}_{\ell,\mathbf{m}} = 0 \,.
\ee

\subsection{Scalar modes}
We next analyze the action for the scalar modes. For the sake of this, we introduce an auxiliary $2$-d scalar field $\Psi^{\left(\text{S}\right)}_{\ell,\mathbf{m}}\left(t,r\right)$ and consider the action~\cite{Hui:2020xxx}
\be\ba
	\tilde{S}^{\left(\text{S}\right)}_{\ell,\mathbf{m}} = \int d^2x\sqrt{-g^{\left(2\right)}} &\bigg[\frac{1}{2}\sqrt{\ell\left(\ell+d-3\right)}\,r^{\left(d-4\right)/2}\text{Re}\left\{\bar{\Psi}^{\left(\text{S}\right)}_{\ell,\mathbf{m}}\varepsilon_{ab}\mathcal{F}^{ab}_{\ell,\mathbf{m}}\right\} \\
	&- \frac{1}{2}\frac{\ell\left(\ell+d-3\right)}{r^2}\left(\left|\Psi^{\left(\text{S}\right)}_{\ell,\mathbf{m}}\right|^2 + r^{d-2}\bar{\mathcal{A}}_{a;\ell,\mathbf{m}}\mathcal{A}^{a}_{\ell,\mathbf{m}}\right)\bigg] \,.
\ea\ee
Classically, this is equivalent to the original action $S^{\left(\text{S}\right)}_{\ell,\mathbf{m}}$ for the scalar modes as can be seen by putting the auxiliary field on-shell,
\be
	\Psi^{\left(\text{S}\right)}_{\ell,\mathbf{m}} = \frac{1}{2}\frac{r^{d/2}}{\sqrt{\ell\left(\ell+d-3\right)}}\varepsilon_{ab}\mathcal{F}^{ab}_{\ell,\mathbf{m}} \,.
\ee

The upshot of this alternative action is that it can be recast in a form similar to the scalar field modes and the gauge field vector modes. This is achieved by integrating out $\mathcal{A}^{a}_{\ell,\mathbf{m}}$ in $\tilde{S}^{\left(\text{S}\right)}_{\ell,\mathbf{m}}$,
\be
	\mathcal{A}^{a}_{\ell,\mathbf{m}} = \frac{r^{-\left(d-4\right)/2}}{\sqrt{\ell\left(\ell+d-3\right)}}\left[\varepsilon^{ab}D_{b} - t^{a}\frac{d-4}{2r}\sqrt{\frac{f_{r}}{f_{t}}}\right]\Psi^{\left(\text{S}\right)}_{\ell,\mathbf{m}} \,.
\ee
As a result,
\be\ba
	\tilde{S}^{\left(\text{S}\right)}_{\ell,\mathbf{m}} &= \int d^2x\sqrt{-g^{\left(2\right)}}\left[-\frac{1}{2}D_{a}\bar{\Psi}^{\left(\text{S}\right)}_{\ell,\mathbf{m}}D^{a}\Psi^{\left(\text{S}\right)}_{\ell,\mathbf{m}} - \frac{1}{2}V^{\left(\text{S}\right)}_{\ell}\left(r\right)\left|\Psi^{\left(\text{S}\right)}_{\ell,\mathbf{m}}\right|^2\right] \\
	&= \int dtdr_{\ast}\left[\frac{1}{2}\left|\partial_{t}\Psi^{\left(\text{S}\right)}_{\ell,\mathbf{m}}\right|^2 -\frac{1}{2}\left|\partial_{r_{\ast}}\Psi^{\left(\text{S}\right)}_{\ell,\mathbf{m}}\right|^2 - \frac{1}{2}f_{t}\left(r\right)V^{\left(\text{S}\right)}_{\ell}\left(r\right)\left|\Psi^{\left(\text{S}\right)}_{\ell,\mathbf{m}}\right|^2\right] \,,
\ea\ee
with the potential given by
\be\ba\label{eq:V1S}
	V^{\left(\text{S}\right)}_{\ell}\left(r\right) &= \frac{\ell\left(\ell+d-3\right)}{r^2} + \frac{\left(d-2\right)\left(d-4\right)}{4r^2}r_{a}r^{a} - \frac{d-4}{2r}D_{a}r^{a} \\
	&= \frac{\ell\left(\ell+d-3\right)}{r^2} + \frac{\left(d-2\right)\left(d-4\right)}{4r^2}f_{r} - \frac{d-4}{2r}\frac{\left(f_{t}f_{r}\right)^{\prime}}{2f_{t}} \,.
\ea\ee

\section{Equations of motion for $p$-form perturbations}
\label{sec:EOM_Seqp}

In higher spacetime dimensions, it is also possible to have $p$-form perturbations, generated by a completely antisymmetric gauge field tensor $A_{\mu_1\mu_2\dots\mu_{p}}$ of rank $p\le d-3$ or, in $p$-form notation, the object
\be
	\mathbf{A}^{\left(p\right)} = \frac{1}{p!}A_{\mu_1\mu_2\dots\mu_{p}}\,\text{d}x^{\mu_1}\wedge\text{d}x^{\mu_2}\wedge\dots\wedge\text{d}x^{\mu_{p}} \,.
\ee
Focusing to spherically symmetric black hole backgrounds with no charge under the $p$-form, the task now is to study the $p$-form extension of the Maxwell action
\be
	S^{\left(p\right)} = -\frac{1}{2}\int \mathbf{F}^{\left(p+1\right)}\wedge \star\mathbf{F}^{\left(p+1\right)} \,,
\ee
where $\mathbf{F}^{\left(p+1\right)}=\text{d}\mathbf{A}^{\left(p\right)}$ is the $\left(p+1\right)$-form field strength tensor, built from the exterior derivative\footnote{The exterior derivative $\text{d}$ of $p$-form $A_{\left(p\right)}$ is defined as
\begin{equation*}
	\text{d}\mathbf{A}^{\left(p\right)} = \frac{1}{p!}\partial_{\mu_1}A_{\mu_2\dots\mu_{p+1}}\,\text{d}x^{\mu_1}\wedge\text{d}x^{\mu_2}\wedge\dots\wedge\text{d}x^{\mu_{p+1}}
\end{equation*}
and has the crucial property of being nilpotent, $\text{d}^2=0$.} of the $p$-form gauge field, and $\star$ is the Hodge dual operation\footnote{The Hodge dual operation of a $p$-form $A^{\left(p\right)}$ in a $d$-dimensional spacetime is the $\left(d-p\right)$-form
\begin{equation*}
	\star\mathbf{A}^{\left(p\right)} = \frac{\sqrt{\left|g\right|}}{p!\left(d-p\right)!}A_{\mu_1\dots\mu_{p}}\epsilon^{\mu_1\dots\mu_{p}}_{\quad\quad\,\,\,\,\mu_{p+1}\dots\mu_{d}}\,\text{d}x^{\mu_{p+1}}\wedge\dots\wedge\text{d}x^{\mu_{d}} \,.
\end{equation*}}. In the traditional index notation,
\be
	F_{\mu_1\mu_2\dots\mu_{p+1}} = \left(p+1\right)\partial_{[\mu_1}A_{\mu_2\dots\mu_{p+1}]}
\ee
and the $p$-form action reads
\be
	S^{\left(p\right)} = -\frac{1}{2\left(p-1\right)!}\int d^{d}x\sqrt{-g}\,F_{\mu_1\mu_2\dots\mu_{p+1}}F^{\mu_1\mu_2\dots\mu_{p+1}} \,.
\ee

The $2+\left(d-2\right)$ decomposition of the $p$-form gauge field into irreducible representations of $SO\left(d-1\right)$ is now achieved via the Hodge decomposition on $\mathbb{S}^{d-2}$. We will adopt the notation of~\cite{Yoshida:2019tvk} and distinguish a $p$-form on $\mathbb{S}^{d-2}$ from a $p$-form on the full manifold by hatting it. For instance,
\be
	\hat{\mathbf{A}}^{\left(p\right)} \equiv \frac{1}{p!}A_{A_1 A_2\dots A_{p}}\,\text{d}\theta^{A_1}\wedge\text{d}\theta^{A_2}\wedge\dots\wedge\text{d}\theta^{A_{p}} \,.
\ee
The spacetime $p$-form gauge field $\mathbf{A}^{\left(p\right)}$ can then at first step be tensorially decomposed as~\cite{Yoshida:2019tvk}
\be
	\mathbf{A}^{\left(p\right)} = \frac{1}{2}\text{d}x^{a}\wedge\text{d}x^{b}\wedge\hat{\mathbf{T}}_{ab}^{\left(p-2\right)} + \text{d}x^{a}\wedge\hat{\mathbf{V}}_{a}^{\left(p-1\right)} + \hat{\mathbf{X}}^{\left(p\right)} \,,
\ee
where the components of the forms $\hat{\mathbf{T}}_{ab}^{\left(p-2\right)}$, $\hat{\mathbf{V}}_{a}^{\left(p-1\right)}$ and $\hat{\mathbf{X}}^{\left(p\right)}$ on the sphere have been identified with the relevant components of the spacetime $p$-form gauge field, that is,
\be
	\left(T_{ab}\right)_{A_1\dots A_{p-2}} \equiv A_{abA_1\dots A_{p-2}} \,,\quad \left(V_{a}\right)_{A_1\dots A_{p-1}} \equiv A_{aA_1\dots A_{p-1}} \,,\quad X_{A_1\dots A_{p}} \equiv A_{A_1\dots A_{p}} \,.
\ee
The Hodge decomposition on $\mathbb{S}^{d-2}$ now tells us that a general $p$-form $\hat{\mathbf{A}}^{\left(p\right)}$ on the sphere can be decomposed into a ``longitudinal'' $\left(p-1\right)$-form $\hat{\mathbf{A}}^{\left(p-1\right)}$ and a ``transverse'' $p$-form $\hat{\mathcal{A}}^{\left(p\right)}$. More specifically,
\be
	\hat{\mathbf{A}}^{\left(p\right)} = \hat{\text{d}}\hat{\mathbf{A}}^{\left(p-1\right)} + \hat{\mathcal{A}}^{\left(p\right)} \,,
\ee
with $\hat{\mathcal{A}}^{\left(p\right)}$ a co-exact $p$-form on $\mathbb{S}^{d-2}$, that is\footnote{The coderivative operator of a $p$-form on a $d$-dimensional spacetime is defined as $\delta\equiv\left(-1\right)^{d\left(p+1\right)+1}\star\text{d}\star$ and, like the exterior derivative, it is also nilpotent, i.e. $\delta^2=0$.} $\hat{\delta}\hat{\mathcal{A}}^{\left(p\right)} = 0$. In components form, this is indeed the transversality condition,
\be
	\hat{\delta}\hat{\mathcal{A}}^{\left(p\right)} = 0 \Leftrightarrow D^{A_1}\mathcal{A}_{A_1 A_2\dots A_{p+1}} = 0 \,.
\ee
The longitudinal mode can be further Hodge decomposed as $\hat{\mathbf{A}}^{\left(p-1\right)} = \hat{\text{d}}\hat{\mathbf{A}}^{\left(p-2\right)} + \hat{\mathcal{A}}^{\left(p-1\right)}$, with $\hat{\mathcal{A}}^{\left(p-1\right)}$ a co-exact $\left(p-1\right)$-form on the sphere. The nilpotency of the exterior derivative then implies that a general form on $\mathbb{S}^{d-2}$ is expressible only by co-exact form fields. The $2+\left(d-2\right)$ decomposition of the $p$-form gauge field into irreducible representations of $SO\left(d-1\right)$ is therefore the following
\be\ba
	\mathbf{A}^{\left(p\right)} &= \frac{1}{2}\text{d}x^{a}\wedge\text{d}x^{b}\wedge\left(\hat{\text{d}}\hat{\mathcal{T}}_{ab}^{\left(p-3\right)}+\hat{\mathcal{T}}_{ab}^{\left(p-2\right)}\right) \\
	&\quad+ \text{d}x^{a}\wedge\left(\hat{\text{d}}\hat{\mathcal{V}}_{a}^{\left(p-2\right)}+\hat{\mathcal{V}}_{a}^{\left(p-1\right)}\right) \\
	&\quad+ \hat{\text{d}}\hat{\mathcal{X}}^{\left(p-1\right)} + \hat{\mathcal{X}}^{\left(p\right)} \,,
\ea\ee
where all the forms that appear are now co-exact on $\mathbb{S}^{d-2}$.

The $p$-form action is invariant under the gauge transformations $\delta_{\Lambda}\mathbf{A}^{\left(p\right)} = \text{d}\Lambda^{\left(p-1\right)}$. After decomposing the gauge parameter $\left(p-1\right)$-form into co-exact forms on the sphere, we can work out that
\be\ba
	\text{d}\Lambda^{\left(p-1\right)} &= \frac{1}{2}\text{d}x^{a}\wedge\text{d}x^{b}\wedge\left(2\,\hat{\text{d}}\hat{\varLambda}_{ab}^{\left(p-3\right)}+D_{a}\hat{\varLambda}_{b}^{\left(p-2\right)}\right) \\
	&\quad+ \text{d}x^{a}\wedge\left(-\hat{\text{d}}\hat{\varLambda}_{a}^{\left(p-2\right)} + D_{a}\hat{\varLambda}^{\left(p-1\right)}\right) + \hat{\text{d}}\hat{\varLambda}^{\left(p-1\right)} \,.
\ea\ee
As a result, the $SO\left(d-1\right)$-decomposed components of the $p$-form gauge field transform according to
\be
	\begin{gathered}
		\delta_{\Lambda}\hat{\mathcal{T}}_{ab}^{\left(p-3\right)} = \hat{\varLambda}_{ab}^{\left(p-3\right)} \,,\quad \delta_{\Lambda}\hat{\mathcal{V}}_{a}^{\left(p-2\right)} = -\hat{\varLambda}_a^{\left(p-2\right)} \,,\quad \delta_{\Lambda}\hat{\mathcal{X}}^{\left(p-1\right)} = \hat{\varLambda}^{\left(p-1\right)} \,, \\
		\delta_{\Lambda}\hat{\mathcal{T}}_{ab}^{\left(p-2\right)} = 2D_{[a}\hat{\varLambda}_{b]}^{\left(p-2\right)} \,,\quad \delta_{\Lambda}\hat{\mathcal{V}}_{a}^{\left(p-1\right)} = D_{a}\hat{\varLambda}^{\left(p-1\right)} \,,\quad \delta_{\Lambda}\hat{\mathcal{X}}^{\left(p\right)} = 0 \,.
	\end{gathered}
\ee
The first line shows that the longitudinal modes are pure gauge. Instead of fixing the gauge, however, we will directly work with gauge invariant combinations. In particular, we can rearrange the independent degrees of freedom into the gauge invariant co-exact $p$-form on the sphere, $\hat{\mathcal{X}}^{\left(p\right)}$, plus the following gauge invariant combinations
\be
	\hat{\mathcal{H}}_{ab}^{\left(p-2\right)} = \hat{\mathcal{T}}_{ab}^{\left(p-2\right)} + 2D_{[a}\hat{\mathcal{V}}_{b]}^{\left(p-2\right)} \,,\quad \hat{\mathcal{A}}_{a}^{\left(p-1\right)} = \hat{\mathcal{V}}_{a}^{\left(p-1\right)} - D_{a}\hat{\mathcal{X}}^{\left(p-1\right)} \,.
\ee
In terms of these, the field strength $\left(p+1\right)$-form is written as
\be\ba
	\mathbf{F}^{\left(p+1\right)} &= \frac{1}{2}\text{d}x^{a}\wedge\text{d}x^{b}\wedge\left(2D_{a}\hat{\mathcal{A}}_{b}^{\left(p-1\right)} + \left(p-2\right)!\,\hat{\text{d}}\hat{\mathcal{H}}_{ab}^{\left(p-2\right)}\right) \\
	&\quad+ \text{d}x^{a}\wedge\left(D_{a}\hat{\mathcal{X}}^{\left(p\right)} - \hat{\text{d}}\hat{\mathcal{A}}_{a}^{\left(p-1\right)}\right) + \hat{\text{d}}\hat{\mathcal{X}}^{\left(p\right)} \,.
\ea\ee

We can now expand into co-exact $p$-form spherical harmonics $Y^{\left(\text{T}\right)A_1 \dots A_{p}}_{\ell,\mathbf{m}}\left(\theta\right)$ on the sphere,
\be\ba
	\left(\mathcal{H}^{ab}\right)^{A_1\dots A_{p-2}}\left(x\right) &= \sum_{\ell,\mathbf{m}}\mathcal{H}^{ab}_{\ell,\mathbf{m}}\left(t,r\right)Y^{\left(\text{T}\right)A_1 \dots A_{p-2}}_{\ell,\mathbf{m}}\left(\theta\right) \,, \\
	\left(\mathcal{A}^{a}\right)^{A_1\dots A_{p-1}}\left(x\right) &= \sum_{\ell,\mathbf{m}}\mathcal{A}^{a}_{\ell,\mathbf{m}}\left(t,r\right)Y^{\left(\text{T}\right)A_1 \dots A_{p-1}}_{\ell,\mathbf{m}}\left(\theta\right) \,, \\
	\mathcal{X}^{A_1\dots A_{p}}\left(x\right) &= \sum_{\ell,\mathbf{m}}\mathcal{X}_{\ell,\mathbf{m}}\left(t,r\right)Y^{\left(\text{T}\right)A_1 \dots A_{p}}_{\ell,\mathbf{m}}\left(\theta\right) \,.
\ea\ee
The important property of the co-exact $p$-form spherical harmonics besides their transversality, $D_{A_1}Y^{\left(\text{T}\right)A_1 \dots A_{p}}_{\ell,\mathbf{m}}=0$, is that they satisfy the eigenvalue problem\footnote{More information on the co-exact $p$-form spherical harmonics can be found in~\cite{Yoshida:2019tvk,Camporesi1994}.}
\be
	D_{B}D^{B}Y^{\left(\text{T}\right)A_1 \dots A_{p}}_{\ell,\mathbf{m}} = -\left[\ell\left(\ell+d-3\right)-p\right]Y^{\left(\text{T}\right)A_1 \dots A_{p}}_{\ell,\mathbf{m}} \,.
\ee

The $p$-form action after this expansion then reduces to
\be
	\begin{gathered}
		S^{\left(p\right)} = \sum_{\ell,\mathbf{m}}\left(S^{\left(p\right)}_{\ell,\mathbf{m}} + S^{\left(p-1\right)}_{\ell,\mathbf{m}} + S^{\left(p-2\right)}_{\ell,\mathbf{m}}\right) \,, \\
		\ba
			S^{\left(p\right)}_{\ell,\mathbf{m}} &= \int d^2x\sqrt{-g^{\left(2\right)}}\,r^{d-2p-2} \left[-\frac{1}{2p!}D_{a}\bar{\mathcal{X}}_{\ell,\mathbf{m}}D^{a}\mathcal{X}_{\ell,\mathbf{m}} - \frac{1}{2p!}\frac{\left(\ell+p\right)\left(\ell+d-p-3\right)}{r^2}\left|\mathcal{X}_{\ell,\mathbf{m}}\right|^2 \right] \,, \\
			S^{\left(p-1\right)}_{\ell,\mathbf{m}} &= \int d^2x\sqrt{-g^{\left(2\right)}}\,r^{d-2p} \bigg[-\frac{1}{4\left(p-1\right)!}\bar{\mathcal{F}}_{ab;\ell,\mathbf{m}}\mathcal{F}^{ab}_{\ell,\mathbf{m}} \\
			&\qquad\qquad\qquad\qquad\qquad\quad- \frac{1}{2\left(p-1\right)!}\frac{\left(\ell+p-1\right)\left(\ell+d-p-2\right)}{r^2}\bar{\mathcal{A}}_{a;\ell,\mathbf{m}}\mathcal{A}^{a}_{\ell,\mathbf{m}}\bigg] \,, \\
			S^{\left(p-2\right)}_{\ell,\mathbf{m}} &= \int d^2x\sqrt{-g^{\left(2\right)}}\,r^{d-2p+2} \left[- \frac{\left(p-2\right)!}{4}\frac{\left(\ell+p-2\right)\left(\ell+d-p-1\right)}{r^2}\mathcal{H}_{ab;\ell,\mathbf{m}}\mathcal{H}^{ab}_{\ell,\mathbf{m}}\right] \,,
		\ea
	\end{gathered}
\ee
where we have defined
\be
	\mathcal{F}^{ab}_{\ell,\mathbf{m}} \equiv D^{a}\mathcal{A}^{b}_{\ell,\mathbf{m}} - D^{b}\mathcal{A}^{a}_{\ell,\mathbf{m}} \,.
\ee
The first thing to observe is that the $\left(p-2\right)$-form sector generated by the spherical harmonic modes $\mathcal{H}^{ab}_{\ell,\mathbf{m}}$ is trivial,
\be
	\mathcal{H}^{ab}_{\ell,\mathbf{m}} = 0 \,.
\ee

\subsection{$p$-form modes}
Similar to the case of spin-$1$ perturbations, we start with the simplest $p$-form sector. Performing the field redefinition
\be
	\mathcal{X}_{\ell,\mathbf{m}}\left(t,r\right) = \sqrt{p!}\frac{\Psi^{\left(p\right)}_{\ell,\mathbf{m}}\left(t,r\right)}{r^{\left(d-2p-2\right)/2}} \,,
\ee
the reduced action for $p$-form modes takes the suggestive form
\be
	S_{\ell,\mathbf{m}}^{\left(p\right)} = \int d^2x\sqrt{-g^{\left(2\right)}}\left[-\frac{1}{2}D_{a}\bar{\Psi}^{\left(p\right)}_{\ell,\mathbf{m}}D^{a}\Psi^{\left(p\right)}_{\ell,\mathbf{m}} - \frac{1}{2}V^{\left(p\right)}_{\ell}\left(r\right)\left|\Psi^{\left(p\right)}_{\ell,\mathbf{m}}\right|^2\right]
\ee
with potential
\be\label{eq:Vpp}
	V^{\left(p\right)}_{\ell}\left(r\right) = \frac{\left(\ell+p\right)\left(\ell+d-p-3\right)}{r^2} + \frac{\left(d-2p-2\right)\left(d-2p-4\right)}{4r^2}r_{a}r^{a} + \frac{d-2p-2}{2r}D_{a}r^{a} \,.
\ee

\subsection{$\left(p-1\right)$-form modes}
For the analysis of the $\left(p-1\right)$-form sector, we introduce an auxiliary $2$-d scalar field $\Psi^{\left(\tilde{p}\right)}_{\ell,\mathbf{m}}\left(t,r\right)$ and consider the action
\be\ba
	\tilde{S}^{\left(p-1\right)}_{\ell,\mathbf{m}} = \int d^2x\sqrt{-g^{\left(2\right)}} &\bigg[\frac{1}{2}\sqrt{\frac{\left(\ell+p-1\right)\left(\ell+d-p-2\right)}{\left(p-1\right)!}}\,r^{\left(d-2p-2\right)/2}\text{Re}\left\{\bar{\Psi}^{\left(\tilde{p}\right)}_{\ell,\mathbf{m}}\varepsilon_{ab}\mathcal{F}^{ab}_{\ell,\mathbf{m}}\right\} \\
	&- \frac{1}{2}\frac{\left(\ell+p-1\right)\left(\ell+d-p-2\right)}{r^2}\left(\left|\Psi^{\left(\tilde{p}\right)}_{\ell,\mathbf{m}}\right|^2 + \frac{r^{d-2p}}{\left(p-1\right)!}\bar{\mathcal{A}}_{a;\ell,\mathbf{m}}\mathcal{A}^{a}_{\ell,\mathbf{m}}\right)\bigg] \,,
\ea\ee
which is classically equivalent to the original action $S^{\left(p-1\right)}_{\ell,\mathbf{m}}$; a fact that becomes obvious after putting the auxiliary field on-shell,
\be
	\Psi^{\left(\tilde{p}\right)}_{\ell,\mathbf{m}} = \frac{1}{2}\frac{r^{\left(d-2p+2\right)/2}}{\sqrt{\left(p-1\right)!\left(\ell+p-1\right)\left(\ell+d-p-2\right)}}\varepsilon_{ab}\mathcal{F}^{ab}_{\ell,\mathbf{m}} \,.
\ee

Integrating out $\mathcal{A}^{a}_{\ell,\mathbf{m}}$ in $\tilde{S}^{\left(p-1\right)}_{\ell,\mathbf{m}}$ instead,
\be
	\mathcal{A}^{a}_{\ell,\mathbf{m}} = \sqrt{\frac{\left(p-1\right)!}{\left(\ell+p-1\right)\left(\ell+d-p-2\right)}}r^{-\left(d-2p-2\right)/2}\left[\varepsilon^{ab}D_{b} - t^{a}\frac{d-2p-2}{2r}\sqrt{\frac{f_{r}}{f_{t}}}\right]\Psi^{\left(\tilde{p}\right)}_{\ell,\mathbf{m}} \,.
\ee
results to the canonically normalized reduced action
\be
	\tilde{S}^{\left(p-1\right)}_{\ell,\mathbf{m}} = \int d^2x\sqrt{-g^{\left(2\right)}}\left[-\frac{1}{2}D_{a}\bar{\Psi}^{\left(\tilde{p}\right)}_{\ell,\mathbf{m}}D^{a}\Psi^{\left(\tilde{p}\right)}_{\ell,\mathbf{m}} - \frac{1}{2}V^{\left(\tilde{p}\right)}_{\ell}\left(r\right)\left|\Psi^{\left(\tilde{p}\right)}_{\ell,\mathbf{m}}\right|^2\right] \,,
\ee
with the potential given by
\be\ba\label{eq:Vpph}
	V^{\left(\tilde{p}\right)}_{\ell}\left(r\right) = \frac{\left(\ell+p-1\right)\left(\ell+d-p-2\right)}{r^2} + \frac{\left(d-2p\right)\left(d-2p-2\right)}{4r^2}r_{a}r^{a} - \frac{d-2p-2}{2r}D_{a}r^{a} \,.
\ea\ee
In fact, this is the same as the potential \eqref{eq:Vpp} for the $p$-form modes after replacing the rank of the $p$-form with its dual,
\be
	p \rightarrow \tilde{p} = d-p-2 \,.
\ee
This is the reason behind our labeling of the master variable for the $\left(p-1\right)$-form sector as $\Psi^{\left(\tilde{p}\right)}_{\ell,\mathbf{m}}$. More collectively, the $p$-form perturbation potential can be rewritten as
\be\label{eq:Vpj}
	V^{\left(j\right)}_{\ell}\left(r\right) = \frac{\left(\ell+j\right)\left(\ell+d-j-3\right)}{r^2} + \frac{\left(d-2j-2\right)\left(d-2j-4\right)}{4r^2}r_{a}r^{a} + \frac{d-2j-2}{2r}D_{a}r^{a} \,,
\ee
where the index $j$ is equal to either $p$, for the $p$-form perturbation modes, or $\tilde{p}=d-p-2$, for the $\left(p-1\right)$-form perturbation modes. This nicely also captures the spin-$1$ vector ($j=1$) and scalar ($j=d-3$) sectors of Eq.~\eqref{eq:V1V} and Eq.~\eqref{eq:V1S} respectively, as well as the spin-$0$ scalar ($j=0$) sector of Eq.~\eqref{eq:V0S}.

\section{Equations of motion for spin-$2$ perturbations}
\label{sec:EOM_Seq2}

Before writing down the relevant action for gravitational (spin-$2$) perturbations, let us first study the decomposition of the metric perturbations $h_{\mu\nu}$ into irreducible $SO\left(d-1\right)$ representation and the construction of gauge invariants. The $\frac{d\left(d+1\right)}{2}$ components of $h_{\mu\nu}$ are rearranged according to
\be\ba
	h_{ab}\left(x\right) &= H_{ab}\left(x\right) \,, \\
	h_{aA}\left(x\right) &= D_{A}H^{\left(\text{S}\right)}_{a}\left(x\right) + h^{\left(\text{V}\right)}_{aA}\left(x\right) \,, \\
	h_{AB}\left(x\right) &= r^2\left(K\left(x\right)\Omega_{AB} + D_{\langle A}D_{B\rangle}G\left(x\right) + D_{(A}h^{\left(\text{V}\right)}_{B)}\left(x\right) + h_{AB}^{\left(\text{TT}\right)}\left(x\right) \right)
\ea\ee
into seven $SO\left(d-1\right)$ scalars
\be
	H_{ab}\left(x\right) \,,\quad H^{\left(\text{L}\right)}_{a}\left(x\right) \,,\quad K\left(x\right) \,\quad\text{and} \,\quad G\left(x\right) \,,
\ee
three $SO\left(d-1\right)$ transverse vectors carrying $d-3$ degrees of freedom each,
\be
	\begin{gathered}
		h^{\left(\text{V}\right)}_{aA}\left(x\right) \,\quad\text{and}\,\quad h^{\left(\text{V}\right)}_{A}\left(x\right) \,, \\
		D^{A}h_{aA}^{\left(\text{V}\right)}\left(x\right) = 0 \,,\quad  D^{A}h_{A}^{\left(\text{V}\right)}\left(x\right) = 0 \,,
	\end{gathered}
\ee
and one $SO\left(d-1\right)$ transverse symmetric tracefree tensor carrying $\frac{\left(d-1\right)\left(d-4\right)}{2}$ degrees of freedom,
\be
	h^{\left(\text{TT}\right)}_{AB}\left(x\right) \,,\quad D^{A}h^{\left(\text{TT}\right)}_{AB}\left(x\right) = 0 \,,\quad \Omega^{AB}h^{\left(\text{TT}\right)}_{AB}\left(x\right) = 0 \,.
\ee
In four-spacetime dimensions, there is no analogue of $h^{\left(\text{TT}\right)}_{AB}\left(x\right)$, which vanishes identically.

Under infinitesimal diffeomorphisms $x^{\mu}\rightarrow x^{\mu}+\xi^{\mu}\left(x\right)$, the metric perturbations transform according to
\be
	\delta_{\xi}h_{\mu\nu} = \nabla_{\mu}\xi_{\nu} + \nabla_{\nu}\xi_{\mu} \,.
\ee
Decomposing the $d$ gauge parameters $\xi_{\mu}$ into $SO\left(d-1\right)$ irreducible representations to three scalars, $\xi_{a}$ and $\xi^{\left(\text{S}\right)}$, and one transverse vector, $\xi^{\left(\text{V}\right)}_{A}$, $D^{A}\xi^{\left(\text{V}\right)}_{A}=0$,
\be
	\xi_{a}\left(x\right) \,,\quad \xi_{A}\left(x\right) = D_{A}\xi^{\left(\text{S}\right)}\left(x\right) + \xi^{\left(\text{V}\right)}\left(x\right) \,,
\ee
the gauge transformation properties of the various $SO\left(d-1\right)$-decomposed components of $h_{\mu\nu}$ can be read to be
\be\ba
	\delta_{\xi}H_{ab} &= D_{a}\xi_{b} + D_{b}\xi_{a} \,,\quad \delta_{\xi}H^{\left(\text{S}\right)}_{a} = \xi_{a} + D_{a}\xi^{\left(\text{S}\right)} - \frac{2}{r}r_{a}\xi^{\left(\text{S}\right)} \,, \\
	\delta_{\xi}K &= \frac{2}{r}r^{a}\xi_{a} + \frac{2}{d-2}\frac{1}{r^2}D_{A}D^{A}\xi^{\left(\text{S}\right)} \,,\quad \delta_{\xi}G = \frac{2}{r^2}\xi^{\left(\text{S}\right)} \,, \\
	\delta_{\xi}h^{\left(\text{V}\right)}_{aA} &= D_{a}\xi^{\left(\text{V}\right)}_{A} - \frac{2}{r}r_{a}\xi^{\left(\text{V}\right)}_{A} \,,\quad \delta_{\xi}h^{\left(\text{V}\right)}_{A} = \frac{2}{r^2}\xi^{\left(\text{V}\right)}_{A} \,, \\
	\delta_{\xi}h^{\left(\text{TT}\right)}_{AB} &= 0 \,.
\ea\ee
One sees, in particular, that the transverse symmetric tracefree tensor is gauge invariant, while $H_{a}^{\left(\text{S}\right)}$, $G$ and $h_{a}^{\left(\text{V}\right)}$ are redundant degrees of freedom. Instead of fixing the gauge, let us work with gauge invariant quantities. For tensor modes, this is just the transverse symmetric tracefree tensor $h^{\left(\text{TT}\right)}_{AB}$. For vector modes, the gauge invariant combination is
\be
	\mathcal{H}^{\left(\text{V}\right)}_{aA} = h^{\left(\text{V}\right)}_{aA} - \frac{1}{2}r^2D_{a}h^{\left(\text{V}\right)}_{A} \,.
\ee
Last, for scalar modes, there are two sets of gauge invariant combinations,
\be\ba
	\mathcal{H}_{ab} &= H_{ab} - 2D_{(a}H^{\left(\text{S}\right)}_{b)} + D_{(a}\left(r^2D_{b)}G\right) \,, \\
	\mathcal{K} &= K - \frac{1}{d-2}D_{A}D^{A}G + rr^{a}D_{a}G - \frac{2}{r}r^{a}H^{\left(\text{S}\right)}_{a} \,.
\ea\ee

In performing the $2+\left(d-2\right)$ decomposition of the field, we expand in scalar, transverse vector and transverse symmetric tracefree tensor spherical harmonics~\cite{Hui:2020xxx,Chodos:1983zi,Higuchi:1986wu},
\be\ba
	\mathcal{H}_{ab}\left(x\right) &= \sum_{\ell,\mathbf{m}}\mathcal{H}_{ab;\ell,\mathbf{m}}\left(t,r\right)Y_{\ell,\mathbf{m}}\left(\theta\right) \,, \\
	\mathcal{K}\left(x\right) &= \sum_{\ell,\mathbf{m}}\mathcal{K}_{\ell,\mathbf{m}}\left(t,r\right)Y_{\ell,\mathbf{m}}\left(\theta\right) \,, \\
	\mathcal{H}^{\left(\text{V}\right)}_{aA}\left(x\right) &= \sum_{\ell,\mathbf{m}}\mathcal{H}_{a;\ell,\mathbf{m}}\left(t,r\right)Y^{\left(\text{T}\right)}_{A;\ell,\mathbf{m}}\left(\theta\right) \,, \\
	h^{\left(\text{TT}\right)}_{AB}\left(x\right) &= \sum_{\ell,\mathbf{m}}h^{\left(\text{T}\right)}_{\ell,\mathbf{m}}\left(t,r\right)Y^{\left(\text{TT}\right)}_{AB;\ell,\mathbf{m}}\left(\theta\right) \,.
\ea\ee

We now look at an explicit action. Solely on the premises of working with equations of motion that are at most second-order in the derivatives, the most general such local theory of gravity is Lovelock gravity~\cite{Lovelock:1971yv}. Treating General Relativity as a low-energy effective field theory, one can write down an infinite number of higher-order curvature corrections in the gravity action~\cite{Donoghue:2017pgk}. As an elementary analysis though, we will focus here to General Relativity, described by the Einstein-Hilbert action. Perturbations around an asymptotically flat vacuum background will then be described by the massless Fierz-Pauli action,
\be
	S^{\left(\text{gr}\right)} = \int d^{d}x\sqrt{-g}\left[-\frac{1}{2}\nabla_{\rho}h_{\mu\nu}\nabla^{\rho}h^{\mu\nu} + \nabla_{\rho}h_{\mu\nu}\nabla^{\nu}h^{\mu\rho} - \nabla_{\mu}h\nabla_{\nu}h^{\mu\nu} + \frac{1}{2}\nabla_{\mu}h\nabla^{\mu}h\right] \,,
\ee
where we are using canonical variables, i.e. the perturbed metric around a background $g_{\mu\nu}$ is $g_{\mu\nu}^{\text{full}} = g_{\mu\nu} + \sqrt{32\pi G}h_{\mu\nu}$. In the presence of matter and other radiation fields, e.g. for a charged black hole, one should furthermore add the corresponding perturbations in the above action, which will also involve coupling of background stress energy-momentum tensor to gravitational perturbations.

Let us ignore for the moment other fields in the system and focus to this pure gravity quadratic action. These other fields would ultimately modify the potentials we will present below by additive pieces. Inserting the spherical harmonic expansions of the metric perturbations as described above and after a few manipulations, we find the following decoupling of the tensor (``$\left(\text{T}\right)$''), vector (``$\left(\text{RW}\right)$'') and scalar (``$\left(\text{Z}\right)$'') modes
\be
	\begin{gathered}
		S^{\left(\text{gr}\right)} = \sum_{\ell,\mathbf{m}}\left(S^{\left(\text{T}\right)}_{\ell,\mathbf{m}} + S^{\left(\text{RW}\right)}_{\ell,\mathbf{m}} + S^{\left(\text{Z}\right)}_{\ell,\mathbf{m}}\right) \\
		\ba
			S^{\left(\text{T}\right)}_{\ell,\mathbf{m}} &= \int d^2x\sqrt{-g^{\left(2\right)}}\,r^{d-2} \bigg[-\frac{1}{2}D_{a}\bar{h}^{\left(\text{T}\right)}_{\ell,\mathbf{m}}D^{a}h^{\left(\text{T}\right)}_{\ell,\mathbf{m}} - -\frac{1}{r}r^{a}D_{a}\left|h^{\left(\text{T}\right)}_{\ell,\mathbf{m}}\right|^2 \\
			&\quad\quad\quad\quad\quad\quad\quad\quad\quad\quad- \frac{1}{2}\frac{\ell\left(\ell+d-3\right)+2\left(d-3\right)}{r^2}\left|h^{\left(\text{T}\right)}_{\ell,\mathbf{m}}\right|^2\bigg] \,, \\
			S^{\left(\text{RW}\right)}_{\ell,\mathbf{m}} &= \int d^2x\sqrt{-g^{\left(2\right)}}\,2r^{d-4} \bigg[ -\frac{1}{4}\mathcal{F}_{ab;\ell,\mathbf{m}}\mathcal{F}^{ab}_{\ell,\mathbf{m}}-\frac{2}{r}r_{a}\text{Re}\left\{\bar{\mathcal{H}}^{b}_{\ell,\mathbf{m}}D_{b}\mathcal{H}^{a}_{\ell,\mathbf{m}}\right\} \\
			&\quad\quad\quad\quad-\frac{1}{2}\left(\frac{\left(\ell+1\right)\left(\ell+d-4\right)}{r^2}\bar{\mathcal{H}}_{a;\ell,\mathbf{m}}\mathcal{H}^{a}_{\ell,\mathbf{m}}-4\left|r_{a}\mathcal{H}^{a}_{\ell,\mathbf{m}}\right|^2\right) \bigg] \,,
		\ea
	\end{gathered}
\ee
\be\ba
	{}&S^{\left(\text{Z}\right)}_{\ell,\mathbf{m}} = \int d^2x\sqrt{-g^{\left(2\right)}}\,r^{d-2} \bigg[ -\frac{1}{2}D_{c}\bar{\mathcal{H}}_{ab;\ell,\mathbf{m}}D^{c}\mathcal{H}^{ab}_{\ell,\mathbf{m}} + D_{c}\bar{\mathcal{H}}_{ab;\ell,\mathbf{m}}D^{b}\mathcal{H}^{ac}_{\ell,\mathbf{m}} \\
	&- \text{Re}\left\{D_{a}\bar{\mathcal{H}}_{\ell,\mathbf{m}}D_{b}\mathcal{H}^{ab}_{\ell,\mathbf{m}}\right\} + \frac{1}{2}D_{a}\bar{\mathcal{H}}_{\ell,\mathbf{m}}D^{a}\mathcal{H}_{\ell,\mathbf{m}} + \frac{\left(d-2\right)\left(d-3\right)}{2}D_{a}\bar{\mathcal{K}}_{\ell,\mathbf{m}}D^{a}\mathcal{K}_{\ell,\mathbf{m}} \\
	&- \left(d-2\right)\text{Re}\left\{D_{a}\bar{\mathcal{K}}_{\ell,\mathbf{m}}\left(D_{b}\mathcal{H}^{ab}_{\ell,\mathbf{m}}-D^{a}\mathcal{H}_{\ell,\mathbf{m}}\right)\right\} \\
	&-\frac{d-2}{r}\text{Re}\left\{\left(D_{a}\bar{\mathcal{H}}_{\ell,\mathbf{m}}+\left(d-4\right)D_{a}\bar{\mathcal{K}}_{\ell,\mathbf{m}}\right)\left(r_{b}\mathcal{H}^{ab}_{\ell,\mathbf{m}}-r^{a}\mathcal{K}_{\ell,\mathbf{m}}\right)\right\} \\
	&-\frac{1}{2}\frac{\ell\left(\ell+d-3\right)}{r^2}\bigg( \bar{\mathcal{H}}_{ab;\ell,\mathbf{m}}\mathcal{H}^{ab}_{\ell,\mathbf{m}} - \left|\mathcal{H}_{\ell,\mathbf{m}}\right|^2 - \left(d-3\right)\left(d-4\right)\left|\mathcal{K}_{\ell,\mathbf{m}}\right|^2 \\
	&\quad\quad\quad\quad\quad\quad\quad\quad- 2\left(d-3\right)\text{Re}\left\{\bar{\mathcal{H}}_{\ell,\mathbf{m}}\mathcal{K}_{\ell,\mathbf{m}}\right\} \bigg) \bigg] \,.
\ea\ee
In the above expressions, we have introduced the notation
\be
	\mathcal{F}^{ab}_{\ell,\mathbf{m}} \equiv D^{a}\mathcal{H}^{b}_{\ell,\mathbf{m}}-D^{b}\mathcal{H}^{a}_{\ell,\mathbf{m}} \,,\quad \mathcal{H}_{\ell,\mathbf{m}}\equiv g_{ab}\mathcal{H}^{ab}_{\ell,\mathbf{m}} \,.
\ee

\subsection{Tensor modes}
We begin with the easier case of the tensor modes and perform the field redefinition
\be
	h^{\left(\text{T}\right)}_{\ell,\mathbf{m}} = \frac{\Psi^{\left(\text{T}\right)}_{\ell,\mathbf{m}}}{r^{\left(d-2\right)/2}} \,.
\ee
The resulting action after integration by parts takes the canonical form
\be
	S^{\left(\text{T}\right)}_{\ell,\mathbf{m}} = \int d^2x\sqrt{-g^{\left(2\right)}}\left[ -\frac{1}{2}D_{a}\bar{\Psi}^{\left(\text{T}\right)}_{\ell,\mathbf{m}}D^{a}\Psi^{\left(\text{T}\right)}_{\ell,\mathbf{m}} - \frac{1}{2}V^{\left(\text{T}\right)}_{\ell}\left(r\right)\left|\Psi^{\left(\text{T}\right)}_{\ell,\mathbf{m}}\right|^2 \right] \,,
\ee
with the tensor modes potential given by
\be\ba\label{eq:V2T}
	V^{\left(\text{T}\right)}_{\ell}\left(r\right) &= \frac{\ell\left(\ell+d-3\right)+2\left(d-3\right)}{r^2}+\frac{d^2-14d+32}{4r^2}r_{a}r^{a}+\frac{d-6}{2r}D_{a}r^{a} \\
	&= \frac{\ell\left(\ell+d-3\right)+2\left(d-3\right)}{r^2}+\frac{d^2-14d+32}{4r^2}f_{r}+\frac{d-6}{2r}\frac{\left(f_{t}f_{r}\right)^{\prime}}{2f_{t}} \,.
\ea\ee

\subsection{Vector (Regge-Wheeler) modes}
Next, for the vector modes we follow a procedure similar to the scalar modes for the spin-$1$ perturbations. We introduce an auxiliary Regge-Wheeler variable $\Psi^{\left(\text{RW}\right)}_{\ell,\mathbf{m}}$ and consider the following action~\cite{Hui:2020xxx}
\be\ba
	\tilde{S}^{\left(\text{RW}\right)}_{\ell,\mathbf{m}} = \int d^2x\sqrt{-g^{\left(2\right)}}&\bigg[ \sqrt{\frac{F_{\ell}\left(r\right)}{2}}r^{\left(d-6\right)/2}\text{Re}\left\{\bar{\Psi}^{\left(\text{RW}\right)}_{\ell,\mathbf{m}}\left(\varepsilon_{ab}\mathcal{F}^{ab}_{\ell,\mathbf{m}}-\frac{4}{r}\sqrt{\frac{f_{r}}{f_{t}}}\,t_{a}\mathcal{H}^{a}_{\ell,\mathbf{m}}\right)\right\} \\
	&-\frac{1}{2}\frac{F_{\ell}\left(r\right)}{r^2}\left(\left|\Psi^{\left(\text{RW}\right)}_{\ell,\mathbf{m}}\right|^2 + 2r^{d-4}\bar{\mathcal{H}}_{a;\ell,\mathbf{m}}\mathcal{H}^{a}_{\ell,\mathbf{m}}\right) \bigg] \,,
\ea\ee
with
\be
	F_{\ell}\left(r\right) \equiv \left(\ell+1\right)\left(\ell+d-4\right) - 2\left(d-3\right)r_{a}r^{a}-2r D_{a}r^{a} \,.
\ee
For Schwarzschild black holes, $F_{\ell}\left(r\right) = \left(\ell-1\right)\left(\ell+d-2\right)$ becomes a constant.

This alternative action retrieves the original action $S^{\left(\text{RW}\right)}_{\ell,\mathbf{m}}$ for the vector modes after integrating out the auxiliary field,
\be
	\Psi^{\left(\text{RW}\right)}_{\ell,\mathbf{m}} = \frac{r^{\left(d-2\right)/2}}{\sqrt{2F_{\ell}\left(r\right)}}\left[\varepsilon_{ab}\mathcal{F}^{ab}_{\ell,\mathbf{m}}-\frac{4}{r}\sqrt{\frac{f_{r}}{f_{t}}}\,t_{a}\mathcal{H}^{a}_{\ell,\mathbf{m}}\right] \,.
\ee
By integrating out the fields $\mathcal{H}^{a}_{\ell,\mathbf{m}}$ instead,
\be
	\mathcal{H}^{a}_{\ell,\mathbf{m}} = \frac{r^{-\left(d-6\right)/2}}{\sqrt{2F_{\ell}\left(r\right)}}\left[\varepsilon^{ab}D_{b} - t^{a}\left(\frac{d-2}{2r}+\frac{F_{\ell}^{\prime}\left(r\right)}{2F_{\ell}\left(r\right)}\right)\sqrt{\frac{f_{r}}{f_{t}}}\right]\Psi^{\left(\text{RW}\right)}_{\ell,\mathbf{m}} \,,
\ee
we end up with a canonically normalized action for the field $\Psi^{\left(\text{RW}\right)}_{\ell,\mathbf{m}}$,
\be
	\tilde{S}^{\left(\text{RW}\right)}_{\ell,\mathbf{m}} = \int d^2x\sqrt{-g^{\left(2\right)}}\left[-\frac{1}{2}D_{a}\bar{\Psi}^{\left(\text{RW}\right)}_{\ell,\mathbf{m}}D^{a}\Psi^{\left(\text{RW}\right)}_{\ell,\mathbf{m}} - \frac{1}{2}V^{\left(\text{RW}\right)}_{\ell}\left(r\right)\left|\Psi^{\left(\text{RW}\right)}_{\ell,\mathbf{m}}\right|^2\right] \,,
\ee
with the Regge-Wheeler potential given by
\be\ba\label{eq:V2RW}
	V^{\left(\text{RW}\right)}_{\ell}\left(r\right) &= \frac{\left(\ell+1\right)\left(\ell+d-4\right)}{r^2} + \frac{\left(d-4\right)\left(d-6\right)}{4r^2}r_{a}r^{a} - \frac{d+2}{2r}D_{a}r^{a} \\
	&+\left[\frac{F_{\ell}^{\prime}}{2F_{\ell}}\left(\frac{d-4}{r}+\frac{F_{\ell}^{\prime}}{2F_{\ell}}\right)-\left(\frac{F_{\ell}^{\prime}}{2F_{\ell}}\right)^{\prime}\right]r_{a}r^{a} - \frac{F_{\ell}^{\prime}}{2F_{\ell}}D_{a}r^{a} \,.
\ea\ee
For Schwarzschild black holes, the terms in the second line are zero.

\subsection{Scalar (Zerilli) modes}
The first observation to find the master variable relevant for the gravitoelectric response of the black hole is that the trace $\mathcal{H}_{\ell,\mathbf{m}}$ of the scalar modes $\mathcal{H}^{ab}_{\ell,\mathbf{m}}$ is an auxiliary field,
\be\ba
	{}&S^{\left(\text{Z}\right)}_{\ell,\mathbf{m}} = \int d^2x\sqrt{-g^{\left(2\right)}}\,r^{d-2} \bigg[ -\frac{1}{2}D_{c}\bar{\mathcal{H}}_{\braket{ab};\ell,\mathbf{m}}D^{c}\mathcal{H}^{\braket{ab}}_{\ell,\mathbf{m}} + D_{c}\bar{\mathcal{H}}_{\braket{ab};\ell,\mathbf{m}}D^{b}\mathcal{H}^{\braket{ac}}_{\ell,\mathbf{m}} \\
	&+ \frac{\left(d-2\right)\left(d-3\right)}{2}D_{a}\bar{\mathcal{K}}_{\ell,\mathbf{m}}D^{a}\mathcal{K}_{\ell,\mathbf{m}} - \left(d-2\right)\text{Re}\left\{D_{a}\bar{\mathcal{K}}_{\ell,\mathbf{m}}\left(D_{b}\mathcal{H}^{\braket{ab}}_{\ell,\mathbf{m}}-\frac{1}{2}D^{a}\mathcal{H}_{\ell,\mathbf{m}}\right)\right\} \\
	&-\frac{d-2}{r}\text{Re}\left\{\left(D_{a}\bar{\mathcal{H}}_{\ell,\mathbf{m}}+\left(d-4\right)D_{a}\bar{\mathcal{K}}_{\ell,\mathbf{m}}\right)\left(r_{b}\mathcal{H}^{\braket{ab}}_{\ell,\mathbf{m}}-r^{a}\left(\mathcal{K}_{\ell,\mathbf{m}}-\frac{1}{2}\mathcal{H}_{\ell,\mathbf{m}}\right)\right)\right\} \\
	&-\frac{1}{2}\frac{\ell\left(\ell+d-3\right)}{r^2}\bigg( \bar{\mathcal{H}}_{\braket{ab};\ell,\mathbf{m}}\mathcal{H}^{\braket{ab}}_{\ell,\mathbf{m}} - \frac{1}{2}\left|\mathcal{H}_{\ell,\mathbf{m}}\right|^2 - \left(d-3\right)\left(d-4\right)\left|\mathcal{K}_{\ell,\mathbf{m}}\right|^2 \\
	&\quad\quad\quad\quad\quad\quad\quad\quad- 2\left(d-3\right)\text{Re}\left\{\bar{\mathcal{H}}_{\ell,\mathbf{m}}\mathcal{K}_{\ell,\mathbf{m}}\right\} \bigg) \bigg] \,,
\ea\ee
and hence it is possible to directly integrate it out. Furthermore, by writing
\be
	\mathcal{H}^{\braket{ab}}_{\ell,\mathbf{m}} = A_{\ell,\mathbf{m}}I_{\left(1\right)}^{\braket{ab}}+B_{\ell,\mathbf{m}}I_{\left(2\right)}^{\braket{ab}}
\ee
with
\be
	I_{\left(1\right)}^{\braket{ab}} = \frac{1}{f_{t}}t^{a}t^{b}+\frac{1}{f_{r}}r^{a}r^{b} \,,\quad I_{\left(2\right)}^{\braket{ab}} = \frac{2}{\sqrt{f_{t}f_{r}}}t^{(a}r^{b)}
\ee
the two independent STF rank-2 tensors in $2$-d, satisfying
\be
	\begin{gathered}
		I_{\left(1\right)\braket{ab}}I_{\left(1\right)}^{\braket{ac}} = \delta_{b}^{c} \,,\quad I_{\left(1\right)\braket{ab}}I_{\left(2\right)}^{\braket{ac}} = \varepsilon_{b}^{\,\,\,c} \,,\quad I_{\left(2\right)\braket{ab}}I_{\left(2\right)}^{\braket{ac}} = -\delta_{b}^{c} \,, \\
		D^{c}I_{\left(1\right)}^{\braket{ab}} = -\frac{f_{t}^{\prime}}{f_{t}}\sqrt{\frac{f_{r}}{f_{t}}}\,t^{c}I^{\braket{ab}}_{\left(2\right)} \,,\quad D^{c}I_{\left(2\right)}^{\braket{ab}} = -\frac{f_{t}^{\prime}}{f_{t}}\sqrt{\frac{f_{r}}{f_{t}}}\,t^{c}I^{\braket{ab}}_{\left(1\right)} \,,
	\end{gathered}
\ee
then
\be
	-\frac{1}{2}D_{c}\bar{\mathcal{H}}_{\braket{ab};\ell,\mathbf{m}}D^{c}\mathcal{H}^{\braket{ab}}_{\ell,\mathbf{m}} + D_{c}\bar{\mathcal{H}}_{\braket{ab};\ell,\mathbf{m}}D^{b}\mathcal{H}^{\braket{ac}}_{\ell,\mathbf{m}} \supset -2\varepsilon_{ab}\text{Re}\left\{D^{a}\bar{A}_{\ell,\mathbf{m}}D^{b}B_{\ell,\mathbf{m}}\right\} \,,
\ee
which shows that $A_{\ell,\mathbf{m}}$ and $B_{\ell,\mathbf{m}}$ are also auxiliary variables. More explicitly, the full Zerilli action for the scalar modes reads, after some integrations by parts,
\be
	S^{\left(\text{Z}\right)}_{\ell,\mathbf{m}} = \int d^2x\sqrt{-g^{\left(2\right)}}\,r^{d-2}\left(\mathcal{L}^{\left(AB\right)}_{\ell,\mathbf{m}} + \mathcal{L}^{\left(KK\right)}_{\ell,\mathbf{m}} + \mathcal{L}^{\left(HH\right)}_{\ell,\mathbf{m}} + \mathcal{L}^{\left(ABK\right)}_{\ell,\mathbf{m}} + \mathcal{L}^{\left(ABH\right)}_{\ell,\mathbf{m}} + \mathcal{L}^{\left(KH\right)}_{\ell,\mathbf{m}} \right) \,,
\ee
\be\ba
	\mathcal{L}^{\left(AB\right)}_{\ell,\mathbf{m}} &= -2\varepsilon^{ab}\text{Re}\left\{D_{a}\bar{A}_{\ell,\mathbf{m}}D_{b}B_{\ell,\mathbf{m}}\right\} - \frac{M_{\ell}\left(r\right)}{r^2}\left(\left|A_{\ell,\mathbf{m}}\right|^2 - \left|B_{\ell,\mathbf{m}}\right|^2 \right) \,, \\
	\mathcal{L}^{\left(KK\right)}_{\ell,\mathbf{m}} &= \frac{\left(d-2\right)\left(d-3\right)}{2}D_{a}\bar{\mathcal{K}}_{\ell,\mathbf{m}}D^{a}\mathcal{K}_{\ell,\mathbf{m}} + \frac{L_{\ell}\left(r\right)}{2r^2}\left|K_{\ell,\mathbf{m}}\right|^2 \,, \\
	\mathcal{L}^{\left(HH\right)}_{\ell,\mathbf{m}} &= \frac{G_{\ell}\left(r\right)}{4r^2}\left|\mathcal{H}_{\ell,\mathbf{m}}\right|^2 \,, \\
	\mathcal{L}^{\left(ABK\right)}_{\ell,\mathbf{m}} &= \left(d-2\right)\text{Re}\left\{\left(\bar{A}_{\ell,\mathbf{m}}I_{\left(1\right)}^{\braket{ab}}+\bar{B}_{\ell,\mathbf{m}}I_{\left(2\right)}^{\braket{ab}}\right)\,D_{a}D_{b}\mathcal{K}_{\ell,\mathbf{m}} \right\} \\
	&\quad+\frac{2\left(d-2\right)}{r}\text{Re}\left\{\left(\bar{A}_{\ell,\mathbf{m}}r^{a}+\sqrt{\frac{f_{r}}{f_{t}}}\,\bar{B}_{\ell,\mathbf{m}}t^{a}\right)D_{a}\mathcal{K}_{\ell,\mathbf{m}}\right\} \\
	\mathcal{L}^{\left(ABH\right)} &= -\frac{d-2}{r}\text{Re}\left\{\left(\bar{A}_{\ell,\mathbf{m}}r^{a}+\sqrt{\frac{f_{r}}{f_{t}}}\,\bar{B}_{\ell,\mathbf{m}}t^{a}\right)D_{a}\mathcal{H}_{\ell,\mathbf{m}}\right\} \,, \\
	\mathcal{L}^{\left(KH\right)}_{\ell,\mathbf{m}} &= \frac{d-2}{2}\text{Re}\left\{\bar{\mathcal{H}}_{\ell,\mathbf{m}}\left[-D_{a}D^{a}\mathcal{K}_{\ell,\mathbf{m}} - \frac{2d-5}{r}r^{a}D_{a}\mathcal{K}_{\ell,\mathbf{m}} + \frac{N_{\ell}\left(r\right)}{r^2}\mathcal{K}_{\ell,\mathbf{m}}\right] \right\} \,,
\ea\ee
where we have defined
\be\ba
	M_{\ell}\left(r\right) &\equiv \ell\left(\ell+d-3\right) - \left[\left(d-2\right)\frac{rf_{t}^{\prime}}{f_{t}} + r^2\left(\frac{f_{t}^{\prime}}{f_{t}}\right)^{\prime}\right]r_{a}r^{a} - \frac{rf_{t}^{\prime}}{f_{t}}rD_{a}r^{a} \,, \\
	G_{\ell}\left(r\right) &\equiv \ell\left(\ell+d-3\right) + \left(d-2\right)\left(d-3\right)r_{a}r^{a} + \left(d-2\right)rD_{a}r^{a} \,, \\
	L_{\ell}\left(r\right) &\equiv \left(d-4\right)\left[\left(d-2\right)\ell\left(\ell+d-3\right) - G_{\ell}\left(r\right)\right] \,, \\
	N_{\ell}\left(r\right) &\equiv \frac{1}{d-2}\left[\left(2d-5\right)\ell\left(\ell+d-3\right) - G_{\ell}\left(r\right)\right] \,.
\ea\ee
By integrating out the auxiliary variables $A_{\ell,\mathbf{m}}$, $B_{\ell,\mathbf{m}}$ and $\mathcal{H}_{\ell,\mathbf{m}}$, it is then possible to write down a Schr\"{o}dinger-like equation of motion for a master variable built from $\mathcal{K}_{\ell,\mathbf{m}}$ and its derivatives. The procedure is quite cumbersome and not enlightening for generic $f_{t}$ and $f_{r}$ so we will just write down the results for the Schwarzschild black hole, for which $f_{t} = f_{r} = 1 -\left(r_{s}/r\right)^{d-3} \equiv f\left(r\right)$ and the above functions reduce to constants,
\be
	\begin{gathered}
		M_{\ell} = \ell\left(\ell+d-3\right) \,,\quad L_{\ell} =\left(d-3\right)\left(d-4\right)\left[\ell\left(\ell+d-3\right)-\left(d-2\right)\right] \,, \\
		G_{\ell} = \ell\left(\ell+d-3\right) + \left(d-2\right)\left(d-3\right) \,,\quad N_{\ell} = 2\ell\left(\ell+d-3\right) - \left(d-3\right) \,.
	\end{gathered}
\ee
After all, this is the only relevant case according to our current setup for the gravitoelectric response of an asymptotically flat and electrically neutral general-relativistic black hole in vacuum.

The Zerilli master variable $\Psi_{\ell,\mathbf{m}}^{\left(\text{Z}\right)}$ is constructed as~\cite{Hui:2020xxx,Ishibashi:2003ap,Kodama:2003jz,Kodama:2003kk,Ishibashi:2011ws}
\be
	\Psi^{\left(\text{Z}\right)}_{\ell,\mathbf{m}} = \frac{4fr^{\frac{d-4}{2}}}{H_{\ell}\left(r\right)}\sqrt{\left(d-2\right)\left(d-3\right)\lambda_{\ell}\ell\left(\ell+d-3\right)}\,\mathcal{V}_{\ell,\mathbf{m}} \,,
\ee
\be
	\mathcal{V}_{\ell,\mathbf{m}} = -\frac{r\mathcal{K}_{\ell,\mathbf{m}}}{2\sqrt{f_{t}f_{r}}}-\frac{\left(d-2\right)r}{2M_{\ell}}\sqrt{\frac{f_{r}}{f_{t}}}\left[A_{\ell,\mathbf{m}}+\frac{1}{2}\mathcal{H}_{\ell,\mathbf{m}}-\frac{1}{f_{r}}r^{a}D_{a}\left(r\mathcal{K}_{\ell,\mathbf{m}}\right)+\frac{rf_{t}^{\prime}}{2f_{t}}\mathcal{K}_{\ell,\mathbf{m}}\right] \,,
\ee
where we have defined
\be
	\lambda_{\ell} \equiv \left(\ell-1\right)\left(\ell+d-2\right)
\ee
and
\be\ba
	H_{\ell}\left(r\right) &= 2\ell\left(\ell+d-3\right) - 2\left(d-2\right)f\left(r\right) + \left(d-2\right)rf^{\prime}\left(r\right) \\
	&= 2\lambda_{\ell} + \left(d-1\right)\left(d-2\right)\left(\frac{r_{s}}{r}\right)^{d-3} \,.
\ea\ee
It satisfies a Schr\"{o}dinger-like equation,
\be
	\left[\partial_{r_{\ast}}^2-\partial_{t}^2-f\left(r\right)V_{\ell}^{\left(\text{Z}\right)}\right]\Psi_{\ell,\mathbf{m}}^{\left(\text{Z}\right)} = 0 \,,
\ee
with the Zerilli potential given by~\cite{Hui:2020xxx,Ishibashi:2003ap,Kodama:2003jz,Kodama:2003kk,Ishibashi:2011ws}
\be
	V_{\ell}^{\left(\text{Z}\right)}\left(r\right) = V_{\ell}^{\left(0\right)}\left(r\right) - \frac{2f^{\prime}\left(r\right)}{r}\frac{\left[2\lambda_{\ell}+\left(d-1\right)\left(d-2\right)\right]\left[H_{\ell}\left(r\right)+2\left(d-3\right)\lambda_{\ell}\right]}{H_{\ell}^2\left(r\right)} \,,
\ee
where $V^{\left(0\right)}\left(r\right)$ is the scalar field potential in Eq.~\ref{eq:V0S}.

\section{Schwarzschild black hole Love numbers}
\label{sec:LNs_SchwarzschildDd}

We know focus to general-relativistic black holes, starting with the case of the higher-dimensional Schwarzschild black hole for which
\be
	f_{t}\left(r\right) = f_{r}\left(r\right) = 1 - \left(\frac{r_{s}}{r}\right)^{d-3} \equiv f\left(r\right) \,.
\ee

\subsection{$p$-form Love numbers}
We will first consider $p$-form perturbations, which are captured by the master variables $\Psi^{\left(j\right)}$ with $j$ labeling the $SO\left(d-1\right)$ sector of the perturbation. In particular, $j=p$ for $p$-form modes and $j=\tilde{p}=d-p-2$ for $\left(p-1\right)$-form modes. Equivalently, $j$ is equal to $n$ dualizations of the rank $p$ of the $p$-form for the $\left(p-n\right)$-form modes. In this notation, the cases of scalar field and spin-$1$ perturbations can also be incorporated via
\be\ba
	s=0&: \Psi^{\left(j=0\right)}_{\ell,\mathbf{m}} = \Psi^{\left(0\right)}_{\ell,\mathbf{m}} \\
	s=1&: \Psi^{\left(j=1\right)}_{\ell,\mathbf{m}} = \Psi^{\left(\text{V}\right)}_{\ell,\mathbf{m}} \,,\quad \Psi^{\left(j=d-3\right)}_{\ell,\mathbf{m}} = \Psi^{\left(\text{S}\right)}_{\ell,\mathbf{m}} \,,
\ea\ee
with no analogues of $\left(p-1\right)$-form modes for the scalar ($p=0$) field. Performing the field redefinition
\be
	\Phi^{\left(j\right)}_{\ell,\mathbf{m}} = \frac{\Psi^{\left(j\right)}_{\ell,\mathbf{m}}}{r^{\frac{d-2}{2}}} \,,
\ee
introducing the variable $\rho=r^{d-3}$ and defining $\Delta=\rho^2 f = \rho\left(\rho-\rho_{s}\right)$, the radial equations of motion for $p$-form perturbations can be rewritten as
\be\label{eq:FullpEOMSchwarzschild}
	\begin{gathered}
		\mathbb{O}^{\left(j\right)}_{\text{full}}\Phi^{\left(j\right)}_{\ell,\mathbf{m}} = \hat{\ell}(\hat{\ell}+1)\Phi^{\left(j\right)}_{\ell,\mathbf{m}} \,, \\
		\mathbb{O}^{\left(j\right)}_{\text{full}} = \partial_{\rho}\,\Delta\,\partial_{\rho} - \frac{r^2\rho^2}{\left(d-3\right)^2\Delta}\partial_{t}^2 + \frac{\rho_{s}}{\rho}\hat{j}^2 \,,
	\end{gathered}
\ee
where we have also introduced the rescaled orbital number and $j$-index
\be
	\hat{\ell} \equiv \frac{\ell}{d-3} \,,\quad \hat{j} \equiv \frac{j}{d-3} \,.
\ee

Let us now solve these to extract the Schwarzschild black hole Love numbers at leading order in the near-zone expansion. There are two near-zone splittings that are of particular interest, controlled by a sign $\sigma=\pm1$,
\be\label{eq:Vp_SchwarzschildD}
	\begin{gathered}
		\mathbb{O}^{\left(j\right)}_{\text{full}} = \partial_{\rho}\,\Delta\,\partial_{\rho} + V_0^{\left(\sigma\right)} + \epsilon\,V_1^{\left(\sigma\right)} \,, \\
		\ba
			V_0^{\left(\sigma\right)} &= -\frac{\rho_{s}^2}{4\Delta}\beta^2\partial_{t}^2 + \frac{\rho_{s}}{\rho}\hat{j}\left(\sigma\beta\,\partial_{t} + \hat{j}\right) \,, \\
			V_1^{\left(\sigma\right)} &= -\frac{r^2\rho^2-r_{s}^2\rho_{s}^2}{\left(d-3\right)^2\Delta}\partial_{t}^2 - \sigma\frac{\rho_{s}}{\rho}\hat{j}\beta\,\partial_{t} \,,
		\ea
	\end{gathered}
\ee
where $\beta = \frac{2r_{s}}{d-3}$ is the inverse Hawking temperature of the $d$-dimensional Schwarzschild black hole. Note that we have introduced a $\partial_{t}$ term; even though this was not present in the original equations of motion, it is still subleading in the near-zone expansion since it does not alter the near-horizon behavior of the solution. This might look like making the equations we want to solve more complicated than necessary, but introducing this term actually makes the problem simpler in the sense that we can now analytically solve the leading order near-zone equations of motion in terms of hypergeometric functions just as in the $d=4$ case.

Indeed, after separating the variables,
\be
	\Phi^{\left(j\right)}_{\omega\ell,\mathbf{m}}\left(t,\rho\right) = e^{-i\omega t}R^{\left(j\right)}_{\omega\ell,\mathbf{m}}\left(\rho\right) \,,
\ee
and introducing the dimensionless radial distance from the event horizon,
\be
	x = \frac{\rho-\rho_{s}}{\rho_{s}} \,,
\ee
the leading order near-zone radial equation of motion reads
\be\label{eq:NZRadialScharzschildDd}
	\left[\frac{d}{dx}\,x\left(1+x\right)\frac{d}{dx} + \frac{\beta^2\omega^2}{4x} - \frac{\left(\beta\omega+2i\sigma\hat{j}\right)^2}{4\left(1+x\right)} \right]R^{\left(j\right)}_{\omega\ell,\mathbf{m}} = \hat{\ell}(\hat{\ell}+1)R^{\left(j\right)}_{\omega\ell,\mathbf{m}}
\ee
and the solution satisfying ingoing boundary conditions at the future event horizon (see Eq.~\eqref{eq:NHbehavior_SphericalSymmetric}) can be analytically found to be 
\be\ba
	R^{\left(j\right)}_{\omega\ell,\mathbf{m}} &= \bar{R}^{\left(j\right)\text{in}}_{\ell,\mathbf{m}}\left(\omega\right)\left(\frac{x}{1+x}\right)^{-i\beta\omega/2} \\
	&\quad \times\left(1+x\right)^{-\sigma\hat{j}}{}_2F_1\left(\hat{\ell}+1-\sigma\hat{j},-\hat{\ell}-\sigma\hat{j};1-i\beta\omega;-x\right) \,.
\ea\ee
Expanding around large distances reveals then that the response coefficients at leading order in the near-zone expansion are
\be\label{eq:ResponseCoefficientspSchwarzschildDd}
	k^{\left(j\right)}_{\ell}\left(\omega\right) = \frac{\Gamma(-2\hat{\ell}-1)\Gamma(\hat{\ell}+1-\sigma\hat{j})\Gamma(\hat{\ell}+1+\sigma\hat{j}-i\beta\omega)}{\Gamma(2\hat{\ell}+1)\Gamma(-\hat{\ell}-\sigma\hat{j})\Gamma(-\hat{\ell}+\sigma\hat{j}-i\beta\omega)} \,.
\ee
At this point, let us remark that the matching has been done directly at the level of the master variables $\Phi^{\left(j\right)}_{\ell,\mathbf{m}}$ which are built at most from derivatives of the actual fields in terms of which the response problem is defined. Therefore, analytically continuing the orbital number $\ell$ or the spacetime dimensionality $d$ is sufficient to unambiguously perform the source/response split. The above coefficients in front of the decaying branches of the master variables $\Phi^{\left(j\right)}_{\ell,\mathbf{m}}$ are then equal to the actual response coefficients we are looking for, up to overall non-zero matching normalization constants. For scalar and electromagnetic perturbations, for instance,
\be
	\begin{gathered}
		k^{\left(j=0\right)}_{\ell}\left(\omega\right) = k^{\left(0\right)}_{\ell}\left(\omega\right) \,, \\
		k^{\left(j=1\right)}_{\ell}\left(\omega\right) = k^{\mathcal{B}\left(1\right)}_{\ell}\left(\omega\right) \,,\quad k^{\left(j=d-3\right)}_{\ell}\left(\omega\right) = -\frac{\ell+d-3}{\ell}k^{\mathcal{E}\left(1\right)}_{\ell}\left(\omega\right) \,.
	\end{gathered}
\ee
For the more general case of $p$-form perturbations, there is a $p$-form sector and $\left(p-1\right)$-form sector, serving as the $p>1$ generalizations of the magnetic and electric sectors respectively that one encounters for electromagnetic perturbations. For these,
\be
	\begin{gathered}
		k^{\left(j=p\right)}_{\ell}\left(\omega\right) = k^{\mathcal{B},\left(p\right)}_{\ell}\left(\omega\right) \,,\quad k^{\left(j=d-p-2\right)}_{\ell}\left(\omega\right) = -\frac{\ell+d-p-2}{\ell+p-1}k^{\mathcal{E},\left(p\right)}_{\ell}\left(\omega\right) \,,
	\end{gathered}
\ee
where the superscripts ``$\mathcal{B}$'' and ``$\mathcal{E}$'' here refer to these extensions of the magnetic-type and electric-type perturbations, although this is just a convention of labeling things; these are not actual magnetic or electric in nature\footnote{To be more concrete, following the notation of Chapter~\ref{ch:TLNsDefinition}, the Love numbers for the $p$-form and $\left(p-1\right)$-form sectors are defined in the worldline EFT as, for a spherically symmetric body,
\begin{align*}
	S_{\text{Love}}^{\left(p\right)} &= \sum_{\ell=1}^{\infty}\frac{1}{p}\int \frac{d\omega}{2\pi}\frac{C_{\text{el},\ell}^{\left(p\right)}\left(\omega\right)}{2\ell!}\,\mathcal{E}_{L|b_1\dots b_{p-1}}^{\left(p\right)}\left(-\omega\right)\mathcal{E}^{\left(p\right)L|b_1\dots b_{p-1}}\left(\omega\right) \\
	&+ \sum_{\ell=1}^{\infty}\frac{1}{p+1}\int \frac{d\omega}{2\pi}\frac{C_{\text{mag},\ell}^{\left(p\right)}\left(\omega\right)}{2\ell!}\,\mathcal{B}_{L|b_1\dots b_{p}}^{\left(p\right)}\left(-\omega\right)\mathcal{B}^{\left(p\right)L|b_1\dots b_{p}}\left(\omega\right) \,,
\end{align*}
respectively, with
\begin{align*}
	\mathcal{E}_{L|b_1\dots b_{p-1}}^{\left(p\right)} &= e_{a_1}^{\mu_1}\dots e_{a_{\ell-1}}^{\mu_{\ell-1}}\nabla_{\langle \mu_1}\dots \nabla_{\mu_{\ell-1}}E_{a_{\ell}\rangle b_1\dots b_{p-1}} \,,\quad E_{a_1a_2\dots a_{p}} = u^{\mu_1}e_{a_1}^{\mu_2}\dots e_{a_{p}}^{\mu_{p+1}}F_{\mu_1\mu_2\dots\mu_{p+1}} \,, \\
	\mathcal{B}_{L|b_1\dots b_{p}}^{\left(p\right)} &= e_{a_1}^{\mu_1}\dots e_{a_{\ell-1}}^{\mu_{\ell-1}}\nabla_{\langle \mu_1}\dots \nabla_{\mu_{\ell-1}}B_{a_{\ell}\rangle b_1\dots b_{p}} \,,\quad B_{a_1a_2\dots a_{p+1}} = e_{a_1}^{\mu_1}e_{a_2}^{\mu_2}\dots e_{a_{p+1}}^{\mu_{p+1}}F_{\mu_1\mu_2\dots\mu_{p+1}} \,.
\end{align*}
These are matched onto the Love part of the response coefficients as defined in the asymptotic expansion, in frequency space,
\begin{align*}
	\mathcal{A}_{a_1\dots a_{p-1}}^{\mathcal{E},\left(p\right)}\left(\omega,\mathbf{x}\right) &\xrightarrow{r\rightarrow\infty} \sum_{\ell=1}^{\infty}\frac{\left(\ell-1\right)!}{\ell!}\left[1+k_{\ell}^{\mathcal{E},\left(p\right)}\left(\omega\right)\left(\frac{\mathcal{R}}{r}\right)^{2\ell+d-3}\right]\bar{\mathcal{A}}_{L|a_1\dots a_{p-1}}^{\mathcal{E},\left(p\right)}\left(\omega\right)x^{L} \,, \\
	\mathcal{A}_{a_1\dots a_{p}}^{\mathcal{B},\left(p\right)}\left(\omega,\mathbf{x}\right) &\xrightarrow{r\rightarrow\infty} \sum_{\ell=1}^{\infty}\frac{\left(\ell-1\right)!}{\ell!}\left[1+k_{\ell}^{\mathcal{B},\left(p\right)}\left(\omega\right)\left(\frac{\mathcal{R}}{r}\right)^{2\ell+d-3}\right]\bar{\mathcal{A}}_{L|a_1\dots a_{p}}^{\mathcal{B},\left(p\right)}\left(\omega\right)x^{L} \,,
\end{align*}
where $\mathcal{A}_{a_1\dots a_{p-1}}^{\mathcal{E},\left(p\right)}=u^{\mu_1}e^{\mu_2}_{a_1}\dots e^{\mu_{p}}_{a_{p-1}}A_{\mu_1\mu_2\dots \mu_{p}}$ and $\mathcal{A}_{a_1\dots a_{p}}^{\mathcal{B},\left(p\right)}=e^{\mu_1}_{a_1}e^{\mu_2}_{a_2}\dots e^{\mu_{p}}_{a_{p}}\mathcal{A}_{\mu_1\mu_2\dots \mu_{p}}^{\left(p\right)}$, according to Eq.~\eqref{eq:WTesnorConsResp}, with $N_{\text{prop}}^{\left(p\right),\mathcal{E}}=-1$ and $N_{\text{prop}}^{\left(p\right),\mathcal{B}}=+1$, understood such that $\braket{\mathcal{A}^{\mathcal{E}/\mathcal{B}}_{a_1\dots a_{n}}\mathcal{A}^{\mathcal{E}/\mathcal{B}|b_1\dots b_{n}}}\left(p\right)=N_{\text{prop}}^{\left(p\right),\mathcal{E}/\mathcal{B}}\delta_{[a_1}^{b_1}\dots \delta^{b_{n}}_{a_{n}]}\frac{-i}{p^2}$}. For the sake of simplicity, however, we will keep referring to \eqref{eq:ResponseCoefficientspSchwarzschildDd} as the response coefficients associated with each type of perturbation.

In the static limit, the response coefficients \eqref{eq:ResponseCoefficientspSchwarzschildDd} become purely real and correspond to the static Love numbers for $p$-form perturbations,
\be\ba\label{eq:StaticpLNs_SchwarzschildDd}
	k^{\left(j\right)\text{Love}}_{\ell}\left(\omega=0\right) &= \frac{\Gamma^2(\hat{\ell}+1-\hat{j})\Gamma^2(\hat{\ell}+1+\hat{j})}{\pi\,\Gamma(2\hat{\ell}+1)\Gamma(2\hat{\ell}+2)}\frac{\sin\pi(\hat{\ell}-\hat{j})\sin\pi(\hat{\ell}+\hat{j})}{\sin2\pi\hat{\ell}} \\
	&= \frac{\Gamma^2(\hat{\ell}+1-\hat{j})\Gamma^2(\hat{\ell}+1+\hat{j})}{2\pi\,\Gamma(2\hat{\ell}+1)\Gamma(2\hat{\ell}+2)}\left[\tan\pi\hat{\ell}\cos^2\pi\hat{j}-\cot\pi\hat{\ell}\sin^2\pi\hat{j}\right] \,.
\ea\ee
Ignoring at the moment the specific values of $\hat{j}$, the static Love numbers appear to exhibit the expected behavior. Namely, as discussed in Section~\ref{sec:PowerCounting}, the power counting argument applied for General Relativity shows that the static Love numbers should be non-zero and non-running for generic $2\hat{\ell}\notin\mathbb{N}$, while they are expected to exhibit a logarithmic running for $2\hat{\ell}\in\mathbb{N}$. However, taking into consideration the explicit possible values of $\hat{j}$ we see a very rich structure depending on the values of the rank of the $p$-form gauge field. First of all, for the static scalar, static magnetic and static electric susceptibilities
\be\ba\label{eq:Static01LNs_SchwarzschildDd}
	{}&k^{\left(j=0\right)}_{\ell} = \frac{\Gamma^{4}(\hat{\ell}+1)}{2\pi\,\Gamma(2\hat{\ell}+1)\Gamma(2\hat{\ell}+2)}\tan\pi\hat{\ell} \,, \\
	{}&k^{\left(j=1\right)}_{\ell} = \frac{\Gamma^2(\hat{\ell}+1-\frac{1}{d-3})\Gamma^2(\hat{\ell}+1+\frac{1}{d-3})}{\pi\,\Gamma(2\hat{\ell}+1)\Gamma(2\hat{\ell}+2)}\frac{\sin\pi(\hat{\ell}-\frac{1}{d-3})\sin\pi(\hat{\ell}+\frac{1}{d-3})}{\sin2\pi\hat{\ell}} \,, \\
	{}&\text{and} \quad k^{\left(j=d-3\right)}_{\ell} = \frac{\Gamma^{2}(\hat{\ell})\Gamma^{2}(\hat{\ell}+2)}{2\pi\,\Gamma(2\hat{\ell}+1)\Gamma(2\hat{\ell}+2)}\tan\pi\hat{\ell}
\ea\ee
respectively. Hence, there are still hints of fine-tuning coming from the vanishing of the static scalar susceptibilities ($j=0$) and the static electric susceptibilities ($j=d-3$) whenever $\hat{\ell}\in\mathbb{N}$. We also get the opportunity to see how the electric/magnetic duality is no longer present in $d>4$, namely, the static magnetic susceptibilities ($j=1$) vanish under the \textit{different} resonant conditions $\hat{\ell}\pm\frac{1}{d-3}\in\mathbb{N}$, which are always non-overlapping with the $\hat{\ell}\in\mathbb{N}$ cases in $d>4$, see Table~\ref{tbl:Staticp3LNs_SchwarzschildDd}.

For generic $0< p\le d-3$, we can break things down into three categories, after also noting that $0<\hat{j}\le1$. The first class of $p$-form perturbations is when $\hat{j}$ is an integer, i.e. when $\hat{j}=1$. This corresponds to $p=d-3$ for the $p$-form $SO\left(d-1\right)$ sector or $p=1$ for the $\left(p-1\right)$-form $SO\left(d-1\right)$ sector. The latter is simply the electric-type electromagnetic response we saw above. The former is a new category of magnetic-like-type perturbations that emerges in $d>4$ and whose Love numbers are again identical to the static electric susceptibilities. These perturbations are just the Hodge dual version of the electric-type electromagnetic perturbations and they are merely a reflection of the Hodge duality symmetry, $\mathbf{F}^{\left(p+1\right)}\rightarrow \star\mathbf{F}^{\left(p+1\right)}$, of the $p$-form action. The qualitative behavior of static responses under $p$-form perturbations in this class is demonstrated in Table~\ref{tbl:Staticp1LNs_SchwarzschildDd}.

The second class of $p$-form perturbations is when $\hat{j}$ is a half-integer, i.e. when $\hat{j}=\frac{1}{2}$. This occurs only for odd spacetime dimensionalities, $d=5,7,\dots$, and now corresponds to $p=\frac{d-3}{2}$ for the $p$-form $SO\left(d-1\right)$ sector or $p=\frac{d-1}{2}$ for the $\left(p-1\right)$-form $SO\left(d-1\right)$ sector, the two types of perturbations again being related by Hodge duality. The static Love numbers for these cases read
\be
	k^{\left(j=\left(d-3\right)/2\right)}_{\ell} = -\frac{\Gamma^2\left(\hat{\ell}+\frac{1}{2}\right)\Gamma^2\left(\hat{\ell}+\frac{3}{2}\right)}{2\pi\,\Gamma(2\hat{\ell}+1)\Gamma(2\hat{\ell}+2)}\cot\pi\hat{\ell} = -\frac{1}{2^{3+4\hat{\ell}}}\frac{1}{k_{\ell}^{\left(j=0\right)}} \,,
\ee
where, in the second equality, we used the Legendre duplication formula for the $\Gamma$-function to compare with the static scalar Love numbers. We therefore see that the behavior of the static Love number for this class of $p$-form perturbations is opposite to that of the electric-type Love numbers, namely, they are non-zero and non-running for $2\hat{\ell}\notin\mathbb{N}$, they are logarithmically running for $\hat{\ell}\in\mathbb{N}$ and they are vanishing for $\hat{\ell}\in\mathbb{N}+\frac{1}{2}$, see Table~\ref{tbl:Staticp2LNs_SchwarzschildDd}.

The final, third, class of $p$-form perturbations contains all the other cases, for which $2\hat{j}\notin\mathbb{N}$. From the general expression for the static Love numbers in Eq.~\eqref{eq:StaticpLNs_SchwarzschildDd}, we see that these are non-zero and non-running for generic $\hat{\ell}$, they are logarithmically running for $2\hat{\ell}\in\mathbb{N}$ and are vanishing for $\hat{\ell}\pm\hat{j}\in\mathbb{N}$, see Table~\ref{tbl:Staticp3LNs_SchwarzschildDd}.

\begin{table}[t]
	\centering
	\begin{tabular}{|c||c|}
		\hline
		\Gape[7pt]{Range of parameters} & Behavior of $k_{\ell}^{\left(j\right)}\left(\omega=0\right)$ \\
		\hline\hline
		$\hat{\ell}\in\mathbb{N}$ & \Gape[8pt]{Vanishing} \\
		\hline
		$\hat{\ell}\in\mathbb{N}+\frac{1}{2}$ & \Gape[9pt]{Running} \\
		\hline
		$2\hat{\ell}\notin\mathbb{N}$ & \Gape[10pt]{Non-vanishing and Non-running} \\
		\hline
	\end{tabular}
	\caption[Behavior of static Love numbers for the first class of $p$-form perturbations of the higher-dimensional Schwarzschild black hole.]{Behavior of static Love numbers for the first class of $p$-form perturbations of the higher-dimensional Schwarzschild black hole, for which $\hat{j}=\frac{j}{d-3}$ is an integer. This class contains the static scalar susceptibilities ($j=0$) and the static electric susceptibilities as well as the Hodge dual $p$-form $SO\left(d-1\right)$ sector of $p$-form perturbations with $p=d-3$ ($j=d-3$). For generic orbital number, the static Love numbers for $p$-form perturbations in this class are non-zero and non-running. They are zero for integer $\hat{\ell}=\frac{\ell}{d-3}$ and they exhibit a classical RG flow for half-integer $\hat{\ell}$. As we will see later, the static electric-type and the static tensor-type tidal Love numbers also behave as prescribed here.}
	\label{tbl:Staticp1LNs_SchwarzschildDd}
\end{table}

\begin{table}[t]
	\centering
	\begin{tabular}{|c||c|}
		\hline
		\Gape[7pt]{Range of parameters} & Behavior of $k_{\ell}^{\left(j\right)}\left(\omega=0\right)$ \\
		\hline\hline
		$\hat{\ell}\in\mathbb{N}$ & \Gape[8pt]{Running} \\
		\hline
		$\hat{\ell}\in\mathbb{N}+\frac{1}{2}$ & \Gape[9pt]{Vanishing} \\
		\hline
		$2\hat{\ell}\notin\mathbb{N}$ & \Gape[10pt]{Non-vanishing and Non-running} \\
		\hline
	\end{tabular}
	\caption[Behavior of static Love numbers for the second class of $p$-form perturbations of the higher-dimensional Schwarzschild black hole.]{Behavior of static Love numbers for the second class of $p$-form perturbations of the higher-dimensional Schwarzschild black hole, for which $j=\frac{d-3}{2}$. This class exists for odd spacetime dimensionalities and contains the static magnetic susceptibilities in $d=5$ ($j=1$). For generic orbital number, the static Love numbers for $p$-form perturbations in this class are non-zero and non-running. They are zero now for half-integer $\hat{\ell}=\frac{\ell}{d-3}$ and they exhibit a classical RG flow for integer $\hat{\ell}$.}
	\label{tbl:Staticp2LNs_SchwarzschildDd}
\end{table}

\begin{table}[t]
	\centering
	\begin{tabular}{|c||c|}
		\hline
		\Gape[7pt]{Range of parameters} & Behavior of $k_{\ell}^{\left(j\right)}\left(\omega=0\right)$ \\
		\hline\hline
		$\hat{\ell}+\hat{j}\in\mathbb{N}$ OR $\hat{\ell}-\hat{j}\in\mathbb{N}$ & \Gape[8pt]{Vanishing} \\
		\hline
		$2\hat{\ell}\in\mathbb{N}$ AND $\hat{\ell}\pm\hat{j}\notin\mathbb{N}$ & \Gape[9pt]{Running} \\
		\hline
		$2\hat{\ell}\notin\mathbb{N}$ AND $\hat{\ell}\pm\hat{j}\notin\mathbb{N}$ & \Gape[10pt]{Non-vanishing and Non-running} \\
		\hline
	\end{tabular}
	\caption[Behavior of static Love numbers for the third class of $p$-form perturbations of the higher-dimensional Schwarzschild black hole.]{Behavior of static Love numbers for the third class of $p$-form perturbations of the higher-dimensional Schwarzschild black hole, for which $\hat{j}=\frac{j}{d-3}$ is neither an integer nor a half-integer. This class contains the static magnetic susceptibilities ($j=1$) in $d\ge6$. For generic $\hat{\ell}=\frac{\ell}{d-3}$, the static Love numbers for $p$-form perturbations in this class are non-zero and non-running, while they exhibit a classical RG flow for $2\hat{\ell}\in\mathbb{N}$. They are now zero along the two branches of non-integer $\hat{\ell}$ cases $\hat{\ell}\pm\hat{j}\in\mathbb{N}$ or $\hat{\ell}\pm\hat{j}\in\mathbb{N}$. As we will see later, the static magnetic-type tidal Love numbers also behave as prescribed here, with the resonant conditions for vanishing Love numbers mimicking those for the static magnetic susceptibilities for which ($j=1$).}
	\label{tbl:Staticp3LNs_SchwarzschildDd}
\end{table}

\subsection{Tidal Love numbers}
For the gravitational (spin-$2$) response of the Schwarzschild black hole, most of the analysis turns out to be exactly the same as the $p$-form perturbation analysis above for particular values of $p$. More specifically, after performing the field redefinitions
\be
	\Phi^{\left(\text{T}\right)}_{\ell,\mathbf{m}} = \frac{\Psi^{\left(\text{T}\right)}_{\ell,\mathbf{m}}}{r^{\frac{d-2}{2}}}\,,\quad \Phi^{\left(\text{RW}\right)}_{\ell,\mathbf{m}} = \frac{\Psi^{\left(\text{RW}\right)}_{\ell,\mathbf{m}}}{r^{\frac{d-2}{2}}} \,,
\ee
the equation of motion for the tensor modes becomes identical to the equation of motion for the scalar field perturbations, Eq.~\eqref{eq:FullpEOMSchwarzschild} with $j=0$. The equation of motion for the spin-$2$ magnetic-type (Regge-Wheeler) modes also takes the form of the $p$-form perturbations equations of motion Eq.~\eqref{eq:FullpEOMSchwarzschild}, now corresponding to the value $j=d-2$. Interestingly, the magnetic-type and tensor-type gravitational perturbations of the higher-dimensional Schwarzschild black hole obey the same equations of motion as the $p$-form and $\left(p-1\right)$-form modes, respectively, for $p=d-2$. The corresponding static magnetic-type and static tensor-type Love number are therefore captured by the general expression in Eq.~\ref{eq:StaticpLNs_SchwarzschildDd} with $j=d-2$ and $j=0$ respectively.

Consequently, the tensor-type tidal Love numbers and the scalar Love numbers of the higher-dimensional Schwarzschild black hole are exactly the same and behave the same way as the Love number of the first class of $p$-form perturbations, see Table~\ref{tbl:Staticp1LNs_SchwarzschildDd}. Similarly, the magnetic-type Love numbers of the higher-dimensional Schwarzschild black hole behave the same way as the second class of $p$-form perturbations for $d=5$ and the same way as the third class of $p$-form perturbations for $d\ge6$, see Table~\ref{tbl:Staticp2LNs_SchwarzschildDd} and Table~\ref{tbl:Staticp3LNs_SchwarzschildDd} respectively.

Let us also note that, since the master variables entering the gravitational perturbations equations of motion are built at most from derivatives of the actual fields in terms of which the response problem is defined, the response coefficients in front of the decaying branches of these master variables are proportional to the actual response coefficients we are looking for, namely,
\be
	k^{\left(\text{T}\right)}_{\ell}\left(\omega\right) = k^{\mathcal{T}\left(2\right)}_{\ell}\left(\omega\right) \,,\quad k^{\left(\text{RW}\right)}_{\ell}\left(\omega\right) = -\frac{\ell+d-2}{\ell-1}k^{\mathcal{B}\left(2\right)}_{\ell}\left(\omega\right) \,.
\ee

Unfortunately, we have not been able to find a useful near-zone truncation of the Zerilli equation of motion. At least for static perturbations, the Zerilli equation has been shown in~\cite{Kodama:2003jz,Ishibashi:2003ap} to reduce to a hypergeometric differential equation after performing a particular Darboux transformation, see also~\cite{Hui:2020xxx,Glampedakis:2017rar}. Using this fact, the authors in~\cite{Hui:2020xxx} have been able to extract the corresponding static electric-type tidal Love numbers to be~\cite{Hui:2020xxx}
\be
	k^{\left(\text{Z}\right)}_{\ell} = \frac{\hat{\ell}}{\hat{\ell}+1}\frac{\Gamma^{2}(\hat{\ell})\Gamma^{2}(\hat{\ell}+2)}{2\pi\,\Gamma(2\hat{\ell}+1)\Gamma(2\hat{\ell}+2)}\tan\pi\hat{\ell} = \frac{\left(\ell+d-3\right)\left(\ell+d-2\right)}{\ell\left(\ell-1\right)}k_{\ell}^{\mathcal{E},\left(2\right)} \,
\ee
where in the second equality we have demonstrated how the static response coefficients of the Zerilli modes are related to the actual response coefficients associated with fields in terms of which we have defined the response problem in Chapter~\ref{ch:TLNsDefinition}.

\section{Scalar and tensor Love numbers of Reissner-Nordstr\"{o}m black holes}
\label{sec:LNs_RNDd}

Next, we consider the higher-dimensional electrically charged Reissner-Nordstr\"{o}m black hole, Eq.~\eqref{eq:RNDdGeometry}. To avoid dealing with coupled differential equations, we will focus to spin-$0$ scalar and spin-$2$ tensor perturbations. The equations of motion for the scalar field perturbations have the same form \eqref{eq:V0S}. As for the gravitational tensor modes, even though there are no tensor modes for the gauge field perturbations to couple to the gravitational tensor modes, one needs to supplemented with the contribution of the background electromagnetic field which comes from the Maxwell action,
\be\ba
	S^{\left(1\right)}_{\text{full}} &= \int d^{d}x\,\sqrt{-g^{\text{full}}}\left[-\frac{1}{4}F^{\text{full}}_{\mu\nu}F^{\text{full}\,\mu\nu}\right] \\
	&\supset \int d^{d}x\,\sqrt{-g}\,\left[2\pi GF_{\mu\nu}F^{\mu\nu}\left(h_{\mu\nu}h^{\mu\nu}-\frac{1}{2}h^2\right)\right] \\
	&\supset \sum_{\ell,\mathbf{m}}\int d^2x\,\sqrt{-g^{\left(2\right)}}\,r^{d-2}\left[ 2\pi G F_{\mu\nu}F^{\mu\nu}\left|h^{\left(\text{T}\right)}_{\ell,\mathbf{m}}\right|^2 \right] \,,
\ea\ee
as well as an additional contribution from the Einstein-Hilbert action due to a non-zero background energy-momentum tensor,
\be\ba
	S^{\left(\text{gr}\right)}_{\text{full}} &= \int d^{d}x\,\sqrt{-g^{\text{full}}}\left[\frac{1}{16\pi G}R^{\text{full}}\right] \\
	&\supset \int d^{d}x\,\sqrt{-g}\,\left[-\frac{R}{2}\left(h_{\mu\nu}h^{\mu\nu}-\frac{1}{2}h^2\right)\right] \\
	&\supset \sum_{\ell,\mathbf{m}}\int d^2x\,\sqrt{-g^{\left(2\right)}}\,r^{d-2}\left[ -2\pi G\frac{d-4}{d-2} F_{\mu\nu}F^{\mu\nu}\left|h^{\left(\text{T}\right)}_{\ell,\mathbf{m}}\right|^2 \right] \,.
\ea\ee
Taking these into account, the equations of motion for the spin-$0$ scalar and spin-$2$ tensor modes in the background of a Reissner-Nordstr\"{o}m black hole turn out to be exactly the same~\cite{Ishibashi:2011ws,Kodama:2003kk,Pereniguez:2021xcj}. We can then follow the same footsteps as for the higher-dimensional Schwarzschild black hole. We employ the near-zone splitting analogous to the $j=0$ Eq.~\eqref{eq:Vp_SchwarzschildD},
\be\label{eq:V0V1_RNDd}
	\begin{gathered}
		\mathbb{O}^{\left(0\right)}_{\text{full}} = \partial_{\rho}\,\Delta\,\partial_{\rho} + V_0 + \epsilon\,V_1 \,, \\
		V_0 = -\frac{\left(\rho_{+}-\rho_{-}\right)^2}{4\Delta}\beta^2\partial_{t}^2 \,,\quad V_1 = -\frac{r^2\rho^2-r_{+}^2\rho_{+}^2}{\left(d-3\right)^2\Delta}\partial_{t}^2 \,,
	\end{gathered}
\ee
where we have again introduced $\rho = r^{d-3}$ and $\beta = \frac{2r_{+}}{d-3}\frac{\rho_{+}}{\rho_{+}-\rho_{-}}$ is the inverse Hawking temperature of the $d$-dimensional Reissner-Nordstr\"{o}m black hole. The leading order near-zone radial solution that is ingoing at the future event horizon has the same form,
\be
	R^{\left(0\right)}_{\omega\ell,\mathbf{m}} = \bar{R}^{\left(0\right)\text{in}}_{\ell,\mathbf{m}}\left(\omega\right)\left(\frac{x}{1+x}\right)^{-i\beta\omega/2}{}_2F_1\left(\hat{\ell}+1,-\hat{\ell};1-i\beta\omega;-x\right) \,,
\ee
where now $x=\frac{\rho-\rho_{+}}{\rho_{+}-\rho_{-}}$, and the corresponding dissipative response coefficients and Love numbers are extracted to be
\be\label{eq:ResponseCoefficientsRND}
	\begin{gathered}
		k^{\left(0\right)}_{\ell}\left(\omega\right) = k^{\left(0\right)\text{Love}}_{\ell}\left(\omega\right) + ik^{\left(0\right)\text{diss}}_{\ell}\left(\omega\right) \,, \\\\
		\ba
			k^{\left(0\right)\text{diss}}_{\ell}\left(\omega\right) &= A_{\ell}\left(\omega\right)\sinh\pi\beta\omega \,, \\
			k^{\left(0\right)\text{Love}}_{\ell}\left(\omega\right) &= A_{\ell}\left(\omega\right)\tan\pi\hat{\ell}\cosh\pi\beta\omega \,,
		\ea \\\\
		A_{\ell}\left(\omega\right) = \frac{\Gamma^2(\hat{\ell}+1)\left|\Gamma(\hat{\ell}+1-i\beta\omega)\right|^2}{2\pi\Gamma(2\hat{\ell}+1)\Gamma(2\hat{\ell}+2)}\left(\frac{\rho_{+}-\rho_{-}}{\rho_{s}}\right)^{2\hat{\ell}+1} \,.
	\end{gathered}
\ee
Again, these vanish for integer $\hat{\ell}$, now even beyond the static limit but always at leading order in the near-zone expansion, while for other values of the orbital number they exhibit the expected behavior based on power counting arguments, i.e. for half-integer $\hat{\ell}$ they logarithmically run and for $2\hat{\ell}\notin\mathbb{N}$ they are non-zero and non-running.

\section{Love symmetries for spin-$s$ and $p$-form perturbations in higher spacetime dimensions}
\label{sec:LoveSymmetryDd}

Despite the more intricate structure of the black hole Love numbers in higher spacetime dimensions, Love symmetry turns out to still exist independently of the value of the rescaled orbital number $\hat{\ell}$. There are now two sets of Love symmetry generators, one for each sign $\sigma=+1$ or $\sigma=-1$ that characterizes the near-zone split in Eq.~\eqref{eq:Vp_SchwarzschildD}. The two Love symmetries are generated by
\be\label{eq:SL2R_SchwarzschildDd}
	\begin{gathered}
		L_0^{\left(\sigma,j\right)} = -\beta\,\partial_{t} - \sigma\hat{j} \,, \\
		L_{\pm1}^{\left(\sigma,j\right)} = e^{\pm t/\beta}\left[\mp\sqrt{\Delta}\,\partial_{\rho} + \partial_{\rho}\left(\sqrt{\Delta}\right)\beta\,\partial_{t} + \sigma\hat{j}\sqrt{\frac{\rho-\rho_{+}}{\rho-\rho_{-}}} \right] \,,
	\end{gathered}
\ee
and satisfy the $\SL$ algebra, while the corresponding Casimir is given by
\be
	\mathcal{C}_2^{\left(\sigma,j\right)} = \partial_{\rho}\,\Delta\,\partial_{\rho} - \frac{\left(\rho_{+}-\rho_{-}\right)^2}{4\Delta}\beta^2\partial_{t}^2 + \hat{j}\frac{\rho_{+}-\rho_{-}}{\rho-\rho_{-}}\left(\sigma\beta\,\partial_{t} + \hat{j}\right) \,,
\ee
which exactly matches the leading order near-zone radial operators for the various cases encountered up until now in this chapter, i.e. with Eq.~\eqref{eq:Vp_SchwarzschildD} for the spin-$0$, $p$-form and spin-$2$ perturbations of the Schwarzschild black hole and with Eq.~\eqref{eq:V0V1_RNDd} for the spin-$0$ scalar and spin-$2$ tensor mode perturbations of the Reissner-Nordstr\"{o}m black hole. To investigate the regularity of the generators \eqref{eq:SL2R_SchwarzschildDd} at the future or the past event horizon, we need to study their near-horizon behavior after transitioning to advanced ($+$) or retarded ($-$) null coordinates $\left(t_{\pm},r,\theta^{A}\right)$ respectively. Although there is no useful closed form for the null coordinates in generic spacetime dimensionality $d\ge4$, we can still study the near-horizon behavior thanks to \eqref{eq:AdvancedRetarded_SchwazrschildDd}-\eqref{eq:Tortoise_SchwazrschildDd}. Doing this, one then immediately sees that the generators \eqref{eq:SL2R_SchwarzschildDd} are indeed regular. This time, in contrast to the $d=4$ case in Eq.~\eqref{eq:SL2RsKerrNewman}, we see that the scalar pieces are also regular. This is of course related to the fact that we are not employing any singular tetrad choice; rather, the master variables $\Phi^{\left(j\right)}_{\ell,\mathbf{m}}$ are related to scalar, electromagnetic and gravitational perturbations through transformations that are manifestly regular at the horizon.

A separable solution $\Phi^{\left(j\right)}_{\omega\ell,\mathbf{m}}$ then furnishes a representation of the Love symmetry,
\be
	\mathcal{C}_2^{\left(\sigma,j\right)}\Phi^{\left(j\right)}_{\omega\ell,\mathbf{m}} = \hat{\ell}(\hat{\ell}+1)\Phi^{\left(j\right)}_{\omega\ell,\mathbf{m}} \,,\quad L_0^{\left(\sigma,j\right)}\Phi^{\left(j\right)}_{\omega\ell,\mathbf{m}} = \left(i\beta\omega-\sigma\hat{j}\right)\Phi^{\left(j\right)}_{\omega\ell,\mathbf{m}} \,.
\ee
The static solution, in particular has an $L_0$-eigenvalue
\be
	L_0^{\left(\sigma,j\right)}\Phi^{\left(j\right)}_{\omega=0,\ell,\mathbf{m}} = -\sigma\hat{j}\,\Phi^{\left(j\right)}_{\omega=0,\ell,\mathbf{m}} \,.
\ee
One important difference compared to the four-dimensional case is that the Casimir is now in general non-integer unless $\hat{\ell}\in\mathbb{N}$. Furthermore, the weight of the static solution is also not integer unless $j=0$ or $j=d-3$.

\subsection{Scalar/tensor perturbations of Reissner-Nordstr\"{o}m black holes}
\begin{figure}[t]
	\centering
	\begin{tikzpicture}
		\node at (0,0) (uml4) {$\upsilon_{-\hat{\ell},2\hat{\ell}}$};
		\node at (0,1) (uml3) {$\upsilon_{-\hat{\ell},\hat{\ell}}$};
		\node at (0,2) (uml2) {$\upsilon_{-\hat{\ell},2}$};
		\node at (0,3) (uml1) {$\upsilon_{-\hat{\ell},1}$};
		\node at (0,4) (uml0) {$\upsilon_{-\hat{\ell},0}$};
		
		\draw [snake=zigzag] (1,-0.1) -- (5,-0.1);
		\draw (1,0) -- (5,0);
		\draw (1,1) -- (5,1);
		\node at (3,1.5) (up) {$\vdots$};
		\node at (3,0.5) (um) {$\vdots$};
		\draw (1,2) -- (5,2);
		\draw (1,3) -- (5,3);
		\draw (1,4) -- (5,4);
		\draw [snake=zigzag] (1,4.1) -- (5,4.1);
		
		\draw[blue] [->] (2.5,2) -- node[left] {$L_{+1}$} (2.5,3);
		\draw[blue] [->] (2,3) -- node[left] {$L_{+1}$} (2,4);
		\draw[red] [<-] (4,3) -- node[right] {$L_{-1}$} (4,4);
		\draw[red] [<-] (3.5,2) -- node[right] {$L_{-1}$} (3.5,3);
	\end{tikzpicture}
	\caption[The finite-dimensional highest-weight representation of $\SL$ whose elements solve the leading order near-zone equations of motion for a massless scalar field in the $d$-dimensional Schwarzschild black hole background with integer rescaled multipolar index $\hat{\ell}=\frac{\ell}{d-3}$ and contains the regular static solution.]{The finite-dimensional highest-weight representation of $\SL$ whose elements solve the leading order near-zone equations of motion for a massless scalar field in the $d$-dimensional Schwarzschild black hole background with integer rescaled multipolar index $\hat{\ell}=\frac{\ell}{d-3}$ and contains the regular static solution.}
	\label{fig:HWSL2R0_SchwarzschildDd}
\end{figure}
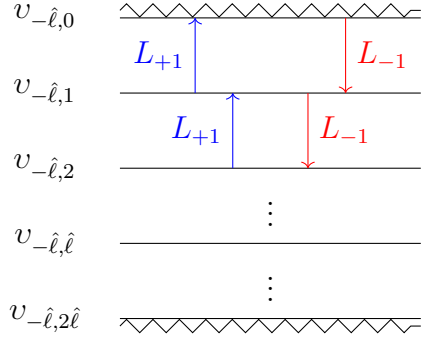

Let us begin with the cases where $j=0$. These capture the $s=0$ scalar and $s=2$ tensor mode perturbations of the Reissner-Nordstr\"{o}m black hole. The vector fields generating the Love $\SL$ symmetry simplify to
\be
	L_0 = -\beta\,\partial_{t} \,,\quad L_{\pm1} = e^{\pm t/\beta}\left[\mp\sqrt{\Delta}\,\partial_{\rho} + \partial_{\rho}\left(\sqrt{\Delta}\right)\beta\,\partial_{t} \right] \,,
\ee
which have the exact same form as the $d=4$ versions \eqref{eq:SL2RSchwarzschild4d}, but with $\rho=r^{d-3}$ appearing in place of $r$~\cite{Bertini:2011ga}. Analogously to the $d=4$ examples, let us construct the highest-weight representation with weight $h=-\hat{\ell}$, starting from the primary state $\upsilon_{-\hat{\ell},0}$, satisfying
\be
	L_{+1}\upsilon_{-\hat{\ell},0} = 0 \,,\quad L_0\upsilon_{-\hat{\ell},0} = -\hat{\ell}\,\upsilon_{-\hat{\ell},0} \quad\Rightarrow\quad \upsilon_{-\hat{\ell},0} = \left(-e^{+t/\beta}\sqrt{\Delta}\right)^{\hat{\ell}} \,,
\ee
where the spherical symmetry of the background geometry has allowed us to focus on axisymmetric ($\mathbf{m}=\mathbf{0}$) perturbations without loss of generality. This state is always regular at the future event horizon, while it is regular at the past event horizon as long as $e^{+t/\beta}\sqrt{\Delta}\sim e^{t_{-}/\beta} \left(r-r_{+}\right)$ is not raised to any negative power. The descendants,
\be
	\upsilon_{-\hat{\ell},n} = \left(L_{-1}\right)^{n} \upsilon_{-\hat{\ell},0} \,,
\ee
are also always regular at the future event horizon and have $L_0$-eigenvalues
\be
	L_0\upsilon_{-\hat{\ell},n} = (n-\hat{\ell})\,\upsilon_{-\hat{\ell},n} \,.
\ee
We see, therefore, the qualitative new feature in $d>4$ that the static solution does not in general belong to a highest-weight representation of the Love $\SL$ symmetry. In particular, the static scalar/tensor mode solution $\Phi_{\omega=0,\ell,\mathbf{m}}$ that is regular at the horizon is an element of the above highest-weight representation \text{if and only if}
\be
	\hat{\ell} \in \mathbb{N} \,,
\ee
which indeed captures the resonant conditions for which the static scalar, and tensor-type tidal, Love numbers of the Reissner-Nordstr\"{o}m black hole vanish. In these cases one can run an argument completely identical to the four-dimensional case, which proves that the static solution is a finite polynomial in $\rho$ (see Figure~\ref{fig:HWSL2R0_SchwarzschildDd}),
\be
	\text{If $\hat{\ell}\in\mathbb{N}$ }\Rightarrow \Phi^{\left(0\right)}_{\omega=0,\ell,\mathbf{m}=\mathbf{0}} \propto \upsilon_{-\hat{\ell},\hat{\ell}} = \sum_{n=0}^{\hat{\ell}}c_{n}\rho^{n} = c_{\hat{\ell}}r^{\ell} + \dots + c_0 \,.
\ee
The absence of terms $\propto r^{-\left(\ell+d-3\right)}$ implies that the corresponding static Love numbers are zero, which reproduces the result based on the explicit regular solution of the static Klein-Gordon equation according to which the series in the r.h.s. of the above equation corresponds to the Legendre polynomial of degree $\hat{\ell}$~\cite{Kol:2011vg}.

Let us also briefly comment on the role of the $\SL$ Love symmetry regarding the running of Love numbers. In $d=4$, for either Schwarzschild or Kerr-Newman black holes, the static Love numbers vanish at all scales. In $d>4$ on the other hand, the static scalar/tensor Love numbers for the Reissner-Nordstr\"{o}m black hole have a more intricate structure~\cite{Kol:2011vg,Hui:2020xxx},
\be
	k_{\ell}^{\left(0\right)} =
	\begin{cases}
		\alpha_{\ell} & \text{for $2\hat{\ell}\notin\mathbb{N}$} \\
		\alpha_{\ell} - \beta_{\ell}\log \frac{r-r_{+}}{L} & \text{for $\hat{\ell}\in\mathbb{N}+\frac{1}{2}$} \\
		0 & \text{for $\hat{\ell}\in\mathbb{N}$}
	\end{cases} \,,
\ee
where $\alpha_{\ell}$ are the renormalized static scalar Love numbers at length scale $L$ and $\beta_{\ell}$ are the associated $\beta$-functions, which turn out to be non-vanishing only in the case of half-integer $\hat{\ell}$. As we have shown here, Love symmetry explains the vanishing of the Love numbers whenever $\hat{\ell}$ is an integer. The absence of running in these cases stems from the fact that the regular and singular static solutions belong to different, locally distinguishable, representations; these are the same as for the four-dimensional Schwarzschild black hole case displayed in Figure~\ref{fig:HWSL2RSchwarzschild} and Figure~\ref{fig:SingSL2RSchwarzschild}. In the nomenclature of Refs.~\cite{Miller1968,Miller1970}, Figure~\ref{fig:HWSL2RSchwarzschild} and Figure~\ref{fig:SingSL2RSchwarzschild} are the representations $D(\,2\hat{\ell}\,)$ and $D^{+-}(\,2\hat{\ell}\,)$ respectively, while, in the notation of Ref.~\cite{Howe1992}, these are the type ``$[\circ]$'' and ``$\circ]\circ[\circ$'' representations $U(-\hat{\ell},-\hat{\ell}\,)$ and $U(\hat{\ell}+1,\hat{\ell}+1)$ respectively.

In all other cases with $\hat{\ell}\notin\mathbb{N}$, however, regular and singular static solutions belong to the same standard $\SL$ representations $D(\hat{\ell},0)$ (\cite{Miller1968,Miller1970}) or $W(4\hat{\ell}(\hat{\ell}+1),0)$ (\cite{Howe1992}) and the Love symmetry does not offer any local criteria from which to infer the absence of running. While this is consistent with the case of half-integer $\hat{\ell}$, it fails to capture the vanishing RG flow for the generic case $2\hat{\ell}\notin\mathbb{N}$. In other words, the Love symmetry $\SL$ representation theory implies a necessary, but not sufficient, condition that an RG flow is expected whenever $\hat{\ell}$ is not an integer. In addition, one also needs the external input of power counting arguments (see Section~\ref{sec:PowerCounting}) which independently implies that a necessary condition for running Love numbers is that $2\hat{\ell}\in\mathbb{N}$. These two necessary conditions together reduce to the prediction of a non-vanishing RG flow only for half-integer $\hat{\ell}$ which is indeed what is found by the explicit computations in this chapter~\cite{Kol:2011vg,Hui:2020xxx}. This is somewhat analogous to the appearance of logarithmic running in conformal perturbation theory, which takes place only if certain resonant conditions are satisfied, c.f.~\cite{Zamolodchikov:1989hfa,Konechny:2003yy}. Curiously, the appearance of logarithms in the degenerate hypergeometric function case is also known as~``resonance'', see e.g.~\cite{Opdam:2001wrq}.

Last, the peculiar result of vanishing scalar/tensor Love numbers for any frequency at leading order in the near-zone approximation whenever $\hat{\ell}\in\mathbb{N}$ can be addressed algebraically via the finite-frequency extension of the Love symmetry generators we presented in Section~\ref{sec:SL2R_Starobinsky}, with the according adjustments regarding the conditions for belonging to a highest-weight representation we saw here.

\subsection{First class of $p$-form perturbations of Schwarzschild black holes}
We now consider the case of the first class of $p$-form perturbations of the Schwarzschild black hole, for which $j=d-3$ and $\rho_{-}=0$, $\rho_{+}=\rho_{s}$. The Love symmetries generators \eqref{eq:SL2R_SchwarzschildDd} then read
\be\label{eq:SL2Rp1_SchwarzschildDd}
	\begin{gathered}
		L_0^{\left(\sigma,j=d-3\right)} = -\beta\,\partial_{t} - \sigma \,, \\
		L_{\pm1}^{\left(\sigma,j=d-3\right)} = e^{\pm t/\beta}\left[\mp\sqrt{\Delta}\,\partial_{\rho} + \partial_{\rho}\left(\sqrt{\Delta}\right)\beta\,\partial_{t} + \sigma\sqrt{\frac{\rho-\rho_{s}}{\rho}} \right] \,.
	\end{gathered}
\ee
The fact that the $L_0$-eigenvalues only get shifted by integer amounts allows to carry the previous analysis in exactly the same way. The primary state of the highest-weight representation with weight $h=-\hat{\ell}$ is given by
\be
	\begin{gathered}
		L_{+1}^{\left(\sigma,j=d-3\right)}\upsilon_{-\hat{\ell},0}^{\left(\sigma,j=d-3\right)} = 0 \,,\quad L_0^{\left(\sigma,j=d-3\right)}\upsilon_{-\hat{\ell},0}^{\left(\sigma,j=d-3\right)} = -\hat{\ell}\,\upsilon_{-\hat{\ell},0}^{\left(\sigma,j=d-3\right)} \,, \\
		\Rightarrow \quad \upsilon_{-\hat{\ell},0}^{\left(\sigma,j=d-3\right)} = \rho^{\sigma}\left(-e^{+t/\beta}\sqrt{\Delta}\right)^{\hat{\ell}-\sigma} \,.
	\end{gathered}
\ee
and, along with its descendants,
\be
	\upsilon_{-\hat{\ell},n}^{\left(\sigma,j=d-3\right)} = \left(L_{-1}^{\left(\sigma,j=d-3\right)}\right)^{n} \upsilon_{-\hat{\ell},0}^{\left(\sigma,j=d-3\right)} \,,
\ee
they furnish a representation of the Love $\SL$ symmetry spanned by states that are regular at the future event horizon. For the regular static solution to belong to this representation, we must therefore have
\be
	\hat{\ell}-\sigma \in\mathbb{N} \Leftrightarrow \hat{\ell}\in\mathbb{N} \,,
\ee
where we used the fact that $\ell\ge1$ for $p\ge1$~\cite{Yoshida:2019tvk,Camporesi1994}. These are again the exact resonant conditions for which the static electric-type Love numbers of the Schwarzschild black hole vanish. The regular static solution is the $(\hat{\ell}-\sigma)$'th descendant and the highest-weight property
\be
	\left(L_{+1}^{\left(\sigma,j=d-3\right)}\right)^{\hat{\ell}-\sigma+1}\Phi_{\omega=0,\ell,\mathbf{m}}^{\left(\sigma,j=d-3\right)} = 0 \quad\text{if $\hat{\ell}\in\mathbb{N}$} \,,
\ee
immediately implies the following polynomial form
\be
	\text{If $\hat{\ell}\in\mathbb{N}$ }\Rightarrow \Phi_{\omega=0,\ell,\mathbf{m}=\mathbf{0}}^{\left(\sigma,j=d-3\right)} \propto \upsilon_{-\hat{\ell},\hat{\ell}-\sigma}^{\left(\sigma,j=d-3\right)} = \rho^{\sigma}\sum_{n=0}^{\hat{\ell}-\sigma}c_{n}\rho^{n} = c_{\hat{\ell}-\sigma}r^{\ell} + \dots + c_0 r^{\sigma\left(d-3\right)} \,,
\ee
with no relevant response modes present and, hence, vanishing static responses.

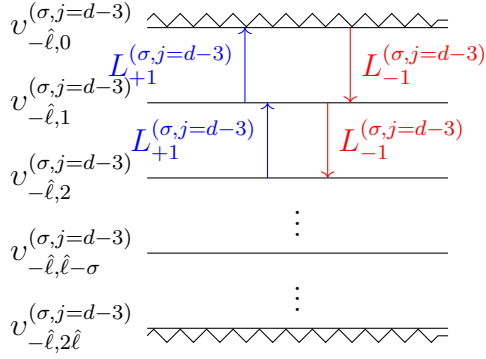
\begin{figure}[t]
	\centering
	\begin{tikzpicture}
		\node at (0,0) (uml4) {$\upsilon_{-\hat{\ell},2\hat{\ell}}^{\left(\sigma,j=d-3\right)}$};
		\node at (0,1) (uml3) {$\upsilon_{-\hat{\ell},\hat{\ell}-\sigma}^{\left(\sigma,j=d-3\right)}$};
		\node at (0,2) (uml2) {$\upsilon_{-\hat{\ell},2}^{\left(\sigma,j=d-3\right)}$};
		\node at (0,3) (uml1) {$\upsilon_{-\hat{\ell},1}^{\left(\sigma,j=d-3\right)}$};
		\node at (0,4) (uml0) {$\upsilon_{-\hat{\ell},0}^{\left(\sigma,j=d-3\right)}$};
		
		\draw [snake=zigzag] (1,-0.1) -- (5,-0.1);
		\draw (1,0) -- (5,0);
		\draw (1,1) -- (5,1);
		\node at (3,1.5) (up) {$\vdots$};
		\node at (3,0.5) (um) {$\vdots$};
		\draw (1,2) -- (5,2);
		\draw (1,3) -- (5,3);
		\draw (1,4) -- (5,4);
		\draw [snake=zigzag] (1,4.1) -- (5,4.1);
		
		\draw[blue] [->] (2.6,2) -- node[left] {$L_{+1}^{\left(\sigma,j=d-3\right)}$} (2.6,3);
		\draw[blue] [->] (2.3,3) -- node[left] {$L_{+1}^{\left(\sigma,j=d-3\right)}$} (2.3,4);
		\draw[red] [<-] (3.7,3) -- node[right] {$L_{-1}^{\left(\sigma,j=d-3\right)}$} (3.7,4);
		\draw[red] [<-] (3.4,2) -- node[right] {$L_{-1}^{\left(\sigma,j=d-3\right)}$} (3.4,3);
	\end{tikzpicture}
	\caption[The finite-dimensional highest-weight representation of $\SL$ whose elements solve the leading order near-zone equations of motion for a first class $p$-form perturbation of the $d$-dimensional Schwarzschild black hole with integer rescaled multipolar index $\hat{\ell}=\frac{\ell}{d-3}$ and contains the regular static solution.]{The finite-dimensional highest-weight representation of $\SL$ whose elements solve the leading order near-zone equations of motion for a first class $p$-form perturbation of the $d$-dimensional Schwarzschild black hole with integer rescaled multipolar index $\hat{\ell}=\frac{\ell}{d-3}$ and contains the regular static solution.}
	\label{fig:HWSL2Rp1_SchwarzschildDd}
\end{figure}

As before, studying the lowest-weight representation with weight $\bar{h}=+\hat{\ell}$ that is spanned by ascendants,
\be
	\bar{\upsilon}_{+\hat{\ell},n}^{\left(\sigma,j=d-3\right)}= \left(-L_{+1}^{\left(\sigma,j=d-3\right)}\right)^{n}\bar{\upsilon}_{+\hat{\ell},0}^{\left(\sigma,j=d-3\right)} \,, \\
\ee
of the lowest-weight vector
\be
	\begin{gathered}
		L_{-1}^{\left(\sigma,j=d-3\right)}\bar{\upsilon}_{+\hat{\ell},0}^{\left(\sigma,j=d-3\right)} = 0 \,,\quad L_0^{\left(\sigma,j=d-3\right)}\bar{\upsilon}_{+\hat{\ell},0}^{\left(\sigma,j=d-3\right)} = +\hat{\ell}\,\bar{\upsilon}_{+\hat{\ell},0}^{\left(\sigma,j=d-3\right)} \,, \\
		\Rightarrow \quad \bar{\upsilon}_{+\hat{\ell},0}^{\left(\sigma,j=d-3\right)} = \rho^{-\sigma}\left(+e^{-t/\beta}\sqrt{\Delta}\right)^{\hat{\ell}+\sigma} \,,
	\end{gathered}
\ee
reveals the static regular solution with vanishing static Love numbers is also the $(\hat{\ell}+\sigma)$'th ascendant and, therefore, this representation is in fact the finite $(2\hat{\ell}+1)$-dimensional type-``$[\circ]$'' representation of the Love $\SL$ symmetry (see Figure~\ref{fig:HWSL2Rp1_SchwarzschildDd}), while the singular static solution belongs to the locally distinguishable type-``$\circ]\circ[\circ$'' representation of Figure~\ref{fig:SingSL2RSchwarzschild}.

\subsection{Second class of $p$-form perturbations of Schwarzschild black holes}
Next, for the case of the second class of $p$-form perturbations of the Schwarzschild black hole, for which $j=\frac{d-3}{2}$ and which only emerges in odd spacetime dimensionalities, the Love symmetries generators \eqref{eq:SL2R_SchwarzschildDd} become
\be\label{eq:SL2Rp2_SchwarzschildDd}
	\begin{gathered}
		L_0^{\left(\sigma,j=d-3\right)} = -\beta\,\partial_{t} - \frac{\sigma}{2} \,, \\
		L_{\pm1}^{\left(\sigma,j=d-3\right)} = e^{\pm t/\beta}\left[\mp\sqrt{\Delta}\,\partial_{\rho} + \partial_{\rho}\left(\sqrt{\Delta}\right)\beta\,\partial_{t} + \frac{\sigma}{2}\sqrt{\frac{\rho-\rho_{s}}{\rho}} \right] \,.
	\end{gathered}
\ee
The $L_0$-eigenvalues now only get shifted by half-integer amounts. The previous analysis can then be applied in an exactly analogous manner to reveal that the static solution regular at the future event horizon now belongs to a highest-weight representation \textit{if and only if} $\hat{\ell}$ is half-integer which captures all the resonant condition of vanishing static Love numbers for this class of perturbations, a result inferred by the polynomial form of the solution implied by the highest-weight property. Similarly to the first class of $p$-form perturbations, this particular highest-weight representation is in fact the finite $(2\hat{\ell}+1)$-dimensional type-``$[\circ]$'' representation of the Love $\SL$ symmetry obtained from the one in Figure~\ref{fig:HWSL2Rp1_SchwarzschildDd} after replacing $\sigma\rightarrow\frac{\sigma}{2}$, while the singular static solution belongs to the locally distinguishable type-``$\circ]\circ[\circ$'' representation of the corresponding form shown in Figure~\ref{fig:SingSL2RSchwarzschild}.

\subsection{Third class of $p$-form perturbations of Schwarzschild black holes}
Last, for the third class of $p$-form perturbations of the Schwarzschild black hole, for which $2\hat{j}\notin\mathbb{N}$, things are a bit more interesting. Explicitly, the Love symmetries generators \eqref{eq:SL2R_SchwarzschildDd} are given by
\be
	\begin{gathered}
		L_0^{\left(\sigma,j\right)} = -\beta\,\partial_{t} - \sigma\hat{j} \,, \\
		L_{\pm1}^{\left(\sigma,j\right)} = e^{\pm t/\beta}\left[\mp\sqrt{\Delta}\,\partial_{\rho} + \partial_{\rho}\left(\sqrt{\Delta}\right)\beta\,\partial_{t} + \sigma\hat{j}\sqrt{\frac{\rho-\rho_{s}}{\rho}} \right] \,,
	\end{gathered}
\ee
and we see that the $L_0$-eigenvalues now get shifted by non-integer amounts. The highest-weight representation with weight $h=-\hat{\ell}$ is spanned by descendants,
\be
	\upsilon_{-\hat{\ell},n}^{\left(\sigma,j\right)} = \left(L_{-1}^{\left(\sigma,j\right)}\right)^{n} \upsilon_{-\hat{\ell},0}^{\left(\sigma,j\right)} \,,
\ee
of the primary state $\upsilon_{-\hat{\ell},0}^{\left(\sigma,j\right)}$ relevant for the third class $p$-form perturbations, satisfying
\be
	\begin{gathered}
		L_{+1}^{\left(\sigma,j\right)}\upsilon_{-\hat{\ell},0}^{\left(\sigma,j\right)} = 0 \,,\quad L_0^{\left(\sigma,j\right)}\upsilon_{-\hat{\ell},0}^{\left(\sigma,j\right)} = -\hat{\ell}\,\upsilon_{-\hat{\ell},0}^{\left(\sigma,j\right)} \,, \\
		\Rightarrow \quad \upsilon_{-\hat{\ell},0}^{\left(\sigma,j\right)} = \rho^{\sigma\hat{j}}\left(-e^{+t/\beta}\sqrt{\Delta}\right)^{\hat{\ell}-\sigma\hat{j}} \,.
	\end{gathered}
\ee
These states are always regular at the future event horizon and their $L_0$-eigenvalues are given by
\be
	L_0^{\left(\sigma,j\right)}\upsilon_{-\hat{\ell},n}^{\left(\sigma,j\right)} = (n-\hat{\ell})\,\upsilon_{-\hat{\ell},n}^{\left(\sigma,j\right)} \,.
\ee
As before, we encounter the new feature in $d>4$ that the static solution does not in general belong to a highest-weight representation of the Love $\SL$ symmetries. For this to happen, there are some resonant conditions that need to be satisfied. In particular, the static solution $\Phi^{\left(j\right)}_{\omega=0,\ell,\mathbf{m}}$ that is regular at the horizon is an element of the above highest-weight representation \text{if and only if}
\be
	\hat{\ell}-\sigma\hat{j} \in \mathbb{N} \,.
\ee
In $d>4$, this only covers one branch of the resonant conditions for which the static Love numbers of this class of $p$-form perturbations vanish (see Table~\ref{tbl:Staticp3LNs_SchwarzschildDd}). Nevertheless, the second branch of these resonant conditions is captured by the second Love symmetry, corresponding to the opposite sign $\sigma$. However, we will see momentarily that the second branch also arises from the lowest-weight representation.

Similar to the $d=4$ cases, the highest-weight property implies a polynomial form. In particular, from the fact that, for arbitrary purely radial functions $F\left(\rho\right)$,
\be
	\left(L_{+1}^{\left(\sigma,j\right)}\right)^{n}\left[\left(\rho-\rho_{-}\right)^{\sigma\hat{j}}F\left(\rho\right)\right] = \left(-e^{+t/\beta}\sqrt{\Delta}\right)^{n}\left(\rho-\rho_{-}\right)^{\sigma\hat{j}}\frac{d^{n}}{d\rho^{n}}F\left(\rho\right) \,,
\ee
the annihilation condition $(L_{+1}^{\left(\sigma,j\right)})^{\hat{\ell}-\sigma\hat{j}+1}\Phi^{\left(j\right)}_{\omega=0,\ell,\mathbf{m}}\bigg|_{\hat{\ell}-\sigma\hat{j}\in\mathbb{N}} = 0$ implies
\be
	\text{If $\hat{\ell}-\sigma\hat{j}\in\mathbb{N}$ }\Rightarrow \Phi_{\omega=0,\ell,\mathbf{m}=\mathbf{0}}^{\left(\sigma,j\right)} \propto \upsilon_{-\hat{\ell},\hat{\ell}-\sigma\hat{j}}^{\left(\sigma,j\right)} = \rho^{\sigma\hat{j}}\sum_{n=0}^{\hat{\ell}-\sigma\hat{j}}c_{n}\rho^{n} = c_{\hat{\ell}-\sigma\hat{j}}r^{\ell} + \dots + c_0 r^{\sigma j} \,,
\ee
which indeed has the appropriate polynomial form from which to infer the vanishing of the static Love numbers by the absence of a response mode.

As for the lowest-weight representation of the Love $\SL$ symmetry for sign $\sigma$ with weight $\bar{h} = +\hat{\ell}$, the lowest-weight vector $\bar{\upsilon}_{+\hat{\ell},0}^{\left(\sigma,j\right)}$ is found to be
\be
	\begin{gathered}
		L_{-1}^{\left(\sigma,j\right)}\bar{\upsilon}_{+\hat{\ell},0}^{\left(\sigma,j\right)} = 0 \,,\quad L_0^{\left(\sigma,j\right)}\bar{\upsilon}_{+\hat{\ell},0}^{\left(\sigma,j\right)} = +\hat{\ell}\,\bar{\upsilon}_{+\hat{\ell},0}^{\left(\sigma,j\right)} \,, \\
		\Rightarrow \quad \bar{\upsilon}_{+\hat{\ell},0}^{\left(\sigma,j\right)} = \rho^{-\sigma\hat{j}}\left(+e^{-t/\beta}\sqrt{\Delta}\right)^{\hat{\ell}+\sigma\hat{j}} \,,
	\end{gathered}
\ee
and is always regular at the past event horizon, while it is regular at the future event horizon as long as $e^{-t/\beta}\sqrt{\Delta}\sim e^{-t_{+}/\beta} \left(r-r_{+}\right)$ is not raised to any negative power. Its ascendants,
\be
	\bar{\upsilon}_{+\hat{\ell},n}^{\left(\sigma,j\right)} = \left(-L_{+1}^{\left(\sigma,j\right)}\right)^{n} \bar{\upsilon}_{+\hat{\ell},0}^{\left(\sigma,j\right)} \,,
\ee
share the same boundary conditions and their charge under $L_0$ is
\be
	L_0^{\left(\sigma,j\right)}\bar{\upsilon}_{+\hat{\ell},n}^{\left(\sigma,j\right)} = (\hat{\ell}-n)\bar{\upsilon}_{+\hat{\ell},n}^{\left(\sigma,j\right)} \,.
\ee
For the static solution regular at the horizon to belong to this representation, we must therefore have
\be
	\hat{\ell}+\sigma\hat{j}\in\mathbb{N} \,,
\ee
and the lowest-weight property implies an analogous polynomial form of the solution with no decaying mode.

\begin{figure}[t]
	\centering
	\begin{subfigure}[b]{0.49\textwidth}
		\centering
		\begin{tikzpicture}
			\node at (0,1) (uml3) {$\upsilon_{-\hat{\ell},\hat{\ell}-\sigma\hat{j}}$};
			\node at (0,2) (uml2) {$\upsilon_{-\hat{\ell},2}^{\left(\sigma,j\right)}$};
			\node at (0,3) (uml1) {$\upsilon_{-\hat{\ell},1}^{\left(\sigma,j\right)}$};
			\node at (0,4) (uml0) {$\upsilon_{-\hat{\ell},0}^{\left(\sigma,j\right)}$};
			
			\draw (1,1) -- (5,1);
			\node at (3,1.5) (up) {$\vdots$};
			\node at (3,0.5) (um) {$\vdots$};
			\draw (1,2) -- (5,2);
			\draw (1,3) -- (5,3);
			\draw (1,4) -- (5,4);
			\draw [snake=zigzag] (1,4.1) -- (5,4.1);
			
			\draw[red] [<-] (2.5,2) -- node[left] {$L_{-1}^{\left(\sigma,j\right)}$} (2.5,3);
			\draw[red] [<-] (2,3) -- node[left] {$L_{-1}^{\left(\sigma,j\right)}$} (2,4);
			\draw[blue] [->] (4,3) -- node[right] {$L_{+1}^{\left(\sigma,j\right)}$} (4,4);
			\draw[blue] [->] (3.5,2) -- node[right] {$L_{+1}^{\left(\sigma,j\right)}$} (3.5,3);
		\end{tikzpicture}
		\caption{The highest-weight $\SL$ representation that contains the regular static solution along the $\hat{\ell}-\sigma\hat{j}\in\mathbb{N}$ branch.}
	\end{subfigure}
	\hfill
	\begin{subfigure}[b]{0.49\textwidth}
		\centering
		\begin{tikzpicture}
			\node at (0,3) (upll) {$\bar{\upsilon}_{+\hat{\ell},\hat{\ell}+\sigma\hat{j}}$};
			\node at (0,2) (upl2) {$\bar{\upsilon}_{+\hat{\ell},2}^{\left(\sigma,j\right)}$};
			\node at (0,1) (upl1) {$\bar{\upsilon}_{+\hat{\ell},1}^{\left(\sigma,j\right)}$};
			\node at (0,0) (upl0) {$\bar{\upsilon}_{+\hat{\ell},0}^{\left(\sigma,j\right)}$};
			
			\draw [snake=zigzag] (1,-0.1) -- (5,-0.1);
			\draw (1,0) -- (5,0);
			\draw (1,1) -- (5,1);
			\draw (1,2) -- (5,2);
			\node at (3,2.5) (up) {$\vdots$};
			\draw (1,3) -- (5,3);
			\node at (3,3.5) (um) {$\vdots$};
			
			\draw[blue] [->] (2.5,0) -- node[left] {$L_{+1}^{\left(\sigma,j\right)}$} (2.5,1);
			\draw[blue] [->] (2,1) -- node[left] {$L_{+1}^{\left(\sigma,j\right)}$} (2,2);
			\draw[red] [<-] (4,1) -- node[right] {$L_{-1}^{\left(\sigma,j\right)}$} (4,2);
			\draw[red] [<-] (3.5,0) -- node[right] {$L_{-1}^{\left(\sigma,j\right)}$} (3.5,1);
		\end{tikzpicture}
		\caption{The lowest-weight $\SL$ representation that contains the regular static solution along the $\hat{\ell}+\sigma\hat{j}\in\mathbb{N}$ branch.}
	\end{subfigure}
	\caption[The infinite-dimensional highest-weight and lowest-weight representations of $\SL$ whose elements solve the leading order near-zone equations of motion for the third class $p$-form perturbations of the $d$-dimensional Schwarzschild black hole with rescaled orbital numbers satisfying $\hat{\ell}\pm\hat{j}\in\mathbb{N}$ and contain the regular static solution.]{The infinite-dimensional highest-weight and lowest-weight representations of $\SL$ whose elements solve the leading order near-zone equations of motion for the third class $p$-form perturbations of the $d$-dimensional Schwarzschild black hole with rescaled orbital numbers satisfying $\hat{\ell}\pm\hat{j}\in\mathbb{N}$ and contain the regular static solution.}
	\label{fig:HWSL2p3_SchwarzschildDd}
\end{figure}
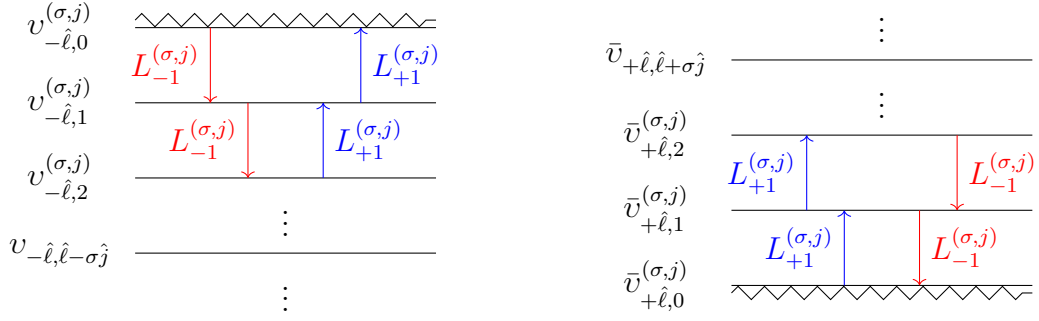

We now see an interesting new feature compared to the case of the first and second classes of $p$-form perturbations of the higher-dimensional Schwarzschild black hole. To begin with, for the first and second classes of $p$-form perturbations, for which $2\hat{j}$ is an integer, the highest-weight representation encountered before turned out to in fact be the finite $(2\hat{\ell}+1)$-dimensional type-``$[\circ]$'' representation whenever the static solution was one of its elements. As for the third class $p$-form perturbations, for which $2\hat{j}$ is not an integer in $d>4$, the lowest-weight representation captures the second branch of resonant conditions associated with vanishing static Love numbers, namely, the branch with $\hat{\ell}+\sigma\hat{j}\in\mathbb{N}$, see Figure~\ref{fig:HWSL2p3_SchwarzschildDd}. The highest-weight and lowest-weight representations associated with the current third class of $p$-form perturbations are non-overlapping and become infinite-dimensional Verma modules. In contrast to the Verma modules encountered in the four-dimensional Kerr-Newman Love multiplets, the regular at the horizon static solution is capable of belonging to either of the two, highest-weight or lowest-weight, modules, depending on which branch of the resonant conditions, $\hat{\ell}-\hat{j}\in\mathbb{N}$ or $\hat{\ell}+\hat{j}\in\mathbb{N}$ respectively, is encountered. The corresponding singular static solutions, however, still belong to the locally distinguishable type-``$\circ]\circ[\circ$'' representation.

\section{Beyond general-relativistic black holes}
\label{sec:LoveSymmetryModGR}

As a last investigation, we will perform a study similar to the Riemann-cubed paradigm in four spacetime dimensions and compute the static scalar Love numbers for some higher-derivative theories of gravity. We will focus to the $\alpha^{\prime}$-corrected gravitational actions of string theory and extract the leading order static scalar susceptibilities for the simplest case of the corresponding modified Schwarzschild black holes. These consist of the Callan-Myers-Perry black hole of bosonic/heterotic string theory~\cite{Callan:1988hs,Myers:1998gt} and the type-II superstring theory $\alpha^{\prime3}$-corrections to the Schwarzschild black hole~\cite{Myers:1987qx}.

The full radial equation of motion for the static scalar field spherical harmonics modes $\Phi_{\ell,\mathbf{m}}\left(r\right)$ reads
\be
	\left[f_{r}\partial_{x}^2 + \frac{x^{2\left(d-4\right)}}{2f_{t}}\partial_{x}\left(\frac{f_{t}f_{r}}{x^{2\left(d-4\right)}}\right)\partial_{x}\right]\Phi_{\ell,\mathbf{m}} = \frac{\ell\left(\ell+d-3\right)}{x^2}\Phi_{\ell,\mathbf{m}} \,,
\ee
where we have introduced the variable
\be
	x = \frac{r_{\text{h}}}{r} \,,
\ee
with $r_{\text{h}}$ the radial location of the event horizon at all values of $\alpha^{\prime}$. We will treat this equation perturbatively around $\alpha^{\prime}=0$, with the order parameter being denoted by $\lambda$ and which is proportional to the appropriate power of $\alpha^{\prime}$ for each situation we will examine. The scalar field is expanded as
\be
	\Phi_{\ell,\mathbf{m}}\left(x\right) = r_{\text{h}}^{\ell}\bar{\mathcal{E}}_{\ell,\mathbf{m}}\left[\Phi_{\ell}^{\left(0\right)}\left(x\right) + \lambda\,\Phi^{\left(1\right)}_{\ell} + \mathcal{O}\left(\lambda^2\right) \right] \,,
\ee
with the zeroth order solution regular at the horizon $x=1$ given by the general-relativistic static scalar field profile,
\be
	\Phi_{\ell}^{\left(0\right)}\left(x\right) = \frac{\Gamma^2(\hat{\ell}+1)}{\Gamma(2\hat{\ell}+1)}\,{}_2F_1\left(\hat{\ell}+1,-\hat{\ell};1;1-\frac{1}{x^{d-3}}\right) \,,
\ee
and with the higher-order terms chosen to grow at infinity slower than the leading order solution,
\be
	\lim_{x\rightarrow0}x^{\ell}\Phi_{\ell}^{\left(n\right)} = 0 \quad \text{for $n>0$} \,.
\ee

\subsection{Bosonic/Heterotic string theory Callan-Myers-Perry black hole}
The Callan-Myers-Perry black hole describes the leading stringy corrections to the Schwarzschild black hole in heterotic/bosonic string theory~\cite{Callan:1988hs,Myers:1998gt}, see also~\cite{Moura:2006pz,Moura:2011rr}. The gravitational action is $\alpha^{\prime}$-corrected by a Riemann-squared term,
\be\label{eq:BosonicHeteroticStringAction}
	\begin{gathered}
		S_{\text{gr}} = \frac{1}{16\pi G}\int d^{d}x\,\sqrt{-g}\left[R - \frac{4}{d-2}\left(\partial\phi\right)^2 + \lambda\,e^{-4\phi/\left(d-2\right)}Y(\tilde{R})\right] \,, \\
		Y\left(R\right) = \frac{1}{2}R_{\mu\nu\rho\sigma}R^{\mu\nu\rho\sigma} \,,\quad \tilde{R}_{\mu\nu}^{\quad\rho\sigma} = R_{\mu\nu}^{\quad\rho\sigma} - \delta_{[\mu}^{[\rho}\nabla_{\nu]}\nabla^{\sigma]}\phi \,,
	\end{gathered}
\ee
where we also included the dilaton term and we are working in the Einstein-frame. The string coupling parameter above is equal to $\lambda=\frac{\alpha^{\prime}}{2}$ for the bosonic and $\lambda=\frac{\alpha^{\prime}}{4}$ for the heterotic string theory. The $d$-dimensional Callan-Myers-Perry black hole geometry has a constant dilaton and is given by~\cite{Callan:1988hs,Myers:1998gt}
\be
	\begin{gathered}
		ds^2 = -f\left(r\right)dt^2 + \frac{dr^2}{f\left(r\right)} + r^2d\Omega_{d-2}^2 \,, \\
		f\left(r\left(x\right)\right) = \left(1-x^{d-3}\right)\left[ 1 - 	\frac{\left(d-3\right)\left(d-4\right)}{2}\frac{\lambda}{r_{\text{h}}^2}\,x^{d-3}\frac{1-x^{d-1}}{1-x^{d-3}} \right] + \mathcal{O}\left(\lambda^2\right) \,.
	\end{gathered}
\ee
The event horizon $r_{\text{h}}$ is related to the ADM mass $M$, as encoded in the Schwarzschild radius $r_{s}$, of the black hole according to
\be
	r_{\text{h}} = r_{s}\left(1+\frac{d-4}{2}\frac{\lambda}{r_{s}^2}\right) + \mathcal{O}\left(\lambda^2\right) \,.
\ee
The black hole solution built perturbatively in $\alpha^{\prime}$ is valid only in regions where $r^2\gg\alpha^{\prime}$. For our purposes, it is sufficient to require that the gravitational radius of the black hole is much bigger than the string length, $r_{s}^2\gg\alpha^{\prime}$.

Let us now look at some specific examples for the $\alpha^{\prime}$-corrected static scalar field profile and the corresponding corrections to the coefficients $\varkappa_{\ell}^{\left(1\right)}$ that appear in front the decaying terms that go like $x^{\ell+d-3} \sim r^{-\left(\ell+d-3\right)}$ for the Callan-Myers-Perry black hole. For $d=5$ and $\ell=2$ or $\ell=4$, we find
\be
	\begin{gathered}
		\ba
			\Phi^{\left(1\right)}_{\ell=2} &= -\frac{7}{2} + x^2 + 8\ln x + 2\left(1-\frac{2}{x^2}\right)\left(\text{Li}_2\left(1-x^2\right)-\frac{\pi^2}{6}\right) \,, \\
			\Phi^{\left(1\right)}_{\ell=4} &= -\frac{35}{x^2} + \frac{80}{3} + x^2 - 36\left(1-\frac{2}{x^2}\right)\ln x \\
			&\quad- 6\left(1-\frac{6}{x^2}+\frac{6}{x^4}\right)\left(\text{Li}_2\left(1-x^2\right)-\frac{\pi^2}{6}\right) \,,
		\ea \\\\
		\varkappa^{\left(1\right)}_{\ell=2} = -\lambda\left(\frac{1}{18} + \frac{2}{3}\ln x\right) \,,\quad \varkappa^{\left(1\right)}_{\ell=4} = -\lambda\left(\frac{43}{3600} + \frac{2}{5}\ln x\right) \,.
	\end{gathered}
\ee
We see that these cases give rise to logarithmically running Love numbers, the value of the constant in front of the logarithms being identified with the $\beta$-function. An example of non-running static scalar Love numbers is the $d=6$, $\ell=3$ case
\be\ba
	\Phi^{\left(1\right)}_{\ell=3} &= \frac{9}{x^2} - \frac{45}{x} + \frac{63}{2} \\
	&\quad- 15\left(1-\frac{2}{x^3}\right)\left(\ln\left(1-x^3\right) + \frac{3}{2}x^2{}_2F_1\left(1,\frac{2}{3};\frac{5}{3};x^3\right)\right) \,, \\
	\varkappa^{\left(1\right)}_{\ell=3} &= -\lambda\frac{5}{2} \,.
\ea\ee
The running/non-running is in fact in accordance with power counting arguments. Indeed, following the arguments used in Section~\ref{sec:PowerCounting}, one expects to find a non-vanishing RG flow if
\be
	1 \le 2\hat{\ell}+1-\frac{2}{d-3} \in \mathbb{N} \,,
\ee
otherwise, the natural expectation is some non-zero and non-running scalar Love numbers. This indeed in accordance with our above results, i.e. for $d=5$ and $\ell=2$ or $\ell=4$ the above condition is satisfied, while, for $d=6$ and $\ell=3$, it is not. It appears, therefore, that the $\alpha^{\prime}$-corrected Riemann-squared action for the bosonic/heterotic string theory does not exhibit any seemingly fine-tuning behavior with respect to the black hole response problem.

\subsection{Type-II superstring theory black holes}
Next, we consider the $\alpha^{\prime3}$-corrected black hole in type-II superstring theory. The type-II superstring theory effective action arising from tree-level amplitude for four-graviton scattering, up to and including terms at eighth order in the graviton and dilaton momenta, is quartic in the Riemann tensor and in the Einstein frame is given by~\cite{Myers:1987qx} (see also~\cite{Chen:2021qrz})
\be\label{eq:TypeIISuperstringAction}
	\begin{gathered}
		S = \frac{1}{16\pi G}\int d^{d}x\,\sqrt{-g}\left[R - \frac{4}{d-2}\left(\partial\phi\right)^2 + \lambda\,e^{-12\phi/\left(d-2\right)}Y(\tilde{R})\right] \,, \\
		Y\left(R\right) = 2R_{\mu\nu\rho\sigma}R_{\kappa\quad\lambda}^{\,\,\,\,\,\nu\rho}R^{\mu\alpha\beta\kappa}R^{\lambda\quad\sigma}_{\,\,\,\,\alpha\beta} + R_{\mu\nu\rho\sigma}R_{\kappa\lambda}^{\quad\rho\sigma}R^{\mu\alpha\beta\kappa}R^{\lambda\quad\nu}_{\,\,\,\,\alpha\beta} \,,
	\end{gathered}
\ee
where $\lambda = \frac{1}{16}\zeta\left(3\right)\alpha^{\prime3}$ is the string coupling parameter and $\tilde{R}_{\mu\nu}^{\quad\rho\sigma}$ is given in \eqref{eq:BosonicHeteroticStringAction}.

The asymptotically flat and electrically neutral black hole solution of this theory now has a non-constant dilaton and its geometry in the Einstein frame reads~\cite{Myers:1987qx,Chen:2021qrz}
\be
	\begin{gathered}
		ds^2 = -f_{t}\left(r\right)dt^2 + \frac{dr^2}{f_{r}\left(r\right)} + r^2d\Omega_{d-2}^2 \,, \\
		f_{t}\left(r\left(x\right)\right) = \left(1-x^{d-3}\right)\left[1+2\frac{\lambda}{r_{\text{h}}^6}\mu\left(x\right)\right] \,, \quad f_{r}\left(r\left(x\right)\right) = \left(1-x^{d-3}\right)\left[1-2\frac{\lambda}{r_{\text{h}}^6}\varepsilon\left(x\right)\right] \,, \\
		\mu\left(x\right) = -\varepsilon\left(x\right) - C_{d}\,x^{3\left(d-1\right)} \,,\quad \varepsilon\left(x\right) = D_{d}\,x^{3\left(d-1\right)} + E_{d}\,x^{d-3}\frac{1-x^{2d}}{1-x^{d-3}} \,,
	\end{gathered}
\ee
where the constants $C_{d}$, $D_{d}$ and $E_{d}$ are given by
\be\ba
	C_{d} &= \frac{2}{3}\left(d-1\right)\left(d-3\right)\left(2\,d^3-10\,d^2+6\,d+15\right) \,, \\
	D_{d} &= -\frac{1}{24}\left(d-3\right)\left(52\,d^4-375\,d^3+758\,d^2-117\,d-570\right) \,, \\
	E_{d} &= \frac{1}{24}\left(d-3\right)\left(20\,d^4-225\,d^3+946\,d^2-1779\,d+1290\right) \,.
\ea\ee

The power counting arguments of Section~\ref{sec:PowerCounting} now imply that one expects logarithmically running scalar Love numbers whenever
\be
	3\le2\hat{\ell}+1-\frac{6}{d-3}\in\mathbb{N}
\ee
at $\alpha^{\prime3}$ order. This means that logarithms appear first at orbital number $\ell=d$. To avoid cumbersome expressions at very high multipolar orders, we focus here to $d=4$. Then, one expects to find non-running and non-vanishing static Love numbers for $\ell=2,3$ but, for $\ell\ge4$, one should be faced with RG-flowing static responses. Indeed, for $\ell=2,3$, we find no logs,
\be
	\begin{gathered}
		\ba
			\Phi^{\left(1\right)}_{\ell=2} &= -\frac{5}{2x} + \frac{5}{6} - \frac{1619}{4200}x^3 - \frac{1619}{2800}x^4 - \frac{1619}{2450}x^5 - \frac{1619}{2352}x^6 - \frac{3153}{784}x^7 + \frac{71}{32}x^8 \,, \\
			\Phi^{\left(1\right)}_{\ell=3} &= -\frac{15}{4x^2} + \frac{3}{x} - \frac{3}{8} - \frac{6693}{19600}x^4 - \frac{6693}{9800}x^5 - \frac{4533}{784}x^6 + \frac{12861}{1960}x^7 - \frac{213}{160}x^8 \,, \\
		\ea \\\\
		\varkappa^{\left(1\right)}_{\ell=2} = -\lambda\frac{1619}{4200} \,,\quad \varkappa^{\left(1\right)}_{\ell=3} = -\lambda\frac{6693}{19600} \,,
	\end{gathered}
\ee
while, for $\ell=4$,
\be
	\begin{gathered}
		\ba
			{}&\Phi^{\left(1\right)}_{\ell=4} = \frac{39195}{x^3} - \frac{480155}{7 x^2} + \frac{2214665}{63 x} -\frac{306241}{63} + 28 x - \frac{14}{9} x^2 + \frac{4}{9}x^3 - \frac{1}{2}x^4 \\
			&- \frac{102407}{12348}x^5 + \frac{327629}{24696} x^6 - \frac{7901}{1372}x^7 + \frac{71}{112}x^8 + \frac{1400}{3x^2}\left(1-\frac{2}{x}\right)\left(5-\frac{42}{x}+\frac{42}{x^2}\right)\ln x \\
			&-560\left(1-\frac{20}{x}+\frac{90}{x^2}-\frac{140}{x^3}+\frac{70}{x^4}\right)\left(\text{Li}_2\left(x\right)+\ln\left(1-x\right)\ln x\right) \,,
		\ea \\\\
		\varkappa^{\left(1\right)}_{\ell=4} = -\frac{3947599}{555660}+\frac{8}{9}\ln x \,,
	\end{gathered}
\ee
as expected. Consequently, similar to the Callan-Myers-Perry black hole, the $\alpha^{\prime3}$-corrected Schwarzschild black hole of type-II superstring theory does not seem to exhibit any fine-tuned scalar Love numbers.

\subsection{A sufficient geometric constraint for the existence of near-zone symmetries}
As we just saw, neither the Callan-Myers-Perry black hole of bosonic/heterotic string theory nor the $\alpha^{\prime3}$-corrected Schwarzschild black hole of type-II superstring theory demonstrate any superficially unnatural black hole scalar Love numbers. This is expected to be accompanied with the absence of a Love symmetry structure.

Having these results as explicit counterexamples, we will attempt now to extract sufficient geometric conditions for the existence of Love symmetry beyond general-relativistic black hole configurations by studying a massless scalar field in the background of a generalized spherically symmetric black hole geometry \eqref{eq:SphericallySymemtricBHDd} with $f_{t}\left(r\right)\ne f_{r}\left(r\right)$. After the field redefinition $\Phi_{\ell,\mathbf{m}}=\Psi_{\ell,\mathbf{m}}^{\left(0\right)}/r^{\frac{d-2}{2}}$, the full massless Klein-Gordon equation derived in Section~\ref{sec:EOM_Seq0} becomes
\be
	\begin{gathered}
		\mathbb{O}_{\text{full}}^{\left(0\right)}\Phi_{\ell,\mathbf{m}} = \hat{\ell}(\hat{\ell}+1)\Phi_{\ell,\mathbf{m}} \,, \\
		\mathbb{O}_{\text{full}}^{\left(0\right)} = \partial_{\rho}\,\Delta_{r}\,\partial_{\rho} + \frac{\Delta_{r}^2}{2\Delta_{t}}\left(\frac{\Delta_{t}}{\Delta_{r}}\right)^{\prime}\partial_{\rho} - \frac{r^2\rho^2}{\left(d-3\right)^2\Delta_{t}}\,\partial_{t}^2 \,,
	\end{gathered}
\ee
where $\rho=r^{d-3}$ and $\hat{\ell}=\ell/\left(d-3\right)$ as before, $\Delta_{t}\equiv\rho^2f_{t}$, $\Delta_{r}\equiv\rho^2f_{r}$ and primes denote derivatives with respect to $\rho$.

Similar to the corresponding analysis in four dimensions presented in Section~\ref{sec:LoveBeyondGR4d}, the following near-zone approximation turns out to be the only possible candidate for enjoying a globally defined $\SL$ symmetry,
\be\label{eq:NZModGRDd}
	\mathbb{O}_{\text{NZ}}^{\left(0\right)} = \partial_{\rho}\,\Delta_{r}\,\partial_{\rho} + \frac{\Delta_{r}^2}{2\Delta_{t}}\left(\frac{\Delta_{t}}{\Delta_{r}}\right)^{\prime}\partial_{\rho} - \frac{r_{\text{h}}^{2\left(d-2\right)}}{\left(d-3\right)^2\Delta_{t}}\,\partial_{t}^2 \,.
\ee
The associated vector fields generating the near-zone $\SL$ algebra,
\be\label{eq:SL2RModGRDd}
	\begin{gathered}
		L_0 = -\beta\,\partial_{t} \,,\quad
		L_{\pm1} = e^{\pm t/\beta}\left[\mp\sqrt{\Delta_{r}}\,\partial_{\rho} + \sqrt{\frac{\Delta_{r}}{\Delta_{t}}}\partial_{\rho}\left(\sqrt{\Delta_{t}}\right)\beta\,\partial_{t}\right] \,,
	\end{gathered}
\ee
with $\beta$ the inverse Hawking temperature \eqref{eq:betaDd}, are regular at both the future and past event horizons and give rise to a Casimir operator that matches this near-zone truncation of the Klein-Gordon operator if and only if\footnote{This is the solution to a particular differential equation that is outputted by the requirement that the Love symmetry vector fields satisfy the $\SL$ algebra and that their Casimir produces a consistent near-zone truncation of the massless Klein-Gordon operator. This equation is actually the same as the four-dimensional analogue \eqref{eq:SL2RModGRConstraint} after replacing $r$ with $\rho$.}
\be\label{eq:SL2RModGRConstraintDd}
	\Delta_{r}\left(\rho\right) = \Delta_{t}\left(\rho\right)\frac{4\Delta_{t}\left(\rho\right)+\left(\frac{\beta_{s}}{\beta}\rho_{\text{h}}\right)^2}{\Delta_{t}^{\prime2}\left(\rho\right)} \,,
\ee
where we have defined the inverse Hawking temperature for the Schwarzschild black hole $\beta_{s}=\frac{2r_{\text{h}}}{d-3}$, or, at the level of the functions $f_{t}\left(r\right)$ and $f_{r}\left(r\right)$ themselves,
\be
	f_{r}\left(r\right) = f_{t}\left(r\right) \frac{\left(d-3\right)^2r^{2\left(d-4\right)}}{\left(r^{2\left(d-3\right)}f_{t}\left(r\right)\right)^{\prime2}} \left[4r^{2\left(d-3\right)}f_{t}\left(r\right) + \left(\frac{\beta_{s}}{\beta}\right)^2r_{\text{h}}^{2\left(d-3\right)}\right] \,.
\ee
For the case of $f_{t}=f_{r}$, the above condition tells us that the general-relativistic Reissner-Nordstr\"{o}m geometry is the only acceptable black hole solution, which already rules out the Callan-Myers-Perry black hole solution. Furthermore, plugging in the explicit $\alpha^{\prime3}$-corrections to the Schwarzschild black hole in type-II superstring theory reveals that the above condition is again not satisfied, in accordance with the explicit computations of the static scalar Love numbers. Of course, this does not rule out all black hole solutions of string theory. An explicit counterexample is the STU black hole of supergravity~\cite{Cvetic:1996kv} which satisfies the above geometric constraint and Love symmetry has indeed been shown to exist for the more general rotating STU black hole configuration~\cite{Cvetic:2021vxa}. Furthermore, even though the geometric condition derived here sets a sufficient constraint on the existence of Love symmetry, this needs not be a necessary constraint as well. In particular, we have only examined the case of a massless scalar field minimally coupled to pure gravity which may very well not be a good representative of the modified theory of gravity under study.

One can also check that the above near-zone $\SL$ implies the vanishing of static Love numbers when $\hat{\ell}\in\mathbb{N}$. Using the same symmetry argument of the regular static solution being an element of a highest-weight representation of this $\SL$, we obtain $\left(L_{+1}\right)^{\hat{\ell}+1}\Phi_{\omega=0,\ell,\mathbf{m}} = 0$ if and only if $\hat{\ell}$ is an integer. From the fact that
\be
	\left(L_{+1}\right)^{n}F\left(\rho\right) = \left(-e^{t/\beta}\sqrt{\Delta_{t}}\right)^{n}\left[\sqrt{\frac{\Delta_{r}}{\Delta_{t}}}\frac{d}{d\rho}\right]^{n}F\left(\rho\right)
\ee
we see that the corresponding static solution is a polynomial but this time in the variable $\tilde \rho$, defined as
\be\label{eq:tilderho}
	d\tilde{\rho} \equiv \sqrt{\frac{\Delta_{t}}{\Delta_{r}}} \,d\rho \Rightarrow \tilde{\rho} = \sqrt{\Delta_{t}+\left(\frac{\beta_{s}}{2\beta}\rho_{\text{h}}\right)^2} +\tilde{\rho}_{\text{h}} - \frac{\beta_{s}}{2\beta}\rho_{\text{h}} \,,
\ee
where $\tilde{\rho}_{\text{h}}$ is an integration constant indicating the location of the event horizon in this new radial coordinate,
\be
	\text{If $\hat{\ell}\in\mathbb{N}$: }\Rightarrow \Phi_{\omega=0,\ell,\mathbf{m}}\left(r\right) = \sum_{n=0}^{\hat{\ell}}c_{n}^{\left(\mathbf{m}\right)}\tilde{\rho}^{n}\left(r\right)
\ee
Asymptotically, $\tilde{\rho}\rightarrow\rho$ due to the asymptotic flatness of $f_{t}$. Expanding this polynomial in $\tilde{\rho}$ at large distance in the initial radial variable $\rho$, one would observe the appearance of an $\rho^{-\hat{\ell}-1}=r^{-\ell-d+3}$ term. However, this term is a relativistic correction in the profile of the ``source'' part of the solution, rather than a response effect from induced multipole moments. Indeed, if the geometric condition \eqref{eq:SL2RModGRConstraintDd} for the existence of a near-zone $\SL$ symmetry is satisfied, we arrive at a situation practically identical to the case of $s=0$ scalar mode perturbations of the Reissner-Nordstr\"{o}m black hole, Eq.~\eqref{eq:V0V1_RNDd}, when working with the variable $\tilde{\rho}$. More explicitly, the full radial Klein-Gordon operator reads,
\be
	\mathbb{O}_{\text{full}}^{\left(0\right)} = \partial_{\tilde{\rho}}\,\Delta_{t}\,\partial_{\tilde{\rho}} - \frac{r^{2\left(d-2\right)}}{\left(d-3\right)^2\Delta_{t}}\,\partial_{t}^2 \,,
\ee
and $\Delta_{t}$ is a quadratic polynomial in $\tilde{\rho}$,
\be
	\Delta_{t} = \left(\tilde{\rho}-\tilde{\rho}_{+}\right)\left(\tilde{\rho}-\tilde{\rho}_{-}\right) \,,
\ee
where we have denoted the locations of the outer (event), and inner (Cauchy) horizons as
\be
	\tilde{\rho}_{\pm} = \tilde{\rho}_{\text{h}} - \frac{1\mp1}{2}\frac{\beta_{s}}{\beta}\rho_{\text{h}} \,.
\ee
Matching onto the worldline EFT is equivalent to solving the equations motion after analytically continuing the orbital number to perform the source/response split of the scalar field, and only in the end sending $\ell$ to take its physical integer values. Doing this, we see that the ``response'' part of the static scalar field is singular at the horizon when $\hat{\ell}\in\mathbb{N}$ and is therefore absent, while the ``source'' part becomes a polynomial of degree $\hat{\ell}$ in $\tilde{\rho}$. Consequently, the corresponding static Love numbers vanish identically and we see again how a polynomial form of the solution is indicative of this vanishing. For generic $\hat{\ell}$, the procedure just described gives the following static scalar Love numbers,
\be
	k_{\ell}^{\left(0\right)} = \frac{\Gamma^4(\hat{\ell}+1)}{2\pi\,\Gamma(2\hat{\ell}+1)\Gamma(2\hat{\ell}+2)}\tan\pi\hat{\ell}\,\left(\frac{\beta_{s}}{\beta}\frac{\rho_{\text{h}}}{\rho_{s}}\right)^{2\hat{\ell}+1} \,,
\ee
where the last factor in the parenthesis is just $\left(\tilde{\rho}_{+}-\tilde{\rho}_{-}\right)/\rho_{s}$. These are exactly the same as the static scalar Love numbers for the higher-dimensional Reissner-Nordstr\"{o}m black hole obtained in Section~\ref{sec:LNs_RNDd}.

One can also apply the same analysis for the $p$-form perturbations equations of motion \eqref{eq:Vpj}. In fact, the equations of motion in the background of a generic electrically neutral black hole geometry can be collectively written as
\be
	\begin{gathered}
		\mathbb{O}_{\text{full}}^{\left(j\right)}\Phi^{\left(j\right)}_{\ell,\mathbf{m}} = \hat{\ell}(\hat{\ell}+1)\Phi^{\left(j\right)}_{\ell,\mathbf{m}} \,, \\
		\mathbb{O}_{\text{full}}^{\left(j\right)} = \partial_{\rho}\,\Delta_{r}\,\partial_{\rho} + \frac{\Delta_{r}^2}{2\Delta_{t}}\left(\frac{\Delta_{t}}{\Delta_{r}}\right)^{\prime}\partial_{\rho} - \frac{r^2\rho^2}{\left(d-3\right)^2\Delta_{t}}\,\partial_{t}^2 + U^{\left(j\right)}\left(\rho\right) \,, \\
	\end{gathered}
\ee
with the reduced potential given by
\be
	U^{\left(j\right)}\left(\rho\right) = \frac{\hat{j}}{d-3}rD_{a}r^{a} - \hat{j}(1-\hat{j})\left(1-r_{a}r^{a}\right) \,.
\ee
Introducing the coordinate $\tilde{\rho}$ in the same way as before, i.e. $d\tilde{\rho} = \sqrt{\frac{\Delta_{t}}{\Delta_{r}}}\,d\rho$, the radial operator is brought to the suggestive form
\be
	\mathbb{O}_{\text{full}}^{\left(j\right)} = \partial_{\tilde{\rho}}\,\Delta_{t}\,\partial_{\tilde{\rho}} - \frac{r^2\rho^2}{\left(d-3\right)^2\Delta_{t}}\,\partial_{t}^2 + U^{\left(j\right)}\left(\rho\right)
\ee
regardless of what the geometry is.

Motivated by the results for the Schwarzschild black hole in Section~\ref{sec:LoveSymmetryDd}, we expect that a non-zero spin will not affect the vector part of the candidate near-zone symmetry generators. In other words, the previous geometric condition is expected to still be an outcome of this analysis. Assuming this is indeed the case, one can go ahead and see whether there is any additional geometric constraint that arises for $p\ne0$. It turns out that there is one additional geometric constraint which will completely fix the geometry. To see this, it is instructive to first express everything using the independent variable $\tilde{\rho}$, namely, rewrite
\be
	f_{r} = \Delta_{t}\left(\frac{1}{\rho}\frac{d\rho}{d\tilde{\rho}}\right)^2 \,.
\ee
Introducing the variable
\be
	u\left(\tilde{\rho}\right) = \left[\rho\left(\tilde{\rho}\right)\right]^{-\hat{j}} \,,
\ee
the reduced potential function then takes the form
\be
	U^{\left(j\right)} = -\left[\frac{1}{u}\frac{d}{d\tilde{\rho}}\left(\Delta_{t}\frac{du}{d\tilde{\rho}}\right) + \hat{j}(1-\hat{j})\right] \,.
\ee
At this point we have made on assumption on the explicit form of $\Delta_{t}$. Assuming that the vector part of the candidate Love symmetry generators is the same as for the scalar response problem, i.e. that $\Delta_{t}$ is a quadratic polynomial in $\tilde{\rho}$, the conditions that these generators form an $\SL$ algebra whose quadratic Casimir operator produces a consistent near-zone truncation of the equations of motion then primarily imply that
\be
	\begin{gathered}
		L_0 = -\beta\,\partial_{t} - \gamma \,, \\
		L_{\pm1} = e^{\pm t/\beta}\left[\mp\sqrt{\Delta_{t}}\,\partial_{\tilde{\rho}} + \partial_{\tilde{\rho}}\left(\sqrt{\Delta_{t}}\right)\beta\,\partial_{t} + \gamma\sqrt{\frac{\tilde{\rho}-\tilde{\rho}_{+}}{\tilde{\rho}-\tilde{\rho}_{-}}}\right] \,, \\
		U^{\left(j\right)}_{\SL} = \frac{\tilde{\rho}_{+}-\tilde{\rho}_{-}}{\tilde{\rho}-\tilde{\rho}_{-}}\gamma^2 \,,
	\end{gathered}
\ee
for some constant $\gamma$. Matching the two potentials then gives a differential equation for $u\left(\tilde{\rho}\right)$,
\be
	\left[\frac{d}{d\tilde{\rho}}\Delta_{t}\frac{d}{d\tilde{\rho}} +\frac{\tilde{\rho}_{+}-\tilde{\rho}_{-}}{\tilde{\rho}-\tilde{\rho}_{-}}\gamma^2\right] u = -\hat{j}(1-\hat{j})\,u \,.
\ee
This can be analytically solved in terms of Euler's hypergeometric functions. In fact, after introducing the variable $x=\frac{\tilde{\rho}-\tilde{\rho}_{+}}{\tilde{\rho}_{+}-\tilde{\rho}_{-}}$, this differential equation is exactly the same as the static problem for $p$-form perturbations of the higher-dimensional Schwarzschild black hole after the replacements $\hat{j}\rightarrow\gamma$ and $\hat{\ell}\rightarrow-\hat{j}$ in Eq.~\eqref{eq:NZRadialScharzschildDd}. The general solution is, therefore,
\be\ba
	u\left(\tilde{\rho}\right) &= c_1\left(\frac{\tilde{\rho}-\tilde{\rho}_{-}}{\tilde{\rho}_{+}-\tilde{\rho}_{-}}\right)^{+\gamma}{}_2F_1\left(1-\hat{j}+\gamma,\hat{j}+\gamma;1;-\frac{\tilde{\rho}-\tilde{\rho}_{+}}{\tilde{\rho}_{+}-\tilde{\rho}_{-}}\right) \\
	&+c_2\left(\frac{\tilde{\rho}-\tilde{\rho}_{-}}{\tilde{\rho}_{+}-\tilde{\rho}_{-}}\right)^{-\gamma}{}_2F_1\left(1-\hat{j}-\gamma,\hat{j}-\gamma;1;-\frac{\tilde{\rho}-\tilde{\rho}_{+}}{\tilde{\rho}_{+}-\tilde{\rho}_{-}}\right) \,.
\ea\ee
Expanding around large distances,
\be\ba
	u\left(\tilde{\rho}\right) \xrightarrow{\tilde{\rho}\rightarrow\infty} \left(c_1+c_2\right)&\bigg[\frac{\Gamma(2\hat{j}-1)}{\Gamma(\hat{j}+\gamma)\Gamma(\hat{j}-\gamma)}\left(\frac{\tilde{\rho}-\tilde{\rho}_{+}}{\tilde{\rho}_{+}-\tilde{\rho}_{-}}\right)^{\hat{j}-1} \\
	&+ \frac{\Gamma(1-2\hat{j})}{\Gamma(1-\hat{j}+\gamma)\Gamma(1-\hat{j}-\gamma)}\left(\frac{\tilde{\rho}-\tilde{\rho}_{+}}{\tilde{\rho}_{+}-\tilde{\rho}_{-}}\right)^{-\hat{j}}\bigg] \,,
\ea\ee
we see that the asymptotic flatness condition, which implies $\tilde{\rho}\rightarrow\rho$, along with the fact that $u=\rho^{-\hat{j}}$, fixes the constant $\gamma$ to be
\be
	\gamma = \pm\hat{j} \,.
\ee
One then observes that the Love symmetry generators are exactly the same as for the general-relativistic Schwarzschild black hole in Eq.~\ref{eq:SL2R_SchwarzschildDd}. More explicitly, the relation between $\rho$ and $\tilde{\rho}$ is required to be
\be
	\rho = \tilde{\rho}-\tilde{\rho}_{-} \,,
\ee
which immediately implies
\be
	f_{r}=f_{t} = 1 - \frac{\rho_{\text{h}}}{\rho} \,.
\ee
In other words, the Schwarzschild black hole is the \textit{only} possible isolated asymptotically flat and electrically neutral black hole that exhibits the Love symmetry beyond the scalar response problem, at least within the current assumptions. Based on this, it is tempting to conclude that the most general theory of gravity whose black hole response problem admits the Love symmetry beyond the scalar response problem has an action of the form
\be
	S^{\left(\text{gr}\right)} = \frac{1}{16\pi G}\int d^{d}x\sqrt{-g}\,R\,f\left(R_{\rho\sigma\mu\nu}\right)
\ee
for arbitrary\footnote{A minimal requirement here is that $f\left(R_{\rho\sigma\mu\nu}=0\right)$ is finite, i.e. that flat Minkowski spacetime is a solution of this theory and, hence, asymptotically flat solutions exist.} functions $f\left(R_{\rho\sigma\mu\nu}\right)$ of the Riemann tensor. This is the most general class of theories of gravity that admits Ricci-flat vacuum solutions, a special subclass of which is $f\left(R\right)$ gravity, but which does not include Lovelock gravity beyond General Relativity.

\section{Summary of Chapter~\ref{ch:LoveSymmetryDd}}
In this chapter, we have seen how the black hole Love numbers in higher spacetime dimensions have a much richer structure, by focusing to spherically symmetric static configurations. To begin with, we saw that, even for general-relativistic black holes, Love numbers are in general non-zero and exhibit no running in accordance with power counting arguments. However, there exist discrete sets of resonant conditions regarding the values of the rescaled orbital number $\hat{\ell}=\frac{\ell}{d-3}$ with respect to which the static Love numbers of general-relativistic black holes do vanish. The characteristic example is the $\hat{\ell}\in\mathbb{N}$ condition for electric-type responses of Schwarzschild black hole~\cite{Kol:2011vg,Hui:2020xxx}.

We have also showed that Love symmetry still manifests itself within the near-zone-approximated equations of motion for linearized perturbations of the Schwarzschild and Reissner-Nordstr\"{o}m black holes~\cite{Charalambous:2022rre,Charalambous:2024tdj}. Interestingly, Love symmetry exists regardless of the orbital number of the perturbation but is only for those resonant conditions for which the Love numbers vanish that the corresponding solution regular at the event horizon belongs to a highest-weight representation. We also saw the new behavior where the selection rule of vanishing magnetic-type Love numbers for one of the two branches of the corresponding resonant conditions is outputted as a consequence of a lowest-weight property, rather than a highest-weight property; this extends to the larger third class of $p$-form perturbations of the higher-dimensional Schwarzschild black hole for which $2\hat{j}\notin\mathbb{N}$. Unfortunately, we have not been able to find a useful near-zone truncation of the equations of motion for gravitoelectric perturbations of the Schwarzschild black hole due to the intricate structure of the higher-dimensional Zerilli equation. As already mentioned, we suspect that there exists a Darboux transformation that maps a near-zone truncation of the Zerilli equation onto an equation of the form of Eq.~\eqref{eq:Vp_SchwarzschildD}. We leave this task of completing the interpretation of vanishing static Love numbers for higher spin perturbations of the higher-dimensional Schwarzschild black hole, by supplementing with the case of gravitoelectric perturbations, for future work.

Furthermore, we have performed an analysis analogous to that of Section~\ref{sec:LoveBeyondGR4d} to extract a sufficient geometric condition for the existence of Love symmetry. Neither the Callan-Myers-Perry geometry of bosonic/heterotic string theory~\cite{Callan:1988hs,Myers:1998gt} nor the $\alpha^{\prime3}$-corrected Schwarzschild black hole of type-II superstring theory~\cite{Myers:1987qx} turned out to satisfy this condition which is in accordance with the explicit computations of the corresponding static scalar Love numbers we presented here. It would be particularly interesting to better study what kind of theories of gravity support black hole geometries that satisfy this geometric constraint. We also gave hints that the Love symmetry appears to be quite special to General Relativity in the sense that, ignoring possible field redefinitions involving derivatives of the master variables, Love symmetry exists for \textit{all} $p$-form perturbations only for general-relativistic black holes. This does not necessarily mean that Love symmetry does not exist in other theories of gravity but it does hint that either these modified theories gravity admit general-relativistic black hole solutions or that Love symmetry might arise only for specific type of perturbations or black hole solutions rather than being a generic property of the theory.

Even though we have not presented it here explicitly, the Love symmetry of the higher-dimensional Schwarzschild and Reissner-Nordstr\"{o}m black holes still acquires a geometric interpretation as generating an $\SL$ isometry subgroup of subtracted geometries, see Section~\ref{sec:SubtractedGeometries4d}. In fact, the extremal limit of the Love symmetry vector fields for the scalar field again recovers the correct Killing vectors generating the enhanced $\SL$ isometries of the $\text{AdS}_2$ near-horizon throat of the extremal Reissner-Nordstr\"{o}m black hole in exactly the same way as prescribed in Chapter~\ref{ch:NHE}.
	\newpage
\chapter{Scalar Love numbers and Love symmetries for the $5$-dimensional Myers-Perry black hole}
\label{ch:5dMP}

We have seen that Love symmetry exists for general-relativistic black holes in higher spacetime dimensions as well as for higher spin perturbations. In the previous chapter, we only focused to the simpler case of spherically symmetric black holes. However, there is no reason other than technicalities not to investigate higher-dimensional rotating black holes. A first observation is that, in $d=1+\left(d-1\right)$ spacetime dimensions, there are more than one planes of rotation; namely, there are $N=\left\lfloor \frac{d-1}{2}\right\rfloor$ orthogonal planes of rotation and, hence, such a black hole would be characterized by $N$ angular momenta. Higher-dimensional asymptotically flat and rotating black holes solving the vacuum Einstein field equations are described by the Myers-Perry geometry~\cite{Myers:1986un}. These posse a $U\left(1\right)^{N}$ isometry subgroup associated with azimuthal rotations around the $N$ planes of rotation, which is the $d$-dimensional generalization of the $U\left(1\right)$ axisymmetry group of the four-dimensional Kerr black hole geometry.

Similar to $d=4$, the Myers-Perry black hole is equipped with hidden symmetries arising from higher-rank Killing tensors which are associated with its complete integrability~\cite{Frolov:2006dqt,Kubiznak:2006kt,Page:2006ka,Krtous:2006qy,Krtous:2007xf,Frolov:2007cb,Frolov:2008jr}. In fact, the Myers-Perry black hole is the most general higher-dimensional asymptotically flat black hole geometry that solves the vacuum Einstein field equations and is equipped with a rank-$2$ conformal Killing-Yano tensor~\cite{Houri:2007uq,Houri:2007xz}. This integrability ensures the remarkable separability of equations of motion for scalar~\cite{Frolov:2006pe} as well as electromagnetic~\cite{Lunin:2017drx} and $p$-form~\cite{Lunin:2019pwz} perturbations. Another different type of enhanced symmetries for the Myers-Perry black hole is its $U\left(1\right)^{N}\rightarrow U\left(N\right)$ enhancement for equi-rotating configurations, where all angular momenta have the same magnitude, that occurs in odd spacetime dimensions~\cite{Myers:1986un}.

This chapter is based on our work in Ref.~\cite{Charalambous:2023jgq}, where we focus to the simplest non-trivial $d>4$ example of scalar responses of the doubly rotating $5$-dimensional Myers-Perry black hole. We will first explicitly present the geometry describing this general-relativistic black hole solution and then proceed to solve the massless Klein-Gordon equation in such a background. As instructed by the worldline EFT, we will extract the scalar Love numbers of the $5$-dimensional Myers-Perry black hole by working in a near-zone expansion. In particular, we will employ two different near-zone truncations and extract the corresponding scalar response coefficients. In contrast to the $5$-dimensional Schwarzschild black hole, we will find that the scalar Love numbers are in general non-zero and exhibit a classical RG flow, even in the static limit. However, if the black hole is equi-rotating or if the perturbation has azimuthal numbers along the two planes of rotation that are equal in magnitude, then the static Love numbers do indeed exhibit the same behavior as for the $5$-dimensional Schwarzschild black hole, i.e. they vanish whenever the rescaled orbital number $\hat{\ell}=\frac{\ell}{2}$ is integer ($\ell$ is even). This will be explained by the existence of two Love symmetries, one for each of the employed two near-zone truncations.

Similarly to the four-dimensional Kerr-Newman black holes, each of the globally defined near-zone $\SL$ symmetry acquires a geometric interpretation as an isometry subgroup of a subtracted geometry. Furthermore, each Love symmetry can be extended to a local near-zone $\SL\times\SL$ symmetry. In fact, these local $\SL\times\SL$ symmetries have precisely the structure needed to infer the non-running/vanishing of all the resonant conditions associated with the leading order near-zone scalar Love numbers, even the ones involving $\omega$-dependence or the unphysical ones accompanied by conical deficits.

We will see that the two globally defined Love symmetries can be realized as subalgebras of an infinite-dimensional $\SL\ltimes\hat{U}\left(1\right)_{\mathcal{V}}^2$ extension. This infinite extension also contains particular subalgebras whose extremal limit can be used to recover the enhanced isometries of the extremal $5$-dimensional Myers-Perry black hole.

\section{Geometry of $5$-dimensional Myers-Perry black hole}
\label{sec:5dMPGeometry}

The Myers-Perry geometry describes an electrically neutral, rotating black hole in higher-dimensional Ricci-flat spacetimes~\cite{Myers:1986un}. In the present chapter, we focus to the $5$-dimensional case with the geometry in Boyer-Lindquist coordinates given by
\be\ba\label{eq:Metric_5dMP}
	ds^2 = -dt^2 &+ \frac{r_{s}^2}{\Sigma}\left(dt-a\sin^2\theta\,d\phi - b\cos^2\theta\,d\psi\right)^2 + \frac{r^2\Sigma}{\Delta}\,dr^2 + \Sigma\,d\theta^2 \\
	&+ \left(r^2+a^2\right)\sin^2\theta\,d\phi^2 + \left(r^2+b^2\right)\cos^2\theta\,d\psi^2 \,,
\ea\ee
where
\be
	\begin{gathered}
		\Sigma = r^2 + a^2\cos^2\theta + b^2\sin^2\theta \,, \\
		\Delta = \left(r^2+a^2\right)\left(r^2+b^2\right)-r_{s}^2r^2 \,,
	\end{gathered}
\ee
and $\theta\in\left[0,\frac{\pi}{2}\right]$ is a direction cosine angle, while $\phi\in\left[0,2\pi\right)$ and $\psi\in\left[0,2\pi\right)$ are periodically identified azimuthal angles. This geometry describes an asymptotically flat black hole with mass $M$ and angular momenta $J_{\phi}$ and $J_{\psi}$ along the two orthogonal planes of rotation, related to the parameters $r_{s}$, $a$ and $b$ appearing in the line element above according to
\be
	M = \frac{3\pi}{8 G}r_{s}^2\,\,\,,\,\,\,J_{\phi} = \frac{2}{3}Ma\,\,\,,\,\,\,J_{\psi} = \frac{2}{3}Mb \,.
\ee

The horizons correspond to the roots of the discriminant $\Delta$ and are located at radial distances $r=r_{\pm}$, with
\be
	r_{\pm}^2 = \frac{1}{2}\left[r_{s}^2-a^2-b^2\pm\sqrt{\left(r_{s}^2-a^2-b^2\right)^2-4a^2b^2}\right] \,.
\ee
The absence of a naked singularity imposes the inequality
\be
	\left|a\right|+\left|b\right| \le r_{s} \,,
\ee
with its saturation indicating the extremality condition. The event horizon $r_{\text{h}}=r_{+}$ and the Cauchy horizon $r_{\text{C}}=r_{-}$ are Killing horizons relative to the Killing vectors
\be
	K^{\left(\pm\right)} = \partial_{t} + \Omega_{\phi}^{\left(\pm\right)}\,\partial_{\phi} + \Omega_{\psi}^{\left(\pm\right)}\,\partial_{\psi}
\ee
respectively, where $\Omega_{\phi}^{\left(\pm\right)} = \frac{a}{r_{\pm}^2+a^2}$ and $\Omega_{\psi}^{\left(\pm\right)} = \frac{b}{r_{\pm}^2+b^2}$. In particular, $\Omega_{\phi}^{\left(+\right)}\equiv\Omega_{\phi}$ and $\Omega_{\psi}^{\left(+\right)}\equiv\Omega_{\psi}$ are the black hole angular velocities as realized by a static observer in the exterior. The inverse metric components can be extracted to be
\be\ba
	g^{\mu\nu}\partial_{\mu}\partial_{\nu} = \frac{1}{\Sigma} &\bigg\{ -\Sigma\,\partial_{t}^2 - \frac{\left(r^2+a^2\right)\left(r^2+b^2\right)r_{s}^2}{\Delta}\left(\partial_{t} + \frac{a}{r^2+a^2}\,\partial_{\phi} + \frac{b}{r^2+b^2}\,\partial_{\psi}\right)^2 \\
	& + \left[\frac{1}{\sin^2\theta} - \frac{a^2-b^2}{r^2+a^2}\right]\partial_{\phi}^2 + \left[\frac{1}{\cos^2\theta} - \frac{b^2-a^2}{r^2+b^2}\right]\partial_{\psi}^2 + \frac{\Delta}{r^2}\,\partial_{r}^2 + \partial_{\theta}^2\bigg\} \,,
\ea\ee
while the only additional term needed to construct the massless Klein-Gordon operator involves the Christoffel symbols,
\be
	g^{\mu\nu}\Gamma_{\mu\nu}^{\sigma}\partial_{\sigma} = -\frac{1}{\Sigma}\bigg\{\frac{1}{r}\frac{d}{dr}\left(\frac{\Delta}{r}\right)\,\partial_{r} + \frac{1}{\sin\theta\cos\theta}\frac{d}{d\theta}\left(\sin\theta\cos\theta\right)\,\partial_{\theta}\bigg\} \,.
\ee

To investigate regularity at the future or the past event horizon, one employs advanced ($+$) or retarded ($-$) coordinates respectively, related to the Boyer-Lindquist coordinates according to
\be\ba
	dt_{\pm} &= dt \pm \frac{\left(r^2+a^2\right)\left(r^2+b^2\right)}{\Delta}dr \,, \\
	d\varphi_{\pm} &= d\phi \pm a\frac{r^2+b^2}{\Delta}dr \,, \\
	dy_{\pm} &= d\psi \pm b\frac{r^2+a^2}{\Delta}dr \,.
\ea\ee
Explicitly, they are given by
\be\ba\label{eq:NullCoordinates_5dMP}
	t_{\pm} &= t \pm \left\{r + \frac{1}{2} \frac{r_{s}^2}{r_{+}^2-r_{-}^2}\left[r_{+}\ln\left|\frac{r-r_{+}}{r+r_{+}}\right| - r_{-}\ln\left|\frac{r-r_{-}}{r+r_{-}}\right|\right]\right\} \,, \\
	\varphi_{\pm} &= \phi \pm \frac{1}{2}\frac{a}{r_{+}^2-r_{-}^2}\left[\frac{r_{+}^2+b^2}{r_{+}}\ln\left|\frac{r-r_{+}}{r+r_{+}}\right| - \frac{r_{-}^2+b^2}{r_{-}}\ln\left|\frac{r-r_{-}}{r+r_{-}}\right|\right] \,, \\
	y_{\pm} &= \psi \pm \frac{1}{2}\frac{b}{r_{+}^2-r_{-}^2}\left[\frac{r_{+}^2+a^2}{r_{+}}\ln\left|\frac{r-r_{+}}{r+r_{+}}\right| - \frac{r_{-}^2+a^2}{r_{-}}\ln\left|\frac{r-r_{-}}{r+r_{-}}\right|\right] \,,
\ea\ee
where the integration constants have been fixed to ensure the asymptotic behaviors
\be
	t_{\pm}\xrightarrow{r\rightarrow\infty} t \pm r \,,\quad \,\varphi_{\pm}\xrightarrow{r\rightarrow\infty} \phi \,,\quad \,y_{\pm}\xrightarrow{r\rightarrow\infty} \psi \,,
\ee
as well as a smooth extremal limit,
\be\ba
	t_{\pm} &\xrightarrow{r_{-}\rightarrow r_{+}} t \pm \left\{r-\frac{r_{s}}{2}\left[\frac{r_{s}r}{r^2-r_{+}^2} + \frac{r_{s}}{2r_{+}}\log\left|\frac{r-r_{+}}{r+r_{+}}\right|\right]\right\} \,, \\
	\varphi_{\pm} &\xrightarrow{r_{-}\rightarrow r_{+}} \phi \mp \frac{a}{2r_{+}}\left[\frac{b^2+r_{+}^2}{r_{+}^2}\frac{r_{+}r}{r^2-r_{+}^2} + \frac{b^2-r_{+}^2}{2r_{+}^2}\log\left|\frac{r-r_{+}}{r+r_{+}}\right|\right] \,, \\
	y_{\pm} &\xrightarrow{r_{-}\rightarrow r_{+}} \psi \mp \frac{b}{2r_{+}}\left[\frac{a^2+r_{+}^2}{r_{+}^2}\frac{r_{+}r}{r^2-r_{+}^2} + \frac{a^2-r_{+}^2}{2r_{+}^2}\log\left|\frac{r-r_{+}}{r+r_{+}}\right|\right] \,.
\ea\ee

\section{Equations of motion for spin-$0$ perturbations}
\label{sec:EOM0_5dMP}
The equations of motion for a massless scalar field in the background of the $5$-dimensional Myers-Perry black hole turn out to be separable,
\be
	\nabla^2 \Phi = \frac{4}{\Sigma}\left[\mathbb{O}_{\text{full}} - \mathbb{P}_{\text{full}}\right]\Phi = 0 \,,
\ee
with the radial and angular operators given by, after introducing the variable $\rho=r^2$,
\be\label{eq:FullEOM_5dMP}
	\begin{gathered}
		\ba
			\mathbb{O}_{\text{full}} &\equiv \partial_{\rho}\,\Delta\,\partial_{\rho} - \frac{\left(a^2-b^2\right)}{4}\left(\frac{1}{\rho+a^2}\,\partial_{\phi}^2 - \frac{1}{\rho+b^2}\,\partial_{\psi}^2\right)  - \frac{1}{4}\left(\rho+a^2+b^2\right)\partial_{t}^2	\\
			&\quad- \frac{\left(\rho+a^2\right)\left(\rho+b^2\right)r_{s}^2}{4\Delta}\left(\partial_{t} + \frac{a}{\rho+a^2}\,\partial_{\phi} + \frac{b}{\rho+b^2}\,\partial_{\psi}\right)^2 \,,
		\ea
		\\
		\mathbb{P}_{\text{full}} \equiv -\frac{1}{4}\left[ \Delta_{\mathbb{S}^3}^{\left(0\right)} + \left(a^2\sin^2\theta+b^2\cos^2\theta\right)\partial_{t}^2 \right] \,.
	\end{gathered}
\ee
In the above expression, we have also identified the scalar ($s=0$) Laplace-Beltrami operator on $\mathbb{S}^3$ in the current direction cosine angular coordinates,
\be
	\Delta_{\mathbb{S}^3}^{\left(0\right)} \equiv \frac{1}{\sin\theta\cos\theta}\partial_{\theta}\left(\sin\theta\cos\theta\,\partial_{\theta}\right) + \frac{1}{\sin^2\theta}\,\partial_{\phi}^2 + \frac{1}{\cos^2\theta}\,\partial_{\psi}^2 \,.
\ee
which is used to define a modified spherical harmonics expansion in the static case from which the Love numbers are more naturally extracted as we will shortly see.

After separating the variables,
\be\label{eq:PhiSepVar_5dMP}
	\Phi_{\omega\ell m j}\left(t,\rho,\theta,\phi,\psi\right) = e^{-i\omega t}e^{im\phi}e^{ij\psi}R_{\omega\ell m j}\left(\rho\right) S_{\omega\ell m j}\left(\theta\right) \,,
\ee
this $5$-dimensional scalar Teukolsky equation decomposes to
\be\ba
	\mathbb{O}_{\text{full}}\Phi_{\omega\ell m j} &= \hat{\ell}(\hat{\ell}+1)\Phi_{\omega\ell m j} \,, \\
	\mathbb{P}_{\text{full}}\Phi_{\omega\ell m j} &= \hat{\ell}(\hat{\ell}+1)\Phi_{\omega\ell m j} \,,
\ea\ee
with
\be
	\hat{\ell} \equiv \frac{\ell}{d-3} = \frac{\ell}{2} \,,
\ee
and $\ell$ an effective orbital number which is in general non-integer for $\omega\ne0$. The azimuthal numbers $m$ and $j$, however, are always integers by virtue of the periodicity of the azimuthal angles.

\subsection{Near-zone approximation}
In order to match observables onto the $1$-body worldline EFT as prescribed in~\ref{sec:RCsGR}, Eq.~\eqref{eq:1ptNewtonianMatching}, we should solve the massless Klein-Gordon equation after employing a near-zone perturbation scheme~\cite{Chia:2020yla,Charalambous:2021kcz,Starobinsky:1973aij,Starobinskil:1974nkd,Castro:2010fd,Maldacena:1997ih}. In the near-zone region, one imposes the asymptotic boundary condition
\be\label{eq:Asympbc_5dMP}
	R_{\omega\ell m j} \xrightarrow{r\rightarrow\infty} \bar{\mathcal{E}}_{\ell m j}\left(\omega\right)\,r^{\ell} = \bar{\mathcal{E}}_{\ell m j}\left(\omega\right)\,\rho^{\hat{\ell}} \,,
\ee
indicating the presence of a source at large distances with multipole moments $\bar{\mathcal{E}}_{\ell m j}\left(\omega\right)$, along with the ingoing boundary condition at the event horizon. Ingoing boundary condition at the future/past event horizon is imposed by requiring ingoing waves at the horizon in advanced ($+$)/retarded ($-$) coordinates,
\be
	\Phi_{\omega\ell m j} \xrightarrow{r\rightarrow r_{+}} T_{\ell m j}^{\left(\pm\right)}\left(\omega\right)\,e^{-i\omega t_{\pm}}e^{im\varphi_{\pm}}e^{ijy_{\pm}}S_{\omega\ell m j}\left(\theta\right) \,,
\ee
with $T_{\ell m j}^{\left(\pm\right)}\left(\omega\right)$ the transmission amplitudes. The relation between advanced/retarded coordinates $\left(t_{\pm},r,\theta,\varphi_{\pm},y_{\pm}\right)$ and Boyer-Lindquist coordinates  $\left(t,r,\theta,\phi,\psi\right)$ is given in Eq.~\eqref{eq:NullCoordinates_5dMP} and implies
\be\label{eq:NHbc_5dMP}
	\begin{gathered}
		R_{\omega\ell m j} \sim \left(\rho-\rho_{+}\right)^{\pm iZ_{+}\left(\omega\right)}\left(\rho-\rho_{-}\right)^{\pm iZ_{-}\left(\omega\right)}\,\,\,\,\,\,\, \text{as $\rho\rightarrow \rho_{+}$} \,, \\
		Z_{\pm}\left(\omega\right) \equiv \pm\frac{r_{\pm}}{2}\frac{\rho_{s}}{\rho_{+}-\rho_{-}}\left(m\Omega_{\phi}^{\left(\pm\right)}+j\Omega_{\psi}^{\left(\pm\right)}-\omega\right) \,.
	\end{gathered}
\ee
Physically, of course, we are only interested in regularity at the future event horizon and $\Omega_{\phi}^{\left(+\right)}\equiv\Omega_{\phi}$ and $\Omega_{\psi}^{\left(+\right)}\equiv\Omega_{\psi}$ measure the black hole angular velocities with respect to the two rotation planes according to an asymptotic observer.

Let us remind here that the near-zone approximation extends beyond the near-horizon or the low frequency regimes. In particular, not only does it preserve the near-horizon dynamics in the radial operator for any frequency $\omega$, but it also overlaps with the asymptotically flat far-zone region $r\gg r_{+}$ where outgoing boundary conditions are imposed. The overlapping intermediate region $r_{+}\ll r \ll \omega^{-1}$ then serves as a matching region that probes the response of the centered body in the outgoing waves that are detected at infinity. Furthermore, the near-zone expansion is not unique as there are infinitely many ways to near-zone-split the equations of motion as long as the ``subleading'' terms can be treated perturbatively in a near-zone expansion. In practice, the splitting is done such that the equations of motion are exactly solvable in terms of elementary functions at leading order in the near-zone expansion. There are two particular near-zone splittings of interest in the current problem, controlled by a sign $\sigma=\pm$. We split the radial operator as
\be\label{eq:NZRadial1_5dMP}
	\mathbb{O}_{\text{full}} = \partial_{\rho}\,\Delta\,\partial_{\rho} + V_0^{\left(\sigma\right)} + \epsilon\,V_1^{\left(\sigma\right)} \,,
\ee
with $\epsilon$ a formal order parameter which is equal to unity for the full equations of motion and equal to zero for the near-zone approximation and
\be\ba\label{eq:NZRadial2_5dMP}
	V_0^{\left(\sigma\right)} = &- \frac{\rho_{s}^2\rho_{+}}{4\Delta}\left(\partial_{t}+\Omega_{\phi}\,\partial_{\phi}+\Omega_{\psi}\,\partial_{\psi}\right)^2 - \frac{a^2-b^2}{4\left(\rho-\rho_{-}\right)}\left(\partial_{\phi}^2-\partial_{\psi}^2\right) \\
	&- \frac{\rho_{s}^2\rho_{+}}{2\left(\rho_{+}-\rho_{-}\right)\left(\rho-\rho_{-}\right)} \left(\Omega_{\phi}-\sigma\,\Omega_{\psi}\right)\partial_{t}\left(\partial_{\phi}-\sigma\,\partial_{\psi}\right) \,,
\ea\ee
\be\ba\label{eq:NZRadial3_5dMP}
	V_1^{\left(\sigma\right)} &= \frac{\rho_{s}\left(a-\sigma b\right)\left[\rho_{s}-\left(a+\sigma b\right)^2\right]}{4\left(\rho_{+}-\rho_{-}\right)\left(\rho-\rho_{-}\right)}\partial_{t}\left(\partial_{\phi}-\sigma\,\partial_{\psi}\right) \\
	&- \frac{\rho_{s}}{4\left(\rho-\rho_{-}\right)}\left[\rho_{s}\,\partial_{t}+\left(a+\sigma b\right)\left(\partial_{\phi}+\sigma\,\partial_{\psi}\right)\right]\partial_{t} - \frac{1}{4}\left(\rho+a^2+b^2+\rho_{s}\right)\partial_{t}^2 \,.
\ea\ee

For the angular operator, we use the splitting
\be\label{eq:NZAngular_5dMP}
	\mathbb{P}_{\text{full}} = -\frac{1}{4}\left[ \Delta_{\mathbb{S}^3}^{\left(0\right)} + \epsilon\left(a^2\sin^2\theta+b^2\cos^2\theta\right)\partial_{t}^2 \right] \,,
\ee
that is, we are near-zone approximating it with the static angular operator. In the static limit, the angular problem is solved by $S_{\ell m j}\left(\theta\right)$, given in Appendix~\ref{app:SphericalHarmonics}, from which $\ell$ is set to be an orbital number ranging in the set of whole numbers, $\ell\in\mathbb{N}$, $m$ is identified as a spherical harmonics integer azimuthal number $\left|m\right|\le\ell$ and $j$ is a second integer azimuthal number ranging from $-\left(\ell-\left|m\right|\right)$ up to $\ell-\left|m\right|$, but with step $2$.

\section{From Love tensors to Love Numbers of $5$-dimensional rotating bodies}
\label{sec:LTs_LNs5d}

The Newtonian matching condition presented in Section~\ref{sec:RCsGR} refers to matching the response tensors $k_{LL^{\prime}}^{\left(s\right)}$. The microscopic computation as formulated above, however, will naturally output the harmonic response coefficients arising after performing a harmonic expansion thanks to the $1$-to-$1$ correspondence between spatial STF tensors and spherical harmonics. As we saw in Section~\ref{sec:LTs_LNs}, extracting an isolating prescription that separates the conservative and dissipative parts of the harmonic response coefficients is in general non-trivial, but for integrable configurations, such as the current remarkably separable perturbations of $5$-dimensional rotating black holes, this decomposition allows to identify the Love numbers as the real part of the harmonic response coefficients, while the imaginary part is assigned to dissipative effects.

While this can be done in $d=1+3$ spacetime dimensions by performing an expansion into spherical harmonics over $\mathbb{S}^2$~\cite{LeTiec:2020bos}, it fails to work for a general rotating body in higher spacetime dimensions. For $d>4$, an expansion into spherical harmonics on $\mathbb{S}^{d-2}$ allows us to extract a simple isolating prescription of the conservative part of the response coefficients only for spherically symmetric and non-rotating bodies. For axisymmetric distributions, one should instead perform a modified harmonic expansion over $\left[\mathbb{S}^1\right]^{N}\subset \mathbb{S}^{d-2}$, with $N=\left\lfloor\frac{d-1}{2}\right\rfloor$ factors of $\mathbb{S}^1$, appropriate for the isometry subgroup $U\left(1\right)^{N}\subset SO\left(d-1\right)$ of such configurations. In $d=5$ spacetime dimensions, this is a modified harmonic expansion over the $\mathbb{S}^1\times\mathbb{S}^1$ part of $\mathbb{S}^3$ in accordance with the $U\left(1\right)\times U\left(1\right)$ azimuthal symmetries. This modified spherical harmonics basis for $d=5$ is introduced and analyzed in Appendix~\ref{app:SphericalHarmonics}.

Following Section~\ref{sec:LTs_LNs}, we start by expanding the $4$-dimensional spatial STF tensors $\bar{\mathcal{E}}_{L}^{\left(s\right)}$ into modified spherical harmonics of integer orbital number $\ell\in\mathbb{N}$,
\be
	\bar{\mathcal{E}}^{\left(s\right)L} = \sum_{m,j}\bar{\mathcal{E}}_{\ell m j}^{\left(s\right)}\mathcal{Y}_{\ell m j}^{L\ast} \,,
\ee
where the constant STF tensors $\mathcal{Y}_{\ell m j}^{L}$ are given by
\be
	\mathcal{Y}_{\ell m j}^{L} = \frac{\left(2\ell+2\right)!!}{4\pi^2\ell!}\oint_{\mathbb{S}^3}d\Omega_3\,n^{\left\langle L \right\rangle}\tilde{Y}_{\ell m j}^{\ast}\left(\mathbf{n}\right) \,,
\ee
with $\tilde{Y}_{\ell m j}\left(\mathbf{n}\right)\equiv \tilde{Y}_{\ell m j}\left(\theta,\phi,\psi\right)$ the modified spherical harmonics on $\mathbb{S}^3$ and $n^{i}\equiv x^{i}/r$. For future reference, the explicit limits of the sums over the azimuthal numbers $m$ and $j$ are
\be
	\sum_{m,j}\left(\dots\right)\equiv \sum_{m=-\ell}^{\ell}\left(\sum_{j=-\left(\ell-\left|m\right|\right),2}^{\ell-\left|m\right|}\left(\dots\right)\right) \,.
\ee
We note that the $j$-sum is being performed with a step $2$. This is merely a convention chosen such that the azimuthal number $m$ resembles the usual azimuthal number of scalar spherical harmonics on $\mathbb{S}^2$.

Then, the response coefficients $k_{\ell m j;\ell^{\prime} m^{\prime} j^{\prime}}^{\left(s\right)}\left(\omega\right)$ are related to the response tensor $k_{LL^{\prime}}^{\left(s\right)}\left(\omega\right)$ according to
\be\label{eq:RNsToRTs5d}
	k_{\ell m j;\ell^{\prime} m^{\prime} j^{\prime}}^{\left(s\right)}\left(\omega\right) = \frac{4\pi^2\ell!}{\left(2\ell+2\right)!!}k_{LL^{\prime}}^{\left(s\right)}\left(\omega\right)\mathcal{Y}_{\ell m j}^{L}\mathcal{Y}_{\ell^{\prime} m^{\prime} j^{\prime}}^{L^{\prime}\ast} \,.
\ee
From the fact that $k_{LL^{\prime}}^{\left(s\right)\ast}\left(\omega\right)=k_{LL^{\prime}}^{\left(s\right)}\left(-\omega\right)$, following from the reality of the source and induced multipole moments in position space, the complex conjugacy relation of the modified spherical harmonics, $\tilde{Y}_{\ell m j}^{\ast}=\tilde{Y}_{\ell,-m,-j}$, implies the following complex conjugacy relation for the response coefficients
\be\label{eq:klmCC5d}
	k_{\ell m j;\ell^{\prime} m^{\prime} j^{\prime}}^{\left(s\right)\ast}\left(\omega\right) = k_{\ell,-m,-j;\ell^{\prime},-m^{\prime},-j^{\prime}}^{\left(s\right)}\left(-\omega\right) \,.
\ee

The conservative/dissipative decomposition of the response tensor in Eq.~\eqref{eq:RTsConsDiss} can then be translated at the level of the response coefficients $k_{\ell mj;\ell^{\prime}m^{\prime}j^{\prime}}^{\left(s\right)}\left(\omega\right)$. The definition \eqref{eq:RNsToRTs5d} and the complex conjugacy relation \eqref{eq:klmCC5d} immediately imply
\be\ba\label{eq:RCsConsDissGen5d}
	k_{\ell mj;\ell^{\prime}m^{\prime}j^{\prime}}^{\left(s\right)\text{Love}}\left(\omega\right) &= \frac{1}{2}\left(k_{\ell mj;\ell^{\prime}m^{\prime}j^{\prime}}^{\left(s\right)}\left(\omega\right)+k_{\ell^{\prime}m^{\prime}j^{\prime};\ell mj}^{\left(s\right)\ast}\left(\omega\right)\right) \,, \\
	k_{\ell mj;\ell^{\prime}m^{\prime}j^{\prime}}^{\left(s\right)\text{diss}}\left(\omega\right) &= \frac{1}{2i}\left(k_{\ell mj;\ell^{\prime}m^{\prime}j^{\prime}}^{\left(s\right)}\left(\omega\right)-k_{\ell^{\prime}m^{\prime}j^{\prime};\ell mj}^{\left(s\right)\ast}\left(\omega\right)\right) \,,
\ea\ee
such that
\be
	k_{\ell mj;\ell^{\prime}m^{\prime}j^{\prime}}^{\left(s\right)}\left(\omega\right) = k_{\ell mj;\ell^{\prime}m^{\prime}j^{\prime}}^{\left(s\right)\text{Love}}\left(\omega\right) + ik_{\ell mj;\ell^{\prime}m^{\prime}j^{\prime}}^{\left(s\right)\text{diss}}\left(\omega\right) \,.
\ee
We note, however, that $k_{\ell mj;\ell^{\prime}m^{\prime}j^{\prime}}^{\left(s\right)\text{Love}}\left(\omega\right)$ and $k_{\ell mj;\ell^{\prime}m^{\prime}j^{\prime}}^{\left(s\right)\text{diss}}\left(\omega\right)$ are in general complex numbers.

We now focus to a case relevant for the current scalar perturbations of the $5$-dimensional Myers-Perry black hole. The axisymmetry of the background implies the decoupling of $m$-modes and $j$-modes, while we enhanced integrability of the configuration reveals that there is no $\ell$-mode mixing either. Then,
\be
	k_{\ell m j;\ell^{\prime} m^{\prime} j^{\prime}}^{\left(s\right)}\left(\omega\right) = k_{\ell m j}^{\left(s\right)}\left(\omega\right)\delta_{\ell\ell^{\prime}}\delta_{mm^{\prime}}\delta_{jj^{\prime}} \,,
\ee
and the frequency space potential perturbation harmonic modes in the Newtonian limit simplify to
\be
	\delta\Phi^{\left(s\right)}_{\ell m j}\left(\omega,r\right) = \frac{\left(\ell-s\right)!}{\ell!}\left[1 + k_{\ell m j}^{\left(s\right)}\left(\omega\right)\left(\frac{\mathcal{R}}{r}\right)^{2\ell+2}\right]r^{\ell}\bar{\mathcal{E}}_{\ell m j}^{\left(s\right)}\left(\omega\right) \,.
\ee

These response coefficients $k_{\ell m j}^{\left(s\right)}\left(\omega\right)$ will in general be analytic functions in the angular momenta of the rotating body as well as the frequency $\omega$ with respect to an inertial observer. The complex conjugacy relation, which now reads $k_{\ell m j}^{\left(s\right)\ast}\left(\omega\right) = k_{\ell,-m,-j}^{\left(s\right)}\left(-\omega\right)$, then allows to explicitly separate the $m$- and $j$-dependencies of the response coefficients as
\be\label{eq:TLNs5d_mjExpansion}
	k_{\ell m j}^{\left(s\right)}\left(\omega\right) = k_{\ell}^{\left(0\right)}\left(\omega\right) + \chi\sum_{n_{\phi}=1}^{\infty}\sum_{n_{\psi}=1}^{\infty}k_{\ell}^{(n_{\phi},n_{\psi})}\left(\omega,\chi_{\phi},\chi_{\psi}\right) \left(im\right)^{n_{\phi}}\left(ij\right)^{n_{\psi}} \,,
\ee
with $\chi_{\phi}$ and $\chi_{\psi}$ the dimensionless spin parameters associated with the $J_{\phi}$ and $J_{\psi}$ angular momenta and the overall formal $\chi$ is to separate the non-spinning part $k_{\ell}^{\left(0\right)}\left(\omega\right)$. All $k_{\ell}^{(n_{\phi},n_{\psi})}\left(\omega,\chi_{\phi},\chi_{\psi}\right)$ are smooth functions of $\chi_{\phi}$ and $\chi_{\psi}$, satisfying the complex conjugation relation $k_{\ell}^{(n_{\phi},n_{\psi})\ast}\left(\omega,\chi_{\phi},\chi_{\psi}\right)=k_{\ell}^{(n_{\phi},n_{\psi})}\left(-\omega,\chi_{\phi},\chi_{\psi}\right)$.

Let us now extract a necessary condition for such a decoupling to occur. This analysis is the $d=5$ version of~\cite{LeTiec:2020bos}. Starting from the physically relevant part of the response tensor,
\be
	k_{L\left\langle L^{\prime} \right\rangle}^{\left(s\right)}\left(\omega\right) = \frac{4\pi^2\ell!}{\left(2\ell+2\right)!!}\sum_{m,j}k_{\ell m j}^{\left(s\right)}\left(\omega\right)\mathcal{Y}^{\ell m j\ast}_{L}\mathcal{Y}^{\ell m j}_{L^{\prime}} \,,
\ee
with $\ell^{\prime}=\ell$ understood, the expansion \eqref{eq:TLNs5d_mjExpansion} implies
\be\ba
	{}&k_{L\left\langle L^{\prime} \right\rangle}^{\left(s\right)}\left(\omega\right) = k_{\ell}^{\left(0\right)}\left(\omega\right)\delta_{LL^{\prime}} + \chi\sum_{n_{\phi}=1}^{\infty}\sum_{n_{\psi}=1}^{\infty} \left(-1\right)^{n_{\phi}+n_{\psi}} \\
	&\times\bigg[k_{\ell}^{(2n_{\phi},2n_{\psi})}\left(\omega,\chi_{\phi},\chi_{\psi}\right)R_{LL^{\prime}}^{(2n_{\phi},2n_{\psi})} - k_{\ell}^{(2n_{\phi}-1,2n_{\psi}-1)}\left(\omega,\chi_{\phi},\chi_{\psi}\right)R_{LL^{\prime}}^{(2n_{\phi}-1,2n_{\psi}-1)} \\
	&+ k_{\ell}^{(2n_{\phi}-1,2n_{\psi})}\left(\omega,\chi_{\phi},\chi_{\psi}\right)I_{LL^{\prime}}^{(2n_{\phi}-1,2n_{\psi})} + k_{\ell}^{(2n_{\phi},2n_{\psi}-1)}\left(\omega,\chi_{\phi},\chi_{\psi}\right)I_{LL^{\prime}}^{(2n_{\phi},2n_{\psi}-1)}\bigg] \,,
\ea\ee
and the tensorial structure of $k_{L\left\langle L^{\prime} \right\rangle}^{\left(s\right)}\left(\omega\right)$ is completely determined by two real-valued symmetric and two real-valued antisymmetric STF tensors,
\be\ba
	R_{LL^{\prime}}^{(2n_{\phi},2n_{\psi})} &\equiv \frac{8\pi^2\ell!}{\left(2\ell+2\right)!!}\sum_{m=1}^{\ell}\left(\sum_{j=-\left(\ell-m\right),2}^{\ell-m}m^{2n_{\phi}}j^{2n_{\psi}}\text{Re}\left\{\mathcal{Y}^{\ell mj\ast}_{L}\mathcal{Y}^{\ell mj}_{L^{\prime}}\right\}\right) \,, \\
	R_{LL^{\prime}}^{(2n_{\phi}-1,2n_{\psi}-1)} &\equiv \frac{8\pi^2\ell!}{\left(2\ell+2\right)!!}\sum_{m=1}^{\ell}\left(\sum_{j=-\left(\ell-m\right),2}^{\ell-m}m^{2n_{\phi}-1}j^{2n_{\psi}-1}\text{Re}\left\{\mathcal{Y}^{\ell mj\ast}_{L}\mathcal{Y}^{\ell mj}_{L^{\prime}}\right\}\right) \,, \\
	I_{LL^{\prime}}^{(2n_{\phi}-1,2n_{\psi})} &\equiv \frac{8\pi^2\ell!}{\left(2\ell+2\right)!!}\sum_{m=1}^{\ell}\left(\sum_{j=-\left(\ell-m\right),2}^{\ell-m}m^{2n_{\phi}-1}j^{2n_{\psi}}\text{Im}\left\{\mathcal{Y}^{\ell mj\ast}_{L}\mathcal{Y}^{\ell mj}_{L^{\prime}}\right\}\right) \,, \\
	I_{LL^{\prime}}^{(2n_{\phi},2n_{\psi}-1)} &\equiv \frac{8\pi^2\ell!}{\left(2\ell+2\right)!!}\sum_{m=1}^{\ell}\left(\sum_{j=-\left(\ell-m\right),2}^{\ell-m}m^{2n_{\phi}}j^{2n_{\psi}-1}\text{Im}\left\{\mathcal{Y}^{\ell mj\ast}_{L}\mathcal{Y}^{\ell mj}_{L^{\prime}}\right\}\right) \,.
\ea\ee
\be\ba
	R_{LL^{\prime}}^{(2n_{\phi},2n_{\psi})} = +R_{L^{\prime}L}^{(2n_{\phi},2n_{\psi})} \,,\quad R_{LL^{\prime}}^{(2n_{\phi}-1,2n_{\psi}-1)} = +R_{L^{\prime}L}^{(2n_{\phi}-1,2n_{\psi}-1)} \,, \\
	I_{LL^{\prime}}^{(2n_{\phi}-1,2n_{\psi})} = -I_{L^{\prime}L}^{(2n_{\phi}-1,2n_{\psi})} \,, \quad I_{LL^{\prime}}^{(2n_{\phi},2n_{\psi}-1)} = -I_{L^{\prime}L}^{(2n_{\phi},2n_{\psi}-1)} \,.
\ea\ee

Finally, let us write the conservative/dissipative decomposition of the response coefficients \eqref{eq:RCsConsDissGen5d} for the current special configuration, which is also the main result of interest of this analysis,
\be\ba\label{eq:RCsConsDiss5d}
	k_{\ell mj}^{\left(s\right)\text{Love}}\left(\omega\right) &= \text{Re}\left\{ k_{\ell mj}^{\left(s\right)}\left(\omega\right) \right\} \,, \\
	k_{\ell mj}^{\left(s\right)\text{diss}}\left(\omega\right) &= \text{Im}\left\{ k_{\ell mj}^{\left(s\right)}\left(\omega\right) \right\} \,.
\ea\ee
The Love numbers are therefore just the real part of the response coefficients, while the imaginary part encodes all the dissipative effects. We remark here that dissipative effects can survive even in the static limit due to frame dragging~\cite{Chia:2020yla,Charalambous:2021mea,Ivanov:2022hlo}.

\section{Scalar Love numbers of $5$-dimensional Myers-Perry black holes}
\label{sec:TLNsComputation}

We will now apply the tools presented in the previous sections to compute the scalar susceptibilities of $5$-dimensional Myers-Perry black holes at leading order in the near-zone expansion and identify the corresponding scalar ($s=0$) Love numbers. After separating the variables as in \eqref{eq:PhiSepVar_5dMP} and introducing
\be
	x \equiv \frac{r^2-r_{+}^2}{r_{+}^2-r_{-}^2} = \frac{\rho-\rho_{+}}{\rho_{+}-\rho_{-}} \,,
\ee
the near-zone equation of motion for the radial wavefunction can be massaged into
\be
	\left[ \frac{d}{dx}\,x\left(1+x\right)\frac{d}{dx} + \frac{Z_{+}^2\left(\omega\right)}{x}-\frac{\tilde{Z}_{-}^{\left(\sigma\right)2}\left(\omega\right)}{1+x} \right] R_{\omega\ell m j} = \hat{\ell}\left(\hat{\ell}+1\right)R_{\omega\ell m j} \,,
\ee
with $Z_{+}\left(\omega\right)$ given in \eqref{eq:NHbc_5dMP} and dictating the near-horizon behavior of the solution and
\be\label{eq:NZZetaMinus_5dMP}
	\tilde{Z}_{-}^{\left(\sigma\right)}\left(\omega\right) \equiv -\frac{r_{+}}{2}\frac{\rho_{s}}{\rho_{+}-\rho_{-}}\left(m\Omega_{\psi}+j\Omega_{\phi}-\sigma\omega\right) \,.
\ee
We note in particular that $\tilde{Z}_{-}^{\left(\sigma\right)}\left(\omega=0\right)=Z_{-}\left(\omega=0\right)$ in \eqref{eq:NHbc_5dMP} reflecting how the near-zone approximation becomes exact in the static limit. The above differential equation has three regular singular points at $x=0$, $x=-1$ and $x\rightarrow\infty$ with the characteristic exponents given in Table~\ref{tbl:CharacteristicExpnents5dMP}.

\begin{table}[t]
	\centering
	\begin{tabular}{|c||c|c|c|}
		\hline
		singular point & $x=0$ & $x=-1$ & $x\rightarrow\infty$ \\
		\hline
		characteristic exponent \#1 & $+iZ_{+}\left(\omega\right)$ & $+i\tilde{Z}_{-}^{\left(\sigma\right)}\left(\omega\right)$ & $\hat{\ell}$ \\
		\hline
		characteristic exponent \#2 & $-iZ_{+}\left(\omega\right)$ & $-i\tilde{Z}_{-}^{\left(\sigma\right)}\left(\omega\right)$ & $-\hat{\ell}-1$ \\
		\hline
	\end{tabular}
	\caption[Characteristic exponents of the near-zone truncated radial Klein-Gordon operator in the background of the $5$-dimensional Myers-Perry black hole.]{Characteristic exponents of the near-zone truncated radial Klein-Gordon operator in the background of the $5$-dimensional Myers-Perry black hole.}
	\label{tbl:CharacteristicExpnents5dMP}
\end{table}

The differential equation can be solved analytically in terms of Euler's hypergeometric functions. For future convenience, we introduce the parameters
\be\ba\label{eq:GammaPlusMinus}
	\Gamma^{\left(\sigma\right)}_{\pm\sigma}\left(\omega\right) &\equiv Z_{+}\left(\omega\right) \mp \sigma\, \tilde{Z}_{-}^{\left(\sigma\right)}\left(\omega\right) \\
	&= \frac{r_{+}}{2}\frac{\rho_{s}}{\rho_{+}-\rho_{-}}\left[\left(m\pm \sigma j\right)\left(\Omega_{\phi}\pm\sigma\Omega_{\psi}\right)-\omega\left(1\pm1\right)\right] \,.
\ea\ee
The solution that is regular at the future event horizon then reads
\be\ba\label{eq:NZRadialSolution_5dMP}
	{}&R_{\omega\ell m j} = \bar{R}_{\ell m j}\left(\omega\right)\,\left(\frac{x}{1+x}\right)^{iZ_{+}\left(\omega\right)} \\
	&\left(1+x\right)^{i\Gamma_{\pm\sigma}^{\left(\sigma\right)}\left(\omega\right)}\,{}_2F_1\left(\hat{\ell}+1+i\Gamma^{\left(\sigma\right)}_{\pm\sigma}\left(\omega\right),-\hat{\ell}+i\Gamma^{\left(\sigma\right)}_{\pm\sigma}\left(\omega\right);1+2iZ_{+}\left(\omega\right);-x\right) \,,
\ea\ee
where $\bar{R}_{\ell m j}\left(\omega\right)$ is fixed from the asymptotic boundary condition \eqref{eq:Asympbc_5dMP} to be proportional to the source moments harmonic modes $\bar{\mathcal{E}}_{\ell mj}\left(\omega\right)$ according to
\be
	\bar{R}_{\ell mj}\left(\omega\right) = \bar{\mathcal{E}}_{\ell mj}\left(\omega\right) \left(\rho_{+}-\rho_{-}\right)^{\hat{\ell}}\frac{\Gamma\left(\hat{\ell}+1+i\Gamma^{\left(\sigma\right)}_{+\sigma}\left(\omega\right)\right)\Gamma\left(\hat{\ell}+1+i\Gamma^{\left(\sigma\right)}_{-\sigma}\right)}{\Gamma\left(2\hat{\ell}+1\right)\Gamma\left(1+2iZ_{+}\left(\omega\right)\right)} \,.
\ee

Up to this point in solving the leading order near-zone problem, we have made no use of the fact that the orbital number $\ell$ is an integer. This is also the instructive prescription of performing the Newtonian matching with the worldline EFT from which the Love numbers are defined~\cite{Charalambous:2021mea,Ivanov:2022hlo}. Namely, in order to extract the response coefficients from the above microscopic computation, one should first analytically continue $\ell$ to be a real number, then expand around large $r$ to read the coefficient in front of the $r^{-\ell-2}$ term and then send $\ell$ to its physical integer values at the end. Doing this, we find, before sending $\ell$ to take its physical values,
\be\label{eq:NZSLNs_5dMP}
	k_{\ell m j}\left(\omega\right) = \frac{\Gamma\left(-2\hat{\ell}-1\right)\Gamma\left(\hat{\ell}+1+i\Gamma^{\left(\sigma\right)}_{+\sigma}\left(\omega\right)\right)\Gamma\left(\hat{\ell}+1+i\Gamma^{\left(\sigma\right)}_{-\sigma}\right)}{\Gamma\left(2\hat{\ell}+1\right)\Gamma\left(-\hat{\ell}+i\Gamma^{\left(\sigma\right)}_{+\sigma}\left(\omega\right)\right)\Gamma\left(-\hat{\ell}+i\Gamma^{\left(\sigma\right)}_{-\sigma}\right)}\left(\frac{\rho_{+}-\rho_{-}}{\rho_{s}}\right)^{2\hat{\ell}+1} \,,
\ee
which can be massaged using the mirror formula for the $\Gamma$-functions into the more transparent result
\be\ba
	k_{\ell m j}&\left(\omega\right) = A_{\ell m j}\left(\omega\right)\times\bigg\{- i\sinh2\pi Z_{+}\left(\omega\right) \\	&+\tan\pi\hat{\ell}\cosh\pi\Gamma^{\left(\sigma\right)}_{+\sigma}\left(\omega\right)\cosh\pi\Gamma^{\left(\sigma\right)}_{-\sigma} - \cot\pi\hat{\ell}\sinh\pi\Gamma^{\left(\sigma\right)}_{+\sigma}\left(\omega\right)\sinh\pi\Gamma^{\left(\sigma\right)}_{-\sigma} \bigg\} \,,
\ea\ee
where $A_{\ell m j}\left(\omega\right)$ is a real constant given by
\be
	A_{\ell m j}\left(\omega\right) \equiv \frac{\left|\Gamma\left(\hat{\ell}+1+i\Gamma^{\left(\sigma\right)}_{+\sigma}\left(\omega\right)\right)\right|^2\left|\Gamma\left(\hat{\ell}+1+i\Gamma^{\left(\sigma\right)}_{-\sigma}\right)\right|^2}{2\pi\,\Gamma\left(2\hat{\ell}+1\right)\Gamma\left(2\hat{\ell}+2\right)}\left(\frac{\rho_{+}-\rho_{-}}{\rho_{s}}\right)^{2\hat{\ell}+1} \,.
\ee
The conservative/dissipative decomposition \eqref{eq:RCsConsDiss5d} then implies
\be\label{eq:5dMPRCsDiss}
	k_{\ell mj}^{\text{diss}}\left(\omega\right) = -A_{\ell mj}\left(\omega\right)\sinh2\pi Z_{+}\left(\omega\right) \,,
\ee
\be\ba\label{eq:5dMPLove}
	k_{\ell mj}^{\text{Love}}\left(\omega\right) = A_{\ell mj}\left(\omega\right)\times&\bigg\{\tan\pi\hat{\ell}\cosh\pi\Gamma^{\left(\sigma\right)}_{+\sigma}\left(\omega\right)\cosh\pi\Gamma^{\left(\sigma\right)}_{-\sigma} \\
	&- \cot\pi\hat{\ell}\sinh\pi\Gamma^{\left(\sigma\right)}_{+\sigma}\left(\omega\right)\sinh\pi\Gamma^{\left(\sigma\right)}_{-\sigma} \bigg\} \,.
\ea\ee
At this point, we would like stress out that the above results should be trusted only for small values of $\omega$. Indeed, the near-zone approximation is accurate only for low frequencies. In particular, in the near-zone split \eqref{eq:NZRadial1_5dMP}-\eqref{eq:NZAngular_5dMP}, we are already approximating at order $\mathcal{O}\left(\omega a,\omega b,\omega^2\right)$~\cite{Charalambous:2022rre}. Nevertheless, the above result is accurate in the static limit which is also the case of interest. The qualitative behavior of the static scalar Love numbers can be found in Table~\ref{tbl:StaticSLNs_5dMP}. In the rest of this chapter, we will be using the above $\omega$-dependent expressions but it should always be kept in mind that they are accurate only in the static limit for non-zero spin parameters. The corresponding qualitative behavior of the leading order near-zone dynamical Love numbers is indicated in Table~\ref{tbl:WSLNs_5dMP}.

\begin{table}[t]
	\centering
	\begin{tabular}{|l|l||l|}
		\hline
		\multicolumn{2}{|c||}{\Gape[6pt]{Range of parameters}} & \multicolumn{1}{c|}{Behavior of $k^{\text{Love}}_{\ell mj}\left(\omega=0\right)$} \\
		\hline\hline
		\multicolumn{2}{|c||}{$\ell\in2\mathbb{N}+1$} & \multicolumn{1}{c|}{\Gape[6pt]{Running}} \\
		\hline
		\multirow{2}{*}{$\ell\in2\mathbb{N}$} & \multicolumn{1}{c||}{$\left|a\right|=\left|b\right|$ OR $\left|m\right|=\left|j\right|$} & \multicolumn{1}{c|}{\Gape[6pt]{Vanishing}} \\
		\cline{2-3}
		& \multicolumn{1}{c||}{$\left|a\right|\ne\left|b\right|$ AND $\left|m\right|\ne\left|j\right|$} & \multicolumn{1}{c|}{\Gape[6pt]{Running}} \\
		\hline
	\end{tabular}
	\caption[Behavior of static scalar Love numbers of the $5$-dimensional Myers-Perry black hole as a function of the black hole angular momenta and the scalar field perturbation orbital and azimuthal numbers.]{Behavior of static scalar Love numbers as a function of the $5$-dimensional Myers-Perry black hole angular momenta $a$ and $b$ and the scalar field perturbation orbital number $\ell$ and azimuthal numbers $m$ and $j$. For generic angular momenta and azimuthal and orbital numbers, the static scalar Love numbers exhibit a classical RG flow. The only exception is when the orbital number is even ($\hat{\ell}$ is integer) and the angular momenta or the azimuthal numbers are equal in magnitude in which case the static Love numbers turn out to vanish.}
	\label{tbl:StaticSLNs_5dMP}
\end{table}

\begin{table}[t]
	\centering
	\begin{tabular}{|l|l||l|}
		\hline
		\multicolumn{2}{|c||}{\Gape[6pt]{Range of parameters}} & \multicolumn{1}{c|}{Behavior of $k^{\text{Love}}_{\ell mj}\left(\omega\ne0\right)$} \\
		\hline\hline
		\multicolumn{2}{|c||}{\Gape[6pt]{$i\Gamma^{\left(\sigma\right)}_{\pm}\left(\omega\right)-\hat{\ell} \notin \mathbb{Z}$}} & \multicolumn{1}{c|}{Running} \\
		\hline
		\multirow{2}{*}{$\begin{matrix} i\Gamma^{\left(\sigma\right)}_{+}\left(\omega\right)-\hat{\ell} = k\in\mathbb{Z} \\ \text{OR} \\ i\Gamma^{\left(\sigma\right)}_{-}\left(\omega\right)-\hat{\ell} = k\in\mathbb{Z} \end{matrix}$} & \multicolumn{1}{c||}{\Gape[9pt]{$-\ell \le k \le 0$}} & \multicolumn{1}{c|}{Vanishing} \\
		\cline{2-3}
		& \multicolumn{1}{c||}{\Gape[10pt]{$k>0$ OR $k<-\ell$}} & \multicolumn{1}{c|}{Non-running and Non-vanishing} \\
		\hline
	\end{tabular}
	\caption[Behavior of the dynamical scalar Love numbers of the $5$-dimensional Myers-Perry black hole at leading order in the near-zone expansion as a function of the black hole angular momenta and the scalar field perturbation frequency and orbital and azimuthal numbers.]{Behavior of $\omega$-dependent scalar Love numbers of the $5$-dimensional Myers-Perry black hole at leading order in the near-zone expansion as a function of the parameters $\Gamma^{\left(\sigma\right)}_{\pm}\left(\omega\right)$, given in \eqref{eq:GammaPlusMinus}, and the generalized orbital number $\hat{\ell}=\frac{\ell}{2}$ of the perturbation. For generic angular momenta and azimuthal and orbital numbers, the static scalar Love numbers exhibit a classical RG flow. However, there is a discrete series of imaginary-valued $\Gamma^{\left(\sigma\right)}_{\pm}$'s for which the near-zone Love numbers do not run. For the $\Gamma^{\left(\sigma\right)}_{-\sigma}$ branch, these are unphysical, accompanied by conical singularities in the scalar field profile. For the $\Gamma^{\left(\sigma\right)}_{+\sigma}\left(\omega\right)$ branch, the $k<0$ modes acquire the interpretation of Total Transmission Modes.}
	\label{tbl:WSLNs_5dMP}
\end{table}

\subsection{Running Love}
We begin analyzing our above result for the scalar Love numbers by first addressing the divergent behavior. For general non-zero spin parameters $a$ and $b$ and general azimuthal numbers $m$ and $j$ and frequency $\omega$, i.e. for general non-zero $\Gamma^{\left(\sigma\right)}_{\pm\sigma}\left(\omega\right)$, the scalar Love numbers \textit{always} diverge, either as $\cot\pi\hat{\ell}$ for integer $\hat{\ell}$ (even $\ell$) or as $\tan\pi\hat{\ell}$ for half-integer $\hat{\ell}$ (odd $\ell$). More specifically, in the limit $\varepsilon\rightarrow0$ where $2\hat{\ell}=n-\varepsilon$ approaches a whole number $n\in\mathbb{N}$, the response coefficients \eqref{eq:NZSLNs_5dMP} develop a simple pole due to the diverging $\Gamma(-2\hat{\ell}-1)$. From the residue of the $\Gamma$-function near negative integers, $\Gamma\left(-n+\varepsilon\right) = \frac{\left(-1\right)^{n}}{n!\,\varepsilon} + \mathcal{O}\left(\varepsilon^0\right)$, the developed pole can be worked out to be
\be\label{eq:kRes_5dMP}
	k_{\ell m j}\left(\omega\right) = -\frac{\left(-1\right)^{n}}{n!\,\varepsilon}\frac{\Gamma\left(\frac{n}{2}+1+i\Gamma^{\left(\sigma\right)}_{+\sigma}\left(\omega\right)\right)\Gamma\left(\frac{n}{2}+1+i\Gamma^{\left(\sigma\right)}_{-\sigma}\right)}{\Gamma\left(-\frac{n}{2}+i\Gamma^{\left(\sigma\right)}_{+\sigma}\left(\omega\right)\right)\Gamma\left(-\frac{n}{2}+i\Gamma^{\left(\sigma\right)}_{-\sigma}\right)\left(n+1\right)!}\left(\frac{\rho_{+}-\rho_{-}}{\rho_{s}}\right)^{n+1} + \mathcal{O}\left(\varepsilon^{0}\right) \,.
\ee

The full solution, however, is regular due to a compensating divergence in the ``source'' part of the scalar field profile. More specifically, the source/response split is performed prior to sending the orbital number to take its physical values with the end result
\be\ba\label{eq:SourceReponseSplit_5dMP}
	R_{\omega\ell m j}\left(\rho\right) &= \bar{\mathcal{E}}_{\ell m j}\left(\omega\right)\,\rho^{\hat{\ell}}\left[Z_{\omega\ell m j}^{\text{source}}\left(\rho\right) + k_{\ell m j}\left(\omega\right)\,\left(\frac{\rho_{s}}{\rho}\right)^{2\hat{\ell}+1} Z_{\omega\ell m j}^{\text{response}}\left(\rho\right)\right] \,, \\
	Z_{\omega\ell m j}^{\text{source}}\left(\rho\right) &= \left(1-\frac{\rho_{+}}{\rho}\right)^{\hat{\ell}} \left(\frac{\rho-\rho_{-}}{\rho-\rho_{+}}\right)^{\mp i\sigma\tilde{Z}_{-}^{\left(\sigma\right)}\left(\omega\right)} \\
	&\times\,{}_2F_1\left(-\hat{\ell}+i\Gamma^{\left(\sigma\right)}_{\mp\sigma}\left(\omega\right),-\hat{\ell}-i\Gamma^{\left(\sigma\right)}_{\pm\sigma}\left(\omega\right);-2\hat{\ell};\frac{\rho_{+}-\rho_{-}}{\rho_{+}-\rho}\right) \,, \\
	Z_{\omega\ell m j}^{\text{response}}\left(\rho\right) &= \left(1-\frac{\rho_{+}}{\rho}\right)^{-\hat{\ell}-1} \left(\frac{\rho-\rho_{-}}{\rho-\rho_{+}}\right)^{\mp i\sigma\tilde{Z}_{-}^{\left(\sigma\right)}\left(\omega\right)} \\
	&\times\,{}_2F_1\left(\hat{\ell}+1+i\Gamma^{\left(\sigma\right)}_{\mp\sigma}\left(\omega\right),\hat{\ell}+1-i\Gamma^{\left(\sigma\right)}_{\pm\sigma}\left(\omega\right);2\hat{\ell}+2;\frac{\rho_{+}-\rho_{-}}{\rho_{+}-\rho}\right) \,.
\ea\ee
In the limit $\varepsilon\rightarrow0$ where $2\hat{\ell}=n-\varepsilon$ approaches a whole number $n\in\mathbb{N}$, $\rho^{\hat{\ell}}Z_{\omega\ell m j}^{\text{source}}$ also develops a simple pole. The diverging component of the ``source'' part of the solution can be obtained from the residue formula \eqref{eq:2F1Residue} to be
\be\ba
	Z_{\omega\ell m j}^{\text{source}} &= -\frac{1}{n!\,\varepsilon}\frac{\Gamma\left(\frac{n}{2}+1+i\Gamma^{\left(\sigma\right)}_{+\sigma}\left(\omega\right)\right)\Gamma\left(\frac{n}{2}+1-i\Gamma^{\left(\sigma\right)}_{-\sigma}\right)}{\Gamma\left(-\frac{n}{2}+i\Gamma^{\left(\sigma\right)}_{+\sigma}\left(\omega\right)\right)\Gamma\left(-\frac{n}{2}-i\Gamma^{\left(\sigma\right)}_{-\sigma}\right)\left(n+1\right)!}\left(\frac{\rho_{+}-\rho_{-}}{\rho_{s}}\right)^{n+1} \\
	&\times \,\left(\frac{\rho_{s}}{\rho}\right)^{n+1}Z_{\omega\ell m j}^{\text{response}} + \mathcal{O}\left(\varepsilon^0\right) \,,
\ea\ee
which exactly cancels with the divergence in the scalar Love numbers whenever $n$ is an integer.

Similar to the scalar perturbations of the four-dimensional Kerr-Newman black seen in Section~\ref{sec:SLNs_KerrNewman}, this diverging behavior of the scalar Love numbers is interpreted as a classical RG flow from the worldline EFT perspective. More specifically, the power counting arguments of Section~\ref{sec:PowerCounting} reveal that whenever $2\hat{\ell}+1\in\mathbb{N}$, the Love numbers get renormalized from an overlapping with the PN corrections of the ``source'' part of the $1$-point function, namely, from the following type of diagrams~\cite{Kol:2011vg,Charalambous:2022rre,Ivanov:2022hlo}
\be
	\Phi_{\omega\ell m j} \supset \vcenter{\hbox{\begin{tikzpicture}
			\begin{feynman}
				\vertex[dot] (a0);
				\vertex[below=1cm of a0] (p1);
				\vertex[above=1cm of a0] (p2);
				\vertex[right=0.4cm of a0, blob] (gblob){};							
				\vertex[right=1.5cm of p1] (b1);
				\vertex[right=1.5cm of p2] (b2);
				\vertex[right=1.19cm of p2] (b22){$\times$};
				\vertex[above=0.7cm of a0] (g1);
				\vertex[above=0.4cm of a0] (g2);
				\vertex[right=0.05cm of a0] (gdtos){$\vdots$};
				\vertex[below=0.7cm of a0] (gN);
				\vertex[left=0.2cm of gN] (gN1);
				\vertex[left=0.2cm of g1] (g11);
				\diagram*{
					(p1) -- [double,double distance=0.5ex] (p2),
					(g1) -- [photon] (gblob),
					(g2) -- [photon] (gblob),
					(gN) -- [photon] (gblob),
					(b1) -- (gblob) -- (b2),
				};
				\draw [decoration={brace}, decorate] (gN1.south west) -- (g11.north west)
				node [pos=0.55, left] {\(2\hat{\ell}+1\)};
			\end{feynman}
	\end{tikzpicture}}} \,.
\ee
Moreover, whenever $2\hat{\ell}+1\in\mathbb{N}$, the the characteristic exponents of the near-zone radial differential equation near $x\rightarrow\infty$ differ by an integer number and Fuchs' theorem ensures that one of the independent solutions will contain logarithms. More explicitly, the solution regular at the future event horizon is still given by \eqref{eq:NZRadialSolution_5dMP}, but its analytic continuation at large distances must be taken as a limiting case with the end result being
\be\ba
	{}&R_{\omega\ell m j} = \bar{\mathcal{E}}_{\ell m j}\left(\omega\right) \rho_{s}^{\hat{\ell}}\left(\frac{\rho-\rho_{-}}{\rho-\rho_{+}}\right)^{i\tilde{Z}_{-}^{\left(\sigma\right)}\left(\omega\right)} \\
	&\times\bigg\{ \left(\frac{\rho-\rho_{+}}{\rho_{s}}\right)^{\hat{\ell}}\sum_{k=0}^{\ell}\frac{\left(-\hat{\ell}+i\Gamma^{\left(\sigma\right)}_{+\sigma}\left(\omega\right)\right)_{k}}{\left(-k+\hat{\ell}+1+i\Gamma^{\left(\sigma\right)}_{-\sigma}\right)_{k}}\frac{\left(\ell-k\right)!}{\ell!\,k!}\left(-x\right)^{-k} \\
	&\,\,\,+\underset{\varepsilon\rightarrow0}{\text{Res}}\left\{k_{\ell-\varepsilon, m j}\left(\omega\right)\right\} \left(\frac{\rho-\rho_{+}}{\rho_{s}}\right)^{-\hat{\ell}-1}\sum_{k=0}^{\infty}\frac{\left(\hat{\ell}+1+i\Gamma^{\left(\sigma\right)}_{+\sigma}\left(\omega\right)\right)_{k}}{\left(-k-\hat{\ell}+i\Gamma^{\left(\sigma\right)}_{-\sigma}\right)_{k}}\frac{\left(\ell+1\right)!}{\left(k+\ell+1\right)!\,k!}\left(+x\right)^{-k} \\
	{}&\quad\times\bigg[\log x + \psi\left(k+1\right) + \psi\left(k+\ell+2\right) \\
	&\quad\quad\quad- \psi\left(k+\hat{\ell}+1+i\Gamma^{\left(\sigma\right)}_{+\sigma}\left(\omega\right)\right) - \psi\left(-k-\hat{\ell}+i\Gamma^{\left(\sigma\right)}_{-\sigma}\right)\bigg] \bigg\} \,,
\ea\ee
where we have identified the coefficient multiplying the second term as the residue \eqref{eq:kRes_5dMP} of the response coefficients as $\ell$ approaches an integer number. From the EFT point of view, this residue is precisely the $\beta$-function dictating the classical RG flow of the Love numbers~\cite{Kol:2011vg,Ivanov:2022hlo,Charalambous:2022rre},
\be
	L\frac{d k_{\ell mj}}{dL} = -\frac{\left(-1\right)^{\ell}}{\ell!}\frac{\Gamma\left(\frac{\ell}{2}+1+i\Gamma^{\left(\sigma\right)}_{+\sigma}\left(\omega\right)\right)\Gamma\left(\frac{\ell}{2}+1+i\Gamma^{\left(\sigma\right)}_{-\sigma}\right)}{\Gamma\left(-\frac{\ell}{2}+i\Gamma^{\left(\sigma\right)}_{+\sigma}\left(\omega\right)\right)\Gamma\left(-\frac{\ell}{2}+i\Gamma^{\left(\sigma\right)}_{-\sigma}\right)\left(\ell+1\right)!}\left(\frac{\rho_{+}-\rho_{-}}{\rho_{s}}\right)^{\ell+1} \,.
\ee
This $\beta$-function is evidently real and is therefore entirely associated with the running of the Love numbers, while the dissipative response exhibits no RG flow.

\subsection{Vanishing static Love}
We now turn to the possible resonant conditions for which the $\beta$-function associated with the near-zone Love numbers is zero. Let us start with the case of static scalar Love numbers, which is after all the only regime within which the near-zone approximation is accurate for generic spin parameters. The static scalar Love numbers vanish only if $\hat{\ell}\in\mathbb{N}$ ($\ell$ is an even integer) and
\be\label{eq:VanishingLove_IntL_Static}
	\begin{gathered}
		\Gamma^{\left(\sigma\right)}_{+\sigma}\left(\omega=0\right) = 0 \Rightarrow \frac{j+\sigma m}{2}\left(\Omega_{\psi}+\sigma\Omega_{\phi}\right) = 0 \,, \\
		\text{OR} \\
		\Gamma^{\left(\sigma\right)}_{-\sigma} = 0 \Rightarrow \frac{j-\sigma m}{2}\left(\Omega_{\psi}-\sigma\Omega_{\phi}\right) = 0 \,.
	\end{gathered}
\ee
In other words, these conditions are satisfied if either $\left|a\right|=\left|b\right|$ or $\left|m\right|=\left|j\right|$. The first case describes an equi-rotating Myers-Perry black hole, which has the property of being equipped with an enhanced isometry group $U\left(1\right) \times U\left(1\right) \rightarrow U\left(2\right)$, while the second case can be regarded as a higher-dimensional generalization of ``axisymmetric'' perturbations~\cite{Pani:2015hfa,Gurlebeck:2015xpa,LeTiec:2020bos} that actually includes here non-axisymmetric cases ($m,j\ne0$). This situation is also similar to the higher-dimensional Schwarzschild black holes~\cite{Kol:2011vg}: for integer $\hat{\ell}$, the static scalar Love numbers vanish, while for half-integer $\hat{\ell}$ they are non-zero and exhibit a classical RG flow discussed above. The corresponding static scalar field profile for $\hat{\ell}\in\mathbb{N}$ is given by
\be\ba\label{eq:StaticRadialVanishingLove}
	R_{\omega=0,\ell m j}\bigg|_{\Gamma^{\left(\sigma\right)}_{\pm\sigma}=0} &= \bar{R}_{\ell m j}\left(\omega=0\right)\,\left(\frac{x}{1+x}\right)^{i\Gamma^{\left(\sigma\right)}_{\mp\sigma}/2} \,{}_2F_1\left(\hat{\ell}+1,-\hat{\ell};1+i\Gamma^{\left(\sigma\right)}_{\mp\sigma};-x\right) \\
	&= \bar{R}_{\ell m j}\left(\omega=0\right)\,\left(\frac{x}{1+x}\right)^{i\Gamma^{\left(\sigma\right)}_{\mp\sigma}/2} \sum_{n=0}^{\hat{\ell}} \frac{\left(\hat{\ell}+n\right)!}{\left(\hat{\ell}-n\right)!} \frac{1}{\left(1+i\Gamma^{\left(\sigma\right)}_{\mp\sigma}\right)_{n}}\frac{x^{n}}{n!} \,,
\ea\ee
where we have used the polynomial form of the hypergeometric function whose one of first two parameters is a negative integer and $\left(a\right)_{n}$ is the Pochhammer symbol.

The vanishing of static Love numbers raises naturalness concerns from the point of view of the worldline EFT~\cite{tHooft:1979rat,Porto:2016zng}. In the absence of any selection rules imposed by an enhanced symmetry structure, the Love numbers are expected to be $\mathcal{O}\left(1\right)$ numbers and exhibit running. In contrast, we find here situations where the static Love numbers vanish at all scales and call upon a symmetry explanation to be presented in the next section. Related to this, for generic real values of the orbital number, the ``source'' and ``response'' parts are given by two infinite series in inverse powers of $\rho$, see Eq.~\eqref{eq:SourceReponseSplit_5dMP}. These two series overlap when $\ell$ takes its physical integer values. When the static Love numbers vanish, the final result after summing these two series is the quasi-polynomial form shown above. However, the ``source'' and ``response'' parts are still given by two infinite series which now conspire to give the quasi-polynomial radial solution. In particular, it is these infinite cancellations resulting in the quasi-polynomial form that need to be addressed by a symmetry argument.

\subsection{Non-running Love}
It is also interesting to investigate other situations where $k_{\ell mj}^{\text{Love}}\left(\omega\right)$ are fine tuned. For even $\ell$ (integer $\hat{\ell}$), we find that the scalar Love numbers \eqref{eq:NZSLNs_5dMP} exhibit no RG flow under the conditions
\be\label{eq:VanishingLove_IntL}
	\begin{gathered}
		i\Gamma^{\left(\sigma\right)}_{+\sigma}\left(\omega\right) \in \mathbb{Z} \Rightarrow \omega_{k} = \frac{j+\sigma m}{2}\left(\Omega_{\psi}+\sigma\Omega_{\phi}\right) + \frac{i}{\beta}k \,, \\
		\text{OR} \\
		i\Gamma^{\left(\sigma\right)}_{-\sigma} \in \mathbb{Z} \Rightarrow \frac{j-\sigma m}{2}\left(\Omega_{\psi}-\sigma\Omega_{\phi}\right) = -\frac{i}{\beta}k \,,
	\end{gathered}
\ee
where $k\in\mathbb{Z}$ and we have introduced the inverse Hawking temperature of the $5$-dimensional Myers-Perry black hole,
\be\label{eq:betaMP}
	\beta = \frac{1}{2\pi T_{H}} = \frac{\rho_{s}r_{+}}{\rho_{+}-\rho_{-}} \,.
\ee
In particular, for $-\hat{\ell}\le k\le\hat{\ell}$ the scalar Love numbers vanish identically, while for $k\ge\hat{\ell}+1$ or $k\le-\hat{\ell}-1$, they are non-zero but still exhibit no running.

Similarly, for odd $\ell$ (half-integer $\hat{\ell}$), the conditions read
\be\label{eq:VanishingLove_HIntL}
	\begin{gathered}
		i\Gamma^{\left(\sigma\right)}_{+\sigma}\left(\omega\right) \in \mathbb{Z} + \frac{1}{2} \Rightarrow \omega_{k} = \frac{j+\sigma m}{2}\left(\Omega_{\psi}+\sigma\Omega_{\phi}\right) + \frac{i}{\beta}\left(k+\frac{1}{2}\right) \,, \\
		\text{OR} \\
		i\Gamma^{\left(\sigma\right)}_{-\sigma} \in \mathbb{Z} + \frac{1}{2} \Rightarrow \frac{j-\sigma m}{2}\left(\Omega_{\psi}-\sigma\Omega_{\phi}\right) = -\frac{i}{\beta}\left(k+\frac{1}{2}\right) \,,
	\end{gathered}
\ee
with vanishing Love numbers whenever $-\hat{\ell}\le k+\frac{1}{2}\le \hat{\ell}$.

Regarding the conditions on $\Gamma^{\left(\sigma\right)}_{-\sigma}$, these are in general accompanied by conical singularities in the scalar field profile because they imply imaginary azimuthal numbers which break the periodicity of the scalar field with respect to azimuthal rotations. The only situation where this does not happen is when $\Gamma^{\left(\sigma\right)}_{-\sigma}=0$ for $\hat{\ell}\in\mathbb{N}$.

Despite their unphysical nature in certain cases, the vanishing/non-running of scalar Love numbers beyond the static limit and for all the situations in \eqref{eq:VanishingLove_IntL}-\eqref{eq:VanishingLove_HIntL} is still an interesting result. The corresponding near-zone radial wavefunction for integer $\hat{\ell}$ takes the form
\be\ba\label{eq:NZRadialVanishingLove}
	R_{\omega\ell m j}\bigg|_{i\Gamma^{\left(\sigma\right)}_{\pm\sigma}\left(\omega\right)=k} &= \bar{R}_{\ell m j}\left(\omega\right)\,\left(\frac{x}{1+x}\right)^{i\Gamma^{\left(\sigma\right)}_{\mp\sigma}\left(\omega\right)} \\
	&\left[x\left(1+x\right)\right]^{k/2}\,{}_2F_1\left(\hat{\ell}+1+k,-\hat{\ell}+k;1+k+i\Gamma^{\left(\sigma\right)}_{\mp\sigma}\left(\omega\right);-x\right) \,.
\ea\ee
For $k\le\hat{\ell}$, this again takes a particular quasi-polynomial form to be addressed in the next section via symmetry arguments as well. As we will see, it is the highest-weight property that dictates this quasi-polynomial form. Nevertheless, for $k<-\hat{\ell}$ the polynomial starts developing $\rho^{-\hat{\ell}-1}$ terms. The absence of logarithms indicates that this is not due to an overlapping with PN corrections to the Newtonian source and therefore these are interpreted as non-vanishing and non-running Love numbers.

As we will see in the next section, the highest-weight representation relevant for the properties of the near-zone Love numbers is actually an indecomposable $\SL$ representation of type-``$\circ[\circ[\circ$''. The states ``sandwiched'' between the two highest-weight modes are the ones for which the near-zone Love numbers vanish, while the states spanning the irreducible (lower) highest-weight representation have non-vanishing and non-running near-zone Love numbers. As for the vectors spanning the upper ladder above the reducible (higher) highest-weight representation, these will be spanned by the states characterized by the resonant conditions with $k\ge\hat{\ell}+1$. For half-integer $\hat{\ell}$ for which $i\Gamma^{\left(\sigma\right)}_{+}\left(\omega\right)$ or $i\Gamma^{\left(\sigma\right)}_{-}\left(\omega\right)$ is a half-integer, the same conclusions are drawn after replacing $k\rightarrow k+\frac{1}{2}$.

\section{Love symmetries}
\label{sec:SL2R}

In Chapter~\ref{ch:LoveSymmetry4d}, we have demonstrated the emergence of an enhanced $\SL$ symmetry, dubbed ``Love symmetry'', of the near-zone equations of motion for black hole perturbations. In the context of this symmetry, the vanishing of Love numbers appears as a constraint imposed by the highest-weight Love symmetry representation structure, to which the relevant black hole perturbation solutions belong~\cite{Charalambous:2021kcz,Charalambous:2022rre}. In particular, the highest-weight property dictates a (quasi-)polynomial form of the regular radial wavefunctions which is the behavior indicative of the vanishing of the Love numbers. The intricate structure of the scalar Love numbers for $5$-dimensional Myers-Perry black holes extracted in the previous section sets a good example of examining this hypothesis. In this section, we will demonstrate the existence of $\SL$ structures of the near-zone Klein-Gordon equation for all values of the spin parameters and azimuthal numbers. As we will see, it is only when the scalar Love numbers vanish that the corresponding regular scalar field solution belongs to a highest-weight representation of the corresponding $\SL$ and the vanishing of the scalar Love numbers will be immediately implied through the highest-weight property.

To demonstrate the enhanced symmetry structure of the near-zone equations of motion, it is convenient to introduce the following sum/difference azimuthal angles,
\be
	\psi_{\pm}=\psi\pm\phi \,. 
\ee
The corresponding angular velocities and azimuthal numbers with respect to these two directions are then
\be
	\Omega_{\pm} = \Omega_{\psi} \pm \Omega_{\phi} \,,\quad m_{\pm}= \frac{j\pm m}{2} \,.
\ee

The two near-zone splits \eqref{eq:NZRadial1_5dMP}-\eqref{eq:NZRadial2_5dMP} have the property of each being equipped with an $\SL$ structure. Indeed, for each sign $\sigma=\pm$, we can find a set of three vector fields satisfying the $\SL$ algebra,
\be
	\left[L_{m}^{\left(\sigma\right)},L_{n}^{\left(\sigma\right)}\right] = \left(m-n\right)L_{m+n}^{\left(\sigma\right)} \,,\quad m,n=0,\pm1 \,.
\ee
These generators are given by the vector fields
\be\label{eq:LoveGen}
	\begin{gathered}
		L_0^{\left(\sigma\right)} = -\beta\left(\partial_{t} + \Omega_{\sigma}\,\partial_{\sigma}\right) \,, \\
		L_{\pm1}^{\left(\sigma\right)} = e^{\pm t/\beta}\left[\mp\sqrt{\Delta}\,\partial_{\rho} + \partial_{\rho}\left(\sqrt{\Delta}\right)\,\beta\left(\partial_{t} + \Omega_{\sigma}\,\partial_{\sigma}\right) + \frac{\rho_{+}-\rho_{-}}{2\sqrt{\Delta}} \beta\Omega_{-\sigma}\,\partial_{-\sigma}\right] \,,
	\end{gathered}
\ee
with $\partial_{\pm}\equiv\partial_{\psi_{\pm}}=\left(\partial_{\psi}\pm\partial_{\phi}\right)/2$ and $\beta$ the inverse Hawking temperature \eqref{eq:betaMP} of the $5$-dimensional Myers-Perry black hole. The corresponding Casimirs,
\be\ba\label{eq:LoveCasimir}
	{}&\mathcal{C}_2^{\left(\sigma\right)} = L_0^{\left(\sigma\right)2} - \frac{1}{2}\left(L_{+1}^{\left(\sigma\right)}L_{-1}^{\left(\sigma\right)}+L_{-1}^{\left(\sigma\right)}L_{+1}^{\left(\sigma\right)}\right) \\
	&= \partial_{\rho}\,\Delta\,\partial_{\rho} - \frac{\rho_{s}^2\rho_{+}}{4\Delta}\left(\partial_{t}+\Omega_{+}\,\partial_{+}+\Omega_{-}\,\partial_{-}\right)^2 - \frac{\rho_{+}-\rho_{-}}{\rho-\rho_{-}}\beta^2\,\left(\partial_{t}+\Omega_{\sigma}\,\partial_{\sigma}\right)\Omega_{-\sigma}\,\,\partial_{-\sigma} \,,
\ea\ee
are precisely equal to the two near-zone truncations of the radial operator \eqref{eq:NZRadial1_5dMP}-\eqref{eq:NZRadial2_5dMP} when working in the initial Boyer-Lindquist azimuthal coordinates $\left(\phi,\psi\right)$. We will denote the individual algebras generated for each sign $\sigma$ as $\SL_{\left(\sigma\right)}$. A crucial property of these generators is that they are regular at both the future and the past event horizons as can be seen by switching to the advanced and retarded null coordinates respectively in Eq.~\ref{eq:NullCoordinates_5dMP}.

Solutions of the near-zone equations of motion then form representations of the corresponding $\SL_{\left(\sigma\right)}$ symmetry, labeled by their Casimir eigenvalues, which are equal to the angular eigenvalues $\hat{\ell}(\hat{\ell}+1)$, and the $L_0$-weights~\cite{Miller1968,Howe1992},
\be
	\mathcal{C}_2^{\left(\sigma\right)}\Phi_{\omega\ell mj} = \hat{\ell}(\hat{\ell}+1)\Phi_{\omega\ell mj} \,,\quad L_0^{\left(\sigma\right)}\Phi_{\omega\ell mj} = h^{\left(\sigma\right)}\Phi_{\omega\ell mj} \,.
\ee
We remark, in particular, that
\be
	h^{\left(\sigma\right)}=i\beta\left(\omega-m_{\sigma}\Omega_{\sigma}\right) = -i\Gamma^{\left(\sigma\right)}_{+\sigma}\left(\omega\right) \,.
\ee

\subsection{Highest-weight banishes static Love}
We saw in the previous section that the static scalar Love numbers vanish whenever $m_{\sigma}\Omega_{\sigma}=0$ for $\sigma=+$ or $\sigma=-$ and only for integer $\hat{\ell}$ (see Eq.~\eqref{eq:VanishingLove_IntL_Static}). We will show here that the vanishing of the static Love numbers follows from the fact that the relevant solution of the near-zone equations of motion belongs to a highest-weight representation of $\SL_{\left(\sigma\right)}$.

Let us construct the highest-weight representation of $\SL_{\left(\sigma\right)}$ with highest-weight $h^{\left(\sigma\right)}_{-\hat{\ell},0}=-\hat{\ell}$~\cite{Miller1968,Howe1992,Charalambous:2021kcz,Charalambous:2022rre}. The primary state $\upsilon_{-\hat{\ell},0}^{\left(\sigma\right)}$ satisfies,
\be
	L_0^{\left(\sigma\right)}\upsilon_{-\hat{\ell},0}^{\left(\sigma\right)}=-\hat{\ell}\upsilon_{-\hat{\ell},0}^{\left(\sigma\right)} \,,\quad L_{+1}^{\left(\sigma\right)}\upsilon_{-\hat{\ell},0}^{\left(\sigma\right)} = 0 \,.
\ee
Supplementing with the condition of definite azimuthal numbers,
\be
	J_0^{\left(\pm\right)}\upsilon_{-\hat{\ell},0}^{\left(\sigma\right)} = m_{\pm }\upsilon_{-\hat{\ell},0}^{\left(\sigma\right)} \,,
\ee
where $J_0^{\left(\pm\right)}=-i\partial_{\pm}$ are the two $\mathfrak{so}\left(3\right)$ $J_0$-generators of the rotation group algebra $\mathfrak{so}\left(4\right)\simeq \mathfrak{so}\left(3\right)\oplus\mathfrak{so}\left(3\right)$ (see Appendix~\ref{app:SphericalHarmonics}), we find, up to an overall normalization constant,
\be\label{eq:SL2RHighestL}
	\upsilon_{-\hat{\ell},0}^{\left(\sigma\right)} = \mathcal{F}_{\sigma}\left(\rho\right) e^{im_{\sigma}\left(\psi_{\sigma}-\Omega_{\sigma}t\right)}e^{im_{-\sigma}\psi_{-\sigma}} \left[e^{t/\beta}\sqrt{\Delta}\right]^{\hat{\ell}} \,,
\ee
with the form factor given by
\be
	\mathcal{F}_{\sigma}\left(\rho\right) \equiv \left(\frac{\rho-\rho_{+}}{\rho-\rho_{-}}\right)^{im_{-\sigma}\beta\Omega_{-\sigma}/2} = \left(\frac{\rho-\rho_{+}}{\rho-\rho_{-}}\right)^{i\Gamma^{\left(\sigma\right)}_{-\sigma}/2} \,.
\ee
This highest-weight vector is regular at the future event horizon but singular at the past event horizon. From the regularity of the generators, all the descendants will also be regular at the future event horizon and singular at the past one. For generic parameters, the highest-weight representation is an infinite-dimensional Verma module and is spanned by the vectors~\cite{Howe1992,Miller1968,Miller1970}
\be
	\upsilon_{-\hat{\ell},n}^{\left(\sigma\right)} = \left[L_{-1}^{\left(\sigma\right)}\right]^{n}\upsilon_{-\hat{\ell},0}^{\left(\sigma\right)} \,,\quad n\ge0\,,
\ee
whose charge under $L_0^{\left(\sigma\right)}$ is
\be
	h_{-\hat{\ell},n}^{\left(\sigma\right)} = n-\hat{\ell} \,.
\ee

Let us compare with the properties of the regular static solution with $m_{\sigma}\Omega_{\sigma}=0$ and $\hat{\ell}\in\mathbb{N}$ which corresponds to vanishing static Love numbers. This is a null state of $L_0^{\left(\sigma\right)}$ regular at the future event horizon and is therefore identified as the $n=\hat{\ell}$ descendant in the highest-weight multiplet,
\be
	\Phi_{\omega=0,\ell mj}\bigg|_{m_{\sigma}\Omega_{\sigma}=0} \propto \upsilon_{-\hat{\ell},\hat{\ell}}^{\left(\sigma\right)} = \left[L_{-1}^{\left(\sigma\right)}\right]^{\hat{\ell}}\upsilon_{-\hat{\ell},0}^{\left(\sigma\right)} \,.
\ee
Noticing that, for generic $\upsilon\left(\rho\right)$,
\be\ba\label{eq:Lp1n}
	\left[L_{+1}^{\left(\sigma\right)}\right]^{n}&\left(\mathcal{F}_{\sigma}\left(\rho\right)e^{im_{\sigma}\left(\psi_{\sigma}-\Omega_{\sigma}t\right)}e^{im_{-\sigma}\psi_{-\sigma}}\upsilon\left(\rho\right)\right) = \\
	&\mathcal{F}_{\sigma}\left(\rho\right)e^{im_{\sigma}\left(\psi_{\sigma}-\Omega_{\sigma}t\right)}e^{im_{-\sigma}\psi_{-\sigma}}\left[-e^{t/\beta}\sqrt{\Delta}\right]^{n}\frac{d^{n}}{d\rho^{n}}\upsilon\left(\rho\right) \,,
\ea\ee
we see that the highest-property,
\be
	\left[L_{+1}^{\left(\sigma\right)}\right]^{\hat{\ell}+1}\Phi_{\omega=0,\ell mj}\bigg|_{m_{\sigma}\Omega_{\sigma}=0} = 0 \,,
\ee
dictates a quasi-polynomial form of the near-zone radial wavefunction,
\be
	R_{\omega=0,\ell mj}\bigg|_{m_{\sigma}\Omega_{\sigma}=0} = \mathcal{F}_{\sigma}\left(\rho\right) \sum_{n=0}^{\hat{\ell}}c_{n}\rho^{n} \,.
\ee
This is precisely the quasi-polynomial form \eqref{eq:StaticRadialVanishingLove} that we wanted to address. In conclusion, we have seen how the vanishing of static scalar Love numbers of the $5$-dimensional Myers-Perry black hole is automatically outputted as a selection rule following from the fact that corresponding solution belongs to a highest-weight representation of the near-zone $\SL_{\left(\sigma\right)}$. On the opposite route, the conditions for the regular at the future event horizon static solution to belong to a highest-weight representation are that $\hat{\ell}\in\mathbb{N}$ and that $m_{\sigma}\Omega_{\sigma}=0$, which are precisely the conditions of vanishing static scalar Love numbers.

\begin{figure}[t]
	\centering
	\begin{subfigure}[b]{0.49\textwidth}
		\centering
		\begin{tikzpicture}
			\node at (0,1) (uml3) {$\upsilon^{\left(\sigma\right)}_{-\hat{\ell},\hat{\ell}}$};
			\node at (0,2) (uml2) {$\upsilon^{\left(\sigma\right)}_{-\hat{\ell},2}$};
			\node at (0,3) (uml1) {$\upsilon^{\left(\sigma\right)}_{-\hat{\ell},1}$};
			\node at (0,4) (uml0) {$\upsilon^{\left(\sigma\right)}_{-\hat{\ell},0}$};
			
			\draw (1,1) -- (5,1);
			\node at (3,1.5) (up) {$\vdots$};
			\node at (3,0.5) (um) {$\vdots$};
			\draw (1,2) -- (5,2);
			\draw (1,3) -- (5,3);
			\draw (1,4) -- (5,4);
			\draw [snake=zigzag] (1,4.1) -- (5,4.1);
			
			\draw[red] [<-] (2.5,2) -- node[left] {$L_{-1}^{\left(\sigma\right)}$} (2.5,3);
			\draw[red] [<-] (2,3) -- node[left] {$L_{-1}^{\left(\sigma\right)}$} (2,4);
			\draw[blue] [->] (4,3) -- node[right] {$L_{+1}^{\left(\sigma\right)}$} (4,4);
			\draw[blue] [->] (3.5,2) -- node[right] {$L_{+1}^{\left(\sigma\right)}$} (3.5,3);
		\end{tikzpicture}
		\caption{The highest-weight representation spanned by states $\{\upsilon^{\left(\sigma\right)}_{-\hat{\ell},n}|n\in\mathbb{N}\}$ which are regular (singular) at the future (past) event horizon and have weights $h^{\left(\sigma\right)}_{-\hat{\ell},n}=n-\hat{\ell}$.}
	\end{subfigure}
	\hfill
	\begin{subfigure}[b]{0.49\textwidth}
		\centering
		\begin{tikzpicture}
			\node at (0,3) (upll) {$\bar{\upsilon}^{\left(\sigma\right)}_{+\hat{\ell},\hat{\ell}}$};
			\node at (0,2) (upl2) {$\bar{\upsilon}^{\left(\sigma\right)}_{+\hat{\ell},2}$};
			\node at (0,1) (upl1) {$\bar{\upsilon}^{\left(\sigma\right)}_{+\hat{\ell},1}$};
			\node at (0,0) (upl0) {$\bar{\upsilon}^{\left(\sigma\right)}_{+\hat{\ell},0}$};
			
			\draw [snake=zigzag] (1,-0.1) -- (5,-0.1);
			\draw (1,0) -- (5,0);
			\draw (1,1) -- (5,1);
			\draw (1,2) -- (5,2);
			\node at (3,2.5) (up) {$\vdots$};
			\draw (1,3) -- (5,3);
			\node at (3,3.5) (um) {$\vdots$};
			
			\draw[blue] [->] (2.5,0) -- node[left] {$L_{+1}^{\left(\sigma\right)}$} (2.5,1);
			\draw[blue] [->] (2,1) -- node[left] {$L_{+1}^{\left(\sigma\right)}$} (2,2);
			\draw[red] [<-] (4,1) -- node[right] {$L_{-1}^{\left(\sigma\right)}$} (4,2);
			\draw[red] [<-] (3.5,0) -- node[right] {$L_{-1}^{\left(\sigma\right)}$} (3.5,1);
		\end{tikzpicture}
		\caption{The lowest-weight representation spanned by states $\{\bar{\upsilon}^{\left(\sigma\right)}_{+\hat{\ell},n}|n\in\mathbb{N}\}$ which are singular (regular) at the future (past) event horizon and have weights $\bar{h}^{\left(\sigma\right)}_{+\hat{\ell},n}=\hat{\ell}-n$.}
	\end{subfigure}
	\caption[Infinite-dimensional highest-weight and lowest-weight representations of $\SL_{\left(\sigma\right)}$ containing near-zone solutions for a massless scalar field in the $5$-dimensional Myers-Perry black hole background with multipolar index $\ell$. Whenever $\hat{\ell}\in\mathbb{N}$ and $m_{\sigma}\Omega_{\sigma}=0$, the solution regular (singular) at the future event horizon is the $\hat{\ell}$'th descendant (ascendant) in the highest-weight (lowest-weight) representation.]{Infinite-dimensional highest-weight and lowest-weight representations of $\SL_{\left(\sigma\right)}$ containing near-zone solutions for a massless scalar field in the $5$-dimensional Myers-Perry black hole background with multipolar index $\ell$. Whenever $\hat{\ell}\in\mathbb{N}$ and $m_{\sigma}\Omega_{\sigma}=0$, the solution regular (singular) at the future event horizon is the $\hat{\ell}$'th descendant (ascendant) in the highest-weight (lowest-weight) representation.}
	\label{fig:SL2R_HW_LW}
\end{figure}
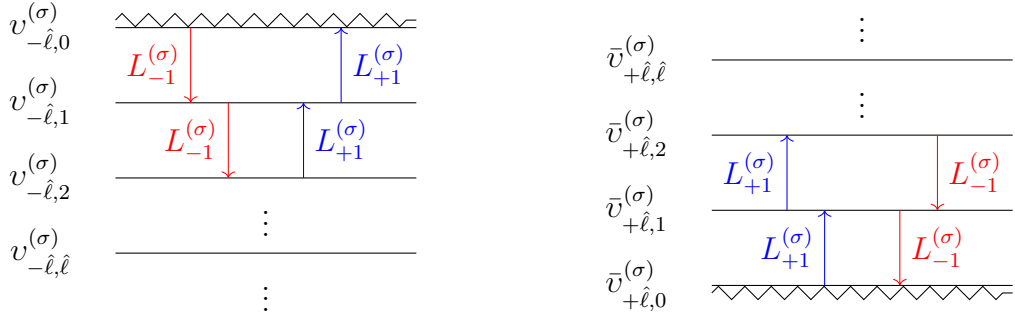

Let us briefly comment on the structure of the lowest-weight representation of $\SL_{\left(\sigma\right)}$, spanned by ascendants $\bar{\upsilon}_{+\hat{\ell},n}^{\left(\sigma\right)}$ of a lowest-weight state $\bar{\upsilon}_{+\hat{\ell},0}^{\left(\sigma\right)}$,
\be
	\bar{\upsilon}_{+\hat{\ell},n}^{\left(\sigma\right)} = \left[-L_{+1}^{\left(\sigma\right)}\right]^{n}\bar{\upsilon}_{+\hat{\ell},0}^{\left(\sigma\right)} \,,
\ee
satisfying
\be
	L_0^{\left(\sigma\right)}\bar{\upsilon}_{+\hat{\ell},0}^{\left(\sigma\right)}=+\hat{\ell}\bar{\upsilon}_{+\hat{\ell},0}^{\left(\sigma\right)} \,,\quad L_{-1}^{\left(\sigma\right)}\bar{\upsilon}_{+\hat{\ell},0}^{\left(\sigma\right)} = 0 \,,
\ee
and having definite azimuthal numbers. We find
\be\label{eq:SL2RLowestL}
	\bar{\upsilon}_{+\hat{\ell},0}^{\left(\sigma\right)} = \bar{\mathcal{F}}_{\sigma}\left(\rho\right) e^{im_{\sigma}\left(\psi_{\sigma}-\Omega_{\sigma}t\right)}e^{im_{-\sigma}\psi_{-\sigma}} \left[e^{-t/\beta}\sqrt{\Delta}\right]^{\hat{\ell}} \,,\quad \bar{\mathcal{F}}_{\sigma}\left(\rho\right) \equiv \left(\frac{\rho-\rho_{+}}{\rho-\rho_{-}}\right)^{-i\Gamma^{\left(\sigma\right)}_{+\sigma}/2} \,.
\ee
This state is a solution of the near-zone equations of motion that is regular at the past event horizon, but singular at the future event horizon. As a result, from the regularity of the generators, all the ascendants will also be regular at the past event horizon and singular at the future one, with $L_0$-charges $\bar{h}^{\left(\sigma\right)}_{\hat{\ell},n}=\hat{\ell}-n$. In particular, when $\hat{\ell}$ is an integer, the $n=\hat{\ell}$ ascendant $\bar{\upsilon}_{+\hat{\ell},\hat{\ell}}^{\left(\sigma\right)}$ will be the static solution of the Klein-Gordon equation with $m_{\sigma}\Omega_{\sigma}=0$ that is singular at the future event horizon. We have just demonstrated that the static solutions regular and singular at the future event horizon belong to different, locally distinguishable, representations of $\SL_{\left(\sigma\right)}$; the highest-weight representation and lowest-weight representation respectively. This is the algebraic manifestation of the absence of RG flow for the static scalar Love numbers in the particular case of $\hat{\ell}\in\mathbb{N}$ and $m_{\sigma}\Omega_{\sigma}=0$. The construction of these highest- and lowest-representations is demonstrated graphically in Figure~\ref{fig:SL2R_HW_LW}.

\subsection{Highest-weight banishes near-zone Love}
Interestingly, the near-zone symmetries can address all the situations \eqref{eq:VanishingLove_IntL}-\eqref{eq:VanishingLove_HIntL}, even the unphysical ones.

If $\hat{\ell}$ is an integer and $i\Gamma^{\left(\sigma\right)}_{+\sigma}\left(\omega\right)=k\in\mathbb{Z}$, then the corresponding near-zone solution has the form
\be
	\Phi_{\omega\ell mj}\bigg|_{i\Gamma^{\left(\sigma\right)}_{+\sigma}\left(\omega\right)=k} \propto e^{kt/\beta}e^{im_{\sigma}\left(\psi_{\sigma}-\Omega_{\sigma}t\right)}e^{im_{-\sigma}\psi_{-\sigma}}R_{\omega\ell mj}\left(\rho\right)\bigg|_{i\Gamma^{\left(\sigma\right)}_{+\sigma}\left(\omega\right)=k} \,,
\ee
where we are suppressing the $\theta$-dependence, and therefore satisfies
\be
	L_0^{\left(\sigma\right)}\Phi_{\omega\ell mj}\bigg|_{i\Gamma^{\left(\sigma\right)}_{+\sigma}\left(\omega\right)=k} = -k\,\Phi_{\omega\ell mj}\bigg|_{i\Gamma^{\left(\sigma\right)}_{+\sigma}\left(\omega\right)=k} \,.
\ee
If $k\le\hat{\ell}$, the regular at the future event horizon solution is then recognized to be the $n=\hat{\ell}-k$ descendant,
\be
	\Phi_{\omega\ell mj}\bigg|_{i\Gamma^{\left(\sigma\right)}_{+\sigma}\left(\omega\right)=k\le\hat{\ell}} \propto \upsilon_{-\hat{\ell},\hat{\ell}-k}^{\left(\sigma\right)} = \left[L_{-1}^{\left(\sigma\right)}\right]^{\hat{\ell}-k}\upsilon_{-\hat{\ell},0}^{\left(\sigma\right)} \,.
\ee
The highest-property,
\be
	\left[L_{+1}^{\left(\sigma\right)}\right]^{\hat{\ell}-k+1}\Phi_{\omega\ell mj}\bigg|_{i\Gamma^{\left(\sigma\right)}_{+\sigma}\left(\omega\right)=k\le\hat{\ell}} = 0 \,,
\ee
does not directly imply any useful quasi-polynomial form itself. Rather, it is the fact that
\be
	\upsilon_{-\hat{\ell},\hat{\ell}-k}^{\left(\sigma\right)} =
	\begin{cases}
		\left(-1\right)^{k}\left[L_{+1}^{\left(\sigma\right)}\right]^{k}\upsilon_{-\hat{\ell},\hat{\ell}}^{\left(\sigma\right)} \quad \text{for }0\le k \le \hat{\ell} \\
		\left[L_{-1}^{\left(\sigma\right)}\right]^{-k}\upsilon_{-\hat{\ell},\hat{\ell}}^{\left(\sigma\right)} \quad\quad\quad\, \text{for }k<0
	\end{cases}
\ee
and the quasi-polynomial form of the static element $\upsilon_{-\hat{\ell},\hat{\ell}}^{\left(\sigma\right)}$ that give a useful expression for the relevant near-zone solution. From \eqref{eq:Lp1n}, and an analogous relation for the action of $L_{-1}^{\left(\sigma\right)}$ we conclude that
\be
	R_{\omega\ell mj}\left(\rho\right)\bigg|_{i\Gamma^{\left(\sigma\right)}_{+\sigma}\left(\omega\right)=k\le\hat{\ell}} = \mathcal{F}_{\sigma}\left(\rho\right) \Delta^{k/2} \sum_{n=0}^{\hat{\ell}-k}c_{n}\rho^{n} \,,
\ee
which has the exact quasi-polynomial form of \eqref{eq:NZRadialVanishingLove} with $k\le\hat{\ell}$.

Another remark here is there are in general two possible highest-weight representations of $\SL$; one has $h^{\left(\sigma\right)}=-\hat{\ell}$, which is the one we saw, and the other has $h^{\left(\sigma\right)}=+\hat{\ell}+1$. Both highest-weight representations contain solutions that are regular at the future event horizon. Since the weights differ by the integer amount $2\hat{\ell}+1$, we see that the primary state with $h^{\left(\sigma\right)}=+\hat{\ell}+1$ is actually a descendant of the primary state with $h^{\left(\sigma\right)}=-\hat{\ell}$,
\be
	\upsilon_{+\hat{\ell}+1,0} = \left[L_{-1}^{\left(\sigma\right)}\right]^{2\hat{\ell}+1}\upsilon_{-\hat{\ell},0} \,.
\ee
Therefore, the highest-weight Verma module of $\SL_{\left(\sigma\right)}$ we have been working with so far is actually of type-``$[\circ[\circ$''. With this fact in hand, we see an interesting algebraic interpretation of the elements of this representation: The irreducible (lower) highest-weight is spanned by states with non-vanishing but also non-running near-zone scalar Love numbers, while the quotient representation sandwiched between the two primary states is spanned by all the possible near-zone solutions regular at the future event horizon that have vanishing scalar Love numbers.

Let us now supplement with the case where $k\ge\hat{\ell}+1$ for which the scalar Love numbers are not zero but still exhibit no running. For the sake of this, we need to look into states $\{\upsilon^{\left(\sigma\right)}_{-\hat{\ell},-\left(n+1\right)}|n\in\mathbb{N}\}$ that span the ladder above this highest-weight representation. These states are ascendants of the state $\upsilon^{\left(\sigma\right)}_{-\hat{\ell},-1}$,
\be
	\upsilon^{\left(\sigma\right)}_{-\hat{\ell},-\left(n+1\right)} = \frac{1}{n!\left(2\hat{\ell}+2\right)_{n}}\left[L_{+1}^{\left(\sigma\right)}\right]^{n}\upsilon^{\left(\sigma\right)}_{-\hat{\ell},-1} \,,
\ee
which satisfies
\be
	L_0^{\left(\sigma\right)}\upsilon^{\left(\sigma\right)}_{-\hat{\ell},-1} = -(\hat{\ell}+1)\upsilon^{\left(\sigma\right)}_{-\hat{\ell},-1} \,,\quad L_{-1}^{\left(\sigma\right)}\upsilon^{\left(\sigma\right)}_{-\hat{\ell},-1} = \upsilon^{\left(\sigma\right)}_{-\hat{\ell},0} \,.
\ee
The resulting first-order inhomogeneous differential equation can be solved to get
\be\ba
	\upsilon^{\left(\sigma\right)}_{-\hat{\ell},-1} &= \frac{\mathcal{F}_{\sigma}\left(\rho\right)e^{im_{\sigma}\left(\psi_{\sigma}-\Omega_{\sigma}t\right)}e^{im_{-\sigma}\Omega_{-\sigma}}}{\left(\rho_{+}-\rho_{-}\right)\left(\hat{\ell}+1+i\Gamma^{\left(\sigma\right)}_{-\sigma}\right)}\left(e^{t/\beta}\sqrt{\Delta}\right)^{\hat{\ell}+1} \\
	&\quad\quad\times{}_2F_1\left(1,2\hat{\ell}+2;2+\hat{\ell}+i\Gamma^{\left(\sigma\right)}_{-\sigma};-\frac{\rho-\rho_{+}}{\rho_{+}-\rho_{-}}\right) \,,
\ea\ee
which is regular at the future event horizon and singular at the past one. Consequently, all the ascendants $\upsilon^{\left(\sigma\right)}_{-\hat{\ell},-\left(n+1\right)}$, with $n\in\mathbb{N}$ will also be regular at the future event horizon near-zone solutions, with $L_0$-eigenvalues $h^{\left(\sigma\right)}_{-\hat{\ell},-\left(n+1\right)}=-(n+\hat{\ell}+1)$. From this, we therefore identify
\be
	\Phi_{\omega\ell mj}\bigg|_{i\Gamma^{\left(\sigma\right)}_{+\sigma}\left(\omega\right)=k\ge\hat{\ell}+1} \propto \upsilon_{-\hat{\ell},-\left(k-\hat{\ell}\right)}^{\left(\sigma\right)} \propto \left[L_{+1}^{\left(\sigma\right)}\right]^{k-\hat{\ell}-1}\upsilon_{-\hat{\ell},-1}^{\left(\sigma\right)} \,.
\ee
As already discussed around \eqref{eq:NZRadialVanishingLove}, we do not expect to find any conspiring quasi-polynomial solution for $k\ge\hat{\ell}+1$. However, we have supplemented with the algebraic property of these remaining states to span the \textit{entire} type-``$\circ[\circ[\circ$'' representation of $\SL_{\left(\sigma\right)}$ for which the highest-weight state has weight $h^{\left(\sigma\right)}_{-\hat{\ell},0}=-\hat{\ell}$ (see Chapter~\ref{ch:LoveSymmetry4d} or Ref.~\cite{Howe1992} for our notation).

As for the near-zone solutions singular at the future event horizon, these can be similarly worked out to span the entire type-``$\circ]\circ]\circ$'' representation of $\SL_{\left(\sigma\right)}$ for which the lowest-weight state has weight $\bar{h}^{\left(\sigma\right)}_{+\hat{\ell},0}=+\hat{\ell}$, thus providing us with an algebraic argument of the absence of RG flow of the Love numbers. These constructions are shown graphically in Figure~\ref{fig:SL2R_oHW_oLW}.

\begin{figure}[t]
	\centering
	\begin{subfigure}[b]{0.49\textwidth}
		\centering
		\begin{tikzpicture}
			\node at (0,-1) (uml4) {$\upsilon^{\left(\sigma\right)}_{-\hat{\ell},2\hat{\ell}+2}$};
			\node at (0,0) (uml3) {$\upsilon^{\left(\sigma\right)}_{-\hat{\ell},2\hat{\ell}+1}$};
			\node at (0,1) (uml2) {$\upsilon^{\left(\sigma\right)}_{-\hat{\ell},2\hat{\ell}}$};
			\node at (0,2) (uml1) {$\upsilon^{\left(\sigma\right)}_{-\hat{\ell},1}$};
			\node at (0,3) (uml0) {$\upsilon^{\left(\sigma\right)}_{-\hat{\ell},0}$};
			\node at (0,4) (umlm1) {$\upsilon^{\left(\sigma\right)}_{-\hat{\ell},-1}$};
			\node at (0,5) (umlm2) {$\upsilon^{\left(\sigma\right)}_{-\hat{\ell},-2}$};
			
			\node at (3,-1.5) (um) {$\vdots$};
			\draw (1,-1) -- (5,-1);
			\draw [snake=zigzag] (1,0.1) -- (5,0.1);
			\draw (1,0) -- (5,0);
			\draw (1,1) -- (5,1);
			\node at (3,1.5) (up) {$\vdots$};
			\draw (1,2) -- (5,2);
			\draw (1,3) -- (5,3);
			\draw (1,4) -- (5,4);
			\draw [snake=zigzag] (1,3.1) -- (5,3.1);
			\draw (1,5) -- (5,5);
			\node at (3,5.5) (up) {$\vdots$};
			
			\draw[red] [<-] (3,3) -- node[left] {$L_{-1}^{\left(\sigma\right)}$} (3,4);
			\draw[red] [<-] (2,2) -- node[left] {$L_{-1}^{\left(\sigma\right)}$} (2,3);
			\draw[red] [<-] (3,0) -- node[left] {$L_{-1}^{\left(\sigma\right)}$} (3,1);
			\draw[red] [<-] (2,-1) -- node[left] {$L_{-1}^{\left(\sigma\right)}$} (2,0);
			\draw[blue] [->] (4,-1) -- node[right] {$L_{+1}^{\left(\sigma\right)}$} (4,0);
			\draw[blue] [->] (4,2) -- node[right] {$L_{+1}^{\left(\sigma\right)}$} (4,3);
			\draw[red] [<-] (2,4) -- node[left] {$L_{-1}^{\left(\sigma\right)}$} (2,5);
			\draw[blue] [->] (4,4) -- node[right] {$L_{+1}^{\left(\sigma\right)}$} (4,5);
		\end{tikzpicture}
		\caption{The type ``$\circ[\circ[\circ$'' representation spanned by states $\{\upsilon^{\left(\sigma\right)}_{-\hat{\ell},j}|j\in\mathbb{Z}\}$ which are regular (singular) at the future (past) event horizon and have weights $h^{\left(\sigma\right)}_{-\hat{\ell},j}=j-\hat{\ell}$.}
	\end{subfigure}
	\hfill
	\begin{subfigure}[b]{0.49\textwidth}
		\centering
		\begin{tikzpicture}
			\node at (0,4) (upl4) {$\bar{\upsilon}^{\left(\sigma\right)}_{+\hat{\ell},2\hat{\ell}+2}$};
			\node at (0,3) (upl3) {$\bar{\upsilon}^{\left(\sigma\right)}_{+\hat{\ell},2\hat{\ell}+1}$};
			\node at (0,2) (upl2) {$\bar{\upsilon}^{\left(\sigma\right)}_{+\hat{\ell},2\hat{\ell}}$};
			\node at (0,1) (upl1) {$\bar{\upsilon}^{\left(\sigma\right)}_{+\hat{\ell},1}$};
			\node at (0,0) (upl0) {$\bar{\upsilon}^{\left(\sigma\right)}_{+\hat{\ell},0}$};
			\node at (0,-1) (uplm1) {$\bar{\upsilon}^{\left(\sigma\right)}_{+\hat{\ell},-1}$};
			\node at (0,-2) (uplm2) {$\bar{\upsilon}^{\left(\sigma\right)}_{+\hat{\ell},-2}$};
			
			\draw [snake=zigzag] (1,-0.1) -- (5,-0.1);
			\draw (1,0) -- (5,0);
			\draw (1,1) -- (5,1);
			\node at (3,1.5) (up) {$\vdots$};
			\draw (1,2) -- (5,2);
			\node at (3,4.5) (um) {$\vdots$};
			\draw [snake=zigzag] (1,2.9) -- (5,2.9);
			\draw (1,3) -- (5,3);
			\draw (1,4) -- (5,4);
			\draw (1,-1) -- (5,-1);
			\draw (1,-2) -- (5,-2);
			\node at (3,-2.5) (up) {$\vdots$};
			
			\draw[blue] [->] (3,-1) -- node[left] {$L_{+1}^{\left(\sigma\right)}$} (3,0);
			\draw[blue] [->] (2,0) -- node[left] {$L_{+1}^{\left(\sigma\right)}$} (2,1);
			\draw[blue] [->] (3,2) -- node[left] {$L_{+1}^{\left(\sigma\right)}$} (3,3);
			\draw[blue] [->] (2,3) -- node[left] {$L_{+1}^{\left(\sigma\right)}$} (2,4);
			\draw[red] [<-] (4,3) -- node[right] {$L_{-1}^{\left(\sigma\right)}$} (4,4);
			\draw[red] [<-] (4,0) -- node[right] {$L_{-1}^{\left(\sigma\right)}$} (4,1);
			\draw[blue] [->] (2,-1) -- node[left] {$L_{+1}^{\left(\sigma\right)}$} (2,-2);
			\draw[red] [<-] (4,-1) -- node[right] {$L_{-1}^{\left(\sigma\right)}$} (4,-2);
		\end{tikzpicture}
		\caption{The type ``$\circ]\circ]\circ$'' representation spanned by states $\{\bar{\upsilon}^{\left(\sigma\right)}_{+\hat{\ell},j}|j\in\mathbb{Z}\}$ which are singular (regular) at the future (past) event horizon and have weights $\bar{h}^{\left(\sigma\right)}_{+\hat{\ell},j}=\hat{\ell}-j$.}
	\end{subfigure}
	\caption[The type-``$\circ\mathrm{[}\circ\mathrm{[}\circ$'' and type-``$\circ\mathrm{]}\circ\mathrm{]}\circ$'' representations of $\SL_{\left(\sigma\right)}$ containing all the near-zone solutions for a massless scalar field in the $5$-dimensional Myers-Perry black hole background that have vanishing/non-running Love numbers regarding the conditions on $\Gamma^{\left(\sigma\right)}_{+\sigma}$.]{The type ``$\circ[\circ[\circ$'' and type ``$\circ]\circ]\circ$'' representations of $\SL_{\left(\sigma\right)}$ containing all the near-zone solutions for a massless scalar field in the $5$-dimensional Myers-Perry black hole background that have vanishing/non-running Love numbers regarding the conditions on $\Gamma^{\left(\sigma\right)}_{+\sigma}$ (see \eqref{eq:VanishingLove_IntL}-\eqref{eq:VanishingLove_HIntL}). For integer $\hat{\ell}$, the relevant condition is $i\Gamma^{\left(\sigma\right)}_{+\sigma}=k\in\mathbb{Z}$ and is depicted above. For half-integer $\hat{\ell}$, the condition becomes $i\Gamma^{\left(\sigma\right)}_{+\sigma}=k+\frac{1}{2}$, $k\in\mathbb{Z}$, and the structure of the representations is as in the above figures after replacing $k\rightarrow k+\frac{1}{2}$.}
	\label{fig:SL2R_oHW_oLW}
\end{figure}

Last, if $\hat{\ell}$ is a half-integer and $i\Gamma^{\left(\sigma\right)}_{+\sigma}\left(\omega\right)=k+\frac{1}{2}$, with $k\in\mathbb{Z}$, then we are looking at near-zone solutions of the form
\be
	\Phi_{\omega\ell mj}\bigg|_{i\Gamma^{\left(\sigma\right)}_{+\sigma}\left(\omega\right)=k+\frac{1}{2}} \propto e^{\left(k+1/2\right)t/\beta}e^{im_{\sigma}\left(\psi_{\sigma}-\Omega_{\sigma}t\right)}e^{im_{-\sigma}\psi_{-\sigma}}R_{\omega\ell mj}\left(\rho\right)\bigg|_{i\Gamma^{\left(\sigma\right)}_{+\sigma}\left(\omega\right)=k+\frac{1}{2}} \,,
\ee
which satisfy
\be
	L_0^{\left(\sigma\right)}\Phi_{\omega\ell mj}\bigg|_{i\Gamma^{\left(\sigma\right)}_{+\sigma}\left(\omega\right)=k+\frac{1}{2}} = -\left(k+\frac{1}{2}\right)\Phi_{\omega\ell mj}\bigg|_{i\Gamma^{\left(\sigma\right)}_{+\sigma}\left(\omega\right)=k+\frac{1}{2}} \,.
\ee
The above analysis is then carried away identically, after replacing $k\rightarrow k+\frac{1}{2}$.

\subsection{Local near-zone $\SL\times\SL$}
Somewhat surprisingly, we can address the remaining situations where $i\Gamma^{\left(\sigma\right)}_{-\sigma}\in\mathbb{Z}$ for integer $\hat{\ell}$ or $i\Gamma^{\left(\sigma\right)}_{-\sigma}\in\mathbb{Z}+\frac{1}{2}$ for half-integer $\hat{\ell}$, even for $\omega\ne0$. This is ought to the observation that the particular near-zone truncations are equipped with a larger $\SL_{\left(\sigma\right),\text{L}}\times\SL_{\left(\sigma\right),\text{R}}$ structure. The first $\SL$ factor is the Love symmetry $\SL$,
\be
	\SL_{\left(\sigma\right),\text{L}}=\SL_{\left(\sigma\right)} \,,
\ee
generated by the globally defined vector fields \eqref{eq:LoveGen}, $L_{m}^{\left(\sigma\right),\text{L}}=L_{m}^{\left(\sigma\right)}$, $m=0,\pm1$. The second, $\SL_{\left(\sigma\right),\text{R}}$, factor is generated by the following vector fields
\be\label{eq:LoveGenSpBroken}
	\begin{gathered}
		L_0^{\left(\sigma\right),\text{R}} = -\beta\Omega_{-\sigma}\,\partial_{-\sigma} \,, \\
		L_{\pm1}^{\left(\sigma\right),\text{R}} = e^{\pm\psi_{-\sigma}/\left(\beta\Omega_{-\sigma}\right)}\left[\mp\sqrt{\Delta}\partial_{\rho}+\partial_{\rho}\left(\sqrt{\Delta}\right)\beta\Omega_{-\sigma}\,\partial_{-\sigma}+\frac{\rho_{+}-\rho_{-}}{2\sqrt{\Delta}}\beta\left(\partial_{t}+\Omega_{\sigma}\,\partial_{\sigma}\right)\right] \,.
	\end{gathered}
\ee
The Casimirs of the two commuting $\SL$'s are exactly the same,
\be
	\mathcal{C}_2^{\left(\sigma\right),\text{R}} = \mathcal{C}_2^{\left(\sigma\right),\text{L}} = \mathcal{C}_2^{\left(\sigma\right)} \,.
\ee
In addition, the sets of vector fields generating the two $\SL$'s are regular at both the future and the past event horizons with respect to the radial variable, i.e. they do not develop poles as $\rho\rightarrow \rho_{+}$. However, $\SL_{\left(\sigma\right),\text{R}}$ is spontaneously broken down to $U\left(1\right)_{\left(\sigma\right),\text{R}}$ by the periodic identification of the azimuthal angles, $\psi_{\pm}\sim\psi_{\pm}+2\pi$, under which $L_{\pm1}^{\left(\sigma\right),\text{R}}$ are singular.

Despite this breaking of $\SL_{\left(\sigma\right),\text{R}}$ by the periodic identification of the azimuthal angles, it can still be used to explain the vanishing/non-running corresponding to the situations where $i\Gamma^{\left(\sigma\right)}_{-\sigma}\in\mathbb{Z}$ for integer $\hat{\ell}$ or $i\Gamma^{\left(\sigma\right)}_{-\sigma}\in\mathbb{Z}+\frac{1}{2}$ for half-integer $\hat{\ell}$ in a similar fashion as in the previous subsection. Solutions of the near-zone equations of motion will furnish representations labeled by the Casimir and $L_0$-eigenvalues,
\be
	\mathcal{C}_2^{\left(\sigma\right),\text{R}}\Phi_{\omega\ell mj}=\hat{\ell}(\hat{\ell}+1)\Phi_{\omega\ell mj} \,,\quad L_0^{\left(\sigma\right),\text{R}}\Phi_{\omega\ell mj} = h^{\left(\sigma\right),\text{R}}\Phi_{\omega\ell mj} \,,
\ee
with
\be
	h^{\left(\sigma\right),\text{R}} = -i\beta m_{-\sigma}\Omega_{-\sigma} = -i\Gamma^{\left(\sigma\right)}_{-\sigma} \,.
\ee
Let us construct the analogous highest-weight representation of $\SL_{\left(\sigma\right),\text{R}}$. The primary state with highest-weight $h_{-\hat{\ell},0}^{\left(\sigma\right),\text{R}}=-\hat{\ell}$, satisfying
\be
	L_0^{\left(\sigma\right),\text{R}}\upsilon_{-\hat{\ell},0}^{\left(\sigma\right),\text{R}} = -\hat{\ell}\upsilon_{-\hat{\ell},0}^{\left(\sigma\right),\text{R}} \,,\quad 	L_{+1}^{\left(\sigma\right),\text{R}}\upsilon_{-\hat{\ell},0}^{\left(\sigma\right),\text{R}} = 0 \,,
\ee
and having definite azimuthal numbers and frequency is given by, up to an overall normalization constant,
\be
	\upsilon_{-\hat{\ell},0}^{\left(\sigma\right),\text{R}} = \mathcal{F}_{\sigma}^{\text{R}}\left(\rho\right)e^{-i\omega t}e^{im_{\sigma}\psi_{\sigma}}e^{\hat{\ell}\psi_{-\sigma}/\left(\beta\Omega_{-\sigma}\right)}\Delta^{\hat{\ell}/2} \,, \quad \mathcal{F}_{\sigma}^{\text{R}}\left(\rho\right) \equiv \left(\frac{\rho-\rho_{+}}{\rho-\rho_{-}}\right)^{i\Gamma^{\left(\sigma\right)}_{+\sigma}\left(\omega\right)/2} \,.
\ee
This state is singular at the past event horizon and has a conical singularity as we go around the azimuthal circles, but develops no pole at the future event horizon with respect to the radial variable. The descendants,
\be
	\upsilon_{-\hat{\ell},n}^{\left(\sigma\right),\text{R}} = \left[L_{+1}^{\left(\sigma\right),\text{R}}\right]^{n}\upsilon_{-\hat{\ell},0}^{\left(\sigma\right),\text{R}} \,,
\ee
share the same boundary conditions, with the conical singularity measured by their charge under $L_0^{\left(\sigma\right),\text{R}}$,
\be
	h_{-\hat{\ell},n}^{\left(\sigma\right),\text{R}} = n-\hat{\ell} \,.
\ee

If $\hat{\ell}\in\mathbb{N}$, there exists a particular descendant that has no conical singularity and is therefore truly regular at the future event horizon. This is the $n=\hat{\ell}$ descendant which is a null state under $L_0^{\left(\sigma\right),\text{R}}$ and corresponds to the regular near-zone solution with $\Gamma^{\left(\sigma\right)}_{+\sigma}=0$. Noticing that
\be\ba
	\left[L_{+1}^{\left(\sigma\right)}\right]^{n}&\left(\mathcal{F}_{\sigma}^{\text{R}}\left(\rho\right)e^{i\omega t}e^{im_{\sigma}\psi_{\sigma}}\upsilon\left(\rho\right)\right) = \\
	&\mathcal{F}_{\sigma}^{\text{R}}\left(\rho\right)e^{i\omega t}e^{im_{\sigma}\psi_{\sigma}}\left[-e^{\psi_{-\sigma}/\left(\beta\Omega_{-\sigma}\right)}\sqrt{\Delta}\right]^{n}\frac{d^{n}}{d\rho^{n}}\upsilon\left(\rho\right) \,,
\ea\ee
for generic $\upsilon\left(\rho\right)$, we see that the highest-weight property,
\be
	\left[L_{+1}^{\left(\sigma\right),\text{R}}\right]^{\hat{\ell}+1}\Phi_{\omega\ell mj}\bigg|_{\Gamma^{\left(\sigma\right)}_{-\sigma}=0} = 0 \,,
\ee
implies the following quasi-polynomial form for the radial wavefunction
\be
	R_{\omega\ell mj}\bigg|_{\Gamma^{\left(\sigma\right)}_{-\sigma}=0} = \mathcal{F}_{\sigma}^{\text{R}}\left(\rho\right)\sum_{n=0}^{\hat{\ell}}c_{n}\rho^{n} \,,
\ee
which exactly matches the one from \eqref{eq:StaticRadialVanishingLove} we wanted to address. The absence of RG flow is also encoded in the representation theory analysis of $\SL_{\left(\sigma\right),\text{R}}$, with the solution singular at the future event horizon (and regular at the past event horizon) being the $n=\hat{\ell}$ ascendant of the locally distinguishable lowest-weight representation of $\SL_{\left(\sigma\right),\text{R}}$ with lowest-weight $\bar{h}_{+\hat{\ell},0}^{\left(\sigma\right),\text{R}}=+\hat{\ell}$.

For the other situations of vanishing/non-running scalar Love numbers that regard the conditions on $\Gamma^{\left(\sigma\right)}_{-\sigma}$ for which the relevant near-zone solutions develop conical singularities, we can follow the same procedure as in the previous subsection and show that all the relevant regular (singular) at the future event horizon near-zone solutions span the entire representation of type-``$\circ[\circ[\circ$'' (type-``$\circ]\circ]\circ$'').

Last, if $\hat{\ell}$ is a half-integer and $i\Gamma^{\left(\sigma\right)}_{+\sigma}\left(\omega\right)=k+\frac{1}{2}$, with $k\in\mathbb{Z}$, then we are looking at near-zone solutions of the form
\be
	\Phi_{\omega\ell mj}\bigg|_{i\Gamma^{\left(\sigma\right)}_{+\sigma}\left(\omega\right)=k+\frac{1}{2}} \propto e^{\left(k+1/2\right)t/\beta}e^{im_{\sigma}\left(\psi_{\sigma}-\Omega_{\sigma}t\right)}e^{im_{-\sigma}\psi_{-\sigma}}R_{\omega\ell mj}\left(\rho\right)\bigg|_{i\Gamma^{\left(\sigma\right)}_{+\sigma}\left(\omega\right)=k+\frac{1}{2}} \,,
\ee
which satisfy
\be
	L_0^{\left(\sigma\right)}\Phi_{\omega\ell mj}\bigg|_{i\Gamma^{\left(\sigma\right)}_{+\sigma}\left(\omega\right)=k+\frac{1}{2}} = -\left(k+\frac{1}{2}\right)\Phi_{\omega\ell mj}\bigg|_{i\Gamma^{\left(\sigma\right)}_{+\sigma}\left(\omega\right)=k+\frac{1}{2}} \,.
\ee
The above analysis is then carried away identically, after replacing $k\rightarrow k+\frac{1}{2}$.

\section{Properties}
\label{sec:Properties}

In this section we will present a number of interesting properties of the Love symmetries. First, we will show how the near-zone $\SL$ symmetries acquire a geometric interpretation as isometries of effective geometries within the framework of subtracted geometries~\cite{Cvetic:2011dn,Cvetic:2011hp}. Then, we will demonstrate how both near-zone symmetries can be realized as subalgebras of a larger, infinite-dimensional $\SL\ltimes \hat{U}\left(1\right)^2_{\mathcal{V}}$ extension, which is interpreted as the $5$-dimensional version of the $\SL\ltimes \hat{U}\left(1\right)_{\mathcal{V}}$ infinite extension of the Love symmetry for Kerr-Newman black holes encountered in Chapter~\ref{ch:Properties}~\cite{Charalambous:2021kcz,Charalambous:2022rre}. We will close this section with another interesting generalization of the near-zone symmetries which will exhaust all the possible near-zone truncations of the equations of motion that are equipped with an enhanced $\SL$ symmetry and acquire a subtracted geometry interpretation.

\subsection{Near-zone symmetries as isometries of subtracted geometries}
To introduce the notion of subtracted geometries for $5$-dimensional black holes, we begin by writing the geometry of the $5$-dimensional Myers-Perry black hole as a fibration over a $4$-d base space~\cite{Chong:2006zx,Cvetic:2011hp},
\be
	\begin{gathered}
		ds^2 = -\Delta_0^{-2/3}G\left(dt+\mathcal{A}\right)^2 + \Delta_0^{1/3}ds_4^2 \,, \\
		ds_4^2 = \frac{d\rho^2}{4X} + \,d\theta^2 + \frac{1}{4}\sum_{i,j=1}^{2}\gamma_{ij}d\phi^{i}d\phi^{j} \,,
	\end{gathered}
\ee
where small Latin indices label the azimuthal angles, with $\phi^1\equiv\phi$ and $\phi^2\equiv\psi$. In the notation more conventionally used to write down the line element of the geometry in Eq.~\eqref{eq:Metric_5dMP},
\be
	\begin{gathered}
		X = \Delta \,,\quad \Delta_0=\Sigma^3 \,,\quad G=\Sigma\left(\Sigma-\rho_{s}\right) \,, \\
		\mathcal{A} = \frac{\rho_{s}\Sigma}{G}\left(a\sin^2\theta\,d\phi + b\cos^2\theta\,d\psi\right) \,, \\
		\frac{1}{4}\sum_{i,j}\gamma_{ij}d\phi^{i}d\phi^{j} = \frac{\rho_{s}}{G}\left(a\sin^2\theta\,d\phi+b\cos^2\theta\,d\psi\right)^2 + \frac{\rho+a^2}{\Sigma}d\phi^2+\frac{\rho+b^2}{\Sigma}d\psi^2 \,.
	\end{gathered}
\ee

Let us forget for a moment what the explicit expressions for $\Delta_0$, $G$, $\mathcal{A}$, $X$ and $\gamma_{ij}$ are. The generic effective geometry then describes a stationary, axisymmetric black hole whose event horizon is the larger root of the function $X$. The function $G$ captures the static limit at the surface $G=0$, setting the boundaries of the ergoregion. The thermodynamic properties of the black hole are completely independent of the warp-factor $\Delta_0$, while they only depend on the near-horizon behavior of the function $G$, the angular potential $\mathcal{A}$ and the induced metric $\gamma_{ij}$~\cite{Cvetic:2011hp}. The warp-factor $\Delta_0$ can therefore be interpreted as to encode information about the environment surrounding the black hole, rather than its internal structure. A subtracted geometry is then a geometry obtained by a modification of the warp-factor, while keeping all the other metric functions fixed~\cite{Cvetic:2011hp,Cvetic:2011dn}.

To reveal a connection to the Love symmetries, it is more convenient to look at the inverse metric,
\be
	\begin{gathered}
		g^{\mu\nu}\partial_{\mu}\partial_{\nu} = \frac{4}{\Delta_0^{1/3}}\left\{X\partial_{\rho}^2 + \frac{1}{4}\,\partial_{\theta}^2 - \frac{\Delta_0}{4G}\partial_{t}^2 + \sum_{i,j}\gamma^{ij}D_{i}D_{j}\right\} \,,\quad D_{i}\equiv \partial_{i} - \mathcal{A}_{i}\partial_{t} \,,
	\end{gathered}
\ee
with $\gamma^{ij}$ the components of the inverse of the induced azimuthal metric $\gamma_{ij}$, $\gamma^{ik}\gamma_{kj}=\delta^{i}_{j}$. The Love symmetry $\SL_{\left(\sigma\right)}$ can then be realized as an isometry of the effective geometry with inverse metric
\be
	\tilde{g}^{\mu\nu}_{\left(\sigma\right)}\partial_{\mu}\partial_{\nu} = \frac{4}{\tilde{\Delta}_{0}^{\left(\sigma\right)1/3}}\left\{X\partial_{\rho}^2 + \frac{1}{4}\,\partial_{\theta}^2 - \frac{\tilde{\Delta}_{0}^{\left(\sigma\right)}}{4G}\partial_{t}^2 + \sum_{i,j}\gamma^{ij}\tilde{D}_{i}^{\left(\sigma\right)}\tilde{D}_{j}^{\left(\sigma\right)}\right\} \,,\quad \tilde{D}_{i}^{\left(\sigma\right)} = \partial_{i} - \tilde{\mathcal{A}}_{i}^{\left(\sigma\right)}\partial_{t} \,,
\ee
where
\be
	\begin{gathered}
		\tilde{\Delta}_{0}^{\left(\sigma\right)} = 16\rho_{+}\rho_{s}^2\left[1+\beta^2\left(\Omega_{\phi}-\sigma\Omega_{\psi}\right)^2\right] \,, \\
		\tilde{\mathcal{A}}^{\left(\sigma\right)} = -\frac{\rho_{+}\rho_{s}^2}{4\Delta}\left(\Omega_{\phi}\,\partial_{\phi}+\Omega_{\psi}\,\partial_{\psi}\right) - \frac{\rho_{+}-\rho_{-}}{\rho-\rho_{-}}\frac{\beta^2}{4}\left(\Omega_{\phi}-\sigma\Omega_{\psi}\right)\left(\partial_{\phi}-\sigma\partial_{\psi}\right) \,.
	\end{gathered}
\ee
More importantly, this effective geometry has exactly the same $4$-d base space $ds_4^2$ and preserves the entire form of the function $G$ which captures information about the static limit. The only difference relevant to the original definition of subtracted geometries~\cite{Cvetic:2011hp,Cvetic:2011dn} is that the angular potential is itself modified, but in such a way that the thermodynamic properties of the black hole remain unaltered.

As a side note, we remark here that it might be possible to realize subtracted geometries by a scaling limit of the full geometry, see e.g.~\cite{Cvetic:2012tr}. It would be interesting to investigate whether the effective geometries associated with the Love symmetries can be manifested as similar scaling limits of the full $5$-dimensional Myers-Perry black hole geometry. We leave such an analysis for future work.

\subsection{Infinite-dimensional extension}
In the previous section, we presented how two different near-zone truncations of the massless Klein-Gordon equation admit two different $\SL$ symmetries, $\SL_{\left(\sigma\right)}$, $\sigma=\pm$. Both of these $\SL$ algebras can be realized as subalgebras of the semi-direct product $\SL_{\left[0,0\right]}\ltimes U\left(1\right)^2_{\mathcal{V}}$, where $\SL_{\left[0,0\right]}$ is generated by the following vector fields
\be
	\begin{gathered}
		L_0^{\left[0,0\right]} = -\beta\,\partial_{t} \,, \\
		L_{\pm1}^{\left[0,0\right]} = e^{\pm t/\beta}\left[\mp\sqrt{\Delta}\,\partial_{\rho} + \partial_{\rho}\left(\sqrt{\Delta}\right)\beta\,\partial_{t} + \frac{\rho_{+}-\rho_{-}}{2\sqrt{\Delta}}\beta\left(\Omega_{+}\,\partial_{+}+\Omega_{-}\,\partial_{-}\right)\right] \,.
	\end{gathered}
\ee
Each of the $U\left(1\right)_{\mathcal{V}}$ factors is generated by vector fields of the form $\upsilon\,\beta\Omega_{\sigma}\,\partial_{\sigma}$ with $\upsilon$ belonging to a representation $\mathcal{V}$ of $\SL_{\left[0,0\right]}$, for each sign $\sigma=+,-$ respectively.

Let us construct one such representation $\mathcal{V}=\left\{\upsilon_{0,k},k\in\mathbb{Z}\right\}$. We first specify $\upsilon_{0,0}=-1$, which belongs to the singleton representation of $\SL_{\left[0,0\right]}$, satisfying $L_{\pm1}^{\left[0,0\right]}\upsilon_{0,0}=0$, and has vanishing azimuthal numbers, therefore further satisfying $L_0^{\left[0,0\right]}\upsilon_{0,0}=0$. We then built the $\upsilon_{0,\pm1}$ states, under the conditions that they can reach the singleton state via the action of $L_{\mp1}^{\left[0,0\right]}$ and that they have weights $h=\mp1$,
\be
	L_0^{\left[0,0\right]}\upsilon_{0,\pm1} = \mp \upsilon_{0,\pm1} \,,\quad L_{\mp1}^{\left[0,0\right]}\upsilon_{0,\pm1} = \mp\upsilon_{0,0} \,.
\ee
Solving these, we arrive at the following basic states of the representation $\mathcal{V}$
\be
	\upsilon_{0,0}=-1 \,,\quad \upsilon_{0,\pm1}=e^{\pm t/\beta}\sqrt{\frac{\rho-\rho_{+}}{\rho-\rho_{-}}} \,,
\ee
which are automatically regular at both the future and the past event horizon. The rest of the representation $\mathcal{V}$ can then be constructed by climbing up or down the ladder,
\be
	\upsilon_{0,\pm n} = \left[L_{\pm1}^{\left[0,0\right]}\right]^{n-1}\upsilon_{0,\pm1} = \left(\pm1\right)^{n-1}\left(n-1\right)!\,e^{\pm nt/\beta}\left(\frac{\rho-\rho_{+}}{\rho-\rho_{-}}\right)^{n/2} \,,
\ee
with integer $n\ge1$. This construction is depicted in Figure~\ref{fig:VSL2R}.

\begin{figure}
	\centering
	\begin{tikzpicture}
		\node at (0,0) (um3) {$\upsilon_{0,-3}$};
		\node at (0,1) (um2) {$\upsilon_{0,-2}$};
		\node at (0,2) (um1) {$\upsilon_{0,-1}$};
		\node at (0,3) (u0) {$\upsilon_{0,0}$};
		\node at (0,4) (up1) {$\upsilon_{0,+1}$};
		\node at (0,5) (up2) {$\upsilon_{0,+2}$};
		\node at (0,6) (up3) {$\upsilon_{0,+3}$};
		
		\node at (3,-0.4) (um) {$\vdots$};
		\draw (1,0) -- (5,0);
		\draw (1,1) -- (5,1);
		\draw (1,2) -- (5,2);
		\draw [snake=zigzag] (1,2.9) -- (5,2.9);
		\draw (1,3) -- (5,3);
		\draw [snake=zigzag] (5,3.1) -- (1,3.1);
		\draw (1,4) -- (5,4);
		\draw (1,5) -- (5,5);
		\draw (1,6) -- (5,6);
		\node at (3,6.4) (up) {$\vdots$};
		
		\draw[blue] [->] (2,0) -- node[left] {$L_{+1}^{\left[0,0\right]}$} (2,1);
		\draw[blue] [->] (2.5,1) -- node[left] {$L_{+1}^{\left[0,0\right]}$} (2.5,2);
		\draw[blue] [->] (3,2) -- node[left] {$L_{+1}^{\left[0,0\right]}$} (3,3);
		\draw[blue] [->] (2.5,4) -- node[left] {$L_{+1}^{\left[0,0\right]}$} (2.5,5);
		\draw[blue] [->] (2,5) -- node[left] {$L_{+1}^{\left[0,0\right]}$} (2,6);
		\draw[red] [<-] (4,0) -- node[right] {$L_{-1}^{\left[0,0\right]}$} (4,1);
		\draw[red] [<-] (3.5,1) -- node[right] {$L_{-1}^{\left[0,0\right]}$} (3.5,2);
		\draw[red] [<-] (3,3) -- node[right] {$L_{-1}^{\left[0,0\right]}$} (3,4);
		\draw[red] [<-] (3.5,4) -- node[right] {$L_{-1}^{\left[0,0\right]}$} (3.5,5);
		\draw[red] [<-] (4,5) -- node[right] {$L_{-1}^{\left[0,0\right]}$} (4,6);
	\end{tikzpicture}
	\caption[A representation $\mathcal{V}$ of $\SL_{\left[0,0\right]}$ used to construct the $\SL_{\left[0,0\right]}\ltimes\hat{U}\left(1\right)^2_{\mathcal{V}}$ extension of the Love symmetries of the $5$-dimensional Myers-Perry black hole.]{A representation $\mathcal{V}$ of $\SL_{\left[0,0\right]}$ used to construct the $\SL_{\left[0,0\right]}\ltimes\hat{U}\left(1\right)^2_{\mathcal{V}}$ extension of the Love symmetries of the $5$-dimensional Myers-Perry black hole.}
	\label{fig:VSL2R}
\end{figure}

Consequently, we can extend $\SL_{\left[0,0\right]}$ into $\SL_{\left[0,0\right]}\ltimes\hat{U}\left(1\right)^2_{\mathcal{V}}$, via the $U\left(1\right)^2_{\mathcal{V}}$ elements
\be
	\upsilon = \sum_{k\in\mathbb{Z}}\upsilon_{0,k}\left(\alpha_{k,+}\,\beta\Omega_{+}\,\partial_{+} + \alpha_{k,-}\,\beta\Omega_{-}\,\partial_{-}\right) \,.
\ee

Within this infinite extension lies a particular $2$-parameter family of $\SL$ subalgebras,
\be
	\SL_{\left[\alpha_{+},\alpha_{-}\right]}\subset \SL_{\left[0,0\right]}\ltimes U\left(1\right)^2_{\mathcal{V}} \,,
\ee
generated by the vector fields
\be
	L_{m}^{\left[\alpha_{+},\alpha_{-}\right]} = L_{m}^{\left[0,0\right]} + \upsilon_{0,m}\left(\alpha_{+}\,\beta\Omega_{+}\,\partial_{+} + \alpha_{-}\,\beta\Omega_{-}\,\partial_{-}\right) \,,\quad m=0,\pm1 \,.
\ee
The corresponding Casimir operator is given by
\be\ba
	{}&\mathcal{C}_2^{\left[\alpha_{+},\alpha_{-}\right]} = \partial_{\rho}\,\Delta\,\partial_{\rho}- \frac{\rho_{s}^2\rho_{+}}{4\Delta}\left(\partial_{t} + \Omega_{+}\,\partial_{+} + \Omega_{-}\,\partial_{-}\right)^2 \\
	&+ \frac{\rho_{+}-\rho_{-}}{\rho-\rho_{-}}\beta^2\left(\partial_{t} + \alpha_{+}\,\Omega_{+}\,\partial_{+} + \alpha_{-}\,\Omega_{-}\,\partial_{-}\right)\left[\left(\alpha_{+}-1\right)\Omega_{+}\,\partial_{+} + \left(\alpha_{-}-1\right)\Omega_{-}\,\partial_{-}\right] \,.
\ea\ee
Note that for an arbitrary pair $\left(\alpha_{+},\alpha_{-}\right)$, these Casimirs do not correspond to any
consistent physical near-zone truncation of the Klein-Gordon equation in the background of the $5$-dimensional Myers-Perry black hole, except in two cases;
\be
	\alpha^{\text{NZ}}_{\sigma} = 1 \,\quad \text{AND} \quad\, \alpha^{\text{NZ}}_{-\sigma} = 0 \,,
\ee
for $\sigma=+$ or $\sigma=-$. These are precisely our two $\SL$ symmetries of the near-zone truncations \eqref{eq:NZRadial1_5dMP}-\eqref{eq:NZRadial2_5dMP} which are now realized as subalgebras of the same larger structure. In the current notation,
\be
	\SL_{\left(+\right)} = \SL_{\left[1,0\right]} \,,\quad \SL_{\left(-\right)} = \SL_{\left[0,1\right]} \,.
\ee
However, the Casimirs with generic $\alpha_{+}$ and $\alpha_{-}$ have another remarkable property: they are the most general globally defined and time-reversal symmetric truncations of the equations of motion which preserve the characteristic exponents in the vicinity of the event horizon (see Appendix~\ref{app:SL2RGenerators}).

\subsection{Infinite zones of Love from local time translations}
Beyond the infinite extension described above involving subtracted geometry truncations of the radial Klein-Gordon operator equipped with an enhanced $\SL$ symmetry, there is another, different type of generalization that gives rise to all the possible near-zone truncations of the equations of motion such that a globally defined $\SL$ symmetry emerges. The corresponding generators make up two towers of near-zone $\SL$'s and are given by
\be
	\begin{gathered}
		L_0^{\left(\sigma\right)}\left[g\left(\rho\right)\right] = L_0^{\left(\sigma\right)}\left[0\right] \,, \\
		L_{\pm1}^{\left(\sigma\right)}\left[g\left(\rho\right)\right] = e^{\pm g\left(\rho\right)/\beta}L_{\pm1}^{\left(\sigma\right)}\left[0\right] \pm e^{\pm \left(t+g\left(\rho\right)\right)/\beta}\sqrt{\Delta}\left(\partial_{\rho}g\left(\rho\right)\right)\partial_{t} \,,
	\end{gathered}
\ee
where $L_{m}^{\left(\sigma\right)}\left[0\right]$ are the Love symmetries generators \eqref{eq:LoveGen} and $g\left(\rho\right)$ is an arbitrary radial function which is regular and non-vanishing at the event horizon. All of these $\SL$ algebras, however, can be realized as cousins of the Love symmetries, corresponding to local $\rho$-dependent temporal translations,
\be
	t \rightarrow \tilde{t} = t + g\left(\rho\right) \,.
\ee
Indeed, when writing the generators using this time coordinate, they acquire the same form as the Love symmetries generators,
\be
	\begin{gathered}
		L_0^{\left(\sigma\right)}\left[g\left(\rho\right)\right] = -\beta\left(\partial_{\tilde{t}} + \Omega_{\sigma}\,\partial_{\sigma}\right) \,, \\
		L_{\pm1}^{\left(\sigma\right)}\left[g\left(\rho\right)\right] = e^{\pm \tilde{t}/\beta}\left[\mp\sqrt{\Delta}\,\partial_{\rho} + \partial_{\rho}\left(\sqrt{\Delta}\right)\,\beta\left(\partial_{\tilde{t}} + \Omega_{\sigma}\,\partial_{\sigma}\right) + \frac{\rho_{+}-\rho_{-}}{2\sqrt{\Delta}} \beta\Omega_{-\sigma}\,\partial_{-\sigma}\right] \,.
	\end{gathered}
\ee
The associated Casimir is then given by \eqref{eq:LoveCasimir} with $t$ replaced $\tilde{t}$. Furthermore, the argument of vanishing static Love numbers remains unaltered even when using these generalized Love symmetries. Namely, when the conditions for vanishing static Love numbers are satisfied, the corresponding regular at the future event horizon static solution is a descendant in the highest-weight representation of these generalized near-zone $\SL$'s and the highest-weight property dictates the (quasi-)polynomial form of the solution.

\section{Relation to NHE isometries}
\label{sec:NHE}

We have seen that the near-zone $\SL$ symmetries can be realized as particular subalgebras of the infinite extension $\SL\ltimes U\left(1\right)^2_{\mathcal{V}}$ and that all of these $\SL$ subalgebras of the infinite extension are approximate symmetries of the $5$-dimensional Myers-Perry black hole, in the sense that they are isometries of geometries that preserve the internal structure of the black hole. Here, we will relate these approximate symmetries with the exact isometries of the near-horizon region of the extremal $5$-dimensional Myers-Perry black hole.

We will start with a brief review of how an enhanced $\SL$ symmetry for the extremal $5$-dimensional Myers-Perry black hole arises in the near-horizon region~\cite{Bardeen:1999px,Kunduri:2007vf,Galajinsky:2012vh,Galajinsky:2013mla,Hakobyan:2017qee,Figueras:2008qh}. Then, we will demonstrate how to take an appropriate extremal limit of the approximated $\SL$ symmetries which will precisely recover the $\SL$ Killing vectors of the near-horizon extremal geometry.

\subsection{Enhanced symmetries for NHE Myers-Perry black hole}
Consider the extremal $5$-dimensional Myers-Perry black hole geometry. The extremality condition reads
\be
	\left|a\right|+\left|b\right| = r_{s} \,,
\ee
with the degenerate horizon located at
\be
	\rho_{+} = \rho_{-} = \frac{\rho_{s}-a^2-b^2}{2} = \left|ab\right| \,.
\ee
To obtain the near-horizon geometry, we perform the following change into co-rotating coordinates~\cite{Bardeen:1999px}
\be
	\tilde{\rho} = \frac{\rho-\rho_{+}}{\lambda} \,,\quad \tau=\lambda t \,,\quad \tilde{\phi} = \phi - \Omega_{\phi}\,t \,,\quad  \tilde{\psi} = \psi - \Omega_{\psi}\,t
\ee
and take the scaling limit $\lambda\rightarrow 0$. The resulting near-horizon extremal (NHE) geometry is then given by~\cite{Galajinsky:2012vh,Galajinsky:2013mla,Hakobyan:2017qee,Figueras:2008qh}
\be
	ds^2_{\text{NHE}} = \frac{\Sigma_{+}}{\rho_0}\left[-\left(\frac{\tilde{\rho}}{\rho_0}\right)^2d\tau^2 + \left(\frac{\rho_0}{2\tilde{\rho}}\right)^2\frac{d\tilde{\rho}^2}{\rho_0}+\rho_0\,d\theta^2\right] + \sum_{i,j=1}^{2}\tilde{\gamma}_{ij}D\tilde{\phi}^{i}D\tilde{\phi}^{j} \,.
\ee
In the above expression, $\rho_0^3=\rho_{+}\rho_{s}^2$, $\Sigma_{+} = \rho_{+}+a^2\cos^2\theta+b^2\sin^2\theta$, small Latin indices label the azimuthal angles with $\tilde{\phi}^1\equiv\tilde{\phi}$ and $\tilde{\phi}^2\equiv\tilde{\psi}$, $\tilde{\gamma}_{ij}$ is the induced metric at the horizon along the azimuthal directions,
\be\ba
	\sum_{i,j=1}^{2}\tilde{\gamma}_{ij}d\tilde{\phi}^{i}d\tilde{\phi}^{j} &= \frac{\rho_{s}}{\Sigma_{+}}\left(a\sin^2\theta\,d\tilde{\phi} + b\cos^2\theta\,d\tilde{\psi}\right)^2 \\
	&+ \left(\rho_{+}+a^2\right)\sin^2\theta\,d\tilde{\phi}^2 + \left(\rho_{+}+b^2\right)\cos^2\theta\,d\tilde{\psi}^2 \,,
\ea\ee
and $D\tilde{\phi}^{i} = d\tilde{\phi}^{i}+k^{i}\tilde{\rho}\,d\tau$ with
\be
	k^{\tilde{\phi}} = \frac{\Omega_{\phi}}{\Sigma_{+}}\left[\cos^2\theta+\frac{b^2}{\rho_{+}}\sin^2\theta\right] \,,\quad k^{\tilde{\psi}} = \frac{\Omega_{\psi}}{\Sigma_{+}}\left[\sin^2\theta+\frac{a^2}{\rho_{+}}\cos^2\theta\right] \,.
\ee
This form of the NHE metric makes the enhanced isometry of the geometry manifest. More explicitly, the NHE geometry of the $5$-dimensional Myers-Perry black hole has an $\SL\times U\left(1\right)^2$ isometry, with $U\left(1\right)^2$ generated by the azimuthal Killing vectors and the $\SL$ Killing vectors given by
\be\label{eq:SL2RKV_BH}
	\begin{gathered}
		\xi_0 = \tau\,\partial_{\tau} - \tilde{\rho}\,\partial_{\tilde{\rho}} \,\,\,,\,\,\, \xi_{+1} = \partial_{\tau} \,, \\
		\xi_{-1} = \left(\frac{\rho_{+}\rho_{s}^2}{4\tilde{\rho}^2} + \tau^2\right)\,\partial_{\tau} - 2\tau\tilde{\rho}\,\partial_{\tilde{\rho}} - \frac{\rho_{+}\rho_{s}}{2\tilde{\rho}}\left(\frac{1}{a}\,\partial_{\tilde{\phi}}+\frac{1}{b}\,\partial_{\tilde{\psi}}\right) \,.
	\end{gathered}
\ee
In the initial $\left(t,\rho,\phi,\psi\right)$ coordinates, the $\SL$ Killing vectors read
\be\label{eq:SL2RKV_BL}
	\begin{gathered}
		\xi_0 = t\,K - \left(\rho-\rho_{+}\right)\partial_{\rho} \,,\quad \xi_{+1} = \lambda^{-1}\,K \,, \\
		\xi_{-1} = \lambda\bigg[ \left(\frac{\rho_{+}\rho_{s}^2}{4\left(\rho-\rho_{+}\right)^2} + t^2 \right)K -2t\left(\rho-\rho_{+}\right)\partial_{\rho} - \frac{\rho_{+}\rho_{s}}{2\left(\rho-\rho_{+}\right)}\left(\frac{1}{a}\,\partial_{\phi} + \frac{1}{b}\,\partial_{\psi}\right) \bigg] \,,
	\end{gathered}
\ee
where $K$ is the Killing vector that becomes null at the event horizon,
\be
	K = \partial_{t} + \Omega_{\phi}\,\partial_{\phi} + \Omega_{\psi}\,\partial_{\psi} = \lambda\,\partial_{\tau} \,.
\ee

The Casimir associated with this $\SL$ algebra is given by
\be\ba\label{eq:SL2RKV_Casimir}
	\mathcal{C}_2^{\SL} &= \partial_{\tilde{\rho}}\,\tilde{\rho}^2\,\partial_{\tilde{\rho}} - \frac{\rho_{+}\rho_{s}^2}{4\tilde{\rho}^2}\,\partial_{\tau}^2 + \frac{\rho_{+}\rho_{s}}{2\tilde{\rho}}\left(\frac{1}{a}\,\partial_{\tilde{\phi}}+\frac{1}{b}\,\partial_{\tilde{\psi}}\right)\partial_{\tau} \\
	&= \partial_{\rho}\left(\rho-\rho_{+}\right)^2\partial_{\rho} -\frac{\rho_{+}\rho_{s}^2}{4\left(\rho-\rho_{+}\right)^2}K^2 + \frac{\rho_{+}\rho_{s}}{2\left(\rho-\rho_{+}\right)}\left(\frac{1}{a}\,\partial_{\phi}+\frac{1}{b}\,\partial_{\psi}\right)K \,,
\ea\ee
and correctly reproduces the full radial operator for the Klein-Gordon equation in the NHE limit after supplementing with the $U\left(1\right)^2$ contributions,
\be
	\mathbb{O}_{\text{full}}^{\left(0\right)} = \mathcal{C}_2^{\SL}-\frac{1}{4}\left(2\rho_{s}-\rho_{+}\right)\left( \Omega_{\phi}\,\partial_{\phi}+\Omega_{\psi}\,\partial_{\psi} \right)^2 + \mathcal{O}\left(\lambda\right) \,.
\ee

\subsection{NHE algebra from infinite extension}
Let us demonstrate now how the NHE $\SL$ Killing vectors can be recovered from the non-extremal $\SL\ltimes U\left(1\right)^2_{\mathcal{V}}$. We parameterize the extremal limit in terms of the Hawking temperature, $T_{H}\rightarrow0$. For example,
\be
	\left|a\right|+\left|b\right| = r_{s}\left(1-4\pi^2T_{H}^2r_{s}^2\right) + \mathcal{O}\left(T_{H}^3\right) \,.
\ee
Consider the $\SL$ subalgebra of $\SL\ltimes U\left(1\right)^2_{\mathcal{V}}$ corresponding to the choices\footnote{We note here that we can always consider a family of such $\SL$ subalgebras for which one adds arbitrary $\mathcal{O}\left(T_{H}^2\right)$ terms in $\alpha_{\pm}$. We also remark that $\alpha_{\sigma_{ab}} = 1 + \frac{r_{+}}{2}\pi T_{H} + \mathcal{O}\left(T_{H}^2\right)$ and $\alpha_{-\sigma_{ab}}=\mathcal{O}\left(T_{H}^2\right)$, where $\sigma_{ab}$ is the sign of the product of the two spin parameters of the black hole.}
\be
	\alpha_{\pm}^{\text{NHE}} = 1 + \frac{r_{+}}{\Omega_{\pm}}\left(\frac{1}{a}\pm\frac{1}{b}\right)\pi T_{H} \,.
\ee
The generators of this algebra are given by
\be
	\begin{gathered}
		L_0^{\text{NHE}} = -\frac{K}{2\pi T_{H}} - \frac{r_{+}}{2}\left(\frac{1}{a}\,\partial_{\phi} + \frac{1}{b}\,\partial_{\psi}\right) \,, \\
		L_{\pm1}^{\text{NHE}} = e^{\pm 2\pi T_{H}t}\left[\mp\sqrt{\Delta}\,\partial_{\rho} + \partial_{\rho}\left(\sqrt{\Delta}\right)\frac{K}{2\pi T_{H}} + \sqrt{\frac{\rho-\rho_{+}}{\rho-\rho_{-}}}\,\frac{r_{+}}{2}\left(\frac{1}{a}\,\partial_{\phi} + \frac{1}{b}\,\partial_{\psi}\right)\right] \,,
	\end{gathered}
\ee
and produce the following Casimir operator
\be\ba
	\mathcal{C}_2^{\text{NHE}} &= \partial_{\rho}\,\Delta\,\partial_{\rho} - \frac{\rho_{+}\rho_{s}^2}{4\Delta}\,K^2 + \frac{\rho_{+}\rho_{s}}{2\left(\rho-\rho_{-}\right)} \left(\frac{1}{a}\,\partial_{\phi} + \frac{1}{b}\,\partial_{\psi}\right)K \\
	&\quad +\frac{\rho_{+}-\rho_{-}}{\rho-\rho_{-}}\frac{\rho_{+}}{4}\left(\frac{1}{a}\,\partial_{\phi} + \frac{1}{b}\,\partial_{\psi}\right)^2 \,.
\ea\ee
Even though this Casimir does not give rise to a near-zone truncation in the non-extremal case, it has the special property of precisely reproducing the Casimir operator \eqref{eq:SL2RKV_Casimir} associated with the NHE $\SL$ Killing vectors when taking the extremal limit,
\be
	\mathcal{C}_2^{\text{NHE}} = \mathcal{C}_2^{\SL} + \mathcal{O}\left(T_{H}\right) \,.
\ee
In fact, by considering the following linear combinations in the extremal limit
\be\ba
	\xi_0 &= \lim_{T_{H}\rightarrow 0} \frac{L_{+1}^{\text{NHE}} - L_{-1}^{\text{NHE}}}{2} \,, \\
	\xi_{+1} &= \lambda^{-1}\lim_{T_{H}\rightarrow0}2\pi T_{H}\,L_0^{\text{NHE}} \,, \\
	\xi_{-1} &= \lambda\lim_{T_{H}\rightarrow0} \frac{L_{+1}^{\text{NHE}} + L_{-1}^{\text{NHE}} +2 L_0^{\text{NHE}}}{2\pi T_{H}} \,,
\ea\ee
we precisely recover the NHE $\SL$ Killing vectors \eqref{eq:SL2RKV_BL} after identifying $\lambda$ with the near-horizon scaling parameter. We see, therefore, that the infinite extension $\SL\ltimes U\left(1\right)^2_{\mathcal{V}}$ contains both the Love symmetries $\SL_{\left(\sigma\right)}$ associated with the non-extremal near-zone truncations as well as a family of $\SL$ subalgebras which in the extremal limit recover the exact $\SL$ Killing vectors of the NHE geometry.

\section{Summary of Chapter~\ref{ch:5dMP}}
\label{sec:Discussion}

In this chapter, we have extended the proposal of the Love symmetry resolution of the seemingly unnatural values of the black hole Love numbers~\cite{Charalambous:2021kcz,Charalambous:2022rre} to higher-dimensional rotating black holes in General Relativity. Namely, we have explored in full the case of static scalar responses of the $5$-dimensional Myers-Perry black hole.

Compared to the examples of Kerr-Newman black holes in $d=4$ spacetime dimensions~\cite{Charalambous:2021mea} and Schwarzschild black holes in $d=5$ spacetime dimensions~\cite{Kol:2011vg}, we find some interesting exact results. To start with, static scalar Love numbers do not in general vanish in $d=5$ for generic spin parameters, not even when $\hat{\ell}\in\mathbb{N}$, as opposed to the Schwarzschild case~\cite{Kol:2011vg}. Beyond vanishing for ``axisymmetric'' perturbations~\cite{Landry:2015zfa,Pani:2015hfa,Gurlebeck:2015xpa}, we also see the new element with no $d=4$ analogue of vanishing for an equi-rotating black hole background. We remark here that the current results can be straightforwardly extended to include the case of $5$-dimensional electrically charged Myers-Perry black holes, mainly due to the fact that the discriminant function remains a quadratic polynomial in $\rho$~\cite{Chong:2005hr,Chong:2006zx}. Scalar Love numbers for $5$-dimensional charged Myers-Perry black holes were also considered in~\cite{Consoli:2022eey}, who however focused on their slowly-rotating limits thus missing the classical RG flow feature.

It appears that the vanishing of static Love numbers for rotating black holes in $d=4$ is the exception rather than the norm. Indeed, as we have demonstrated in this chapter, Love numbers for rotating black holes in $d=5$ are in general non-zero and exhibit running. Regardless, we were still able to find near-zone truncations acquiring $\SL$ Love symmetries just like in $d=4$ Kerr-Newman black holes and $d\ge4$ Reissner-Nordstr\"{o}m black holes~\cite{Bertini:2011ga,Kim:2012mh,Charalambous:2021kcz,Charalambous:2022rre}. In the special situations where Love numbers do vanish, however, it is the highest-property of the corresponding Love symmetry that outputs this vanishing as a selection rule. We see therefore that the existence of near-zone Love symmetries appears to be routed in black holes in General Relativity, rather than only with background geometries and perturbations with vanishing Love numbers.

At the same time, we have demonstrated here that the highest-weight representation of the near-zone $\SL$'s, along with its full extension into the representation of type ``$\circ[\circ[\circ$'', plays a special role regarding the scalar response problem: it is entirely spanned by near-zone solutions with vanishing/non-running Love numbers. These properties are also shared with the Love symmetry presented in Chapter~\ref{ch:LoveSymmetry4d} for the $d=4$ Kerr-Newman black hole~\cite{Charalambous:2021kcz,Charalambous:2022rre}. These two features, the existence of near-zone $\SL$ symmetries and the vanishing/non-running of Love numbers, appear therefore to be mutually inclusive, with the solutions of vanishing Love numbers furnishing a quotient representation of the highest-weight Verma module of the near-zone $\SL$. We remind here, though, that only the static results can be trusted within the near-zone regime.

On that account, it is interesting to further study this hypothesis. On the one hand, it is instructive to extend the analysis to other general-relativistic black holes. The obvious next candidate to analyze is the higher-dimensional Myers-Perry black holes whose scalar field perturbations are still separable~\cite{Frolov:2006pe,Frolov:2008jr}. A technical obstacle in this approach, however, is the fact that angular eigenvalues in $d>5$ are not known in closed form, but can be obtained as an expansion in spin parameters ratios, see e.g.~\cite{Cho:2011yp}. It would be interesting, in particular, to analyze the fate of scalar Love numbers for equi-rotating Myers-Perry black holes in odd spacetime dimensions which have the enhanced isometry subgroup $U\left(1\right)^{N}\rightarrow U\left(N\right)$. Moreover, it only deems appropriate to extend to higher-spin fields, namely, electromagnetic and gravitational perturbations. At least for spin-$1$ perturbations, this should be very similar to the work done here thanks to the separability of electromagnetic perturbations in the background of Myers-Perry black holes~\cite{Lunin:2017drx}.

On the other hand, it is still an open question whether Love symmetry exists in theories of gravity beyond general relativity. A preliminary analysis around this was done in Section~\ref{sec:LoveBeyondGR4d} and Section~\ref{sec:LoveSymmetryModGR}, where a sufficient geometric condition was extracted for spherically symmetric black holes~\cite{Charalambous:2022rre}. It would be interesting to supplement that analysis with sufficient \text{and} necessary constraints, investigate what type of theories of gravity support such geometries and whether the corresponding Love symmetries live up to their names, i.e. whether they can address the potential vanishing of Love numbers. As a counterexample, it was shown in~\cite{Charalambous:2022rre} that Love symmetry does not exist for the case of Riemann-cubed modifications of General Relativity, see also~\cite{Cai:2019npx,Cardoso:2018ptl,DeLuca:2022tkm}. This nicely fitted with the corresponding computation of static scalar Love numbers which were found to be non-zero and exhibit the expected RG flow.
	\newpage
\chapter{Conclusion and Discussion}
\label{ch:Discussion}

After going through the relativistic definition of the response problem in Chapter~\ref{ch:TLNsDefinition}, we came across a very interesting result in Chapter~\ref{ch:TLNsBlackHoles4d}: General-relativistic black holes in $d=4$ spacetime dimensions exhibit a vanishing conservative static response under either scalar, electromagnetic or gravitational perturbations, i.e. their static Love numbers vanish identically~\cite{Damour:2009vw,Binnington:2009bb,Gurlebeck:2015xpa,Bicak:1977,Bicak:1976,Poisson:2014gka,Landry:2015zfa,Pani:2015hfa,LeTiec:2020spy,LeTiec:2020bos,Chia:2020yla,Charalambous:2021mea}. The vanishing of static Love numbers can be traced back to a particular (quasi-)polynomial form of the relevant solutions of linearized black hole perturbations which, from the worldline EFT point of view, appears as a tower of magic zeroes arising from cancellations between the profiles of the ``source'' and ``response'' modes~\cite{Charalambous:2022rre}. This challenges the naturalness of General Relativity and calls upon an enhanced symmetry explanation~\cite{tHooft:1979rat,Porto:2016zng}. We saw in Chapter~\ref{ch:LoveSymmetry4d} that, once the response problem is appropriately formulated in terms of a near-zone expansion, an enhanced symmetry acting on black hole perturbations manifests itself: the $\SL$ Love symmetry~\cite{Charalambous:2021kcz,Charalambous:2022rre}. The Love symmetry revives the theory's naturalness via representation theory arguments, namely, the fact that the regular static solution belongs to a highest-weight representation precisely outputs the conspiring (quasi-)polynomial form of the solution, from which to infer the vanishing of static Love numbers, as a selection rule~\cite{Charalambous:2021kcz,Charalambous:2022rre}.

There are some unconventional features of the Love symmetries regarding their property to offer IR selection rules. In particular, they have the feature of mixing IR and UV modes as can be seen from the fact that representations of the near-zone $\SL$ have non-zero frequencies and, thus, they are not directly manifested at the level of the worldline EFT. However, the action of the Love symmetry remains compatible with the near-zone conditions in the near-extreme limit $M/\beta\ll1$. For the highest-weight representation, the corresponding frequencies have the same form as the ``near-horizon'' modes presented in~\cite{Zimmerman:2011dx,Cvetic:2013lfa}. More interestingly, the purely imaginary spacing is precisely equal to the universal QNMs level spacing as extracted from Padmanabhan's argument~\cite{Padmanabhan:2003fx}. Another possible connection to the QNM spectrum has been suggested in~\cite{Charalambous:2022rre}, where the complex frequencies of highest-weight elements were contrasted to highly dumped QNMs and total transmission modes~\cite{Cook:2016fge,Cook:2016ngj}. Still, it is somewhat unclear whether this should be considered as a triumph of naturalness in the sense of ’t Hooft~\cite{tHooft:1979rat}, or rather an example of a ``UV miracle''. It remains to be seen whether this unconventional example may provide useful lessons for other famous hierarchy problems.

In Chapter~\ref{ch:Properties}, we presented a number of properties and generalizations of the Love symmetry in four spacetime dimensions, notably its infinite-dimensional extension into $\SL\ltimes\hat{U}\left(1\right)_{\mathcal{V}}$~\cite{Charalambous:2021kcz,Charalambous:2022rre}. This infinite extension contains three interesting $\SL$ subalgebras: two of them are the Love symmetry algebra and a second globally defined $\SL$ algebra associated with a different near-zone splitting of the equations of motion, the Starobinsky near-zone approximation~\cite{Starobinsky:1973aij,Starobinskil:1974nkd}. The third $\SL$ subalgebra is actually a family of $\SL$ subalgebras which have the remarkable property of recovering another well-known $\SL$ structure associated with black holes: the enhanced isometry of the near-horizon geometry of extremal black holes~\cite{Bardeen:1999px,Amsel:2009et}.

A popular slogan is that ``black holes are the hydrogen atom of the 21st century'', see, e.g., \cite{tHooft:2016sdu,Cardoso:2017cfl,EHT2019}. We see that this comparison is actually accurate in a very concrete technical sense. Low-energy dynamics of both systems is governed by an emergent integrable algebraic structure. It is still natural to wonder who ordered these structures. What are the reasons for the $SO\left(4\right)$ Laplace-Runge-Lentz symmetry of the hydrogen atom from the viewpoint of the full quantum electrodynamics and for the Love symmetry of black holes from the viewpoint of the full general relativity? We are not aware of a good answer in the hydrogen case, but it looks plausible that, for black holes, the horizon is the culprit. In fact, near-zone $\SL$ symmetries are suspected to be remnants of the enhanced symmetry structure of extremal black holes. This is supported by the fact that the near-zone symmetries acquire the interpretation as isometries of subtracted geometries~\cite{Cvetic:2011hp,Cvetic:2011dn} which have the property of filtering out the environment of the black hole while preserving its near-horizon structure. Furthermore, we already saw in Chapter~\ref{ch:LoveSymmetryDd} and Chapter~\ref{ch:5dMP} that non-zero static Love numbers for higher-dimensional black holes~\cite{Kol:2011vg,Hui:2020xxx,Charalambous:2023jgq} do not signal the loss of symmetry~\cite{Charalambous:2023jgq,Charalambous:2024tdj}.

However, once we depart from general-relativistic configurations, the Love symmetry does not in general exist, in accordance with the corresponding computations of black hole Love numbers~\cite{Cai:2019npx,Cardoso:2018ptl,Charalambous:2022rre,DeLuca:2022tkm}. Nevertheless, the near-horizon enhanced isometry group appears to be a generic feature of degenerate black holes~\cite{Kunduri:2007vf}. This statement can be easily checked at least for a generic spherically symmetric black hole with at most two degenerate horizons. The crucial difference with black hole geometries such as those of General Relativity is that there are no near-horizon modes of extremal black holes that can propagate in the ``far-horizon'' region. For general-relativistic black holes, on the other hand, we saw in Section~\ref{sec:NHEReview4d} that static axisymmetric modes do survive beyond the near-horizon regime. This accidental robustness of the near-horizon symmetry of extremal black holes can be traced to the fact that the full discriminant function determining the locations of the horizons remains a quadratic polynomial. In fact, the sufficient geometric conditions for the existence of Love symmetry for scalar perturbations of spherically symmetric black holes in a generic theory of gravity that we derived in Section~\ref{sec:LoveBeyondGR4d} and Section~\ref{sec:LoveSymmetryModGR} precisely imply this form of the discriminant function. Related to this, it would be particularly interesting to supplement that analysis with sufficient \textit{and necessary} constraints, investigate what type of theories of gravity support such geometries and whether the corresponding Love symmetries live up to their names, i.e. whether they can address the potential vanishing of Love numbers. Furthermore, this robustness of the near-horizon symmetry of extremal general-relativistic black holes appears to be reflected in the accidental symmetry found in Ref.~\cite{Porfyriadis:2021psx}, see also~\cite{Hadar:2020kry,Horowitz:2022leb,Horowitz:2023xyl}, which maps perturbations of exactly extremal black holes to perturbations of near-extremal black holes. It would be instructive to seek a connection between this accidental symmetry of extremal black holes in General Relativity and our globally defined near-zone symmetries of non-extremal black holes.

Although there is currently no rigorous proof of the interpretation of near-zone symmetries as relics of near-horizon extremal isometries, another interesting route would be to study perturbations of black objects with non-spherical horizons, namely, black $p$-branes~\cite{Duff:1993ye,Emparan:2001wn,Emparan:2006mm}. One would then attempt to identify near-zone truncations admitting an $SO\left(p+1,2;\mathbb{R}\right)$ symmetry, appropriate for the corresponding enhanced symmetry of the near-horizon geometry of extremal black $p$-branes. We leave this for future work~\cite{InProgress_pBranes}. If such an analysis turns out to yield affirmative results, then the Love symmetry may shine more light on the potential holographic descriptions of asymptotically flat general-relativistic black holes~\cite{Bardeen:1999px,Guica:2008mu,Castro:2010fd,Lu:2008jk}.

At the same time, it is important to remember that the study of black hole responses is far from being a pure theorist's exercise. These effects contribute to gravitational waveforms of binary inspirals, and the corresponding Wilson coefficients will be probed by the forthcoming gravitational wave observations~\cite{Porto:2016zng,Cardoso:2017cfl}. An approximate enhanced symmetry provides an extremely valuable addition and a useful organizing principle to the effective field theory toolbox. Chiral symmetry of pion interactions is one of the most famous and successful illustrations of this. Similar to the pion case, it is important to systematically work out all consequences of the Love symmetry, including the ones beyond the strict static limit. To achieve this, it should be fruitful to replace low frequency expansion with the near-zone expansion. By treating the symmetry breaking parameters as spurions under the Love symmetry, it should be possible to obtain analogues of the Gell-Mann-Okubo relations~\cite{Gell-Mann:1961omu,Okubo:1961jc} for finite frequency responses and quasinormal modes. A potential approach along these lines would be to identify the effective black hole geometries, for which Love symmetries are isometries, as scaling limits of the full asymptotically flat black hole solution, see e.g.~\cite{Cvetic:2012tr}. In fact, based on the results of Refs.~\cite{Mano:1996vt,Mano:1996mf,Mano:1996gn,Sasaki:2003xr}, it should be possible to reconstruct the full solution to the Teukolsky equation as a series over hypergeometric functions furnishing representations of the Love $\SL$ symmetry.

It would be unfair not to acknowledge other attempts to infer the vanishing of Love numbers via enhanced symmetry arguments, notably the ladder symmetry structure of the static response problem~\cite{Hui:2021vcv,Hui:2022vbh,Berens:2022ebl,Katagiri:2022vyz} whose origins can be traced back to the notion of ``mass ladder operators'' constructible for spacetimes admitting a closed conformal Killing vector~\cite{Cardoso:2017qmj}. Another attempt has been through the manifestation of a Schr\"{o}dinger symmetry at the level of the static problem~\cite{Achour:2021dtj,BenAchour:2022uqo,BenAchour:2022fif,BenAchour:2023dgj}. These all have the attractive feature of addressing the vanishing of the static Love numbers in a more conventional way, directly at the IR regime. At least for the ladder symmetry structure of Ref.~\cite{Hui:2021vcv}, it seems to be related to the interesting geometric properties of the subtracted geometries associated with the near-zone symmetries, namely, the fact that they are conformally flat~\cite{Hui:2022vbh} and, hence, acquire closed conformal Killing vectors~\cite{Cardoso:2017qmj}. It remains to be seen whether there is a deeper connection between these and our Love symmetry argument.

Closing, a promising further study of the enhanced symmetries of non-extremal black holes would be through Celestial holography techniques. It was shown in~\cite{Donnay:2015abr}, for example, that near-horizon asymptotic symmetries of non-extremal black holes are reminiscent of the BMS algebra~\cite{Bondi:1962px,Sachs:1962zza}. Furthermore, the geometry of an event horizon is described by Carrollian physics, i.e. ``vanishing-speed-of-light'' physics, and vector fields preserving the Carrollian geometry of the horizon have been studied in more detail in~\cite{Donnay:2019jiz}. It was also recently suggested that Carrollian and Celestial holography are secretly equivalent descriptions~\cite{Donnay:2022aba,Donnay:2022wvx}. Interestingly, Carroll symmetry was actually the first attempt to infer vanishing Love numbers via symmetry arguments~\cite{Penna:2018gfx}. In the end, it might actually be constructive to fall victims to the ``history repeats'' statement, but this time in hopes of bridging seemingly unconnected concepts labeling the same structure towards a potentially novel holographic description of asymptotically flat black holes.

	\newpage
	\cleardoublepage
	\appendix
	\phantomsection

	\begin{appendices}
	\newpage
\appchapter{List of symbols and conventions}
\label{app:Conventions}

\appsection{List of abbreviations}
\begin{abbrv}
	\item[LIGO] Laser Interferometer Gravitational-wave Observatory \\
	\item[LISA] Laser Interferometer Space Antenna \\
	\item[NS] Neutron Star \\
	\item[BH] Black Hole \\
	\item[EFT] Effective Field Theory \\
	\item[PN] Post-Newtonian \\
	\item[IR] Infrared \\
	\item[UV] Ultraviolet \\
	\item[STF] Symmetric Tracefree \\
	\item[RG] Renormalization Group (flow) \\
	\item[NP] Newman-Penrose (formalism) \\
	\item[GHP] Geroch-Held-Penrose (formalism) \\
	\item[NDA] Naive Dimensional Analysis \\
	\item[w.r.t.] with respect to \\
	\item[w.l.o.g.] without loss of generality \\
	\item[NHE] Near-horizon Extremal \\
	\item[AdS] Anti de Sitter (spacetime) \\
	\item[BMS] Bondi-Metzner-Sachs \\
\end{abbrv}

\appsection{Notation and conventions}

We will employ geometrized units with $c=1$ and work with the mostly-positive metric Lorentzian signature, $\left(\eta_{\mu\nu}\right) = \text{diag}\left(-1,+1,+1,\dots\right)$. In Chapters~\ref{ch:Intro}-\ref{ch:TLNsDefinition}, we will keep Newton's gravitational constant $G$ explicit, but we will set it to unity in the remaining chapters. For the Riemann curvature tensor, we will be using the sign convention
\[
	R^{\rho}_{\,\,\,\,\,\sigma\mu\nu} = \partial_{\mu}\Gamma^{\rho}_{\nu\sigma} - \partial_{\nu}\Gamma^{\rho}_{\mu\sigma} + \Gamma^{\rho}_{\mu\lambda}\Gamma^{\lambda}_{\nu\sigma} - \Gamma^{\rho}_{\nu\lambda}\Gamma^{\lambda}_{\mu\sigma} \,,
\]
with $\Gamma^{\rho}_{\mu\nu} = \frac{1}{2}g^{\rho\sigma}\left(\partial_{\mu}g_{\nu\sigma}+\partial_{\nu}g_{\sigma\mu}-\partial_{\sigma}g_{\mu\nu}\right)$ the Christoffel symbols. The corresponding sign convention for the Einstein-Hilbert action and gravitational stress-energy-momentum tensor will be
\[
	S_{\text{EH}} = +\frac{1}{16\pi G}\int d^{d}x\,\sqrt{-g}\,R \,,\quad T_{\mu\nu} = -\frac{2}{\sqrt{-g}}\frac{\delta\left(\sqrt{-g}\mathscr{L}_{\text{m}}\right)}{\delta g^{\mu\nu}} \,,
\]
where $S_{\text{m}} = \int d^{d}x\,\sqrt{-g}\,\mathscr{L}_{\text{m}}$ denotes the non-gravitational, ``matter'', action, such that the Einstein field equations in asymptotically flat spacetimes read ${ R_{\mu\nu}-\frac{1}{2}g_{\mu\nu}R = 8\pi G T_{\mu\nu} }$.

Small Latin letters from the beginning of the alphabet, $a,b\dots$, will denote spatial indices running from $1$ to $d-1$ for a $d$-dimensional spacetime, while small Greek letters, $\mu,\nu,\dots$, will denote spacetime indices running from $0$ to $d-1$, with $x^0$ the temporal coordinate, and repeated indices will be summed over. 
We will also be employing the multi-index notation $a_1\dots a_{\ell}\equiv L$, within which $x^{a_1}\dots x^{a_{\ell}}\equiv x^{L}$ and $\partial_{a_1}\dots\partial_{a_{\ell}}\equiv\partial_{L}$.

The symmetric or antisymmetric part of a tensor $T_{a_1a_2\dots}$ with respect to a set of indices $\left\{\mu_1,\mu_2,\dots\right\}$ will be denoted by enclosing the indices within round parentheses or brackets respectively:
\begin{align*}
	T_{\left(\mu_1\mu_2\dots \mu_{n}\right)\nu_1\nu_2\dots}\quad &\text{: Symmetric part of tensor $T$ with respect to indices $\left\{\mu_1,\mu_2,\dots,\mu_{n}\right\}$} \,, \\
	T_{\left[\mu_1\mu_2\dots \mu_{n}\right]\nu_1\nu_2\dots}\quad &\text{: Antisymmetric part of tensor $T$ with respect to indices $\left\{\mu_1,\mu_2,\dots,\mu_{n}\right\}$} \,.
\end{align*}
These are normalized such that symmetric (antisymmetric) part of a purely symmetric (antisymmetric) tensor is equal to the tensor itself. For instance, for rank-$2$ tensors,
\begin{equation*}
	T_{\left(\mu\nu\right)} = \frac{1}{2}\left(T_{\mu\nu}+T_{\nu\mu}\right) \,,\quad T_{\left[\mu\nu\right]} = \frac{1}{2}\left(T_{\mu\nu}-T_{\nu\mu}\right) \,,
\end{equation*}
while, for rank-$n$ tensors,
\begin{equation*}
	T_{\left(\mu_1\mu_2\dots\mu_{n}\right)} = \frac{1}{n!}\sum_{\sigma\in\mathcal{P}_{n}}T_{\mu_{\sigma_1}\mu_{\sigma_2}\dots\mu_{\sigma_{n}}} \,,\quad T_{\left[\mu_1\mu_2\dots\mu_{n}\right]} = \frac{1}{n!}\delta_{\mu_1\mu_2\dots\mu_{n}}^{\nu_1\nu_2\dots\nu_{n}}T_{\nu_1\nu_2\dots\nu_{n}} \,,
\end{equation*}
with $\mathcal{P}_{n}$ the set of permutations of $\left\{1,2,\dots,n\right\}$ and $\delta_{\mu_1\mu_2\dots\mu_{n}}^{\nu_1\nu_2\dots\nu_{n}}$ the rank-$n$ generalized Kronecker symbol.

In this thesis, we will casually encounter symmetric trace-free (STF) spatial tensors. We will denote the symmetric trace-free part of a spatial tensor with respect to a set of indices $\left\{a_1,a_2,\dots\right\}$ by enclosing the indices within angular brackets:
\begin{equation*}
	T_{\vev{a_1a_2\dots  a_{\ell}} b_1b_2\dots}\quad \text{: STF part of tensor $T$ with respect to indices $\left\{a_1,a_2,\dots,a_{\ell}\right\}$} \,.
\end{equation*}
For example, for a rank-$2$ tensor in $d-1$ spatial dimensions,
\begin{equation*}
	T_{\vev{ab}} = T_{\left(ab\right)} - \frac{1}{d-1}\gamma_{ab}T^{\,\,\,c}_{c} \,,
\end{equation*}
where spatial indices are lowered and raised through the induced spatial metric, $\gamma_{ab}$, and its inverse, $\gamma^{ab}$. For higher-rank tensors, the expressions get cumbersome, but in the case the tensor is already symmetric one can write down the following expansion in the STF basis:
\begin{equation*}
	T_{\vev{a_1a_2\dots a_{\ell}}} = \sum_{p=0}^{\left\lfloor\frac{\ell}{2}\right\rfloor}\left(-1\right)^{p}\frac{\left(2\ell-2p+d-5\right)!!}{\left(2\ell+d-5\right)!!}\,\ell!\,\gamma_{(a_1a_2}\dots\gamma_{a_{2p-1} a_{2p}}T_{a_{2p+1}\dots a_{\ell})b_1b_2\dots b_{p}}^{\qquad\qquad\qquad b_1b_2\dots b_{p}} \,.
\end{equation*}
In all applications in this thesis, the spatial metric is for all purposes the flat space metric, $\gamma_{ab}=\delta_{ab}$.

\appsection{List of symbols}
\begin{abbrv}
	\item[$n\text{PN}$] $n$'th Post-Newtonian order \\
	\item[$x^{L}$] $x^{a_1}x^{a_2}\dots x^{a_{\ell}}$ \\
	\item[$\partial_{L}$] $\partial_{a_1}\partial_{a_2}\dots\partial_{a_{\ell}}$ \\
	\item[$\mathcal{E}_{L}^{\left(s\right)}$] Electric-type moments of rank-$\ell$ for spin-$s$ field \\
	\item[$\mathcal{B}_{L,a}^{\left(s\right)}$] Magnetic-type moments of rank-$\ell$ for spin-$s$ field \\
	\item[$\mathcal{T}_{L,ab}^{\left(s\right)}$] Tensor-type moments of rank-$\ell$ for spin-$s$ field \\
	\item[$\bar{\mathcal{E}}_{L}^{\left(s\right)}$] Background source electric-type moments of rank-$\ell$ for spin-$s$ field \\
	\item[$\bar{\mathcal{B}}_{L,a}^{\left(s\right)}$] Background source magnetic-type moments of rank-$\ell$ for spin-$s$ field \\
	\item[$\bar{\mathcal{T}}_{L,ab}^{\left(s\right)}$] Background source tensor-type moments of rank-$\ell$ for spin-$s$ field \\
	\item[$Q_{L}^{\left(s\right),\mathcal{E}}$] Electric-type multipole moments of rank-$\ell$ for spin-$s$ field \\
	\item[$Q_{L,a}^{\left(s\right),\mathcal{B}}$] Magnetic-type multipole moments of rank-$\ell$ for spin-$s$ field \\
	\item[$Q_{L,ab}^{\left(s\right),\mathcal{T}}$] Tensor-type multipole moments of rank-$\ell$ for spin-$s$ field \\
	\item[$\mathbb{S}^{n}$] Unit-radius $n$-sphere \\
	\item[$\Delta_{\mathbb{S}^{n}}$] Laplace-Beltrami operator on $\mathbb{S}^{n}$ \\
	\item[$\epsilon_{\mu_1\mu_2\dots\mu_{n}}$] Levi-Civita symbol in $n$ dimensions \\
	\item[$\varepsilon_{\mu_1\mu_2\dots\mu_{n}}$] Levi-Civita tensor in $n$ dimensions \\
	\item[$T_{H}$] Hawking temperature \\
	\item[$\kappa$] Surface gravity \\
	\item[$\beta$] Inverse surface gravity \\
\end{abbrv}

%
	\newpage
\appchapter{Useful formulae involving the $\Gamma$-function and Euler's hypergeometric function}
\label{app:2F1Gamma}

In this Appendix, we enumerate a number of useful formulae relevant in solving the near-zone equations of motion and extracting the conservative and dissipative pieces of the response coefficients. All of these formulae can be found in the NIST Digital Library of Mathematical Functions~\cite{NIST}.

\appsection{$\Gamma$-function}
\label{sec:GammaFunction}

We begin with the (complete) $\Gamma$-function, defined by Euler's integral,
\be
	\Gamma\left(z\right) = \int_0^{\infty}dt\,t^{z-1}e^{-t} \,,\quad \text{Re}\left\{z\right\}>0 \,,
\ee
and serves as an extension of the familiar factorial function, satisfying the recurrence relation $\Gamma\left(z+1\right)=z\Gamma\left(z\right)$. For positive integer arguments, it is just the usual factorial offset by one unit,
\be
	\Gamma\left(n\right) = \left(n-1\right)! \,,\quad n=1,2,\dots \,.
\ee
The $\Gamma$-function can also be analytically continued to $\text{Re}\left\{z\right\}\le0$. For example, this can be done by the mirror/reflection formula
\be\label{eq:GammaMirrorFormula}
	\Gamma\left(z\right)\Gamma\left(1-z\right) = \frac{\pi}{\sin\pi z} \,,
\ee
which is particularly useful when studying the behavior of the response coefficients as it allows to explicitly reveal the vanishing or running of the Love numbers when sending the orbital number to range in its physical integer values.

The $\Gamma$-function is a meromorphic function with no roots and with simple poles at non-positive integers, with residue
\be
	\underset{z=-n}{\text{Res}}\Gamma\left(z\right) = \frac{\left(-1\right)^{n}}{n!} \,,\quad n=0,1,2,\dots \,.
\ee
Its logarithmic derivative defines the digamma or $\psi$-function,
\be
	\psi\left(z\right) = \frac{\Gamma^{\prime}\left(z\right)}{\Gamma\left(z\right)}
\ee
which is a meromorpic function with simple poles of residue $-1$ at semi-negative integers, satisfying the recursion relation $\psi\left(z+1\right) = \psi\left(z\right) + z^{-1}$. 

Another useful $\Gamma$-function identity is the Legendre duplication formula,
\be
	\Gamma\left(z\right)\Gamma\left(z+\frac{1}{2}\right) = 2^{1-2z}\sqrt{\pi}\Gamma\left(2z\right) \,,
\ee
which is a special case of the Gauss multiplication formula,
\be
	\prod_{k=1}^{n}\Gamma\left(z+\frac{k-1}{n}\right) = n^{\frac{1}{2}-nz}\left(2\pi\right)^{\frac{n-1}{2}}\Gamma\left(nz\right) \,.
\ee
The Legendre duplication formula helps comparing the expressions of the response coefficients written in this thesis with other works in the literature. Last, it is sometimes convenient to employ the identity
\be
	\left|\Gamma\left(n+1+ix\right)\right|^2 = \frac{\pi x}{\sin\pi x}\prod_{k=1}^{n}\left(k^2+x^2\right) \,,\quad n\in\mathbb{N}\,,\quad x\in\mathbb{R}
\ee
to write the Love numbers in a more practical manner.

\appsection{Hypergeometric function}
\label{sec:2F1Function}

Euler's hypergeometric function is characterized by $2+1$ parameters, $a$, $b$ and $c$, and is defined on the disk $\left|z\right|<1$ by the series
\be
	{}_2F_1\left(a,b;c;z\right) = \sum_{k=0}^{\infty}\frac{\left(a\right)_{k}\left(b\right)_{k}}{\left(c\right)_{k}}\frac{z^{k}}{k!} \,,
\ee
where $\left(a\right)_{k} = \frac{\Gamma\left(a+k\right)}{\Gamma\left(a\right)}$ is the Pochhammer symbol, sometimes also referred to as the rising factorial. It is one of the independent solutions expandable as a Frobenius series around $z=0$ of the hypergeometric differential equation
\be
	\left[z\left(1-z\right)\frac{d^2}{dz^2}+\left[c-\left(a+b+1\right)z\right]\frac{d}{dz} - ab\right]\,y\left(z\right) = 0 \,,
\ee
given that $c$ is not a non-positive integer. Useful transformation properties within the principal branch $\left|\text{Arg}\left(1-z\right)\right|<\pi$ involve Euler's transformation,
\be
	{}_2F_1\left(a,b;c;z\right) = \left(1-z\right)^{c-a-b}{}_2F_1\left(c-a,c-b;c;z\right) \,,
\ee
and the two Pfaff transformations,
\be\ba
	{}_2F_1\left(a,b;c;z\right) &= \left(1-z\right)^{-a}{}_2F_1\left(a,c-b;c;\frac{z}{z-1}\right) \\
	&= \left(1-z\right)^{-b}{}_2F_1\left(c-a,b;c;\frac{z}{z-1}\right) \,.
\ea\ee
The hypergeometric function can be analytically continued to $\left|z\right|>1$ via
\be\ba\label{eq:2F1LargeX}
	{}_2F_1\left(a,b;c;z\right) &= \frac{\Gamma\left(c\right)\Gamma\left(b-a\right)}{\Gamma\left(b\right)\Gamma\left(c-a\right)}\left(-z\right)^{-a}{}_2F_1\left(a,a-c+1;a-b+1;\frac{1}{z}\right) \\
	&+\frac{\Gamma\left(c\right)\Gamma\left(a-b\right)}{\Gamma\left(a\right)\Gamma\left(c-b\right)}\left(-z\right)^{-b}{}_2F_1\left(b,b-c+1;b-a+1;\frac{1}{z}\right) \,,
\ea\ee
which is valid for $\left|\text{Arg}\left(-z\right)\right|<\pi$, e.g. for negative real arguments such as the hypergeometric functions encountered in this thesis. This analytic continuation formula is particularly useful when extracting the source/response splitting of the profiles of the black hole perturbations and, subsequently, the response coefficients.

The hypergeometric function is analytic for all $a,b\in\mathbb{C}$ but does not exist for non-positive integer values of the parameter $c$ due to the development of simple poles. Nevertheless, the following limit is well defined
\be
	\lim\limits_{c\rightarrow-n}\frac{{}_2F_1\left(a,b;c;z\right)}{\Gamma\left(c\right)} = \frac{\left(a\right)_{n+1}\left(b\right)_{n+1}}{\left(n+1\right)!}z^{n+1}{}_2F_1\left(a+n+1,b+n+1;n+2;z\right) \,.
\ee
This is relevant when discussing the seemingly diverging behavior of the Love numbers which is compensated by a divergence of the above form in the ``source'' part of the solution with end result being a regular solution profile involving logarithms that reflect the classical RG flow of the Love numbers.

Last, when $a$ or $b$ is a non-positive integer, the hypergeometric function reduces to a polynomial,
\be
	{}_2F_1\left(-n,b;c;z\right) = \sum_{k=0}^{n}\left(-1\right)^{k}\binom{n}{k}\frac{\left(b\right)_{k}}{\left(c\right)_{k}}\frac{z^{k}}{k!} \,,\quad n=0,1,2,\dots \,,
\ee
as long as $c$ is not a negative integer larger than $n$.

	\newpage
\appchapter{Some properties of $\SL$ representations}
\label{app:SL2RRepresentations}

The lowest-weight and highest-weight representations are two standard reducible representations of $\SL$. Here we present some basic properties of these representations (see~\cite{Howe1992} for more detail\footnote{In the notation of that book we have $L_{\pm1}^{\text{here}} = \mp e^{\mp,\text{there}}$, $L_0^{\text{here}} = h^{\text{there}}/2$, $\mathcal{C}_2^{\text{here}}=\mathcal{C}_2^{\text{there}}/4$, $h^{\text{here}} = \lambda^{\text{there}}/2$.}). The highest-weight module $V_{h}$ has a basis of $L_0$-eigenvectors $\upsilon_j$ ($j={0,1,2,...}$) that satisfy
\be
	\begin{split}
		& L_0 \upsilon_{j}=\left(h + j\right)\upsilon_{j} \,,\quad L_{-1}\upsilon_{j} = \upsilon_{j+1} \,, \\
		& L_{+1} \upsilon_{j} = j\left(2h+j -1\right)\upsilon_{j-1} \,,\quad L_{+1} \upsilon_0 = 0 \,, \\
		& \mathcal{C}_2\upsilon_{j} = h\left(h-1\right)\upsilon_{j} \,.
	\end{split} 
\ee
The lowest-weight module $\bar V_{\bar{h}}$ has a basis of $L_0$-eigenvectors $\bar{\upsilon}_j$ ($j={0,1,2,...}$) that satisfy
\be
	\begin{split}
		& L_0 \bar{\upsilon}_{j} = \left(\bar{h} - j\right)\bar{\upsilon}_{j} \,,\quad L_{+1} \bar{\upsilon}_{j} = -\bar{\upsilon}_{j+1} \,, \\
		& L_{-1} \bar{\upsilon}_{j} = j\left(2\bar{h} - j + 1\right)\bar{\upsilon}_{j-1} \,,\quad L_{-1} \bar{\upsilon}_0 = 0 \,, \\
		& \mathcal{C}_2\bar{\upsilon}_{j} = \bar{h}\left(\bar{h}+1\right)\bar{\upsilon}_{j} \,.
	\end{split} 
\ee

Note that Ref.~\cite{Miller1970} offers a somewhat different classification of reducible representations of $\SL$. In particular, the representation relevant for the $d=4$ Schwarzschild case is the finite-dimensional representation $D(2\ell)$,
whereas the generic highest/lowest-weight Verma modules correspond to the representations $D^-(2\ell)/D^+(2\ell)$, respectively.

For completeness, we also present here the three additional standard modules $W\left(\mu,\lambda\right)$, $\bar{W}\left(\mu,\lambda\right)$ and $U\left(\nu^+,\nu^-\right)$, spanned by infinite sets of vectors $\upsilon_{j}$ ($j\in \mathbb{Z}$), which satisfy
\be
	\begin{split}
		W\left(\mu,\lambda\right): \quad & L_0\upsilon_{j} = \left(\frac{\lambda}{2} + j\right)\upsilon_{j} \,,\quad L_{-1}\upsilon_{j} = \upsilon_{j+1} \,,\quad \mathcal{C}_2\upsilon_{j} = \left(\mu/4\right)\upsilon_{j} \,, \\
		& L_{+1}\upsilon_{j} = -\frac{1}{4}\left[\mu-\left(\lambda+2j-1\right)^2+1\right]\upsilon_{j-1} \,, \\
		\bar{W}\left(\mu,\lambda\right): \quad & L_0\bar{\upsilon}_{j} = \left(\frac{\lambda}{2} + j\right)\bar{\upsilon}_{j} \,, \quad L_{+1}\bar{\upsilon}_{j} = -\bar{\upsilon}_{j-1} \,,\quad \mathcal{C}_2\bar{\upsilon}_{j} = \left(\mu/4\right) \bar{\upsilon}_{j} \,,\\
		& L_{-1}\bar{\upsilon}_{j} = \frac{1}{4}\left[\mu-\left(\lambda+2j+1\right)^2+1\right]\bar{\upsilon}_{j+1} \,, \\
		U\left(\nu^{+},\nu^{-}\right): \quad & L_0\upsilon_{j} = \left(\frac{\nu^{+}-\nu^{-}}{2} + j\right)\upsilon_{j} \,,\quad L_{\mp1}\upsilon_{j} = \left(j\pm\nu^{\pm}\right)\upsilon_{j\pm1} \,, \\ 
		& \mathcal{C}_2\upsilon_{j} = \frac{\nu^{+} + \nu^{-}}{2}\left(\frac{\nu^{+} + \nu^{-}}{2} - 1\right)\upsilon_{j} \,.
	\end{split}
\ee
	\newpage
\appchapter{Derivation of $\SL$ generators}
\label{app:SL2RGenerators}

In Chapter~\ref{ch:LoveSymmetry4d} we presented the existence of near-zone truncations of the Teukolsky equation, governing the dynamics of perturbations of the Kerr-Newman black hole, equipped with an $\SL$ symmetry structure. In this appendix, we will present the derivation of the form of the generators of these symmetries. We will do this by starting with a generic ansatz for the form of the $\SL$ generators under the assumption that separable solutions of the Teukolsky equation furnish $\SL$ representations. We will impose the preliminary constraints that the associated Casimir operators yield operators that preserve the characteristic exponents of the full equations of motion in the vicinity of the black hole event horizon. We will end up with an infinite number of $\SL$ algebras, most of which are not consistent near-zone truncations and also not globally defined.

The upshot of using this approach is that finding the most general truncation that preserves the near-horizon characteristic exponents also ensures that we will find all the possible near-zone truncations admitting $\SL$ symmetries as a subset. As we will see, there will be two towers of near-zone $\SL$ symmetries controlled by an arbitrary parameter which spontaneously breaks the $\SL$ symmetry down to $U\left(1\right)$, in the sense that these are locally defined and singular as we go around the azimuthal circle. The only possible globally defined near-zone $\SL$ symmetries will then correspond to setting this symmetry breaking parameter to zero and will precisely correspond to the Love and Starobinsky near-zone symmetries presented in Chapter~\ref{ch:LoveSymmetry4d}, up to local $r$-dependent translations in time as in noticed in Section~\ref{sec:SL2RtTranslations4d}. We will also investigate the situations where the $\SL$ symmetry of the truncations preserving the near-horizon characteristic exponents can be enhanced to full $2$-d conformal structure $\SL\times\SL$. This will allow to interpret the characteristic exponents near the inner and the outer horizons of the black hole as the $\text{CFT}_2$ spin weight and scaling dimension of the perturbation respectively.

The analysis in this Appendix has the feature of generalizing some already known results in the literature, such as the local $\SL\times\SL$ originally found in~\cite{Castro:2010fd}. We focus on the non-extremal case but, as we discuss in Chapter~\ref{ch:NHE}, there is a particular subset of these near-horizon-monodromy-preserving truncations of the Teukolsky operator equipped with $\SL$ symmetries for non-extremal black holes that can be utilized to get the correct spin-weighted $\SL$ Killing vectors of the near horizon extremal black hole configuration as well. Furthermore, the steps followed here are generic nature and can be applied for the derivation of all the near-zone $\SL$ symmetries associated with perturbations of other black hole configurations, e.g. the scalar, electromagnetic and gravitational perturbations of higher-dimensional Schwarzschild black holes and the scalar perturbations of the $5$-dimensional Myers-Perry black hole.

\appsection{Truncated radial Teukolsky operator preserving the near-horizon monodromy}
\label{sec:TeukolskyTruncations}

The full master Teukolsky operator \eqref{eq:teuk}-\eqref{eq:teukExplicit} for the spin-$s$ NP scalar has the form,
\be\ba
	\mathbb{T}^{\left(s\right)}_{\text{full}}\Psi_{s} &= \frac{1}{\Sigma}\left[G^{\mu\nu}_{\text{full}}\left(r,\theta\right)\nabla_{\mu}\nabla_{\nu} + s\left(\gamma^{\mu}_{\text{full}}\nabla_{\mu}\left(r,\theta\right) + F_{\text{full},0}\left(r,\theta\right)\right)\right]\Psi_{s} \\
	&= \frac{1}{\Sigma}\left[\mathbb{O}^{\left(s\right)}_{\text{full}} - \mathbb{P}^{\left(s\right)}_{\text{full}}\right]\Psi_{s}\,,
\ea\ee
where $G^{\mu\nu}_{\text{full}}\equiv\Sigma g^{\mu\nu}_{\text{full}}$ is the rescaled Kerr-Newman metric and $\gamma^{\mu}_{\text{full}}$ and $F_{\text{full},0}$ are tetrad-dependent modifications of the Klein-Gordon operator due a non-zero spin weight. All of these functions depend only on the radial and polar coordinates by virtue of the $\mathbb{R}_{t}\times U_{\phi}\left(1\right)$ isometry group of the background geometry. In the second equality above, we have dubbed as $\mathbb{O}^{\left(s\right)}_{\text{full}}$ and $\mathbb{P}^{\left(s\right)}_{\text{full}}$ the full radial and angular Teukolsky operators respectively to stress out the separability of the Teukolsky equation. The important point here is that the second derivatives terms in the equations of motion always come from the inverse metric.

We wish to explore the possibility of enhanced symmetries arising in the radial problem after employing a truncation such that we preserve the near-horizon boundary conditions for any perturbation frequency. The most general truncation of the radial operator that preserves the characteristic exponents as we approach the event horizon at $r=r_{+}$ has the form
\be\ba
	\mathbb{O}_{\text{trunc}} &= \Delta^{-s}\partial_{r}\,\Delta^{s+1}\,\partial_{r} - \frac{\left(r_{+}-r_{-}\right)^2}{4\Delta}\beta^2\left(\partial_{t}+\Omega\,\partial_{\phi}-\frac{2s}{\beta}\right)\left(\partial_{t}+\Omega\,\partial_{\phi}\right) \\
	&\quad + s\left(s+1\right) + \delta g^{\mu\nu}\,\partial_{\mu}\partial_{\nu} + \delta\gamma^{\mu}\,\partial_{\mu} + \delta F_0 \,,
\ea\ee
where $\delta g^{\mu\nu}$, $\delta\gamma^{\mu}$ and $\delta F_0$ are terms that are subleading in the vicinity of the event horizon, $\Delta=\left(r-r_{+}\right)\left(r-r_{-}\right)$ is the discriminant function for the Kerr-Newman black hole and $\beta = 2\frac{r_{+}^2+a^2}{r_{+}-r_{-}}$ is the inverse Hawking temperature of the black hole. For example, the form of $\delta g^{\mu\nu}$ that preserves the background $\mathbb{R}_{t}\times U\left(1\right)_{\phi}$ symmetries as well as the separability of the Teukolsky equation in the current Boyer-Lindquist coordinates is
\be
	\begin{gathered}
		\delta g^{rr} = \Delta^2f_{rr}\left(r\right) \,,\quad \delta g^{tt} = f_{tt}\left(r\right) \,, \\
		\delta g^{t\phi} = \Omega\,f_{t\phi}\left(r\right) \,,\quad \delta g^{\phi\phi} = \Omega^2\,f_{\phi\phi}\left(r\right) \,, \\
		\delta g^{tr} = \Delta\,f_{tr}\left(r\right) \,,\quad \delta g^{r\phi} = \Omega\,\Delta\,f_{r\phi}\left(r\right) \,, \\
		\delta g^{\mu\theta} = 0 \,,
	\end{gathered}
\ee
with all $f_{\mu\nu}$ being purely radial functions that are regular at $r=r_{+}$ and the $\Omega$ insertions are to remind us that these functions are absent for non-rotating black holes. We remark here that we are also allowing the generation of the non-diagonal terms $\delta g^{tr}$ and $\delta g^{r\phi}$, as well as the offset diagonal term $\delta g^{rr}$. These are zero in the full background geometry but are in general allowed in the current treatment, as long as they decay sufficiently fast on the horizon. All these terms, however, can be removed by performing appropriate $r$-dependent ``far-horizon'' coordinate transformations. More explicitly, introducing coordinates $(\tilde{t},\tilde{r},\tilde{\phi})$, related to $\left(t,r,\phi\right)$ according to
\be
	\begin{gathered}
		\frac{d\tilde{r}}{\sqrt{\Delta\left(\tilde{r}\right)}} = \frac{dr}{\sqrt{\Delta\left(r\right)+\delta g^{rr}\left(r\right)}} \,, \\
		d\tilde{t} = dt - \frac{\delta g^{tr}\left(r\right)}{\Delta\left(r\right)+\delta g^{rr}\left(r\right)}\,dr \,,\quad d\tilde{\phi} = d\phi - \frac{\delta g^{r\phi}\left(r\right)}{\Delta\left(r\right) + \delta g^{rr}\left(r\right)}\,dr \,,
	\end{gathered}
\ee
we can set
\be
	\delta\tilde{g}^{\tilde{r}\tilde{r}} = 0 \,,\quad \delta\tilde{g}^{\tilde{t}\tilde{r}} = 0 \,,\quad \delta\tilde{g}^{\tilde{r}\tilde{\phi}} = 0 \,.
\ee

In the rest of this appendix, we will be working in the $(\tilde{t},\tilde{r},\tilde{\phi},\theta)$ coordinates but we will drop the tildes to ease our notation. Nevertheless, in the final expressions, we will always have in mind to supplement with the more general results corresponding to the replacements,
\be\label{eq:t_r_phi_Rep}
	r \rightarrow r + \Delta^2 g_{r}\left(r\right) \,\,\,,\,\,\, t\rightarrow t + g_{t}\left(r\right) \,\,\,,\,\,\, \phi\rightarrow \phi + g_{\phi}\left(r\right) \,,
\ee
with $g_{r}$, $g_{t}$ and $g_{\phi}$ radial functions that are regular near the horizon, giving rise to non-zero $\delta g^{rr}$, $\delta g^{tr}$ and $\delta g^{r\phi}$ respectively.

The fall-offs of $\delta\gamma^{\mu}$ and $\delta F_0$ are similarly dictated by
\be
	\begin{gathered}
		\delta\gamma^{r} = \Delta\,f_{r}\left(r\right) + s\,\Delta\,F_{r}\left(r\right) \,,\\
		\delta\gamma^{t} = f_{t}\left(r\right) + s\,F_{t}\left(r\right) \,, \\
		\delta\gamma^{\phi} = \Omega\,f_{\phi}\left(r\right) + s\,\Omega\,F_{\phi}\left(r\right) \,, \\
		\delta\gamma^{\theta} = 0 \,, \\
		\delta F_0 = s\,\Delta\,F_0\left(r\right) \,,
	\end{gathered}
\ee
with all $f_{\mu}$, $F_{\mu}$ and $F_0$ again being purely radial function regular at the event horizon. We note that we have assigned $\delta F_0$ entirely to a non-zero spin-weight to reflect the condition that, whatever enhanced symmetry we find, we want its generators to act as Lie derivatives onto a scalar ($s=0$) perturbation. We have accordingly separated the $\delta\gamma^{\mu}$ components. We can now utilize the additional gauge actions, associated with local Lorentz transformations acting on the tetrad vectors, to further simplify the above operator in a general manner. Given that the Teukolsky operator acts on the perturbed NP scalars $\Psi_{s}$ which have boost weights equal to their spin weights, $b=s$, and since the Teukolsky operator itself is built from background quantities, that is, it is invariant under gauge transformations of the perturbations, we can always perform a ``far-horizon'' local boost transformation,
\be\label{eq:s_Rep}
	\Psi_{s} \xrightarrow{\text{boost}} e^{s\,\Delta^{\varepsilon}\,\eta\left(r\right)} \Psi_{s} \,,
\ee
with $\eta\left(r\right)$ a purely radial function that is regular at the horizon, chosen such that we remove the $s\,F^{r}\partial_{r}$ term, while the exponent $\varepsilon$ must be such that the near-horizon characteristic exponents of the equations of motion are simultaneously preserved. Indeed, choosing, $\eta\left(r\right)$ to satisfy
\be
	2\left[\Delta^{\varepsilon}\,\eta\left(r\right)\right]^{\prime} = -F_{r} \,,
\ee
removes the term that goes like $s\,\partial_{r}$, while the modifications in the scalar part are always subleading in the vicinity of the horizon as long as $\varepsilon>1$. We will adopt this particular gauge where $F^{r}=0$ but always have in mind to supplement with the more general results corresponding to the above local boost transformation. All in all, the most general truncated radial Teukolsky operator that preserves the near-horizon characteristic exponents is given by, after some rearrangement and up to the $r$-dependent coordinate transformations and local boosts transformations described above,
\be\ba
	\mathbb{O}_{\text{trunc}} = \mathcal{C}_2^{\text{Star}} &+ f_{tt}\left(r\right)\,\partial_{t}^2 + 2f_{t\phi}\left(r\right)\Omega\,\partial_{t}\partial_{\phi} + f_{\phi\phi}\left(r\right)\Omega^2\,\partial_{\phi}^2 \\
	&+ \Delta\,f_{r}\left(r\right)\,\partial_{r} + f_{t}\left(r\right)\,\partial_{t} + f_{\phi}\left(r\right)\Omega\,\partial_{\phi} \\
	&+ s\left[F_{t}\left(r\right)\,\partial_{t} + F_{\phi}\left(r\right)\,\partial_{\phi} + \Delta F_0\left(r\right)\right] \,,
\ea\ee
where
\be\ba
	\mathcal{C}_2^{\text{Star}} &= \Delta^{-s}\partial_{r}\,\Delta^{s+1}\partial_{r} - \frac{\left(r_{+}^2+a^2\right)^2}{\Delta}\left(\partial_{t}+\Omega\,\partial_{\phi}\right)^2 \\
	&\quad+ s\frac{\left(r_{+}^2+a^2\right)\Delta^{\prime}}{\Delta}\left(\partial_{t}+\Omega\,\partial_{\phi}\right) + s\left(s+1\right)
\ea\ee
is the Starobinsky near-zone truncation introduced here for future convenience.

\appsection{Solving the $\SL$ constraints}
With this setup for the generic truncation of the radial operator, we now investigate the existence of three operators, $L_0$, $L_{+1}$ and $L_{-1}$, generating the $\SL$ algebra,
\be\label{eq:AlgebraConstraints}
	\left[L_{m},L_{n}\right] = \left(m-n\right)L_{m+n} \,,\quad m,n=0,\pm1 \,,
\ee
and whose Casimir precisely recovers a truncation of the radial operator of the type we just described,
\be\label{eq:CasimirConstraints}
	\mathcal{C}_2 = L_0^2 - \frac{1}{2}\left(L_{+1}L_{-1}+L_{-1}L_{+1}\right) \equiv \mathbb{O}_{\text{trunc}} \,.
\ee
Representations of $\SL$ are labeled by $\mathcal{C}_2$ and $L_0$. Using the $\mathbb{R}_{t}\times U\left(1\right)_{\phi}\times U\left(1\right)_{\psi}$ isometry of the background, we therefore make the following generic ansatz for the $\SL$ generators
\be
	\begin{gathered}
		L_0 = -\left(\beta_{t}\,\partial_{t} + \beta_{\phi}\Omega\,\partial_{\phi}\right) + s\,\beta_0 \,, \\
		L_{\pm 1} = \tilde{G}_{\pm}\left(t,r,\phi\right)\,\partial_{r} + \tilde{K}_{\pm}\left(t,r,\phi\right)\,\partial_{t} + \tilde{H}_{\pm}\left(t,r,\phi\right)\Omega\,\partial_{\phi} + s\,\tilde{\Lambda}_{\pm}\left(t,r,\phi\right) \,,
	\end{gathered}
\ee
with $\beta_{t}$, $\beta_{\phi}$ and $\beta_0$ constants and with all scalar parts of the generators assigned to a non-zero spin weight. However, we are not explicitly assuming that the vector part functions do not depend on $s$, even though this will turn out to be the case as we will see.

With this starting ansatz for the generators, we begin applying the $\SL$ algebra constraints. First of all, $\left[L_{\pm 1},L_0\right]=\pm L_{\pm 1}$ implies that,
\be
	\left(\beta_{t}\,\partial_{t} + \beta_{\phi}\Omega\,\partial_{\phi}\right)\tilde{X}_{\pm} = \pm \tilde{X}_{\pm} \,,\quad \tilde{X}_{\pm}\in\left\{\tilde{G}_{\pm},\tilde{K}_{\pm},\tilde{H}_{\pm},\tilde{\Lambda}_{\pm}\right\} \,,
\ee
which can be solved to eliminate the explicit $t$-dependence,
\be
	\tilde{X}_{\pm}\left(t,r,\phi\right) = e^{\pm t/\beta_{t}}X_{\pm}(r,\hat{\phi}) \,,\quad \hat{\phi}\equiv\phi-\frac{\beta_{\phi}}{\beta_{t}}\,\Omega\,t \,,\quad X_{\pm}\in\left\{G_{\pm},K_{\pm},H_{\pm},\Lambda_{\pm}\right\} \,.
\ee
We are assuming here, of course, that $\beta_{t}\ne0$. If this is not the case, the $\left[L_{\pm 1},L_0\right]=\pm L_{\pm 1}$ algebra constraints are used to eliminate the explicit $\phi$-dependence instead. Nevertheless, the final results based on the $\beta_{t}\ne0$ assumption will turn out to include $\beta_{t}=0$ as a special case so we will proceed here with $\beta_{t}\ne0$. From the above form of the functions $\tilde{X}_{\pm}$, it is instructive to reformulate the problem in the $(t,r,\hat{\phi})$ coordinates, instead of $\left(t,r,\phi\right)$. In these coordinates, the generators read
\be
	\begin{gathered}
		L_0 = -\beta_{t}\,\partial_{t} + s\,\beta_0 \,, \\
		L_{\pm 1} = e^{\pm t/\beta_{t}}\left[G_{\pm}(r,\hat{\phi})\,\partial_{r} + K_{\pm}(r,\hat{\phi})\,\partial_{t} + \hat{H}_{\pm}(r,\hat{\phi})\Omega\,\partial_{\hat{\phi}} + s\,\Lambda_{\pm}(r,\hat{\phi})\right] \,,
	\end{gathered}
\ee
with $\hat{H}_{\pm} = H_{\pm} - \frac{\beta_{\phi}}{\beta_{t}}K_{\pm}$. The remaining algebra constraints $\left[L_{\pm1},L_{\mp1}\right]=\pm2L_0$ then become
\begin{subequations}
	\begin{gather}
		\label{eq:AC1}
		G_{[\pm}\partial_{r}G_{\mp]} + \hat{H}_{[\pm}\Omega\,\partial_{\hat{\phi}}G_{\mp]} = \pm \beta_{t}^{-1}K_{(\pm}G_{\mp)} \,, \\
		\label{eq:AC2}
		G_{[\pm}\partial_{r}\hat{H}_{\mp]} + \hat{H}_{[\pm}\Omega\,\partial_{\hat{\phi}}\hat{H}_{\mp]} = \pm \beta_{t}^{-1}K_{(\pm}\hat{H}_{\mp)} \,, \\
		\label{eq:AC3}
		G_{[\pm}\partial_{r}K_{\mp]} + \hat{H}_{[\pm}\Omega\,\partial_{\hat{\phi}}K_{\mp]} = \pm \beta_{t}^{-1}\left(K_{\pm}K_{\mp}-\beta_{t}^2\right) \,, \\
		\label{eq:AC4}
		G_{[\pm}\partial_{r}\Lambda_{\mp]} + \hat{H}_{[\pm}\Omega\,\partial_{\hat{\phi}}\Lambda_{\mp]} = \pm \beta_{t}^{-1}\left(K_{(\pm}\Lambda_{\mp)} + \beta_0\beta_{t}\right) \,,
	\end{gather}
\end{subequations}
where we have defined the symmetric and antisymmetric operations with respect to the ``$\pm$'' indices $A_{(\pm}B_{\mp)}\equiv \left(A_{\pm}B_{\mp}+A_{\mp}B_{\pm}\right)/2$ and $A_{[\pm}B_{\mp]}\equiv \left(A_{\pm}B_{\mp}-A_{\mp}B_{\pm}\right)/2$. These are supplemented with the Casimir constraints associated with the near-horizon characteristic exponents of the equations of motion,
\begin{subequations}
	\begin{gather}
		\label{eq:CC1}
		-G_{\pm}G_{\mp} = \Delta \,\,\,,\,\,\, G_{(\pm}K_{\mp)} = 0 \,\,\,,\,\,\, G_{(\pm}\hat{H}_{\mp)} = 0 \,, \\
		\label{eq:CC2}
		-\left(K_{\pm}K_{\mp}-\beta_{t}^2\right) = -\frac{\left(r_{+}^2+a^2\right)^2}{\Delta} + f_{tt}\left(r\right) \,, \\
		\label{eq:CC3}
		-K_{(\pm}\hat{H}_{\mp)} = -\frac{\left(r_{+}^2+a^2\right)^2}{\Delta}\left(1-\frac{\beta_{\phi}}{\beta_{t}}\right) + f_{t\hat{\phi}}\left(r\right) \,, \\
		\label{eq:CC4}
		-\hat{H}_{\pm}\hat{H}_{\mp} = -\frac{\left(r_{+}^2+a^2\right)^2}{\Delta}\left(1-\frac{\beta_{\phi}}{\beta_{t}}\right)^2 + f_{\hat{\phi}\hat{\phi}}\left(r\right) \,,
	\end{gather}
\end{subequations}
\begin{subequations}
	\begin{gather}
		\label{eq:CC5}
		\ba
			{}&-\left[\frac{1}{2}\partial_{r}\left(G_{\pm}G_{\mp}\right) + \hat{H}_{(\pm}\Omega\,\partial_{\hat{\phi}}G_{\mp)} \mp \beta_{t}^{-1}K_{[\pm}G_{\mp]} + 2s\, \Lambda_{(\pm}G_{\mp)}\right] \\
			&\qquad\qquad\qquad\qquad\qquad\qquad\qquad\qquad\qquad\qquad= \left(s+1\right)\Delta^{\prime} + \Delta\,f_{r}\left(r\right) \,,
		\ea \\
		\label{eq:CC6}
		\ba
			{}&-\left[G_{(\pm}\partial_{r}K_{\mp)} + \hat{H}_{(\pm}\Omega\,\partial_{\hat{\phi}}K_{\mp)} + 2s \left(\Lambda_{(\pm}K_{\mp)} + \beta_0\beta_{t}\right)\right] \\
			&\qquad\qquad\qquad\qquad\qquad\qquad\qquad\quad= f_{t}\left(r\right) + s\left[\frac{\left(r_{+}^2+a^2\right)\Delta^{\prime}}{\Delta} + F_{t}\left(r\right)\right] \,,
		\ea \\
		\label{eq:CC7}
		\ba
			{}&-\left[G_{(\pm}\partial_{r}\hat{H}_{\mp)} + \frac{1}{2}\Omega\,\partial_{\hat{\phi}}(\hat{H}_{\pm}\hat{H}_{\mp}) \mp \beta_{t}^{-1}K_{[\pm}\hat{H}_{\mp]} + 2s\, \Lambda_{(\pm}\hat{H}_{\mp)}\right] \\
			&\qquad\qquad\qquad\qquad\qquad= f_{\hat{\phi}}\left(r\right) + s\left[\frac{\left(r_{+}^2+a^2\right)\Delta^{\prime}}{\Delta}\left(1-\frac{\beta_{\phi}}{\beta_{t}}\right) + F_{\hat{\phi}}\left(r\right)\right] \,,
		\ea \\
		\label{eq:CC8}
		\ba
			{}&-\left[G_{(\pm}\partial_{r}\Lambda_{\mp)} + \hat{H}_{(\pm}\Omega\,\partial_{\hat{\phi}}\Lambda_{\mp)} \mp \beta_{t}^{-1}K_{[\pm}\Lambda_{\mp]} + s\,\left(\Lambda_{\pm}\Lambda_{\mp} - \beta_0^2\right)\right] \\
			&\qquad\qquad\qquad\qquad\qquad\qquad\qquad\quad\qquad\qquad\qquad\qquad= s+1 + \Delta F_0\left(r\right) \,.
		\ea
	\end{gather}
\end{subequations}
We can now proceed to solve the above algebra and Casimir constraints. The first thing to notice is that the Casimir constraints imply that products of any component of $X_{\pm}$ and any component of $X_{\mp}$ are $\hat{\phi}$-independent and $s$-independent. As such, any $\hat{\phi}$-dependence must come in the form,
\be
	X_{\pm}(r,\hat{\phi}) = e^{\pm A(r,\hat{\phi})}\chi_{\pm}\left(r\right)\,\,\,,\,\,\,\chi_{\pm}\in\left\{g_{\pm}, k_{\pm}, \hat{h}_{\pm}, \lambda_{\pm}\right\}
\ee
with $\chi_{\pm}\left(r\right)$ being independent of $s$ as well.

For the radial components $G_{\pm}(r,\hat{\phi})$, \eqref{eq:CC1} and the differential equation \eqref{eq:AC1} allow to completely fix the radial dependence. In particular, the second Casimir constraint in \eqref{eq:CC1} makes the RHS of \eqref{eq:AC1} zero, while the second term in the algebra constraint \eqref{eq:AC1} also vanishes by virtue of the third Casimir constraint in \eqref{eq:CC1} since, due to the first of \eqref{eq:CC1}, $\hat{H}_{[\pm}\Omega\,\partial_{\hat{\phi}}G_{\mp]} = G_{(\pm}\hat{H}_{\mp)}\Omega\,\partial_{\hat{\phi}}\ln G_{\pm} = 0$. As such, all the $r$-dependence in $A(r,\hat{\phi})$ can be separated with the end result,
\be
	G_{\pm}(r,\hat{\phi}) = e^{\pm A(\hat{\phi})}\,\left[\mp\sqrt{\Delta}\right] \,.
\ee
Here, we note that we also have the freedom of an overall integration constant $c$ that comes in the form $e^{\pm c}$. Such overall constant reciprocal rescalings of $L_{\pm1}$, however, are automorphisms and are, therefore, algebraically trivial. With this extracted function form of $g_{\pm} = \mp\sqrt{\Delta}$, the second and third of the Casimir constraints \eqref{eq:CC1} imply that the radial functions in the $t$- and $\hat{\phi}$-components of the generators $L_{\pm 1}$ are equal,
\be
	k_{\pm}\left(r\right) = k\left(r\right) \,,\quad \hat{h}_{\pm}\left(r\right) = \hat{h}\left(r\right) \,.
\ee
It is then straightforward to conclude that the purely angular function $A(\hat{\phi})$ is linear in $\hat{\phi}$. For example, taking the $\hat{\phi}$-derivative of the algebra constraint \eqref{eq:AC2} or \eqref{eq:AC3}, we get
\be
	A^{\prime\prime}(\hat{\phi}) = 0 \Rightarrow A(\hat{\phi}) = \tau\hat{\phi} \,,
\ee
with $\tau$ an arbitrary integration constant, while we have again ignored algebraically trivial reciprocal rescalings of $L_{\pm 1}$.

Moving on, multiplying \eqref{eq:AC2} with $\tau\Omega$ and adding with it \eqref{eq:AC3}, we end up with a first order non-linear differential equation for the function $\beta_{t}^{-1}k\left(r\right)+\tau\Omega\,\hat{h}\left(r\right)$ which can be solved to find
\be\label{eq:KmH}
	\beta_{t}^{-1}k\left(r\right)+\tau\Omega\,\hat{h}\left(r\right) = \frac{\left(\sqrt{r-r_{+}}+\sqrt{r-r_{-}}\right)^4 + c_1}{\left(\sqrt{r-r_{+}}+\sqrt{r-r_{-}}\right)^4 - c_1} \,,
\ee
with $c_1$ an integration constant. In the mean time, the Casimir constraints \eqref{eq:CC2} and \eqref{eq:CC4} tell us that, up to an overall sign, which falls into the category of automorphisms of the $\SL$ algebra,
\be
	k\left(r\right) \sim \frac{r_{+}^2+a^2}{\sqrt{\Delta}} \quad\text{and}\quad \hat{h}\left(r\right) \sim \left(1-\frac{\beta_{\phi}}{\beta_{t}}\right)\frac{r_{+}^2+a^2}{\sqrt{\Delta}}\quad\text{as}\quad \Delta\rightarrow0 \,,
\ee
where the sign of $\hat{h}\left(r\right)$ relative to $k\left(r\right)$ was also fixed by using the near-horizon behavior of \eqref{eq:CC3}. From these, we infer the near-horizon behavior of \eqref{eq:KmH} which fixes $c_1=\left(r_{+}-r_{-}\right)^2$ regardless of the values of $\beta_{t}$, $\beta_{\phi}$ and $\tau$. This greatly simplifies \eqref{eq:KmH} to
\be
	\beta_{t}^{-1}k\left(r\right)+\tau\Omega\,\hat{h}\left(r\right) = \partial_{r}\left(\sqrt{\Delta}\right) \,,
\ee
and results in the following quite pleasing $s$-independent expressions of $k\left(r\right)$ and $\hat{h}\left(r\right)$ after solving \eqref{eq:AC2} and matching the near-horizon behavior of $\hat{h}\left(r\right)$
\be
	\beta_{t}^{-1}k\left(r\right) = \partial_{r}\left(\sqrt{\Delta}\right) - \tau\Omega\left(1-\frac{\beta_{\phi}}{\beta_{t}}\right)\frac{r_{+}^2+a^2}{\sqrt{\Delta}} \,,\quad \hat{h}\left(r\right) = \left(1-\frac{\beta_{\phi}}{\beta_{t}}\right)\frac{r_{+}^2+a^2}{\sqrt{\Delta}} \,.
\ee
These also imply that $f_{t}=f_{\phi}=0$. Moreover, the near-horizon behavior of $k\left(r\right)$ implies the following relation between $\beta_{t}$, $\beta_{\phi}$ and $\tau$
\be\label{eq:SL2Rbetas}
	\beta_{t}\left(1-\tau\beta\Omega\right) = \beta\left(1-\tau\beta_{\phi}\Omega\right) \,.
\ee
As a result, one can eliminate $\beta_{t}$ or $\beta_{\phi}$, depending on the value of $\tau$. More explicitly,
\be
	\begin{gathered}
		\beta_{t} = \beta\,\frac{1-\tau\beta_{\phi}\Omega}{1-\tau\beta\Omega} \equiv \beta^{\left(\tau\right)} \quad\text{and}\quad \beta_{\phi}\in\mathbb{R} \quad\text{if}\quad \tau\ne\frac{1}{\beta\Omega} \,, \\
		\text{ OR } \\
		\beta_{\phi} = \beta \quad\text{and}\quad\beta_{t}\in\mathbb{R} \quad\text{if}\quad \tau=\frac{1}{\beta\Omega} \,.
	\end{gathered}
\ee

We now move on to the remaining constraints \eqref{eq:AC4} and \eqref{eq:CC5}-\eqref{eq:CC8}. \eqref{eq:AC4} results in a linear first-order inhomogeneous differential equation for $\Lambda_{+}\left(r\right)+\Lambda_{-}\left(r\right)$ which is easily solved to give
\be
	\lambda_{+}\left(r\right) + \lambda_{-}\left(r\right) = -\left[2\beta_0\sqrt{\frac{r-r_{+}}{r-r_{-}}} + \frac{r_{+}-r_{-}}{2\sqrt{\Delta}}\right] \,,
\ee
where the integration constant was fixed by the near-horizon behavior of \eqref{eq:CC6}. Next, \eqref{eq:CC5} simplifies to
\be
	s\left[\sqrt{\Delta}\left(\lambda_{+}\left(r\right)-\lambda_{-}\left(r\right)\right) + \Delta^{\prime}\right] = -\Delta f_{r}\left(r\right)
\ee
and, since $f_{r}\left(r\right)$ is independent of the spin weight, $f_{r}\left(r\right)=0$, leaving us with
\be
	\lambda_{+}\left(r\right)-\lambda_{-}\left(r\right) = -2\partial_{r}\left(\sqrt{\Delta}\right) \,.
\ee
These two finally tell us that
\be
	\lambda_{\pm}\left(r\right) = -\left(1\pm1\right)\partial_{r}\left(\sqrt{\Delta}\right) + \left(1-\beta_0\right)\sqrt{\frac{r-r_{+}}{r-r_{-}}}
\ee
and the remaining constraints fix the spin-weighted ``far-horizon'' corrections $F_{t}\left(r\right)$, $F_{\hat{\phi}}\left(r\right)$ and $F_0\left(r\right)$.

In summary, after some rearrangement, the most general subtracted geometry $\SL$ is generated by
\be\label{eq:SL2RGeneral}
	\begin{gathered}
		L_0 = -\left(\beta_{t}\,\partial_{t} + \beta_{\phi}\Omega\,\partial_{\phi}\right) + s\,\beta_0 \,, \\
		\ba
			L_{\pm1} = e^{\pm \left[t/\beta + \tau\left(\phi-\Omega t\right)\right]}&\bigg[\mp\sqrt{\Delta}\,\partial_{r} + \left(\frac{\beta_{t}}{\beta}\sqrt{\frac{r-r_{+}}{r-r_{-}}} + \frac{r_{+}-r_{-}}{2\sqrt{\Delta}}\right)\beta\,\partial_{t} \\
			&+ \left(\frac{\beta_{\phi}}{\beta}\sqrt{\frac{r-r_{+}}{r-r_{-}}} + \frac{r_{+}-r_{-}}{2\sqrt{\Delta}}\right)\beta\Omega\,\partial_{\phi} \\
			&- s\left(\left(\beta_0\pm1\right)\sqrt{\frac{r-r_{+}}{r-r_{-}}} + \left(1\pm1\right)\frac{r_{+}-r_{-}}{2\sqrt{\Delta}}\right) \bigg] \,,
		\ea
	\end{gathered}
\ee
with $\beta_{t}=\beta^{\left(\tau\right)}$ and generic $\beta_{\phi}$, $\beta_0$ and $\tau$ if $\tau\beta\Omega\ne1$ or $\beta_{\phi}=\beta$ and generic $\beta_{t}$ and $\beta_0$ if $\tau\beta\Omega=1$. The associated Casimir is given by,
\be
	\mathcal{C}_2 = \mathcal{C}_2^{\text{Star}} + \frac{r_{+}-r_{-}}{r-r_{-}}\big[\beta_{t}\,\partial_{t}+\beta_{\phi}\Omega\,\partial_{\phi}+s\left(1-\beta_0\right)\big]\left[\left(\beta_{t}-\beta\right)\partial_{t} + \left(\beta_{\phi}-\beta\right)\Omega\,\partial_{\phi}-s\,\beta_0\right] \,.
\ee

From all of the above Casimirs, only the subset with $(\beta_{\phi},\beta_0)=(0,1)$ or $(\beta_{\phi},\beta_0)=(\beta,0)$ give rise to valid near-zone approximations, for which all the static terms must be kept explicitly, and precisely correspond to $\tau$-generalizations of the Love and Starobinsky near-zones respectively.

Interestingly, all the generators written above are automatically regular at both the future and the past event horizons, as can be seen by working in advanced and retarded null coordinates respectively. This statement of course only holds for the vector, $s=0$, parts, since the scalar $s\ne0$ parts are tetrad-dependent and we are working here in the singular-at-the-horizon Kinnersley tetrad~\cite{Kinnersley1969}. More explicitly,
\be\ba\label{eq:SL2RGeneralAdvanced}
	L_0\big|_{s=0} &= -\left(\beta_{t}\,\partial_{t_{+}} + \beta_{\phi}\Omega\,\partial_{\varphi_{+}}\right) = -\left(\beta_{t}\,\partial_{t_{-}} + \beta_{\phi}\Omega\,\partial_{\varphi_{-}}\right) \,, \\
	L_{\pm1}\big|_{s=0} &= \exp\left\{\pm\left[\frac{t_{+}-r}{\beta}-\tau\left(\varphi_{+}-\Omega\left(t_{+}-r\right)\right)\right]\right\} \left(\frac{r-r_{-}}{r_{+}}\right)^{\mp 2M\left(1+\tau\beta\Omega\right)/\beta} \\
	&\quad\times\bigg[\mp\left(r-r_{\mp}\right)\partial_{r}+\frac{r-r_{\mp}}{r-r_{-}}\left(\left(\beta_{t}\mp\left(r+r_{+}\right)\right)\partial_{t_{+}} + \beta_{\phi}\Omega\,\partial_{\varphi_{+}}\right) \\
	&\quad\quad\quad+ \frac{1\mp1}{2}\frac{r_{+}-r_{-}}{r-r_{-}}\beta\left(\partial_{t_{+}}+\Omega\,\partial_{\varphi_{+}}\right)\bigg] \\
	&= \exp\left\{\pm\left[\frac{t_{-}+r}{\beta}-\tau\left(\varphi_{-}-\Omega\left(t_{-}+r\right)\right)\right]\right\} \left(\frac{r-r_{-}}{r_{+}}\right)^{\pm 2M\left(1+\tau\beta\Omega\right)/\beta} \\
	&\quad\times\bigg[\mp\left(r-r_{\pm}\right)\partial_{r}+\frac{r-r_{\pm}}{r-r_{-}}\left(\left(\beta_{t}\pm\left(r+r_{+}\right)\right)\partial_{t_{-}} + \beta_{\phi}\Omega\,\partial_{\varphi_{-}}\right) \\
	&\quad\quad\quad+ \frac{1\pm1}{2}\frac{r_{+}-r_{-}}{r-r_{-}}\beta\left(\partial_{t_{-}}+\Omega\,\partial_{\varphi_{-}}\right)\bigg] \,,
\ea\ee
in advanced ($+$) and retarded ($-$) null coordinates \eqref{eq:NullCoordiantes_KerrNewman}.

\appsection{Globally defined, time-reversal symmetric $\SL$'s}
In the main text, we are mostly interested in those approximations which preserve all the symmetries of the full geometry, including time-reversal invariance\footnote{Time-reversal invariance here refers to the simultaneous time-reversal transformation $t\rightarrow-t$ \textit{and} the flip of the direction of rotation of the black hole $a\rightarrow-a$.}, and are globally defined. These correspond to $\tau=0$ and make up a $2$-parameter family of subtracted geometry approximations enjoying an $\SL$ symmetry, labeled by the $\beta_{\phi}$ and $\beta_0$ parameters. The generators read,
\be
	\begin{gathered}
		L_0 = -\left(\beta\,\partial_{t} + \beta_{\phi}\Omega\,\partial_{\phi}\right) + s\beta_0 \,, \\
		\ba
			L_{\pm1} = e^{\pm t/\beta}&\left[\mp\sqrt{\Delta}\,\partial_{r} + \partial_{r}\left(\sqrt{\Delta}\right)\beta\,\partial_{t} + \frac{a}{\sqrt{\Delta}}\,\partial_{\phi} - s\,\left(1\pm1\right)\partial_{r}\left(\sqrt{\Delta}\right)\right] \\
			&+\left(e^{\pm t/\beta}\sqrt{\frac{r-r_{+}}{r-r_{-}}}\right)\big(\beta_{\phi}\Omega\,\partial_{\phi} + s\left(1-\beta_0\right)\big) \,,
		\ea
	\end{gathered}
\ee
and the associated Casimir operator is given by,
\be
	\mathcal{C}_2 = \mathcal{C}_2^{\text{Star}} + \frac{r_{+}-r_{-}}{r-r_{-}}\big[\beta\,\partial_{t}+\beta_{\phi}\Omega\,\partial_{\phi}+s\left(1-\beta_0\right)\big]\big[\left(\beta_{\phi}-\beta\right)\Omega\,\partial_{\phi} - s\,\beta_0\big] \,.
\ee
As already mentioned, only the subset with $(\beta_{\phi},\beta_0)=(0,1)$ (Love near-zone) or $(\beta_{\phi},\beta_0)=(\beta,0)$ (Starobinsky near-zone) give rise to valid near-zone approximations. We also remind here that we still have the freedom of performing the coordinate transformations \eqref{eq:t_r_phi_Rep} and the local boost transformation \eqref{eq:s_Rep}. From these, only the temporal translations $t\rightarrow t + g_{t}\left(r\right)$ preserve the near-zone behavior which, however, break the time-reflection symmetry.

A nice remark here is that the $\tau$-generalized generators of the local near-zone $\SL$ can be obtained by the $\tau=0$ globally defined ones after performing a particular $\phi$-dependent temporal translation,
\be
	L_{m}^{\left(\tau=0\right)} \xrightarrow[\tau\beta\Omega\ne1]{t \rightarrow t + \tau\beta\left(\phi-\Omega\,t\right)} L_{m}^{\left(\tau\ne0\right)}
\ee

\appsection{Extension to $\SL\times\SL$}
We will finish with a short investigation on the possibility of extending the above-found $\SL$ symmetries, for which the radial operator truncations preserve the characteristic exponents near the event horizon, into the full $2$-d global conformal structure $\SL\times\SL$.

Consider, therefore, two general such $\SL$ symmetries generated by vector fields $L_{m}$ and $\bar{L}_{m}$ of the form \eqref{eq:SL2RGeneral}. They are characterized by parameters $\{\beta_{t},\beta_{\phi},\beta_0,\tau\}$ and $\{\bar{\beta}_{t},\bar{\beta}_{\phi},\bar{\beta}_0,\bar{\tau}\}$ respectively, with each set of parameters being subject to the relation \eqref{eq:SL2Rbetas}. By working out the requirement that
\be
	\left[L_{m},\bar{L}_{n}\right] = 0 \,,\quad m,n=0,\pm1 \,,
\ee
we extract the following summarizing condition
\be
	L_0 + \bar{L}_0 = -\beta K + s \,,
\ee
where $K=\partial_{t} + \Omega\,\partial_{\phi}$ is the Killing vector with respect to which the event horizon is a Killing horizon. In other words, $\bar{\beta}_{t} = \beta - \beta_{t}$, $\bar{\beta}_{\phi} = \beta - \beta_{\phi}$ and $\bar{\beta}_0 = 1-\beta_0$. The associated Casimirs turn out to be exactly the same and can be written in the suggestive form
\be\ba
	\mathcal{C}_2 &= \bar{\mathcal{C}}_2 = \Delta^{-s}\partial_{r}\,\Delta^{s+1}\,\partial_{r} +s\left(s+1\right) \\
	&\quad- \frac{r_{+}-r_{-}}{r-r_{+}}\left[\left(\frac{L_0+\bar{L}_0}{2}\right)^2-\frac{s^2}{4}\right] + \frac{r_{+}-r_{-}}{r-r_{-}}\left[\left(\frac{L_0-\bar{L}_0}{2}\right)^2-\frac{s^2}{4}\right] \,.
\ea\ee
In a $\text{CFT}_2$ interpretation, this shows that the characteristic exponents near the outer and inner horizons are the (squares of half of the) scaling dimension and spin-weight of the spacetime NP scalar under the action of the $\text{CFT}_2$ dilaton and Lorentz generators respectively, shifted by $-s^2/4$ for non-zero NP spin weight. In this language, the above truncations of the radial operator seek to preserve the scaling dimension but allow to approximate the $\text{CFT}_2$ spin-weight. The remaining generators can similarly be written as
\be\ba
	L_{\pm1} &= e^{\pm\left[t/\beta + \tau\left(\phi-\Omega\,t\right)\right]} \\
	&\quad\times\left[\mp\sqrt{\Delta}\,\partial_{r} - \sqrt{\frac{r-r_{-}}{r-r_{+}}}\frac{L_0+\bar{L}_0 \pm s}{2} - \sqrt{\frac{r-r_{+}}{r-r_{-}}}\frac{L_0-\bar{L}_0 \pm s}{2}\right] \,, \\
	\bar{L}_{\pm1} &= e^{\pm\left[t/\beta + \bar{\tau}\left(\phi-\Omega\,t\right)\right]} \\
	&\quad\times\left[\mp\sqrt{\Delta}\,\partial_{r} - \sqrt{\frac{r-r_{-}}{r-r_{+}}}\frac{L_0+\bar{L}_0 \pm s}{2} - \sqrt{\frac{r-r_{+}}{r-r_{-}}}\frac{\bar{L}_0-L_0 \pm s}{2}\right] \,.
\ea\ee

Last, for the case of near-zone $\SL\times\SL$'s, there is one tower of such enhancements, corresponding to $\left(\beta_{\phi},\beta_0\right)=\left(0,1\right)$ and, thus, $\left(\bar{\beta}_{\phi},\bar{\beta}_0\right)=\left(\beta,0\right)$. One of the outcomes of the current analysis is then that near-zone $\SL\times\SL$'s can \textit{never} be globally defined. The best one can do is to have near-zone $\SL\times\SL$ symmetries spontaneously broken down to $\SL\times U\left(1\right)$ from the periodic identification of the azimuthal angle. This is precisely the enhancements of the Love symmetry presented in Section~\ref{sec:LocalSL2RLove4d}. We remark here that the construction of near-zone $\SL\times\SL$ symmetries described above also contains the Kerr/CFT proposal for the $4$-dimensional Kerr black hole in~\cite{Castro:2010fd} as a special case, corresponding to a different near-zone truncation with $\beta_{t}=\beta\frac{r_{+}-r_{-}}{2r_{+}}$ and $\bar{\beta}_{t}=\beta\frac{r_{+}+r_{-}}{2r_{+}}$ and which has the unique property of preserving the characteristic exponent at the inner horizon as well.
	\newpage
\appchapter{Generalized Lie derivative on GHP tensors}
\label{app:LieDerivative}

In this appendix, we present the construction of the generalized ``spin-weighted'' Lie derivative $\Lstr_{\xi}$ employed in Section~\ref{sec:SpinWeightedGenerators}. In particular, we will construct the most general generalized Lie derivative acting on objects that are spacetime tensors of some particular rank and also carry definite GHP weights. We will refer to such object as GHP tensors, with the ``tensors'' and ``GHP'' parts referring to their homogeneous transformation rules under diffeomorphisms and local boosts and rotations respectively.

\appsection{Constructing the spin-weighted Lie derivative}
\label{sec:SpinWeightedLieDerGeneral}

There are some minimal conditions that such a Lie derivative must obey for it to be well defined. In the following conditions, we denote a general GHP tensor of GHP weights $\left\{p,q\right\}$ and spacetime tensorial structure $\left(k,l\right)$, with $k$ the contravariant rank and $l$ the covariant rank, as $A_{\left\{p,q\right\}}^{\left(k,l\right)}$.
\begin{itemize}
	\item It reduces to the usual unique Lie derivative $\mathcal{L}_{\xi}$ in the case the GHP tensor it acts on has zero GHP weights, i.e. when it acts on pure spacetime tensors,
	\be
		\Lstr_{\xi}A_{\left\{0,0\right\}}^{\left(k,l\right)} = \mathcal{L}_{\xi}A_{\left\{0,0\right\}}^{\left(k,l\right)} \,.
	\ee
	
	\item It obeys the Leibniz rule,
	\be
		\Lstr_{\xi}\left(A_{\left\{p,q\right\}}^{\left(k,l\right)}B_{\left\{p^{\prime},q^{\prime}\right\}}^{\left(k^{\prime},l^{\prime}\right)}\right) = \left(\Lstr_{\xi}A_{\left\{p,q\right\}}^{\left(k,l\right)}\right)B_{\left\{p^{\prime},q^{\prime}\right\}}^{\left(k^{\prime},l^{\prime}\right)} + A_{\left\{p,q\right\}}^{\left(k,l\right)}\left(\Lstr_{\xi}B_{\left\{p^{\prime},q^{\prime}\right\}}^{\left(k^{\prime},l^{\prime}\right)}\right) \,.
	\ee
	
	\item It acts covariantly, leaving the GHP weights and spacetime rank unaltered, i.e. when it acts on a GHP tensor, it outputs a GHP tensor of the same type,
	\be
		\Lstr_{\xi}A_{\left\{p,q\right\}}^{\left(k,l\right)} \xrightarrow{\lambda,\chi}\lambda^{b}e^{is\chi}\Lstr_{\xi}A_{\left\{p,q\right\}}^{\left(k,l\right)} \,,
	\ee
	with the spin and boost weights related to the GHP weights according to $b=\frac{p+q}{2}$ and $s=\frac{p-q}{2}$.
	
	\item It is linear in the vector $\xi^{\mu}$ with respect to which we are Lie dragging.
\end{itemize}

Then, in order to satisfy the first two conditions, one starts with an ansatz that contains at most first spacetime derivatives, such that the Leibniz rule is satisfied,
\be
	\Lstr_{\xi} = \mathcal{L}_{\xi} + \lambda_{0,\xi}\left(p,q;k,l\right) + \lambda_{1,\xi}^{\mu}\left(p,q;k,l\right)\partial_{\mu} \,,
\ee
where $\left(k,l\right)$ and $\left\{p,q\right\}$ are the spacetime rank and GHP weights of the GHP tensor the generalized Lie derivative acts on. The Leibniz rule then further imposes
\be\ba
	\lambda_{0,\xi}\left(p+p^{\prime},q+q^{\prime};k+k^{\prime},l+l^{\prime}\right) &= \lambda_{0,\xi}\left(p,q;k,l\right) + \lambda_{0,\xi}\left(p^{\prime},q^{\prime};k^{\prime},l^{\prime}\right) \,, \\
	\lambda_{1,\xi}^{\mu}\left(p+p^{\prime},q+q^{\prime};k+k^{\prime},l+l^{\prime}\right) &= \lambda_{1,\xi}^{\mu}\left(p,q;k,l\right) = \lambda_{1,\xi}^{\mu}\left(p^{\prime},q^{\prime};k^{\prime},l^{\prime}\right) \,.
\ea\ee
The first constraint is the usual multi-variable Cauchy functional equation implying that $\lambda_{0,\xi}\left(p,q;k,l\right)$ must be linear in $k$, $l$, $p$ and $q$, while the second constraint tells us that $\lambda_{1,\xi}^{\mu}\left(p,q;k,l\right)$ is independent of the spacetime rank and GHP weights. Combining with the first condition of getting the usual Lie derivative $\mathcal{L}_{\xi}$ when acting on pure spacetime tensors with zero GHP weights,
\be
	\lambda_{0,\xi}\left(0,0;k,l\right) = 0\,\,\,\,,\,\,\,\,\lambda_{1,\xi}^{\mu}\left(0,0;k,l\right) = 0 \,,
\ee
we see that so far we have
\be\label{eq:LieIntermediate}
	\Lstr_{\xi} = \mathcal{L}_{\xi} + p\,\alpha_{\xi} + q\,\beta_{\xi} \,,
\ee
with $\alpha_{\xi}$ and $\beta_{\xi}$ some spacetime scalar functions, linear in $\xi$, that are independent of $k$, $l$, $p$ and $q$.

We next impose the third condition of this generalized Lie derivative to act covariantly on GHP tensors. Since the additive modifications on the usual Lie derivative are independent of the spacetime tensorial structure of the GHP tensor it acts on, it is sufficient to consider the case of spacetime scalars of general GHP weights for which $\mathcal{L}_{\xi}=\xi^{\mu}\nabla_{\mu}$. The spacetime covariant derivative is written in terms of the NP directional derivatives,
\be
	D \equiv \ell^{\mu}\nabla_{\mu}\,\,\,\,,\,\,\,\,\triangle \equiv n^{\mu}\nabla_{\mu}\,\,\,\,,\,\,\,\,\delta \equiv m^{\mu}\nabla_{\mu}\,\,\,\,,\,\,\,\,\bar{\delta} \equiv \bar{m}^{\mu}\nabla_{\mu} \,,
\ee
as
\be
	\nabla_{\mu} =-\ell_{\mu}\triangle - n_{\mu}D + m_{\mu}\bar{\delta} + \bar{m}_{\mu}\delta \,,
\ee
so,
\be
	\xi^{\mu}\nabla_{\mu} = -\xi_{\ell}\triangle - \xi_{n}D + \xi_{m}\bar{\delta} + \xi_{\bar{m}}\delta \,.
\ee
It is well known that the NP derivatives do not transform covariantly under local boosts and rotations when acting on NP scalars. Their ``bad'' transformation rules read
\be
	\begin{pmatrix}
		D \\
		\triangle \\
		\delta \\
		\bar{\delta}
	\end{pmatrix} \xrightarrow{\lambda,\chi}
	\begin{pmatrix}
		\lambda\left[D + \frac{1}{2}\left(D\ln\left(\mathscr{B}^{p}\bar{\mathscr{B}}^{q}\right)\right)\right] \\
		\lambda^{-1}\left[\triangle + \frac{1}{2}\left(\triangle\ln\left(\mathscr{B}^{p}\bar{\mathscr{B}}^{q}\right)\right)\right] \\
		e^{i\chi}\left[\delta + \frac{1}{2}\left(\delta\ln\left(\mathscr{B}^{p}\bar{\mathscr{B}}^{q}\right)\right)\right] \\
		e^{-i\chi}\left[\bar{\delta} + \frac{1}{2}\left(\bar{\delta}\ln\left(\mathscr{B}^{p}\bar{\mathscr{B}}^{q}\right)\right)\right]
	\end{pmatrix} \,,
\ee
where we have defined
\be
	\mathscr{B}\equiv \lambda e^{i\chi} \,.
\ee
As a result, the usual Lie derivative transforms as
\be
	\mathcal{L}_{\xi} \xrightarrow{\lambda,\chi}\mathcal{L}_{\xi} + \frac{1}{2}\left(\xi^{\mu}\nabla_{\mu}\ln\left(\mathscr{B}^{p}\bar{\mathscr{B}}^{q}\right)\right) \,,
\ee
and the covariance of the generalized Lie derivative implies the following transformation rules for the unknown scalar functions $\alpha_{\xi}$ and $\beta_{\xi}$
\be\ba
	\alpha_{\xi} &\xrightarrow{\lambda,\chi} \alpha_{\xi} - \frac{1}{2}\xi^{\mu}\nabla_{\mu}\ln\mathscr{B} \,, \\
	\beta_{\xi} &\xrightarrow{\lambda,\chi} \beta_{\xi} - \frac{1}{2}\xi^{\mu}\nabla_{\mu}\ln\bar{\mathscr{B}} \,.
\ea\ee
Comparing with the transformation laws for the ``bad'' spin coefficients~\eqref{eq:NPSpinCoefficientsBad},
\be
	\begin{pmatrix}
		\varepsilon \\
		\gamma \\
		\beta \\
		\alpha
	\end{pmatrix} \xrightarrow{\lambda,\chi}
	\begin{pmatrix}
		\lambda\left(\varepsilon + \frac{1}{2}D\ln\mathscr{B}\right) \\
		\lambda^{-1}\left(\gamma + \frac{1}{2}\triangle\ln\mathscr{B}\right) \\
		e^{i\chi}\left(\beta + \frac{1}{2}\delta\ln\mathscr{B}\right) \\
		e^{-i\chi}\left(\alpha + \frac{1}{2}\bar{\delta}\ln\mathscr{B}\right)
	\end{pmatrix} \,,
\ee
we see that the non-homogeneous part of the transformation laws for the scalar functions $\alpha_{\xi}$ and $\beta_{\xi}$ can be reproduced from
\be
	\alpha_{\xi}^{\text{bad}} = -\xi^{\mu}\zeta_{\mu} \,\,\,\,,\,\,\,\,\beta_{\xi}^{\text{bad}} = -\xi^{\mu}\bar{\zeta}_{\mu} \,,
\ee
where
\be
	\zeta_{\mu} = -\ell_{\mu}\gamma - n_{\mu}\varepsilon + m_{\mu}\alpha + \bar{m}_{\mu}\beta = -\frac{1}{2}\left(n^{\nu}\nabla_{\mu}\ell_{\nu} - \bar{m}^{\nu}\nabla_{\mu}m_{\nu}\right) \,.
\ee
In conclusion, the most general generalized Lie derivative that acts on GHP tensors of GHP weights $\left\{p,q\right\}$ reads,
\be
	\Lstr_{\xi} = \mathcal{L}_{\xi} -\xi^{\mu}\left(p\,\zeta_{\mu} + q\,\bar{\zeta}_{\mu}\right) + p\,\eta_{\xi} + q\,\vartheta_{\xi} \,,
\ee
with $\eta_{\xi}$ and $\vartheta_{\xi}$ two scalar functions independent of $p$ and $q$ that transform covariantly with zero GHP weight,
\be
	\eta_{\xi}\xrightarrow{\lambda,\chi}\eta_{\xi} \,,\quad \vartheta_{\xi}\xrightarrow{\lambda,\chi}\vartheta_{\xi} \,,
\ee
that are also linear in the vector field $\xi^{\mu}$. For the minimal choice $\eta_{\xi}=\vartheta_{\xi}=0$, we retrieve the usual GHP derivative~\cite{Geroch:1973am}. These scalar functions that appear above are arbitrary but part of them can be fixed by uniquely constructing a generalized Lie derivative when Lie dragging along a Killing vector. This was first proposed by Ludwig et al.~\cite{Ludwig2000,Ludwig:2001hx} and is obtained from our above construction by requiring the existence of a Killing vector, satisfying $\mathcal{L}_{\xi}g_{\mu\nu}=0$, to be apparent directly at the level of the tetrad vectors. It is sufficient to impose
\be
	n_{\mu}\Lstr_{\xi}\ell^{\mu} = \bar{m}_{\mu}\mathcal{L}_{\xi}m^{\mu} = 0 \quad\text{when}\quad \mathcal{L}_{\xi}g_{\mu\nu}=0 \,,
\ee
uniquely fixing $\eta_{\xi}$ and $\vartheta_{\xi}$ with the end result for the generalized Lie derivative
\be\label{eq:LieKilling}
	\Lstr_{\xi} = \mathcal{L}_{\xi} + b\,n_{\mu}\mathcal{L}_{\xi}\ell^{\mu} - s\,\bar{m}_{\mu}\mathcal{L}_{\xi}m^{\mu} \quad \text{when}\quad \mathcal{L}_{\xi}g_{\mu\nu}=0 \,.
\ee

\appsection{Preserving the algebra}
\label{sec:PreservingAlgebra}

A further feature we would like the generalized Lie derivative to have is for it to preserve the algebra already satisfied by the usual Lie derivative with respect to some vector field generators of the algebra. In particular, we would like to preserve the identity
\be
	\left[\mathcal{L}_{\xi_1},\mathcal{L}_{\xi_2}\right] = \mathcal{L}_{\left[\xi_1,\xi_2\right]_{\text{LB}}} \,,
\ee
where $\left[\xi_1,\xi_2\right]_{\text{LB}} = \mathcal{L}_{\xi_1}\xi_2$ is the Lie bracket of the two vector fields. At the level of the scalar functions $\alpha_{\xi}$ and $\beta_{\xi}$ in \eqref{eq:LieIntermediate}, prior to requiring the generalized Lie derivative to act covariantly on a general GHP tensor, the above algebra-preserving identity implies
\be\label{eq:LieAlgebraPreserving}
	\left[\Lstr_{\xi_1},\Lstr_{\xi_2}\right] = \Lstr_{\left[\xi_1,\xi_2\right]_{\text{LB}}} \Rightarrow
	\begin{cases}
		\mathcal{L}_{\xi_1}\alpha_{\xi_2} - \mathcal{L}_{\xi_2}\alpha_{\xi_1} = \alpha_{\left[\xi_1,\xi_2\right]_{\text{LB}}} \\
		\mathcal{L}_{\xi_1}\beta_{\xi_2} - \mathcal{L}_{\xi_2}\beta_{\xi_1} = \beta_{\left[\xi_1,\xi_2\right]_{\text{LB}}}
	\end{cases} \,.
\ee
Then, after writing $\alpha_{\xi} = -\xi^{\mu}\zeta_{\mu} + \eta_{\xi}$ and $\beta_{\xi} = -\xi^{\mu}\bar{\zeta}_{\mu} + \vartheta_{\xi}$ to ensure the covariance of the generalized Lie derivative and since the tensorial structure associated with the geometry itself is carried only by the tetrad vectors, the most general form of the scalar function $\eta_{\xi}$ is some linear operator containing at least one spacetime covariant derivative acting on the vector field $\xi^{\mu}$,
\be\ba
	\eta_{\xi} &= \left[H^{\left(\mu\nu\right)} + \frac{1}{2}\left(\ell^{[\mu}n^{\nu]} - m^{[\mu}\bar{m}^{\nu]}\right)\right]\nabla_{\nu}\xi_{\mu} + H^{\left(\mu\nu\right)\rho}\nabla_{\rho}\nabla_{\nu}\xi_{\mu} + \dots \,, \\
	\vartheta_{\xi} &= \left[\Theta^{\left(\mu\nu\right)} + \frac{1}{2}\left(\ell^{[\mu}n^{\nu]} + m^{[\mu}\bar{m}^{\nu]}\right)\right]\nabla_{\nu}\xi_{\mu} + \Theta^{\left(\mu\nu\right)\rho}\nabla_{\rho}\nabla_{\nu}\xi_{\mu} + \dots \,,
\ea\ee
where $H^{\left(\mu\nu\right)}$, $\Theta^{\left(\mu\nu\right)}$, $H^{\left(\mu\nu\right)\rho}$, $\Theta^{\left(\mu\nu\right)\rho}$, $\dots$ are pure spacetime tensors carrying zero GHP weights that are symmetric in their first two indices, while the antisymmetric parts of the rank-$2$ tensors were fixed such that we reproduce the generalized Lie derivative \eqref{eq:LieKilling} when Lie dragging along a Killing vector. We want the algebra preserving identity to be independent of the involved vector fields $\xi_1^{\mu}$ and $\xi_2^{\mu}$. Unfortunately, it can be shown directly from the algebra preserving constraints \eqref{eq:LieAlgebraPreserving} that this is never possible, unless $\xi^{\mu}$ is a Killing vector.
	\newpage
\appchapter{Modified Spherical Harmonics basis on $\mathbb{S}^3$}
\label{app:SphericalHarmonics}

In this appendix, we present the basis used for the modified harmonic expansion, naturally associated with the computation of Love numbers for axisymmetric distributions in $d=5$ spacetime dimensions. For the sake of this, we need to perform a harmonic expansion over the $\mathbb{S}^1\times\mathbb{S}^1$ subpart of $\mathbb{S}^3$, appropriate for the general isometry group factor $U\left(1\right)\times U\left(1\right)$ of such configurations.

\appsection{The basis}

The basis we are looking for is extracted by solving the eigenvalue problem for the Laplace-Beltrami operator on $\mathbb{S}^3$ expressed in the direction cosine angular coordinates appearing in the Myers-Perry black hole line element,
\be\label{eq:LaplaceBeltramiEigenvalue}
	\left[\frac{1}{\sin\theta\cos\theta}\partial_{\theta}\left(\sin\theta\cos\theta\,\partial_{\theta}\right) + \frac{1}{\sin^2\theta}\,\partial_{\phi}^2 + \frac{1}{\cos^2\theta}\,\partial_{\psi}^2\right] \tilde{Y}_{\lambda}\left(\theta,\phi,\psi\right) = \lambda \tilde{Y}_{\lambda}\left(\theta,\phi,\psi\right) \,.
\ee

Let us present first how this is indeed the Laplace-Beltrami operator on $\mathbb{S}^3$. In usual spherical coordinates $\left(r,\vartheta_1,\vartheta_2,\varphi\right)$, the $4$-dimensional spatial position vector with components $\left(x^1,x^2,x^3,x^4\right)$ in Cartesian coordinates is written as
\be\ba
	x^1 &= r\cos\vartheta_1 \,, \\
	x^2 &= r\sin\vartheta_1\cos\vartheta_2 \,, \\
	x^3 &= r\sin\vartheta_1\sin\vartheta_2\cos\varphi \,, \\
	x^4 &= r\sin\vartheta_1\sin\vartheta_2\sin\varphi \,,
\ea\ee
where $x^1$ plays the role of the $3$-dimensional $z$-axis. In these coordinates, $\vartheta_1\in\left[0,\pi\right]$ and $\vartheta_2\in\left[0,\pi\right]$ are two polar angles, while $\varphi\in\left[0,2\pi\right)$ is a periodically identified azimuthal angle. The Laplacian operator acting on scalar functions then reads
\be
	\nabla_{4}^2 = \partial_{1}^2 + \partial_{2}^2 + \partial_{3}^2 + \partial_{4}^2 = \frac{1}{r^3}\partial_{r}\left(r^3\partial_{r}\right) + \frac{1}{r^2}\Delta_{\mathbb{S}^3}^{\left(0\right)} \,,
\ee
with $\Delta_{\mathbb{S}^3}^{\left(0\right)}$ the Laplace-Beltrami operator on $\mathbb{S}^3$ acting on scalar ($s=0$) functions, which in spherical coordinates is given explicitly by
\be
	\Delta_{\mathbb{S}^3}^{\left(0\right)} = \frac{1}{\sin^2\vartheta_1}\left[\partial_{\vartheta_1}\left(\sin^2\vartheta_1\,\partial_{\vartheta_1}\right) + \frac{1}{\sin\vartheta_2}\partial_{\vartheta_2}\left(\sin\vartheta_2\,\partial_{\vartheta_2}\right) + \frac{1}{\sin^2\vartheta_2}\,\partial_{\varphi}^2\right] \,.
\ee
In order to transit to direction cosine coordinates $\left(r,\theta,\phi,\psi\right)$, we split the four Cartesian coordinates $x^{i}$ into two pairs of two and project onto these two planes of rotation,
\be\ba
	x^1 &= r\cos\theta\cos\psi \,, \\
	x^2 &= r\cos\theta\sin\psi \,, \\
	\\
	x^3 &= r\sin\theta\cos\phi \,, \\
	x^4 &= r\sin\theta\sin\phi \,,
\ea\ee
with $\theta\in\left[0,\frac{\pi}{2}\right]$ a direction cosine angle, while $\phi\in\left[0,2\pi\right)$ and $\psi\in\left[0,2\pi\right)$ are two periodically identified azimuthal angles. Then, the Laplace operator on scalar functions has the same form as in spherical coordinates, but with the Laplace-Beltrami operator now identified as
\be
	\Delta_{\mathbb{S}^3}^{\left(0\right)} = \frac{1}{\sin\theta\cos\theta}\partial_{\theta}\left(\sin\theta\cos\theta\,\partial_{\theta}\right) + \frac{1}{\sin^2\theta}\,\partial_{\phi}^2 + \frac{1}{\cos^2\theta}\,\partial_{\psi}^2 \,.
\ee
In particular, the transformation rule between spherical coordinates $\left(\vartheta_1,\vartheta_2,\varphi\right)$ and direction cosine coordinates $\left(\theta,\phi,\psi\right)$ allows to identify $\phi=\varphi$, while $\psi$ and $\theta$ are related to $\vartheta_1$ and $\vartheta_2$ according to
\be
	\sin\theta = \sin\vartheta_1\sin\vartheta_2\,\,\,,\,\,\,\tan\psi = \tan\vartheta_1\cos\vartheta_2 \,.
\ee
The generators of the algebra $\mathfrak{so}\left(4\right)\simeq\mathfrak{so}\left(3\right)\oplus\mathfrak{so}\left(3\right)$ in direction cosine coordinates are more transparent by introducing the sum/difference azimuthal angles $\psi_{\pm}=\psi\pm\phi$ and they are organized in the two commuting $\mathfrak{so}\left(3\right)$'s, labeled by a sign $\sigma=+$ or $\sigma=-$,
\be
	\begin{gathered}
		J_0^{\left(\sigma\right)} = -i\partial_{\sigma} \,, \\
		J_{\pm1}^{\left(\sigma\right)} = e^{\pm i\psi_{\sigma}}\left[\partial_{2\theta} \pm i\cot2\theta\,\partial_{\sigma} \mp\frac{i}{\sin2\theta}\,\partial_{-\sigma}\right] \,, \\
		\left[J_{\pm1}^{\sigma},J_0^{\sigma^{\prime}}\right] = \mp J_{\pm1}^{\left(\sigma\right)}\delta_{\sigma,\sigma^{\prime}} \,\,\,,\,\,\, \left[J_{\pm1}^{\sigma},J_{\mp1}^{\sigma^{\prime}}\right] = \mp 2J_{0}^{\left(\sigma\right)}\delta_{\sigma,\sigma^{\prime}} \,\,\,,\,\,\, \left[J_{\pm1}^{\sigma},J_{\pm1}^{\sigma^{\prime}}\right] = 0 \,.
	\end{gathered}
\ee

Returning to the eigenvalue problem \eqref{eq:LaplaceBeltramiEigenvalue}, this is reduced to a one-dimensional problem after separating the azimuthal angles as
\be
	\tilde{Y}_{\ell m j}\left(\theta,\phi,\psi\right) = S_{\ell m j}\left(\theta\right)\frac{e^{im\phi}}{\sqrt{2\pi}}\frac{e^{ij\psi}}{\sqrt{2\pi}} \,,
\ee
with the azimuthal numbers $m$ and $j$ being integers by virtue of the periodicity of the angles $\phi$ and $\psi$ with period $2\pi$, and $S_{\ell m j}\left(\theta\right)$ satisfying the first order ordinary differential equation
\be
	\left[\frac{1}{\sin\theta\cos\theta}\frac{d}{d\theta}\left(\sin\theta\cos\theta\,\frac{d}{d\theta}\right) - \frac{m^2}{\sin^2\theta} - \frac{j^2}{\cos^2\theta}\right]S_{\ell m j}\left(\theta\right) = -\ell\left(\ell+2\right)S_{\ell m j}\left(\theta\right) \,.
\ee
We remark here that we have set the eigenvalues to $\lambda \equiv -\ell\left(\ell+2\right)$ such that $\ell$ resembles the corresponding orbital quantum number appearing in scalar spherical harmonics on $\mathbb{S}^3$. However, at this point, $\ell$ is not restricted to be a whole number.

This differential equation must now be solved in parallel with the boundary condition of regularity of $S_{\ell m j}\left(\theta\right)$ along the full domain of the direction cosine angle, $\theta\in\left[0,\frac{\pi}{2}\right]$. The unique solution is not hard to extract,
\be
	S_{\ell m j}\left(\theta\right) = N_{\ell m j}\sin^{\left|m\right|}\theta\,\cos^{\left|j\right|}\theta\,{}_2F_1\left(-\frac{\ell-\left|m\right|-\left|j\right|}{2},\frac{\ell+\left|m\right|+\left|j\right|}{2}+1;1+\left|j\right|;\cos^2\theta\right) \,,
\ee
with $N_{\ell m j}$ a normalization constant, while regularity at $\theta=0$ imposes the discretization condition
\be
	\frac{\ell-\left|m\right|-\left|j\right|}{2} \in \mathbb{N}_0 = \left\{0,1,2,\dots\right\} \,.
\ee
We can, thus, assign whole numbers to $\ell$ as in the usual scalar spherical harmonics of $\mathbb{S}^3$, and restrict the domain of the azimuthal numbers $m$ and $j$ such that the above condition is satisfied. This can be achieved, for example, by letting $\left|m\right|\le\ell$ to resemble the corresponding azimuthal number of the usual scalar spherical harmonics, but restricting $j$ to take integer values according to
\be
	j = -\left(\ell-\left|m\right|\right)\,,\,\, -\left(\ell-\left|m\right|\right) + 2\,,\,\, \dots\,,\,\, \left(\ell-\left|m\right|\right) - 2\,,\,\, \left(\ell-\left|m\right|\right) \,,
\ee
where we emphasize the step $2$ in the successively allowed values of $j$.

The eigenfunctions $\tilde{Y}_{\ell m j}$ can be seen to be orthogonal on $\mathbb{S}^3$ in these coordinates using properties of the Jacobi polynomials, while choosing the normalization constant to be
\be
	N_{\ell m j} = \frac{1}{\left(\left|j\right|\right)!}\sqrt{\left(2\ell+2\right)\frac{\left(\frac{\ell+\left|m\right|+\left|j\right|}{2}\right)!\left(\frac{\ell-\left|m\right|+\left|j\right|}{2}\right)!}{\left(\frac{\ell-\left|m\right|-\left|j\right|}{2}\right)!\left(\frac{\ell+\left|m\right|-\left|j\right|}{2}\right)!}} \,,
\ee
ensures the orthonormality identity
\be
	\oint_{\mathbb{S}^3} d\Omega_3\,\tilde{Y}_{\ell m j}^{\ast}\tilde{Y}_{\ell^{\prime}m^{\prime}j^{\prime}} = \delta_{\ell\ell^{\prime}}\delta_{mm^{\prime}}\delta_{jj^{\prime}} \,,
\ee
where asterisks indicate complex conjugation and the $3$-sphere integration measure in the direction cosine coordinates reads
\be
	\oint_{\mathbb{S}^3} d\Omega_3 = \int_{0}^{\frac{\pi}{2}}d\theta\int_{0}^{2\pi}d\phi\int_{0}^{2\pi}d\psi\,\sin\theta\cos\theta \,.
\ee

Two useful properties of these modified spherical harmonic functions are their transformation under complex conjugation, which simply reverses the sign of the azimuthal numbers,
\be
	\tilde{Y}_{\ell m j}^{\ast} = \tilde{Y}_{\ell,-m,-j} \,,
\ee
and under a parity transformation\footnote{In spherical coordinates, parity acts as $\left(\vartheta_1,\vartheta_2,\varphi\right)\rightarrow\left(\pi-\vartheta_1,\pi-\vartheta_2,\pi+\varphi\right)$ which is translated into directed cosine coordinates to $\left(\theta,\phi,\psi\right)\rightarrow\left(\theta,\pi+\phi,\pi+\psi\right)$.} $\left(\theta,\phi,\psi\right)\rightarrow\left(\theta,\pi+\phi,\pi+\psi\right)$, which only adds an overall phase,
\be
	\tilde{Y}_{\ell m j}\left(\theta,\pi+\phi,\pi+\psi\right) = \left(-1\right)^{m+j}\tilde{Y}_{\ell m j}\left(\theta,\psi,\phi\right) \,.
\ee

Let us count how many such basis states exist for a given value of the orbital number $\ell$. This is given by
\be
	\tilde{d}_{\ell} = \sum_{m=-\ell}^{\ell}\left(\sum_{j=-\left(\ell-\left|m\right|\right),2}^{\ell-\left|m\right|}\right) = \sum_{m=-\ell}^{\ell}\left(\ell-\left|m\right|+1\right) = \left(\ell+1\right)^2
\ee
and is exactly the same as the degeneracy of the scalar spherical harmonics\footnote{The scalar spherical harmonics of degree $\ell$ on $\mathbb{S}^{n}$ have a degeneracy,
\begin{equation*}
	d_{\ell}\left(n\right) = \frac{\left(2\ell+n-1\right)\left(\ell+n-2\right)!}{\ell!\left(n-1\right)!} \,.
\end{equation*}
For $n=3$, this gives $d_{\ell}\left(3\right)=\left(\ell+1\right)^2$.} on $\mathbb{S}^3$. Subsequently, the modified spherical harmonics $\tilde{Y}_{\ell m j}$ are equivalent to the usual scalar spherical harmonics on $\mathbb{S}^3$. In particular, the scalar spherical harmonics $Y_{\ell,\ell_2,\mu}$ on $\mathbb{S}^3$, with $\left|\mu\right|\le\ell_2\le\ell$, can always be written as a linear combination of the modified spherical harmonics,
\be
	Y_{\ell,\ell_2,\mu}\left(\vartheta_1,\vartheta_2,\varphi\right) = \sum_{m,j}c_{\ell,\ell_2,\mu;m,j}\tilde{Y}_{\ell m j}\left(\theta,\phi,\psi\right) \,,
\ee
and, thus, they form a complete set of $\tilde{d}_{\ell}$ linearly independent and orthonormal unit vectors. We remark here that the azimuthal numbers $\mu$ and $m$ are \textit{not} the same as they appear in $Y_{\ell,\ell_2,\mu}$ and $Y_{\ell m j}$; while $\mu$ and $m$ have the same range in both spherical harmonics bases, their multiplicities do not match. Nevertheless, in the above expansion it is straightforward to see that $c_{\ell,\ell_2,\mu;m,j}=c_{\ell,\ell_2,\mu;j}\delta_{m,\mu}$. In addition, the complex conjugacy relation $Y_{\ell,\ell_2,\mu}^{\ast} = \left(-1\right)^{\mu}Y_{\ell,\ell_2,-\mu}$ further implies
\be
	c_{\ell,\ell_2,\mu;j}^{\ast} = \left(-1\right)^{\mu} c_{\ell,\ell_2,-\mu;-j}\,.
\ee

One useful consequence of this completeness is that the two spherical harmonics bases obey the same addition theorem, expressed in terms of the Gegenabuer polynomials $C_{\ell}^{\left(1\right)}\left(x\right)$ as
\be
	\sum_{m,j}\tilde{Y}_{\ell m j}\left(\mathbf{n}\right)\tilde{Y}_{\ell m j}^{\ast}\left(\mathbf{n}^{\prime}\right) = \sum_{\ell_2,\mu}Y_{\ell,\ell_2,\mu}\left(\mathbf{n}\right)Y_{\ell,\ell_2,\mu}^{\ast}\left(\mathbf{n}^{\prime}\right) = \frac{\ell+1}{2\pi^2}C_{\ell}^{\left(1\right)}\left(\mathbf{n}\cdot\mathbf{n}^{\prime}\right) \,.
\ee

\appsection{Correspondence with STF tensors}
We will now present the $1$-to-$1$ correspondence between $4$-dimensional spatial STF tensors of rank-$\ell$ and the modified spherical harmonics $\tilde{Y}_{\ell m j}$ with the same orbital number $\ell$. There are two key observations that ensure this correspondence. First, the basic STF tensor of rank-$\ell$ $n^{\left\langle L\right\rangle}$, where $n^{i}=\frac{x^{i}}{r}$ are the projectors along the $i$-th spatial direction $x^{i}$, has eigenvalue $-\ell\left(\ell+2\right)$ under the action of the $4$-dimensional flat space Laplace operator, i.e. the same eigenvalue as all $\tilde{Y}_{\ell m j}$ with the same $\ell$.

Second, the number of independent components of a rank-$\ell$ STF tensor in  $n+1$ spatial dimensions is the number of degrees of freedom of a rank-$\ell$ symmetric tensor minus the number of traces that need to be removed,
\be
	d_{\ell}^{\text{STF}}\left(n\right) =
	\begin{pmatrix}
		n+\ell \\
		\ell
	\end{pmatrix} -
	\begin{pmatrix}
		n+\ell-2 \\
		\ell-2
	\end{pmatrix}
	= \frac{\left(2\ell+n-1\right)\left(\ell+n-2\right)!}{\ell!\left(n-1\right)!} \,.
\ee
For $n=3$, this gives that the number of independent components of a $4$-dimensional spatial STF tensor of rank-$\ell$ is equal to $\left(\ell+1\right)^2$, i.e. the same as the number $\tilde{d}_{\ell}$ of basis function $\tilde{Y}_{\ell m j}$ with orbital number $\ell$. Consequently, any $4$-dimensional spatial STF tensor of rank-$\ell$ can be written as a linear combination of $\tilde{Y}_{\ell m j}$ for all the possible values of the azimuthal numbers $m$ and $j$. In particular, for the basic STF tensor $n^{\left\langle L \right\rangle}$,
\be
	n^{\left\langle L \right\rangle} = A_{\ell}\sum_{m,j}\mathcal{Y}^{L}_{\ell m j}\tilde{Y}_{\ell m j}\left(\mathbf{n}\right) \,,
\ee
where $\mathbf{n}$ is a shorthand for the direction cosine coordinates $\left(\theta,\phi,\psi\right)$ associated with the position vector $x^{i}$ from which $n^{i}$ is constructed, the constant STF tensors $\mathcal{Y}^{L}_{\ell m j}$ are given by
\be
	\mathcal{Y}^{L}_{\ell m j} = \frac{1}{A_{\ell}}\oint_{\mathbb{S}^3} d\Omega_{3}\,n^{\left\langle L \right\rangle}\tilde{Y}_{\ell m j}^{\ast}\left(\mathbf{n}\right) \,,
\ee
and $A_{\ell}$ is a real normalization constant chosen such that\footnote{Using the addition theorem of scalar spherical harmonics, which is also an addition theorem for the modified spherical harmonics, this normalization constant can be found to be
\begin{equation*}
	A_{\ell} = \frac{4\pi^2\ell!}{\left(2\ell+2\right)!!} \,.
\end{equation*}}
\be
	\tilde{Y}_{\ell m j}\left(\mathbf{n}\right) = \mathcal{Y}^{L\ast}_{\ell m j}n_{\left\langle L \right\rangle} \,.
\ee

A general $4$-dimensional spatial STF tensor $\mathcal{E}^{L}=\mathcal{E}^{\left\langle L \right\rangle}$ can then be expanded into its modified multipole moments $\mathcal{E}_{\ell m j}$ according to
\be
	\mathcal{E}^{L} = \sum_{m,j}\mathcal{E}_{\ell m j}\mathcal{Y}^{L\ast}_{\ell m j} \,,
\ee
with
\be
	\mathcal{E}_{\ell m j} = A_{\ell} \mathcal{E}_{L}\mathcal{Y}^{L}_{\ell m j} = \mathcal{E}_{L}\oint_{\mathbb{S}^3} d\Omega_{3}\,n^{\left\langle L \right\rangle}\tilde{Y}_{\ell m j}^{\ast}\left(\mathbf{n}\right) \,,
\ee
such that
\be
	\mathcal{E}_{L}n^{L} = \sum_{m,j}\mathcal{E}_{\ell m j}\tilde{Y}_{\ell m j} \,.
\ee

Last, from the complex conjugacy relation of the modified spherical harmonics, we can see that
\be
	\mathcal{Y}^{L\ast}_{\ell m j} = \mathcal{Y}^{L}_{\ell, -m, -j} \,,\quad \mathcal{E}_{\ell m j}^{\ast} = \mathcal{E}_{\ell, -m, -j} \,.
\ee
	\end{appendices}

	\clearpage
	\phantomsection
	\addcontentsline{toc}{chapter}{\bibtitlename}
	\bibliographystyle{JHEP}
	\bibliography{TLNsAll_References_Inspirehep}

\end{document}